\documentclass[12pt,twoside,vi]{mitthesis}
\usepackage{lgrind}
\pdfoutput=1
\usepackage{cmap}
\usepackage[T1]{fontenc}

\usepackage{mycommands}

\def\all{all}
\ifx\files\all  \else

\begin{document}
\pagenumbering{roman} 
%
%
%
%
%
%
%
%
%
%
%

\title{QCD Beyond Leading Power}

\author{Gherardo Vita}
\department{Department of Physics}

 \degree{Doctor of Philosophy \and Master of Science}

\degreemonth{September}
\degreeyear{2020}
\thesisdate{August 13, 2020}


\supervisor{Iain W. Stewart}{Director, Center for Theoretical Physics}

\chairman{Nergis Mavalvala}{Associate Department Head of Physics}

\maketitle



\cleardoublepage
\begin{abstractpage}
In this thesis I study the infrared limits of Quantum Chromodynamics (QCD) beyond leading power by developing effective quantum field theory techniques.
I apply these formal developments to deepen our understanding of the infrared structure of gauge theory amplitudes and cross sections in the soft, collinear and Regge limit, as well as to improve predictions for collider observables. 

Using and extending the framework of Soft and Collinear Effective Theory (SCET), I explore the ingredients of factorization beyond leading power constructing subleading hard scattering operator and radiative jet and soft functions for processes such as Higgs boson production as well as the production of electroweak gauge bosons and their decay.

I introduce new subleading power gauge invariant objects, the \emph{$\theta$-jet} and \emph{$\theta$-soft functions}, which arise in the renormalization of non local gauge invariant objects beyond leading power. I use them to achieve, for the first time, the resummation of collinear and soft logarithms beyond leading power for a collider observable in QCD. 

I study the perturbative power corrections to differential distributions for color singlet production at the Large Hadron Collider (LHC). I explore the implications of retaining the full dependence on the kinematics of the color singlet particles and I highlight and solve the subtleties related to the regularization of rapidity divergences beyond leading power for which I propose a new regulator, the \emph{pure rapidity regulator}.

I present a new method to employ cutting edge multiloop techniques for the computation of the expansions of cross sections in the collinear limit. This allows the extraction of universal ingredients arising in the collinear limit of QCD to an unprecedented level of precision in perturbation theory. It also provides a powerful method to construct systematically improvable analytic approximations to differential distributions at an order beyond what is currently feasible in full kinematics.

I also examine factorization and resummation in the Regge limit at the amplitude and the cross section level. I develop a Lagrangian formalism for the treatment of fermion mediated forward scattering processes in QCD, which are generated at subleading power in the forward limit, and apply it to obtain the quark reggeization as well as the BFKL resummation of small-$x$ logarithms for di-photon production at the LHC.

\end{abstractpage}


\cleardoublepage


\pagestyle{plain}
\tableofcontents
\newpage
\listoffigures
\newpage
\listoftables

\pagenumbering{arabic} 
\chapter{Introduction}
The description of a vast range of natural phenomena that have complex microscopic interactions can often be dramatically simplified by developing effective theories which capture the macroscopic dynamics of the physical system under study.
One captures the dependence on the microscopic degrees of freedom via a small set of parameters that can be fixed either by calculations or by experiments. Take for example the description of the propagation of a perturbation inside a fluid. Its dynamics  in principle could be described by studying the interactions between each of the molecules composing the fluid, but given the number of degrees of freedom at play, this is practically infeasible. However, we know that if we look at the fluid dynamics at long distances it exhibits a very simple behavior. The small perturbation is described by a wave equation that depends on the molecular details of the fluid through a single parameter, the acoustic velocity. Such a description is valid because of the large separation, or \emph{hierarchy}, between the scale at which we are observing the fluid and the microscopic distances at which the molecular interactions take place. This same strategy of formulating \emph{effective field theories} has wide application in virtually every area of physics, from hydrodynamics to gravity and to relativistic quantum field theories of elementary particles and nuclei, where it has perhaps been most systematically and thoroughly developed.

Given an effective theory it is important to give a precise and quantitative measure of its range of validity and applicability. Moreover, as precision in theoretical predictions keep up with the advances in experimental measurements, it is imperative to devise methods to estimate theoretical uncertainties, which can be done systematically using an effective theory, and to systematically improve such theories in order to to reduce them.

In the last fifty years, the quest to understand the fundamental laws of nature at the most microscopic scales has received enormous contributions from experiments at particle accelerators. As the technology has been developed to build larger and larger colliders, physicists are able to probe higher and higher energy scales and therefore interactions and processes that happen at smaller and smaller length scales.

\begin{figure*}
 \centering
 \includegraphics[width=0.49\textwidth]{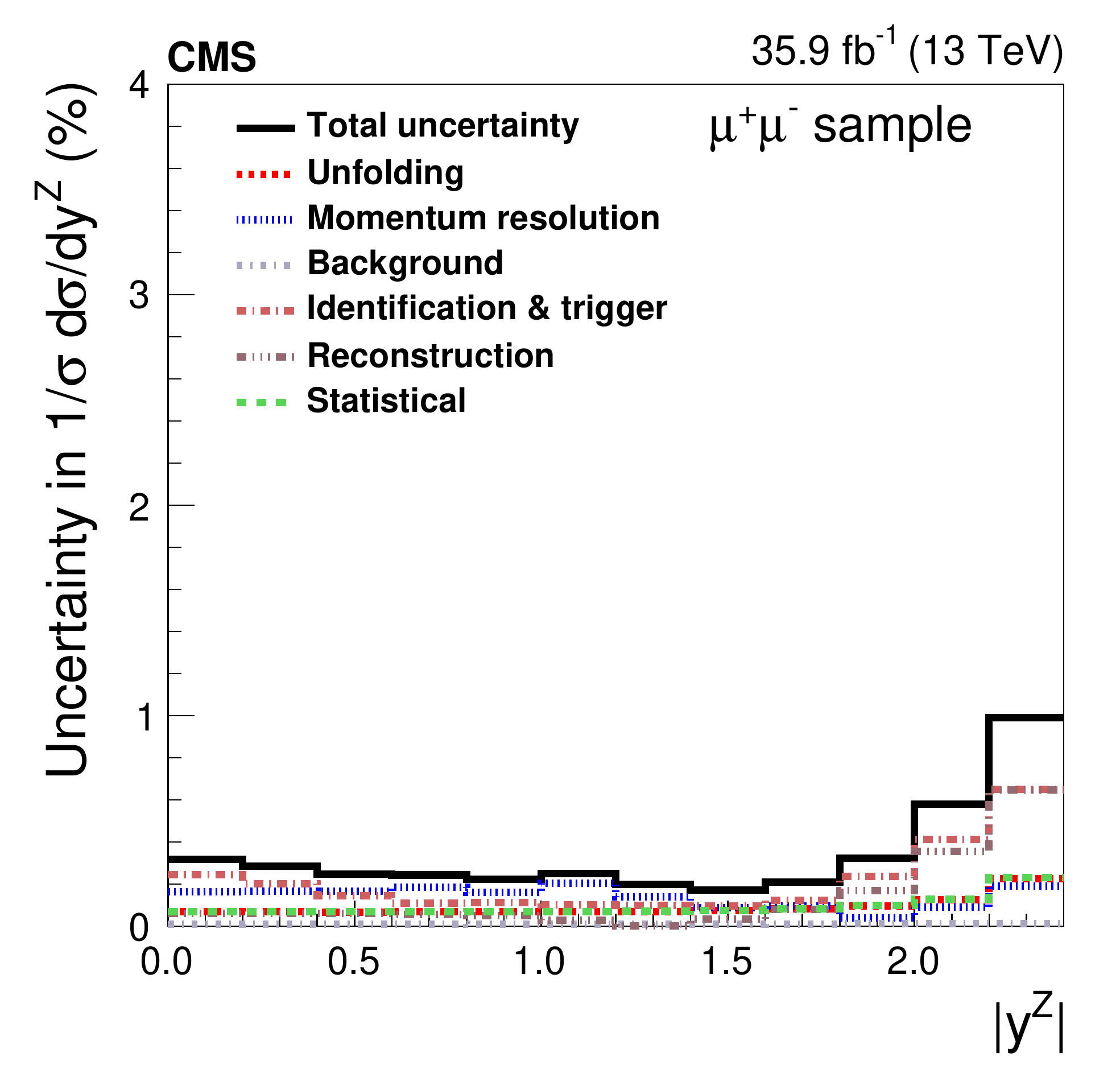}
 \hfill
 \includegraphics[width=0.49\textwidth]{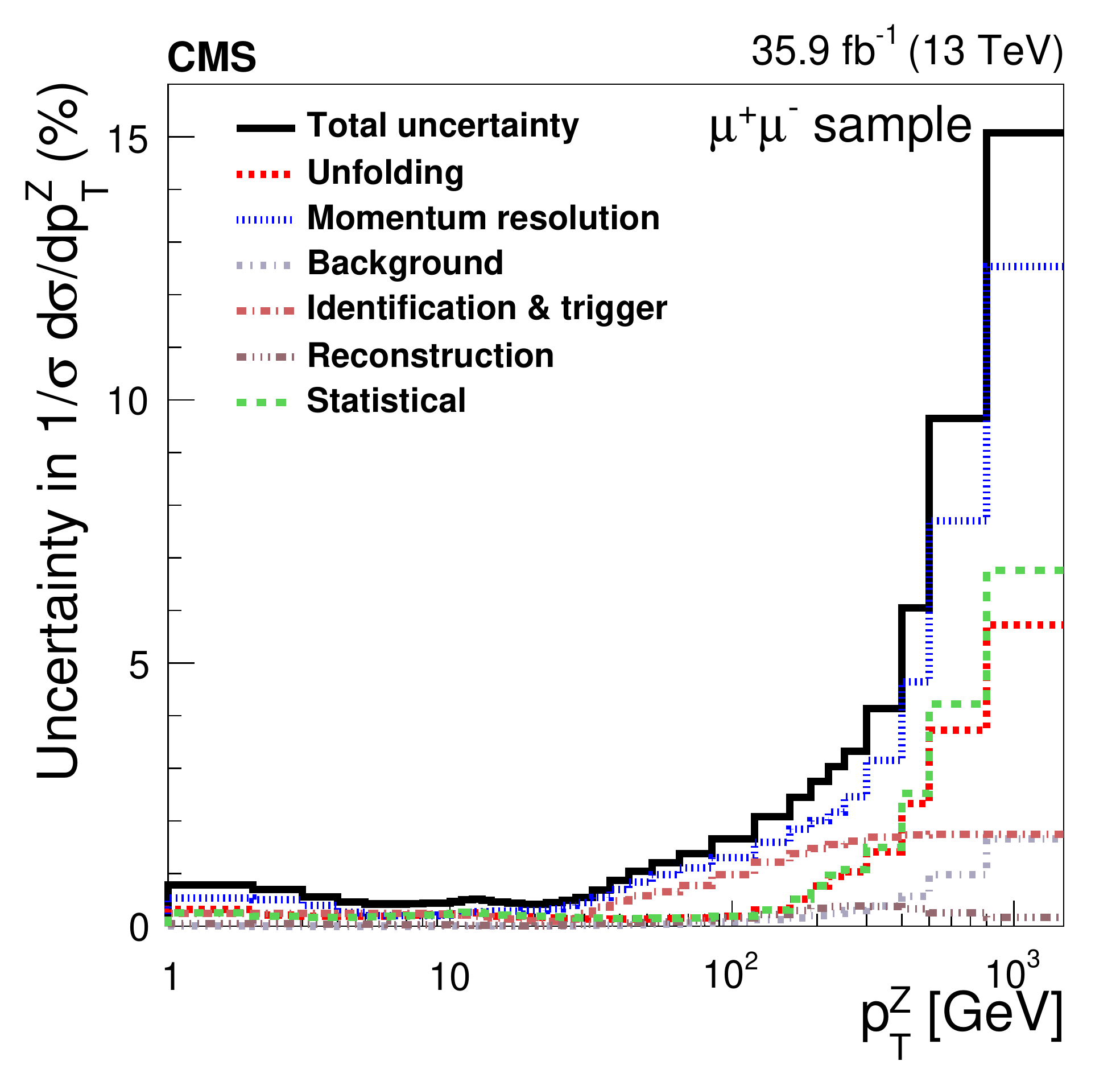}
 \caption{The relative sources of uncertainty in the normalized rapidity (left) and transverse momentum (right) spectrum measurements for the $Z$ boson in the muon sample as measured by the CMS experiment at the LHC\cite{Sirunyan:2019bzr}. We observe that the uncertainty is at (or below) the one percent level for a large range of the spectra.}
 \label{fig:DYmeasurements}
\end{figure*}

Nowadays at the Large Hadron Collider (LHC), key benchmark processes like the rapidity and transverse momentum spectra of electroweak gauge bosons in the Drell-Yan process are measured with an astonishing level of precision, see \fig{DYmeasurements}. The precision of the experimental measurements at the LHC will improve even further in the coming years as more data gets collected and improvements in the analysis techniques take place. 

It is therefore of utmost importance that the theoretical understanding of the interactions underlying the processes happening at the LHC is deepened and broadened adequately in order to provide theoretical predictions that at least match the level of precision of experiments.

In this thesis we will consider effective theories for relativistic non-abelian quantum field theories with a focus on developing new techniques and concepts for the study of Quantum Chromodynamics (QCD) in its perturbative regime beyond leading power.

Among the fundamental interactions of nature, QCD, the strong interaction, plays the dominant role for a large variety of processes at colliders and it constitutes the main source of corrections and background to virtually any processes taking place at the energy scales probed by modern colliders.

Unfortunately, making precise theoretical predictions for QCD processes is extremely challenging. Even after leveraging the power of perturbation theory, the non-abelian nature of the theory and its relatively large coupling require the inclusion of a plethora of subprocesses to obtain control of the predictions.

Moreover, because of the large hierarchy of scales that enter in a high-energy scattering process, there are many interesting observables that one can probe in collider experiments whose cross sections are enhanced by events at the corners of the phase space of the scattering particles. The population of these kinematic regions spoils the convergence of the perturbation series in the strong coupling constant, requiring an all-order understanding of the QCD interactions in such regions. 

For example, let's take the case of the production of a color singlet particle $h$ in proton-proton collision, which is the case of the production of a Higgs boson or also an electroweak gauge boson at the LHC. The perturbative description of hadronic color singlet cross sections for infrared and collinear safe observables becomes inadequate when the value of the observable forces hadronic radiation produced on top of the colorless final state to be in the infrared or collinear regime.
As a matter of fact when imposing an infrared-sensitive measurement $\Obs$ in a scattering process with characteristic hard scale $Q$, the cross section $\sigma$ receives contributions of up to two logarithms $L = \ln(Q/\Obs)$ per order of the coupling constant, \emph{i.e.}\ $\sigma \sim \as^n L^{2n}$.
In the limit of $\Obs \to 0$ the presence of such logarithms signals the sensitivity to infrared physics and truncated perturbation theory does not yield an accurate description of the observable, as the large logarithms $L$ can overcome the suppression in $\as$.
The ability to include infinite towers of these large logarithms in the theoretical predictions is therefore crucial in order to obtain reliable predictions for physical observables that are sensitive to these kinematic regions.
It is therefore important to devise different methods to control these logarithms when calculating cross sections.

In order to understand the organization of the expansion of cross sections for observables at high energy colliders, like the LHC, the Future Circular Collider (FCC), the B-factories, the Electron-Ion Collider (EIC), etc. and more generally of perturbative gauge theories, let's take for concreteness the cross section  for the transverse momentum distribution of the Higgs boson at the LHC. 
The two main parameters that we can use to perform an expansion of such cross section are the coupling constant $\alpha_s$ and the dimensionless ratio $\Obs\equiv q^2_T/m^2_H$ between $q_T$, the transverse momentum of the Higgs, and the hard scale of the process $m_H$, the Higgs mass.
A common approach to calculate cross sections is to perform a perturbative expansion in the coupling constant. For the case of transverse momentum distribution $\frac{\df \sigma}{\df \Obs}$ we can schematically sketch its fixed order expansion as
\be\label{eq:asexp}
	\frac{\df \sigma}{\df \Obs} = c_0(\Obs) + c_1(\Obs) \,\alpha_s + c_2(\Obs)\, \alpha_s^2 +  c_3(\Obs)\, \alpha_s^3 +  \dots \,
\ee
where each coefficient of the expansion $c_i(\Obs)$ is a full function of $\Obs$ and therefore of the transverse momentum. For processes at energies much larger than $\lqcd$, the coupling constant is indeed a small parameter ($\alpha_s(m_H) \sim 0.1$) and therefore we could in principle obtain precise predictions by truncating the series in \eq{asexp} at a high enough order in $\alpha_s$.
However, as previously mentioned, the series coefficient $c_i(\Obs)$ will contain singularities as $\Obs \to 0$ spoiling the convergence of the series. We can therefore take another approach and organize the expansion in terms of $\Obs$ itself rather than $\alpha_s$ 
\be\label{eq:tauexp}
	\frac{\df \sigma}{\df \Obs} = \frac{d\sigma^{(0)}}{\df \Obs}+\frac{d\sigma^{(1)}}{\df \Obs}+\frac{d\sigma^{(2)}}{\df \Obs}+\cdots  \,,
\ee
where the first term $\frac{\df \sigma^{(0)}}{\df \Obs}$ is referred to as the \emph{leading power} term and the subsequent terms represents higher and higher terms in the $\Obs$ expansion $\frac{d\sigma^{(i)}}{\df \Obs} \sim \Obs^i\, \frac{\df \sigma^{(0)}}{\df \Obs} $ represent the terms beyond leading power. Note that each term in the series of \eq{tauexp} is at all orders in the coupling constant $\alpha_s$ and has a strict organization of the logarithmic divergences. If we look in more detail, \eq{tauexp} has the form
\begin{align}\label{eq:tauexp_full}
	\frac{\df \sigma}{\df \Obs} &=\sum\limits_{n=0}^{\infty} \left(  \frac{ \alpha_s}{\pi} \right)^n\left[ c^{(0)}_{n,-1}\delta(\Obs) + \sum \limits_{m=0}^{2n-1} c_{n,m}^{(0)} \left( \frac{\log^m (\Obs)}{\Obs}  \right)_+\right] 
	\hspace{5mm}\leftarrow\,\text{\small Leading Power (LP)}
\nn \\&  		
	+\sum\limits_{n=1}^{\infty} \left(  \frac{\alpha_s}{\pi} \right)^n \sum \limits_{m=0}^{2n-1} c_{n,m}^{(1)}~  \log^m \Obs
	\hspace{26mm}\leftarrow\,\text{\small Next to Leading Power (NLP)}
\nn \\ &
	+  \sum\limits_{n=1}^{\infty} \left(  \frac{\alpha_s}{\pi} \right)^n \sum \limits_{m=0}^{2n-1} c_{n,m}^{(2)}~ \Obs \log^m \Obs
	\hspace{10mm}\leftarrow\,\text{{\footnotesize Next-to-Next to Leading Power} {\small (NNLP)}} 
\nn \\ &
+\cdots 
\end{align}
where in general the coefficients $c_{n,m}^{(i)}$ maybe function of other kinematic variables and are not simple constants. For the case of the Higgs $q_T$ they involve dependence on other parameters like the energy of the collider, the rapidity of the Higgs and on other measurements or constraints imposed on the Higgs phase space through experimental cuts. Moreover they will contain non-perturbative functions like the Parton Distribution Functions.

\section{Motivation For Studying Subleading Power}\label{sec:motivations}
Having presented the organization of the expansion of cross sections, in this section we want to present a set of motivations to pursue the study of gauge theories beyond leading power.

\subsection{A New Direction To Explore Gauge Theories}\label{sec:newdirection}
\begin{figure}
\begin{center}
		{\includegraphics[width=14cm]{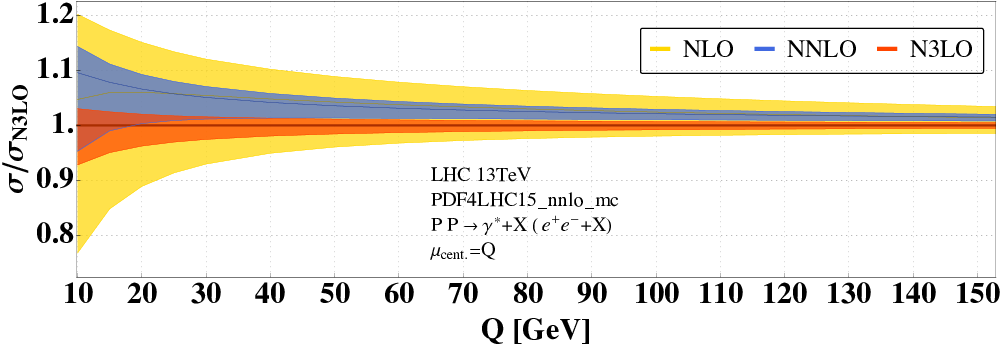}}
\end{center}
\vspace{-0.4cm}
\caption{An example of state-of-the-art fixed order cross section (N$^3$LO).The Drell-Yan cross section for a virtual photon order by order in perturbation theory up to N$^3$LO as a function of the virtuality $Q^2$ of the photon with uncertainty bands, normalized by the N$^3$LO result \cite{Duhr:2020seh}.} \label{fig:N3LO}
\end{figure}

Using the organization of \eq{tauexp_full}, we see that the determining the full cross section amounts to determine the coefficients $c_{n,m}^{(i)}$ in \eq{tauexp_full}
\footnote{Technically that is not true due to issues with the convergence of such expansions. It is known that the perturbative expansions of gauge theories are not convergent but rather asymptotic expansions and one could equally worry about the convergence of the power expansion in \eq{tauexp}. As is commonly done, we will ignore the fact that we are dealing with asymptotic expansions assuming that the issue of converge will become relevant at much higher order than the ones that we will treat in this thesis. The analysis of uncertainties at a given order provides a cross check on the validity of this hypothesis.}.
We can therefore identify three main directions that determine the accuracy of amplitudes and cross sections: 
\begin{itemize}
	\item $n$ is the perturbative order in the $\alpha_s$ expansion.
	\item $m$ gives the logarithmic accuracy.
	\item $i$ determines the order of the power expansion in $\Obs$.
\end{itemize}
Our first motivation is that while there is great control and understanding of cross sections at \emph{fixed order} in $\alpha_s$ and/or at leading power in $\Obs$, much less is understood beyond leading power.
As a matter of fact, at fixed order, the state of the art for cross sections in hadron-hadron collisions has now reached the impressive precision of N$^3$LO, i.e. the inclusion of all the terms suppressed up to $\alpha_s^3$ with respect to the leading order which correspond to the $c_{3,m}^{(i)}$ coefficients in \eq{tauexp_full}. So far this has been achieve for inclusive cross sections for Drell-Yan, see \fig{N3LO}, and Higgs production~\cite{Anastasiou:2015ema,Mistlberger:2018etf,Duhr:2019kwi,Duhr:2020seh,Duhr:2020sdp}.

Also, since fixed order expansion of distributions for a large class of observables, like transverse momentum distribution of the Higgs or DY, are not reliable and the resummation of logarithmic enhanced terms is needed in order to obtain precise predictions, a huge effort in the phenomenology community has been taken to obtain resummed cross sections. The number of logarithms that are predicted by the resummation is dictated by the logarithmic accuracy, denoted by N$^k$LL. Explicitly, for the first few orders, a leading power resummation at N$^k$LL can be used to predict all the terms $c^{(0)}_{n,m}$ in \eq{tauexp_full}, satisfying
\begin{align} \label{eq:whichms_intro}
\text{LL predicts}: m=2n-1\,, \\
\text{NLL predicts}: m\geq 2n-2\,, \nn \\
\text{NNLL predicts}: m\geq 2n-4\,, \nn \\
\text{N$^3$LL predicts}: m\geq 2n-6\,, \nn 
\end{align}
The use of effective field theory techniques and in particular factorization theorems in SCET has dramatically improved the ability to perform resummation at higher levels of accuracy for a variety of processes. The state of the art is leading power N$^3$LL accuracy which has been achieved for a number of $e^+e^-$ event shapes \cite{Becher:2008cf,Abbate:2010xh,Chien:2010kc,Hoang:2014wka,Moult:2018jzp} as well as for transverse momentum distributions of color singlets at the LHC \cite{Chen:2018pzu, Bizon:2018foh,Wiesemann:2020gbm}. Note that thanks to the very recent result of ours in \cite{Ebert:2020yqt}, which is based on the new method presented in \chap{collinear_expansion_chapter}, we now have the full leading power structure of $q_T$ distributions at N$^3$LO and therefore all the ingredients to push the leading power resummation of $q_T$-dependent distributions like Drell-Yan and Higgs production at the LHC to N$^3$LL$^\prime$.
\begin{figure}
\begin{center}	 
		{\includegraphics[width=7.1cm,trim=0mm 0mm 0mm 1cm, clip]{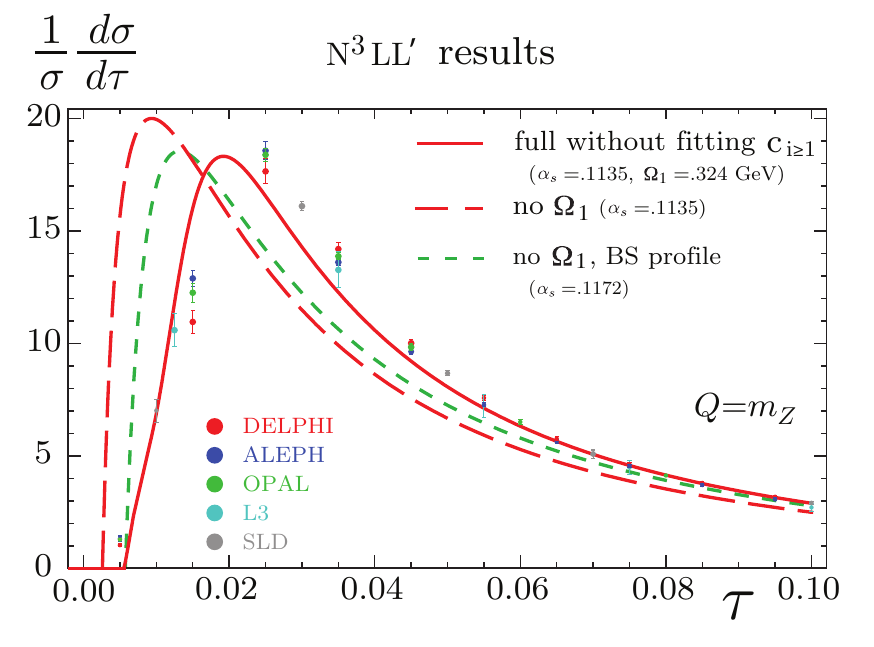}}	 
\qquad 
		{\includegraphics[width=6cm]{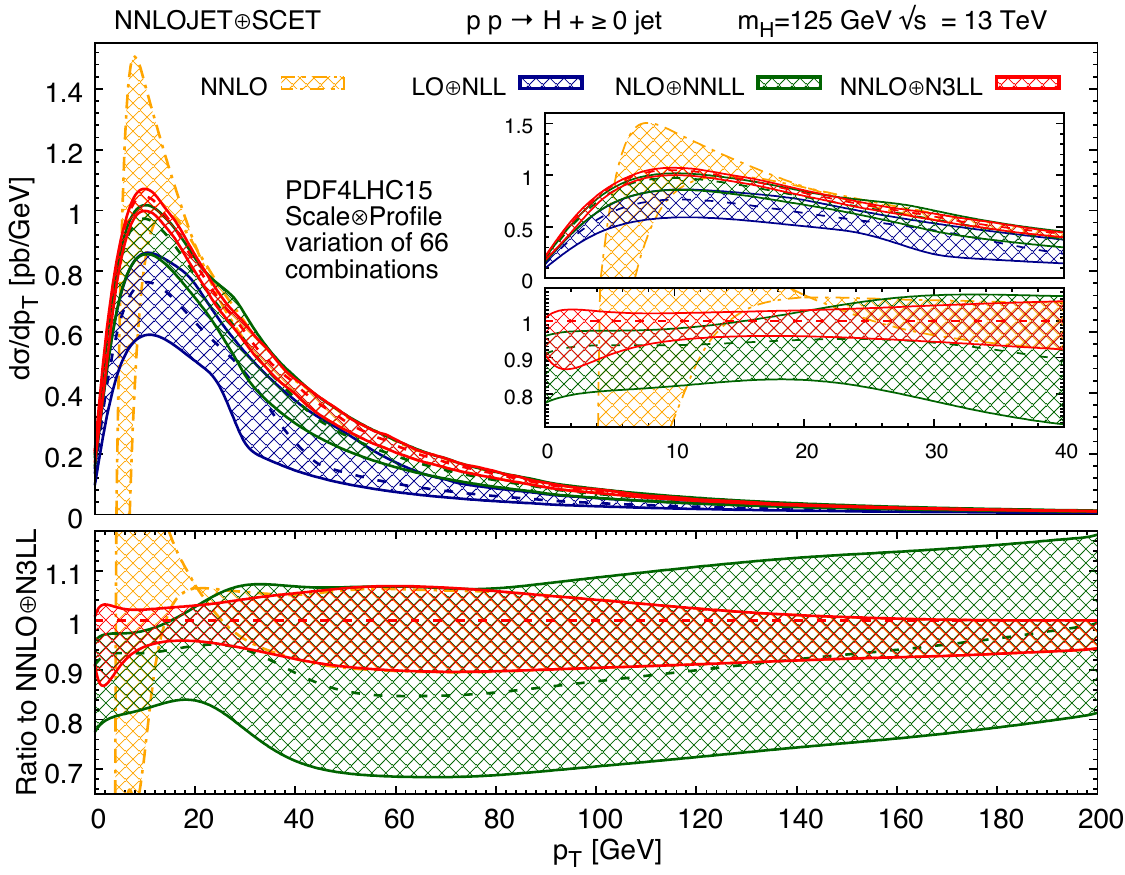}}
\end{center}
\vspace{-0.4cm}
\caption{Two examples of state-of-the-art resummation (N$^3$LL). The thrust event shape spectrum in $e^+ e^-$ annihilation (left) \cite{Abbate:2010xh} and the Higgs transverse momentum spectrum at the LHC \cite{Chen:2018pzu}(right).} \label{fig:N3LLresumm}
\end{figure}

While huge progress in exploring higher orders in fixed order calculations and leading power resummation has been made, much less is understood at subleading power.
\begin{figure}
\begin{center}
\includegraphics[width=0.50\columnwidth]{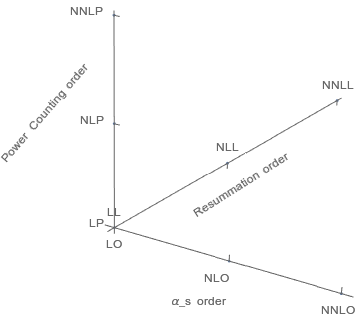} 
\end{center}
\vspace{-0.4cm}
\caption{The expansion of perturbative cross sections can be pictured as a volume in a 3D plane characterized by the perturbative, logarithmic, and power counting order to which it is computed} \label{fig:3dexpansion}
\end{figure}
We can represent pictorially the theoretical control on a cross section or amplitude, by the volume in the 3D space of \fig{3dexpansion} where the edge of the volume are determined by the largest value of the indices $i,m,n$ of the coefficients $c_{nm}^{(i)}$ that we can calculate.
Therefore, studying cross sections and amplitudes beyond leading power opens up an entire new direction in the understanding of gauge theories which is orthogonal to the ones that have been intensively explored in the last 40 years.
As always happens when starting exploring new directions, many things change, new questions arise and new tools are needed in order to proceed.
As an example, we can consider how much the understanding of gauge theories has been pushed by going beyond leading order in perturbation theory. 
We can also think about how many of concepts that are valid at leading logarithmic approximation fail to generalize when going to higher logarithmic accuracy. 
By analogy, it then becomes clear that many of the aspects that are successful in deriving factorization and resummation at leading power will fail when going beyond leading power.
Some of the fundamental questions that need to be answered are "what is the structure of factorization theorems?", "How much of the behavior of gauge theories beyond leading power is directly related (and predictable) from their leading power?", "What is new?", "How do we regulate, renormalize, and  resum matrix elements and cross sections beyond leading power?"
It is also interesting to explore and identify universal objects regulating factorization and resummation at subleading power, analogously to how concepts like the cusp anomalous dimension of Wilson lines dictates many aspects of the leading power resummation.

Moreover, symmetries that dictate the physics when we only allow for leading power interactions to occur may no longer hold beyond leading power.
This can give rise to new phenomena that cannot happen at leading power, similarly to how flavor changing in neutral current processes don't occur in the Standard Model at tree level. An important example, phenomena associated to soft quarks, will be analyzed in this thesis.

\subsection{Pushing The Boundaries of Slicing Methods}
\begin{figure}
\begin{center}	 
	\includegraphics[height=5cm]{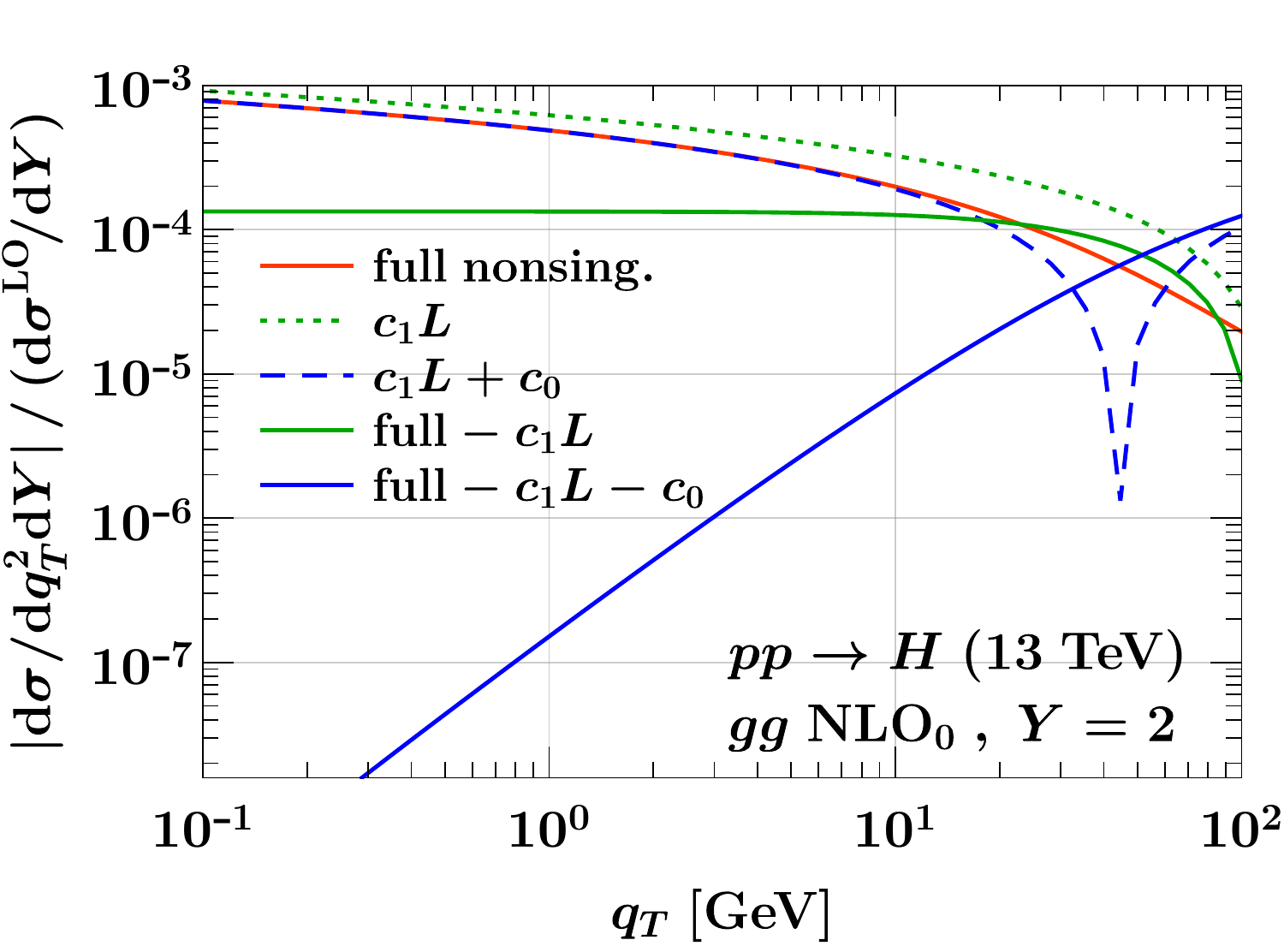}
	\includegraphics[height=5cm]{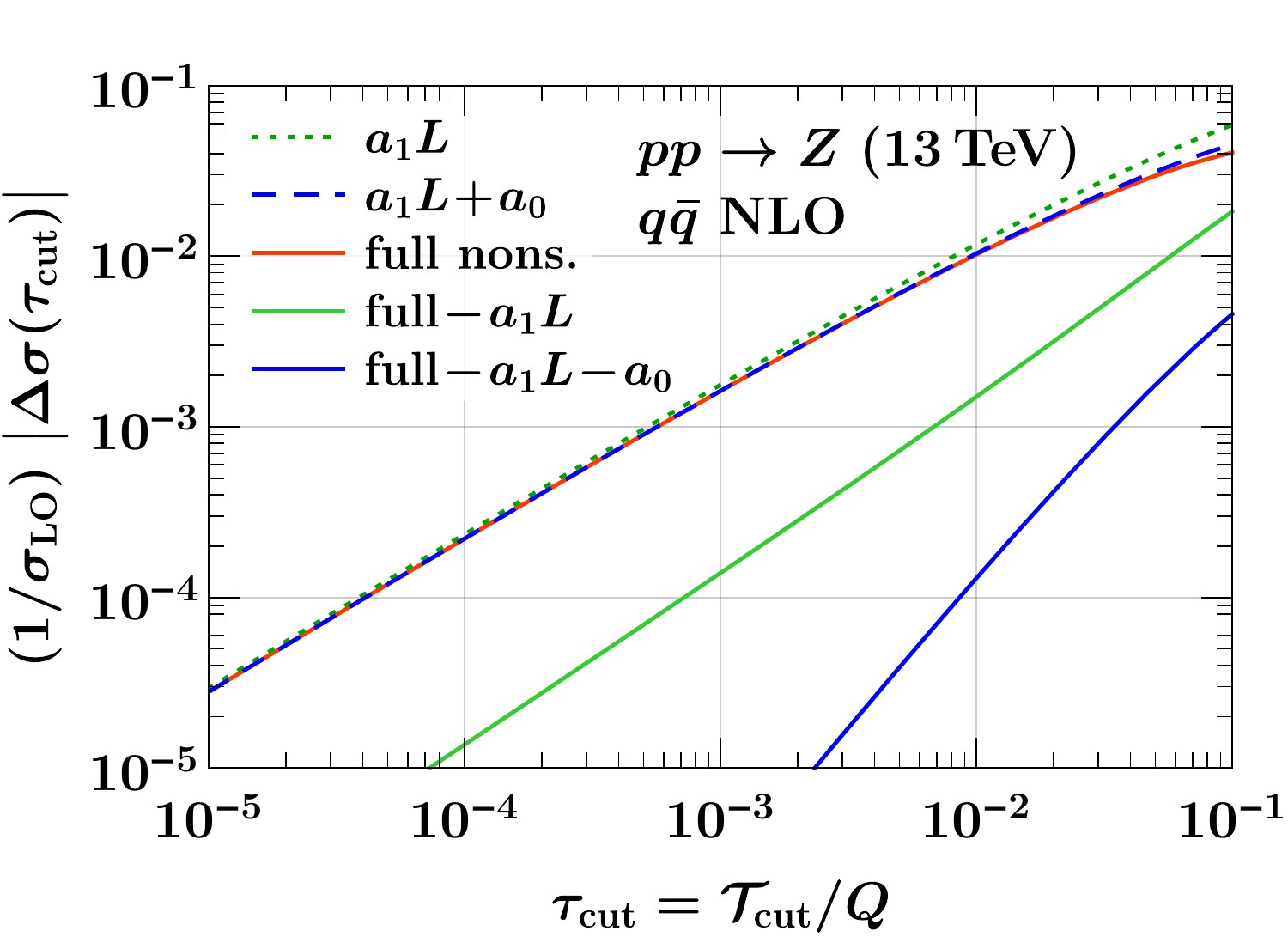}
\end{center}
\vspace{-0.4cm}
\caption{Improvements due to the inclusion of next-to-leading power (NLP) terms in slicing methods. The solid lines represent the residual error after the inclusion of: no power corrections (red), leading log NLP terms (green), NLL NLP terms (blue). Plots for $pp \to H$ with the $q_T$ subtraction scheme (left) and Z production with $N$-Jettiness (right). The left plot is from \cite{Ebert:2018gsn} and its derivation will be presented in \chap{ptNLL}. The right plot is from \cite{Ebert:2018lzn}.} \label{fig:slicing}
\end{figure}

An important application of studying QCD beyond leading power is the calculation of subleading power corrections to event shape observables, such as $0$-jettiness \cite{Stewart:2010tn} or $q_T$.
Recently, there has been considerable interest in the use of event shape observables for performing NNLO fixed order subtractions using the $q_T$ \cite{Catani:2007vq} or $N$-jettiness \cite{Boughezal:2015aha,Gaunt:2015pea} subtraction schemes. 
Schematically, in a \emph{slicing} method one uses an event shape observable to split the calculation of a cross section $\sigma(X)$ in two regions, by slicing the phase space in order to circumvent the issue of the cancellation of infrared divergences between virtual and real contributions
\be 
	\sigma(X)=\int\limits_0 d \Obs  \frac{d\sigma(X)}{d\Obs}	= \underbrace{ \int \limits_0^{\Obs_\text{cut}} d \Obs \frac{d\sigma(X)}{d \Obs} }_\text{below the cut} +\underbrace{\int\limits_{\Obs_\text{cut}} d \Obs \frac{d\sigma(X)}{d\Obs}}_\text{above the cut}
\ee
The region "below the cut" contains the infrared sensitivity, but effective field theory techniques, like SCET, allow the analytic calculation of the singular structure of the distribution in the event shape. In the region "above the cut" one emission is resolved by the slicing parameter $\Obs_\text{cut}$ (for example a non-zero $q_T$ in Higgs production implies that the Higgs is recoiling against some resolved radiation) and it is therefore free\footnote{More generally, the cross section above the cut is not necessarily free of IR divergences, but it is  a calculation at lower loop level for a higher point amplitude whose IR divergences are much easier to deal with.} of infrared divergences and can be calculated numerically.
These  have been extremely successful applied to color singlet production \cite{Catani:2009sm,Ferrera:2011bk,Catani:2011qz,Grazzini:2013bna,Cascioli:2014yka,Ferrera:2014lca,Gehrmann:2014fva,Grazzini:2015nwa,Grazzini:2015hta,Campbell:2016yrh,Boughezal:2016wmq}, to the production of a single jet in association with a color singlet particle \cite{Boughezal:2015aha,Boughezal:2015dva,Boughezal:2016isb,Boughezal:2016dtm,Campbell:2019gmd}, to inclusive photon production \cite{Campbell:2016lzl}, and recently also to top pair production \cite{Catani:2019hip,Catani:2020tko}. 
Progress is also being done in extending these methods to N$^3$LO~\cite{Cieri_2019,Billis:2019vxg} and 
in particular, thanks to a very recent result of ours~\cite{Ebert:2020yqt} which is based on the new method presented in \chap{collinear_expansion_chapter}, all the ingredients for $q_T$ subtraction at N$^3$LO are now available.
By choosing the cut parameter to be small the region below the cut can be approximated by a power expansion like \eq{tauexp}
\be\label{eq:exp_slicing}
	\int \limits_0^{\Obs_\text{cut}} d \Obs \frac{d\sigma}{d \Obs} ~ \overset{\Obs_\text{cut} \to 0}{\sim}~ \int \limits_0^{\Obs_\text{cut}} d \Obs \frac{d\sigma^{(0)}}{\df \Obs}+\int \limits_0^{\Obs_\text{cut}} \frac{d\sigma^{(1)}}{\df \Obs}+\cO(\Obs_\text{cut}^2)
\ee
Usually since only the leading power is understood, as we said in \sec{newdirection}, it is common to include only the first term in \eq{exp_slicing}. However, this forces us to take $\Obs_\text{cut}$ extremely small in order to minimize the residual error induced by neglected higher powers terms, and this slows down tremendously the numerical calculation of the region above the cut.
For both $N$-jettiness and $q_T$ subtractions the inclusion of the terms beyond leading power dramatically improves the numerical accuracy, and thereby the computational efficiency, of the method. For example for color-singlet production the efficiency improves by up to an order of magnitude per logarithmic order calculated~\cite{Moult:2016fqy, Moult:2017jsg, Ebert:2018lzn,Ebert:2018gsn}, see \fig{slicing}.
The inclusion of subleading terms is even more important at higher orders due to the presence of stronger and stronger logarithmic singularities and therefore they are important to extend the application of slicing methods to the calculation of fully differential cross sections at higher orders. 
Moreover, for complicated process the calculation above the cut becomes numerically instable already for values of the slicing parameter that are too large to sufficiently reduce the residual error. 
In this case the inclusion of terms beyond leading power are not just important, but rather crucial, since they open the door to the calculation of cross sections which would otherwise be unfeasible.

\subsection{New Phenomena Arising Beyond Leading Power}
To conclude this section let us present a final motivation for studying subleading power. 
As we mentioned in \sec{newdirection}, in the same way that processes involving flavor changing in neutral current cannot be described in the Standard Model using only tree level interactions, there are QCD processes that vanish in the leading kinematic limit and thus cannot be described by looking only at the leading power interactions in the soft, collinear and forward regime.
For example, as we will see in \chap{gluon_ops} and \chap{radiative}, while at leading power the emission of soft radiation from energetic collinear particles is constrained to be only eikonal, at subleading power we can have emission of non-eikonal soft particles.
The relaxation of the constraint on the polarization of the soft radiation emitted allows, for example, processes that are forbidden at leading power by helicity selection rules to take place beyond leading power.
Observables of this class are, for example, certain angular coefficients for Drell-Yan~\cite{Bacchetta:2019qkv,Ebert:fiducial}, for which only fixed order calculations are available \cite{Gauld:2017tww} as well as transverse momentum distributions of quarkonia in particular spin configurations \cite{Fleming:2019pzj}

In \chap{qregge} we will study another process that starts beyond leading power, which is the forward scattering mediated by a quark exchange\cite{Moult:2017xpp}.
At leading power in the forward limit only processes mediated by gluons are allowed. 
This implies that at leading power no forward scattering process with fermion number flipping in a given angular direction can take place. 
However, going beyond leading power it becomes possible to mediate the forward scattering by a quark, allowing the study, for example, of color neutral particle pair production in the forward limit at the LHC.

\subsection{Bootstrapping QCD Amplitudes and Cross Sections}
Another interesting theoretical motivation for the study of subleading power, is the potential to provide more constraints on the structure of amplitudes and cross sections. This would be fundamental to the application of bootstrapping techniques to QCD.
The bootstrap program aims at fully reconstruct amplitudes by making ansatzes of their functional form and then constraining and cross checking them by using general symmetry relations and their behavior in different kinematic limits.
It has been applied with great success for amplitudes in $\cN=4$ SYM (see, for example, \cite{Dixon:2014iba,Dixon:2014xca,Caron-Huot:2016owq,Caron-Huot:2019bsq,Caron-Huot:2019vjl}) where the large number of symmetries imposes strong constraints on the structure of amplitudes, especially in the planar case where super-conformal invariance dramatically constrains the functional dependence on the kinematic invariants.
To give an idea of how powerful these techniques are in the context of planar $\cN=4$, recently the 6 gluon amplitude has been bootstrapped up to 7 loop orders~\cite{Caron-Huot:2019vjl}.
Bootstrapping QCD amplitudes or cross sections is significantly more challenging and beyond what is currently feasible. However, bootstrapping techniques have been recently employed to determine QCD matrix elements and cross sections in the soft and collinear limits~\cite{Li:2016axz,Li:2016ctv,Ebert:2020yqt,Ebert:ThingsToCome}. 
An interesting application for these techniques in QCD at the cross section level is the bootstrap of the Energy-Energy Correlation (EEC)~\cite{Basham:1978bw}.
The EEC is a particularly simple observable since it is a differential distribution that depends only on a single variable. 
Also, it has singularities sitting only at the two end-points and momentum sum rules that provide powerful constraints~\cite{Korchemsky:2019nzm,Chen:2020vvp}.
While at fixed order the EEC has only been recently calculated at $\cO(\alpha_s^2)$ in QCD \cite{Dixon:2018qgp,Luo:2019nig}, its perturbative structure at the end points is much better understood\cite{Moult:2018jzp,Dixon:2019uzg}.
Moreover, much progress has been achieved in understanding the EEC in $\cN=4$ SYM \cite{Hofman:2008ar,Belitsky:2013xxa,Belitsky:2013bja,Belitsky:2013ofa,Henn:2019gkr,Moult:2019vou} and its relation to the behavior of energy correlators in QCD~\cite{Korchemsky:2019nzm,Dixon:2019uzg,Chicherin:2020azt,Henn:2020omi,Chen:2020uvt}, thus providing an interesting playground for the extension of bootstrapping techniques to QCD at the amplitude and cross section level. 
Going forward, given that a solid knowledge of the infrared limits of QCD beyond leading power provides very powerful constraints, the topics presented in this thesis and, more generally, the progress in the understanding of QCD beyond leading power are going to be of fundamental importance for the success of the bootstrap program in QCD.
To give a practical example of how the subject of this thesis can contribute to this program, we note that in a work of mine published in \cite{Moult:2019vou}, but not included in this manuscript,  we obtained the leading logarithmic resummation of the EEC in $\cN=4$ SYM in the back-to-back limit beyond leading power by extending the techniques presented in this thesis in \chap{subRGE} and \chap{ptNLL} for the resummation of subleading power event shape observables and treatment of rapidity divergences beyond leading power. This work constitutes the first step towards an all order understanding of the EEC in QCD beyond leading power.

\section{Kinematic Limits of QCD}\label{sec:expansions_intro}
In order to understand cross sections in their infrared and collinear kinematic regimes, it is necessary to identify and characterize these regimes.
Simply speaking, one can classify two particles with momenta $p_i$ and $p_j$ as collinear to each other when $p_i \cdot p_j \to 0$, while a particle with momentum $p_i$ is considered soft when $p_i^\mu \to 0$.
More precisely, particles should be classified as soft and collinear \emph{relative to the hard scale} of the process.
\begin{align} \label{eq:IR}
 p_i~\mathrm{collinear~to}~p_j \,:\quad p_i \cdot p_j \ll Q^2
\,,\qquad
 p_i~\mathrm{soft} \,:\quad p_i^\mu \ll Q
\,.\end{align}
More generally, in order to study these kinematic limits, it is convenient to work with light-cone coordinates. We take two light-like vectors $n^\mu$ and $\bn^\mu$ which we can identify, for example, as the directions of the incoming protons in an LHC collision  
\be
n^\mu = (1,0,0,1)\, \qquad \bn^\mu = (1,0,0,-1) \,.
\ee 
With the lightcone directions at hand, we can decompose any generic momentum $p^\mu$ as
\begin{align}\label{eq:lcdef_intro}
 p^\mu &= p^+ \frac{\bn^\mu}{2} + p^- \frac{n^\mu}{2} + p_\perp^\mu \equiv (p^+, p^-, p_\perp)
\,,\end{align}
where the $p^\pm$ components are explicitly given by
\begin{align}\label{eq:lcdef2_intro}
 p^- &= \bn \cdot p = p^0 + p^z \,,\quad p^+ = n \cdot p = p^0 - p^z
\,,\end{align}
and $p_\perp$ is the remaining transverse component. It is straightforward to generalize the above decomposition to the case of $N$ collinear directions. We just need $N$ pairs of lightlike vectors, $\{n_i^\mu,\bn_i^\mu\}_{i=1}^N$, one pair for each direction.

Employing the lightcone notation just introduced in \eq{lcdef_intro}, we can thus classify a momentum $p^\mu = (p^+, p^-, p_\perp)$ in the different dominant kinematic regions as
\begin{alignat}{3} 
\label{eq:modes}
 &\text{hard}:          \qquad &&p^\mu \sim Q \, (1, 1, 1)
\,,\nn\\
 &n\text{-collinear}:   \qquad &&p^\mu \sim Q \, (\lambda^2, 1, \lambda)
\,,\nn\\
 &\bn\text{-collinear}: \qquad &&p^\mu \sim Q \, (1, \lambda^2, \lambda)
\,,\nn\\
 &\text{soft}:          \qquad &&p^\mu \sim Q \, (\lambda^m, \lambda^m, \lambda^m)   \,,\qquad m=1,2
\,.\end{alignat}
Here, $\lambda \ll 1$ is an auxiliary power counting parameter indicating the suppression of the different modes relative to the hard scale $Q$.
Let us discuss \eq{modes} in more detail:
\begin{itemize}
 \item The hard region describes momenta directly associated with the production of the hard probe of the process, $h$.
       Taking $h$ to have invariant mass $p_h^2 = Q^2$, parametrically hard momenta also have virtuality $p^2 \sim Q^2$.
       For example, virtual corrections to the partonic process are sensitive to this scaling.
 \item The $n$-collinear region describes a momentum where $n \cdot p \ll \bn \cdot p$, and hence, from \eq{lcdef_intro}, $p$ is aligned with the $n$-direction.
       The scaling of the transverse component follows by noting that for on-shell particles $p^+p^- \sim p_{\perp}^2 \sim \lambda^2 Q^2$.
 \item The soft region describes low-energetic, but isotropic radiation, as is manifest from the homogeneous scaling in \eq{modes}.
       The choice of $m$ in \eq{modes} depends on whether the observable $\Obs$ under consideration is sensitive to the lightcone momenta only ($m=2$) or also to transverse momenta ($m=1$).
       In the SCET literature, these two cases are referred to as ultrasoft and soft, respectively.
\end{itemize}
Note that in more general cases, such as also measuring final-state jets or complicated observables, more modes may arise, and this has given rise to a plethora of \emph{scaling hierarchies} in the literature, see for example \refscite{Seymour:1997kj,Dasgupta:2001sh,Cheung:2009sg, Bauer:2011uc,vonManteuffel:2013vja,Larkoski:2014gra,Larkoski:2015zka, Larkoski:2015kga, Procura:2014cba,Becher:2015hka,Chien:2015cka,Ellis:2010rwa,Banfi:2010pa,Kelley:2012kj,Pietrulewicz:2016nwo,Hornig:2017pud,Neill:2018yet,Lee:2019lge,Chien:2019osu}.
For sufficiently inclusive observables as considered in this thesis it suffices to  consider \eq{modes} as well as Glauber modes, which will be the focus of \sec{glaubers} and play a crucial part in describing the \emph{Regge limit}\footnote{In the literature many different names can be associated with this limit, such as forward limit, high-energy limit, small-$x$ limit, or BFKL regime.} of cross sections and amplitudes. 

Above, we have only given heuristic arguments for the observation that the scalings in \eq{modes} are the only relevant ones.
In SCET~\cite{Bauer:2000ew, Bauer:2000yr, Bauer:2001ct, Bauer:2001yt, Bauer:2002nz}, an effective field theory to describe QCD in the infrared limit the separation of modes corresponding to \eq{modes} it is performed at the Lagrangian level for quark and gluon fields. In this thesis I will use and extend the SCET framework to study the behavior of gauge theories beyond leading power, for this reason \sec{scet} will be dedicated to a review of different aspects of SCET.

Before concluding this section, it is important to note that the modes stated in \eq{modes}
also arise in proofs of QCD factorization in what is sometimes referred to as the \emph{direct} QCD approach. These proofs are based on the insight that there is a one-to-one correspondence between the momentum regions giving rise to the large $Q$ behavior and mass divergences in massless perturbation theory~\cite{Sterman:1978bi,Libby:1978bx}.
These mass divergences arise at pinch-singular surfaces, \emph{i.e.}~momentum regions when loop momenta can not be deformed away from singularities in the appearing propagators~\refscite{Collins:1989gx,Sterman:1995fz,Collins:1350496}.
Similarly, these singular surfaces also arise when analyzing Feynman diagrams using the method of regions~\cite{Beneke:1997zp}.

\section{Overview}
\label{sec:outline}
In this section I present an overview of the work performed in the thesis. Other than explaining how this manuscript is organized, I will frame the work of this thesis in context with respect to the literature that has appeared on the subject.\\

In \chap{scet} I give a review of the SCET framework with a focus on the ingredients necessary for the study of gauge theories beyond leading power. 
I will explain what is the structure of factorization theorems, how subleading power corrections from different sources can be organized and I clarify some aspects of factorization breaking beyond leading power as well as the role of Glauber interactions in subleading power factorization.\\

In \chap{gluon_ops} and \chap{radiative}, I present the derivation of the ingredients of factorization beyond leading power in SCET, with a focus on event shape observables.
In \chap{gluon_ops} I construct subleading hard scattering operator for Higgs production and decay in gluon fusion leveraging the power of helicity selection rules thanks to the use of gauge invariant helicity operators. 
In \chap{radiative} I describe how to achieve factorization for event shapes in $e^+ e^-$ beyond leading power in the presence of subleading Lagrangian insertions that give rise to radiative jet and soft functions.

The results of these two chapters have been published in \cite{Moult:2017rpl,Moult:2019mog}. A related work for this thesis, that has not been included in this manuscript, has been published in \cite{Chang:2017atu} where we derived the subleading power hard scattering operators for the scalar current, which mediates Higgs decay and production in $b \bar{b}$ annihilation. 
The framework of helicity SCET operators has also been used in the literature to derive the subleading power operators for the vector current \cite{Feige:2017zci}.\\

In \chap{subRGE}, I discuss the leading logarithmic resummation of thrust in gluon induced Higgs decay at subleading power.
In order to do so I introduce new subleading power gauge invariant objects, the \emph{$\theta$-jet} and \emph{$\theta$-soft functions}, which arise in the renormalization of subleading soft and jet functions beyond leading power.
I present the subleading power factorization formula for this observable at leading logarithmic accuracy, and derive and solve the renormalization group equations (RGE) for the hard, jet and soft functions appearing in the factorization, including running coupling effects. 
This work has been published in \cite{Moult:2018jjd} and constitutes the first example of the resummation of collinear and soft logarithms beyond leading power for an event shape observable in QCD.

While the results in this chapter are derived for Higgs thrust, many of the results derived in this work constitute universal elements of renormalization and resummation for massless gauge theories at subleading power. 
In a work for the thesis not included in this manuscript, we show that the same framework allows for the resummation of soft fermion emissions in $\cN=1$ SUSY QCD \cite{Moult:2019uhz}.
As an other example, the $\theta$-soft function and its mixing under RG evolution with subleading power soft functions, as well as the solution of the subleading RGE equations derived in our work was later  used in the literature by an independent group to derive the resummation of threshold logarithms beyond leading power for Drell-Yan and Higgs production at the LHC \cite{Beneke:2018gvs,Beneke:2019mua}.\\

In \chap{ptNLL}, I present the study of rapidity logarithms, and the associated rapidity divergences, at subleading order in the power expansion.
I discuss the structure of subleading-power rapidity divergences and how to consistently regulate them.
I introduce a new pure rapidity regulator and a corresponding $\overline{\rm MS}$-like scheme, which handles rapidity divergences while maintaining the homogeneity of the power expansion.
I find that power-law rapidity divergences appear at subleading power, which give rise to derivatives of parton distribution functions. As a concrete application, I consider the $q_T$ spectrum for color-singlet production, for which I compute the complete $q_T^2/Q^2$ suppressed power corrections at $\mathcal{O}(\alpha_s)$, including both logarithmic and non-logarithmic terms and retaining the full dependence on the kinematics of the color singlet particle.

As explained in \sec{motivations}, the calculation of perturbative power corrections is important to improve slicing methods for fully differential fixed order calculations and having full dependence on the kinematic of the color singlet particle allows the use of such power corrections in calculations involving fiducial cuts on the color singlet phase space.
The work presented in this chapter has been published in \cite{Ebert:2018gsn} and constitutes the first calculation of subleading terms for transverse momentum distributions at the LHC and therefore for the $q_T$-subtraction scheme~\cite{Catani:2007vq}.
An analogous calculation of the analytic power corrections for $q_T$ subtractions has been later reproduced for the case of inclusive color singlet cross sections in \cite{Cieri:2019tfv}, while the inclusion of corrections to $q_T$ distributions due to the application of fiducial cuts to the decays of the color singlet for Higgs and Drell-Yan production has been studied in \refscite{Ebert:2019zkb,Ebert:fiducial}. 
Analogous corrections due to radiation from massive final states have been considered in \refcite{Buonocore:2019puv}.\\

In \chap{collinear_expansion_chapter}, I present a new method to employ cutting edge multiloop techniques for the computation of the expansions of cross sections in the collinear limit. This work has been published in \cite{Ebert:2020lxs}. This method provides a powerful way to construct systematically improvable analytic approximations to differential distributions. I illustrate this with an example by calculating the rapidity distribution of the Higgs boson at the LHC at NNLO in this approximation. I show that with just two orders in the collinear expansion, I obtain a smaller than 2\% discrepancy with respect to the known exact NNLO result for the entire range the rapidity spectrum.

This work also allows the extraction of universal ingredients arising in the collinear limit of QCD to an unprecedented level of precision in perturbation theory. 
In a work of mine not included in this manuscript, we demonstrate this by using the methods developed in \chap{collinear_expansion_chapter} to calculate for the first time at N$^3$LO both the $N$-Jettiness beam functions~\cite{Ebert:2020unb} as well as the transverse momentum depends PDFs ($q_T$ dependent beam functions) for gluon and quarks \cite{Ebert:2020yqt}.
Ref.~\cite{Ebert:2020yqt} constitutes the last missing ingredient to determine the full singular structure of an infrared observable ($q_T$) at N$^3$LO in QCD. This allows the $q_T$ subtraction to be extended to N$^3$LO and opens the door to fully differential calculations at N$^3$LO for Drell-Yan and Higgs production at the LHC. 
Since it determines the full log-independent structure of $q_T$ at N$^3$LO, it allows the resummation of $q_T$ distributions at N$^3$LL$^\prime$ and it provides the boundaries needed for N$^4$LL resummation. 
Similarly, the $N$-jettiness beam functions that we calculated in \refcite{Ebert:2020unb} are one of the cornerstones of $N$-jettiness subtraction at N$^3$LO. They are also crucial to extend the resummation of $\Tau_N$ to N$^3$LL$^\prime$ and N$^4$LL accuracy, and for matching N$^3$LO calculations to parton showers based on $\Tau_0$ resummation, like \texttt{GENEVA}~\cite{Alioli:2012fc,Alioli:2015toa}.\\

In \chap{qregge}, I study factorization and resummation in the Regge limit both at the amplitude and the cross section level. In this chapter, which has been published in \cite{Moult:2017xpp}, I develop a Lagrangian formalism for the treatment of fermion mediated forward scattering processes in QCD. This constitutes an example of  a process that only takes place beyond leading power, as it is forbidden by the symmetries of QCD at leading power. Using the operators of the derived Lagrangians, I derived the quark Regge trajectory at 1 loop as well as the BFKL resummation of small-$x$ logarithms for diphoton production at the LHC.

I conclude in \chap{conclusions_thesis}.\\

During the thesis I have worked on other papers that have not been included in this manuscript, besides the ones already mentioned throughout this outline. Therefore, to frame these works in the context of the study of QCD beyond leading power that I have carried out in the thesis, let me finish this section by briefly presenting them.

As we explain in \chap{ptNLL}, the structure of corrections for $q_T$ distribution is complicated by the presence of rapidity divergences at subleading power. 
In \chap{ptNLL} we study them at fixed order, but an all-order understanding necessary for carry out resummation of $q_T$ distributions is significantly more challenging to obtain. 
In the work published in \cite{Moult:2019vou}, we tackle this problem and carry out the resummation of the Energy-Energy Correlation (EEC) in the back-to-back limit
\footnote{The EEC in the back-to-back limit is intimately related to $q_T$ factorization, so much so that its logarithmic structure at leading power can be completely predicted recycling the ingredients necessary for $q_T$ resummation~\cite{Moult:2018jzp}. }
at subleading power in $\cN=4$ super-Yang-Mills. 
This work constitutes the first resummation at subleading power for an observable involving rapidity logarithms and includes the introduction of \emph{rapidity identity operators}, that will generically appear at subleading power in problems involving both rapidity and virtuality scales. This the case, for example, of the $p_{T}$ spectrum for color singlet boson production at hadron colliders and the resummation of power suppressed logarithms in the Regge limit.

I have also worked on power correction for the $N$-jettiness subtraction scheme~\cite{Boughezal:2015aha,Gaunt:2015pea}.
In the work published in \cite{Ebert:2018lzn}, we calculated the complete $\Tau$ suppressed power corrections at $\mathcal{O}(\alpha_s)$, including both logarithmic and non-logarithmic terms and retaining the full dependence on the kinematics of the color singlet particle. 
This work extended part of the results of \cite{Moult:2016fqy, Moult:2017jsg} to next-to-leading logarithmic accuracy and clarified in detail the necessity of retaining the full kinematic dependence on the color singlet phase space and thus of including power corrections induced by the additional measurements on the color singlet, which where not included in the calculations of \refscite{Boughezal:2016zws,Boughezal:2018mvf}.
In the work published in \cite{Bhattacharya:2018vph}, we have initiated the study of power corrections for LHC processes involving a jet in the final state. We have developed a procedure to efficiently expand spinor-helicity amplitudes in the soft and in the collinear limit to arbitrary power.
This is important not only because expressions for higher multiplicity amplitudes, as the ones that are needed for the study of power corrections to color singlet processes in association to one or more jets, are efficiently expressed in the spinor-helicity formalism, but also to facilitate the study of the universal behavior of amplitudes in the collinear limit beyond leading power, which has been recently object of study in the literature~\cite{Stieberger:2015kia,Nandan:2016ohb}. 
Finally, we apply this method to the case of Higgs production in association with a jet presenting the leading logarithmic contribution due to subleading power matrix elements for certain channels.

\chapter{SCET Beyond Leading Power}\label{sec:scet}

In this chapter we present an overview of the framework of Soft and Collinear Effective Theory which will be used in this thesis. 
In particular, the goal of this chapter is to give a coherent picture for the understanding of massless gauge theories beyond leading power in an effective field theory framework.
Note that, while many of the ingredients that we will introduce in this chapter had already appeared in the SCET literature, a substantial portion of the concepts and tools presented in this chapter has been developed for this thesis (all of which are published) and related works.
Therefore, we wish to provide here a self contained discussion hoping that it puts into context the role of the different works contained in this thesis to subleading power factorization more generally.

SCET is an effective field theory of QCD describing the interactions of collinear and soft particles in the presence of a hard interaction \cite{Bauer:2000ew, Bauer:2000yr, Bauer:2001ct, Bauer:2001yt, Bauer:2002nz}. 
SCET is constructed as an expansion about the light cone in powers of $\lambda$, following the notation of \eq{lcdef_intro}. Momenta are expanded into \emph{label} and \emph{residual} components
\begin{equation} \label{eq:label_dec_rf}
p^\mu = p_\ell^\mu + k^\mu = \bn \cdot p_\ell \, \frac{n^\mu}{2} + \lp_{\ell\perp}^\mu + k^\mu\,,
\,\end{equation}
where, $\bn \cdot p_\ell  \sim Q$ and $p_{\ell\perp} \sim \la Q$ are the large label momentum components, with $Q$ a characteristic scale of the hard interaction, while $k\sim \la^2 Q$ is a small residual momentum. A multipole expansion is then performed to obtain fields with momenta of definite scaling, namely collinear quark and gluon fields for each collinear direction, as well as soft quark and gluon fields.
Independent gauge symmetries are enforced for each set of collinear or soft fields.  As a consequence of the multipole expansion all fields and their derivatives acquire a definite power counting \cite{Bauer:2001ct}, shown in \Tab{tab:PC_intro}. 
The detailed structure of the fields will be described shortly. 

\begin{table}
\begin{center}
\begin{tabular}{| l | c | c |c |c|c|c |c|r| }
  \hline                       
  Operator & $\cB_{n\perp}^\mu$ & $\bar\cP^{}$ & $\cP_\perp^\mu$& $\bar n\cdot\cB_{us}$ &$\cB_{us\,\perp}^\mu$ & $\bar n\cdot \partial_{us}$ & $ n\cdot \partial_{us}$ & $ \partial_{us}^\perp$ \\
  Power Counting & $\lambda$  &  $\lambda^0$ & $\lambda$& $\lambda^2$& $\lambda^2$& $\lambda^2$& $\lambda^2$& $\lambda^2$ \\
  \hline  
\end{tabular}
\end{center}
\caption{
Power counting for building block operators in $\text{SCET}_\text{I}$.
}
\label{tab:PC_intro}
\end{table}

\section{Hard, Dynamical and Glauber Lagrangians}\label{sec:scet_lagr}

Given that each operator and derivative has a well defined power counting and that Lagrangians are made of operators, it is clear that the action of the theory can be written via Lagrangians with definite power counting. The SCET Lagrangian is expanded as
\begin{align} \label{eq:SCETLagExpand_RF}
\cL_{\text{SCET}}=\sum_{i\geq0} \cL_{\text{SCET}}^{(i)}\,,
\end{align}
with each term having a definite power counting ${\cal O}(\lambda^i)$.
The leading power SCET Lagrangian $\cL_{\text{SCET}}^{(0)}$ can be organized as
\be\label{eq:SCETLagLP}
	\cL_{\text{SCET}}^{(0)} =  \cL_\hard^{(0)}+ \cL_\dyn^{(0)} +\cL_G^{(0)} = \cL_\hard^{(0)}+ \Big(\sum_{\{n_i\}} \cL_{n_i}^{(0)} + \cL_s^{(0)} \Big)+\cL_G^{(0)} \,.
\ee
More generally, we can organize the expansion of the Lagrangians beyond leading power as 
\begin{align} 
\cL_{\text{SCET}}=\cL_\hard+\cL_\dyn= \sum_{i\geq0} \cL_\hard^{(i)}+\sum_{i\geq0} \cL_\dyn^{(i)} +\cL_G^{(0)} \,,
\end{align}
As written, the SCET Lagrangian is divided into three different contributions:\\

The \emph{hard scattering Lagrangian}, $ \cL_\hard^{(i)}$, contains local operators and is derived by a matching calculation. It describes operators from the SM Lagrangian necessary to mediate the hard scattering interactions. 
For example, the hard scattering operators mediating the Born process $q \bar{q} \to Z$ for Drell Yan will be part of this Lagrangian. It is important to notice that while this Lagrangian is referred as ``hard", the fields entering in this Lagrangians are not hard, but rather collinear%
\footnote{When considering QCD beyond leading power there are hard scattering operators containing also soft fields and soft derivatives. We will see explicit example of this in \sec{nnlp_soft}}.
There are no active hard degrees of freedom in SCET.

One could pictorially interpret the operators in $\cL_\hard$ as the ones answering the question "how do the color charged particles interact at the hard scale?", "How do collinear quarks and gluons interact at the hard scale to produce the desired final state?". All the hard modes are integrated out and their contribution is contained in the hard matching coefficients (the Wilson coefficients) of the operators.
The work in \chap{gluon_ops} is dedicated to derive both the hard scattering operators as well as their Wilson coefficients for $ \cL_\hard$ up to $\cO(\lambda^2)$ for the case Higgs production in gluon fusion.\\

The \emph{dynamical Lagrangians}, $\cL_\dyn^{(i)}$, describe the long wavelength dynamics in the effective theory of collinear modes in each collinear sector $n$ as well as that of soft modes.
At leading power, the dynamical Lagrangian can be further split in a term involving only soft fields $\cL_s^{(0)}$ and a sum of purely collinear Lagrangians $\cL_{n_i}^{(0)}$, one for each non overlapping collinear directions $n_i$.%
\footnote{It is important to note that "overlapping" reference vectors $n$ and $n'$, with $n\cdot n' \sim \ord{\lambda^2}$ provide equivalent descriptions. This enforces a symmetry on the effective theory known as reparametrization invariance (RPI) \cite{Manohar:2002fd,Chay:2002vy}. We will often use this symmetry throughout this thesis to simplify our description, for example by choosing that the total $\perp$ momentum of a particular collinear sector vanishes.}. 

At subleading power the distinction between collinear and soft dynamical Lagrangians is less meaningful since many operators in the subleading power dynamical Lagrangians contain both soft and collinear fields as we will see in \chap{radiative}.
The dynamical Lagrangians dictate how the color charged particles interact after, or before, the hard scattering interaction takes place. It describes, for example, how a $n$-collinear gluon can split into an $n$-collinear quark-antiquark pair. As a matter of fact, the Altarelli-Parisi splitting as well as the splitting amplitudes can be completely derived via $\cL^{(0)}_n$. 
Note that since they carry no information about the hard scattering process, the dynamical Lagrangians are process independent and therefore enjoy a degree of universality.
We will present the SCET dynamical Lagrangians up to $\cO(\lambda^2)$ in \sec{subleading_lagrangians}.\\

The \emph{Glauber Lagrangian}~\cite{Rothstein:2016bsq}, $\cL_G^{(0)}$, describes leading power interactions between soft and collinear modes in the form of non-local potentials, which break factorization, unless they can be shown to cancel. 
Note that for the Glauber Lagrangian we only have singled out the leading power $\cL_G^{(0)}$ and no power correction $\sum_{i>0}\cL_G^{(i)}$.
This is because the distinction between Glaubers and other types of non local operators  beyond leading power ceases to be transparent and meaningful as operators beyond leading power cannot break factorization while the main characteristic of $\cL^{(0)}_G$ is that it does. 
We will provide more details of the Glauber interactions in SCET and in particular on this subtlety of the classification of Glaubers beyond leading power in \sec{glaubers}. Note also that a subset of subleading power operators involving Glaubers which describes quark reggeization \cite{Moult:2017xpp} is the focus of \chap{qregge}.

\section{Gauge Invariant Fields and Soft Decoupling}\label{sec:scet_fields}

In this section we review some of the basic SCET concepts that we will use in the rest of the thesis.
We will write the SCET fields for $n_i$-collinear quarks and gluons, $\xi_{n_i,\lp}(x)$ and $A_{n_i,\lp}(x)$,  in position space with respect to the residual momentum and in momentum space with respect to the large momentum components. The large momentum $\lp$, and the collinear direction then act as labels for the fields. Derivatives acting on the fields give the residual momentum dependence, $i \partial^\mu \sim k \sim \la^2 Q$, while label momentum operators  $\cP_{n_i}^\mu$ give the label momentum $\cP_{n_i}^\mu\, \xi_{n_i,\lp} = \lp^\mu\, \xi_{n_i,\lp}$.  An important feature of the multipole expansion is that the propagator for collinear fields 
\begin{align}\label{eq:prop}
\frac{1}{\bar n \cdot \tilde p~ n \cdot p_r + \tilde p_{\perp} ^2}\,,
\end{align}
is independent of $\bar n \cdot p_r$ and $p_{r\perp}^\mu$ and hence is local in the residual $x^-$, $x^\perp$ components. This will allow for factorized expressions involving convolutions to be reduced to a single variable convolution in the position along the lightcone, and will play an important role in our definitions of the \emph{radiative functions} in \chap{radiative}.

Soft degrees of freedom are described in the effective theory by separate quark and gluon fields, $q_{us}(x)$ and $A_{us}(x)$. We will assume that we are working in the SCET$_\text{I}$ theory where these soft degrees of freedom are referred to as ultrasoft, indicated by the subscript "us". These fields do not carry label momentum, and have $i \partial^\mu \sim \la^2Q$.

The use of gauge invariant soft and collinear SCET fields play a central role in achieving factorization for QCD amplitudes and cross sections, since it allows us to systematically construct gauge invariant non-local operators. Collinear gauge invariant quark and gluon fields are defined as \cite{Bauer:2000yr,Bauer:2001ct}
\begin{align} \label{eq:chiB_RF}
\chi_{{n_i},\w}(x) &= \Bigl[\delta(\w - \bnP_{n_i})\, W_{n_i}^\dagger(x)\, \xi_{n_i}(x) \Bigr]
\,,\\
\cB_{{n_i}\perp,\w}^\mu(x)
&= \frac{1}{g}\Bigl[\delta(\w + \bnP_{n_i})\, W_{n_i}^\dagger(x)\,i  D_{{n_i}\perp}^\mu W_{n_i}(x)\Bigr]
 \,, \nn
\end{align}
where
\begin{equation}
i  D_{{n_i}\perp}^\mu = \cP^\mu_{{n_i}\perp} + g A^\mu_{{n_i}\perp}\,,
\end{equation}
is the collinear covariant derivative and
\begin{equation} \label{eq:Wn_RF}
W_{n_i}(x) = \biggl[~\sum_\text{perms} \exp\Bigl(-\frac{g}{\bnP_{n_i}}\,\bn\sdt A_{n_i}(x)\Bigr)~\biggr]\,,
\end{equation}
is a Wilson line of ${n_i}$-collinear gluons in label momentum space, and the label operators in \eqs{chiB_RF}{Wn_RF} only act inside the square brackets. The collinear Wilson line $W_{n_i}(x)$ is localized with respect to the residual position $x$, and we can therefore treat
$\chi_{{n_i},\w}(x)$ and $\cB_{{n_i},\w}^\mu(x)$ as local quark and gluon fields from the perspective of ultrasoft derivatives $\partial^\mu$ that act on $x$. 

Since also soft particles carry color charge, there are \emph{soft} gauge transformation under which they transform \cite{iain_notes}. It is therefore equivalently important to define gauge invariant ultrasoft quark and gluon fields. In order to do so we first need to decouple the collinear fields from the ultrasoft particles. In the SCET framework, this is achieved at the level of the Lagrangian through the BPS field redefinition \cite{Bauer:2002nz} of the collinear fields
\be \label{eq:BPSfieldredefinition_v2}
\cB^{a\mu}_{n\perp}\to \cY_n^{ab} \cB^{b\mu}_{n\perp} , \qquad \chi_n^\alpha \to Y_n^{\alpha \bbeta} \chi_n^\beta\,,
\ee
which is performed in each collinear sector. Here $Y_n$, $\cY_n$ are fundamental and adjoint ultrasoft Wilson lines, where for a general color $SU(3)$ representation, $r$, the ultrasoft Wilson line is defined by
\be
Y^{(r)}_n(x)=\bold{P} \exp \left [ ig \int\limits_{0}^\infty ds\, n\cdot A^a_{us}(x+sn)  T_{(r)}^{a}\right]\,,
\ee
where $\bold P$ denotes path ordering.  The BPS field redefinition decouples the ultrasoft degrees of freedom from the leading power collinear Lagrangian \cite{Bauer:2002nz}, so that they appear only in the hard scattering vertex. In other contexts this is sometimes referred as leading power eikonalization.
To define gauge invariant ultrasoft fields, we can group all Wilson lines arising from the BPS field redefinition with fields to form gauge invariant combinations. In particular, we can define an ultrasoft gauge invariant quark field as
\begin{align} \label{eq:usgaugeinvdef}
\psi_{us(i)}=Y^\dagger_{n_i} q_{us}\,,
\end{align}
Similarly, we can group Wilson lines with gauge covariant derivatives in an arbitrary representation, $r$,
\begin{align}\label{eq:soft_gluon_RF}
Y^{(r)\,\dagger}_{n_i} i D^{(r)\,\mu}_{us} Y^{(r)}_{n_i }&=i \partial^\mu_{us} + [Y_{n_i}^{(r)\,\dagger} i D^{(r)\,\mu}_{us} Y^{(r)}_{n_i}]=i\partial^\mu_{us}+T_{(r)}^{a} g \cB^{a\mu}_{us(i)}\,, \nn \\
Y^{(r)\,\dagger}_{n_i} i \overleftarrow{D}^{(r)\,\mu}_{us} Y^{(r)}_{n_i }&=i \overleftarrow{\partial}^\mu_{us} + [Y_{n_i}^{(r)\,\dagger} i \overleftarrow{D}^{(r)\,\mu}_{us} Y^{(r)}_{n_i}]=i \overleftarrow{\partial}^\mu_{us}-T_{(r)}^{a} g \cB^{a\mu}_{us(i)}\,,
\end{align}
to obtain ultrasoft derivatives that only act on gauge invariant products of fields, and gauge invariant ultrasoft gluon fields
\begin{align} \label{eq:softgluondef_RF}
g \cB^{a\mu}_{us(i)}= \left [   \frac{1}{in_i\cdot \partial_{us}} n_{i\nu} i G_{us}^{b\nu \mu} \cY^{ba}_{n_i}  \right] \,.
\end{align}
Here the square brackets indicate that the derivatives acts only on the Wilson lines within the brackets.

Gauge invariant ultrasoft operators will be necessarily non-local at the ultrasoft scale, involving ultrasoft Wilson lines. 
However, the form of this non-locality is completely determined by the BPS field redefinition. 
Matching calculations from QCD to the effective theory are performed prior to the BPS field redefinition, when the theory is local at the hard scale, and then the non-locality arises later from the BPS field redefinition. 
However, one can alternatively take a bottom-up approach and consider gauge invariant soft gluon fields as building blocks of the theory. This approach has been used in \cite{Feige:2017zci,Moult:2017rpl,Chang:2017atu} to construct the operator bases at subleading power, as we will see in \chap{gluon_ops}. In \sec{subleading_lagrangians} we will show that the subleading power Lagrangians of SCET can be rewritten after BPS field redefinition purely in terms of gauge invariant ultrasoft quark and gluon building blocks and collinear fields \cite{Moult:2019mog}. This will play an important role, allowing for the definition of gauge invariant radiative functions in \chap{radiative}.

In both \Eqs{eq:usgaugeinvdef}{eq:soft_gluon_RF}, the fields are ultrasoft gauge invariant for an arbitrary lightlike direction, $n_i$, and the soft fields themselves are not naturally associated with a given direction.
Due to \Eq{eq:BPSfieldredefinition_v2}, this direction is usually naturally taken to coincide with that of a collinear direction, for example the direction of the collinear particles emitting the soft quark or gluon. Note that the ultrasoft gauge invariant gluon field is the analogue of the gauge invariant collinear gluon field of \Eq{eq:chiB_RF}, which can also be written
\begin{align}  
g\cB_{n_i\perp}^{A\mu} =\left [ \frac{1}{\bar \cP}    \bar n_{i\nu} i G_{n_i}^{B\nu \mu \perp} \cW^{BA}_{n_i}         \right]\,.
\end{align}

The gauge invariance of the ultrasoft fields in \Eqs{eq:usgaugeinvdef}{eq:soft_gluon_RF}, is enforced by the presence of the soft Wilson line. This also implies that it has Feynman rules describing an arbitrary number of soft emissions. For example, expanded up to two emissions, we have
\begin{align}\label{eq:twogluonBus}
g\cB^\mu_{us(n)}&=g\left(   A^{\mu a}_{us}(k)\, T^a -k^\mu \frac{n \cdot A^a_{us}(k)\, T^a}{n \cdot k} \right)+g^2 [T^a, T^b] \frac{n \cdot A^a_{us}(k_1)}{n \cdot k_1} A^{b\,\mu }_{us}(k_2) + \nn \\[0.2cm]
&+\frac{g^2}{2}\left( \frac{k^\mu_{1} +k^\mu_{2}}{n \cdot k_1+n \cdot k_2}\right) \left(   \frac{T^a T^b}{n \cdot k_1}+\frac{T^b T^a}{n \cdot k_2 } \right)n \cdot A_{us}^a(k_1)\, n \cdot A^b_{us}(k_2) +\cdots\,.
\end{align}
\section{Components of Factorization Beyond Leading Power}
\label{sec:sub-fact}

In this section we discuss the different components contributing to the formulation of a subleading power factorization theorem in SCET, extending the brief discussion provided in \Sec{sec:scet_lagr}. 
We will take as a concrete example an SCET$_{\text{I}}$ event shape in $e^+e^-\to $ dijets, however, the considerations are more general. Throughout this section, except for \sec{glaubers}, we will not consider possible contributions from leading power Glauber modes, which we assume decouple from the soft and collinear modes. For the case of $e^+e^-$ this is reasonable, since all QCD particles are in the final state, where we expect leading Glauber effects can be absorbed into the direction of soft Wilson lines \cite{Rothstein:2016bsq}.

Consider a dimensionless SCET$_\text{I}$ event shape observable, $\tau$, in $e^+e^-\to $ dijets, which is chosen to vanish in the dijet limit. As a concrete example, one can consider $\tau=1-T$, where $T$ is the thrust observable \cite{Farhi:1977sg}. Analogously to what we have seen in \sec{motivations}, we can write the cross section as a power expansion
\begin{align}\label{eq:xsec_homog}
\frac{\df\sigma}{\df \tau} &=\frac{\df\sigma^{(0)}}{\df \tau} +\frac{\df\sigma^{(1)}}{\df \tau} +\frac{\df\sigma^{(2)}}{\df \tau}+\frac{\df\sigma^{(3)}}{\df \tau} +{\cal O}(\tau)\,,
\end{align}
where $\df\sigma^{(n)}/\df \tau\sim \tau^{n/2-1}$ due to the scaling relation $\lambda \sim \sqrt{\tau}$. We wish to find factorized expressions for the non-zero contributions in this series, in terms of hard, jet and soft functions of the schematic form
\begin{align} \label{eq:sec5facsigma}
&\hspace{-0.25cm}\frac{\df\sigma^{(n)}}{\df \tau} =
 Q\sigma_0 
\sum_{j}  H^{(n_{Hj})}_{j} \otimes J^{(n_{Jj})}_{j} \otimes S_j^{(n_{Sj})}  
\,.\end{align}
Here $H^{(n_{Hj})}_{j}$ describes matching coefficients, while $J^{(n_{Jj})}_{j}$ and $S_j^{(n_{Sj})}$ are field theoretic matrix elements involving only collinear or soft fields, and $Q$ is the center of mass energy. Here the superscripts denote the power suppression, and we have
\begin{align}
n=n_{Hj}+n_{Jj}+n_{Sj}\,.
\end{align}
In general, there will be multiple distinct hard, jet and soft functions at each power, indicated by the sum over $j$.

In the effective theory approach, the first step towards this goal is to write the expression for the differential cross section in terms of full theory QCD matrix elements,
\begin{align}
\frac{d\sigma}{\df \tau} = \frac{1}{2Q^2} \sum_X \tilde \delta^{(4)}_q  
    \bra{L} O (0) \ket{X}\bra{X} O(0) \ket{L} \delta\big( \tau - \tau(X) \big)\,,
\end{align}
where for  $e^+e^-\to$ dijets through a virtual photon, $O=\mathcal{J}^\mu L_\mu$, where $L_\mu$ is the leptonic current which includes the $e^+e^-$, photon propagator and couplings, and $\mathcal{J}^\mu=\bar q \gamma^\mu q$.  Here we use the shorthand notation $\tilde \delta^{(4)}_q=(2\pi)^4\delta^4(q-p_X)$ for the momentum conserving delta function. The summation over all final states, $X$, includes phase space integrations. Here $|L\rangle$ denotes the $e^+e^-$ leptonic initial state. The measurement of the dijet observable is enforced by $ \delta\big( \tau - \tau(X) \big) $, where $\tau(X)$, returns the value of the observable $\tau$ as measured on the final state $X$.

For $\tau \ll 1$, we are in the dijet limit and  can match onto SCET hard scattering operators with two collinear sectors
\begin{align}\label{eq:match_intro}
O=\sum_{\lambda_l,j,k} \cC_{\lambda_l\,j}^{(k)} O_{\lambda_l\,j}^{(k)}\,.
\end{align}
Here the sum is over powers in $\lambda$ indicated by the superscript $(k)$, and at each power distinct operators are labeled by $\lambda_l$ and $j$ which include helicity and color labels.  Our labels are split such that $\lambda_l = \pm$ indicates the helicity of the lepton current and the index $j$ denotes all helicity and color labels of the QCD component of the current. The $\cC_{\lambda_l j}^{(k)}$ coefficients include the electromagnetic coupling and charges.

We work to all orders in the strong coupling, $\alpha_s$, but to leading order in the electroweak couplings. We can therefore factorize out the leptonic component, $J_{\lambda l}$ of the hard scattering operators
\begin{align}
O_{\lambda_l\,j}^{(k)}=J_{\lambda_l}\tilde O_j^{(k)}\,.
\end{align}
Evaluating the tree level matrix element involving the external electron states, the expression for the cross section can be written in terms of matrix elements in the effective theory  as
\begin{align}\label{eq:pre_expand}
&\frac{d\sigma}{d \tau} = N \sum_X    \tilde \delta^{(4)}_q  \sum_{\lambda_l}
 \Big \l 0 \Big| \sum_{j,k} \cC_{\lambda_l\,j}^{(k)} \tO_j^{(k)}  \Big| X \Big \r_{\cL_{\text{dyn}} }     
 \Big \l X \Big |\sum_{j,k} \cC_{\lambda_l\,j}^{(k)} \tO_j^{(k)} \Big |0\Big \r_{\cL_{\dyn}}   
 \delta\Big( \tau - \sum_l \tau^{(l)}(X) \Big)
 \,.
\end{align}
After having calculated the leptonic matrix element we are left with a normalization factor $N$, whose explicit form is not relevant for the current discussion.

To achieve an expression with homogeneous power counting, as in \Eq{eq:xsec_homog}, we must systematically expand \Eq{eq:pre_expand} in $\lambda$, working to all orders in $\alpha_s$. At leading power, assuming that the action of the measurement function factorizes, this is simple. The BPS field redefinition decouples leading power soft and collinear interactions so that the Hilbert space factorizes, and the state $|X\r$ can be written
\begin{align}
|X\r =|X_n \r |X_\bn \r |X_{us} \r\,.
\end{align}
Algebraic manipulations can then be used to organize  \Eq{eq:pre_expand} into a form involving separate matrix elements of soft and collinear fields, and hence derive a bare factorization formula. 

\begin{figure}
\begin{center}
\includegraphics[width=0.23\columnwidth]{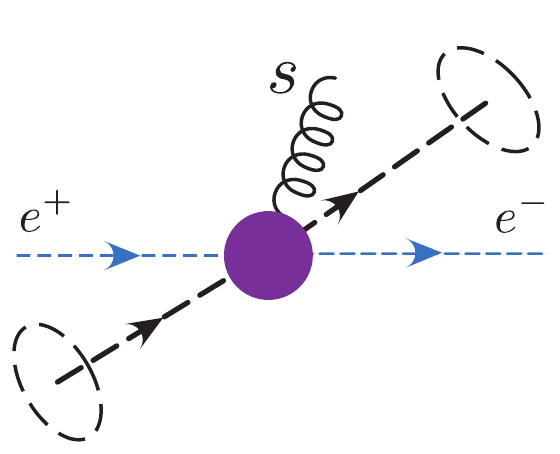} 
\hspace{0.1cm}
\includegraphics[width=0.23\columnwidth]{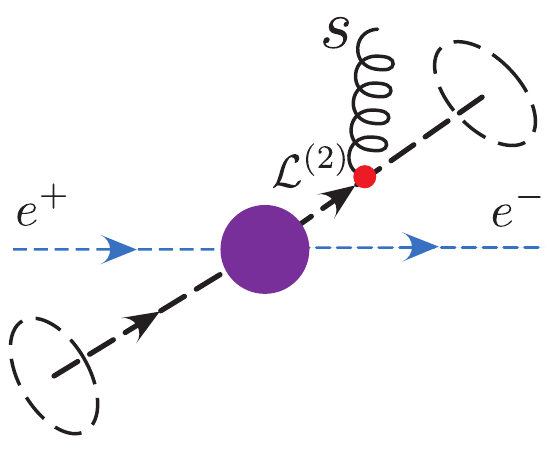} 
\hspace{0.1cm}
\includegraphics[width=0.23\columnwidth]{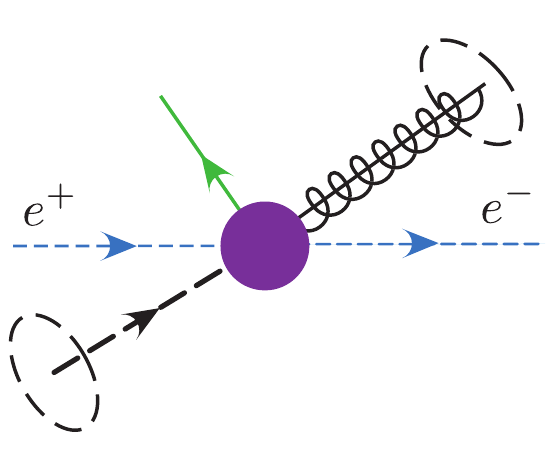} 
\hspace{0.1cm}
\includegraphics[width=0.23\columnwidth]{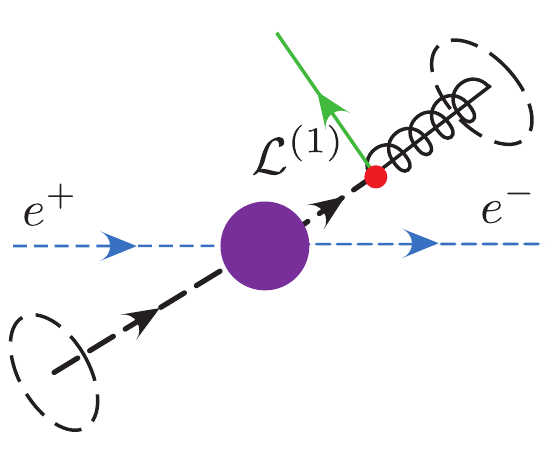} 
\raisebox{0cm}{ \hspace{-0.2cm} 
  $a$)\hspace{3.4cm}
  $b$)\hspace{3.4cm} 
  $c$)\hspace{3.4cm}
  $d$)\hspace{4cm} } 
\\[-25pt]
\end{center}
\vspace{-0.4cm}
\caption{ 
Subleading power contributions from the emission of a soft quark or gluon in a hard scattering. The subleading emission can either be from a local hard scattering operator, as shown in a), c), or from a radiative contribution from the energetic partons, as in b), d). }\label{fig:subleadingamp_intro} 
\end{figure}

In the effective field theory organization it is then evident from \Eq{eq:pre_expand} that there are three sources of power corrections
\begin{enumerate}
\item Subleading power hard scattering operators.
\item Subleading power corrections to the measurement function.
\item Subleading power Lagrangian insertions.
\end{enumerate}
We will discuss the structure of each of these sources of power corrections.

\subsection{Hard Scattering Operators}\label{sec:hard_fact}

The matching of QCD onto SCET gives rise to hard scattering operators, see \Eq{eq:match_intro}. 
These operators are local at the scale of the matching, as shown schematically in \Fig{fig:subleadingamp_intro}. 
Subleading power hard scattering operators with two collinear directions are discussed in detail in \chap{gluon_ops} and published in \Refs{Feige:2017zci,Moult:2017rpl,Chang:2017atu} where complete bases were derived for $\bar q \Gamma q$ and $gg$ currents using the approach of helicity operators \cite{Moult:2015aoa,Kolodrubetz:2016uim}. 
The leading order matching was also performed. In addition, operator bases for $N$-jet configurations were studied in \cite{Beneke:2017ztn,Beneke:2018rbh}.

Hard scattering operators at subleading power are similar to those at leading power in that they are formed from products of the SCET operator building blocks of \Tab{tab:PC_intro}.
These building blocks provide a complete basis of building blocks to all powers, as can be proven by the use of equations of motion and operator relations~\cite{Marcantonini:2008qn}. 
The difference between leading and subleading power comes from additional collinear or ultrasoft fields, or $\cP_\perp$ operators which are inserted into the hard scattering operator to give the power suppression. 
For example, leading power hard scattering operators for more inclusive processes typically have a single collinear field in each collinear sector. 
For concreteness one can take the case of Higgs production in gluon fusion (which will be thoroughly presented in \chap{gluon_ops}) where the leading power operator has this property
\be\label{eq:LPggop}
\includegraphics[valign=c,width=0.2\columnwidth]{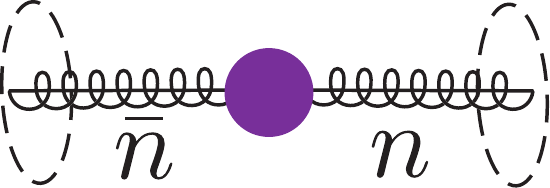}\quad.
\ee
However, at subleading power, operators can have multiple collinear fields in a single sector, as for example
\be
\includegraphics[valign=c,width=0.2\columnwidth]{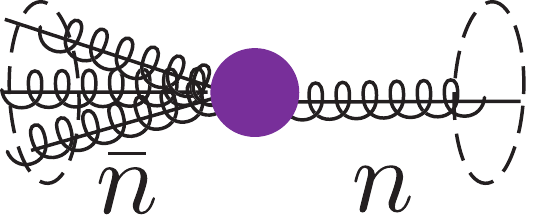}\quad.
\ee
While it may be tempting to think that subleading power operators can be seen as some "dressing" or correction of leading power operators, it is important to note that that  is not true. At subleading power, intrinsically new operators which cannot be obtained just by tweaking or correcting the leading power operator appear. For example, beyond leading power there are operators with different fermion number per collinear sector w.r.t. the leading power.  Again, taking an example from the case of $gg \to H$, an operator that contributes to the cross section at subleading power is
\be\label{eq:NLPgqop}
{\includegraphics[valign=c,width=0.2\columnwidth]{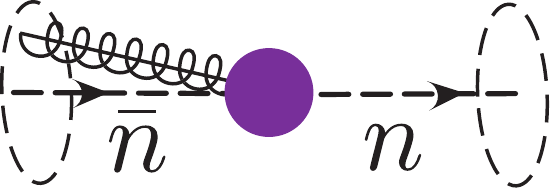}}\quad,
\ee
where we have non-zero fermion number in each collinear sector, while the leading power operator for the same process has zero fermion number in each sector, see \eq{LPggop}.
It is important to understand that here we have chosen the case of $gg \to H$ to illustrate these features of hard scattering operators beyond leading power just for convenience, as $gg \to H$ is a standard process of significant interest, but these features are general and can be seen also in the hard scattering operators of other processes, like the Drell-Yan process or Higgs decay to $b \bar{b}$ \cite{Feige:2017zci,Chang:2017atu}.

Note that the operator in \eq{NLPgqop} contributes already at $\cO(\alpha_s)$ to the cross section as will be shown in detail in \sec{matching_nlp}.


If the entire power suppression comes from the hard scattering operator then the factorization proceeds similar to at leading power. 
In particular, in this case only the leading power Lagrangian is required since subleading power Lagrangian insertions would induce additional power suppression, and therefore factorization formulae can be derived through the BPS field redefinition.
However, since subleading power hard scattering operators can have multiple fields per collinear sector, the final factorization formulas typically not only have a richer convolution structure, but they will also include jet, beam or soft functions with additional fields and very different renormalization properties.
The richer structure of the factorization formulas due to subleading hard scattering operators for $gg \to H$ is discussed in detail in \sec{discussion_gluonops} and summarized in \Tab{tab:fact_func}.
The presence of a soft function involving additional fields due to a subleading hard scattering operator is crucial for the leading log resummation of event shape observables beyond leading power and its subtle renormalization is one of the main challenges solved by my work published in \refcite{Moult:2018jjd}, which is presented in \chap{subRGE}.


\subsection{Measurement Function Factorization}\label{sec:obs_fact}

The action of the measurement function $\tau(X)$, which is a function of the soft and collinear momenta, must also be expanded homogeneously in the soft and collinear limits.  As shown in \Eq{eq:pre_expand}, we expand the measurement function as
\begin{align} \label{eq:measurement_exp}
 \tau(X) &= \tau^{(0)}(X) + \tau^{(2)}(X) + \ldots \,.
\end{align}
Here we have assumed that any $\mathcal{O}(\lambda) = \mathcal{O}(\sqrt{\tau})$ corrections to the measurement function vanish.\footnote{This is true for most observables in the SCET formulation of \Ref{Bauer:2000ew, Bauer:2000yr, Bauer:2001ct, Bauer:2001yt} in which label and residual momentum are exactly conserved. In the alternate approach of \Ref{Freedman:2011kj}, where momentum is not strictly conserved, an $\mathcal{O}(\lambda)$ contribution to the measurement function does appear. As shown in \Ref{Freedman:2013vya} this $\mathcal{O}(\lambda)$ contribution to the measurement function contributes as a product with an $\cO(\lambda)$ operator arising from the expansion of momentum conserving delta functions, and contributes at $\lambda^2$.} This was shown explicitly for the case of thrust in \cite{Feige:2017zci}. 

The measurement function enters  \Eq{eq:pre_expand} as a delta function constraint on the final state. This constraint can be expanded as
\begin{align}\label{eq:meas_expand}
\delta\big( \tau - \tau(X) \big) &=
\delta\big( \tau - \tau^{(0)}(X) - \tau^{(2)}(X) - \ldots \big) 
\\&= \delta\big( \tau - \tau^{(0)}(X) \big) - \tau^{(2)}(X)\: \delta' \big( \tau - \tau^{(0)}(X) \big) + \ldots
 \,,\nn
\end{align}
where the dots represent higher derivatives of delta functions.

To achieve a factorization, one must show that the measurement operators at each power can be factorized into contributions from soft or collinear degrees of freedom. To be specific, we restrict ourselves to what we have referred to as ``pseudo-additive observables'' \cite{Feige:2017zci} which we define as those observables with measurement functions that can be factorized into contributions from collinear and ultrasoft modes at each order in the power expansion in the form
\begin{align}\label{eq:fact_meas}
 \tau^{(i)}(X)= \tau_n^{(i)}(X_n,G_\bn,G_s)+ \tau_\bn^{(i)}(X_\bn,G_n,G_s) + \tau_{us}^{(i)}(X_{us},G_n,G_\bn)\,.
\end{align}
The factors $G_n$, $G_\bn$, $G_{us}$, which can enter the measurement function for any sector, are global properties of a sector, and must be defined independent of the order in perturbation theory.\footnote{While the factors $G_n, G_\bn, G_{us}$ are often trivial, an example where they are not is the factorization for the ``soft haze'' region of \Ref{Larkoski:2015kga,Larkoski:2017iuy,Larkoski:2017cqq}, describing the factorization in endpoint region of energy correlation function based jet substructure observables \cite{Larkoski:2014gra,Moult:2016cvt}.}  In this case one can define field theoretic measurement functions $\widehat \cM_n^{(i)}$, $\widehat \cM_\bn^{(i)}$, and $\widehat \cM_{us}^{(i)}$ from the energy momentum tensor of the theory \cite{Lee:2006nr,Sveshnikov:1995vi,Korchemsky:1997sy,Bauer:2008dt,Belitsky:2001ij}. The measurement functions act as 
\begin{align}
\widehat \cM_n^{(i)} |X_n\r &= \delta(\tau - \tau_n^{(i)}(X_n))|X_n\r \,, \nn \\
\widehat \cM_{\bar n}^{(i)} |X_{\bar n}\r &= \delta(\tau - \tau_{\bar n}^{(i)}(X_{\bar n}))|X_{\bar n}\r \,, \nn \\
\widehat \cM_{us}^{(i)} |X_{us}\r &= \delta(\tau - \tau_{us}^{(i)}(X_{us}))|X_{us}\r \,.  
\end{align}
The subleading power measurement function has been derived for the thrust observable in \cite{Freedman:2013vya,Feige:2017zci} to $\cO(\lambda^2)$.

Note that subleading power corrections can also arise for measurement functions of observables, such as kinematic factors, that are not small in the $\lambda \ll 1$ limit. This is particularly important when multiple measurements are performed on the final states as occurs for \emph{Born measurements} in fully differential cross sections at hadron colliders. 

As an example, let us consider the beam thrust \cite{Stewart:2010tn} event shape $\Tau_0$ or the $q_T$ spectrum in color singlet production at the LHC. To obtain distributions that are fully differential in the momentum of the color singlet, one needs to include not only a measurement function for the observable $\Tau_0$ or $q_T$, but also a measurement $\hat\delta_Y \equiv  \delta(Y -\hat{Y}(X))$ for the rapidity and one for the invariant mass of the color singlet $\hat\delta_Q \equiv  \delta(Q^2 -\hat{Q^2}(X))$. To be precise, if we call $q^\mu$ the 4-momentum of the color singlet in the hadronic center of mass frame, the rapidity and the invariant mass measurements take the form
\be\label{eq:bornMeas}
	\hat\delta_Y = \delta\left(Y - \frac{1}{2}\log\frac{q^-}{q^+}\right)\,,\qquad\hat\delta_Q = \delta\Bigl(Q^2 - q^+q^- - {\vec{q}}_\perp^{\,\,2}\Bigr)\,.
\ee
At Born level the $\Tau_0$ or $q_T$ measurement gives $\delta(\Tau_0)$ or $\delta(q_T)$, however the observables defined in \eq{bornMeas} are in general non trivial already at the Born level. Hence, they are referred to as \emph{Born measurements}.
As shown in the fixed order calculations of \cite{Moult:2016fqy,Moult:2017jsg,Ebert:2018lzn,Ebert:2018gsn} and explained in detail in \cite{Ebert:2018lzn}, the power corrections to the Born measurements contribute significantly to the power correction of the differential distribution. In particular they introduce new non-perturbative functions, namely derivatives of the Parton Distribution Functions (PDFs), which do not appear at leading power.
We will see a detailed example of this in \sec{qT} when deriving the perturbative power corrections to the Higgs and DY transverse momentum spectra \cite{Ebert:2018gsn} at the LHC.

Since we are interested in deriving factorization to $\cO(\lambda^2)$, and the first subleading power correction to the measurement function appears at $\cO(\lambda^2)$, contributions to the cross section whose power suppression arises from the measurement functions can be factorized just like at leading power by using the BPS field redefinition. Any insertion of subleading power Lagrangians or hard scattering operators would lead to further power suppression.

\subsection{Factorization with Lagrangian Insertions}\label{sec:subl_insert}

The most non-trivial aspect of subleading power factorization is the factorization of the subleading power Lagrangians. At leading power this is achieved in SCET through the BPS field redefinition, however, the BPS field redefinition does not decouple soft and collinear interactions beyond leading power. The Lagrangian governing the dynamics of the effective theory has the power expansion
\begin{align} 
\cL_\dyn=\sum_{i\geq0} \cL^{(i)} \,.
\end{align}
When working to any fixed power in $\lambda$ only a finite number of insertions of $\cL^{(i)}$, $i\geq1$ need to be considered. Explicitly, if we consider a time-ordered product ($T$-product) in the effective theory and we are interested in its expansion to $\cO(\lambda^2)$, we have
{\small
\begin{align}
&  \langle 0| T\{ \tilde O_j^{(k)}(0) \text{exp}[i \mbox{$\int$} d^4x \cL_\dyn]  \} |X \rangle        \\
&= \langle 0| T\{ \tilde O_j^{(k)}(0)  \text{exp}[i \mbox{$\int$} d^4x (\cL^{(0)} +\cL^{(1)} +\cL^{(2)}+\cdots )        ]  \} |X\rangle    \nn \\
&=\left \langle 0 \left| T\bigg\{ \tilde O_j^{(k)}(0)  \text{exp}[i \mbox{$\int$} d^4x \cL^{(0)}] \bigg (1 \! +\!  i \mbox{$\int$}d^4y\cL^{(1)}\!+\! \frac{1}{2}    \big( i\mbox{$\int$} d^4y\cL^{(1)} \big)  \big( i\mbox{$\int$} d^4z \cL^{(1)} \big) \!+\! i\mbox{$\int$} d^4z\cL^{(2)}   \!+\! \cdots  \bigg)          \bigg\} \right |X \right\rangle   \nn \\
&=  \left \langle 0 \left | T\bigg\{ \tilde O_j^{(k)}(0)   \bigg(1+i\mbox{$\int$} d^4y\cL^{(1)}+\frac{1}{2} \big( i\mbox{$\int$} d^4y\cL^{(1)} \big)  \big( i\mbox{$\int$} d^4z \cL^{(1)} \big) +i\mbox{$\int$} d^4z\cL^{(2)} \bigg)           \bigg\} \right |X \right\rangle_{\cL^{(0)}} +\cdots\,,  \nn
\end{align}}
where the dots represent higher power corrections.
In the final expression  all matrix elements are evaluated using the leading power SCET Lagrangian, and the subleading power Lagrangians appear only a finite number of times. From now on we will drop the ${\cal L}^{(0)}$ subscript. This expression highlights that to achieve factorization of the dynamics at any finite power in the power expansion, it is sufficient to show a decoupling of the leading power interactions. The insertions of the subleading power Lagrangians in the matrix elements will lead to the radiative functions which are the focus of \chap{radiative}.

The leading power interactions of ultrasoft and collinear degrees of freedom can be decoupled at the Lagrangian level using the BPS field redefinition of \Eq{eq:BPSfieldredefinition_v2}. After the BPS field redefinition, the leading power SCET Lagrangian decomposes as
\begin{align}
\cL^{(0)}= \sum_{n_i} \cL^{(0)}_{n_i} +\cL^{(0)}_{us}\,,
\end{align}
where the sum is over distinct collinear sectors. Since the leading power Lagrangian defines the time evolution,
states in the Hilbert space can then also be factorized as
\begin{align}
|X\r =|X_n \r |X_\bn \r |X_{us} \r\,.
\end{align}
Here we work in the interaction picture defined by the leading power Lagrangian, and considering perturbations in the power expansion. Note that these perturbations are in $\lambda$, unlike the interaction picture defining the perturbative expansion in $\alpha_s$. Here corrections in $\alpha_s$ are kept to all orders. 

This allows hard-soft-collinear factorization to be achieved to any power in the effective field theory. Deriving the explicit structure of the factorization in the case of subleading power Lagrangian insertions will be the main focus of \chap{radiative}, and will give rise to radiative functions.

\subsection{Factorized Cross Section to $\cO(\lambda^2)$}\label{sec:sum_fact}

To show how the different sources of power corrections in the effective theory presented so far enter the calculation of a physical observable, we will now combine these contributions to achieve a homogenous power expansion for the $e^+e^-\to$ dijet cross section and give expressions at each order in the power expansion in terms of matrix elements of hard scattering operators, Lagrangian insertions, and measurement functions. Here we consider only the terms which arise up to $\mathcal{O}(\lambda^2)$.  At $\mathcal{O}(\lambda^0)$ we have the simple expression in terms of the different helicity configurations of the  leading power operator
\begin{align}\label{eq:xsec_lam0}
\frac{d\sigma}{d\tau}^{(0)}\!\! = N  \sum_X    \tilde \delta^{(4)}_q \sum_{\lambda_l}    
 \bra{0}   \sum_{\lambda_q} C^{(0)}_{(\lambda_l;\lambda_q)} \tO^{(0)\dg}_{(\lambda_q)}(0) \ket{X}
 \bra{X} \sum_{\lambda_q}  C^{(0)}_{(\lambda_l;\lambda_q)}  \tO^{(0)}_{(\lambda_q)}(0) \ket{0} 
 \, \delta\big( \tau - \tau^{(0)}(X) \big)\,.
\end{align}
Since all matrix elements are now evaluated with the leading power Lagrangian, there are no interactions between soft and collinear degrees of freedom, and the factorization into collinear and soft matrix elements is simply an algebraic exercise, leading to the well known factorization for the thrust observable \cite{Korchemsky:1999kt,Fleming:2007qr,Schwartz:2007ib}. We will review this factorization in more detail in \Sec{sec:fact_RadiativeFunction}.

At $\mathcal{O}(\lambda^1)$ we have potential contributions from $\mathcal{O}(\lambda^1)$ hard scattering operators, as well as subleading Lagrangian insertions,
{\begin{small}
\begin{align}\label{eq:xsec_lam_RF}
\frac{d\sigma}{d\tau}^{(1)} &= N \sum_{X,i}  \tilde \delta^{(4)}_q  \bra{0} C_i^{(1)*} \tO_i^{(1) \dagger}(0) \ket{X}\bra{X} C^{(0)} \tO^{(0)}(0) \ket{0}   \delta\big( \tau - \tau^{(0)}(X) \big) +\text{h.c.} 
 \\
&\quad+ N \sum_X    \tilde \delta^{(4)}_q  \int d^4x \bra{0}  \ATO \big(-i\cL^{(1)}(x) \b
ig)  C^{(0)*} \tO^{(0)\dagger}(0) \ket{X}\bra{X}  C^{(0)} \tO^{(0)}0) \ket{0} \delta\big( \tau - \tau^{(0)}(X) \big)+\text{h.c.} \nn \\
&=0
\,. \nn
\end{align}
\end{small}}%
where here and in the following $\TO$($\ATO$) denotes (anti-)time ordering.
The vanishing of $\frac{d\sigma}{d\tau}^{(1)}$ from hard scattering operators was explained for thrust in \cite{Feige:2017zci}, and in \Sec{sec:RF_thrust} we will discuss the analogous explanation for the vanishing of Lagrangian insertion contributions at this order

At $\mathcal{O}(\lambda^2)$ all three sources of power corrections contribute
\begin{align}
\frac{d\sigma}{d\tau}^{(2)} =\frac{d\sigma}{d\tau}^{(2),\text{hard}} +\frac{d\sigma}{d\tau}^{(2),\text{measure}}+\left(\frac{d\sigma}{d\tau}^{(2),\text{radiative}}+\frac{d\sigma}{d\tau}^{(2),\text{mixed}}\right)\,,
\end{align}
or more explicitly
{\begin{footnotesize}
\begin{align}\label{eq:xsec_lam2_RF}
&\frac{d\sigma}{d\tau}^{(2)} = N \sum_{X,i}  \tilde \delta^{(4)}_q  \bra{0} C_i^{(2)*} \tO_i^{(2)\dagger}(0) \ket{X}\bra{X} C^{(0)} \tO^{(0)}(0) \ket{0} \delta\big( \tau - \tau^{(0)}(X) \big) +\text{h.c.} \\
&+ N \sum_{X,i,j}  \tilde \delta^{(4)}_q  \bra{0} C_i^{(1)*} \tO_i^{(1)\dagger}(0) \ket{X}\bra{X} C_j^{(1)} \tO_j^{(1)}(0) \ket{0}   \delta\big( \tau - \tau^{(0)}(X) \big)   \nn\\
&- N \sum_X  \tilde \delta^{(4)}_q \bra{0} C^{(0)*} \tO^{(0)\dagger}(0) \ket{X}\bra{X}  C^{(0)} \tO^{(0)}(0) \ket{0}   \tau^{(2)}(X)\, \delta'\big( \tau - \tau^{(0)}(X) \big)\nn\\
&+ N \sum_{X,i}  \tilde \delta^{(4)}_q   \int d^4x  \bra{0} C_i^{(1)*} \tO_i^{(1)\dagger}(0) \ket{X}\bra{X}\TO ( i \cL^{(1)}(x)) C^{(0)} \tO^{(0)}(0) \ket{0}   \delta\big( \tau - \tau^{(0)}(X) \big)+\text{h.c.}    \nn\\
&+ N \sum_{X,i}  \tilde \delta^{(4)}_q   \int d^4x  \bra{0} \ATO( -i \cL^{(1)}(x)) C_i^{(1)*} \tO_i^{(1)\dagger}(0) \ket{X}\bra{X}  C^{(0)} \tO^{(0)}(0) \ket{0}   \delta\big( \tau - \tau^{(0)}(X) \big)+\text{h.c.}    \nn\\
&+ N \sum_X  \tilde \delta^{(4)}_q \int d^4x  \bra{0}  C^{(0)*} \tO^{(0)\dagger}(0) \ket{X}\bra{X} \TO( i \cL^{(2)}(x)) C^{(0)} \tO^{(0)}(0) \ket{0}   \delta\big( \tau - \tau^{(0)}(X) \big)  +\text{h.c.} \nn\\
&-\frac{N}{2} \sum_X    \tilde \delta^{(4)}_q  \int d^4x \int d^4y \bra{0}\ATO\cL^{(1)}(x) \cL^{(1)}(y)  C^{(0)*} \tO^{(0)\dagger}(0) \ket{X}\bra{X}  C^{(0)} \tO^{(0)}0) \ket{0}   \delta\big( \tau - \tau^{(0)}(X) \big)  +\text{h.c.} \nn\\
&-\frac{N}{2} \sum_X   \tilde \delta^{(4)}_q  \int d^4x  \int d^4y \bra{0}\ATO\cL^{(1)}(x)  C^{(0)*} \tO^{(0) \dagger}(0) \ket{X}\bra{X}\TO \cL^{(1)}(y) C^{(0)} \tO^{(0)}0) \ket{0}  \delta\big( \tau - \tau^{(0)}(X) \big) \,.\nn
\end{align}
\end{footnotesize}}%
Unlike the ${\cal O}(\lambda)$ power correction, the $\mathcal{O}(\lambda^2)$ correction to the cross section does not vanish. The $\mathcal{O}(\lambda^2)$ power correction for thrust was computed at fixed order to $\cO(\alpha_s)$ and to $\cO(\alpha_s^2)$ using SCET in \cite{Freedman:2013vya} and \cite{Moult:2016fqy,Moult:2017jsg}, respectively. Since the interactions between soft and collinear degrees of freedom have been decoupled, by algebraic manipulation of \Eq{eq:xsec_lam2_RF}, the contributions to the cross section at each order in the power expansion can be expressed as a sum of vacuum matrix elements, involving a measurement function insertion, and each containing only collinear $n$, collinear $\bar n$, or ultrasoft fields.  To do this, we write the constraint on the final state as a sum of the measurement operators
\begin{align}
\delta(\tau-\hat \tau)= \int d\tau_n d\tau_{\bar n} d\tau_{us} \delta(\tau-\tau_n -\tau_{\bar n}-\tau_{us}) \delta(\tau_n -\hat \tau_n) \delta(\tau_{\bar n} -\hat \tau_{\bar n}) \delta(\tau_{us}-\hat \tau_{us})\,.\nn
\end{align}
We can then perform the sum over the $|X_n\rangle \langle X_n|$,  $|X_{\bar n}\rangle \langle X_{\bar n}|$, and $|X_{us}\rangle \langle X_{us}|$ states to simplify all the matrix elements to vacuum matrix elements. The Lorentz, Dirac, and color structure can be simplified using Fierz relations, and the symmetry properties of the vacuum matrix elements, such that each matrix element is a scalar, and there are no index contractions between the soft and collinear functions, namely a completely factorized form. 

In order to better summarize all the ingredients presented so far in this section, we want to conclude by presenting pictorially the structure of the factorized cross section for thrust in $e^+e^-$ annihilation at next to leading power, \eq{xsec_lam2_RF}, in terms of leading and subleading hard, jet and soft functions
\begin{align}\label{eq:factorization_ee}
&\frac{1}{\sigma_0}\frac{d\sigma^{(2),e^+e^-}}{d\tau}=  \\
&\underbrace{\left [ \left| \fd{1.5cm}{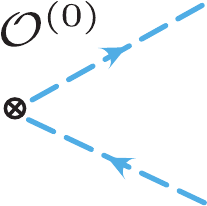} ~ \right|^2\cdot\int d r_2^+ \fd{3cm}{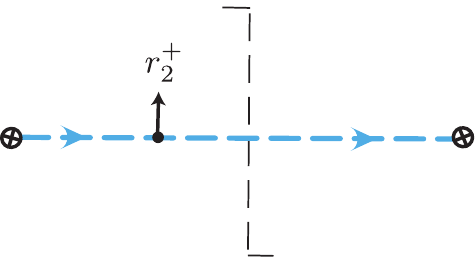} \otimes \fd{3cm}{figures_b/1quark_jetfunction_low}\otimes  \fd{3cm}{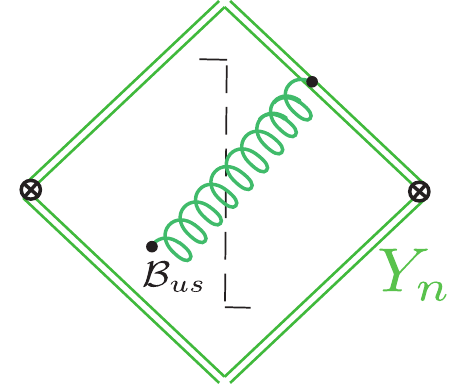} \right. }_{\text{Soft Gluon Correction}}\nn \\
+&\underbrace{ \left. \int d\omega_1 \Re\left ( \fd{1.5cm}{figures_b/collinear_category1_hard_low}\cdot \fd{1.5cm}{figures_b/2quark_hardfunc_low.pdf}^\dagger \right )  \otimes\fd{3cm}{figures_b/gq_jetfunction_low}\otimes \fd{3cm}{figures_b/1quark_jetfunction_low} \otimes \fd{3cm}{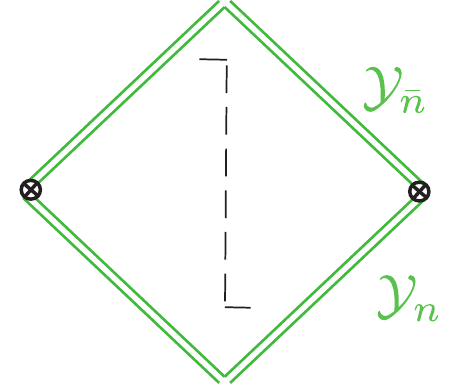} \right ]}_{\text{Collinear Gluon Correction}}\nn \\
+&\underbrace{ \left [  \left| \fd{1.5cm}{figures_b/2quark_hardfunc_low.pdf}~ \right|^2  \cdot   \int dr_2^+ dr_3^+ \fd{3cm}{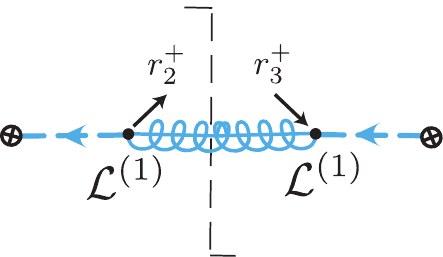}\otimes \fd{3cm}{figures_b/1quark_jetfunction_low} \otimes  \fd{3cm}{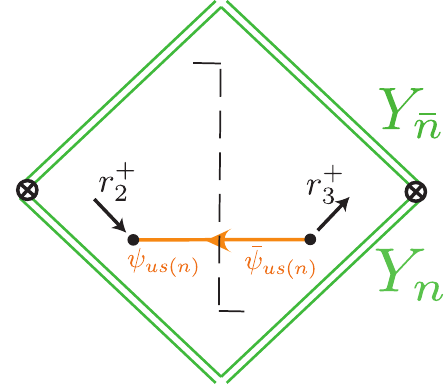}  \right. }_{\text{Soft Quark Correction}}\nn \\
 +&\underbrace{ \left. \int d \omega_1 d \omega_2 \left| \fd{1.5cm}{figures_b/collinear_category2_hard_low}~ \right|^2\otimes\fd{3cm}{figures_b/2quark_jetfunc_low}\otimes \fd{3cm}{figures_b/1gluon_jetfunc_low}\otimes \fd{3cm}{figures_b/eikonal_factor_gluon_low.pdf} \right]}_{\text{Collinear Quark Correction}}\,. \nn
\end{align}
In \eq{factorization_ee} we have included only the terms that contribute to this subleading power contribution at leading logarithmic order~\cite{Feige:2017zci,Moult:2019mog,Moult:2019uhz}. We can see that the first and third lines are terms where the power suppression comes from the presence of subleading hard scattering operators, i.e. the class of corrections presented in \sec{hard_fact}. On the other hand in the second and fourth line we have terms where the suppression comes from the insertion of subleading power Lagrangians giving rise to radiative functions. We will discuss in detail the factorization of the radiative function contributions appearing in \eq{factorization_ee} in \chap{radiative} and in particular in \sec{RF_thrust}.
\section{Factorization, Glaubers and Factorization Breaking}
\label{sec:glaubers}
The proof of cancellation of Glauber modes for a cross section is an open problem in QCD for many relevant cross sections.
It is an extremely difficult task since it must be carried out at all orders in the coupling constant. 
So far it has only been shown for inclusive color singlet cross sections and transverse momentum distributions\cite{Bodwin:1984hc,Collins:1984kg,Collins:1985ue,Collins:1988ig,Collins:1989gx,Collins:1350496} using direct QCD methods. 
SCET and its recent extension of SCET with Glaubers \cite{Rothstein:2016bsq} offers a promising framework to organize factorization proofs and identifying factorization breaking effects.
In this section we will clarity what can and what cannot break factorization even when considering QCD beyond leading power, by leveraging  the power of the SCET framework.

After the BPS field redefinition, the soft and collinear leading power dynamical Lagrangians decouple. 
Given the leading power decomposition of the SCET Lagrangian of \eq{SCETLagLP}, this implies that the only way to couple collinear and soft fields at leading power is via the Glauber Lagrangian $\cL_G^{(0)}$.
While it is true that both hard scattering operators as well as subleading Lagrangian insertions do couple soft and collinear fields beyond leading power, as explained in \sec{subl_insert}, they can do it only a finite number of times. 

In order to understand factorization in the presence of Glaubers and/or subleading power contributions, it is important to remember that factorization theorems for hadronic cross sections are well defined concepts only order by order in the power counting. This means that
one assigns a power counting to a scale entering the cross section, expands the cross section and thus obtains a homogeneous expansion such as \eq{xsec_homog}
\begin{align}\label{eq:xsec_homog_rep}
\frac{\df\sigma}{\df \tau} &=\frac{\df\sigma^{(0)}}{\df \tau} +\frac{\df\sigma^{(1)}}{\df \tau} +\frac{\df\sigma^{(2)}}{\df \tau}+\frac{\df\sigma^{(3)}}{\df \tau} +{\cal O}(\tau)\,.
\end{align}
One then derives separate factorization theorems for each term in this expansion.
Now, leading power factorization theorems in QCD for an IRC safe observable typically involve one term, such that it is sometimes referred as "the" factorization theorem for such observable. As we have seen in \eq{xsec_lam0}, the leading power factorization theorem for an event shape observable at a lepton collider takes the form
\begin{align} \label{eq:xsec_lam0_schem}
&\hspace{-0.25cm}\frac{\df\sigma^{(0)}}{\df \tau} \sim
 H J_n^{(0)} \otimes J_\bn^{(0)} \otimes S^{(0)}  
\,.\end{align}
However, the fact that the factorization theorem involves only one term is only due to the simplicity of QCD at leading power for this class of observables.
As we have seen in \sec{sub-fact}, beyond leading power this is no longer the case. The factorization theorem in \eq{xsec_lam2_RF} is a different and independent object w.r.t. the leading power one.

Now that we have clarified this simple but sometimes subtle point, it is now easy to understand that the insertion of whatever subleading power operator that couples soft and collinear fields automatically cannot be a source of factorization violation. 
This is because each insertion of such operator changes the order in the power counting and therefore simply generates a new term in \emph{another} factorization theorem, i.e. a factorization theorem at a different order in the power counting.

The uniqueness of the leading power Glauber Lagrangian $\cL_G^{(0)}$ is that it is the only Lagrangian that couples soft and collinear fields (as well as collinear fields in different collinear directions) at leading power. 
For this reason it is possible to insert an infinite number of soft and collinear interactions via $\cL^{(0)}_G$ while staying at the same order in the power counting therefore breaking factorization.
Therefore, it is clear why subleading power Glauber Lagrangians, i.e. subleading power operators involving Glauber potentials, again cannot break factorization. 
Carrying a power suppression, they cannot break factorization for the same reason that other subleading power operators do not. 
Hence, there is no particular reason to either treat or classify them differently from the rest of the subleading power dynamical Lagrangians.

The property of Glaubers to couple collinear and soft fields at leading power has to immediately worry us, as it invalidates the factorization of the Hilbert space of the theory which is one of the cornerstones of the factorization proofs.
The solution to this problem in SCET is to show that the Glauber operator contributions from $\cL^{(0)}_G$ (and those from $\cL^{(0)}_G$ only) cancel. 

It is important to notice that the fact that only the leading power $\cL^{(0)}_G$ can break factorization, does not imply that we should only prove that Glaubers cancel at leading power, i.e. for the leading power factorization theorem. 
The cancellation of Glaubers is a property of the factorization theorem (or more precisely of the ingredients contributing to the cross section at a given order in the power counting) and thus it has to be shown order by order in the power counting, as we have very different factorization theorems at each order. 
If we take a process and an observable in which all the insertions of $\cL^{(0)}_G$ cancel at leading power, it is definitely not guaranteed that this will continue to be true beyond leading power. Hence, one can have an observable that factorizes at leading power but doesn't factorize beyond leading power.
This is due to the fact that at each order in the power counting we have very different ingredients entering the factorization formulas. 
In particular, we have intrinsically new hard scattering operators as well as subleading Lagrangian insertions appearing beyond leading power and their $T-$products with arbitrary numbers of $\cL^{(0)}_G$ can in principle yield new Glauber interactions which are completely independent from the ones at leading power, thus requiring a dedicated analysis to prove whether they cancel or not.
This further motivates the study of QCD beyond leading power.

\chapter{A Subleading Operator Basis and Matching for $gg\to H$}\label{sec:gluon_ops}

\section{Introduction}\label{sec:intro_gluonops}

Factorization theorems play an important role in understanding the all orders behavior of observables in Quantum Chromodynamics (QCD).
While typically formulated at leading power, the structure of subleading power corrections is of significant theoretical and practical interest. 
A convenient formalism for studying factorization in QCD is the Soft Collinear Effective Theory (SCET) \cite{Bauer:2000ew, Bauer:2000yr, Bauer:2001ct, Bauer:2001yt}, an effective field theory describing the soft and collinear limits of QCD. SCET allows for a systematic power expansion in $\lambda \ll 1$ at the level of the Lagrangian, and simplifies many aspects of factorization proofs \cite{Bauer:2002nz}. 
SCET has been used to study power corrections at the level of the amplitude \cite{Larkoski:2014bxa} and to derive factorization theorems at subleading power for $B$ decays \cite{Lee:2004ja,Beneke:2004in,Hill:2004if,Bosch:2004cb,Beneke:2004rc,Paz:2009ut,Benzke:2010js}. 
More recently, progress has been made towards understanding subleading power corrections for event shape observables \cite{Freedman:2013vya,Freedman:2014uta,Moult:2016fqy,Feige:2017zci}.

In this chapter, we focus on the power suppressed hard scattering operators describing the gluon initiated production (or decay) of a color singlet scalar. 
We present a complete operator basis to $\cO(\lambda^2)$ in the SCET power expansion using operators of definite helicity \cite{Moult:2015aoa,Kolodrubetz:2016uim,Feige:2017zci}, and discuss how helicity selection rules simplify the structure of the basis. We also classify all operators which can contribute at the cross section level at $\cO(\lambda^2)$, and discuss the structure of interference terms between different operators in the squared matrix element.
We then perform the tree level matching onto our operators. 
These results can be used to study subleading power corrections either in fixed order, or resummed perturbation theory, and compliment our recent analysis for the case of $q\bar q$ initiated production \cite{Feige:2017zci}.

We will consider the production of a color singlet final state, which we take for concreteness to be the Higgs, with the underlying hard Born process
\begin{equation} \label{eq:interaction}
g_a(q_a)\, g_b(q_b) \to H(q_1)
\,,\end{equation}
where $g_{a,b}$ denote the colliding gluons, and $H$ the outgoing Higgs particle. 
We work in the Higgs effective theory, with an effective Higgs gluon coupling
\begin{align}\label{eq:heft_intro}
\cL_{\text{hard}}=\frac{C_1(m_t, \alpha_s)}{12\pi v}G^{\mu \nu}G_{\mu \nu} H\,,
\end{align}
obtained from integrating out the top quark. Here $v=(\sqrt{2} G_F)^{-1/2}=246$ GeV, and the matching coefficient is known to $\cO(\alpha_s^3)$ \cite{Chetyrkin:1997un}.

The active-parton exclusive jet cross section corresponding to \eq{interaction} can be proven to factorize for a variety of jet resolution variables. 
For concreteness we will take the case of beam thrust, $\tau_B$. 
The leading power factorized expression for the beam thrust cross section can be written schematically in the form \cite{Stewart:2010tn}
\begin{align} \label{eq:sigma}
\frac{\df\sigma^{(0)}}{\df \tau_B} &=
\int\!\df x_a\, \df x_b\, \df \Phi(q_a \!+ q_b; q_1)\, M(\{q_1\})\
 H_g^{(0)}(\{q_i\})\: 
\Bigl[ B_g^{(0)} B_g^{(0)}  \Bigr]\otimes S_g^{(0)}  
\,,\end{align}
where the $x_{a,b}$ denote the momentum fractions of the incoming partons, $\df \Phi$ denotes the Lorentz-invariant phase space for the Born process in \eq{interaction}, and $M(\{q_i\})$ denotes the measurement made on the color singlet final state.
\footnote{By referring to active-parton factorization we imply that this formula ignores contributions from proton spectator interactions~\cite{Gaunt:2014ska} that occur through the Glauber Lagrangian of Ref.~\cite{Rothstein:2016bsq}. There are also perturbative corrections at ${\cal O}(\alpha_s^4)$ that are described by a single function $B_{gg}$ in place of $B_gB_g$~\cite{Zeng:2015iba,Rothstein:2016bsq}.}
The dependence on the underlying hard interaction is encoded in the hard function $\hH(\{q_i\})$ and the trace is over color.  
The soft function $\hS$ describes soft radiation, and the beam functions $B_i$ describe energetic initial-state radiation along the beam directions \cite{Stewart:2009yx}. 
The factorization theorem of \Eq{eq:sigma} allows logarithms of $\tau_B$ to be resummed to all orders through the renormalization group evolution of the hard, beam and soft functions.

The factorization formula in \Eq{eq:sigma} captures all terms in the cross section scaling as $\tau_B^{-1}$, including delta function terms. More generally the cross section can be expanded in powers of $\tau_B$ as,
\begin{align}\label{eq:cross_expand}
\frac{\df\sigma}{\df\tau_B} &=\frac{\df\sigma^{(0)}}{\df\tau_B} +\frac{\df\sigma^{(1)}}{\df\tau_B} +\frac{\df\sigma^{(2)}}{\df\tau_B}+\frac{\df\sigma^{(3)}}{\df\tau_B} +{\cal O}(\tau)\,.
\end{align}
Here the superscript refers to the suppression in powers of $\sqrt{\tau_B}$ relative to the leading power cross section.
This particular convention is chosen due to the power expansion in SCET, where one typically takes the SCET power counting parameter $\lambda$ to scale like $\lambda^2 \sim \tau_B$. 
Odd orders in \Eq{eq:cross_expand} are expected to vanish, and we will show this explicitly for $\df\sigma^{(1)}/\df\tau_B$. 
The first non-vanishing power correction to the cross section then arises from $\df\sigma^{(2)}/\df\tau_B$, which contains all terms that scale like $\cO(\tau_B^0)$.  

It is generally expected that the power corrections in \Eq{eq:cross_expand} obey a factorization formula similar to that of \Eq{eq:sigma}. 
Schematically, 
\begin{align} \label{eq:sigma_sub}
&\hspace{-0.25cm}\frac{\df\sigma^{(n)}}{\df\tau_B} =
\int\!\df x_a\, \df x_b\, \df \Phi(q_a \!+ q_b; q_1)\,M(\{q_1\})\
\sum_{j}   H^{(n_{Hj})}_{j} \otimes 
  \Big[ B^{(n_{Bj})}_{j}  B^{(n'_{Bj})}_{j}\Big] \otimes S_j^{(n_{Sj})}  
,\end{align}
where $j$ sums over the multiple contributions that appear at each order, $n_{Hj}+n_{Bj}+n_{Bj}'+n_{Sj}=n$, and $\otimes$ denotes a set of convolutions, whose detailed structure has not been specified and is known to be more complicated than typical leading power factorization theorems.
We also let $\otimes$ include nontrivial color contractions. 
The derivation of such a formula would enable for the resummation of subleading power logarithms using the renormalization group evolution of the different functions appearing in \Eq{eq:sigma_sub}, allowing for an all orders understanding of power corrections to the soft and collinear limits.

To derive a factorization theorem in SCET, QCD is matched onto SCET, which consists of hard scattering operators in $\cL_\hard$ and a Lagrangian $\cL_\dyn$ describing the dynamics of soft and collinear radiation
\begin{align} 
\cL_{\text{SCET}}=\cL_\hard+\cL_\dyn \,.
\end{align}
The dynamical Lagrangian can be divided into two parts
\begin{align}
\cL_\dyn=\cL_{\text{fact}}+\cL_{G}^{(0)} \,.
\end{align}
Here $\cL_{G}^{(0)}$ is the leading power Glauber Lagrangian determined in Ref.~\cite{Rothstein:2016bsq} which couples together soft and collinear fields in an apriori non-factorizable manner, and $\cL_{\text{fact}}$ includes both the leading interactions which can be factorized into independent soft and collinear Lagrangians, and subleading power interactions which are factorizable as products of soft and collinear fields.
Our focus here is on determining the subleading power $\cL_\hard$ for $gg\to H$, and $\cL_\dyn$ only plays a minor role when we carry out explicit matching calculations (and $\cL_{G}^{(0)}$ does not play a role at all since these matching calculations are tree level). 

The hard scattering operators are process dependent, while the Lagrangian $\cL_\dyn$ is universal and the relevant terms for our analysis are known in SCET to $\cO(\lambda^2)$ in the power expansion \cite{Manohar:2002fd,Chay:2002vy,Beneke:2002ni,Beneke:2002ph,Pirjol:2002km,Bauer:2003mga}. 
A field redefinition can be performed in the effective theory \cite{Bauer:2002nz} which allows for the decoupling of leading power soft and collinear interactions in $\cL_{\text{fact}}$. 
If $\cL_{G}^{(0)}$ is proven to be irrelevant, then the Hilbert spaces for the soft and collinear dynamics are factorized, and a series of algebraic manipulations can be used to write the cross section as a product of squared matrix elements, each involving only collinear or soft fields. 
This provides a field theoretic definition of each of the functions appearing in \Eq{eq:sigma_sub} in terms of hard scattering operators and Lagrangian insertions in SCET. 
Since the Lagrangian insertions are universal, the remaining ingredient which is required to derive a subleading power factorization theorem for the $gg\to H$ process is a complete basis of subleading power hard scattering operators.
The derivation of a basis, which is the goal of this chapter, provides the groundwork for a systematic study of power corrections for color singlet production through gluon fusion.

An outline of this chapter is as follows. 
In \Sec{sec:review_gluonops} we provide a brief review of SCET and of the helicity building blocks required for constructing subleading operators in SCET. In \Sec{sec:basis} we present a complete basis of operators to $\cO(\lambda^2)$ for the gluon initiated production of a color singlet, and carefully classify which operators can contribute to the cross section at $\cO(\lambda^2)$. 
In \Sec{sec:matching} we perform the tree level matching to the relevant operators. 
We conclude and discuss directions for future study in \Sec{sec:conclusions_gluon_ops}.

\section{Helicity Operators in SCET}\label{sec:review_gluonops}

In this section we review the use of helicity operators in SCET, while for more general features of SCET we refer the reader to \chap{scet}.
Reviews of SCET can be found in \Refs{iain_notes,Becher:2014oda}, and more detailed discussions on the use of helicity operators can be found in \Refs{Moult:2015aoa,Kolodrubetz:2016uim,Feige:2017zci}.

\subsection{SCET}\label{sec:review_scet}

SCET is an effective field theory of QCD describing the interactions of collinear and soft particles in the presence of a hard interaction \cite{Bauer:2000ew, Bauer:2000yr, Bauer:2001ct, Bauer:2001yt, Bauer:2002nz}. 
Collinear particles  are characterized by a large momentum along a particular light-like direction, while soft particles are characterized by having a small momentum with homogenous scaling of all its components. 
For each jet direction present in the problem we define two light-like reference vectors $n_i^\mu$ and $\bn_i^\mu$ such that $n_i^2 = \bn_i^2 = 0$ and $n_i\sdt\bn_i = 2$. 
We can then write any four-momentum $p$ as
\begin{equation} \label{eq:lightcone_dec}
p^\mu = \bn_i\sdt p\,\frac{n_i^\mu}{2} + n_i\sdt p\,\frac{\bn_i^\mu}{2} + p^\mu_{n_i\perp}\
\,.\end{equation}
A particle with momentum $p$ close to the $\vec{n}_i$ direction will be referred to as $n_i$-collinear. 
In lightcone coordinates its momenta scale like $(n_i\!\cdot\! p, \bn_i \!\cdot\! p, p_{n_i\perp}) \sim \bn_i\!\cdot\! p$ $\,(\la^2,1,\la)$. 
Here $\la \ll 1$ is a formal power counting parameter determined by the measurements or kinematic restrictions imposed on the QCD radiation. 
The choice of reference vectors is not unique, and any two reference vectors, $n_i$ and $n_i'$, with $n_i\cdot n_i' \sim \ord{\lambda^2}$ describe the same physics. 
The freedom in the choice of $n_i$ is represented in the effective theory as a symmetry known as reparametrization invariance (RPI) \cite{Manohar:2002fd,Chay:2002vy}. 
More explicitly, there are three classes of RPI transformations under which the EFT is invariant
\begin{alignat}{3}\label{eq:RPI_def}
&\text{RPI-I} &\qquad &  \text{RPI-II}   &\qquad &  \text{RPI-III} \nn \\
&n_{i \mu} \to n_{i \mu} +\Delta_\mu^\perp &\qquad &  n_{i \mu} \to n_{i \mu}   &\qquad & n_{i \mu} \to e^\alpha n_{i \mu} \nn \\
&\bar n_{i \mu} \to \bar n_{i \mu}  &\qquad &  \bar n_{i \mu} \to \bar n_{i \mu} +\epsilon_\mu^\perp  &\qquad & \bar n_{i \mu} \to e^{-\alpha} \bar n_{i \mu}\,.
\end{alignat}
The transformation parameters are assigned the power counting $\Delta^\perp \sim \lambda$, $\epsilon^\perp \sim \lambda^0$, and $\alpha\sim \lambda^0$. 
Additionally, while $\alpha$ can be a finite parameter, the parameters $\Delta^\perp$ and $\epsilon^\perp$ are infinitesimal, and satisfy $n_i\cdot \Delta^\perp=\bar n_i\cdot \Delta^\perp=n_i \cdot \epsilon^\perp=\bar n_i \cdot \epsilon^\perp=0$. 
RPI symmetries can be used to relate operators at different orders in the power expansion, and will be used in this chapter to relate the Wilson coefficients of several subleading power operators to the leading power Wilson coefficients for the $gg\to H$ process.
Furthermore, the RPI-III symmetry will constrain the form of the Wilson coefficients of our subleading power operators. 
At tree level the Wilson coefficients are simply rational functions of the large momentum components of the fields appearing in the operator, which must satisfy the rescaling symmetries of RPI-III.

SCET is constructed by decomposing momenta into label and residual components
\begin{equation} \label{eq:label_dec}
p^\mu = \lp^\mu + k^\mu = \bn_i \sdt\lp\, \frac{n_i^\mu}{2} + \lp_{n_i\perp}^\mu + k^\mu\,.
\,\end{equation}
The momenta $\bn_i \cdot\lp \sim Q$ and $\lp_{n_i\perp} \sim \la Q$ are referred to as the label components, where $Q$ is a typical scale of the hard interaction, while $k\sim \la^2 Q$ is a small residual momentum describing fluctuations about the label momentum.
Fields with momenta of definite scaling are obtained by performing a multipole expansion. 
Explicitly, the effective theory consists of collinear quark and gluon fields for each collinear direction, as well as soft quark and gluon fields. 
Independent gauge symmetries are enforced for each set of fields, which have support for the corresponding momenta carried by that field \cite{Bauer:2003mga}. The leading power gauge symmetry is exact, and is not corrected at subleading powers.

In SCET, fields for $n_i$-collinear quarks and gluons, $\xi_{n_i,\lp}(x)$ and $A_{n_i,\lp}(x)$, are labeled by their collinear direction $n_i$ and their large momentum $\lp$. 
The collinear fields are written in a mixed representation, namely they are written in position space with respect to the residual momentum and in momentum space with respect to the large momentum components. 
Derivatives acting on collinear fields give the residual momentum dependence, which scales as $i \partial^\mu \sim k \sim \la^2 Q$, whereas the label momentum operator $\cP^\mu$ gives the label momentum component. 
It acts on a collinear field as $\cP^\mu\, \xi_{n_i,\lp} = \lp^\mu\, \xi_{n_i,\lp}$. Note that we do not need an explicit $n_i$ label on the label momentum operator, since it is implied by the field that the label momentum operator is acting on. 
We will use the shorthand notation $\bnP = \bn_i\sdt\cP$.  
We will often suppress the explicit momentum labels on the collinear fields, keeping only the label of the collinear sector, ${n_i}$. 
Of particular relevance for the construction of subleading power operators is the $\cP_\perp$ operator, which identifies the $\cO(\lambda)$ perp momenta between two collinear fields within a collinear sector.

Soft degrees of freedom are described in SCET by quark and gluon fields $q_{us}(x)$ and $A_{us}(x)$.
In this chapter we will restrict ourselves to the SCET$_\text{I}$ theory where the soft degrees of freedom are referred to as ultrasoft so as to distinguish them from the soft modes of SCET$_\text{II}$ \cite{Bauer:2002aj}. 
The operators we construct are also applicable in the SCET$_\text{II}$ theory, but additional soft operators would be required. 
For a more detailed discussion see \Ref{Feige:2017zci}. 
The ultrasoft fields carry residual momenta, $i \partial^\mu \sim \la^2Q$, but do not carry label momenta, since they are not associated with any collinear direction. 
Correspondingly, they also do not carry a collinear sector label. 
The ultrasoft fields are able to exchange residual momenta between distinct collinear sectors while remaining on-shell.

SCET is constructed such that manifest power counting in the expansion parameter $\la$ is maintained at every stage of a calculation. 
All fields have a definite power counting \cite{Bauer:2001ct}, shown in \Tab{tab:PC}, and the SCET Lagrangian is expanded as a power series in $\lambda$
\begin{align} \label{eq:SCETLagExpand}
\cL_{\text{SCET}}=\cL_\hard+\cL_\dyn= \sum_{i\geq0} \cL_\hard^{(i)}+ 
 {\cal L}_G^{(0)} + \sum_{i\geq0} \cL^{(i)} \,.
\end{align}
Here $(i)$ denotes objects at ${\cal O}(\lambda^i)$ in the power counting. 
The Lagrangians $ \cL_\hard^{(i)}$ contain the hard scattering operators $O^{(i)}$, and are determined by an explicit matching calculation. 
The hard scattering operators encode all process dependence, while the $\cL^{(i)}$ describe the dynamics of ultrasoft and collinear modes in the effective theory, and are universal. 
The terms we need are explicitly known to $\mathcal{O}(\lambda^2)$, and can be found in a summarized form in \cite{iain_notes}. 
Finally, ${\cal L}_G^{(0)} $ is the leading power Glauber Lagrangian \cite{Rothstein:2016bsq}, which describes the leading power coupling of soft and collinear degrees of freedom through potential operators.

\begin{table}
\begin{center}
\begin{tabular}{| l | c | c |c |c|c| r| }
  \hline                       
  Operator & $\cB_{n_i\perp}^\mu$ & $\chi_{n_i}$& $\cP_\perp^\mu$&$q_{us}$&$D_{us}^\mu$ \\
  Power Counting & $\lambda$ &  $\lambda$& $\lambda$& $\lambda^3$& $\lambda^2$ \\
  \hline  
\end{tabular}
\end{center}
\caption{
Power counting for building block operators in $\text{SCET}_\text{I}$.
}
\label{tab:PC}
\end{table}

In this chapter we will be interested in subleading power hard scattering operators, in particular, $\cL_\hard^{(1)}$ and $\cL_\hard^{(2)}$.
The hard effective Lagrangian at each power is given by a product of hard scattering operators and Wilson coefficients,
\begin{align} \label{eq:Leff_sub_explicit}
\cL^{(j)}_{\text{hard}} = \sum_{\{n_i\}} \sum_{A,\cdot\cdot} 
  \bigg[ \prod_{i=1}^{\ell_A} \int \! \! \df \omega_i \bigg] \,
& \vO^{(j)\dagger}_{A\,\lotsdots}\big(\{n_i\};
   \omega_1,\ldots,\omega_{\ell_A}\big) \nn\\
& \times
\vC^{(j)}_{A\,\lotsdots}\big(\{n_i\};\omega_1,\ldots,\omega_{\ell_A} \big)
\,.
\end{align}
The appropriate collinear sectors $\{n_i\}$ are determined by directions found in the collinear states of the hard process being considered. 
If there is a direction $n_1'$ in the state then we sum over the cases where each of $n_1$, $\ldots$, $n_4$ is set equal to this $n_1'$.\footnote{Technically the $n_i$ in $\{n_i\}$ are representatives of an equivalence class determined by demanding that distinct classes $\{n_i\}$ and $\{n_j\}$ have $n_i\cdot n_j\gg \lambda^2$.} 
The sum over $A,\cdot\cdot$ in \eq{Leff_sub_explicit} runs over the full basis of operators that appear at this order, which are specified by either explicit labels $A$ and/or helicity labels $\cdot\cdot$ on the operators and coefficients.  
The $\vC^{(j)}_{A}$ are also vectors in the color subspace in which the $\mathcal{O}(\lambda^j)$ hard scattering operators $\vec O_A^{(j)\dagger}$ are decomposed.
Explicitly, in terms of color indices, we follow the notation of \Ref{Moult:2015aoa} and have
\begin{align} \label{eq:Opm_color}
\vO^\dagger_\lotsdots  &= O_\lotsdots^{a_1\dotsb \alpha_n}\, \vT^{\, a_1\dotsb \alpha_n}
 \,, \nn\\
C_{\lotsdots}^{a_1\dotsb\alpha_n}
 &= \sum_k C_{\lotsdots}^k T_k^{a_1\dotsb\alpha_n}
\equiv \vT^{ a_1\dotsb\alpha_n} \vC_{\lotsdots}
\,.\end{align}
Here $\vT^{\, a_1\dotsb\alpha_n}$ is a row vector of color structures that spans the color conserving subspace. 
The $a_i$ are adjoint indices and the $\alpha_i$ are fundamental indices. 
The color structures do not necessarily have to be independent, but must be complete. 

Hard scattering operators involving collinear fields are constructed out of products of fields and Wilson lines that are invariant under collinear gauge transformations~\cite{Bauer:2000yr,Bauer:2001ct}. 
The field building blocks for these operators are collinear gauge-invariant quark and gluon fields, defined as
\begin{align} \label{eq:chiB}
\chi_{{n_i},\w}(x) &= \Bigl[\delta(\w - \bnP_{n_i})\, W_{n_i}^\dagger(x)\, \xi_{n_i}(x) \Bigr]
\,,\\
\cB_{{n_i}\perp,\w}^\mu(x)
&= \frac{1}{g}\Bigl[\delta(\w + \bnP_{n_i})\, W_{n_i}^\dagger(x)\,i  D_{{n_i}\perp}^\mu W_{n_i}(x)\Bigr]
 \,. \nn
\end{align}
For this particular definition of $\chi_{{n_i},\w}$, we have $\w > 0$ for an incoming quark and $\w < 0$ for an outgoing antiquark. 
For $\cB_{{n_i},\w\perp}$, $\w > 0$ ($\w < 0$) corresponds to outgoing (incoming) gluons.
The covariant derivative in \eq{chiB} is given by,
\begin{equation}
i  D_{{n_i}\perp}^\mu = \cP^\mu_{{n_i}\perp} + g A^\mu_{{n_i}\perp}\,,
\end{equation}
and the collinear Wilson line is defined as
\begin{equation} \label{eq:Wn}
W_{n_i}(x) = \biggl[~\sum_\text{perms} \exp\Bigl(-\frac{g}{\bnP_{n_i}}\,\bn\sdt A_{n_i}(x)\Bigr)~\biggr]\,.
\end{equation}
The emissions summed in the Wilson lines are $\ord{\lambda^0}$ in the power counting. 
The square brackets indicate that the label momentum operators act only on the fields in the Wilson line. 
The collinear Wilson line, $W_{n_i}(x)$, is localized with respect to the residual position $x$, so that
$\chi_{{n_i},\w}(x)$ and $\cB_{{n_i},\w}^\mu(x)$ can be treated as local quark and gluon fields from the perspective of the ultrasoft degrees of freedom. 

All operators in the theory must be invariant under ultrasoft gauge transformations.
Collinear fields transform under ultrasoft gauge transformations as background fields of the appropriate representation. 
Dependence on the ultrasoft degrees of freedom enters the operators through the ultrasoft quark field $q_{us}$, and the ultrasoft covariant derivative $D_{us}$, defined as 
\begin{equation}
i  D_{us}^\mu = i  \partial^\mu + g A_{us}^\mu\,.
\end{equation}
Other operators, such as the ultrasoft gluon field strength, can be constructed from the ultrasoft covariant derivative. The power counting for these operators is shown in \Tab{tab:PC}.

The complete set of collinear and ultrasoft building blocks is summarized in \Tab{tab:PC}. 
These can be combined, along with Lorentz and Dirac structures, to construct a basis of hard scattering operators at any order in the SCET power counting.
All other field and derivative combinations can be reduced to this set by the use of equations of motion and operator relations~\cite{Marcantonini:2008qn}.
As shown in \Tab{tab:PC}, both the collinear quark and collinear gluon building block fields scale as ${\cal O}(\lambda)$. 
Therefore, while for most jet processes only a single collinear field appears in each sector at leading power, subleading power operators can involve multiple collinear fields in the same collinear sector, as well as $\cP_\perp$ insertions. 
The scaling of an operator is simply obtained by adding up the powers for the building blocks it contains. 
This implies that at higher powers hard scattering operators involve more and more fields, or derivative insertions, leading to any increasingly complicated structure.
Furthermore, to ensure that the effective theory completely reproduces all IR limits of the full theory, as well as to guarantee that the renormalization group evolution of the operators is closed, it is essential that operator bases in SCET are complete, namely all operators consistent with the symmetries of the problem must be included. 
Enumerating a minimal basis of operators becomes difficult at subleading power, and it is essential to be able to efficiently identify independent operators, as well as to make manifest all symmetries of the problem.

\subsection{Helicity Operators}\label{sec:review_helicity}

An efficient approach to simplify operator bases in SCET is to use operators of definite helicity \cite{Moult:2015aoa,Kolodrubetz:2016uim,Feige:2017zci}. 
This general philosophy is well known from the study of on-shell scattering amplitudes, where it leads to compact expressions, removes gauge redundancies, and makes symmetries manifest. 
The use of helicities is also natural in SCET since the effective theory is formulated as an expansion about identified light like directions with respect to which helicities are naturally defined, and collinear fields carry these directions as labels. 
Furthermore, since SCET is formulated in terms of collinear gauge invariant fields, see \Eq{eq:chiB}, one can naturally project onto physical polarizations.
SCET helicity operators were introduced in \cite{Moult:2015aoa} where they were used to study leading power processes with high multiplicities.
This was extended to subleading power in \cite{Kolodrubetz:2016uim} where it was shown that the use of helicity operators is also convenient when multiple fields appear in the same collinear sector.
In this section we briefly review SCET helicity operators, since we will use them to simplify the structure of the subleading power basis for $gg\to H$.
We will follow the notation and conventions of~\cite{Moult:2015aoa,Kolodrubetz:2016uim,Feige:2017zci}.
A summary of the complete set of operators that we will use is given in \Tab{tab:helicityBB}. 

We define collinear gluon and quark fields of definite helicity as
\begin{subequations}
	\label{eq:cBpm_quarkhel_def}
\begin{align} 
\label{eq:cBpm_def}
\cB^a_{i\pm} &= -\ve_{\mp\mu}(n_i, \bn_i)\,\cB^{a\mu}_{n_i\perp,\w_i}
\,, \\
\label{eq:quarkhel_def}
 \chi_{i \pm}^\alpha &= \frac{1\,\pm\, \gamma_5}{2} \chi_{n_i, - \omega_i}^\alpha
\,,\qquad\quad
\bar{\chi}_{i \pm}^\balpha =  \bar{\chi}_{n_i, - \omega_i}^\balpha \frac{1\,\mp\, \gamma_5}{2}\,.
\end{align}
\end{subequations}
Here $a$, $\alpha$, and $\balpha$ are adjoint, $3$, and $\bar 3$ color indices respectively, and the $\omega_i$ labels on both the gluon and quark building blocks are taken to be outgoing, which is also used for our helicity convention. 
Using the standard spinor helicity notation (see e.g. \cite{Dixon:1996wi} for an introduction) 
\begin{align} \label{eq:braket_def}
|p\rangle\equiv \ket{p+} &= \frac{1 + \ga_5}{2}\, u(p)
  \,,
 & |p] & \equiv \ket{p-} = \frac{1 - \ga_5}{2}\, u(p)
  \,, \\
\bra{p} \equiv \bra{p-} &= \mathrm{sgn}(p^0)\, \bar{u}(p)\,\frac{1 + \ga_5}{2}
  \,, 
 & [p| & \equiv \bra{p+} = \mathrm{sgn}(p^0)\, \bar{u}(p)\,\frac{1 - \ga_5}{2}
  \,, \nn 
\end{align}
with $p$ lightlike, the polarization vector of an outgoing gluon with momentum $p$ can be written
\begin{equation}
 \ve_+^\mu(p,k) = \frac{\mae{p+}{\ga^\mu}{k+}}{\sqrt{2} \langle kp \rangle}
\,,\qquad
 \ve_-^\mu(p,k) = - \frac{\mae{p-}{\ga^\mu}{k-}}{\sqrt{2} [kp]}
\,,\end{equation}
where $k\neq p$ is an arbitrary light-like reference vector, chosen to be $\bn_i$ in \eq{cBpm_def}.  

Since fermions always arise in pairs, we can define currents with definite helicities. 
Here we will restrict to the case of two back to back directions, $n$ and $\bar n$, as is relevant for $gg\to H$. 
A more general discussion can be found in \Refs{Kolodrubetz:2016uim,Feige:2017zci}. 
We define helicity currents where the quarks are in opposite collinear sectors,
 \begin{align} \label{eq:jpm_back_to_bacjdef}
 & h=\pm 1:
 & J_{n \bn \pm}^{\balpha\beta}
 & = \mp\, \sqrt{\frac{2}{\omega_n\, \omega_\bn}}\, \frac{   \ve_\mp^\mu(n, \bn) }{\langle \bn \mp | n \pm\rangle}   \, \bar{\chi}^\balpha_{n\pm}\, \gamma_\mu \chi^\beta_{\bn \pm}
 \,, \\
 & h=0:
 & J_{n \bn 0}^{\balpha\beta}
 & =\frac{2}{\sqrt{\vphantom{2} \omega_n \,\omega_\bn}\,  [n \bn] } \bar \chi^\balpha_{n+}\chi^\beta_{\bn-}
 \,, \qquad
 (J^\dagger)_{n \bn 0}^{\balpha\beta}=\frac{2}{\sqrt{ \vphantom{2} \omega_n \, \omega_\bn}  \langle n  \bn \rangle  } \bar \chi^\balpha_{n-}\chi^\beta_{\bn+}
 \,, \nn
 \end{align}
as well as where the quarks are in the same collinear sector,
\begin{align}\label{eq:coll_subl}
 & h=0:
 & J_{i0}^{\balpha \beta} 
  &= \frac{1}{2 \sqrt{\vphantom{2} \omega_{\bar \chi} \, \omega_\chi}}
  \: \bar \chi^\balpha_{i+}\, \Sl{\bar n}_i\, \chi^\beta_{i+}
   \,,\qquad
   J_{i\bar 0}^{\balpha \beta} 
  = \frac{1}{2 \sqrt{\vphantom{2} \omega_{\bar \chi} \, \omega_\chi}}
  \: \bar \chi^\balpha_{i-}\, \Sl {\bar n}_i\, \chi^\beta_{i-}
 \,, \\[5pt]
  & h=\pm 1:
 & J_{i\pm}^{\balpha \beta}
  &= \mp  \sqrt{\frac{2}{ \omega_{\bar \chi} \, \omega_\chi}}  \frac{\epsilon_{\mp}^{\mu}(n_i,\bar n_i)}{ \big(\l n_i \mp | \bar{n}_i \pm \r \big)^2}\: 
   \bar \chi_{i\pm}^\balpha\, \gamma_\mu \Sl{\bar n}_i\, \chi_{i\mp}^\beta
 \,. \nn
\end{align}
Here $i$ can be either $n$ or $\bar n$. 
All of these currents are manifestly invariant under the RPI-III symmetry of SCET.
The Feynman rules for all currents are very simple, and are given in~\cite{Feige:2017zci}.  
Note that the operators $J_{n \bn \pm}^{\balpha\beta}$, $J_{i0}^{\balpha \beta}$, and $J_{i\bar 0}^{\balpha \beta}$ have quarks of the same chirality, and hence are the ones that will be generated by vector gauge bosons.  

\begin{table}
 \begin{center}
  \begin{tabular}{|c|c|cc|ccc|c|ccc|}
	\hline \phantom{x} & \phantom{x} & \phantom{x} 
	& \phantom{x} & \phantom{x} & \phantom{x} & \phantom{x} 
	& \phantom{x} & \phantom{x} & \phantom{x} & \phantom{x} 
	\\[-13pt]                      
 Field: & 
    $\cB_{i\pm}^a$ & $J_{ij\pm}^{\balpha\beta}$ & $J_{ij0}^{\balpha\beta}$ 
    & $J_{i\pm}^{\balpha \beta}$ 
	& $J_{i0}^{\balpha \beta}$ & $J_{i\bar 0}^{\balpha \beta}$  
    & $\cP^{\perp}_{\pm}$ 
	& $\partial_{us(i)\pm}$ & $\partial_{us(i)0}$ & $\partial_{us(i)\bar{0}}$
	\\[3pt] 
 Power counting: &	
    $\lambda$ &  $\lambda^2$ &  $\lambda^2$
	& $\lambda^2$ & $\lambda^2$& $\lambda^2$ & $\lambda$ 
    & $\lambda^2$ & $\lambda^2$  & $\lambda^2$
	\\
 Equation: & 
   (\ref{eq:cBpm_def}) & \multicolumn{2}{c|}{(\ref{eq:jpm_back_to_bacjdef})} 
     & \multicolumn{3}{c|}{(\ref{eq:coll_subl})} & (\ref{eq:Pperppm}) 
     & \multicolumn{3}{c|}{(\ref{eq:partialus})}
    \\
  \hline  
  \end{tabular}\\
\vspace{.3cm} 
  \begin{tabular}{|c|cc|}
	\hline  \phantom{x} &  \phantom{x} & \phantom{x} 
	\\[-13pt]                        
 Field: & 
 	$\cB^a_{us(i)\pm}$ & 
    \!\!$\cB^a_{us(i)0}$  
	\\[3pt] 
 Power counting: &
 	$\lambda^2$ & $\lambda^2$ 
	\\ 
 Equation: & 
     \multicolumn{2}{c|}{(\ref{eq:Bus})}  
    \\
	\hline
  \end{tabular}
 \end{center}
\vspace{-0.3cm}
\caption{The helicity building blocks in $\text{SCET}_\text{I}$ that will be used to construct a basis of hard scattering operators for $gg\to H$, together with their power counting order in the $\lambda$-expansion, and the equation numbers where their definitions may be found. The building blocks also include the conjugate currents $J^\dagger$ in cases where they are distinct from the ones shown.
} 
\label{tab:helicityBB}
\end{table}

At subleading power one must also consider insertions of the $\cP_{i\perp}^\mu$ operator. 
Note that we can drop the explicit $i$ index on the $\cP_\perp$ operator, as it is implied by the field that the operator is acting on.
The $\cP_{\perp}^\mu$ operator acts on the perpendicular subspace defined by the vectors $n_i, \bar n_i$, so  it is naturally written as
\begin{align} \label{eq:Pperppm}
\cP_{+}^{\perp}(n_i,\bar n_i)=-\epsilon^-(n_i,\bar n_i) \cdot \cP_{\perp}\,, \qquad \cP_{-}^{\perp}(n_i,\bar n_i)=-\epsilon^+(n_i,\bar n_i) \cdot \cP_{\perp}\,.
\end{align} 
The $\cP^\perp_\pm$ operator carry helicity $h=\pm 1$. 
We use square brackets to denote which fields are acted upon by the $\cP^{\perp}_{\pm}$ operator, for example
$\cB_{i+} \left [ \cP^{\perp}_{+}  \cB_{i-}  \right]  \cB_{i-}$,
indicates that the $\cP^{\perp}_{+}$ operator acts only on the middle field, whereas for currents, we use a curly bracket notation
\begin{align}\label{eq:p_perp_notation}
  \big\{ \cP^{\perp}_\lambda J_{i 0 }^{\balpha \beta} \big\}  
  & = \frac{1}{2 \sqrt{\vphantom{2}\omega_{\bar \chi} \, \omega_\chi }} \:
   \Big[  \cP^{\perp}_{\lambda}  \bar \chi^\balpha_{i +}\Big] \Sl {\bar n}_i \chi^\beta_{i+}
  \,, \\
 \big\{ J_{i0 }^{\balpha \beta} (\cP^{\perp}_{\lambda})^\dagger \big\}
  &=  \frac{1}{2\sqrt{\vphantom{2}\omega_{\bar \chi} \, \omega_\chi}} \:
  \bar \chi^\balpha_{i+} \Sl {\bar n}_i \Big[   \chi^\beta_{i+} (\cP^{\perp}_{\lambda})^\dagger \Big]
  \,, \nn
\end{align}
to indicate which of the fields within the current is acted on.

To work with gauge invariant ultrasoft gluon fields, we construct our basis post BPS field redefinition. The BPS field redefinition is defined by \cite{Bauer:2002nz}
\be \label{eq:BPSfieldredefinition}
\cB^{a\mu}_{n\perp}\to \cY_n^{ab} \cB^{b\mu}_{n\perp} , \qquad \chi_n^\alpha \to Y_n^{\alpha \bbeta} \chi_n^\beta,
\ee
and is performed in each collinear sector. Here $Y_n$, $\cY_n$ are fundamental and adjoint ultrasoft Wilson lines. For a general representation, r, the ultrasoft Wilson line is defined by
\be
Y^{(r)}_n(x)=\bold{P} \exp \left [ ig \int\limits_{-\infty}^0 ds\, n\cdot A^a_{us}(x+sn)  T_{(r)}^{a}\right]\,,
\ee
where $\bold P$ denotes path ordering. 
The BPS field redefinition has the effect of decoupling ultrasoft and collinear degrees of freedom at leading power \cite{Bauer:2002nz}, and it accounts for the full physical path of ultrasoft Wilson lines~\cite{Chay:2004zn,Arnesen:2005nk}.

The BPS field redefinition introduces ultrasoft Wilson lines into the hard scattering operators.
These Wilson lines can be arranged with the ultrasoft fields to define ultrasoft gauge invariant building blocks.
In particular, the gauge covariant derivative in an arbitrary representation, $r$, can be sandwiched by Wilson lines and decomposed as
\begin{align}\label{eq:soft_gluon}
Y^{(r)\,\dagger}_{n_i} i D^{(r)\,\mu}_{us} Y^{(r)}_{n_i }=i \partial^\mu_{us} + [Y_{n_i}^{(r)\,\dagger} i D^{(r)\,\mu}_{us} Y^{(r)}_{n_i}]=i\partial^\mu_{us}+T_{(r)}^{a} g \cB^{a\mu}_{us(i)}\,.
\end{align}
Here we have defined the ultrasoft gauge invariant gluon field by
\begin{align} \label{eq:softgluondef}
g \cB^{a\mu}_{us(i)}= \left [   \frac{1}{in_i\cdot \partial_{us}} n_{i\nu} i G_{us}^{b\nu \mu} \cY^{ba}_{n_i}  \right] \,.
\end{align}
In the above equations the derivatives act only within the square brackets.
Note from \eq{softgluondef}, that  $n_i\cdot \cB^{a}_{us(i)}= 0$.
The Wilson lines which remain after this procedure can be absorbed into a generalized color structure, $\vT_{\BPS}$ (see \cite{Kolodrubetz:2016uim} for more details).
Determining a complete basis of color structures is straightforward, and detailed examples are given in~\cite{Feige:2017zci}.

Having defined gauge invariant ultrasoft gluon fields, we can now define ultrasoft gauge invariant gluon helicity fields and derivative operators which mimic their collinear counterparts.
For the ultrasoft gluon helicity fields we define the three building blocks
\begin{equation} \label{eq:Bus}
\cB^a_{us(i)\pm} = -\ve_{\mp\mu}(n_i, \bn_i)\,\cB^{a\mu}_{us(i)},\qquad  \cB^a_{us(i)0} =\bar n_\mu  \cB^{a \mu}_{us(i)}   
\,,\end{equation}
and similarly for the ultrasoft derivative operators
\begin{equation}  \label{eq:partialus}
\partial_{us(i)\pm} = -\ve_{\mp\mu}(n_i, \bn_i)\,\partial^{\mu}_{us},\qquad   \partial_{us(i)0} =\bar n_{i\mu} \partial^{\mu}_{us}, \qquad \partial_{us(i)\bar 0} = n_{i \mu} \partial^{\mu}_{us}
\,.\end{equation}
Unlike for the gauge invariant collinear gluon fields, for the ultrasoft gauge invariant gluon field we use three building block fields to describe the two physical degrees of freedom because the ultrasoft gluons are not fundamentally associated with any direction.
Without making a further gauge choice, their polarization vectors do not lie in the perpendicular space of any fixed external reference vector.
When inserting ultrasoft derivatives into operators we will use the same curly bracket notation defined for the $\cP_\perp$ operators in \Eq{eq:p_perp_notation}. 

Gauge invariant ultrasoft quark fields can also appear explicitly in operator bases at subleading powers.
From \Tab{tab:PC} we see that they power count as $\cO(\lambda^3)$, and are therefore not relevant for our construction of an $\cO(\lambda^2)$ operator basis.
Details on the structure of subleading power helicity operators involving ultrasoft quarks can be found in~\cite{Feige:2017zci}.
It is important to emphasize that although ultrasoft quarks do not appear in the hard scattering operators at $\cO(\lambda^2)$ they do appear in the calculation of cross sections or amplitudes at $\cO(\lambda^2)$ due to subleading power Lagrangian insertions in the effective (examples where they play an important role for factorization in $B$-decays include both exclusive decays~\cite{Bauer:2002aj,Mantry:2003uz,Beneke:2003pa} and inclusive decays~\cite{Bosch:2004cb,Lee:2004ja,Beneke:2004in}).
Such ultrasoft quark contributions also played an important role in the recent subleading power perturbative SCET calculation of \Ref{Moult:2016fqy}.

Finally, we note that the helicity operator basis presented in this section only provides a complete basis in $d=4$, and we have not discussed evanescent operators \cite{Buras:1989xd,Dugan:1990df,Herrlich:1994kh}. 
An extension of our basis to include evanescent operators would depend on the regularization scheme. 
However, in general additional building block fields would be required, for example an $\epsilon$ scalar gluon $\cB^a_{\epsilon}$ to encode the $(-2\epsilon)$ transverse degrees of freedom of the gluon.
As in standard loop calculations, we expect that the evanescent operators at each loop order could be straightforwardly identified and treated. 
Since we do not perform a one-loop matching to our operators, we leave a complete treatment of evanescent operators to future work.

\section{Operator Basis}\label{sec:basis}

In this section we enumerate a complete basis of power suppressed operators up to $\cO(\lambda^2)$  for the process $gg\to H$. The organization of the operator basis in terms of helicity operators will make manifest a number of symmetries arising from helicity conservation, greatly reducing the operator basis. Helicity conservation is particularly powerful in this case due to the spin-$0$ nature of the Higgs. The complete basis of field structures is summarized in \Tab{tab:summary}. In \Sec{sec:discussion_gluonops} we will show which operators contribute to the cross section at $\cO(\lambda^2)$. These operators are indicated with a check mark in the table.

Examining \eq{Leff_sub_explicit} we see that the hard Lagrangian in SCET is written as a sum over label momenta of the hard operators.  For the special case of two back-to-back collinear sectors this reduces to
\begin{align}\label{eq:sum_dir}
\cL_{\text{hard}}^{(j)} = \sum_{n} \sum_{A,\cdot\cdot}  
\bigg[ \prod_{i=1}^{\ell_A} \int \! \! \df \omega_i \bigg] \,
& \vO^{(j)\dagger}_{A\,\lotsdots}\big(n,\bn;
   \omega_1,\ldots,\omega_{\ell_A}\big) \nn\\
& \times
\vC^{(j)}_{A\,\lotsdots}\big(n,\bn;\omega_1,\ldots,\omega_{\ell_A} \big)
\,.
\end{align}
When writing our basis, we therefore do not need to include operators which are identical up to the swap of $n\leftrightarrow \bar n$.  This means that when writing an operator with different field structures in the two collinear sectors we are free to make an arbitrary choice for which is labeled $n$ and which $\bar n$, and this choice can be made independently for each operator. When squaring matrix elements, all possible interferences are properly incorporated by the sum over directions in \Eq{eq:sum_dir}.

As discussed in \Sec{sec:intro_gluonops}, we will work in the Higgs effective theory with a Higgs gluon coupling given by the effective Lagrangian in \Eq{eq:heft_intro}. We therefore do not consider operators generated by a direct coupling of quarks to the Higgs. All quarks in the final state are produced by gluon splittings. The extension to include operators involving quarks coupling directly to Higgs, as relevant for $H\to b\bar b$, is straightforward using the helicity building blocks given in \Sec{sec:review_helicity}.

{
\renewcommand{\arraystretch}{1.4}
\begin{table}[t!]
\hspace{-0.15cm} \scalebox{0.9}{
\begin{tabular}{| c | l | l | c | c | c | }
	\hline 
Order & $\!$Category &  Operators (equation number) 
 & \#$\!$  helicity & \#$\!$ of
 & $\!\sigma_{2j}^{\cO(\lambda^2)}\!\! \ne\! 0\!$
 \\[-8pt]
 & & & configs & \! color\! & 
 \\ \hline 
$\mathcal{O}(\lambda^0)$  
   & $\! H gg$ & $O_{\cB \lambda_1 \lambda_1}^{(0)ab}=\cB^a_{n  \lambda_1} \cB^a_{\bar n  \lambda_1} H$ \,(\ref{eq:hgg})
   & 2  & 1 & $\checkmark$
    \\ \hline
$\mathcal{O}(\lambda)$ 
   & $\! H q \bar{q} g$  
    & $O_{\cB n,\bar n \lambda_1 ( \lambda_i)}^{(1)a\,\balpha\bt}
	=  \cB_{n,\bar n  \lambda_1}^a\, J_{n\bar n\, \lambda_j}^{\balpha\bt}\,H$ \,(\ref{eq:H1_basis},\ref{eq:H1_basis2}) 
    & 4 &  1 &   $\checkmark$ 
    \\ \hline
$\mathcal{O}(\lambda^2)$
   & $\! H q \bar q Q \bar Q$  & $O_{qQ1(\lambda_1;\lambda_2)}^{(2)\balpha\bt\bgamma\delta}
	= J_{(q)n {\lambda_1}\, }^{\balpha\bt}\, J_{(Q) \bar n {\lambda_2}\, }^{\bgamma\delta}\,H$ \,(\ref{eq:Z2_basis_qQ}) 
    & 4  & 2 & 
    \\
	& & $O_{qQ2(\lambda_1;\lambda_1)}^{(2)\balpha\bt\bgamma\delta}
	= J_{(q \bar{Q}) n  \lambda_1\, }^{\balpha\bt}\, J_{(Q \bar{q}) \bar{n}  \, \lambda_1\, }^{\bgamma\delta}\,H $ \,(\ref{eq:Z2_basis_qQ_2})
    & 2  & 2 &  
    \\
    & & $O_{qQ3(\lambda_1;-\lambda_1)}^{(2)\balpha\bt\bgamma\delta}
	= J_{(q) n \bar n \lambda_1\, }^{\balpha\bt}\, J_{(Q) n\bar n -\lambda_1\, }^{\bgamma\delta}\,H$ \,(\ref{eq:Z2_basis_qQ_3})
    & 2 &  2 & 
    \\ \cline{2-6}
   & $\! H q \bar q q \bar q$  & $O_{qq1(\lambda_1;\lambda_2)}^{(2)\balpha\bt\bgamma\delta}
	= J_{(q)n {\lambda_1}\, }^{\balpha\bt}\, J_{(q) \bar n {\lambda_2}\, }^{\bgamma\delta}\,H$ \,(\ref{eq:Z2_basis_qq}) 
    & 3 & 2 &   
    \\
	& & $O_{qq3(\lambda_1;-\lambda_1)}^{(2)\balpha\bt\bgamma\delta}
	= J_{(q) n \bar n \lambda_1\, }^{\balpha\bt}\, J_{(q) n\bar n -\lambda_1\, }^{\bgamma\delta}\,H$ \,(\ref{eq:Z2_basis_qq_3})
    & 1 & 2 & 
    \\ \cline{2-6}
   & $\! H q \bar q g g$  & $O_{\cB1\lambda_1 \lambda_2(\lambda_3)}^{(2)ab\, \balpha\bt}
	=  \cB_{n \lambda_1}^a \cB_{\bar n \lambda_2}^b \, J_{n\,{\lambda_3} }^{\balpha\bt}   \,H $ \,(\ref{eq:Hqqgg_basis3}) 
    & 4  &  3 & $\checkmark$
    \\
	& & $O_{\cB2\lambda_1 \lambda_2(\lambda_3)}^{(2)ab\, \balpha\bt}
	=  \cB_{\bar n \lambda_1}^a \cB_{\bar n \lambda_2}^b \, J_{n\,{\lambda_3} }^{\balpha\bt}   \,H $ \,(\ref{eq:Hqqgg_basis4}) 
    & 2  & 3 & 
    \\ \cline{2-6}
   & $\! H gggg$  & $O_{4g1\lambda_1 \lambda_2 \lambda_3 \lambda_4}^{(2)a b c d}
	= S  \cB^a_{n \lambda_1} \cB^b_{n \lambda_2} \cB^c_{\bn \lambda_3} \cB^d_{\bn \lambda_4} H $ \,(\ref{eq:H_basis_gggg_1}) 
    & 3  & 9 &  
    \\
	&   & $O_{4g2\lambda_1 \lambda_2 \lambda_3 \lambda_4}^{(2)a b c d}
	= S  \cB^a_{n \lambda_1} \cB^b_{\bn \lambda_2} \cB^c_{\bn \lambda_3} \cB^d_{\bn \lambda_4} H$ \,(\ref{eq:H_basis_gggg_2}) 
    & 2  & 9 & $\checkmark$
    \\ \cline{2-6}
   &  $\cP_\perp$   & $O_{\cP\chi \lambda_1 (\lambda_2)[\lambda_{\cP}]}^{(2)a\,\balpha\bt}
	= \cB_{n\lambda_1}^a \, \{J_{\bar n\, {\lambda_2}    }^{\balpha\bt}(\cP_{\perp}^{\lambda_{\cP}})^\dagger\}\,H$  \,(\ref{eq:Hqqgpperp_basis_same})  
    & 4  & 1 & $\checkmark$
    \\
	& & $O_{\cP\cB \lambda_1 \lambda_2 \lambda_3[\lambda_{\cP}]}^{(2)abc}
	= S\, \cB_{n\lambda_1}^a\, \cB_{\bar n \lambda_2}^b \left [\cP_{\perp}^{\lambda_{\cP}} \cB_{\bar n \lambda_3}^c \right ]\! H\! $  \,(\ref{eq:Hgggpperp_basis}) 
    & 4  & 2 & $\checkmark$ 
    \\ \cline{2-6}
	&  	$\!$Ultrasoft $\!\!\!$    & $O_{\chi (us(n))0:(\lambda_1)}^{(2)a\,\balpha\bt} 
	= \cB_{us(n)0}^a \, J_{n\bar n\,\lambda_1}^{\balpha\bt}\,H $ \,(\ref{eq:soft_insert_basis}) 
    & 2 & 1 & 
    \\
	&   & $O_{\chi (us(\bar n))0:(\lambda_1)}^{(2)a\,\balpha\bt} 
	= \cB_{us(\bar n)0}^a \, J_{n\bar n\,\lambda_1}^{\balpha\bt}\,H $ \,(\ref{eq:soft_insert_basis2}) 
    & 2 &  1 & 
    \\
	& & $O_{\partial \chi (us(i))\lambda_1:(\lambda_2)}^{(2)\,\balpha\bt}
	= \{\partial_{us(i)\lambda_1} \, J_{n\bar n\,\lambda_2}^{\balpha\bt}\}\,H $ \,(\ref{eq:soft_derivative_basis}) 
    & 4 & 1 & 
    \\
	& & $O_{\cB (us(n))\lambda_1:\lambda_2 \lambda_3 }^{(2)abc}
	=\cB_{us(n) \lambda_1}^a \,\cB_{n\, \lambda_2}^{b}\,\cB_{\bn\, \lambda_3}^{c}\,H$ \,(\ref{eq:Hgggus}) 
    & 2  & 2 & $\checkmark$
    \\
	& & $O_{\cB (us(\bar n))\lambda_1:\lambda_2 \lambda_3 }^{(2)abc}
	=\cB_{us(\bar n) \lambda_1}^a \,\cB_{n\, \lambda_2}^{b}\,\cB_{\bn\, \lambda_3}^{c}\,H$ \,(\ref{eq:Hgggus_2}) 
	& 2  & 2 & $\checkmark$
    \\
	& & $O_{\partial \cB  (us(i))\lambda_1:\lambda_2 \lambda_3 }^{(2)ab}
	=\left[ \partial_{us(i) \lambda_1} \,\cB_{n\, \lambda_2}\right] \,\cB_{\bn\, \lambda_3}\,H $ \,(\ref{eq:Hdggus}) 
	& 4  & 1 & $\checkmark$ 
    \\ \hline
\end{tabular}}
\vspace{0.1cm}
\caption{Basis of hard scattering operators for $gg\to H$ up to ${\cal O}(\lambda^2)$. The $\lambda_i$ denote helicities, $S$ represents a symmetry factor present for some cases, and detailed lists of operators can be found in the indicated equation.  The number of allowed helicity configurations are summarized in the fourth column. The final column indicates which operators contribute to the cross section up to $\mathcal{O}(\lambda^2)$ in the power expansion, as discussed in Sec. \ref{sec:discussion_gluonops}. Counting the helicity configurations there are a total of 53 operators, of which only 28 
contribute to the cross section at $\mathcal{O}(\lambda^2)$. Of those 28, only 24 have non zero Wilson coefficients at tree level since the operators in \eq{Hqqgpperp_basis_same} are absent at this order. These numbers do not include the number of distinct color configurations which are indicated in the 5th column. 
}
\label{tab:summary}
\end{table}
}

\subsection{Leading Power}\label{sec:lp}

The leading power operators for $gg\to H$ in the Higgs effective theory are well known. Due to the fact that the Higgs is spin zero, the only two operators are
\begin{align}
 \boldsymbol{g_n g_{\bn}:}   {\vcenter{\includegraphics[width=0.18\columnwidth]{figures/Leading_gg_low}}} \nn
\end{align}
\vspace{-0.4cm}
\begin{alignat}{2}\label{eq:hgg}
 &O_{\cB++}^{(0)ab}
=\cB_{n +}^a\, \cB_{\bar n+}^b\,   H
\,, \qquad &&O_{\cB--}^{(0)ab}
=\cB_{n-}^a\, \cB_{ \bn -}^b\, H\,.
\end{alignat}
Here the purple circled denotes that this is a hard scattering operator in the effective theory, while the dashed circles indicate which fields are in each collinear sector.
Note that here we have opted not to include a symmetry factor at the level of the operator. We will include symmetry factors in the operator only when there is an exchange symmetry within a given collinear sector. We assume that overall symmetry factors which involve exchanging particles from different collinear sectors are taken into account at the phase space level. The color basis here is one-dimensional, and we take it to be
\begin{align} \label{eq:leading_color}
 \vT^{ab} = \de_{ab}\,, \qquad 
 \vT_{\BPS}^{ab} = \bigl( \cY_{n}^T \cY_{\bar n} \bigr)^{ab} 
                 = \bigl( \cY_{\bn}^T \cY_{n} \bigr)^{ba} 
\,.
\end{align}

\subsection{Subleading Power}\label{sec:nlp}

Due to the spin zero nature of the Higgs, the $\cO(\lambda)$ operators are highly constrained. To simplify the operator basis we will work in the center of mass frame and we will further choose our $n$ and $\bar n$ axes so that the total label $\perp$ momentum of each collinear sector vanishes. This is possible in an SCET$_\text{I}$ theory since the ultrasoft sector does not carry label momentum, and it implies that we do not need to include operators where the $\cP_\perp$ operator acts on a sector with a single collinear field. At $\cO(\lambda)$ the suppression in the operator must therefore come from an explicit collinear field. 

There are two possibilities for the collinear field content of the operators, either three collinear gluon fields, or two collinear quark fields and a collinear gluon field. Interestingly, the helicity selection rules immediately eliminate the possibility of $\cO(\lambda)$ operators with three collinear gluon fields, since they cannot sum to a zero helicity state. We therefore only need to consider operators involving two collinear quark fields and a collinear gluon field. The helicity structure of these operators is also constrained. In particular, to cancel the spin of the collinear gluon field, the collinear quark current must have helicity $\pm1$. Furthermore, the quark-antiquark pair arises from a gluon splitting, since we are considering gluon fusion in the Higgs EFT, and therefore both have the same chirality. Together this implies that the quarks are described by the current $J_{n\bar n\,\pm}^{\balpha\bt}$. The only two operators in the basis at  $\cO(\lambda)$ are
\begin{align}
 \boldsymbol{q_n (\bar qg)_{\bn}:}   {\vcenter{\includegraphics[width=0.18\columnwidth]{figures/SubleadingQG_low}}} \nn
\end{align}
\vspace{-0.4cm}
\begin{alignat}{2} \label{eq:H1_basis}
&O_{\cB\bar n+(+)}^{(1)a\,\balpha\bt}
= \cB_{\bar n+}^a\, J_{n\bar n\,+}^{\balpha\bt} \,H
\,,\qquad &
&O_{\cB\bar n-(-)}^{(1)a\,\balpha\bt}
= \cB_{\bar n-}^a \, J_{n\bar n\,-}^{\balpha\bt}\,H
\,,
\end{alignat}
for the case that the gluon field is in the same sector as the antiquark field, which we have taken to be $\bar n$, and
\begin{align}
 \boldsymbol{(q g)_n \bar q_{\bn}:}   {\vcenter{\includegraphics[width=0.18\columnwidth]{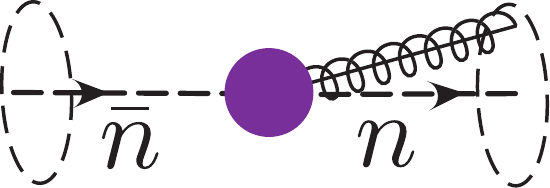}}} \nn
\end{align}
\vspace{-0.4cm}
\begin{alignat}{2} \label{eq:H1_basis2}
&O_{\cB n-(+)}^{(1)a\,\balpha\bt}
= \cB_{n-}^a\, J_{n\bar n\,+}^{\balpha\bt} \,H
\,,\qquad &
&O_{\cB n+(-)}^{(1)a\,\balpha\bt}
= \cB_{n+}^a \, J_{n\bar n\,-}^{\balpha\bt}\,H
\,,
\end{alignat}
for the case that the gluon field is in the same direction as the collinear quark field.  In both cases the color basis is one-dimensional $\vT^{a\, \al\bbeta} = T^a_{\al\bbeta}$. After the BPS field redefinition we have
\begin{align} \label{eq:nlp_color}
 \vT_{\BPS}^{ a \al\bbeta} 
    &= \left (Y_n^\dagger Y_{\bar n} T^a   \right )_{\alpha \bar \beta}
    \,,
& \vT_{\BPS}^{ a \al\bbeta} 
    &= \left (T^a Y_n^\dagger Y_{\bar n}  \right )_{\alpha \bar \beta}
    \,,
\end{align}
for \eqs{H1_basis}{H1_basis2} respectively.


\subsection{Subsubleading Power}\label{sec:nnlp}

At $\cO(\lambda^2)$ the allowed operators can include either additional collinear field insertions, insertions of the $\cP_\perp$ operator, or ultrasoft field insertions. We will treat each of these cases in turn.

\subsubsection{Collinear Field Insertions}\label{sec:nnlp_collinear}

We begin by considering operators involving only collinear field insertions. At $\cO(\lambda^2)$ the operator can have four collinear fields. These operators can be composed purely of collinear gluon fields, purely of collinear quark fields, or of two collinear gluon fields and a collinear quark current. In each of these cases helicity selection rules will restrict the possible helicity combinations of the operators.

\vspace{0.4cm}
\noindent{\bf{Two Quark-Two Gluon Operators:}}

We begin by considering operators involving two collinear quark fields and two collinear gluon fields, which are again severely constrained by the helicity selection rules. Since the two gluons fields can give either helicity $0$ or $2$, the only way to achieve a total spin zero is if the quark fields must be in a helicity zero configuration. Furthermore, since they arise from a gluon splitting they must have the same chirality. This implies that all operators must involve only the currents $J_{n\, 0}^{\balpha\bt}$ or $J_{n\, \bar 0}^{\balpha\bt}$, where we have taken without loss of generality that the two quarks are in the $n$-collinear sector, as per the discussion below \Eq{eq:sum_dir}. The gluons can then either be in opposite collinear sectors, or in the same collinear sector. The color basis before BPS field redefinition is identical for the two cases. It is three dimensional, and we take as a basis
\begin{equation} \label{eq:ggqqll_color}
\vT^{\, ab \alpha\bbeta}
= \Bigl(
(T^a T^b)_{\alpha\bbeta}\,,\, (T^b T^a)_{\alpha\bbeta} \,,\, \tr[T^a T^b]\, \delta_{\alpha\bbeta}
\Bigr)
\,.\end{equation} 

In the case that the two collinear gluons are in opposite collinear sectors a basis of helicity operators is given by
\begin{align}
 \boldsymbol{(g q\bar q)_n (g)_{\bn}:}   {\vcenter{\includegraphics[width=0.18\columnwidth]{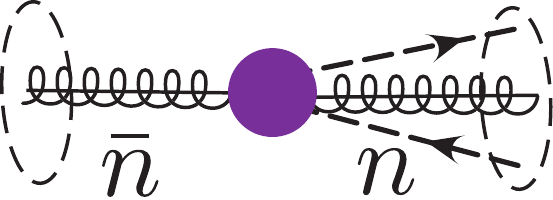}}} \nn
\end{align}
\vspace{-0.4cm}
\begin{alignat}{2} \label{eq:Hqqgg_basis3}
&O_{\cB1++(0)}^{(2)ab\, \balpha\bt}
=   \cB_{n+}^a\, \cB_{\bar n+}^b \, J_{n\,{0} }^{\balpha\bt}  \,H\, , \qquad 
&&O_{\cB1++(\bar 0)}^{(2)ab\, \balpha\bt}
=\cB_{n+}^a\, \cB_{\bar n+}^b  \, J_{n\,{\bar 0} }^{\balpha\bt}   \,H\, ,  \\
&O_{\cB1--(0)}^{(2)ab\, \balpha\bt}
=  \cB_{n-}^a\, \cB_{\bar n-}^b \, J_{n\,{0} }^{\balpha\bt}    \,H\, , \qquad 
&&O_{\cB1--(\bar 0)}^{(2)ab\, \balpha\bt}
= \cB_{ n-}^a\, \cB_{\bar n-}^b  \, J_{n\,{\bar 0} }^{\balpha\bt}   \,H\, . \nn
\end{alignat}
The color basis after BPS field redefinition is given by
\begin{equation}
\vT_{\BPS}^{\, ab \alpha\bbeta}
= \Bigl(
(\cY_n^T \cY_\bn)^{cb}  (T^a T^c)_{\alpha\bbeta} \,,\, 
(\cY_n^T \cY_\bn)^{cb}  (T^c T^a)_{\alpha\bbeta} \,,\, 
T_F (\cY_n^T \cY_\bn)^{ab} \, \delta_{\alpha\bbeta}
\Bigr)
\,,
\end{equation}
where we have used $\tr[T^a T^b] = T_F \delta^{ab}$.

In the case that the two gluons are in the same collinear sector a basis of helicity operators is given by
\begin{align}
& \boldsymbol{(q \bar q)_n (gg)_{\bn}:}{\vcenter{\includegraphics[width=0.18\columnwidth]{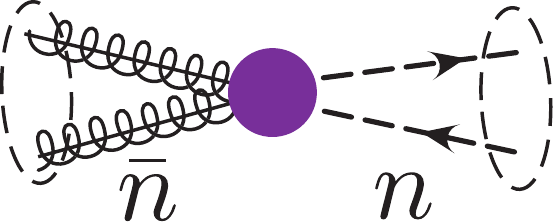}}}  \nn
\end{align}
\vspace{-0.4cm}
\begin{align}\label{eq:Hqqgg_basis4}
&O_{\cB2+-(0)}^{(2)ab\, \balpha\bt}
= \cB_{\bar n+}^a\, \cB_{ \bar n-}^b  \, J_{n\,{0} }^{\balpha\bt}\,   H\,, \qquad 
&O_{\cB2+-(\bar 0)}^{(2)ab\, \balpha\bt}
= \cB_{\bar n+}^a\, \cB_{ \bar n-}^b  \, J_{n\,{\bar 0} }^{\balpha\bt}\,   H\,.
\end{align}
The color basis after BPS field redefinition is
\begin{equation}
\vT_{\BPS}^{\, ab \alpha\bbeta}
= \Bigl(
(Y_n^\dagger Y_\bn T^a T^b Y^\dagger_{\bar n} Y_n )_{\alpha\bbeta}
  \,,\, 
(Y_n^\dagger Y_\bn T^b T^a Y^\dagger_{\bar n} Y_n )_{\alpha\bbeta}
   \,,\, 
\tr[T^a T^b]\, \delta_{\alpha\bbeta} \Bigr)
\,.
\end{equation}

\vspace{0.4cm}
\noindent{\bf{Four Gluon Operators:}}

Operators involving four collinear gluon fields can have either two collinear gluon fields in each sector, or three collinear gluon fields in one sector. A basis of color structures before BPS field redefinition is given by
\begin{equation} \label{eq:gggg_color}
\vT^{ abcd} =
\frac{1}{2}\begin{pmatrix}
\tr[abcd] + \tr[adcb] \\ \tr[acdb] + \tr[abdc] \\ \tr[adbc] + \tr[acbd] \\
\tr[abcd] - \tr[adcb] \\ \tr[acdb] - \tr[abdc] \\ \tr[adbc] - \tr[acbd] \\ 2\tr[ab]\, \tr[cd] \\ 2\tr[ac]\, \tr[db] \\ 2\tr[ad]\, \tr[bc]
\end{pmatrix}^{\!\!\!T}
.\end{equation}
Here we have used a simplified notation, writing only the adjoint indices of the color matrices appearing in the trace.  For example, $\tr[abcd] \equiv \tr[T^a T^b T^c T^d]$.
The color bases after BPS field redefinition will be given separately for each case.
For the specific case of SU($N_c$) with $N_c=3$ we could further reduce the color basis by using the relation
\begin{align}
&\tr[abcd+dcba] + \tr[acdb+bdca] + \tr[adbc+cbda]
\nn\\ & \qquad
= \tr[ab]\tr[cd] + \tr[ac]\tr[db] + \tr[ad]\tr[bc]
\,.\end{align}
We choose not to do this, as it makes the structure more complicated, and because it does not hold for $N_c>3$.

In the case that there are two collinear gluon fields in each collinear sector, a basis of helicity operators is given by
\begin{align}
& \boldsymbol{(gg)_n (gg)_{\bn}:}{\vcenter{\includegraphics[width=0.18\columnwidth]{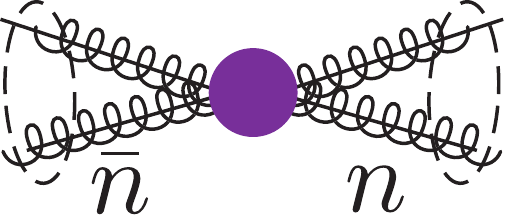}}}  \nn
\end{align}
\vspace{-0.4cm}
\begin{alignat}{2} \label{eq:H_basis_gggg_1}
&O_{4g1++++}^{(2)a b c d}
= \,\frac{1}{4}  \cB^a_{n +} \cB^b_{n +} \cB^c_{\bn +} \cB^d_{\bn +} \,H
\,,\qquad &
&O_{4g1+-+-}^{(2)a b c d}
= \, \cB^a_{n +} \cB^b_{n -} \cB^c_{\bn +} \cB^d_{\bn -} \,H
\,,\\
&O_{4g1----}^{(2)a b c d}
= \,\frac{1}{4}  \cB^a_{n -} \cB^b_{n -} \cB^c_{\bn -} \cB^d_{\bn -} \,H
\,.\qquad &\nn
\end{alignat}
The spin zero nature of the Higgs implies that a number of helicity configurations do not contribute, and therefore are not included in our basis operators here.
The color basis after BPS field redefinition is given by
\begin{equation}
\vT_{\text{BPS}}^{ abcd} =
\frac{1}{2}\begin{pmatrix}
(\tr[ T^{a'}   T^{b'}     T^{c'}   T^{d'}   ] + \tr[  T^{d'}   T^{c'}     T^{b'}   T^{a'}  ]) \cY^{a' a}_{n} \cY^{b' b}_{n} \cY^{c' c}_{\bn} \cY^{d' d}_{\bn}
\\
(\tr[ T^{a'}    T^{c'}   T^{d'}     T^{b'}  ] + \tr[ T^{b'}     T^{d'}   T^{c'}    T^{a'} ] ) \cY^{a' a}_{n} \cY^{b' b}_{n} \cY^{c' c}_{\bn} \cY^{d' d}_{\bn}
\\
(\tr[ T^{a'}    T^{d'}     T^{b'}     T^{c'}   ] + \tr[  T^{c'}     T^{b'}     T^{d'}    T^{a'}  ] ) \cY^{a' a}_{n} \cY^{b' b}_{n} \cY^{c' c}_{\bn} \cY^{d' d}_{\bn}
\\
(\tr[ T^{a'}   T^{b'}     T^{c'}   T^{d'}   ] - \tr[  T^{d'}   T^{c'}     T^{b'}   T^{a'}  ]) \cY^{a' a}_{n} \cY^{b' b}_{n} \cY^{c' c}_{\bn} \cY^{d' d}_{\bn}
\\
(\tr[ T^{a'}    T^{c'}   T^{d'}     T^{b'}  ] - \tr[ T^{b'}     T^{d'}   T^{c'}    T^{a'} ] ) \cY^{a' a}_{n} \cY^{b' b}_{n} \cY^{c' c}_{\bn} \cY^{d' d}_{\bn}
\\
(\tr[ T^{a'}    T^{d'}     T^{b'}     T^{c'}   ] - \tr[  T^{c'}     T^{b'}     T^{d'}    T^{a'}  ] ) \cY^{a' a}_{n} \cY^{b' b}_{n} \cY^{c' c}_{\bn} \cY^{d' d}_{\bn}
\\
\frac{1}{2}  \delta^{ab} \delta^{cd} 
\\
\frac{1}{2} (\cY^T_n \cY_\bn)^{ac} (\cY^T_n \cY_\bn)^{bd} 
\\
\frac{1}{2} (\cY^T_n \cY_\bn)^{ad} (\cY^T_n \cY_\bn)^{bc}
\end{pmatrix}^{\!\!\!T} 
.\end{equation}

The other relevant case has three gluons in one sector, which we take to be the $\bn$ collinear sector. The basis of operators is then given by
\begin{align}
& \boldsymbol{(g)_n (ggg)_{\bn}:}{\vcenter{\includegraphics[width=0.18\columnwidth]{figures/Subleading_4ga_low}}}  \nn
\end{align}
\vspace{-0.4cm}
\begin{alignat}{2} \label{eq:H_basis_gggg_2}
&O_{4g2+++-}^{(2)a b c d}
= \,\frac{1}{2}  \cB^a_{n +} \cB^b_{\bn +} \cB^c_{\bn +} \cB^d_{\bn -} \,H
\,,\qquad &
&O_{4g2-+--}^{(2)a b c d}
= \,\frac{1}{2}  \cB^a_{n -} \cB^b_{\bn +} \cB^c_{\bn -} \cB^d_{\bn -} \,H
\,.
\end{alignat}
In this case, the post-BPS color basis is given by
\begin{equation}
\vT_{\text{BPS}}^{ abcd} =
\frac{1}{2}\begin{pmatrix}
(\tr[ T^{a'}   T^{b'}     T^{c'}   T^{d'}   ] + \tr[  T^{d'}   T^{c'}     T^{b'}   T^{a'}  ]) \cY^{a' a}_{n} \cY^{b' b}_{\bn} \cY^{c' c}_{\bn} \cY^{d' d}_{\bn}
\\
(\tr[ T^{a'}    T^{c'}   T^{d'}     T^{b'}  ] + \tr[ T^{b'}     T^{d'}   T^{c'}    T^{a'} ] ) \cY^{a' a}_{n} \cY^{b' b}_{\bn} \cY^{c' c}_{\bn} \cY^{d' d}_{\bn}
\\
(\tr[ T^{a'}    T^{d'}     T^{b'}     T^{c'}   ] + \tr[  T^{c'}     T^{b'}     T^{d'}    T^{a'}  ] ) \cY^{a' a}_{n} \cY^{b' b}_{\bn} \cY^{c' c}_{\bn} \cY^{d' d}_{\bn}
\\
(\tr[ T^{a'}   T^{b'}     T^{c'}   T^{d'}   ] - \tr[  T^{d'}   T^{c'}     T^{b'}   T^{a'}  ]) \cY^{a' a}_{n} \cY^{b' b}_{\bn} \cY^{c' c}_{\bn} \cY^{d' d}_{\bn}
\\
(\tr[ T^{a'}    T^{c'}   T^{d'}     T^{b'}  ] - \tr[ T^{b'}     T^{d'}   T^{c'}    T^{a'} ] ) \cY^{a' a}_{n} \cY^{b' b}_{\bn} \cY^{c' c}_{\bn} \cY^{d' d}_{\bn}
\\
(\tr[ T^{a'}    T^{d'}     T^{b'}     T^{c'}   ] - \tr[  T^{c'}     T^{b'}     T^{d'}    T^{a'}  ] ) \cY^{a' a}_{n} \cY^{b' b}_{\bn} \cY^{c' c}_{\bn} \cY^{d' d}_{\bn}
\\
\frac{1}{2}  (\cY^T_n \cY_\bn)^{ab} \delta^{cd} 
\\
\frac{1}{2} (\cY^T_n \cY_\bn)^{ac} \delta^{bd} 
\\
\frac{1}{2} (\cY^T_n \cY_\bn)^{ad} \delta^{bc}
\end{pmatrix}^{\!\!\!T} 
.\end{equation}

The helicity basis has made extremely simple the task of writing down a complete and minimal basis of four gluon operators, which would be much more difficult using traditional Lorentz structures. The helicity operators also make it simple to implement the constraints arising from the spin zero nature of the Higgs.

\vspace{0.4cm}
\noindent{\bf{Four Quark Operators:}}

We now consider the case of operators involving four collinear quark fields. These operators are again highly constrained by the helicity selection rules and chirality conservation, since each quark-antiquark pair was produced from a gluon splitting. In particular, these two constraints imply that there are no operators with non-vanishing Wilson coefficients with three quarks in one collinear sector. Therefore, we need only consider the cases where there are two quarks in each collinear sector.

When constructing the operator basis we must also treat separately the case of identical quark flavors $H q \bar q q\bar q$ and distinct quark flavors $H  q \bar q Q\bar Q $. For the case of distinct quark flavors $H q \bar q Q\bar Q $ we will have a $q\leftrightarrow Q$ symmetry for the operators. Furthermore the two quarks of flavor $q$, and the two quarks of flavor $\bar Q$, are necessarily of the same chirality.  In the case that both quarks of the same flavor appear in the same current, the current will be labeled by the flavor. Otherwise, the current will be labeled with ($q \bar{Q}$) or ($Q\bar{q}$) appropriately. For all these cases, the color basis is
\begin{equation} \label{eq:qqqq_color}
\vT^{\, \al\bbeta\ga\bdelta} =
\Bigl(
\de_{\al\bdelta}\, \de_{\ga\bbeta}\,,\, \delta_{\al\bbeta}\, \de_{\ga\bdelta}
\Bigr)
\,.\end{equation}
We will give results for the corresponding $\bar T_{\rm BPS}^{\, \al\bbeta\ga\bdelta}$ basis as we consider each case below.

For the case of operators with distinct quark flavors $H q \bar q Q\bar Q $ and two collinear quarks in each of the $n$ and $\bn$ sectors there are three possibilities. There is either a quark anti-quark pair of the same flavor in each sector (e.g. $(q \bar q)_n(Q \bar Q)_\bn$), a quark and an antiquark of distinct flavors in the same sector (e.g. $(q \bar Q)_n(Q \bar q)_\bn$), or two quarks with distinct flavors in the same sector(e.g. $(q Q)_n(\bar q \bar Q)_\bn$). In the case that there is a quark anti-quark pair of the same flavor in each sector, the basis of helicity operators is
\vspace{0.3cm}\begin{align}
 \boldsymbol{(q \bar q)_n(Q \bar Q)_\bn:}    {\vcenter{\includegraphics[width=0.18\columnwidth]{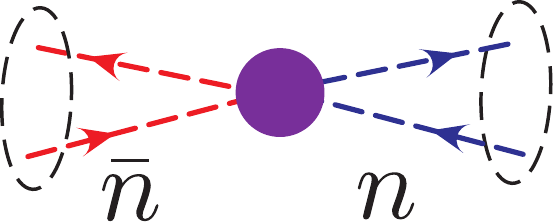}}}
\nn
\end{align}
\vspace{-0.4cm}
\begin{alignat}{2} \label{eq:Z2_basis_qQ}
&O_{qQ1(0;0)}^{(2)\balpha\bt\bgamma\delta}
= \, J_{(q)n 0\, }^{\balpha\bt}\, J_{(Q) \bar n 0\, }^{\bgamma\delta} H
\,,\qquad &
&O_{qQ1(0;\bar 0)}^{(2)\balpha\bt\bgamma\delta}
= \, J_{(q) n  0\, }^{\balpha\bt}\, J_{(Q) \bar n \bar 0\,}^{\bgamma\delta} H
\,,\\
&O_{qQ1(\bar 0;0)}^{(2)\balpha\bt\bgamma\delta}
= \,  J_{(q)n  \bar 0\, }^{\balpha\bt}\, J_{(Q) \bar n 0\, }^{\bgamma\delta} H
\,,\qquad &
&O_{qQ1(\bar 0;\bar 0)}^{(2)\balpha\bt\bgamma\delta}
=\, J_{(q) n \bar 0\, }^{\balpha\bt}\, J_{(Q) \bar n \bar 0\, }^{\bgamma\delta} H
\,,\nn
\end{alignat}
where we have chosen the $q$ quark to be in the $n$ sector. Since all the operators have total helicity $0$ along the $\hat n$ direction, there are only chirality constraints here and no constraints  from angular momentum conservation.  In the case that there is a quark anti-quark of distinct flavors in the same sector, chirality and angular momentum conservation constrains the basis to be 
\begin{align}
\boldsymbol{(q \bar Q)_n(Q \bar q)_\bn:}     {\vcenter{\includegraphics[width=0.18\columnwidth]{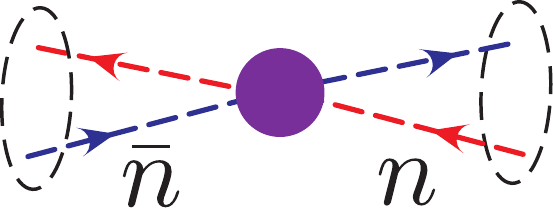}}}
\nn
\end{align}
\vspace{-0.4cm}
\begin{align} \label{eq:Z2_basis_qQ_2}
&O_{qQ2(0;0)}^{(2)\balpha\bt\bgamma\delta}
=\, J_{(q \bar{Q}) n 0\, }^{\balpha\bt}\, J_{(Q\bar{q} ) \bar{n} 0\, }^{\bgamma\delta} H
\,,\qquad 
O_{qQ2(\bar 0;\bar 0)}^{(2)\balpha\bt\bgamma\delta}
=\, J_{(q \bar{Q}) n \bar 0\, }^{\balpha\bt}\, J_{(Q\bar{q} )\bar{n} \bar 0\, }^{\bgamma\delta} H
\,.
\end{align}
For the operators in \eqs{Z2_basis_qQ}{Z2_basis_qQ_2} the color basis after BPS field redefinition is
\begin{align}  \label{eq:TBPS_OqQ12}
\vT_{\BPS}^{ \al\bbeta\ga\bdelta} &=
\left( \Big[ Y_{n}^\dagger Y_{\bar n}  \Big]_{\al\bdelta}\,\Big[ Y_{\bar n}^\dagger Y_{n}  \Big]_{\ga\bbeta}\,,\, \delta_{\al\bbeta}\, \delta_{\ga\bdelta}
\right)
\,.
\end{align}
When there are two quarks of distinct flavors in the same sector the basis of helicity operators is constrained by chirality and reduced further to just two operators by angular momentum conservation, giving 
\begin{align}
\boldsymbol{(q Q)_n(\bar q \bar Q)_\bn:}    {\vcenter{\includegraphics[width=0.18\columnwidth]{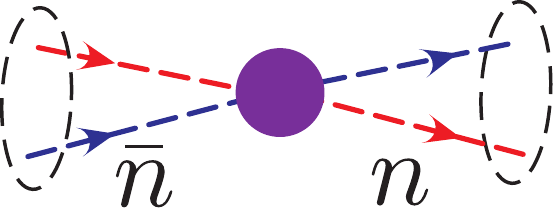}}}
\nn
\end{align}
\vspace{-0.4cm}
\begin{align}\label{eq:Z2_basis_qQ_3}
&O_{qQ3(+;-)}^{(2)\balpha\bt\bgamma\delta}
=\, J_{(q) n \bar n +\, }^{\balpha\bt}\, J_{(Q) n\bar n -\, }^{\bgamma\delta} H
\,,\qquad &
O_{qQ3(-;+)}^{(2)\balpha\bt\bgamma\delta}
= \, J_{(q)n \bar n -\, }^{\balpha\bt}\, J_{(Q) n\bar n +\, }^{\bgamma\delta} H
\,.
\end{align}
For the operators in \eq{Z2_basis_qQ_3} the color basis after BPS field redefinition is
\begin{align} \label{eq:TBPS_OqQ3}
\vT_{\BPS}^{ \al\bbeta\ga\bdelta} &=
\left( \left[ Y_{n}^\dagger Y_{\bar n}  \right ]_{\al\bdelta}\,\left[ Y_{n}^\dagger Y_{\bar n}  \right ]_{\ga\bbeta}
 \,,\, 
 \left[ Y_{n}^\dagger Y_{\bar n}  \right ]_{\al\bbeta}\,\left[ Y_{n}^\dagger Y_{\bar n}  \right ]_{\ga\bdelta}
\right)
\,.
\end{align}

In the cases considered in \eqs{Z2_basis_qQ}{Z2_basis_qQ_2} where there is a quark and antiquark field in the same collinear sector, we have chosen to work in a basis using $J_{i0}^{\balpha\beta}$ and $J_{i\bar 0}^{\balpha\beta}$ which contain only fields in a single collinear sector. One could also construct an alternate form for the basis, for example using the currents $J_{n\bn\lambda}^{\balpha\beta}$. From the point of view of factorization our basis is the most convenient since the fields in the $n$ and $\bar n$-collinear sectors are only connected by color indices, which will simplify later steps of factorization proofs. In the following, we will whenever possible use this logic when deciding between equivalent choices for our basis. 

For identical quark flavors the operators are similar to those in Eqs.~(\ref{eq:Z2_basis_qQ},\ref{eq:Z2_basis_qQ_3}). The distinct operators include
\begin{align}
 \boldsymbol{(q \bar q)_n(q \bar q)_\bn:}    {\vcenter{\includegraphics[width=0.18\columnwidth]{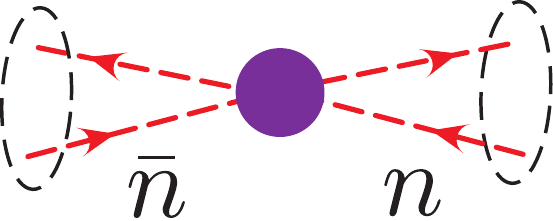}}}
\nn
\end{align}
\vspace{-0.4cm}
\begin{alignat}{2} \label{eq:Z2_basis_qq}
&O_{qq1(0;0)}^{(2)\balpha\bt\bgamma\delta}
= \,\frac14\, J_{(q)n 0\, }^{\balpha\bt}\, J_{(q) \bar n 0\, }^{\bgamma\delta} H
\,,\qquad &
&\phantom{O_{qq1(0;\bar 0)}^{(2)\balpha\bt\bgamma\delta}
= \, J_{(q) n  0\, }^{\balpha\bt}\, J_{(q) \bar n \bar 0\,}^{\bgamma\delta} H
\,,}\\
&O_{qq1(\bar 0;0)}^{(2)\balpha\bt\bgamma\delta}
= \,  J_{(q)n  \bar 0\, }^{\balpha\bt}\, J_{(q) \bar n 0\, }^{\bgamma\delta} H
\,,\qquad &
&O_{qq1(\bar 0;\bar 0)}^{(2)\balpha\bt\bgamma\delta}
=\, \frac14\, J_{(q) n \bar 0\, }^{\balpha\bt}\, J_{(q) \bar n \bar 0\, }^{\bgamma\delta} H
\,,\nn
\end{alignat}
\begin{align}
\boldsymbol{(q q)_n(\bar q \bar q)_\bn:}    {\vcenter{\includegraphics[width=0.18\columnwidth]{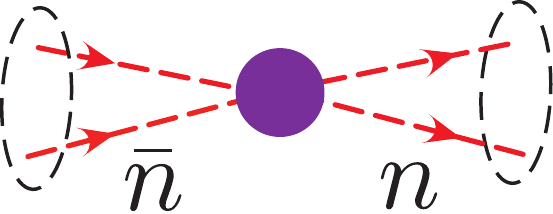}}}
\nn
\end{align}
\vspace{-0.4cm}
\begin{align}\label{eq:Z2_basis_qq_3}
&O_{qq3(+;-)}^{(2)\balpha\bt\bgamma\delta}
=\, J_{(q) n \bar n +\, }^{\balpha\bt}\, J_{(q) n\bar n -\, }^{\bgamma\delta} H
\,.
\end{align}
Note that in \Eq{eq:Z2_basis_qq} there are only three operators due to the equivalence between the two operators
\begin{align}
 \sum_n J_{(q) n  0\, }^{\balpha\bt}\, J_{(q) \bar n \bar 0\,}^{\bgamma\delta} H 
\equiv \,  \sum_n J_{(q)n  \bar 0\, }^{\balpha\bt}\, J_{(q) \bar n 0\, }^{\bgamma\delta} H
\,,
\end{align}
due to the fact that the $n$ label is summed over, as in \Eq{eq:sum_dir}.
We also have the same color bases as in \eqs{TBPS_OqQ12}{TBPS_OqQ3} for $O_{qq1}^{(2)}$ and $O_{qq3}^{(2)}$ respectively.

\subsubsection{$\cP_\perp$ Insertions}\label{sec:nnlp_perp}

Since we have chosen to work in a frame where the total $\perp$ momentum of each collinear sector vanishes, operators involving explicit insertions of the $\cP_\perp$ operator first appear at $\cO(\lambda^2)$. The $\cP_\perp$ operator can act only in a collinear sector composed of two or more fields. At $\cO(\lambda^2)$, there are then only two possibilities, namely that the $\cP_\perp$ operator is inserted into an operator involving two quark fields and a gluon field, or it is inserted into an operator involving three gluon fields. 

In the case that the $\cP_\perp$ operator is inserted into an operator involving two quark fields and a gluon field, the helicity structure of the operator is highly constrained. In particular, the quark fields must be in a helicity zero configuration. Combined with the fact that they must have the same chirality, this implies that all operators must involve only the currents $J_{\bar n\, 0}^{\balpha\bt}$ or $J_{\bar n\, \bar 0}^{\balpha\bt}$. Here we have again taken without loss of generality that the two quarks are in the $\bar n$-collinear sector. A basis of operators is then given by
\begin{align}
&   \boldsymbol{(g)_n (q\bar q\, \cP_\perp)_{\bn}:}{\vcenter{\includegraphics[width=0.18\columnwidth]{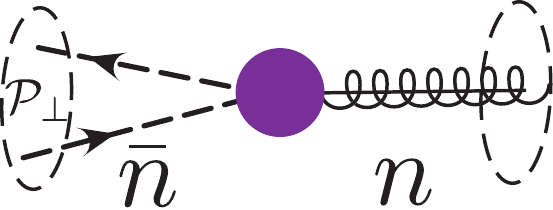}}}  \nn
\end{align}
\vspace{-0.4cm}
\begin{alignat}{2}\label{eq:Hqqgpperp_basis_same}
&O_{\cP \chi + (0)[+]}^{(2)a\,\balpha\bt}
= \cB_{n+}^a\, \big\{ \cP_{\perp}^{+} J_{\bar n\, 0}^{\balpha\bt} \big\}\,  H
\,,\qquad &
&O_{\cP\chi - (0)[-]}^{(2)a\,\balpha\bt}
= \cB_{n-}^a\, \big\{ \cP_{\perp}^{-} J_{\bar n\, 0    }^{\balpha\bt} \big\} \, H 
\,,\\
&O_{\cP\chi + (\bar 0)[+]}^{(2)a\,\balpha\bt}
=  \cB_{n+}^a\,  \big\{ \cP_{\perp}^{+} J_{\bar n\, \bar0    }^{\balpha\bt} \big\}\,  H
\,,\qquad &
&O_{\cP\chi - (\bar 0)[-]}^{(2)a\,\balpha\bt}
= \cB_{n-}^a \, \big\{ \cP_{\perp}^{-} J_{\bar n\, \bar0    }^{\balpha\bt} \big\}  \, H
\,.\nn
\end{alignat}
Since we have assumed that the total $\cP_\perp$ in each collinear sector is zero, integration by parts can be used to make the $\cP_\perp$ operator act only on either the quark, or the antiquark field, which has been used in \Eq{eq:Hqqgpperp_basis_same}. (The additional operators that are needed when we relax this assumption are discussed in Appendix A of \cite{Moult:2017rpl}.)
The color basis is one-dimensional
\begin{equation} \label{eq:nnlp_color_quark_perp}
\vT^{a\, \al\bbeta} = T^a_{\al\bbeta}\,.
\end{equation}
After BPS field redefinition the structure is given by
\begin{align} \label{eq:nnlp_color_quark_perpBPS}
\vT_{\BPS}^{ a \al\bbeta} 
=\left ( Y^\dagger_{\bar n} T^b \cY_n^{ba}   Y_{\bar n}  \right )_{\alpha \bar \beta}
= \bigl( \cY_n^T \cY_\bn \bigr)^{ac}\: T^c_{\alpha \bar \beta} 
\,.
\end{align}

In the case that the $\cP_\perp$ operator is inserted into an operator involving three gluon fields, the helicity selection rules simply imply that the helicities must add to zero.
A basis of operators involving three collinear gluon fields and a $\cP_\perp^\pm$ insertion is given by
\begin{align}
&  \boldsymbol{(g)_n (gg\, \cP_\perp)_{\bn}:}{\vcenter{\includegraphics[width=0.18\columnwidth]{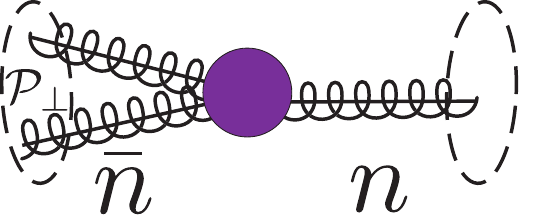}}}  \nn
\end{align}
\vspace{-0.4cm}
\begin{alignat}{2} \label{eq:Hgggpperp_basis}
&O_{\cP\cB +++[-]}^{(2)abc}
=  \cB_{n+}^a\, \cB_{\bar n+}^b\, \left [\cP_{\perp}^{-} \cB_{\bar n+}^c \right ] \,H
\,,\qquad && O_{\cP\cB ---[+]}^{(2)abc}
= \cB_{n-}^a\, \cB_{\bar n-}^b\, \left [\cP_{\perp}^{+} \cB_{\bar n-}^c \right ] \,H\,,
 \nn \\
&O_{\cP\cB ++-[+]}^{(2)abc}
= \, \cB_{n+}^a\, \cB_{ \bar n+}^b\, \left [\cP_{\perp}^{+} \cB_{\bar n-}^c \right ] \,H
\,,\qquad 
&&O_{\cP\cB --+[-]}^{(2)abc}
= \cB_{n-}^a\, \cB_{ \bn -}^b\, \left [\cP_{\perp}^{-} \cB_{\bar n +}^c \right ] \,H
\,.
\end{alignat}
Note that the analogous operators with the helicities $O_{\cP\cB +-+[+]}^{(2)abc}$ and $O_{\cP\cB -+-[-]}^{(2)abc}$ are not eliminated, but instead are equivalent to those in the last row by integrating the $\cP_\perp^\pm$  by parts onto the other $\bn$-collinear field since the total $\cP_\perp$ in each collinear sector is zero. (The additional operators that are needed when we relax this assumption are discussed in Appendix A of \cite{Moult:2017rpl}.)

The basis of color structures here is two dimensional,
\begin{equation} \label{eq:ggg_perp_color}
\vT^{abc} =
\begin{pmatrix}
  i  f^{abc} \\ d^{abc}
\end{pmatrix}
\,,
\qquad
\vT_{\BPS}^{abc} =
\begin{pmatrix}
  i  f^{a'b'c'}\, {\cal Y}_n^{a'a} {\cal Y}_\bn^{b'b} {\cal Y}_{\bn}^{c'c}  \\  
  d^{a'b'c'}\, {\cal Y}_n^{a'a} {\cal Y}_\bn^{b'b} {\cal Y}_{\bn}^{c'c} 
\end{pmatrix}
 =
\begin{pmatrix}
   i f^{bcd} {\cal Y}_{\bn}^{a'd} {\cal Y}_n^{a'a}   \\  
  d^{bcd}\, {\cal Y}_\bn^{a'd}  {\cal Y}_n^{a'a}  
\end{pmatrix}
\,.
\end{equation}
In the BPS redefined color structure we have written it both in a form that makes the structure of the Wilson lines appearing from the field redefinition clear, as well as in a simplified form.

\subsubsection{Ultrasoft Insertions}\label{sec:nnlp_soft}

At $\cO(\lambda^2)$ we have the possibility of operators with explicit ultrasoft insertions. To have label momentum conservation these operators must have a collinear field in each collinear sector. Interestingly, despite the fact that the leading power operator has two collinear gluon fields, for the operators involving an ultrasoft insertion one can have either two collinear quark fields, or two collinear gluon fields.

The construction of an operator basis involving ultrasoft gluons is more complicated due to the fact that they are not naturally associated with a given lightcone direction. There are therefore different choices that can be made when constructing the basis. We will choose to work in a basis where all ultrasoft derivatives acting on ultrasoft Wilson lines are absorbed into $\cB_{us}$ fields. To understand why it is always possible to make this choice, we consider two pre-BPS operators involving two collinear quark fields, and an ultrasoft derivative
\begin{align}
O^\mu_1=\bar \chi_{\bar n} (i D_{us}^\mu) \chi_n\,, \qquad O^\mu_2=\bar \chi_{\bar n} (-i \overleftarrow D_{us}^\mu) \chi_n\,,
\end{align}
where $(-i \overleftarrow D_{us}^\mu)=(i D_{us}^\mu)^\dagger$ and we have not made the contraction of the $\mu$ index explicit, as it is irrelevant to the current discussion. Performing the BPS field redefinition, we obtain 
\begin{align}
O^\mu_{1\text{BPS}}= i \bar \chi_{\bar n}Y_{\bar n}^\dagger D_{us}^\mu Y_n \chi_n\,, \qquad
 O^\mu_{2\text{BPS}}=-i \bar \chi_{\bar n} Y_{\bar n}^\dagger \overleftarrow D_{us}^\mu Y_n \chi_n\,.
\end{align}
If we want to absorb all derivatives acting on Wilson lines into $\cB_{us}$ fields, we must organize the Wilson lines in the operators as
\begin{align}
O^\mu_{1\text{BPS}}= i\bar \chi_{\bar n}Y_{\bar n}^\dagger Y_n (Y_n^\dagger  D_{us}^\mu Y_n) \chi_n\,, \qquad 
O^\mu_{2\text{BPS}}=-i \bar \chi_{\bar n} (Y_{\bar n}^\dagger \overleftarrow D_{us}^\mu Y_{\bar n}) Y_{\bar n}^\dagger  Y_n \chi_n\,.
\end{align}
Using \Eq{eq:soft_gluon} we see that this can be written entirely in terms of $\partial_{us}$ operators acting on collinear fields, and the two ultrasoft gauge invariant gluon fields $\cB_{us(n)}$ and $\cB_{us(\bar n)}$ for $O^\mu_{1\text{BPS}}$ and $O^\mu_{2\text{BPS}}$ respectively. Note, however, that ultrasoft gluon fields defined with respect to both lightcone directions are required. Alternatively, it is possible to work only with $\cB_{us(n)}$, for example, but in this case we see that the ultrasoft derivative must also be allowed to act explicitly on pairs of ultrasoft Wilson lines, for example $[\partial_{us}^\mu (Y_n^\dagger Y_\bn)]$. In constructing our complete basis we will choose to avoid this so that ultrasoft derivatives acting on soft Wilson lines occur only within the explicit $\cB_{us}$ fields. This choice also makes our basis more symmetric.

For the operators involving one ultrasoft gluon and two collinear quarks we have the basis
\begin{align}
&  \boldsymbol{g_{us}(q)_n (\bar q)_{\bn}:}{\vcenter{\includegraphics[width=0.18\columnwidth]{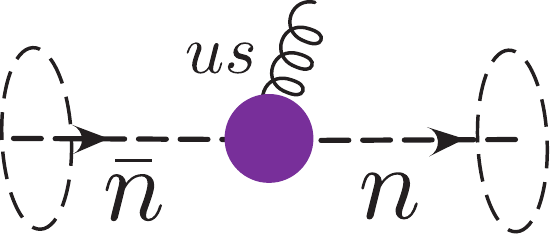}}}  \nn
\end{align}
\vspace{-0.4cm}
\begin{alignat}{2} \label{eq:soft_insert_basis}
&O_{\chi(us(n))-:(+)}^{(2)a\,\balpha\bt}
=  \cB_{us(n)-}^a\, J_{n\bar n\,+}^{\balpha\bt}\, H
\,,\qquad &
&O_{\chi(us(n))+:(-)}^{(2)a\,\balpha\bt}
=\cB_{us(n)+}^a \, J_{n\bar n\, -}^{\balpha\bt}\, H
\,, 
\end{alignat}
with the unique color structure
\begin{equation} 
\vT_{\BPS}^{\,a\, \al\bbeta} = \left ( T^a Y^\dagger_{n} Y_{\bn} \right )_{\alpha \bar\beta}
\,,\end{equation}
and
\begin{alignat}{2} \label{eq:soft_insert_basis2}
&O_{\chi(us(\bar n))+:(+)}^{(2)a\,\balpha\bt}
=  \cB_{us(\bar n)+}^a\, J_{n\bar n\,+}^{\balpha\bt}\, H
\,,\qquad &
&O_{\chi(us)(\bar n))-:(-)}^{(2)a\,\balpha\bt}
=\cB_{us(\bar n)-}^a \, J_{n\bar n\, -}^{\balpha\bt}\, H
\,, 
\end{alignat}
with the unique color structure
\begin{equation} 
\vT_{\BPS}^{\,a\, \al\bbeta}=(Y_n^\dagger Y_\bn T^a)_{\alpha\bbeta}
\,.\end{equation}
Note that the color structures associated with the two different projections of the $\cB_{us}$ field are distinct. All other helicity combinations vanish due to helicity selection rules. The helicity selection rules differ between \Eq{eq:soft_insert_basis} and \Eq{eq:soft_insert_basis2} due to the different choice of reference vector for the ultrasoft field in the two cases.

We also have operators involving two collinear quark fields and a single ultrasoft derivative, 
\begin{align}
&  \boldsymbol{\partial_{us} (q)_n (\bar q)_{\bn}:}{\vcenter{\includegraphics[width=0.18\columnwidth]{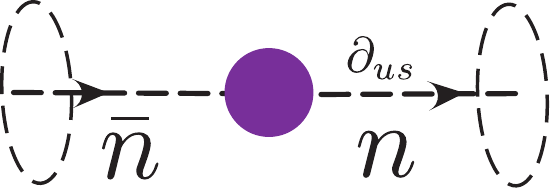}}}  \nn
\end{align}
\vspace{-0.4cm}
\begin{alignat}{2} \label{eq:soft_derivative_basis}
&O_{\partial \chi (us(n))-:(+)}^{(2)\,\balpha\bt}
=  \{\partial_{us(n)-}\, J_{n\bar n\,+}^{\balpha\bt}\}\, H
\,,\qquad &
&O_{\partial \chi (us(n))+:(-)}^{(2)\,\balpha\bt}
=  \{\partial_{us(n)+} \, J_{n\bar n\, -}^{\balpha\bt}\}\, H\,,
\end{alignat}
with the unique color structure given before and after BPS field redefinition by
\begin{align} \label{eq:leading_color_deriv}
 \vT^{\al\bbeta} = (\de_{\al\bbeta})\,, \qquad 
 \vT_{\BPS}^{\al\bbeta} =  \big[Y_n^\dagger Y_{\bar n} \big]_{\al\bbeta} 
\,,
\end{align}
and
\begin{alignat}{2} \label{eq:soft_derivative_basis2}
&O_{\partial^\dagger \chi (us(\bar n))+:(+)}^{(2)\,\balpha\bt}
=  \{ J_{n\bar n\,+}^{\balpha\bt}\,     (i\partial_{us(\bar n)+})^\dagger\}\, H
\,,~ &
&O_{\partial^\dagger \chi (us(\bar n))-:(-)}^{(2)\,\balpha\bt}
=  \{  J_{n\bar n\, -}^{\balpha\bt}\, (i\partial_{us(\bar n)-})^\dagger\}\, H\,,
\end{alignat}
with the unique color structure given before and after BPS field redefinition by
\begin{align} \label{eq:leading_color_deriv2}
 \vT^{\al\bbeta} = (\de_{\al\bbeta})\,, \qquad 
 \vT_{\BPS}^{\al\bbeta} =  \big[Y_n^\dagger Y_{\bar n} \big]_{\al\bbeta} 
\,.
\end{align}
Although the color structure happens to be the same in both cases, we have separated them to highlight the different decompositions of the ultrasoft derivatives in the two cases. Note that the form of the ultrasoft derivatives which appear is constrained by the helicity constraints.

Similarly, we have the corresponding operators involving two collinear gluons. A basis of helicity operators involving two collinear gluons and a single ultrasoft gluon field is given by
\begin{align}
& \boldsymbol{g_{us}(g)_n (g)_{\bn}:}{\vcenter{\includegraphics[width=0.18\columnwidth]{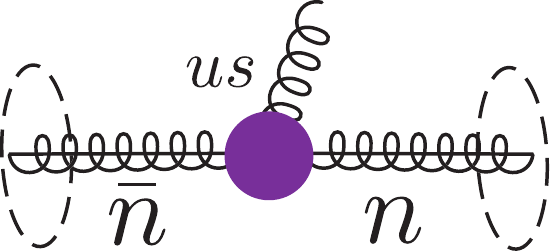}}}  \nn
\end{align}
\vspace{-0.4cm}
\begin{align}\label{eq:Hgggus}
  &O_{\cB (us(n))0:++}^{(2)abc}
  =  \cB_{us(n)0}^a\, \cB_{n+}^b\, \cB_{\bar n+}^c\,  H
  \,, \qquad && O_{\cB (us(n))0:--}^{(2)abc} 
  =   \cB_{ us(n)0}^a\, \cB_{n-}^b\, \cB_{\bar n-}^c\,  H
  \,, 
\end{align}
with the basis of color structures,\footnote{In order to see how the Wilson line structure in \Eq{eq:Z2g_colorus} arises, we look at the object $D_{us}^{ab} \cB_{n}^c \cB_{\bn}^d$ pre-BPS field redefinitions. This object must be contracted with a tensor to make it a singlet under ultrasoft gauge transformations. Each of these resulting forms can be mapped onto the color structures of \Eq{eq:Z2g_colorus} after performing the BPS field redefinition} 
\begin{equation} \label{eq:Z2g_colorus}
\vT_{\BPS}^{abc} =
\begin{pmatrix}
i  f^{abd}\, \big({\cal Y}_n^T {\cal Y}_{\bar n}\big)^{dc} \\
 d^{abd}\, \big({\cal Y}_n^T {\cal Y}_{\bar n}\big)^{dc} 
\end{pmatrix}^T
\,,
\end{equation}
and
\begin{align}\label{eq:Hgggus_2}
  &O_{\cB (us(\bar n))0:++}^{(2)abc}
  =  \cB_{us(\bar n)0}^a\, \cB_{n+}^b\, \cB_{\bar n+}^c\,  H
  \,, \qquad && O_{\cB (us(\bar n))0:--}^{(2)abc} 
  =   \cB_{ us(\bar n)0}^a\, \cB_{n-}^b\, \cB_{\bar n-}^c\,  H
  \,, 
\end{align}
with the basis of color structures,
\begin{equation} \label{eq:Z2g_colorus_2}
\vT_{\BPS}^{abc} =
\begin{pmatrix}
i  f^{abd}\, \big({\cal Y}_{\bar n}^T {\cal Y}_{n}\big)^{dc} \\
 d^{abd}\, \big({\cal Y}_{\bar n}^T {\cal Y}_{n}\big)^{dc} 
\end{pmatrix}^T
\,.
\end{equation}
We have only included the $\vT_{\BPS}^{abc}$ version of the color structure here because the $\cB_{us(n)\lambda}^a$ are generated by BPS field redefinition. 

The Wilson coefficients of the operators that include $\cB_{us(n)0}$ can be related to the Wilson coefficients of the leading power operators using RPI symmetry (see \cite{Larkoski:2014bxa}). In particular, we have 
\begin{align} \label{eq:usRPIrelation}
C^{(2)}_{\cB n(us)0:\lambda_1, \lambda_1}&=-\frac{\partial C^{(0)}_{\lambda_1, \lambda_1} }{\partial \omega_1}  
\,, 
\end{align}
where $C^{(0)}_{\lambda_1, \lambda_1}$ is the Wilson coefficient for the leading power operator of \Eq{eq:hgg}. We will explicitly verify this at the level of tree level matching in \Sec{sec:matching}.

We must also consider operators with an insertion of $\partial_{us(n)}$ with two collinear gluons in different collinear sectors. The gluon equations of motion allow us to eliminate the operators $in\cdot \partial \cB_{n\perp}$ and $i\bar n\cdot \partial \cB_{\bar n\perp}$, which can be rewritten purely in terms of collinear objects~\cite{Marcantonini:2008qn}. Furthermore, we again choose to organize our basis of operators such that ultrasoft derivatives act on ultrasoft Wilson lines only within the $\cB_{us}$ fields, as was done in the quark case. (We also do not include operators where the ultrasoft derivative acts on the Higgs field, since this is moved to the other fields by integration by parts.) The basis of operators involving ultrasoft derivatives is then given by
\begin{align}
& \boldsymbol{\partial_{us}(g)_n (g)_{\bn}:}{\vcenter{\includegraphics[width=0.18\columnwidth]{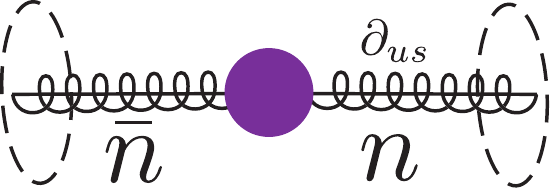}}}  \nn
\end{align}
\vspace{-0.4cm}
\begin{alignat}{2}\label{eq:Hdggus}
&O_{\partial \cB (us(n))\bar 0:++}^{(2)ab}
=  \cB_{n+}^a  \, \left[ \partial_{us(n)\bar 0} \cB_{\bar n+}^b\right] \,H
\,, \,~~ && O_{\partial \cB (us(n))\bar 0:--}^{(2)ab} 
=    \cB_{n-}^a \left[\partial_{us(n)\bar 0} \cB_{\bar n-}^b\right]\, H
\,,
\end{alignat}
with the basis of color structures
\begin{equation} \label{eq:Z2gd_colorus}
\vT_{\BPS}^{ab} =
\big({\cal Y}_n^T {\cal Y}_{\bar n}\big)^{ab} 
\,.
\end{equation}
and
\begin{alignat}{2}\label{eq:Hdggus2}
&O_{\partial \cB  (us(\bar n))0:++}^{(2)ab}
=  \left[ \partial_{us(\bar n)0}\, \cB_{n+}^a\right]\, \cB_{\bar n+}^b\,  H
\,, \,~~ && O_{\partial \cB  (us(\bar n))0:--}^{(2)ab} 
=  \left [ \partial_{us(\bar n)0}\, \cB_{n-}^a \right] \, \cB_{\bar n-}^b\,  H
\,,
\end{alignat}
with the basis of color structures
\begin{equation} \label{eq:Z2gd_colorus2}
\vT_{\BPS}^{ab} =
\big({\cal Y}_{\bar n}^T {\cal Y}_{n}\big)^{ab} 
\,.
\end{equation}

The Wilson coefficients of the operators that include $\partial_{us(n)0}$ can also be related to the Wilson coefficients of the leading power operators using RPI symmetry (see \cite{Larkoski:2014bxa}). In particular, we have 
\begin{align} \label{eq:usRPIrelationb}
C^{(2)}_{\partial \cB (us(\bar n)) 0:\lambda_1 \lambda_1}&=-\frac{\partial C^{(0)}_{\lambda_1,\lambda_1} }{\partial \omega_1}  
\,, 
\end{align}
where $C^{(0)}_{\lambda_1,\lambda_1}$ is the Wilson coefficient for the leading power operator of \Eq{eq:hgg}. We will explicitly show how this arises in the tree level matching in \Sec{sec:matching}.

\subsection{Cross Section Contributions and Factorization}\label{sec:discussion_gluonops}

{
\renewcommand{\arraystretch}{1.6}
	\begin{center}
\begin{table}[t!]
	\begin{center}
		\begin{tabular}{| c | l | c | c | c | c | }
			\hline
			& Operators & Factorization  & Beam $n$ &  Beam $\bar n$ & Soft  \\
			\hline 
			$\!\!\mathbf{\mathcal{O}(\lambda^0)}\!\!$ 
			&$O_{\cB}^{(0)}  O_{\cB}^{(0)} $ 
			& $H_g^{(0)} B_{g}^{(0)} B_{g}^{(0)} S_g^{(0)}$
			& $\cB_n \,\hat\delta\, \cB_n$ 
			& $\cB_\bn \,\hat\delta\, \cB_\bn$
			& $\cY_n^T \cY_\bn \widehat\cM^{(0)}\, \cY_\bn^T \cY_n$ 
			\\
			\hline
			$\!\!\mathbf{\mathcal{O}(\lambda^2)}\!\!$ 
			&$O_{\cB\bar n}^{(1)}  O_{\cB\bar n}^{(1)}$
			& $H_{g1}^{(0)} B_q^{(0)} B_{qgg}^{(2)} S_{q}^{(0)}$  
			& $\bar \chi_n \,\hat\delta\,\chi_n $ 
			& $\bar \chi_\bn \cB_\bn \hat\delta\, \cB_\bn\chi_\bn   $   
			& $Y_\bn^\dagger Y_n \widehat\cM^{(0)}\,   Y_n^\dagger Y_\bn$   
			\\
			\cline{2-6}
			&$O^{(0)}  O_{\cB1}^{(2)} $
			& $H_{g2}^{(0)} B_{gqq}^{(2)} B_{g}^{(0)} S_{g}^{(0)}$  
			& $\!\!\bar \chi_n \cB_n  \chi_n \hat\delta\, \cB_n\!\!  $ 
			& $ \cB_\bn \,\hat\delta\, \cB_\bn $   
			& $\cY_n^T \cY_\bn \widehat\cM^{(0)}\, \cY_\bn^T \cY_n$    
			\\
			\cline{2-6}
			&$O^{(0)}  O_{\cP\chi}^{(2)} $
			& $H_{g3}^{(0)} B_g^{(0)} B_{gq P}^{(2)} S_{g}^{(0)}$  
			& $\cB_n \,\hat\delta\,\cB_n $ 
			& $\!\!\bar \chi_\bn [\cP_\perp \chi_\bn] \hat\delta\, \cB_\bn\!\!   $   
			& $\cY_n^T \cY_\bn \widehat\cM^{(0)}\, \cY_\bn^T \cY_n$   
			\\
			\cline{2-6}
			&$O^{(0)} O_{\cP \cB}^{(2)} $
			& $H_{g4}^{(0)} B_{g}^{(0)} B_{gg P}^{(2)} S_{g}^{(0)}$  
			& $\bar \cB_n \,\hat\delta\, \cB_n  $ 
			& $\!\!\cB_\bn [\cP_\perp \cB_\bn] \hat\delta\, \cB_\bn  \!\! $   
			& $\cY_n^T \cY_\bn \widehat\cM^{(0)}\, \cY_\bn^T \cY_n$   
			\\
			\cline{2-6}
			&$O^{(0)}  O_{4g2}^{(2)}$
			& $H_{g5}^{(0)} B_g^{(0)} B_{gg}^{(2)} S_{g}^{(0)}$  
			& $\cB_n \,\hat\delta\,\cB_n $ 
			& $\!\! \cB_\bn \cB_\bn \cB_\bn \hat\delta\, \cB_\bn \!\! $   
			& $\cY_n^T \cY_\bn \widehat\cM^{(0)}\, \cY_\bn^T \cY_n$
			\\
			\cline{2-6}
			&$\!\! O^{(0)}  O_{\cB (us)0}^{(2)} \!\!\!$
			& $H_{g6}^{(0)} B_g^{(0)} B_{g}^{(0)} S_{g\cB}^{(2)}$  
			&  $\cB_n \,\hat\delta\,\cB_n   $	 
			&$\cB_\bn \,\hat\delta\,\cB_\bn   $  
			& $\!\cB_{us(n) 0}\, \cY_n\cY_\bn \widehat\cM^{(0)}\, \cY_\bn \cY_n\!\!$     
			\\
			\cline{2-6}
			& $\!\!O^{(0)}  O_{\partial(us)0}^{(2)} \!\!\!$
			& $H_{g7}^{(0)} B_g^{(0)} B_{g}^{(0)} S_{g\partial 0}^{(2)}$  
			&  $\cB_n \,\hat\delta\,\cB_n   $	 
			&$\cB_\bn \,\hat\delta\,\cB_\bn   $  
			& $\!\partial_{us(n) 0}\, \cY_n\cY_\bn \widehat\cM^{(0)}\, \cY_\bn \cY_n\!\!$     
			\\	
			\cline{2-6}
			& $\!\!O^{(0)}  O_{\partial(us)\bar 0}^{(2)} \!\!\!$
			& $H_{g8}^{(0)} B_g^{(0)} B_{g}^{(0)} S_{g\partial \bar 0}^{(2)}$  
			&  $\cB_n \,\hat\delta\,\cB_n   $	 
			&$\cB_\bn \,\hat\delta\,\cB_\bn   $  
			& $\! \partial_{us(n) \bar 0}\, \cY_n\cY_\bn \widehat\cM^{(0)}\, \cY_\bn \cY_n\!\!$     
			\\
			\hline
		\end{tabular}
		\caption{Subleading beam and soft functions arising from products of hard scattering operators in the factorization of Higgs with a jet veto, and their field content. Helicity and color structures have been suppressed. We have  not included products of operators whose beam and soft functions are identical to those shown by charge conjugation or $n\leftrightarrow \bn$.}		\label{tab:fact_func}
	\end{center}
\end{table}
	\end{center}
}

While the basis of operators presented in this section is quite large, many of the operators will not contribute to a physical cross section at $\cO(\lambda^2)$. In this section we briefly discuss the helicity operator basis, focusing in particular on understanding which operators can contribute to the cross section for an SCET$_\text{I}$ event shape observable, $\tau_B$, measured on $gg\to H$. In \Sec{sec:contribs_lam}, we show that there are no contributions to the cross section from hard scattering operators at $\cO(\lambda)$, which would correspond to power corrections of $\sqrt{\tau_B}$. Then in \Sec{sec:contribs}, we use helicity selection rules to determine which operators can contribute at $\cO(\lambda^2)=\cO(\tau_B)$. The results are summarized in \Tab{tab:summary}.

Given the set of contributing operators, one can then determine the full subleading power factorization theorem for the related observables with Higgs production. Here we restrict ourselves to determining the structure of the factorization theorem terms arising purely from our subleading hard scattering operators, written in terms of hard, beam and soft functions. A summary of these results is given in \Tab{tab:fact_func}.  In many cases the beam and soft functions which appear in the subleading power factorization formula are identical to those at leading power. For the case of the soft functions this simplification arises due to color coherence, allowing a simplification to the Wilson lines in the soft functions that appear. For gluon-gluon and quark-quark color channels the leading power soft functions are
\begin{align}\label{eq:soft_func_def}
S_g^{(0)}=\frac{1}{(N_c^2-1)}  \tr \big\langle 0 \big| \cY^T_{\bar n} \cY_n \widehat{\cM}^{(0)}\cY_n^T \cY_{\bar n} \big|0 \big\rangle\,, \qquad
S_q^{(0)}=\frac{1}{N_c}  \tr \big\langle 0 \big| Y^\dagger_{\bar n} Y_n \widehat{\cM}^{(0)}Y_n^\dagger Y_{\bar n} \big|0 \big\rangle\,, 
\end{align}
and depend on the kinematic variables probed by the measurement operator $\widehat{\cM}^{(0)}$.
For the beam functions, this simplification occurs since the power correction is often restricted to a single collinear sector. The other collinear sector is then described by the leading power beam functions (incoming jet functions) for gluons and quarks~\cite{Stewart:2009yx,Stewart:2010qs}
\begin{align}\label{eq:beam_func_def}
\frac{\delta^{ab}}{N_c^2-1}\, B_g^{(0)}
  &=- \frac{\omega \, \theta(\omega)}{2\pi}  \int \!\!\frac{dx^-}{2|\omega|}\:  e^{\frac{i}{2} \ell^+ x^-}  \Big\langle p \Big|\, \cB^{\mu a}_{n\perp} \big(x^- \text{\small $\frac{n}{2}$}\big)\, \hat{\delta}\, \left[ \delta(\omega-\bar \cP) \cB^{b}_{n\perp \mu}(0) \right] \,\Big|p\Big\rangle
   \,, \\
\frac{\delta^{\alpha \bbeta}}{N_c}\, B_q^{(0)}
  &= \frac{\theta(\omega)}{2\pi} \int \!\!\frac{dx^-}{2|\omega|}\: e^{\frac{i}{2} \ell^+ x^-} 
  \Big\langle p\Big|\, \chi_{n}^{\alpha} \big(x^- \text{\small $\frac{n}{2}$}\big) \frac{\Sl{\bar n}}{2} \,\hat{\delta}\,  \left[ \delta(\omega-\bar \cP)     \bar \chi_{n}^{\bar\beta}(0) \right]
  \,\Big| p \Big\rangle
\,, \nn
\end{align}
where we take $\ell^+ \gg \Lambda_{\rm QCD}^2/\omega$.
The result for the leading power measurement function $\hat{\delta}$ appearing in these beam functions depends on the factorization theorem being treated. Often the beam functions are inclusive in which case $\hat{\delta}=1$, giving functions of the momentum fraction of the struck parton $x$ and a single invariant mass momentum variable,  $B_{g}^{(0)}(x,\omega \ell^+)$ and $B_{q}^{(0)}(x,\omega \ell^+)$. Here we assume an SCET$_{\text{I}}$ type measurement that does not fix the $\cP_\perp$ of the measured particle. This assumption has been explicitly used in writing the form of the beam functions in \Eq{eq:beam_func_def}, as well as in our construction of the operator basis.

\subsubsection{Vanishing at $\cO(\lambda)$}\label{sec:contribs_lam}

We begin by considering possible contributions to the cross section at $\cO(\lambda)$.
While we will not discuss the factorization of the cross section in detail, the contribution of the hard scattering operators to the cross section at $\cO(\lambda)$ can be written schematically as
{\begin{small}
\begin{align}\label{eq:xsec_lam}
&\frac{d\sigma^{(1)} }{d\tau_B} \supset N \sum_{X,i}  \tilde \delta^{(4)}_q  \bra{P_1 P_2} C_i^{(1)} O_i^{(1)}(0) \ket{X}\bra{X} C^{(0)} O^{(0)}(0) \ket{P_1 P_2}   \delta\big( \tau_B - \tau_B^{(0)}(X) \big) +\text{h.c.} 
\,. 
\end{align}
\end{small}}
Here $N$ is a normalization factor, $P_1, P_2$ denote the incoming hadronic states, and we use the shorthand notation $\tilde \delta^{(4)}_q=(2\pi)^4\delta^4(q-p_X)$ for the momentum conserving delta function. This expression should merely be taken as illustrative of the operator contributions, and in particular, we have not made explicit any color or Lorentz index contractions, nor the treatment of the initial state. The summation over all final states, $X$, includes phase space integrations. The measurement of the observable is enforced by $ \delta\big( \tau_B - \tau_B^{(0)}(X) \big) $, where $\tau_B^{(0)}(X)$, returns the value of the observable $\tau_B$ as measured on the final state $X$. The explicit superscript $(0)$ indicates that the measurement function is expanded to leading power, since here we focus on the power suppression due to the hard scattering operators. 

From \Eq{eq:xsec_lam} we see that hard scattering operators contribute to the $\cO(\lambda)$ cross section through their interference with the leading power operator. The $\cO(\lambda)$ basis of operators is given in \Eqs{eq:H1_basis}{eq:H1_basis2}, each of which involves a single collinear quark field in each collinear sector. Conservation of fermion number then immediately implies that these operators cannot have non-vanishing matrix elements with the leading power operator which consists of a single collinear gluon field in each sector. Therefore, all contributions from hard scattering operators vanish at $\cO(\lambda)$. Although we do not consider them in this chapter, using similar arguments one can show that all other sources of power corrections, such as Lagrangian insertions, also vanish at $\cO(\lambda)$.

\subsubsection{Relevant Operators at $\cO(\lambda^2)$}\label{sec:contribs}

Unlike the $\cO(\lambda)$ power corrections, the power corrections at $\cO(\lambda^2)=\cO(\tau_B)$ will not vanish. Contributions to the cross section at $\cO(\lambda^2)$ whose power suppression arises solely from hard scattering operators take the form either of a product of two $\cO(\lambda)$ operators or as a product of an $\cO(\lambda^2)$ operator and an $\cO(\lambda^0)$ operator
{\begin{small}
\begin{align}\label{eq:xsec_lam2}
\frac{d\sigma^{(2)}}{d\tau_B} &\supset N \sum_{X,i}  \tilde \delta^{(4)}_q  \bra{P_1 P_2} C_i^{(2)} O_i^{(2)}(0) \ket{X}\bra{X} C^{(0)} O^{(0)}(0) \ket{P_1 P_2} \delta\big( \tau_B - \tau_B^{(0)}(X) \big) +\text{h.c.}\nn \\
&+ N \sum_{X,i,j}  \tilde \delta^{(4)}_q  \bra{P_1 P_2} C_i^{(1)} O_i^{(1)}(0) \ket{X}\bra{X} C_j^{(1)} O_j^{(1)}(0) \ket{P_1 P_2}   \delta\big( \tau_B - \tau_B^{(0)}(X) \big)+\text{h.c.} \,.
\end{align}
\end{small}}
For $gg\to H$ the operator basis has only a single operator at $\cO(\lambda)$ (up to helicities and $n\leftrightarrow \bar n$), which was given in \Eqs{eq:H1_basis}{eq:H1_basis2}. This operator will contribute to the cross section at $\cO(\lambda^2)$, as indicated in \Tab{tab:fact_func}.

The contributions from $\cO(\lambda^2)$ hard scattering operators are highly constrained since they must interfere with the leading power operator. We will discuss each possible contribution in turn, and the summary of all operators which can contribute to the $\cO(\lambda^2)$ cross section is given in \Tab{tab:summary}. The schematic structure of the beam and soft functions arising from each of the different operator contributions is shown in \Tab{tab:fact_func}. The subleading beam and soft functions enumerated in this table are universal objects that will appear in processes initiated by different Born level amplitudes (such as $q\bar q$ annihilation), unless forbidden by symmetry. In this initial investigation, we content ourselves with only giving the field content of the beam and soft functions. In \Tab{tab:fact_func}, to save space, we do not write the external vacuum states for the soft functions, or the external proton states for the beam functions, nor do we specify the space-time positions of the fields. We do not present here the full definitions analogous to the leading power definitions given in \Eqs{eq:soft_func_def}{eq:beam_func_def}, but using the field content given in \Tab{tab:fact_func}. Deriving full definitions goes hand in hand with presenting the complete factorization theorems for these contributions, which will be given in future work.

\vspace{0.6cm}
\noindent{\bf{Two Quark-One Gluon Operators:}}

The two quark-one gluon operators, $O_{\cB\bar n}^{(1)}$ can contribute to the cross section by interfering with themselves. These operators are interesting since they effectively have a quark like cusp, instead of a gluon like cusp as is true of the leading power operators.  They contribute with a leading power quark channel soft function $S_q^{(0)}$, a quark beam function $B_q^{(0)}$ and a subleading power beam function $B_{qgg}^{(2)}$ that has fermion number crossing the cut (as indicated by its first $q$ subscript). 

\vspace{0.4cm}
\noindent{\bf{Two Quark-Two Gluon Operators:}}\\
\indent In the case of the two quark-two gluon operators, the only operators that will have a non-vanishing contribution are those that have the two gluons in different collinear sectors, namely $O_{\cB1}^{(2)}$. This gives a gluon beam function $B_g^{(0)}$, soft function $S_g^{(0)}$, and a subleading power beam function with gluon quantum numbers crossing the cut $B_{gqq}^{(2)}$ (with three color contractions). The operator $O_{\cB2}^{(2)}$, which has two quarks in a helicity $0$ configuration in one collinear sector, and two gluons in a helicity $0$ configuration in the other collinear sector does not contribute, since rotational invariance implies that its interference with the leading power operator vanishes. 

\vspace{0.4cm}
\noindent{\bf{Four Gluon Operators:}}

To give a non-vanishing interference with the leading power operator the four gluon operators must have an odd number of collinear gluon fields in each sector. This implies that $O_{4g1}^{(2)}$ does not contribute, while  $O_{4g2}^{(2)}$ does. Once again we can prove that $O_{4g2}^{(2)}$ generates a contribution that enters with simply the leading power gluon soft function $S_g^{(0)}$ (the direct proof of this requires some fairly extensive color algebra). This happens despite the fact that the subleading power beam function $B_{gg}^{(2)}$ has six color contractions.  The contribution from this four gluon operator first enters the cross section at $\cO(\alpha_s^2)$.

\vspace{0.4cm}
\noindent{\bf{Four Quark Operators:}}

For a four quark operator to interfere with the leading power operator, it must have both zero fermion number and a helicity 1 projection in each collinear sector. This eliminates all four quark operators from contributing to the cross section at $\cO(\lambda^2)$.

\vspace{0.4cm}
\noindent{\bf{$\cP_\perp$ Operators:}}

Both the operators involving $\cP_\perp$ insertions have the correct symmetry properties and therefore both $O_{\cP \chi }^{(2)}$ and $O_{\cP \cB }^{(2)}$ can contribute to the $\cO(\lambda^2)$ cross section. Both contributions have a leading power gluon beam function $B_g^{(0)}$ and soft function $S_g^{(0)}$. The operator $O_{\cP \cB }^{(2)}$ has a similar structure to the operator $\cO^{(2)}_{\cP1}$ found in the quark case in \cite{Feige:2017zci}, which contributes a leading log to the thrust (beam thrust) cross section \cite{Moult:2016fqy}. It involves a subleading power beam function $B_{ggP}^{(2)}$ (with two color contractions). On the other hand, we find in \Sec{sec:match_qqg_lam2} that the operator $O_{\cP \chi }^{(2)}$ has a vanishing Wilson coefficient at tree level, so its factorized contribution starts at least at ${\cal O}(\alpha_s^2)$ for the cross section. It has a subleading beam function $B_{gqP}^{(2)}$ with a single color contraction.

\vspace{0.4cm}
\noindent{\bf{Ultrasoft Operators:}}

The ultrasoft operators involving quark fields cannot contribute to the cross section through interference with the leading power operator due to fermion number conservation. Therefore, only the gluon operators of \Eqs{eq:Hgggus}{eq:Hdggus} contribute. They have leading power gluon beam functions $B_g^{(0)}$.

\subsubsection{Comparison with $\bar q\, \Gamma q$}\label{sec:compare}

It is interesting to briefly compare the structure of the operator basis, as well as the contributions to the $\cO(\lambda^2)$ cross section, to the basis for a process with two collinear sectors initiated by the $\bar q \Gamma q$ current as discussed in \cite{Feige:2017zci}. 
The leading power factorization theorems for the two cases are essentially identical, with simply a replacement of quark and gluon jet (beam) functions, as well as the color charges of the Wilson lines in the soft functions. 
However, at subleading power there are interesting differences arising both from the helicity structure of the currents, as well as from the form of the leading power Wilson coefficient.

An interesting feature of $gg\to H$ is that the Wilson coefficient for the leading power operator, which is given in \Sec{sec:matching_lp}, depends explicitly on the large label momenta of the gluons at tree level. 
This is not the case for the $\bar q\, \Gamma q$ current, whose leading power operator has a Wilson coefficient that is independent of the large label momenta at tree level. 
As discussed in \Sec{sec:nnlp_soft}, the Wilson coefficients of hard scattering operators involving insertions of $n \cdot \partial$, $\bar n \cdot \partial$, or $\cB_{us(n)0}$ are related to the derivatives of the leading power Wilson coefficients by RPI. 
This implies that these particular operators vanish at tree level for a $\bar q\, \Gamma q$ current, but are present at tree level for $gg\to H$. 
For the $\bar q \, \Gamma q$ current the power corrections from the ultrasoft sector arise instead only from subleading power Lagrangian insertions. 
Therefore, the nature of power corrections in the two cases is quite different in terms of the organization of the effective theory in the ultrasoft sector. 
However, this does not say anything about their numerical size which would require a full calculation. 
Furthermore, the organization of the collinear hard scattering operators is nearly identical in the two cases.

Despite this difference in the organization of the particular corrections within the ultrasoft sector of the effective theory, there is also much similarity in the way that the subleading power operators contribute to the cross section at $\cO(\lambda^2)$. 
In particular, in both cases, operators involving an additional ultrasoft or collinear gluon field as compared with the leading power operator contribute as an interference of the form $\cO(\lambda^2) \cO(1)$, see \Tab{tab:fact_func}. 
This is guaranteed by the Low-Burnett-Kroll theorem \cite{Low:1958sn,Burnett:1967km}. 
However, the subleading hard scattering operators that have a different fermion number in each sector than the leading power operators contribute as $\cO(\lambda) \cO(\lambda)$. 
For the $gg\to H$ case, this is the $O_{\cB\bar n}^{(1)}$ operator, while for a $q\Gamma \bar q$ current considered in \cite{Feige:2017zci}, it was a hard scattering operator involving two collinear quarks recoiling against a collinear gluon. 
In the NNLO calculation of power corrections for the $q\Gamma \bar q$ case \cite{Moult:2016fqy}, this operator played an important role, as it gave rise to a leading logarithmic divergence not predicted by a naive exponentiation of the one-loop result, and it is expected that the same will be true here. 
We plan to consider this calculation in a future work, and to understand in more detail the relation between the leading logarithmic divergences for the $q\Gamma \bar q$ current, compared with a $gg$ current.

\section{Matching}\label{sec:matching}

In this section we perform the matching to the operators relevant for the calculation of the $\cO(\lambda^2)$ cross section, which were enumerated in \Sec{sec:contribs} and summarized in \Tab{tab:summary}. As discussed in \Sec{sec:intro_gluonops}, we will work in the context of an effective Higgs gluon coupling
\begin{align}\label{eq:heft}
\cL_{\text{hard}}=\frac{C_1(m_t, \alpha_s)}{12\pi v}G^{\mu \nu}G_{\mu \nu} H\,,
\end{align}
obtained from integrating out the top quark. Here $v=(\sqrt{2} G_F)^{-1/2}=246$ GeV, and the matching coefficient is known to $\cO(\alpha_s^3)$ \cite{Chetyrkin:1997un}. Corrections to the infinite top mass can be included in the matching coefficient $C_1$. We use the sign convention
\begin{align}
G_{\mu \nu}^a=\partial_\mu A_\nu^a-\partial_\nu A_\mu^a+g f^{abc} A_{\mu}^b A_\nu^c\,,\qquad iD^\mu=i\partial^\mu+gA^\mu\,.
\end{align}
In the matching, we take all particles as outgoing. However, to avoid a cumbersome notation we use $\epsilon$ instead of $\epsilon^*$ for the polarization of an outgoing gluon. We also restrict to Feynman gauge although we check gauge invariance by enforcing relevant Ward identities. For operators involving collinear gluon fields gauge invariance is guaranteed through the use of the $\cB_\perp$ fields. 

The Higgs effective Lagrangian has Feynman rules for $2$, $3$, and $4$ gluons which are summarized in \App{app:expand_gluon}.  Due to the non-negative powers of momenta appearing in these Feynman rules they give rise to Wilson coefficients which are less singular than those arising from power corrections to the ultrasoft and collinear dynamics of SCET. This will be seen explicitly in the subleading power matching calculations. To simplify the notation throughout this chapter we will suppress the factor of $C_1(m_t, \alpha_s)/(12\pi v)$, and simply write the Feynman rules and matching relations for the operator 
\begin{align}
O^{\text{hard}}=G^{\mu \nu}G_{\mu \nu} H\,.
\end{align}
The dependence on $C_1(m_t, \alpha_s)/(12\pi v)$ is trivially reinstated. 

Throughout the matching, collinear gluons in the effective theory will be indicated in Feynman diagrams as a spring with a line drawn through them, collinear quarks will be indicated by dashed lines, and ultrasoft gluons will be indicated with an explicit ``us". This will distinguish them from their full theory counterparts for which standard Feynman diagram notation for quarks and gluons is used. Furthermore, for the full theory diagrams, we will use the $\otimes$ symbol to denote the vertex of the Higgs effective theory, as compared with the purple circle used to denote a hard scattering operator in the effective theory.

Due to the large number of operators present in our basis, we find it most convenient to express the results of the tree-level matching in the form of the Wilson coefficient multiplying the relevant operator. For this purpose we define a shorthand notation with a caligraphic {\cal O},
\begin{align}
\cO_X^{(i)} = C_X^{\text{tree}} O_X^{(i)}\,,
\end{align}
where as before, the superscript indicates the power suppression, and the subscript is a label that denotes the field and helicity content.
We will write results for $\cO_X^{(i)}$ in a form such that it is trivial to identify the tree level Wilson coefficient $C_X^{\text{tree}}$, so that higher order corrections can be added as they become available.

\subsection{Leading Power Matching}\label{sec:matching_lp}

The leading power matching is of course well known, however, we reproduce it here for completeness and to illustrate the matching procedure. The matching can be performed using a two gluon external state. Since the leading power operator is independent of any $\perp$ momenta, in performing the matching we can take the momenta 
\begin{align}
p_1^\mu =\omega_1 \frac{n^\mu}{2}, \qquad  p_2 =\omega_2 \frac{\bar n^\mu}{2}\,,
\end{align}
and the polarizations to be purely $\perp$, namely $\epsilon_{i}^\mu=\epsilon_{i\perp}^\mu$. All of the operators in \Sec{sec:lp} give a non-vanishing contribution to the two-gluon matrix element for this choice of polarization.

In the two gluon matrix element, this choice of polarization does not remove overlap with any of the operators in \Sec{sec:lp}.
Expanding the QCD result, we find
\begin{align}
\left.\fd{2.0cm}{figures/matching_gg_low} \right |_{\cO(\lambda^0)} &= -2i \delta^{ab} \omega_1 \omega_2 \epsilon_{1\perp} \cdot \epsilon_{2\perp}\,.
\end{align}
This is reproduced by the leading power operator
\begin{align}
\cO^{(0)}_{\cB}= -2\omega_1 \omega_2 \delta^{ab} \cB^a_{\perp \bar n, \omega_2} \cdot \cB^b_{\perp n, \omega_1} H\,,
\end{align}
or in terms of the helicity basis of \Eq{eq:hgg}, we have
\begin{align}
\cO^{(0)}_{\cB++}&=2\omega_1 \omega_2 \delta^{ab} \cB^a_{ \bar n +, \omega_2} \cdot \cB^b_{n +, \omega_1} H\,, \qquad
\cO^{(0)}_{\cB--}=2\omega_1 \omega_2 \delta^{ab} \cB^a_{ \bar n -, \omega_2} \cdot \cB^b_{n -, \omega_1} H\,.
\end{align}

While we focus here on the case where there is zero perp momentum in each collinear sector, we also give the Feynman rule in the case that each sector has a non-zero perp momentum. This will allow us to illustrate the gauge invariance properties of the collinear gluon field $\cB_\perp$. The expansion of the collinear gluon field with an incoming momentum $k$ is given by
\begin{align}
\cB^\mu_{n\perp}&=  A^{\mu a}_{\perp k} T^a -k^\mu_\perp \frac{\bar n \cdot A^a_{nk} T^a}{\bar n \cdot k} +\cdots \,,
\end{align}
where the dots represent terms with multiple gluon fields. The two gluon terms are given in \App{app:expand_gluon}. Gauge invariance therefore dictates the Feynman rule of our operator in the case of generic perp momenta for the two gluon fields,
\begin{align}
\fd{2.0cm}{figures/matching_gg_FeynRule_low} &= -2i \delta^{ab} \omega_1 \omega_2 \left( \epsilon_{1\perp}^{\mu} -p_{1\perp}^\mu \frac{\bar n \cdot \epsilon_1}{ \bar n \cdot p_1}  \right) \left( \epsilon_{2\perp}^{\mu} -p_{2\perp}^\mu \frac{\bar n \cdot \epsilon_2}{ \bar n \cdot p_2}  \right)  \,.
\end{align}
We note that the additional terms are essential to enforce that the required Ward identities are satisfied, and the result is gauge invariant. While this is trivial in this simple leading power example, for the more complicated matching calculations considered in the remainder of the chapter we will often perform the matching for particular kinematic configurations, and the gauge invariance of the collinear gluon fields is an important ingredient to uniquely obtain the full result.

\subsection{Subleading Power Matching}\label{sec:matching_nlp}

We now consider the matching at $\cO(\lambda)$. In \Sec{sec:nlp} we argued that the only $\cO(\lambda)$ operator which can contribute to the cross section at $\cO(\lambda^2)$ has two collinear quark fields in opposite collinear sectors and a collinear gluon field. We can therefore perform the matching using this external state. For concreteness we start with the case with a quark in the $n$-collinear sector, and a gluon and antiquark in the $\bar n$-collinear sector, $(q)_n (\bar q g)_\bn$. Since the power suppression arises from the explicit fields, and all propagators are off shell, we can use the kinematics
\begin{align}
p_1^\mu =\omega_1 \frac{n^\mu}{2}, \qquad  p_2 =\omega_2 \frac{\bar n^\mu}{2}, \qquad p_3 =\omega_3 \frac{\bar n^\mu}{2}\,,
\end{align}
and take the polarization of the gluon to be purely $\perp$, $\epsilon_{i}^\mu=\epsilon_{i\perp}^\mu$. This choice suffices to obtain non-zero matrix elements for all the operators we want to probe, and to distinguish them from one another.

For the matching calculations, we will use the notation
\begin{align}
u_n(i)=P_n u(p_i)\,, \qquad \text{and} \qquad v_n(i)=P_n v(i)\,, \qquad \text{with} \qquad P_n=\frac{\Sl{n} \Sl{\bar n}}{4}\,,
\end{align}
for the projected SCET spinors. Here we have taken the momentum $p_i$ to be $n$-collinear, but similar relations exist for the case that it is $\bar n$-collinear. 
The spinors obey
\begin{align}
u(p_i) = \Big( 1+ \frac{\Sl{p}_{i\perp}}{\bar{n} \cdot p_i} \frac{\Sl{\bar{n}}}{2} \Big)  u_n(i) \,, 
 \qquad
u(p_i) = \Big( 1+ \frac{\Sl{p}_{i\perp}}{n \cdot p_i} \frac{\Sl{n}}{2} \Big) u_{\bar n}(i) \,,
\end{align}
for the $n$-collinear and $\bar n$-collinear cases respectively, with direct analogs for the $v(p_i)$ spinors.

Expanding the QCD diagram to $\cO(\lambda)$, we find
\begin{align}
\left. \fd{2.5cm}{figures/matching_lam_qqg_low}\right |_{\cO(\lambda)} &= \frac{-2ig\omega_3}{\omega_2}  \bar u_n (p_1) \Sl{\epsilon}_{3\perp} T^a v_{\bar n}(p_2)\,.
\end{align}
There are no contributions from time ordered products in the effective theory to this particular matrix element used in the matching. This is due to the fact that there are no $\cO(\lambda^0)$ or $\cO(\lambda^1)$ operators involving just two quark fields, and the collinear Lagrangian insertions in each section preserve the fermion number of each sector, so this particular matrix element can not be obtained from Lagrangian insertions starting from the leading power operator involving two collinear gluons.   Therefore, the result must be reproduced entirely by a hard scattering operator in SCET. This operator is given by
\begin{align}
\cO^{(1)}_{\cB \bar n}= -2g\, \frac{\omega_3}{\omega_2}\,
  \bar \chi_{n,\omega_1} \Sl{\cB}_{\perp \bar n, \omega_3} 
    \chi_{\bar n,-\omega_2} H\,.
\end{align}
or, in terms of the helicity operators of \Eq{eq:H1_basis}
\begin{align}
\cO^{(1)}_{\cB \bar n+(+)}&=4g\frac{\omega_3}{\omega_2} T^a_{\alpha \bbeta} \sqrt{\frac{\omega_1 \omega_2}{2}}  \langle \bar n n \rangle \cB^a_{\bar n+,\omega_3} J^{\balpha \beta}_{n\bar n+} H\,, \nn \\
\cO^{(1)}_{\cB \bar n-(-)}&=-4g\frac{\omega_3}{\omega_2}T^a_{\alpha \bbeta}  \sqrt{\frac{\omega_1 \omega_2}{2}}  [ \bar n n ] \cB^a_{\bar n-,\omega_3} J^{\balpha \beta}_{n\bar n-} H\,.
\end{align}
The Wilson coefficient has a singularity as the energy fraction of the quark in the $\bar n$-collinear sector becomes soft. This operator will therefore contribute to the leading logarithmic divergence at the cross section level at $\cO(\lambda^2)$.  Note that this operator is explicitly RPI-III invariant, with its Wilson coefficient taking the form of a ratio of the large momentum components of the two $\bar n$ collinear fields.

For convenience, we also give the full Feynman rule for this operator
\begin{align}
\fd{3cm}{figures/matching_lam_qqg_Feyn_low}
  &=-2ig T^c \frac{\omega_3 }{\omega_2}   \left( \gamma^\nu_\perp-\frac{\Sl{p}_{3\perp} n^\nu}{\omega_3} \right)\,.
\end{align}
Note that this Feynman rule contains terms that were not present in the matching calculation due to the special choice of kinematics used there. These additional terms are determined by the gauge invariant gluon field, $\cB_{\perp \bar n}$, and it is easy to see that they ensure that this operator satisfies the required Ward identities.

The matching for the operators in the case  $(\bar q)_n (q g)_\bn$ can be easily obtained from the above results by exploiting charge conjugation. This gives
\begin{align}
\cO^{(1)}_{\cB n}= -2g \frac{\omega_3}{\omega_1} \bar \chi_{n,\omega_1} \Sl{\cB}_{\perp n, \omega_3} \chi_{\bn,-\omega_2} H\,,
\end{align}
and for the helicity operators in \eq{H1_basis2} we obtain
\begin{align}
\cO^{(1)}_{\cB n+(-)}
  &=4g\frac{\omega_3}{\omega_1} T^a_{\alpha \bbeta} 
   \sqrt{\frac{\omega_1 \omega_2}{2}}  \langle \bar n n \rangle \cB^a_{n-,\omega_3} J^{\balpha \beta}_{n\bar n+} H
  \,, \nn \\
\cO^{(1)}_{\cB n-(+)}
  &=-4g\frac{\omega_3}{\omega_1}T^a_{\alpha \bbeta}  
  \sqrt{\frac{\omega_1 \omega_2}{2}}  [ \bar n n ] 
  \cB^a_{n+,\omega_3} J^{\balpha \beta}_{n\bar n-} H
  \,.
\end{align}

This concrete matching calculation at subleading power also clearly illustrates the distinction between subleading power hard scattering operators, and the standard picture of leading power factorization in terms of splitting functions. In the leading power factorization for $H\to gq\bar q$, when the $q\bar q$ pair become collinear, the amplitude factorizes into $H\to gg$ multiplied by a universal $g\to q\bar q$ splitting function. This gives rise to a leading power contribution, due to the nearly on-shell propagator of the intermediate gluon that undergoes the splitting. For the operator considered here, the gluon which splits into the $q\bar q$ pair is far off-shell, due to the fact that the $q$ and $\bar q$ are in distinct collinear sectors. Because of this, it is represented in the effective theory by a local contribution (namely a hard scattering operator), and this operator is power suppressed. The hard scattering operators therefore describe precisely the contributions that are not captured by a splitting function type factorization. While this is particularly clear for the operator considered here, this picture remains true for the subleading power hard scattering operators for the more complicated partonic contributions considered at subsubleading power in \Sec{sec:matching_nnlp}. The hard scattering operators describe local contributions, which do not factorize in standard splitting function type picture, and therefore in general have no relation to known splitting functions which appear in the literature.

\subsection{Subsubleading Power Matching}\label{sec:matching_nnlp}

In this section we perform the tree level matching to the $\cO(\lambda^2)$ operators, considering only those which contribute at the cross section level at $\cO(\lambda^2)$, as discussed in \Sec{sec:contribs}. Since there are a number of operators, each with different field content, we will consider each case separately.

\subsubsection{Ultrasoft Derivative}\label{sec:us_deriv}

We begin by performing the matching to the ultrasoft derivative operators of \Sec{sec:nnlp_soft}. To perform the matching we can use a state consisting of two perpendicularly polarized collinear gluons, and we take our momenta as
\begin{align}
p_1^\mu=(\omega_1 +k_1)\frac{n^\mu}{2} +p_{1r}\frac{\bar n^\mu}{2}+p_{1\perp}^\mu\,, \qquad p_2^\mu=(\omega_2 +k_2)\frac{\bar n^\mu}{2} +p_{2r}\frac{n^\mu}{2}+p_{2\perp}^\mu\,. 
\end{align}
Since we have taken non-zero label perp momentum to keep the particles on shell we will have operators contributing that involving the $\cP_\perp$ operator. These operators were not included in our basis, since we assumed zero total perp momentum in each sector. (See Appendix A of \cite{Moult:2017rpl} for the additional operators required in the case that the collinear sectors have non-vanishing perp momentum.) However, these terms are easy to identify. Dropping these terms involving the label perp momentum to identify the contributions relevant for the matching, we find
\begin{align}
\left. \fd{2.0cm}{figures/matching_gg_low}\right |_{\cO(\lambda^2)}&=-2i \delta^{ab} \omega_1 k_2 \epsilon_{1\perp} \cdot \epsilon_{2\perp} -2i\delta^{ab} \omega_2 k_1 \epsilon_{1\perp} \cdot \epsilon_{2\perp}\,.
\end{align}
This result must be completely reproduced by hard scattering operators in the effective theory, since the relevant subleading propagator insertions are proportional to residual components of the $\perp$ momentum, which we have taken to be zero in the matching (see \App{app:expand_gluon}, and in particular \Eq{eq:lam2_gluon_prop}).

The operators given in \Sec{sec:nnlp_soft} were defined post BPS field redefinition, in which case the partial derivative operator $\partial^\mu$ acts on gauge invariant building blocks. While the distinction between pre- and post-BPS field redefinition is not relevant for the calculation of the matrix elements in this particular case, since there are no ultrasoft emissions, it of course determines the form that the operators are written in. For convenience, we give the operators both before and after BPS field redefinition. Note that the collinear gluon field transforms as an adjoint matter field under ultrasoft gauge transformations since the ultrasoft gauge field acts as a background field.

Matching onto pre-BPS field redefinition operators, we find
\begin{align}\label{eq:preBPS_softgluon}
\cO^{(2)}_{n\cdot D}=4\omega_1 \tr \left[ \cB^\mu_{\perp n, \omega_1}   [in \cdot  D_{us}, \cB^\mu_{\perp \bar n, \omega_2} ] \right ] H\,, \quad \cO^{(2)}_{\bar n\cdot D}=4\omega_2 \tr \left[ \cB^\mu_{\perp \bar n, \omega_2}  [i\bar n \cdot D_{us},  \cB^\mu_{\perp n, \omega_1}] \right] H\,,
\end{align}
where the trace is over color. This color structure will be fixed by matching with an additional ultrasoft gluon in \Sec{sec:us_gluon}. To determine the operators post-BPS field redefinition, we can either directly apply the BPS field redefinition, or simply match to the operators of \Sec{sec:nnlp_soft}. We find that the operators where the ultrasoft derivative acts on the gluon fields are given by
\begin{align}
\cO^{(2)ab}_{\partial \cB (us)(0)}=-2\omega_1 \cB^{\mu a}_{\perp n, \omega_1}   i n \cdot  \partial \cB^{\mu b}_{\perp \bar n, \omega_2} H\,, \quad \cO^{(2)ab}_{\partial \cB (us) (\bar 0)}=-2\omega_2 \cB^{\mu a}_{\perp \bar n, \omega_2}  i\bar n \cdot  \partial   \cB^{\mu b}_{\perp n, \omega_1} H\,,
\end{align}
or expanded in terms of the helicity operator basis
\begin{align}
\cO^{(2)ab}_{\partial \cB (us(n)) \bar 0:++}
  &=-2\omega_1 \cB^{\mu a}_{n +, \omega_1}   i  \partial_{us(n)\bar 0} 
  \cB^{\mu b}_{\bar n +, \omega_2} H
  \,, \quad \\
\cO^{(2)ab}_{\partial \cB (us(\bar n))0:++}
  &=-2\omega_2  \cB^{\mu a}_{\bar n +, \omega_2}  i\partial_{us(\bar n) 0}  \cB^{\mu b}_{n +, \omega_1} H
  \,, \nn \\
\cO^{(2)ab}_{\partial \cB (us(n)) \bar 0:--}
  &=-2\omega_1  \cB^{\mu a}_{n -, \omega_1}   i  \partial_{us(n)\bar 0} \cB^{\mu b}_{\bar n -, \omega_2} H
  \,, \quad \nn\\
\cO^{(2)ab}_{\partial \cB (us(\bar n))0:--}
  &=-2\omega_2  \cB^{\mu a}_{\bar n -, \omega_2}  i\partial_{us(\bar n) 0}  \cB^{\mu b}_{n -, \omega_1} H
   \,. \nn
\end{align}
Here the color indices are contracted against the basis of color structures given in \Eq{eq:Z2gd_colorus}.
These operators also give rise, after BPS field redefinition to operators involving $\cB_{us}$. These will be discussed in \Sec{sec:us_gluon}.

As mentioned above \Eq{eq:Hdggus}, using the gluon equations of motion we can eliminate operators involving $\bar n\cdot \partial\cB_{\bar n}$ and $n\cdot \partial \cB_n$ from our basis to all orders in perturbation theory. This structure of the ultrasoft derivative operators is important for the matching at $\cO(\lambda^2)$. In particular, only the $\bar n \cdot \partial$ acts on the $n$-collinear sector, and only the $n\cdot \partial$ acts on the $\bar n$-collinear sector. These correspond to the residual components of the label momenta. In a graph consisting of only collinear particles (i.e. no ultrasoft particles) the residual components of the label momenta can be chosen to vanish, so that these operators do not contribute. In all purely collinear graphs computed in the remainder of this chapter, we will always make this choice, and therefore, these operators will not contribute. However, these operators will contribute, and will play an important role, when ultrasoft particles are present in the graph.

\subsubsection{qqg}\label{sec:match_qqg_lam2}

We now consider the case of the  $\cO(\lambda^2)$ operators involving two collinear quark fields, a collinear gluon field, and a $\cP_\perp$ insertion. In \Sec{sec:nnlp} we argued that the only such operators have both quark fields in the same collinear sector, which we will take to be the $n$-collinear sector. To perform the matching we take the kinematics
\begin{align}\label{eq:qqg_momentum}
p_1^\mu=\omega_1 \frac{n^\mu}{2}+p_\perp^\mu +p_{1r} \frac{\bar n^\mu}{2}\,, \qquad  p_2^\mu=\omega_2 \frac{n^\mu}{2}-p_\perp^\mu +p_{2r} \frac{\bar n^\mu}{2}\,, \qquad p_3^\mu=\omega_3 \frac{\bar n^\mu}{2}\,.
\end{align}
With this choice all subleading Lagrangian insertions in SCET vanish. This can be seen from the explicit subleading Lagrangians and Feynman rules given in \App{app:expand_gluon} by noting that these give contributions to this matrix element the involve residual components of $\perp$ momentum, or residual components of the large label momentum, which are zero for the choice of momentum in \Eq{eq:qqg_momentum}.   The result must therefore be entirely reproduced by hard scattering operators. Expanding the QCD result we find that it vanishes at $\cO(\lambda^2)$
\begin{align}
\left. \fd{2.5cm}{figures/matching_lam2_qqg_low} \right|_{\cO(\lambda^2)}&= 0\,.
\end{align}
This is expected since this diagram involves only collinear dynamics in a single collinear sector, and non-trivial terms will be reproduced by power suppressed Lagrangians.
Therefore, at tree level, the hard scattering operators involving two quarks in the same sector along with a $\cP_\perp$ insertion have  vanishing Wilson coefficients. We do not have an argument that the Wilson coefficients of these operators would continue to vanish at higher orders in perturbation theory, and therefore we do not expect this to be the case.

\subsubsection{ggg}\label{subsec:ggg_match_lam2}

We now consider matching to the $\cO(\lambda^2)$ three gluon operators which have a single $\cP_\perp$. In \Sec{sec:contribs} we have argued that the only such operators that contribute to the cross section at $\cO(\lambda^2)$ have two gluons in the same collinear sector, which we take to be $\bar n$ for concreteness. To perform the matching, we take the kinematics as
\begin{align}
p_1=\omega_1 \frac{n^\mu}{2}\,, \qquad p_2=\omega_2 \frac{\bar n^\mu}{2}+p_\perp^\mu+p_{2r} \frac{n^\mu}{2}\,, \qquad p_3=\omega_3 \frac{\bar n^\mu}{2}-p_\perp^\mu+p_{3r} \frac{n^\mu}{2}\,.
\end{align}
As a further simplification, we can take the polarization vector of the gluon in the $n$-collinear sector to be purely $\perp$,
$\epsilon_{1}^\mu=\epsilon_{1\perp}^\mu\,.$  All of the three gluon operators in our basis give a non-vanishing contribution to the three-gluon matrix element for this choice of polarizations.

In performing the expansion of the QCD diagrams we will obtain all three projections of the polarization vectors, namely $\bar n \cdot \epsilon_{2,3}$, $n \cdot \epsilon_{2,3}$, and $p_{\perp} \cdot \epsilon_{2,3\perp}$. However, all of the operators in our basis are formed from $\cB_{\perp}$, and therefore contain only the $n \cdot \epsilon_{2,3}$ and $p_{\perp} \cdot \epsilon_{2,3\perp}$ components. From the on-shell conditions for the gluon we have the relation
\begin{align}
\omega_2 \frac{\bar n \cdot \epsilon_2}{2}=\frac{p_\perp^2 n\cdot \epsilon_2}{2\omega_2}-p_\perp \cdot \epsilon_\perp\,,
\end{align}
and similarly for $\epsilon_3$. Note that we always use the Minkowski signature for the $\perp$ momenta, i.e. $p_\perp^2=-\vec p_\perp^{~2}$. In performing the matching one can therefore keep track of only the $\perp$ polarizations,  as long as the $\bar n \cdot \epsilon$ polarizations are converted into $n \cdot \epsilon$ and $p_\perp\cdot \epsilon_\perp$ using the above equation. This allows one to simplify the structure of the matching while keeping enough terms to reconstruct operators formed from $\cB_\perp$ gluon fields.

Expanding the QCD diagrams, and keeping only the $\perp$ terms of the polarizations we find
\begin{align}
&\left. \left(\fd{2.25cm}{figures/matching_lam2_ggg4_low} + \fd{2.25cm}{figures/matching_lam2_ggg1_low}\right) \right |_{\cO(\lambda^2)} \nn \\
&\hspace{0cm}= -4gf^{abc} \frac{\omega_3}{\omega_2}\left( \epsilon_{1\perp} \cdot \epsilon_{2\perp} p_{\perp}\cdot \epsilon_{3\perp} -\epsilon_{2\perp} \cdot \epsilon_{3\perp} p_{\perp} \cdot \epsilon_{1\perp}    \right ) -4g f^{abc} \epsilon_{1\perp}\cdot \epsilon_{2\perp} p_{\perp}\cdot \epsilon_{3\perp}  +[(2,b)\leftrightarrow (3,c)]\,, \nn \\
&\left. \fd{2.25cm}{figures/matching_lam2_ggg2_low}\right|_{\cO(\lambda^2)} =0\,, \nn \\
&\left.\fd{2.25cm}{figures/matching_lam2_ggg3_low}\right|_{\cO(\lambda^2)} =-4g f^{abc} \left(p_\perp \cdot \epsilon_{3\perp} \epsilon_{1\perp}\cdot \epsilon_{2\perp} -p_\perp \cdot \epsilon_{1\perp} \epsilon_{2\perp}\cdot \epsilon_{3\perp} +\frac{\omega_2}{\omega_3} p_\perp \cdot \epsilon_{3\perp} \epsilon_{1\perp} \cdot \epsilon_{2\perp}   \right) \nn \\
&\hspace{4cm}+[(2,b)\leftrightarrow (3,c)]\,,
\end{align}
We have shown results for the individual diagrams to emphasize the structure of the contributions, namely that only the diagrams involving an off-shell propagator or the Higgs EFT three gluon vertex contribute.
Simplifying this result, we find that the sum of the QCD diagrams is given by
\begin{align}
&\left. \left(\fd{2.25cm}{figures/matching_lam2_ggg4_low} + \fd{2.25cm}{figures/matching_lam2_ggg1_low}+\fd{2.25cm}{figures/matching_lam2_ggg2_low}+\fd{2.25cm}{figures/matching_lam2_ggg3_low}\right) \right |_{\cO(\lambda^2)} \nn \\
= &4gf^{abc} \left (2+\frac{\omega_3}{\omega_2}+\frac{\omega_2}{\omega_3} \right) \epsilon_{2\perp}\cdot \epsilon_{3\perp} p_\perp\cdot \epsilon_{1\perp} -4gf^{abc} \left( 2+\frac{\omega_3}{\omega_2}+\frac{\omega_2}{\omega_3}  \right)\epsilon_{1\perp} \cdot \epsilon_{2\perp} p_\perp \cdot \epsilon_{3\perp}\nn \\
&  -4gf^{abc} \left( 2+\frac{\omega_3}{\omega_2}+\frac{\omega_2}{\omega_3}  \right)\epsilon_{1\perp} \cdot \epsilon_{3\perp} p_\perp \cdot \epsilon_{2\perp} \,.
\end{align}
For the choice of kinematics and polarizations used in the matching there are no SCET subleading Lagrangian contributions at this power, for similar reasons to the case of $gq\bar q$ discussed above. Therefore, the hard scattering operators must exactly reproduce the QCD result.

We write the operators and their Wilson coefficients both in the helicity basis of \Eq{eq:Hgggpperp_basis}, as well as in a more standard Lorentz structures, as the two may prove useful for different purposes. In terms of standard Lorentz structures the tree level matching gives
\begin{align}
\cO^{(2)}_{\cP \cB1}&=-\left( \frac{1}{2}\right)4g \left(  2+\frac{\omega_3}{\omega_2}+ \frac{\omega_2}{\omega_3}  \right)i f^{abc} \cB^a_{n\perp,\omega_1}\cdot \left[  \cP_\perp \cB^b_{\bar n \perp,\omega_2}\cdot  \right] \cB_{\bar n \perp,\omega_3}^c    H\,, \nn \\
\cO^{(2)}_{\cP \cB2}&=4g\left( 2+\frac{\omega_3}{\omega_2}  + \frac{\omega_2}{\omega_3}\right) if^{abc}\left[ \cP_\perp \cdot \cB_{\bar n \perp,\omega_3}^a \right] \cB^b_{n\perp,\omega_1} \cdot \cB_{\perp \bar n, \omega_2}^c    H\,.
\end{align}
We have written the first operator in this form to incorporate the symmetry factor.
In the helicity basis, we have
\begin{align}\label{eq:3g_hel_match}
&\cO_{\cP\cB +++[-]}^{(2)}
= 4g if^{abc} \left(2+\frac{\omega_3}{\omega_2}+ \frac{\omega_2}{\omega_3}\right) \cB_{n+,\omega_1}^a\, \cB_{\bar n+,\omega_3}^b\, \left [\cP_{\perp}^{-} \cB_{\bar n+,\omega_2}^c \right ] \,H
\,,\nn \\
&\cO_{\cP\cB ---[+]}^{(2)}
=4g if^{abc}\left(2+\frac{\omega_3}{\omega_2}+ \frac{\omega_2}{\omega_3}\right) \cB_{n-,\omega_1}^a\, \cB_{\bar n-,\omega_3}^b\, \left [\cP_{\perp}^{+} \cB_{\bar n-,\omega_2}^c \right ] \,H\,,
 \nn \\
&\cO_{\cP\cB ++-[+]}^{(2)}
= -2g if^{abc}\left(2+\frac{\omega_3}{\omega_2}+ \frac{\omega_2}{\omega_3}\right)\, \cB_{n+,\omega_1}^a\, \cB_{ \bar n-,\omega_3}^b\, \left [\cP_{\perp}^{+} \cB_{\bar n+,\omega_2}^c \right ] \,H
\,,\nn \\
&\cO_{\cP\cB -+-[-]}^{(2)}
= -2g if^{abc}\left(2+\frac{\omega_3}{\omega_2}+ \frac{\omega_2}{\omega_3}\right)\cB_{n-,\omega_1}^a\, \cB_{ \bn -,\omega_3}^b\, \left [\cP_{\perp}^{-} \cB_{\bar n +,\omega_2}^c \right ] \,H
\,.
\end{align}
We therefore see explicitly that the helicity selection rules are realized in the tree level matching. Furthermore, the Wilson coefficient is formed from Bose symmetric combinations of ratios of the large momentum components of the $\bar n$ collinear fields, as required by RPI-III invariance.
For convenience, we also give the Feynman rule of the combined operator with three external gluons
\begin{align}\label{eq:feynrule_ggg3}
\fd{2.75cm}{figures/feynrule_ggg3_low}&\\%
	&\hspace{-2.5cm}=4g f^{abc}\Bigl( 2  + \frac{ \omega_2}{\omega_3} + \frac{ \omega_3}{\omega_2}\Bigr)\biggl[  p_\perp^\mu g_\perp^{\nu \rho}  - p_\perp^\nu g_\perp^{\mu \rho} - p_\perp^\rho g_\perp^{\mu \nu} +\frac{p^2_\perp}{\omega_2 \omega_3}\left( \omega_3 n^\nu g_\perp^{\mu \rho}- \omega_2 n^\rho g_\perp^{\mu \nu}  + p_\perp^\mu n^\nu n^\rho\right) \biggr]
    . \nn
\end{align}
This contains additional terms not present in the earlier matching calculation, due to the particular choice of $\perp$ polarizations used to simplify the matching. One can explicitly check that this operator satisfies the Ward identity, which is gauranteed by the fact that it is written in terms of $\cB_\perp$ fields. 
It is also interesting to  note that the Wilson coefficient of this operator has a divergence as either $\omega_2$, or $\omega_3$ become soft, so that it will give rise to a leading logarithmic divergence in the cross section at $\cO(\lambda^2)$.

\subsubsection{Ultrasoft Gluon}\label{sec:us_gluon}

The operators involving a single ultrasoft insertion were given in \Sec{sec:nnlp_soft}, and it was argued that they were related by RPI to the leading power operator involving two collinear gluons.  In this section we will explicitly perform the tree level matching to verify that this relation holds. The operators in \Sec{sec:nnlp_soft} were given after BPS field redefinition, since it is more convenient when enumerating a complete basis to work with a gauge invariant ultrasoft gluon field. While it is possible to directly match to the post-BPS operators, we will first perform the matching to pre-BPS field redefinition operators involving ultrasoft covariant derivatives, and verify the color structure given in \Eq{eq:preBPS_softgluon}. We will then give the operators after BPS field redefinition.

We perform the matching to a three particle external state, with one collinear gluon in each sector, and a single ultrasoft gluon.
To simplify the matching we take the momenta of the collinear particles as
\begin{align}
p_1^\mu=\omega_1 \frac{n^\mu}{2}\,, \qquad p_2^\mu =\omega_2 \frac{\bar n^\mu}{2}\,,
\end{align}
and the momentum of the ultrasoft particle as
\begin{align}
p_3^\mu=\bar n \cdot p_3 \frac{n^\mu}{2} +n\cdot p_3 \frac{\bar n^\mu}{2} +p_{3\perp}^\mu\,,
\end{align}
where $(n\cdot p_3, \bar n \cdot p_3, p_{3\perp})\sim (\lambda^2, \lambda^2,\lambda^2)$.
The full theory QCD diagrams expanded to $\cO(\lambda^2)$ are given by
\begin{align}
\left. \fd{2.25cm}{figures/matching_gg_soft_low}\right|_{\cO(\lambda^2)}  &= 2g \omega_2 f^{abc} \epsilon_1 \cdot \epsilon_2 \frac{\bar n\cdot p_3}{n\cdot p_3}n\cdot \epsilon_3+4g f^{abc} \omega_2 \epsilon_1 \cdot \epsilon_3 \frac{\epsilon_{2\perp} \cdot p_{3\perp}}{n \cdot p_3} \nn \\
&-4g f^{abc} \omega_2 \epsilon_2\cdot  \epsilon_3 \frac{p_{3\perp} \cdot \epsilon_{1\perp}}{ n \cdot p_3}\,, \nn \\
\left.\fd{2.25cm}{figures/matching_gg_soft_2_low} \right|_{\cO(\lambda^2)} &=-2g \omega_1 f^{abc} \epsilon_1 \cdot \epsilon_2 \frac{n\cdot p_3}{\bar n\cdot p_3}\bar n\cdot \epsilon_3-4g f^{abc} \omega_1 \epsilon_2 \cdot \epsilon_3 \frac{\epsilon_{1\perp} \cdot p_{3\perp}}{\bar n \cdot p_3} \nn \\
&+4g f^{abc} \omega_1 \epsilon_1\cdot \epsilon_3 \frac{p_{3\perp} \cdot \epsilon_{2\perp}}{\bar n \cdot p_3}\,, \nn \\
\left.\fd{2.25cm}{figures/matching_gg_soft_4_low} \right|_{\cO(\lambda^2)} &= 0\,,\nn \\
\left.\fd{2.25cm}{figures/matching_gg_soft_3_low} \right|_{\cO(\lambda^2)} &=2g f^{abc} \omega_1 \epsilon_1 \cdot \epsilon_2 n\cdot \epsilon_3-2g f^{abc} \omega_2 \epsilon_1 \cdot \epsilon_2 \bar n \cdot \epsilon_3 \,. 
\end{align}
In this case there are also contributions from $T$ product diagrams in SCET correcting the emission of an ultrasoft gluon. Once we subtract these terms from the full theory result, the remainder will be localized at the hard scale.
The $\cO(\lambda^2)$ Feynman rule for the emission of a ultrasoft gluon from a collinear gluon is given by (see \App{app:expand_gluon} and e.g.  \cite{Larkoski:2014bxa} for the explicit Feynman rule)
\begin{align}
\fd{3cm}{figures/soft_lam2_scet_low}&=\langle | T \cB^\nu_{n\perp}(0) \cL^{(2)}_{A_n} | \epsilon_n, p_n; \epsilon_s,p_s \rangle=-i f^{abc} \epsilon_{n\mu} \frac{2\epsilon_{s\rho} p_{s\sigma}}{p_n^- n\cdot p_s} \left(  g^{\mu \rho}_\perp g^{\sigma \nu}_\perp -    g^{\mu \sigma}_\perp g^{\rho \nu}_\perp  \right)\,.
\end{align}
The two SCET diagrams involving this Lagrangian insertion are given by
\begin{align}
\fd{2.25cm}{figures/matching_gg_soft_scet_low}  &=\frac{4\omega_2 f^{abc} }{n\cdot p_3} (\epsilon_1 \cdot \epsilon_3 p_{3\perp} \cdot \epsilon_{2\perp} -\epsilon_{1\perp} \cdot p_{3\perp} \epsilon_2 \cdot \epsilon_3)\,, \nn \\
\fd{2.25cm}{figures/matching_gg_soft_scet_2_low}  &=\frac{4\omega_1 f^{abc} }{\bar n\cdot p_3} (\epsilon_1 \cdot \epsilon_3 p_{3\perp} \cdot \epsilon_{2\perp} -\epsilon_{1\perp} \cdot p_{3\perp} \epsilon_2 \cdot \epsilon_3)\,.
\end{align}
Finally we also have contributions from the ultrasoft derivative operators of \Sec{sec:us_deriv}, with a leading power emission of a ultrasoft gluon. For these diagrams we find
\begin{align}
\fd{2.25cm}{figures/matching_gg_soft_scet_deriv_low}&= 2g \omega_2 f^{abc} \epsilon_1 \cdot \epsilon_2 \frac{\bar n\cdot p_3}{n\cdot p_3}n\cdot \epsilon_3\,, \quad
\fd{2.25cm}{figures/matching_gg_soft_scet_deriv_2_low}&= -2g \omega_1 f^{abc} \epsilon_1 \cdot \epsilon_2 \frac{n\cdot p_3}{\bar n\cdot p_3}\bar n\cdot \epsilon_3\,.
\end{align}
The SCET $T$-products therefore exactly reproduce the QCD diagrams, with the exception of the contribution from the three gluon vertex of the Higgs effective theory.
Subtracting the SCET contributions from the expansion of the QCD diagrams, we find that the hard scattering operators are given by
\begin{align}\label{eq:preBPS_softgluon2}
\cO^{(2)}_{n\cdot D}=4\omega_1 \tr \left[ \cB^\mu_{\perp n, \omega_1}   [n \cdot  D_{us}, \cB^\mu_{\perp \bar n, \omega_2} ] \right ] H\,, \quad \cO^{(2)}_{\bar n\cdot D}=4\omega_2 \tr \left[ \cB^\mu_{\perp \bar n, \omega_2}  [\bar n \cdot D_{us},  \cB^\mu_{\perp n, \omega_1}] \right] H\,,
\end{align}
as stated in \Eq{eq:preBPS_softgluon}.
In terms of gauge invariant ultrasoft gluon fields we have
\begin{align}
\cO^{(2)}_{\cB(us(n))}&=\left(i  f^{abd}\, \big({\cal Y}_n^T {\cal Y}_{\bar n}\big)^{dc}\right)  \left (-2g \omega_2  \cB^a_{n\perp, \omega_1} \cdot \cB^b_{\bar n \perp, \omega_2} \cB^c_{us(n)0} \right)\,,  \nn \\
\cO^{(2)}_{\cB(us(\bar n))}&=\left(i  f^{abd}\, \big({\cal Y}_{\bar n}^T {\cal Y}_{n}\big)^{dc}\right)  \left (-2g \omega_1  \cB^a_{n\perp, \omega_1} \cdot \cB^b_{\bar n \perp, \omega_2} \cB^c_{us(\bar n)0} \right)\,,
\end{align}
where the color structures that appear at tree level are the first components of the color basis of \Eqs{eq:Z2g_colorus}{eq:Z2g_colorus_2}.
In terms of helicity operators, 
\begin{align}
\hspace{-0.5cm}\cO_{\cB(us(n)) 0:++}^{(2)}&=-2g \left(i  f^{abd}\, \big({\cal Y}_n^T {\cal Y}_{\bar n}\big)^{dc}\right)  \omega_2  \cB^a_{n+,\omega_1}  \cB^b_{\bar n+,\omega_2}  \cB^c_{us(n)0} H\,,\nn \\
\cO_{\cB(us(n)) 0:--}^{(2)}&=-2g \left(i  f^{abd}\, \big({\cal Y}_n^T {\cal Y}_{\bar n}\big)^{dc}\right)   \omega_2  \cB^a_{n-,\omega_1}  \cB^b_{\bar n-,\omega_2} \cB^c_{us(n) 0} H\,, \nn \\
\hspace{-0.5cm}\cO_{\cB(us(\bar n)) 0:++}^{(2)}&=-2g \left(i  f^{abd}\, \big({\cal Y}_{\bar n}^T {\cal Y}_{n}\big)^{dc}\right)  \omega_1  \cB^a_{n+,\omega_1}  \cB^b_{\bar n+,\omega_2}  \cB^c_{us(\bar n)0} H\,,\nn \\
\cO_{\cB(us(\bar n)) 0:--}^{(2)}&=-2g \left(i  f^{abd}\, \big({\cal Y}_{\bar n}^T {\cal Y}_{n}\big)^{dc}\right)  \omega_1  \cB^a_{n-,\omega_1}  \cB^b_{\bar n-,\omega_2} \cB^c_{us(\bar n) 0} H\,.
\end{align}
This agrees with the relation derived from RPI symmetry, given in \Eq{eq:usRPIrelation}. For convenience, we also give the Feynman rule for the contribution of the hard scattering operators to a single ultrasoft emission both before BPS field redefinition
\begin{align}
\fd{2.75cm}{figures/matching_gg_soft_Feynrule_low}&=2g f^{abc} \omega_1 g^{\mu \nu}_\perp n^\rho -2g f^{abc} \omega_2 g^{\mu \nu}_\perp \bar n^\rho\,,
\end{align}
as well as after BPS field redefinition
\begin{align}
\fd{2.75cm}{figures/matching_gg_soft_Feynrule_BPS_low}&=2g f^{abc} \left[ \omega_1 \left( n^\rho - \frac{ n \cdot p_3}{\bar n \cdot p_3} \bar n^\rho \right) - \omega_2  \left(\bar n^\rho - \frac{\bar n \cdot p_3}{n \cdot p_3} n^\rho \right) \right]  \nn \\
&=2g f^{abc} \left[  
	  n^\rho \left( \omega_1 + \frac{\bar n \cdot p_3}{n \cdot p_3} \omega_2  \right) 
-\bar n^\rho \left( \omega_2 + \frac{ n \cdot p_3}{\bar n \cdot p_3} \omega_1 \right) \right] \,.
\end{align}
Note that the contribution from hard scattering operators before the BPS field redefinition is local, but not gauge invariant, since before BPS field redefinition there are also SCET $T$-product diagrams involving. After BPS field redefinition, the contribution from the hard scattering operators is gauge invariant, but at the cost of locality. However, as emphasized in \cite{Feige:2017zci}, the form of the non-locality is dictated entirely by the BPS field redefinition, and is therefore not problematic. It is therefore advantageous to work in terms of the ultrasoft gauge invariant building blocks, so that the contributions from the hard scattering operators alone are gauge invariant.
Note also that here we have restricted the $\perp$ momentum of the two collinear particles to vanish for simplicity. Furthermore, because of the ultrasoft wilson lines in the color structure of \Eq{eq:Z2g_colorus}, there are also Feynman rules with multiple ultrasoft emissions. This is analogous to the familiar case of the $\cB_\perp$ operator which has Feynman rules for the emission of multiple collinear gluons.

\subsubsection{qqgg}

A basis for the operators involving two collinear quark and two collinear gluon fields was given in \Sec{sec:nnlp_collinear}. In \Sec{sec:contribs} it was argued that the only non-vanishing contributions to the cross section at $\cO(\lambda^2)$ arise from operators with the two collinear quarks and a collinear gluon in one sector, recoiling against a collinear gluon in the other sector.  

In performing the matching to these operators there are potentially $T$-product terms from the three gluon $\cO(\lambda^2)$ operator of \Sec{subsec:ggg_match_lam2}, where one of the gluons splits into a $q\bar q$ pair.
By choosing the momentum
\begin{align}
p_1^\mu&=\omega_1 \frac{n^\mu}{2}+p_\perp^\mu +p_{1r} \frac{\bar n^\mu}{2}\,, \quad p_2^\mu=\omega_2 \frac{n^\mu}{2}-p_\perp^\mu +p_{2r} \frac{\bar n^\mu}{2}\,, \quad
p_3^\mu=\omega_3 \frac{\bar n^\mu}{2}\,, \quad p_4^\mu=\omega_4 \frac{n^\mu}{2}\,,
\end{align}
we see from \Eq{eq:feynrule_ggg3} that all SCET $T$-product contributions vanish, so that the result must be reproduced by hard scattering operators in SCET. 
Expanding the QCD diagrams to $\cO(\lambda^2)$, we find that all the contributions from the two gluon vertex in the Higgs effective theory vanish
\begin{align}
\left. \fd{2.5cm}{figures/matching_lam2_qqgg1_low}\right|_{\cO(\lambda^2)}  &=0\,, \quad \left.\fd{2.5cm}{figures/matching_lam2_qqgg2_low}\right|_{\cO(\lambda^2)}  = 0\,, \quad \left.\fd{2.5cm}{figures/matching_lam2_qqgg3_low}\right|_{\cO(\lambda^2)}   = 0\nn \\
\left.\fd{2.25cm}{figures/matching_lam2_qqgg4_low}\right|_{\cO(\lambda^2)}  &=0\,, \quad \left.\fd{2.5cm}{figures/matching_lam2_qqgg5_low}\right|_{\cO(\lambda^2)}  =0\,.
\end{align}
This result might be anticipated from the structure of the diagrams. However, there is a non-vanishing contribution from the three-gluon vertex in the Higgs effective theory
\begin{align}
\left. \fd{2.5cm}{figures/matching_lam2_qqgg6_low} \right|_{\cO(\lambda^2)}  &=-\frac{4g^2 f^{abc} \omega_4 \epsilon_{3\perp}\cdot \epsilon_{4\perp}}{(\omega_1+\omega_2)^2}  \bar u_n(p_1) T^a \frac{\Sl{\bar n}}{2} v_n(p_2)\,. 
\end{align}
In terms of standard Lorentz and Dirac structures the corresponding hard scattering operator is given by
\begin{align}
\cO^{(2)}_{\cB1}=\frac{4g^2 if^{abc} \omega_4 }{(\omega_1+\omega_2)^2}\cB^b_{n\perp,\omega_4 }\cdot \cB^c_{\bar n \perp,\omega_3} \bar \chi_{n,\omega_1} T^a \frac{\Sl{\bar n}}{2} \chi_{n,-\omega_2}H\,.
\end{align}
Projected onto the helicity operator basis of \Eq{eq:Hqqgg_basis3}, and using the color basis of \Eq{eq:ggqqll_color}, we find 
\begin{align}
\cO^{(2)}_{\cB1++(0)}&=-\frac{4g^2 \omega_4 }{(\omega_1+\omega_2)^2}2\sqrt{\omega_1 \omega_2} \left(  (T^a T^b)_{\alpha \bbeta} -(T^b T^a)_{\alpha \bbeta}  \right) \cB^a_{n+,\omega_4} \cB^b_{\bar n+,\omega_3}  J_{n0}^{\balpha \beta} H\,,\nn \\
\cO^{(2)}_{\cB1--(0)}&=-\frac{4g^2 \omega_4 }{(\omega_1+\omega_2)^2}2\sqrt{\omega_1 \omega_2} \left(  (T^a T^b)_{\alpha \bbeta} -(T^b T^a)_{\alpha \bbeta}  \right) \cB^a_{n-,\omega_4} \cB^b_{\bar n-,\omega_3}  J_{n0}^{\balpha \beta} H\,, \nn\\
\cO^{(2)}_{\cB1++(\bar0)}&=-\frac{4g^2  \omega_4 }{(\omega_1+\omega_2)^2}2\sqrt{\omega_1 \omega_2} \left(  (T^a T^b)_{\alpha \bbeta} -(T^b T^a)_{\alpha \bbeta}  \right) \cB^a_{n+,\omega_4} \cB^b_{\bar n+,\omega_3}  J_{n\bar0}^{\balpha \beta}H\,, \nn \\
\cO^{(2)}_{\cB1--(\bar0)}&=-\frac{4g^2  \omega_4 }{(\omega_1+\omega_2)^2}2\sqrt{\omega_1 \omega_2} \left(  (T^a T^b)_{\alpha \bbeta} -(T^b T^a)_{\alpha \bbeta}  \right) \cB^a_{n-,\omega_4} \cB^b_{\bar n-,\omega_3}  J_{n\bar0}^{\balpha \beta}H\,.
\end{align}
For convenience, we also give the Feynman rule for the operator
\begin{align}
\fd{2.5cm}{figures/matching_lam2_qqgg_Feyn_low}&=-\frac{4g^2 f^{abc} T^a \omega_4}{(\omega_1+\omega_2)^2} \left(  g^{\mu \nu}_\perp -\frac{p^\nu_{4\perp} \bar n^\mu}{\omega_4} \right) \frac{\Sl{\bar n}}{2}\,.
\end{align}
Again, this contains additional terms not present in the matching calculation, and it is straightforward to check that they are necessary to satisfy the required Ward identities.

\subsubsection{gggg}

Finally, we consider the matching to the operators involving four collinear gluon fields. A basis of such operators was given in \Eq{eq:H_basis_gggg_2}. In \Sec{sec:contribs} it was argued that to contribute to the cross section at $\cO(\lambda^2)$, there must be three collinear gluons in the same sector. For concreteness, we take this to be the $\bar n$ sector. The operators with three gluons in the $n$ sector can be obtained by crossing $\bar n \leftrightarrow n$.

To perform the matching we choose the momenta as
\begin{align}\label{eq:momentagggg}
p_1^\mu&=\omega_1 \frac{ n^\mu}{2}\,, \quad p_2^\mu=\omega_2 \frac{\bar n^\mu}{2}\,, \quad p_3^\mu=\omega_3 \frac{\bar n^\mu}{2}-p_\perp^\mu +p_{3r} \frac{n^\mu}{2}\,, \quad p_4^\mu=\omega_4 \frac{\bar n^\mu}{2}+p_\perp^\mu +p_{4r} \frac{ n^\mu}{2}\,.
\end{align}
With this choice, each particle in the $\bar n$ sector is on-shell, but the sum of any two of their momenta is off-shell,
\be
	p_i^2 = 0\,, \qquad (p_1 + p_j)^2 \sim \cO(1)\,, \qquad (p_j + p_k)^2 \sim \cO(\lambda^2) \,, \qquad j,k=2,3,4 \,;~ j\neq k\,,
\ee
which regulates all propagators. This particular choice of momenta is convenient since it simplifies $T$-product contributions from SCET. Furthermore, we take the external polarizations to be purely perpendicular, i.e. $\epsilon_i^\mu = \epsilon_{i\perp}^\mu$. All of the four gluon operators give a non-vanishing contribution to the four-gluon matrix element for this choice of polarization, allowing their Wilson coefficients to be obtained.

In computing the full theory diagrams for the matching it is convenient to separate the diagrams into those involving on-shell propagators, which will be partially reproduced by $T$-product terms in SCET, and diagrams involving only off-shell propagators. Since the four gluon operators obtain their power suppression entirely from the fields, for diagrams involving only off-shell propagators the residual momenta in \Eq{eq:momentagggg} can be ignored, as they contribute only power suppressed contributions. Diagrams with on-shell propagators are regulated by the residual momenta in \Eq{eq:momentagggg}.

We begin by considering the expansion of the full theory diagrams that don't involve any on-shell propagators. In this case, all $\perp$ momenta can be set to zero, and the result will be purely local. The relevant QCD diagrams expanded to $\cO(\lambda^2)$ arise from the four gluon vertex in the Higgs effective theory,\\
\noindent\begin{minipage}{.3\linewidth}
\begin{equation}
\hspace{2cm}\left.\fd{2.25cm}{figures/matching_lam2_gggg1_low}\right|_{\cO(\lambda^2)} \nn
\end{equation}
\end{minipage}%
\begin{minipage}{.7\linewidth}
\begin{align}
=&4ig^2 (f^{eab}f^{ecd}+f^{ead}f^{ecb}) \epsilon_{1\perp} \cdot \epsilon_{3\perp} \epsilon_{2\perp} \cdot \epsilon_{4\perp} \nn \\
&+4ig^2 (f^{eac}f^{ebd} +f^{ead}f^{ebc} ) \epsilon_{1\perp} \cdot \epsilon_{4\perp} \epsilon_{2\perp} \cdot \epsilon_{3\perp} \nn \\
&+4ig^2 ( f^{eab}f^{edc}+f^{eac}f^{edb}) \epsilon_{1\perp} \cdot \epsilon_{2\perp} \epsilon_{3\perp} \cdot \epsilon_{4\perp} \,, 
\end{align}
\end{minipage}\\
\newline
from a splitting off of the three gluon vertex,\\
\noindent\begin{minipage}{.3\linewidth}
\begin{equation}
\hspace{1cm}\left.\left( \fd{2.15cm}{figures/matching_lam2_gggg2_low}+\text{perms} \right)\right|_{\cO(\lambda^2)} \nn
\end{equation}
\end{minipage}%
\begin{minipage}{.7\linewidth}
\begin{align}
=&2ig^2 \left( \frac{\omega_3-\omega_2}{\omega_4}   \right) f^{abe}f^{cde} \epsilon_{1\perp} \cdot \epsilon_{4\perp} \epsilon_{3\perp} \cdot \epsilon_{2\perp} \nn \\
&  +[(2,d)\leftrightarrow (4,b)] +[(3,c)\leftrightarrow (4,b)] \,,
\end{align}
\end{minipage}\\
\newline
and from multiple emissions off of the two gluon vertex, either using the four gluon vertex with a single off-shell propagator\\
\noindent
\raisebox{1cm}{\begin{minipage}{.3\linewidth}
\begin{equation}
\hspace{0.25cm}\left.\left(\fd{2.15cm}{figures/matching_lam2_gggg3a_low}+\text{perms} \right) \right|_{\cO(\lambda^2)} \nn
\end{equation}
\end{minipage}}%
\begin{minipage}{.7\linewidth}
\begin{align}
\hspace{1cm}=& 2ig^2\left( \frac{\omega_2}{\omega_3+\omega_4}\right)\nn \\
& \left[   f^{bae} f^{cde} (\epsilon_{3\perp} \cdot \epsilon_{4\perp} \epsilon_{1\perp} \cdot \epsilon_{2\perp} -\epsilon_{4\perp} \cdot \epsilon_{2\perp} \epsilon_{3\perp} \cdot \epsilon_{1\perp}) \right.\nn \\
&+f^{bce}f^{ade} (\epsilon_{1\perp} \cdot \epsilon_{4\perp} \epsilon_{3\perp} \cdot \epsilon_{2\perp} -\epsilon_{4\perp} \cdot \epsilon_{2\perp} \epsilon_{1\perp} \cdot \epsilon_{3\perp})\nn \\
&\left. +f^{bde}f^{ace}(\epsilon_{1\perp} \cdot \epsilon_{4\perp} \epsilon_{3\perp} \cdot \epsilon_{2\perp}-\epsilon_{3\perp} \cdot \epsilon_{4\perp} \epsilon_{1\perp} \cdot \epsilon_{2\perp}) \right] \nn \\
&\hspace{-0cm}  +[(2,d)\leftrightarrow (3,c)] +[(2,d)\leftrightarrow (4,b)] \,,
\end{align}
\end{minipage}\\
\newline
\noindent or sequential emissions with two off-shell propagators

\noindent\begin{minipage}{.3\linewidth}
\begin{equation}
\left.\left(\fd{2.15cm}{figures/matching_lam2_gggg7_low}+\text{perms}\right) \right|_{\cO(\lambda^2)} \nn
\end{equation}
\end{minipage}%
\begin{minipage}{.7\linewidth}
\begin{align}
=&2ig^2 \frac{\omega_2 \omega_3}{\omega_4(\omega_3+\omega_4)}   \epsilon_{2\perp} \cdot \epsilon_{3\perp} \epsilon_{4\perp} \cdot \epsilon_{1\perp} f^{abe}f^{ecd}\nn \\
&  +[\text{perms}]   \,.
\end{align}
\end{minipage}\\
In the last case we have not explicitly listed the permutations, since all possible permutations are required.

We now consider the expansion of the full theory diagrams involving on-shell propagators. These will generically involve both local and non-local pieces. The non-local pieces will be directly reproduced by $T$-products in the effective theory.  The first class of diagrams involving on-shell propagators are those with all propagators on-shell. Here, at tree level, the dynamics occurs entirely within a single collinear sector. The two relevant QCD diagrams expanded to $\cO(\lambda^2)$ are
\begin{align}
\left.\fd{2.45cm}{figures/matching_lam2_gggg3_low} \right|_{\cO(\lambda^2)} &=0\,, \qquad
\left.\left(\fd{2.00cm}{figures/matching_lam2_gggg5_low} +\text{perms} \right)\right|_{\cO(\lambda^2)} &=0\,,
\end{align}
both of which have vanishing subleading power contributions. 

Next, we consider diagrams involving both on-shell and off-shell propagators. To simplify the results, we will often use the relation
\begin{align}
\frac{p_\perp^2}{(p_2+p_3)^2}=-\frac{\omega_3}{\omega_2}\,,
\end{align}
which will allow us to write the result in terms of a local term, which is just a rational function of the label momenta, and a non-local term, which explicitly contains the on-shell propagator. These non-local terms will be cancelled by the $T$-product diagrams in SCET.  For a first class of diagrams, where we have a nearly on-shell splitting in the $\bar n$-collinear sector, we have both a local term
\begin{align}
\left.\left(\fd{2.25cm}{figures/matching_lam2_gggg4_low}\right) \right|_{\cO(\lambda^2)}& =4ig^2 f^{aed}f^{bce}\frac{(\omega_3 - \omega_4)}{(\omega_3 + \omega_4)}\epsilon_{1\perp} \cdot \epsilon_{2\perp} \epsilon_{3\perp} \cdot \epsilon_{4\perp}\,,
\end{align}
when the splitting is into the particles $3$ and $4$, as well as a term that has both local and non-local pieces
\begin{align}
\left.\left(~\fd{2.25cm}{figures/matching_lam2_gggg4b_low}~ +~ \fd{2.25cm}{figures/matching_lam2_gggg4c_low}~\right) \right|_{\cO(\lambda^2)}& \\
&\hspace{-6cm} =\frac{4ig^2 f^{aeb}f^{dce}}{\omega_4}\left[ \frac{2(\omega_2 + \omega_3)}{(p_2 + p_3)^2}  p_\perp \cdot \epsilon_{1\perp}   p_\perp \cdot \epsilon_{2\perp}  \epsilon_{3\perp}\cdot \epsilon_{4\perp}  \right.\nn\\
&\hspace{-3.2cm}\left. -(2\omega_3 + \omega_4) \epsilon_{1\perp} \cdot \epsilon_{4\perp} \epsilon_{2\perp} \cdot \epsilon_{3\perp} \vphantom{\frac{(a)}{(b)}}\right]  + [3 \leftrightarrow 4, b \leftrightarrow c , p_\perp \to -p_\perp]\,. \nn
\end{align}
As will be discussed in more detail when we consider the corresponding diagrams in the EFT, the first permutation is purely local, since there is no corresponding $T$-product term in the effective theory, and thus it must be fully reproduced by a hard scattering operator. This particular splitting allows a slight simplification in the calculation of the SCET diagrams. For a second class of diagrams, where we have an on-shell splitting emitted from an off-shell leg, we again have a purely local term
\begin{align}
\left.\left(\fd{2.05cm}{figures/matching_lam2_gggg6_low}\right) \right|_{\cO(\lambda^2)}& =0\,, 
\end{align}
as well as non-local contributions,
\begin{align}
\left.\left(~\fd{2.05cm}{figures/matching_lam2_gggg6b_low}~ +~ \fd{2.05cm}{figures/matching_lam2_gggg6c_low}~\right) \right|_{\cO(\lambda^2)}& \\
&\hspace{-6.5cm} =2ig^2 f^{aeb}f^{dce}\left[ \left(\frac{4\omega_4}{(\omega_2 + \omega_3)(p_2 + p_3)^2}\right)   p_\perp \cdot \epsilon_{1\perp} p_\perp \cdot \epsilon_{2\perp} \epsilon_{3\perp}\cdot \epsilon_{4\perp} \right.\nn\\
&\hspace{-4.5cm}\left. -\frac{\omega_3(\omega_2 - \omega_3)(\omega_2 + \omega_3 + \omega_4)^2}{\omega_2\omega_4(\omega_2 + \omega_3)^2} \epsilon_{1\perp} \cdot \epsilon_{4\perp} \epsilon_{2\perp} \cdot \epsilon_{3\perp}  \right]  + [3 \leftrightarrow 4, b \leftrightarrow c , p_\perp \to -p_\perp]\,.\nn
\end{align}
Again, we see the same pattern, that the first permutation gives rise to a purely local term, while the second two permutations give rise to both local and non-local terms.

Finally, we have the diagrams involving the three gluon vertex in the Higgs effective theory. We again have a local contribution
\begin{align}
\left.\left(\fd{2.05cm}{figures/matching_lam2_gggg8_low}\right) \right|_{\cO(\lambda^2)}& =-2ig^2 f^{ade}f^{ebc} \frac{\omega_2(\omega_3-\omega_4)}{(\omega_3 + \omega_4)^2} \epsilon_{1\perp} \cdot \epsilon_{2\perp} \epsilon_{3\perp} \cdot \epsilon_4\,, 
\end{align}
and a non-local contribution
\begin{align}
\left.\left(~\fd{2.05cm}{figures/matching_lam2_gggg8b_low}~ +~ \fd{2.05cm}{figures/matching_lam2_gggg8c_low}~\right) \right|_{\cO(\lambda^2)}&  \\
&\hspace{-6.5cm} =2ig^2 f^{aeb}f^{dce}\left[ \frac{8}{(p_2 + p_3)^2}  p_\perp \cdot \epsilon_{1\perp}  p_\perp \cdot \epsilon_{2\perp}   \epsilon_{3\perp}\cdot \epsilon_{4\perp}  \right.\nn\\
&\hspace{-4.cm}\left. -\biggl\{\frac{ (\omega_3 + \omega_4)^2 - \omega_2\omega_3}{\omega_2\omega_4}\biggr\}   \epsilon_{1\perp} \cdot \epsilon_{4\perp}   \epsilon_{2\perp} \cdot \epsilon_{3\perp}  \right]  + [3 \leftrightarrow 4, b \leftrightarrow c , p_\perp \to -p_\perp]\,. \nn
\end{align}

The non-local terms in the above expansions must be reproduced by $T$-product terms in the effective theory. First, there are potential contributions from $\cO^{(2)}_{\cP\cB}$, with the two gluon Feynman rule for $\cB_{\bar n,\perp}$, which is given in \App{app:expand_gluon}. Such contributions give vanishing overlap for our choice of $\perp$ polarizations. There are however $T$-product contributions arising from the  three gluon $\cO^{(2)}_{\cP\cB}$ operator, with an  $\cL^{(0)}$ insertion.
The three gluon Feynman rule for the $\cO^{(2)}_{\cP\cB}$ vertex was given in \Eq{eq:feynrule_ggg3}.
Since the $\cO^{(2)}_{\cP\cB}$ operator has an explicit $\cP_\perp$ insertion, it vanishes in the case that either of the particles in the $\bar n$ sector has no perpendicular momentum. This is why our particular choice of momenta for the matching simplifies the structure of the $T$-products. The two non-vanishing permutations are given by
\begin{align}\label{eq:ggggSCET}
    & \fd{2.00cm}{figures/matching_gggg_scetb_low} + \fd{2.00cm}{figures/matching_gggg_scetc_low}  \\
	=&-8ig^2 f^{abe}f^{ecd} \frac{(\omega_2 + \omega_3 + \omega_4)^2 }{(\omega_3 + \omega_2)\omega_4}\left[ \frac{\omega_3}{(\omega_2 + \omega_3)} \epsilon_{1\perp}\cdot \epsilon_{4\perp}   \epsilon_{3\perp}\cdot \epsilon_{2\perp} - \frac{ p_\perp\cdot\epsilon_{1\perp} p_\perp\cdot \epsilon_{2\perp} \epsilon_{3\perp} \cdot \epsilon_{4\perp}}{(p_2 + p_3)^2} \right] \nn \\
	&+ [3 \leftrightarrow 4, b \leftrightarrow c , p_\perp \to -p_\perp]\,, \nn 
\end{align}
which consists both of a local and a non-local term. The non-local terms exactly reproduce the ones obtained in the QCD expansion 
\begin{align}
	\hspace{-1cm}
   \left(\fd{1.80cm}{figures/matching_gggg_scetb_low} + \fd{1.80cm}{figures/matching_gggg_scetc_low} \right)_\text{\!\!\!non-loc.} \hspace{-0.5cm}
    & \!\! = \left(\fd{1.90cm}{figures/matching_lam2_gggg4b_low} + \fd{1.90cm}{figures/matching_lam2_gggg6b_low} + \fd{1.90cm}{figures/matching_lam2_gggg8b_low} +\! \text{ perms }\!\!\! \right)_\text{\!\!\!non-loc.} \hspace{-1cm} \nn \\[0.4cm]
	&\hspace{-5cm}=8ig^2    p_\perp\cdot\epsilon_{1\perp}  p_\perp\cdot \epsilon_{2\perp}   \epsilon_{3\perp} \cdot \epsilon_{4\perp} \left( \frac{f^{abe}f^{ecd}} {(p_2 + p_3)^2} \frac{(\omega_2 + \omega_3 + \omega_4)^2 }{(\omega_3 + \omega_2)\omega_4} + [3 \leftrightarrow 4 , b \leftrightarrow c] \right) \,. 
\end{align}
While it is of course necessary that the EFT reproduces all such non-local terms, this is also a highly non trivial cross check of both the three and four gluon matching.

The matching coefficients for the hard scattering operators are given by the remaining local terms. Before presenting the result we briefly comment on the organization of the color structure. All diagrams are proportional to $f^{abe}f^{cde}$, $f^{ace}f^{bde}$ or $f^{ade}f^{bce}$, which are related by the Jacobi identity $f^{abe}f^{cde} = f^{ace}f^{bde} - f^{ade}f^{bce}$. A basis in terms of structure constants can easily be related to the trace basis of \eqref{eq:gggg_color} using
\begin{align}\label{eq:colmatrrel}
f^{ace}f^{bde} &= \tr[abdc] + \tr[acdb] - \tr[acbd] - \tr[adbc] = e_2 - e_3\,,\nn \\
f^{ade}f^{bce} &= \tr[abcd] + \tr[adcb] - \tr[acbd] - \tr[adbc] = e_1 - e_3\,, 
\end{align}
where $e_i$ is the $i$-th element of the basis in \eqref{eq:gggg_color}. We find it most convenient to write the Wilson coefficient in the $(f^{ace}f^{bde},f^{ade}f^{bce})$ basis.
After subtracting the local piece of the SCET $T-$product of \eqref{eq:ggggSCET} from the full theory graphs, and manipulating the result to bring it into a compact form, we find the following operator
\begin{align}
\cO^{(2)}_{4g}= 16 \pi \alpha_s f^{ade}f^{bce}  (\cB^a_{n\perp,\omega_i} \cdot \cB^b_{\bar n \perp, \omega_j})(\cB^c_{\bar n \perp, \omega_k} \cdot \cB_{\bar n \perp, \omega_\ell}^d) \left( 3 + \dfrac{\omega_j^3 + \omega_k^3 + \omega_\ell^3 + \omega_j\omega_k\omega_\ell}{(\omega_j + \omega_k)(\omega_j + \omega_\ell)(\omega_k + \omega_\ell)}\right) \,.
\end{align}
The Wilson coefficient is manifestly RPI-III invariant. When the matrix element of this operator is taken we are forced to sum over permutations which gives the proper Bose symmetric result, as well as inducing terms with other color structures. In terms of the helicity operators of \Eq{eq:H_basis_gggg_2}, we have
\begin{align}
\cO^{(2)}_{4g} 
 &= 16 \pi \alpha_s f^{ade}f^{bce}  \left( 3 + \dfrac{\omega_j^3 + \omega_k^3 + \omega_\ell^3 + \omega_j\omega_k\omega_\ell}{(\omega_j + \omega_k)(\omega_j + \omega_\ell)(\omega_k + \omega_\ell)}\right) \nn \\
 &\ \ \times \Big[ \cB^a_{n+,\omega_i}   \cB^b_{\bar n +, \omega_j} \cB^c_{\bar n +, \omega_k}   \cB_{\bar n -, \omega_\ell}^d + \cB^a_{n+,\omega_i}   \cB^b_{\bar n +, \omega_j} \cB^c_{\bar n -, \omega_k}   \cB_{\bar n +, \omega_\ell}^d  
 \nn \\
 &\qquad +\cB^a_{n-,\omega_i}   \cB^b_{\bar n -, \omega_j} \cB^c_{\bar n +, \omega_k}   \cB_{\bar n -, \omega_\ell}^d + \cB^a_{n-,\omega_i}   \cB^b_{\bar n -, \omega_j} \cB^c_{\bar n -, \omega_k}   \cB_{\bar n +, \omega_\ell}^d \Big] 
 \nn \\
 &= 16 \pi \alpha_s  
  \biggl[ 3 + \dfrac{\omega_j^3 \!+\! \omega_k^3 \!+\! \omega_\ell^3 \!+\! \omega_j\omega_k\omega_\ell}{(\omega_j \!+\! \omega_k)(\omega_j \!+\! \omega_\ell)(\omega_k \!+\! \omega_\ell)}\biggr]
  \Big[ (f^{ade}f^{bce}\! +\! f^{ace}f^{bde}) \cB^a_{n+,\omega_i}   \cB^b_{\bar n +, \omega_j} \cB^c_{\bar n +, \omega_k}   \cB_{\bar n -, \omega_\ell}^d  \nn \\
& \qquad\qquad 
  - (f^{ade}f^{bce} + f^{abe}f^{cde} )\cB^a_{n-,\omega_i}   \cB^b_{\bar n +, \omega_j} \cB^c_{\bar n -, \omega_k}   \cB_{\bar n -, \omega_\ell}^d \Big] 
 \,.
\end{align}
We see that all the helicity selection rules are satisfied in the tree level matching, as expected. We have also checked the result using the automatic FeynArts \cite{Hahn:2000kx} and FeynRules implementation of the HiggsEffectiveTheory \cite{Alloul:2013bka}. For more complicated calculations at subleading power in SCET it would be interesting to fully automate the computation of Feynman diagrams involving power suppressed SCET operators and Lagrangians.

The four gluon operators derived in this section can be used to study $\cO(\alpha_s^2)$  collinear contributions at $\cO(\lambda^2)$. It would be interesting to understand in more detail the universality of collinear splittings at subleading power, as well as collinear factorization properties. For some recent work in this direction from a different perspective, see \cite{Stieberger:2015kia,Nandan:2016ohb}.
The behavior of these Wilson coefficients is also quite interesting. They exhibit a singularity as any pair of collinear particles simultaneously have their energy approach zero. This was also observed in the Wilson coefficients for operators describing the subleading collinear limits of two gluons emitted off of a $q\bar q$ vertex \cite{Feige:2017zci}.

\section{Conclusions}\label{sec:conclusions_gluon_ops}

In this chapter we have presented a complete basis of operators at $\cO(\lambda^2)$ in the SCET expansion for color singlet production of a scalar through gluon fusion, as relevant for $gg\to H$. 
To derive a minimal basis we used operators of definite helicities, which allowed us to significantly reduce the number of operators in the basis. 
This simplification is due to helicity selection rules which are particularly constraining due to the scalar nature of the produced particle. 
We also classified all possible operators which could contribute to the cross section at $\cO(\lambda^2)$. 
In performing this classification the use of a helicity basis again played an important role, allowing us to see from simple helicity selection rules which operators could contribute. 
While the total number of subleading power operators is large, the number that contribute at the cross section level is smaller. 
We compared the structure of the contributions to the case of a quark current, $\bar q \Gamma q$, finding interesting similarities, despite a slightly different organization in the effective theory.

A significant portion of this chapter was devoted to a tree level calculation of the Wilson coefficients of the subleading power operators which can contribute to the cross section at $\cO(\lambda^2)$.
The Wilson coefficients obtained in this matching will allow for a study of the power corrections at NLO and for the study of the leading logarithmic renormalization group structure at subleading power. 
An initial investigation of the renormalization group properties of several subleading power operators relevant for the case of $e^+e^-\to \bar q q$ was considered in \cite{Freedman:2014uta}. 

A number of directions exist for future study, with the goal of understanding factorization at subleading power.
In particular, one would like to combine the hard scattering operators derived in this chapter with the subleading SCET Lagrangians to derive a complete factorization theorem at subleading power for a physical event shape observable. 
Combined with the operators in \cite{Feige:2017zci}, all necessary ingredients are now available to construct such a subleading factorization for thrust for $\bar q q$ or $gg$ dijets in $e^+e^-$ collisions.
This would also allow for a test of the universality of the structure of the subleading factorization.
The operators of this chapter can also be used to study threshold resummation, where power corrections of $\mathcal{O}((1-z)^0)$ have received considerable attention \cite{Dokshitzer:2005bf,Grunberg:2007nc,Laenen:2008gt,Laenen:2008ux,Grunberg:2009yi,Laenen:2010uz,Almasy:2010wn,Bonocore:2014wua,White:2014qia,deFlorian:2014vta,Bonocore:2015esa,Bonocore:2016awd}, particularly for the $q\bar q$ channel, but it would be interesting to extend this to the $gg$ case.

An interesting application of current relevance of the results presented in this chapter is to the calculation of fixed order power corrections for NNLO event shape based subtractions.
Gaining analytic control over power corrections can significantly improve the performance and stability of such subtraction schemes.
This has been studied for $q\bar q$ initiated Drell Yan production to NNLO in \cite{Moult:2016fqy} using a subleading power operator basis in SCET (see also \cite{Boughezal:2016zws} for a direct calculation in QCD).
Combined with the results for the operator basis and matching for $q\bar q$ initiated processes given in \cite{Feige:2017zci}, the operator basis presented in this chapter will allow for the systematic study of power corrections for color singlet production and decay.

\chapter{Subleading Power Factorization with Radiative Functions}\label{sec:radiative}

\section{Introduction}\label{sec:introRadiative}

The simplicity of the soft and collinear limits of gauge theories allows for an all orders understanding of the behavior of amplitudes and cross sections, typically formulated in terms of factorization theorems~\cite{Collins:1985ue,Collins:1988ig,Collins:1989gx}. Unlike for observables which are amenable to a local operator product expansion (OPE)~\cite{Wilson:1969zs}, these general factorization theorems typically involve non-local matrix elements with Wilson lines. While the structure of these matrix elements is well understood at leading power, the structure of power corrections is much less well understood. In general, complicated non-local matrix elements, typically involving operators strung along the light cone dressing the leading power Wilson line structure, are required \cite{Balitsky:1987bk,Balitsky:1990ck,Bauer:2001mh,Bosch:2004cb,Lee:2004ja}.

The soft collinear effective theory (SCET) \cite{Bauer:2000ew, Bauer:2000yr, Bauer:2001ct, Bauer:2001yt}, an effective field theory describing the soft and collinear limits of QCD, provides an operator and Lagrangian based formalism for deriving factorization theorems at subleading power. As an example, SCET has been used to systematically study power corrections to the leading power factorization for $B\to X_s \gamma$, $B\to X_u l \bar \nu$  \cite{Korchemsky:1994jb,Bauer:2001yt} in the shape function region  \cite{Bigi:1993ex,Neubert:1993um,Mannel:1994pm}, and derive subleading factorization theorems in terms of universal non-local operators \cite{Beneke:2002ph,Leibovich:2002ys,Kraetz:2002rv,Neubert:2002yx,Burrell:2003cf,Bosch:2004cb,Lee:2004ja,Beneke:2004in}. In this case, the power corrections take the form of non-local operators describing both soft fluctuations at the scale  $\Lambda_{\text{QCD}}$, in terms of matrix elements of the $B$ meson, as well as the coupling of soft and collinear modes. 

Recently, there has been significant interest in understanding the subleading power soft and collinear limits of perturbative scattering amplitudes and event shape observables. 
This has been motivated at the amplitude level both by their relation to asymptotic symmetries (see e.g. \cite{Strominger:2013jfa,Cachazo:2014fwa,Casali:2014xpa,Cheung:2016iub,Bern:2014oka,He:2014bga,Larkoski:2014bxa,He:2017fsb}), as well as to better understand the structure of amplitudes by studying their limits (see e.g. \cite{Dixon:2011pw,Dixon:2014iba,Dixon:2015iva,Caron-Huot:2016owq,Dixon:2016nkn}).  
At the cross section level an understanding of subleading power corrections will allow for the improved accuracy of perturbative predictions involving resummation, and improvements to next-to-next-to-leading order subtraction schemes \cite{Boughezal:2015dva,Boughezal:2015aha,Gaunt:2015pea} by analytically calculating subleading power corrections \cite{Moult:2016fqy,Boughezal:2016zws,Moult:2017jsg,Boughezal:2018mvf,Ebert:2018lzn,Ebert:2018gsn,Bhattacharya:2018vph}, amongst many other applications. 
From explicit calculations, there are hints for the simplicity of power corrections at higher loop order, for example in splitting functions \cite{Dokshitzer:2005bf},  in the threshold limit \cite{Matsuura:1987wt,Matsuura:1988sm,Hamberg:1990np,DelDuca:2017twk,Dulat:2017prg,Bahjat-Abbas:2018hpv}, for event shape observables \cite{Moult:2016fqy,Boughezal:2016zws,Moult:2017jsg,Balitsky:2017flc,Dixon:2018qgp}, for power corrections in quark masses \cite{Liu:2017vkm,Liu:2018czl}, and in the Regge limit \cite{Bruser:2018jnc}. 
To obtain an all loop understanding, and identify universal structures which persist at subleading powers, it is desirable to formulate subleading power factorization theorems whose renormalization group structure allows the prediction of higher loop results from lower loop data, as has been successful at leading power. 
Recently this was used to derive the first resummation at subleading power for the thrust event shape observable in $H\to gg$ \cite{Moult:2018jjd} and for threshold in \cite{Beneke:2018gvs}.

In this chapter we use SCET to derive an all orders gauge invariant factorization for subleading power soft emissions, focusing in particular on non-local corrections described by so called radiative functions. 
We use the SCET Lagrangian, formulated in terms of non-local gauge invariant quark and gluon fields to provide gauge invariant definitions of the radiative functions for the emission of both soft quarks and gluons.
Gauge invariance is guaranteed by an intricate Wilson line structure, dictated by the symmetries of the effective theory. 
We show how these radiative functions appear in factorization formulas at subleading power, both at the level of the amplitude and the cross section, as multilocal matrix elements with convolutions of operators along the lightcone. 
These operator insertions correct the leading power Wilson line structure.
This completes our derivation of all the required components for subleading power factorization initiated in \cite{Feige:2017zci,Moult:2017rpl}, and we review in detail the complete factorization structure at subleading power, highlighting the role that radiative functions play.

\begin{figure}[t!]
\begin{center}
\includegraphics[width=0.23\columnwidth]{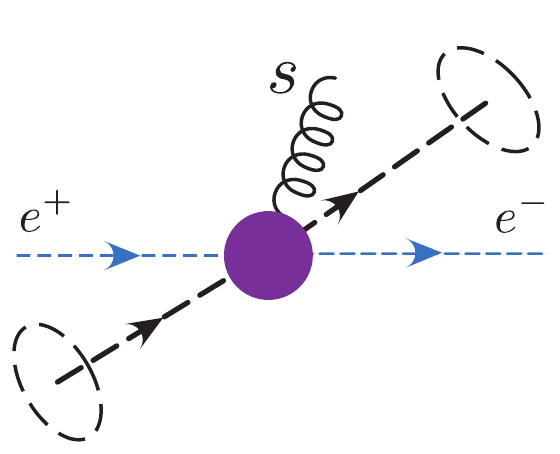} 
\hspace{0.1cm}
\includegraphics[width=0.23\columnwidth]{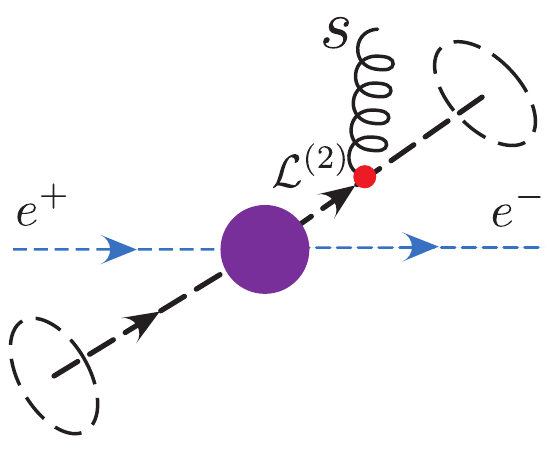} 
\hspace{0.1cm}
\includegraphics[width=0.23\columnwidth]{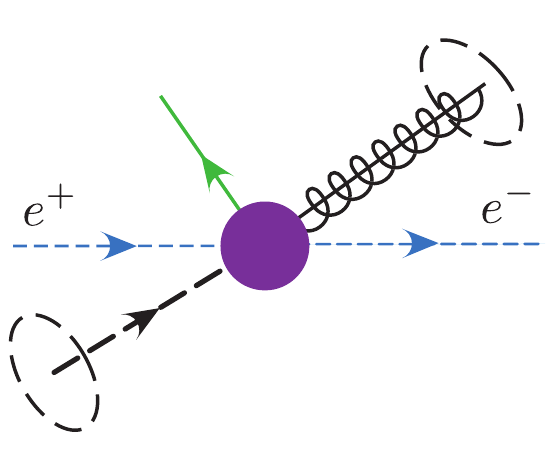} 
\hspace{0.1cm}
\includegraphics[width=0.23\columnwidth]{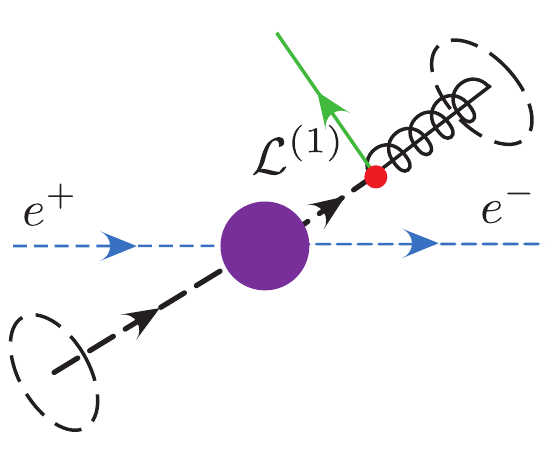} 
\raisebox{0cm}{ \hspace{-0.2cm} 
  $a$)\hspace{3.4cm}
  $b$)\hspace{3.4cm} 
  $c$)\hspace{3.4cm}
  $d$)\hspace{4cm} } 
\\[-25pt]
\end{center}
\vspace{-0.4cm}
\caption{ 
Subleading power contributions from the emission of a soft quark or gluon in a hard scattering. The subleading emission can either be from a local hard scattering operator, as shown in a), c), or from a radiative contribution from the energetic partons, as in b), d). } 
\label{fig:subleadingamp}
\end{figure}

If we consider the subleading power emission of a soft quark or gluon from a hard scattering vertex, there are two potential classes of contributions, as shown in \Fig{fig:subleadingamp}. First, there are contributions where the soft emission localizes to the hard scattering vertex, as shown in \Fig{fig:subleadingamp}. Here the power suppression arises due the lack of a nearly on-shell propagator, and these contributions are described by local hard scattering operators, complete bases of which are known for  $\bar q \Gamma q$  \cite{Feige:2017zci,Chang:2017atu} and $gg$ currents \cite{Moult:2017rpl}, as well as recently for $N$-jet configurations \cite{Beneke:2017ztn,Beneke:2018rbh}. Second, there are contributions from a non-local emission from the energetic parton, which arise from corrections beyond the eikonal limit to the dynamics of the interaction of the soft and collinear particles. Such contributions were studied in the abelian case in the work of Del Duca  \cite{DelDuca:1990gz}, extending the work of Low, Burnett and Kroll (LBK) \cite{Low:1958sn,Burnett:1967km}, and were referred to as radiative jet functions. For the emission of a single soft gluon from an energetic quark, they have been extended to the non-abelian case in \cite{Bonocore:2015esa, Bonocore:2016awd}. They were also studied in \cite{Larkoski:2014bxa} using SCET, where a one-loop expression for soft emission was derived. Our work goes beyond this, by providing explicit all orders factorization in terms of gauge invariant soft and collinear matrix elements. Here we will refer to the general class of such objects as radiative functions. We reserve ``radiative jet function" for the analogous objects at cross section level.

\begin{figure}[t!]
\begin{center}
\includegraphics[width=0.43\columnwidth]{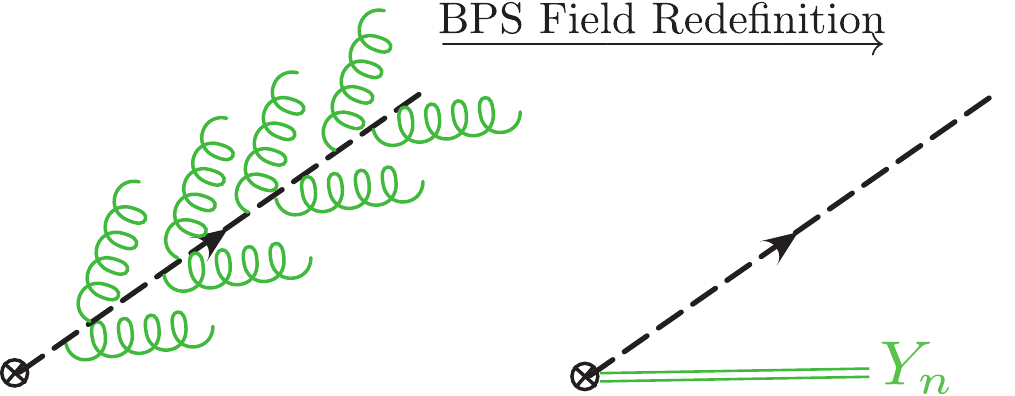} 
\hspace{0.1cm}
\includegraphics[width=0.43\columnwidth]{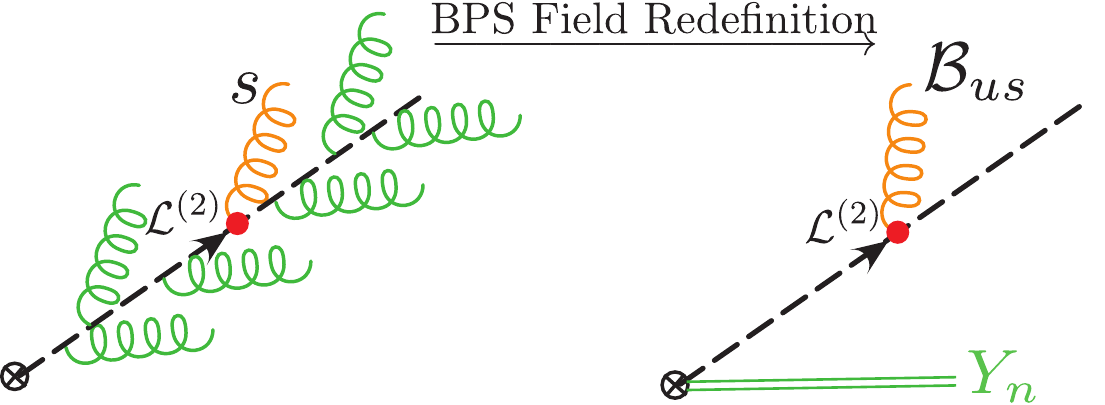} 
\raisebox{0cm}{ \hspace{-0.2cm} 
  $a$)\hspace{7cm}
  $b$)\hspace{3.4cm} 
} 
\\[-25pt]
\end{center}
\vspace{-0.4cm}
\caption{ 
The all orders factorization of soft gluons at leading power a) and next to leading power b). In the leading power case, all soft emissions are absorbed into a Wilson line $Y_n$. In the presence of a next to leading power emission, they are replaced by a Wilson line, and a gauge invariant soft gluon field emitted from the light cone.
 } 
\label{fig:sprig_NLP}
\end{figure}

To provide an all orders description, one must consider a subleading power soft emission in the presence of an arbitrary number of leading power soft, or collinear emissions.  At leading power, the energetic partons emitted from the hard scattering eikonalize, and act as a source for the long wavelength soft radiation. In this limit, the dynamics of the energetic partons can be integrated out, and replaced with a Wilson line along their path. This is shown schematically in \Fig{fig:sprig_NLP}. This leads to the ubiquitous appearance of cusped light-like Wilson lines in the description of the soft and collinear limits of gauge theory amplitudes and cross sections, whose renormalization is controlled by the universal $\Gamma_{\text{cusp}}$ \cite{Korchemsky:1987wg,Korchemsky:1991zp}. Beyond leading power we expect corrections to this picture associated with the breakdown of eikonalization, namely we expect the Wilson lines to be decorated with operators, which we will associate with radiative functions. To achieve subleading power factorization of amplitudes and cross sections, and to understand the universality of these factorizations, we would like to have a systematic approach to the construction of gauge invariant radiative functions in terms of well defined field theoretic objects. This is more difficult due to the nature of the operators, which possess intricate Wilson line structure to ensure gauge invariance in a non-abelian theory.  In QED, this is not an issue since the gauge group is abelian.

To see how radiative functions naturally emerge from the effective theory, we consider the SCET Lagrangian (here we restrict ourselves to the case of \SCETi), which consists of both hard scattering operators, and a dynamical Lagrangian
\begin{align} \label{eq:SCETLagExpand_intro}
\cL_{\text{SCET}}=\cL_\hard+\cL_\dyn= \sum_{i\geq0} \cL_\hard^{(i)}+\sum_{i\geq0} \cL^{(i)}+\cL_G^{(0)} \,,
\end{align}
each of which is a power expansion in $\lambda$. The hard scattering operators, included in $\cL_\hard$, describe all the localized contributions in \Fig{fig:subleadingamp}, while the non-local contributions are described by the dynamical Lagrangians, $\cL^{(i)}$. The dynamical Lagrangian is universal, and known up to $\cO(\lambda^2)$ \cite{Beneke:2002ni,Chay:2002vy,Manohar:2002fd,Pirjol:2002km,Beneke:2002ph,Bauer:2003mga}. Finally, $\cL_G^{(0)}$ is the leading power Glauber Lagrangian \cite{Rothstein:2016bsq}. The SCET Lagrangian is fixed by the symmetries of the theory, namely soft and collinear gauge symmetries and reparametrization invariance \cite{Manohar:2002fd,Chay:2002vy}, and is known to not be renormalized to all orders in $\alpha_s$ \cite{Beneke:2002ph}, which will allow us to prove the universality of the radiative functions.

In the SCET framework, the leading power eikonalization of \Fig{fig:sprig_NLP} is achieved at the level of the Lagrangian through the BPS field redefinition \cite{Bauer:2002nz}
Consider a subleading power emission, in the presence of an arbitrary number of additional soft emissions, as shown in \Fig{fig:sprig_NLP}. The insertion of the subleading power Lagrangian implies that the soft emissions cannot simply be pulled back into a Wilson line at the hard scattering vertex, since they become trapped at the subleading power Lagrangian insertion. The Wilson lines appearing in the BPS field redefinition can be used to sandwich the covariant derivative describing the soft gluon emission. For a derivative in an arbitrary representation, $r$, we have
\begin{align}\label{eq:soft_gluon_intro}
Y^{(r)\,\dagger}_{n_i} i D^{(r)\,\mu}_{us} Y^{(r)}_{n_i }=i \partial^\mu_{us} + [Y_{n_i}^{(r)\,\dagger} i D^{(r)\,\mu}_{us} Y^{(r)}_{n_i}]=i\partial^\mu_{us}+T_{(r)}^{a} g \cB^{a\mu}_{us(i)}\,,
\end{align}
which allows us to define the ultrasoft gauge invariant gluon building block field by
\begin{align} \label{eq:softgluondef_RF_intro}
g \cB^{a\mu}_{us(i)}= \left [   \frac{1}{in_i\cdot \partial_{us}} n_{i\nu} i G_{us}^{b\nu \mu} \cY^{ba}_{n_i}  \right] \,.
\end{align}
Furthermore, there will be a single remaining Wilson line at the hard scattering vertex. This is shown schematically in \Fig{fig:sprig_NLP}. After applying the BPS field redefinition, there are no further interactions between soft and collinear partons. The finite number of subleading power interactions between soft and collinear fields at a given power are represented by gauge invariant operator insertions along the lightcone, which dress the leading power Wilson lines. These give rise to universal non-local string operators appearing in subleading power factorization theorems. A similar picture also applies to soft quark emission. This provides a systematic way to provide gauge invariant operator definitions of radiative functions in the effective theory, which is the goal of this chapter. The use of non-local gauge invariant fields is crucial to achieve factorization for these non-local operators, since it enables gauge invariant definitions of soft and collinear matrix elements tied together by  convolution variables, that can be separately renormalized. This is non-trivial in a non-abelian gauge theory, where the soft emission carries a color charge.

An important result of this chapter is that we will show how to achieve an all orders gauge invariant factorization at the cross section level for the radiative contributions to event shape observables. This will allow us to express the cross section as a convolution of gauge invariant collinear and soft factors, each of which can in principle be separately renormalized. Unlike at leading power, these subleading power factorizations will involve an additional convolution over the gauge invariant momentum (or equivalently position along the light cone) of the insertion of the gauge invariant field. 

As an example, we can consider the factorization involving a radiative contribution at cross section level involving a soft quark field. We will show that such a contribution can be writen as a convolution which can be shown schematically as
\begin{align}
\fd{3cm}{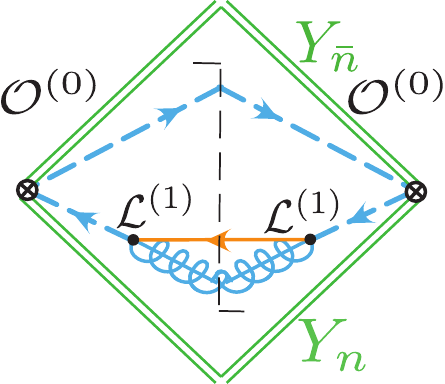} =  \int dr_2^+ dr_3^+ \fd{3cm}{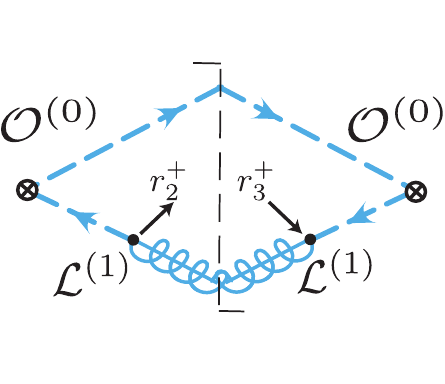} \otimes  \fd{3cm}{figures_b/soft_quark_diagram_wilsonframe_low.pdf}\,.
\end{align}
This involves a factorization into a convolution over a collinear radiative function which emits the soft quark, and a soft function involving a gauge invariant soft quark field, in addition to the standard Wilson lines. This form of factorization allows us to describe systematically, and in a gauge invariant manner, all subleading power corrections to event shape observables.

As noted earlier, non-local gauge invariant  operators describing the coupling of soft and collinear particles have appeared in the SCET literature on subleading power corrections in $B$-physics \cite{Bigi:1993ex,Neubert:1993um,Mannel:1994pm,Korchemsky:1994jb,Bauer:2001yt,Beneke:2002ph,Leibovich:2002ys,Kraetz:2002rv,Neubert:2002yx,Burrell:2003cf,Bosch:2004cb,Lee:2004ja,Beneke:2004in}, many of which have a similar structure to those discussed in this chapter.
Here we provide a simplified approach by determining the subleading power Lagrangians in terms of gauge invariant building blocks, and then we focus on the application to perturbative scattering amplitudes and collider event shape observables.  
We will also emphasize the connection to the radiative functions approach which has been studied in the literature \cite{Bonocore:2015esa, Bonocore:2016awd}.

A complete bases of hard scattering operators is given in \cite{Feige:2017zci,Moult:2017rpl,Chang:2017atu} as well as the expansion of the measurement function \cite{Feige:2017zci}. Here we will focus on the radiative type contributions, namely those term involving additional integrals over the position of Lagrangian insertions, see \sec{subl_insert}. We will formulate the factorization of the radiative contributions to the cross section as products of gauge invariant soft and collinear matrix elements involving either one or two convolutions, corresponding to the one or two Lagrangian insertions which can exist when working to $\cO(\lambda^2)$. 

Explicitly, we can derive a representation of the form
{\small
\begin{align}\label{eq:sec3_convolution}
\frac{1}{\sigma_0}\frac{d\sigma^{(2),\text{radiative}} }{d\tau}&= Q^5\sum_j H_j(Q^2)  \int \frac{dr_1^+}{2\pi Q}  S_{j}(Q\tau_{us}, r_2^+) \otimes  J_{\bar n,j}(Q^2\tau_\bn) \otimes J_{n,j}(Q^2\tau_n, Q r_2^+) \\
&+Q^5\sum_j H_j(Q^2)  \int \frac{dr_1^+}{2\pi Q}   \int \frac{dr_2^+}{2\pi Q}  S_{j}(\tau_{us},r_1^+, r_2^+) \otimes  J_{\bar n,j}(Q^2\tau_\bn) \otimes J_{n,j}(Q^2\tau_n, Qr_1^+,Qr_2^+) \nn \\
&+Q^5\sum_j H_j(Q^2)  \int \frac{dr_1^+}{2\pi Q}   \int \frac{dr_2^+}{2\pi Q}  S_{j}(Q\tau_{us}, r_1^+,r_2^+) \otimes  J_{\bar n,j}(Q^2\tau_\bn,Qr_1^+) \otimes J_{n,j}(Q^2\tau_n, Qr_2^+)\,, \nn
\end{align}}%
where we choose to make the arguments of the soft functions dimension 1 and the arguments of the jet functions dimension 2 analogously to leading power.
Here $\otimes$ denotes the convolution in the thrust variable, $\tau$, 
\be
\int d\tau_n d\tau_{\bar n} d\tau_{us} \delta(\tau-\tau_n -\tau_{\bar n}-\tau_{us}) \,,
\ee
and we have used the symmetry under $n\leftrightarrow \bar n$ to combine several equivalent contributions. The derivation of this factorized form at the cross section level is the main goal of this section. The factorization derived in this section, as well as that derived in \sec{radiative}, will be at the bare level, namely we do not consider the renormalization of the hard, jet and soft functions. To derive a renormalized factorization formula, one must show that the hard, jet and soft functions can be separately renormalized, and that the convolutions in the $r_1^+$ and $r_2^+$ variables are well defined. This is in general non-trivial, and even in simple cases the renormalization of the subleading power jet and soft functions, as we will see in \sec{subRGE} involves mixing with additional operators that do not appear in the matching \cite{Paz:2009ut,Moult:2018jjd,Beneke:2018gvs}, with evanescent operators \cite{Buras:1989xd,Dugan:1990df,Herrlich:1994kh} and possibly with EOM operators \cite{Beneke:2019kgv} (though the particular EOM operators will differ from those found in \cite{Beneke:2019kgv} due to differences in the construction of the subleading power Lagrangians), and the convolutions do not naively converge  \cite{Beneke:2003pa}. However, the derivation of a bare factorization is the first step towards a complete, renormalized factorization. 

In \Sec{sec:fact_RadiativeFunction} we will work out explicitly the structure of the factorization of the matrix elements for those contributions involving Lagrangian insertions, which give rise to the radiative jet functions. This will provide the ingredients needed to construct \Eq{eq:sec3_convolution} explicitly. This in turn yields all the pieces needed to explore the full factorization for subleading power thrust, which we plan to pursue in future work.

An outline of this chapter is as follows. In \Sec{sec:subleading_lagrangians} we derive the form of the subleading power SCET Lagrangians describing the interactions of the non-local gauge invariant quark and gluon fields, using the BPS field redefinition and the equations of motion. In \Sec{sec:RadiativeFunction_intro} we use these Lagrangians to study subleading power factorization at the amplitude level, showing how radiative functions naturally emerge, and deriving in detail their structure. We also compare our radiative functions to those previously discussed in the literature. In \Sec{sec:fact_RadiativeFunction} we consider radiative functions at the level of the cross section for event shape observables, and derive the structure of convolutions between the radiative functions. In \Sec{sec:RF_thrust} we classify all radiative functions contributing to $\SCETi$ type observables in $e^+e^-\to$ dijets. We conclude in \Sec{sec:conclusionsRadiative}.

\section{Subleading Lagrangians for Gauge Invariant Fields}\label{sec:subleading_lagrangians}

As has been emphasized, to achieve factorization into separately gauge invariant soft and collinear factors, it is essential that the radiative functions be formulated in terms of non-local gauge invariant fields, namely $\cB_{us}$ and $\psi_{us}$. We therefore will derive the subleading power Lagrangians describing the interactions of these non-local fields to all orders in $\alpha_s$.

The general form of the subleading power Lagrangians is quite complicated, since they describe the complete dynamics  of the soft and collinear sectors to all orders in $\alpha_s$. Nevertheless, due to the power counting and locality of the effective theory, there are a finite number of terms in each Lagrangian. Operationally, at a fixed order in perturbation theory, the number of terms in the Lagrangian which actually contribute is relatively small since most terms involve higher numbers of fields. Before proceeding to the full derivation of the subleading power Lagrangians, we give the structure of the Lagrangian in terms of field content, ignoring the detailed Dirac, Lorentz, and color structures. This is useful for understanding the general structure of the Lagrangians, and the order in perturbation theory at which different terms can contribute.

 At $\cO(\lambda)$ the field structure of the Lagrangian is given by
\begin{align}\label{eq:L1_fields}
\cL_n^{(1)\text{BPS}}&\sim\frac{1}{\bar \cP}\bar \chi_n \chi_n \cP_\perp \{  \partial_{us} \text{ or }\cB_{us(n)} \}   + \cB_{n\perp} \cB_{n\perp} \cP_\perp \{  \partial_{us} \text{ or }\cB_{us(n)} \}\nn \\
& + \frac{1}{\bar \cP} \bar \chi_n \chi_n \cB_{n\perp} \{  \partial_{us} \text{ or }\cB_{us(n)} \} +    \cB_{n\perp} \cB_{n\perp} \cB_{n\perp} \{  \partial_{us} \text{ or }\cB_{us(n)} \} \nn \\
&+ \frac{1}{\bar \cP}\bar \chi_n \cB_{n\perp} \psi_{us(n)} +\frac{1}{\bar \cP} \bar \psi_{us(n)} \cB_{n\perp}  \chi_n\,,
\end{align}
where we have organized the structure according to the collinear field content.
The number of fields appearing in the Lagrangian is fixed by power counting and locality, and at $\cO(\lambda)$ the Lagrangian involves up to three collinear fields. The operators that involve multiple collinear fields will not contribute at tree level to the emission of a soft parton from a single collinear parton, but are necessary to correctly reproduce the complete subleading power expression at loop level, or for multiple collinear emissions.

At $\cO(\lambda^2)$ the field structure of the Lagrangian is given by
\begin{align}\label{eq:L4_fields}
\cL_n^{(2)\text{BPS}}&\sim \cB_{n\perp} \cB_{n\perp} \{ \partial_{us} \partial_{us} \text{ or } \cB_{us(n)} \cB_{us(n)} \text{ or } \partial_{us} \cB_{us(n)} \} \\
&+\frac{1}{\bar \cP}\bar \chi_n \chi_n \{ \partial_{us} \partial_{us} \text{ or } \cB_{us(n)} \cB_{us(n)} \text{ or } \partial_{us} \cB_{us(n)} \}   \nn \\
&+\frac{1}{\bar \cP} \cB_{n\perp} \cB_{n\perp} \cB_{n\perp} \cB_{n\perp} \{  \partial_{us} \text{ or }\cB_{us(n)} \}  + \frac{1}{\bar \cP} \cB_{n\perp} \cB_{n\perp} \cP_\perp \cB_{n\perp}\{  \partial_{us} \text{ or }\cB_{us(n)} \}  \nn \\
&+  \frac{1}{\bar \cP} \cB_{n\perp} \cB_{n\perp}  \cP_\perp^2 \{  \partial_{us} \text{ or }\cB_{us(n)} \}\nn \\
&+ \frac{1}{\bar \cP^2}\bar \chi_n \chi_n \cB_{n\perp} \cB_{n\perp} \{  \partial_{us} \text{ or }\cB_{us(n)} \}    + \frac{1}{\bar \cP^2}\bar \chi_n \chi_n \cP_\perp \cB_{n\perp} \{  \partial_{us} \text{ or }\cB_{us(n)} \} \nn \\
&  + \frac{1}{\bar \cP^2}\bar \chi_n \chi_n \cP_\perp^2 \{  \partial_{us} \text{ or }\cB_{us(n)} \} + \frac{1}{\bar \cP^3}\bar \chi_n \chi_n \bar \chi_n \chi_n\{  \partial_{us} \text{ or }\cB_{us(n)} \}  \nn \\
&+ \frac{1}{\bar \cP}\bar \chi_n \chi_n \bar \chi_n \psi_{us(n)}   + \frac{1}{\bar \cP} \chi_n \cB_{n\perp}  \cB_{n\perp}    \psi_{us(n)}  + \bar \chi_n \partial_{us} \psi_{us(n)}  + \frac{1}{\bar \cP}\chi_n \cB_{n\perp}  \cP_\perp \psi_{us(n)} +\text{h.c.} \,, \nn
\end{align}
where we have again organized the terms based on their collinear field content, and we see that the $\cO(\lambda^2)$ Lagrangian involves up to four collinear fields.  

In this section we derive the exact form of the Lagrangians given in \Eqs{eq:L1_fields}{eq:L4_fields}.
We begin in \Sec{sec:sum_deriv} by summarizing the notation used in this section and the BPS transformations of different covariant derivative operators, which will allow us to write the subleading power Lagrangians in terms of gauge invariant quark and gluon fields. 
In \Sec{sec:eom} we discuss our reorganization of the Lagrangians using the equations of motion in the effective theory. 
Then, in Secs.~\ref{sec:lp_Lagrangian}-\ref{sec:lam2_Lagrangian} we present our simplified results for the BPS redefined Lagrangians in terms of gauge invariant quark and gluon fields, as well as relevant Feynman rules. 
These will be used to derive the structure of the radiative functions in \Secs{sec:RadiativeFunction_intro}{sec:fact_RadiativeFunction}.

\subsection{Field Redefinitions for Subleading Lagrangians}\label{sec:sum_deriv}

The subleading power Lagrangians in SCET are typically written in a local form, which still involve the interactions of soft and collinear partons \cite{Pirjol:2002km,Manohar:2002fd,Bauer:2003mga}. 
To derive subleading power factorization formulas involving radiative functions, we would like to rewrite them in terms of the non-local gauge invariant quark and gluon fields.
This can be achieved by performing the BPS field redefinition and manipulating the Wilson lines into gauge invariant combinations, which is the goal of this section.

Before BPS field redefinition the subleading power Lagrangians are written in terms of a variety of different covariant derivatives which we summarize here for convenience. The gauge covariant derivatives that we will use are defined by
\begin{align}
iD^\mu_n &= i\partial^\mu_n +g A^\mu_n\,, \qquad
 i \partial^\mu_n = \frac{\bn^\mu}{2} n \cdot \partial + \frac{n^\mu}{2} \overline{\cP} + \cP_\perp^\mu\,, \nn \\
 iD^\mu_{ns} &=i D^\mu_n +\frac{\bn^\mu}{2}gn \cdot A_{us}\,,\qquad
i\partial^\mu_{ns}=i \partial^\mu_n +\frac{\bn^\mu}{2} gn\cdot A_{us}\,, \nn \\
iD_{us}^\mu&=i\partial^\mu+gA^\mu_{us}\,,
\end{align}
and their gauge invariant versions are given by
\begin{align}
i\cD^\mu_{n}&=W_n^\dagger iD^\mu_{n} W_n\,, \qquad i\cD^\mu_{n\perp}=W_n^\dagger iD^\mu_{n\perp} W_n= \cP^\mu_{n\perp}+gB^\mu_{n\perp}\,, \qquad i \cD^\mu_{ns}=W_n^\dagger iD^\mu_{ns} W_n\,.
\end{align}
It is also useful to summarize the transformation of the different derivative operators under the BPS field redefinition. These are all derived using the defining relations of the Wilson line, 
\begin{align}
Y_n^\dagger Y_n=1, \qquad in\cdot D_{us} Y_n=0\,,
\end{align}
which imply the relations
\begin{align}
Y_n^\dagger in\cdot D_{us} Y_n = i n \cdot \partial_{us}\,, \qquad &&Y_n^\dagger gn \cdot A_{us} Y_n = i n \cdot \partial_{us} - Y_n^\dagger in\cdot \partial_{us} Y_n \,.
\end{align}
In addition, the ultrasoft Wilson lines commute with the label momentum operators
\begin{align}
[Y_n,\cP_\perp^\mu] = 0\,, \qquad [Y_n,\bar \cP] = 0\,.
\end{align}
Denoting the BPS transformation of an operator $\hat{O}$ as $\BPS[\hat{O}]$, we then have the following transformations for the derivative operators
\begin{align}
	\BPS[iD^\mu_{n \perp}] &=  Y_n iD^{\mu}_{n \perp} Y_n^\dagger\,, \qquad \BPS[i\cD^\mu_{n\perp}]  = Y_n i\cD^{\mu}_{n \perp} Y_n^\dagger\,, \qquad
    \BPS[i\cD^\mu_{ns}] =Y_n i\cD^{ \mu}_{n } Y_n^\dagger\,.
\end{align}
Additional useful relations are given in \App{app:BPS_identities}.

Given these identities, it is now a straightforward algebraic exercise to compute the BPS field redefinitions of the Lagrangians. By applying the unitarity condition on the ultrasoft Wilson lines, all ultrasoft Wilson lines can either be cancelled, or absorbed into gauge invariant soft quark or gluon fields, as defined in \Eqs{eq:usgaugeinvdef}{eq:softgluondef_RF}. To illustrate explicitly how this works, we consider two simple examples. First, consider a term from the leading power collinear gluon Lagrangian,
\begin{align}
	&\BPS\left[\tr \big\{ ([i \cD^\mu_{ns}, i \cD^\nu_{ns}])^2\big\} \right]    
	= \tr \Big\{ \left([ Y_n i \cD^{(0)\mu}_{n} Y_n^\dagger, Y_n i \cD_n^{(0)\nu} Y_n^\dagger]\right)^2\Big\}  \nn \\
	&\hspace{2cm}= \tr \Big\{ \left(Y_n [ i \cD^{(0)\mu}_{n} ,  i \cD_n^{(0)\nu} ] Y_n^\dagger\right)^2\Big\}
	= \tr \Big\{ \left([i \cD^{(0)\mu}_{n} , i \cD_n^{(0)\nu} ]\right)^2\Big\}\,.
\end{align}
In this case,  all the soft Wilson lines explicitly cancel, decoupling the interactions of the ultrasoft and collinear gluons. As a second example we consider a term from $\cL^{(1)}$ which contains an explicit $\cD_{us}$. Here we find that the ultrasoft gluons do not decouple
\begin{align}
	&\BPS\left[\tr\Big\{ \big[ i {\cal D}_{ns}^\mu,i {\cal D}_{n\perp}^\nu \big]\big[i {\cal D}_{ns\mu},iD^\perp_{us\,\nu} \big]\Big\} \right] \,=\, \tr\Big\{ \big[ Y_n i \cD_{n}^{(0)\mu} Y^\dagger_n,\, Y_n i \cD^{(0)\, \nu}_{n\perp}Y^\dagger_n \big]\big[Y_n i \cD^{(0)}_{n\,\mu} Y^\dagger_n,iD^\perp_{us\,\nu} \big]\Big\}  \nn\\[0.2cm]
	&= \tr\Big\{ \big[  i \cD_{n}^{(0)\mu} ,\, i \cD^{(0)\, \nu}_{n\perp} \big] \big[ i \cD^{(0)}_{n\,\mu},Y^\dagger_n iD^\perp_{us\,\nu} Y_n \big]\Big\} \, = \, \tr\Big\{ \big[  i \cD_{n}^{(0)\mu} ,\, i \cD^{(0)\, \nu}_{n\perp} \big] \big[ i \cD^{(0)}_{n\,\mu},i\partial_{us\, \nu}^\perp + g\cB^\perp_{us\,\nu} \big]\Big\}\,.
\end{align}
In the last step we used the definition of the gauge invariant ultrasoft gluon field. The derivation of the BPS field redefinition for other terms in the Lagrangian proceeds similarly, so in the following sections we will simply state the final results for the BPS redefined Lagrangians.

\subsection{Simplifications Using the Equations of Motion}\label{sec:eom}

In addition to writing the subleading power Lagrangians in terms of the non-local gauge invariant quark and gluon fields, we can also simplify their structure using the equations of motion. Recall that when building bases of hard scattering operators, only the gauge invariant building blocks in \Tab{tab:PC} are required. In particular, for the collinear gluon field, only the two degrees of freedom in $\cB_{n\perp}$ appear explicitly, and not the other components of $\cB_{n}$. In particular, the large components of the gauge field $\bar n \cdot A_n$ appear entirely in Wilson lines, and the small components have been eliminated using the equations of motion. We begin by reviewing how this is achieved, following the results of \cite{Marcantonini:2008qn}, and then apply the same simplifications to the subleading power Lagrangians.

In SCET the collinear gauge invariant covariant derivative is given by
\begin{align}
\cD^\mu_n=W_n^\dagger D^\mu_n W_n\,.
\end{align} 
which can be broken into components as
\begin{align}
i\cD^{\perp \mu}_n &=\cP^\mu_{n\perp}+g\cB^\mu_{n\perp}\,, \qquad i\overleftarrow{\cD}^{\perp \mu}_n =-\cP^{\dagger\mu}_{n\perp}-g\cB^\mu_{n\perp}\,, \nn \\
in\cdot \cD_n&=in\cdot \partial +gn\cdot \cB_n\,,  \qquad in\cdot \overleftarrow{\cD_n}=in\cdot \overleftarrow{\partial} -gn\cdot \cB_n\,, \qquad
i\bar n \cdot \cD_n=\bar \cP\,,
\end{align}
where we have defined the gauge invariant fields for the different components as
\begin{align}
g\cB^\mu_{n\perp}=\left[  \frac{1}{\bar \cP_n}   [i\bar n \cdot \cD_n, i\cD_n^{\perp \mu}]  \right]\,, \qquad gn \cdot \cB_n = \left[  \frac{1}{\bar \cP_n}   [i\bar n \cdot \cD_n, in\cdot \cD_n]  \right]\,.
\end{align}
Here the $\bar \cP_n$ operators act only within the external square brackets.
We can now eliminate the $n\cdot \cB_n$ component of the gluon field using the equation of motion
\begin{align}\label{eq:eomndotb}
\bar n \cdot \cP g n\cdot \cB_n=-2\cP_{\perp} \cdot \cB^{n\perp}_\nu +\frac{4}{\bar n \cdot \cP} g^2 T^A \sum\limits_f  \bar \chi_n^f T^A \frac{\Sl{\bn}}{2} \chi_n^f +\frac{2}{\bar n \cdot \cP}[\cB^\perp_{n\nu},[\bar n \cdot \cP g\cB^\perp_{n\nu}] ]\,.
\end{align}
This allows the Lagrangian to be written entirely in terms of $\cB_{n\perp}$ fields. From the form of \Eq{eq:eomndotb} we can see why this will lead to significant simplifications when studying soft emissions from a single collinear gluon, since all terms on the right hand side involve either a higher number of fields, or the $\cP_\perp$ operator. When studying soft emission at lowest order and lowest multiplicity, any term of the form $\cB_{us} \cB_{n} n\cdot \cB_{n}$ can therefore be dropped, which will simplify our discussion of the radiative functions.

Additionally, it is also possible to eliminate from the Lagrangian all instances of the ultrasoft derivative operator $n\cdot \partial_n$ acting on $n$-collinear fields. This is achieved for the collinear quark field using the equation of motion
\begin{align}
in\cdot \partial_n \chi_n=-g n\cdot \cB_n \chi_n -i \Sl{\cD}_n^\perp \frac{1}{\bar \cP}  i \Sl{\cD}_n^\perp \chi_n\,,
\end{align}
and for the collinear gluon field using
\begin{align} \label{eq:eomndotpartialB}
&\bar \cP [in\cdot \partial_n g\cB^\mu_{n\perp}]=-\left[ \cP^\perp_\nu [g\cB^\mu_{n\perp},g\cB^\nu_{n\perp}]   \right]     -  \left[ \cB^{n\perp}_\nu [g\cP^{[\mu}_\perp,g\cB^{\nu]}_{n\perp}]   \right]    -\left[ \cB^\perp_{n\nu} [g\cB^\mu_{n\perp},g\cB^\nu_{n\perp}]   \right] \\
&+\frac{\bar \cP}{2}[\cP^\mu_\perp g n\cdot \cB_n]   -\left[ \cP^\perp_\nu \cP^{[\mu}_\perp g\cB_n^{\nu ]}    \right]   +\frac{\bar \cP}{2}[\cB^\mu_{n\perp} g n\cdot \cB_n]    -\frac{1}{2}   \left[  gn\cdot \cB_n,[\bar \cP g\cB_{n\perp}^\mu]  \right]  \nn \\
&-g^2 T^A \sum\limits_f \left[ \bar \chi_n^f T^A \gamma^\mu_\perp \frac{1}{\bar \cP^\dagger}  (\Sl{\cP}_\perp^\dagger +g \Sl{\cB}_{n\perp})   \frac{\Sl{\bn}}{2}  \chi_n^f   \right] -g^2 T^A \sum\limits_f   \left[  \bar \chi_n^f \frac{\Sl{\bn}}{2}(\Sl{\cP}_\perp^\dagger +g \Sl{\cB}_{n\perp}) \frac{1}{\bar \cP^\dagger} T^A \gamma_\perp^\mu \chi_n^f  \right]\,. \nn
\end{align}
These equations of motion, particularly for the gluon case are considerably more cumbersome. When writing the full Lagrangian, as well as for performing fixed order calculations, we therefore find it simpler to work with ultrasoft derivatives. However, we note that if we are interested in tree level soft emissions off of a single collinear line, an identical discussion as for $n\cdot B_n$ applies, and we can ignore all appearances of $n \cdot \partial_{us}$ acting on $n$ collinear fields in the Lagrangian. By using these equations of motion, we are therefore able to greatly simplify the structure of the radiative functions we consider.

\subsection{Lagrangian at $\cO(\lambda^0)$}\label{sec:lp_Lagrangian}

For completeness, we begin by considering the leading power SCET Lagrangian. Those familiar with the leading power BPS field redefinition and SCET Lagrangian can skip to the next section. Before BPS field redefinition, the leading power Lagrangian involves interactions between collinear and ultrasoft particles. It can be written as \cite{Bauer:2000ew, Bauer:2000yr, Bauer:2001ct, Bauer:2001yt}
\begin{align} \label{eq:leadingLag_2}
\cL_{\dyn}^{(0)} &= \cL^{(0)}_{n \xi} + \cL^{(0)}_{n g} +  \cL^{(0)}_{us}\,, 
\end{align}
where
\begin{align}
\cL^{(0)}_{n \xi} &= \bar{\xi}_n\big(i n \cdot D_{ns} + i \slashed{D}_{n \perp} W_n \frac{1}{\overline{\cP}_n} W_n^\dagger i \slashed{D}_{n \perp} \big)  \frac{\slashed{\bar{n}}}{2} \xi_n\,, \\
\cL^{(0)}_{n g} &= \frac{1}{2 g^2} \tr \big\{ ([i D^\mu_{ns}, i D^\nu_{ns}])^2\big\} + \zeta \tr \big\{ ([i \partial^\mu_{ns},A_{n \mu}])^2\big\}+2 \tr \big\{\bar{c}_n [i \partial_\mu^{ns}, [i D^\mu_{ns},c_n]]\big\} \,,\nn 
\end{align}
and the ultrasoft Lagrangian, $\cL^{(0)}_{us}$, is simply the QCD Lagrangian. Throughout this chapter, we use a general covariant gauge with gauge fixing parameter $\zeta$ for the collinear gluons, and $c_n$ are the corresponding ghosts.

After performing the BPS field redefinition we have
\begin{align}
\cL^{(0)\text{BPS}} &= \cL^{(0)\text{BPS}}_{n \xi} + \cL^{(0)\text{BPS}}_{n g} +  \cL^{(0)}_{us}\,, 
\end{align}
where the ultrasoft Lagrangian is unchanged. The collinear quark Lagrangian is given by
\begin{align}
\cL^{(0)\text{BPS}}_{n \xi   }&=\bar \chi_n  \left(  i n \cdot \cD_{n}  + i \slashed{\cD}_{n \perp}  \frac{1}{\overline{\cP}_n}  i \slashed{\cD}_{n \perp}  \right)  \frac{\Sl \bn}{2} \chi_n\,,
\end{align}
and the collinear gluon Lagrangian is given by\footnote{Note that $\tr \big\{ ([i \cD^\mu_{n}, i \cD^\nu_{n}])^2\big\} = \tr \big\{ W_n^\dagger([i D^\mu_{n}, i D^\nu_{n}])^2W_n\big\} = \tr \big\{ ([i D^\mu_{n}, i D^\nu_{n}])^2\big\} $ which is the form sometimes used in the literature to write down this term of the collinear leading power Lagrangian.}
\begin{align}
\cL^{(0)\text{BPS}}_{n g} &=  \frac{1}{2 g^2} \tr \big\{ ([i \cD^\mu_{n}, i \cD^\nu_{n}])^2\big\} + \zeta \tr \big\{ ([i \partial^\mu_{n},A_{n \mu}])^2\big\}+2 \tr \big\{\bar{c}_n [i \partial_\mu^{n}, [i D^\mu_{n},c_n]]\big\} \,,
\end{align}
explicitly showing that ultrasoft and collinear interactions have been decoupled to leading power. 

\subsection{Lagrangian at $\cO(\lambda)$}\label{sec:lam_Lagrangian}

Before BPS field redefinition, the $\mathcal{O}(\lambda)$ Lagrangian can be written 
\begin{align}
\cL^{(1)}={\cal L}_{\chi_n}^{(1)}+{\cal L}_{A_n}^{(1)}+{\cal L}_{\chi_n q_{us}}^{(1)} \,,
\end{align}
where \cite{Chay:2002vy,Pirjol:2002km,Manohar:2002fd,Bauer:2003mga}
\begin{align}\label{eq:sublagcollq_2}
{\cal L}_{\chi_n}^{(1)} &= \bar \chi_n \Big(
i \slashed{D}_{us\perp}\frac{1}{ \bar \cP} 
i \slashed{\cal D}_{n\perp}
+i \slashed{\cal D}_{n\perp}\, \frac{1}{\bar \cP} 
i \slashed{D}_{us\perp} 
\Big)\frac{\slashed{\bar{n}}}{2} \chi_n \ ,
 \end{align}
describes the interactions between collinear quarks and gluons, and
\begin{align}
{\cal L}_{A_n}^{(1)}&= \frac{2}{g^2}\text{Tr}\Big(
\big[ i {\cal D}_{ns}^\mu,i {\cal D}_{n\perp}^\nu \big]\big[
i {\cal D}_{ns\mu},iD^\perp_{us\,\nu} 
\big]
\Big)  
+ 2 \zeta \text{Tr} \left(   [i D_{us\perp}^\mu, A_{n\perp \mu}] [i\partial^\nu_{ns},A_{n\nu}] \right) \nn \\
&+2 \text{Tr}   \left( \bar c_n [iD_{us\perp}^\mu, [iD^\perp_{n\mu}, c_n  ]]  \right)          +2 \text{Tr}   \left( \bar c_n [\cP_\perp^\mu,[ W_n iD^\perp_{us\,\mu} W_n^\dagger, c_n   ]]  \right)    \,,  
\end{align}
describes the dynamics of the pure gluon sector, including gauge fixing terms\footnote{Note that the presence of power suppressed gauge fixing Lagrangians is necessary due to the fact that RPI symmetry connects Lagrangians at different orders in the power counting, and would be broken if they were not included.  For example, these subleading power gauge fixing Lagrangians have been shown to give important contributions to the derivation of the LBK theorem for gluons in SCET, see Appendix D of \cite{Larkoski:2014bxa}.}, and
\begin{align}
{\cal L}_{\chi_n q_{us}}^{(1)} 
&= \bar{\chi}_n  g \slashed{\cB}_{n\perp} q_{us}+\text{h.c.,}
\end{align}  
describes the coupling of soft and collinear quarks.

We now wish to express the subleading power Lagrangians in a simplified form in terms of the gauge invariant building blocks, which will be one of the main results of this chapter. This organization of the Lagrangians after BPS field redefinition was also considered in \cite{Larkoski:2014bxa}, although there it was performed schematically. Here we will provide explicit expressions for all components, as well as use the equations of motion to simplify the result so that it can easily be used for subleading power factorization.

After performing the BPS field redefinition, we can perform the same division of the Lagrangian as above,  
\begin{align}
\cL^{(1)\text{BPS}}={\cal L}_{\chi_n}^{(1)\text{BPS}}+{\cal L}_{A_n}^{(1)\text{BPS}}+{\cal L}_{\chi_n q_{us}}^{(1)\text{BPS}} \,,
\end{align}
where the collinear quark Lagrangian is given by
\begin{align} 
{\cal L}_{\chi_n}^{(1) \text{BPS}} &=g \bar \chi_n  \cB^\perp_{us(n)} \cdot \cP_\perp \frac{\Sl \bn}{\bar \cP} \chi_n   +   g \bar \chi_n  \partial^\perp_{us(n)} \cdot \cP_\perp \frac{\Sl \bn}{\bar \cP} \chi_n 
 +\bar \chi_n \left( i\Sl{\partial}_{\perp us}  \frac{1}{\bar \cP} g \Sl{\cB}_{n\perp} +g\Sl{\cB}_{n\perp}\frac{1}{\bar \cP} i\Sl{\partial}_{\perp us}   \right) \frac{\Sl{\bar n}}{2} \chi_n  \nn \\  
&+\bar \chi_n \left( i\Sl{\cB}_{\perp us(n)}  \frac{1}{\bar \cP} g \Sl{\cB}_{n\perp} +g\Sl{\cB}_{n\perp}\frac{1}{\bar \cP} i\Sl{\cB}_{\perp us(n)}   \right) \frac{\Sl{\bar n}}{2} \chi_n \,, 
\end{align}
the collinear gluon Lagrangian is divided into three pieces
\begin{align}
{\cal L}_{A_n}^{(1)\text{BPS}}={\cal L}_{g_n}^{(1) \text{BPS}}  +  {\cal L}_{\text{ghost}}^{(1) \text{BPS}}  +  {\cal L}_{\text{gf}}^{(1) \text{BPS}}\,,
\end{align}
which are given by
\begin{align}
{\cal L}_{g_n}^{(1) \text{BPS}}&=[\cP^{\nu}_{\perp} \cB^{\mu a}_\perp][i\partial^\perp_{us \nu}  \cB^{\mu a}_{n\perp}]-[\cP^\mu_\perp \cB^{\nu a}_\perp][i\partial^\perp_{us \nu} \cB^{\mu a}_{n\perp}] \nn \\
&~~~\,-igf_{abc} \left\{ \cB^{\nu a}_{us \perp} \cB^{\mu b}_{n\perp}[\cP^\mu_\perp \cB^{\nu c}_{n\perp}] -\cB^{\nu a}_{us \perp} \cB^{\mu b}_{n\perp}[\cP^\nu_\perp \cB^{\mu c}_{n\perp}] + \cB^{\mu a}_{n\perp} \cB^{\nu b}_{n\perp}  [i\partial_{us \nu}^\perp g \cB_{n\perp}^{\mu c}] \right\}\nn \\
&~~~\,+g^2 f_{abe} f_{cde} \cB^{\mu a}_{n\perp}   \cB^{\nu b}_{n\perp} \cB^{\nu c}_{us \perp} \cB^{\mu d}_{n\perp}  \nn \\
&~~~\,+ [\bar \cP \cB^{\nu a}_{n\perp}] [i\partial^\perp_{us \nu} n\cdot \cB^a_n] + ig f^{abc}[\bar \cP \cB^{\nu a}_{n\perp}] n\cdot \cB^b_n \cB_{us \perp}^{c\nu}\,,\nn \\
{\cal L}_{\text{ghost}}^{(1) \text{BPS}} &= 2 \text{Tr}  \left(    \bar c_n [ i \partial^\mu_{us\perp},[i D^\perp_{n\mu}, c_n]] \right)    +2 \text{Tr}  \left(    \bar c_n [ T^a,[i D^\perp_{n\mu}, c_n]]        \right)   g \cB_{us(n)}^{a\mu} \,   \nn \\
&+2 \text{Tr}  \left(    \bar c_n  [ \cP^\mu_\perp  , [W_n   i \partial_{us\perp\mu}   W^\dagger_n, c_n] ]   \right)         +2 \text{Tr}  \left(    \bar c_n  [ \cP_{\perp \mu}  , [W_n   T^a   W^\dagger_n, c_n] ]   \right)   g \cB_{us(n)}^{a\mu} \,, \nn \\
{\cal L}_{\text{gf}}^{(1) \text{BPS}} &= 2\zeta \text{Tr}\left([i\partial^\mu_{us\perp},A_{n\perp \mu}][i \partial_n^\nu,A_{n\nu}]    \right)    + 2\zeta \text{Tr}\left( [T^a,A_{n\perp \mu}][i \partial_n^\nu,A_{n\nu}]     \right) g \cB_{us(n)}^{a\mu} \,.
\end{align}
Finally, the interaction of soft quarks is described by the Lagrangian
\begin{align}
{\cal L}_{\chi_n \psi_{us}}^{(1)\text{BPS}} 
&= \bar{\chi}_n g \slashed{\cB}_{n\perp} \psi_{us (n)}+\text{h.c.} \,.
\end{align}

The structure of the $\cO(\lambda)$ Lagrangian is quite complicated, since it describes the complete dynamics of the subleading power corrections to the soft and collinear dynamics, including ghost and gauge fixing terms. In its current form, it also involves multiple polarizations of the collinear gluon field.  To simplify its structure, we use the equations of motion\footnote{Note that the EOM are homogeneus in the power counting $\lambda$, but not in the coupling constant. Therefore the use of the EOM can reshuffle terms among different orders in $g$, but it won't move terms between Lagrangians at different orders.}, as discussed in \Sec{sec:eom}. Simplifying the result to focus only on ultrasoft emissions out of two collinear fields we find the structure
\begin{align}\label{eq:L1_complete}
{\cal L}_{n}^{(1)\text{BPS}} =&  -2g[\cP_\perp^\mu \, \cB_{n\perp }^\nu] [\cB_{n  \nu}^\perp , \cB^{\perp}_{us \,\mu}] +g \bar \chi_n  \cB^\perp_{us(n)} \cdot \cP_\perp \frac{\Sl \bn}{\bar \cP} \chi_n + \bar{\chi}_n g \slashed{\cB}_{n\perp} \psi_{us (n)}+\text{h.c.} \nn \\
&+ \cB_{us} \cdot K^{(1)}_{\cB_{us}} + K^{(1)}_{\partial_{us}}\,,
\end{align}
where $K^{(1)}_{\cB_{us}}$ and $K^{(1)}_{\partial_{us}}$ contain $\geq 3$ collinear fields, and are therefore not relevant for our current analysis. 
In this form, the Lagrangian is written entirely in terms gauge invariant fields, and due to the organization in terms of fields, it is clear at which order in perturbation theory each term can contribute. After performing the BPS field redefinition, and writing the result in terms of collinear and soft gauge invariant fields, the soft and collinear fields are only coupled through Lorentz and color indices, as well as through potential derivative operators. Since each of the building blocks appearing in the Lagrangian is separately gauge invariant, this will allow for a simple factorization into collinear and soft components, tied together through Lorentz and color indices, which will give rise to the radiative functions.

\begin{figure}[t!]
\begin{center}
\begin{align}
	\fd{3cm}{figures/feynman_rule_singlesoft_emission_sub}~&=~\frac{2igT^a}{\bn \cdot p}\left(p_\perp^\mu - \frac{k_s^\perp \cdot p_\perp}{n\cdot k_s} n^\mu\right)\frac{\bnslash}{2}\nn \\
	\fd{3cm}{figures/feynman_rule_trigluon_sub}&=2gf^{abc}\left[  g_\perp^{\mu \nu} p_\perp^\rho -g_\perp^{\mu \nu} p_\perp\cdot k_\perp  \frac{n^\rho}{n\cdot k} \right.  \nn\\
&\hspace{2cm}\left.- \left(p_\perp^\rho - p_\perp\cdot k_\perp  \frac{n^\rho}{n\cdot k} \right)\left( p_\perp^\mu \frac{\bn^\nu}{\bn \cdot p} + p_\perp^\nu \frac{\bn^\mu}{\bn \cdot p} -p_\perp^2 \frac{\bn^\mu\bn^\nu}{(\bn \cdot p)^2}  \right) \right]\nn \\
\fd{3cm}{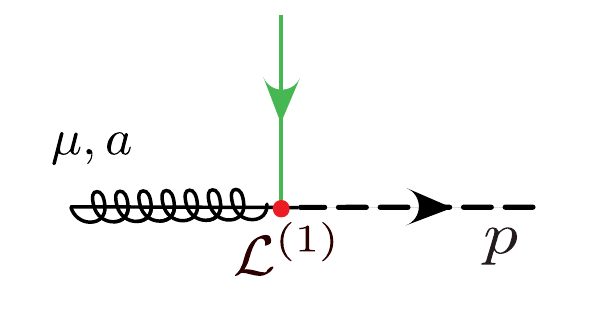}&=igT^a\left( \gamma_\perp^\mu - \frac{\Sl{p}_{\perp}\, \bar{n}^\mu}{\bar n \cdot p} \right)	\nn
\end{align}
\end{center}
\vspace{-0.4cm}
\caption{ Feynman rules for the $\cO(\lambda)$ Lagrangian describing the emission of a soft gluon or quark.
 } 
\label{fig:lam_feynrules}
\end{figure}

The first three terms of \Eq{eq:L1_complete} describe the $\cO(\lambda)$ emission of a soft gluon from a collinear gluon, a soft gluon from a collinear quark, and a soft quark from a collinear quark or gluon, respectively. Using the Lagrangian, we can derive the tree level Feynman rules, which are given in \Fig{fig:lam_feynrules}. Note that in accord with the LBK theorem, the single ultrasoft gluon Feynman rule of $\cL^{(1)}$ vanishes when the label $\perp$ momentum of the collinear leg is set to zero. Unlike for the emission of a soft gluon, the Feynman rule for a soft quark emission does not vanish when the $\cP_\perp$ of the collinear line vanishes.

Since the Lagrangian is defined in terms of gauge invariant soft quark and gluon fields, which involve ultrasoft Wilson lines, they also give the Feynman rules for an arbitrary number of additional leading power soft gluon emissions.  Similarly, the gauge invariant collinear fields also involve collinear Wilson lines, which describe collinear radiative corrections to the above Feynman rules.   
The $K^{(1)}_{\cB_{us}}$ and $K^{(1)}_{\partial_{us}}$ in \Eq{eq:L1_complete} involve additional collinear fields. For a single soft emission from a collinear line, these can first appear at loop level. We will not work out the explicit form of these loop contributions in this initial chapter, however,  we will discuss their contributions in later sections.

\subsection{Lagrangian at $\cO(\lambda^2)$}\label{sec:lam2_Lagrangian}

At $\mathcal{O}(\lambda^2)$ the SCET Lagrangian before BPS field redefinition can be written as \cite{Pirjol:2002km,Manohar:2002fd,Bauer:2003mga}
\begin{align}
	{\cal L}^{(2)} &= {\cal L}_{\chi_n}^{(2)}  + {\cal L}_{A_n}^{(2)} 
	+ {\cal L}_{\chi_n q_{us}}^{(2)} \,,
\end{align}
where for convenience, we further decompose the gluon Lagrangian as
\begin{align}
{\cal L}_{A_n}^{(2)}={\cal L}_{g_n}^{(2) }  +  {\cal L}_{\text{ghost}}^{(2) }  +  {\cal L}_{\text{gf}}^{(2) }\,.
\end{align}
The different components of the Lagrangian are given by
\begin{align}
	\cL_{\chi_n q_{us}}^{(2)}&=  \bar \chi_n \frac{\Sl \bn}{2} [ W_n^\dagger in\cdot D W_n]  q_{us}    + \bar \chi_n \frac{\Sl \bn}{2}  i\Sl \cD_{n\perp}    \frac{1}{\overline{\cP}}   ig \Sl \cB_{n\perp}  q_{us}+ \text{h.c.} \,, \nn \\
	\cL_{\chi_n}^{(2)}&=  \bar \chi_n   \left( i\Sl D_{us\perp} \frac{1}{\overline{\cP}}  i\Sl D_{us\perp}-  i\Sl \cD_{n\perp}   \frac{i\bn \cdot D_{us}}{(\overline{\cP})^2}   i\Sl \cD_{n\perp}   \right)    \frac{\Sl \bn}{2} \chi_n\,, \nn \\
	\cL_{ng}^{(2)}&=\frac{1}{g^2} \text{Tr} \left(  [  i\cD^\mu_{ns} , iD_{us}^{\perp \nu}   ]    [ i \cD_{ns\mu}   ,i D^\perp_{us\nu}   ] \right)   +\frac{1}{g^2} \text{Tr} \left(  [ iD^\mu_{us\perp}  ,i D^\nu_{us\perp}   ]    [ i\cD^\perp_{n\mu}  , i\cD^\perp_{n\nu}   ] \right) \nn \\
	&\hspace{-0.4cm}+ \frac{1}{g^2} \text{Tr} \left(  [ i\cD_{ns}^{\mu}  , i n\cdot \cD_{ns}  ]    [  i\cD_{ns\mu} , i\bn \cdot D_{us}   ] \right)+\frac{1}{g^2} \text{Tr} \left(  [ iD^\mu_{us\perp}  ,  i\cD^\perp_{n\nu}  ]    [  i\cD^\perp_{n\mu}  ,  iD_{us \perp}^\nu  ] \right)\,, \nn \\
	\cL_{\text{gf}}^{(2)}&= \zeta \text{Tr} \left(   [ i D^\mu_{us\perp}, A_{n\perp \mu}]   [i D^\nu_{us\perp}, A_{n\perp \nu}] \right)   + \zeta \text{Tr} \left(   [i \bn \cdot D_{us},n \cdot A_n]   [i \partial^\mu_{ns}, A_{n\mu}] \right) \,, \nn \\
	\cL_{\text{ghost}}^{(2)}&= 2 \text{Tr} \left(  \bar c_n [iD^\mu_{us\perp},[ W_n iD^\perp_{us\mu}W_n^\dagger ,c_n]] \right)+ \text{Tr} \left( \bar c_n [i \bn \cdot D_{us} ,[ i n\cdot D_{ns},c_n]] \right) \nn \\
	&+ \text{Tr} \left( \bar c_n [\overline{\cP},[W_n i \bn \cdot D_{us} W_n^\dagger,c_n]] \right)\,.
\end{align}
After performing the BPS field redefinition, and writing the result in terms of ultrasoft gauge invariant fields, we find that the Lagrangians involving quark fields can be written
\begin{align}
{\cal L}_{\chi_n}^{(2) \text{BPS}} &= \bar \chi_n  \left( i \Sl \partial_{us\perp}  \frac{1}{ \bar \cP} i \Sl \partial_{us\perp}    \right)   \frac{\Sl \bn}{2} \chi_n \nn \\
&-\bar \chi_n  \frac{i\bar n \cdot \partial_{us} }{\bar \cP^2} \cP_\perp^2 \frac{\Sl{\bar n}}{2} \chi_n              -\bar \chi_n g \Sl{\cB}_\perp  \frac{i\bar n \cdot \partial_{us} }{\bar \cP^2} \Sl{\cP}_\perp \frac{\Sl{\bar n}}{2} \chi_n    \nn \\
&-\bar \chi_n  \frac{i\bar n \cdot \partial_{us} }{\bar \cP^2} \Sl{\cP}_\perp \Sl{\cB}_\perp \frac{\Sl{\bar n}}{2} \chi_n              -\bar \chi_n g \Sl{\cB}_\perp  \frac{i\bar n \cdot \partial_{us} }{\bar \cP^2} \Sl{\cB}_\perp \frac{\Sl{\bar n}}{2} \chi_n \nn \\
&+   \bar \chi_n  \left(  T^a \gamma^\mu_\perp \frac{1}{\bar \cP}  i \Sl \partial_{us\perp} - i  {\overleftarrow{\Sl \partial}}_{us\perp} \frac{1}{\bar \cP} T^a \gamma^\mu_\perp    \right)   \frac{\Sl \bn}{2} \chi_n  \,   g \cB_{us(n)}^{a\mu} \nn \\
&-\bar \chi_n T^a \frac{\cP_\perp^2 }{\bar \cP^2}  \frac{\Sl{\bar n}}{2} \chi_n    g \bar n \cdot\cB_{us(n)}^{a}        -\bar \chi_n g \Sl{\cB}_\perp  \frac{T^a }{\bar \cP^2} \Sl{\cP}_\perp \frac{\Sl{\bar n}}{2} \chi_n   g \bn \cdot \cB_{us(n)}^{a} \nn \\
&-\bar \chi_n \Sl{\cP}_\perp  \frac{T^a  }{\bar \cP^2}  \Sl{\cB}_\perp \frac{\Sl{\bar n}}{2} \chi_n      g\bn \cdot \cB_{us(n)}^{a}        -\bar \chi_n g \Sl{\cB}_\perp  \frac{T^a  }{\bar \cP^2} \Sl{\cB}_\perp \frac{\Sl{\bar n}}{2} \chi_n g \bn \cdot \cB_{us(n)}^{a} \nn \\
&+    \bar \chi_n  \left( T^a \gamma^\mu_\perp \frac{1}{\bar \cP} T^b \gamma^\nu_\perp   \right)   \frac{\Sl \bn}{2} \chi_n  \,   g \cB_{us(n)}^{a\mu}   g \cB_{us(n)}^{b\nu}\nn \\
&+\bar \chi_n \frac{\Sl \bn}{2}  in\cdot \cB_n  \psi^{(n)}_{us}    + \bar \chi_n \frac{\Sl \bn}{2}  i\Sl \cD_{n\perp}    \frac{1}{\overline{\cP}}   ig \Sl \cB_{n\perp}  \psi^{(n)}_{us}+ \text{h.c.}\,.
\end{align}

The Lagrangians describing the pure glue sector are more complicated, involving both ghost and gauge fixing terms. We find that they can be written
\begin{align}\label{eq:subsubgluonlagr}
	\cL_{ng}^{(2)\text{BPS}}&=\text{Tr} \left(i n\cdot \partial\cB_{n \mu}^\perp    i \bn\cdot \partial\cB^{\perp\mu}_{n} - [ \cPbar n\cdot \cB_{n}  ]    i\bn \cdot \partial n\cdot\cB_{n} -[\cP_\perp^\mu n \cdot \cB_n ] i\bn \cdot \partial  \cB_{n\mu}^\perp \right) \nn \\
	&+ g \text{Tr} \left( \partial_\perp^{[\mu} \cB_{us \perp}^{\nu]}[ \cB^\perp_{n\mu}  , \cB^\perp_{n\nu}   ] - in\cdot \partial\cB_{n\mu}^\perp [ \cB_{n\mu}^\perp , \bn \cdot \cB^{(n)}_{us}  ]  \right. \nn \\
	&\left.\qquad\quad +\cP_\perp^\mu n \cdot \cB_n [ \cB_{n\mu}^\perp , \bn \cdot \cB^{(n)}_{us}  ] + \cPbar n\cdot \cB_{n}     [  n\cdot\cB_{n} , \bn \cdot \cB^{(n)}_{us}   ]   \right)\nn \\
	&+  g^2\text{Tr} \left([ \cB_{us}^{\perp (n) \mu}  ,\cB_{us}^{\perp (n) \nu}   ]    [ \cB^\perp_{n\mu}  , \cB^\perp_{n\nu}   ]   \right) \,, \nn\\
	\cL_{{\text{gf}}}^{(2)\text{BPS}}&= \zeta \text{Tr} \left(   [ i \partial^\mu_{us\perp}+T^a g \cB^{a\mu}_{us(n)} , A_{n\perp \mu}]   [i \partial^\nu_{us\perp}+T^a g \cB^{a\nu}_{us(n)} , A_{n\perp \nu}] \right)   \nn \\
	&+ \zeta \text{Tr} \left(   [i \bn \cdot \partial_{us} +T^a g \bn\cdot \cB^{a}_{us(n)}  ,n \cdot A_n]   [i \partial^\mu_{n}, A_{n\mu}] \right) \,, \nn \\
	\cL_{\text{ghost}}^{(2)\text{BPS}}&= 
	 2 \text{Tr} \left(  \bar c_n [i\partial^\mu_{us\perp}+T^a g  \cB^{a\mu}_{us(n) \perp},[ W_n (i\partial^\perp_{us\,\mu} +T^b g  \cB^{b\perp}_{us(n)\mu})W_n^\dagger ,c_n]] \right) \nn \\
	&+ \text{Tr} \left( \bar c_n [i \bn \cdot \partial_{us}+ T^a g  \bn \cdot\cB^{a}_{us(n)},[ i n\cdot D_{n},c_n]] \right) \nn \\
	&+ \text{Tr} \left( \bar c_n [\overline{\cP},[W_n( i \bn \cdot \partial_{us}+T^a g \bn\cdot \cB^{a}_{us(n)} ) W_n^\dagger,c_n]] \right)\,.
\end{align}

To make the $\cO(\lambda^2)$ Lagrangian more tractable, we can use the equations of motion to write it entirely in terms of our basis of gauge invariant building blocks. This is a straightforward, but tedious algebraic exercise, and therefore we simply present the final result. Using the equations of motion to rewrite the Lagrangian in terms of our operator basis, we find 
\begin{align}\label{eq:lam2_BPS}
	\cL^{(2)\text{BPS}}&= \text{Tr} \left(i n\cdot \partial\cB_{n \mu}^\perp    i \bn\cdot \partial\cB^{\perp\mu}_{n} - 4[ \cP_{n\perp}\cdot \cB_{n\perp} ]    i\bn \cdot \partial \frac{\cP_{n\perp}\cdot \cB_{n\perp}}{\bar n \cdot \cP}  +2\left[\cP_\perp^\mu\frac{\cP_{n\perp}\cdot \cB_{n\perp}}{\bar n \cdot \cP} \right] i\bn \cdot \partial  \cB_{n\mu}^\perp \right)  \nn \\
	&- \bar \chi_n   \frac{\partial^2_{\perp}}{\overline{\cP}}\frac{\Sl \bn}{2} \chi_n+ \nn \\
	&+g\bar \chi_n \frac{\Sl \bn}{2}  i\Sl \cP_{\perp}    \frac{1}{\overline{\cP}}   \Sl \cB_{n\perp}  \psi_{us(n)} - g\bar \chi_n \frac{\Sl \bn}{2} \frac{2}{\overline{\cP}}\cP_\perp \cdot \cB_n  \psi_{us(n)} + \text{h.c.} \nn \\
	&+ \text{Tr} \left(2g\left[i\partial^{\mu}_\perp  \cB^{\nu}_{us}\right]\,[ \cB^\perp_{n\mu}  , \cB^\perp_{n\nu}   ] +g^2 [ \cB_{us}^{\perp \mu}  ,\cB_{us}^{\perp \nu}   ]    [ \cB^\perp_{n\mu}  , \cB^\perp_{n\nu}   ]  \right)  \nn \\ 
	&+g\bar \chi_n   \left( [i\Sl \partial_{us\perp}  \Sl \cB_{us\perp}] + 2 \cB_{us\perp} \cdot   i{\partial}_{us\perp}  \right) \frac{1}{\overline{\cP}}  \frac{\Sl \bn}{2} \chi_n + g^2 \bar \chi_n  \Sl \cB_{us\perp} \Sl \cB_{us\perp}  \frac{1}{\overline{\cP}}\frac{\Sl \bn}{2} \chi_n \nn \\
	&+ \cB_{us} \cdot K^{(2)}_{\cB_{us}} + K^{(2)}_{\partial_{us}}\,, 
\end{align}
where, as in \Eq{eq:L1_complete}, the $K$ contain $\geq3$ collinear fields, which will not be relevant for the discussion in this chapter. This gives the Lagrangian at $\cO(\lambda^2)$ in terms of gauge invariant soft and collinear quark and gluon fields in such a way that it is clear at which order in perturbation theory each term can contribute. We have used the EOM to write it entirely in terms of the $\cB_{n\perp}$ field, eliminating the other polarizations. For practical applications, we can also apply the EOM of \Eq{eq:eomndotpartialB}, however, this significantly complicates the structure of the Lagrangian, and therefore we have not written it out explicitly. This simplified form of the $\cO(\lambda^2)$ Lagrangian is one of the key results of this chapter. We again emphasize that its highly non-trivial non-local structure, involving a multitude of soft and collinear Wilson lines, is fully determined by the structure of the BPS field redefinition, and the local $\SCETi$ Lagrangians, allowing it to be constructed systematically. This form, in terms of gauge invariant building blocks linked only by Lorentz and color indices, will allow for a straightforward factorization into radiative functions.

\begin{figure}[t!]
\begin{center}
\small
\begin{align}
\fd{5cm}{figures/feynman_rule_singlesoft_emission_subsub}\hspace{-.8cm}&=igT^a\left[ 2 k^\perp_{1 \mu} -\Sl k^\perp_s \gamma^\perp_\mu -\left( 2 k_s^\perp \cdot k_1^\perp -(k_s^\perp)^2 \right) \frac{n_\mu}{n\cdot k_s} \right]\frac{\bnslash}{2\bar n \cdot p}\nn \,,\\
\fd{4cm}{figures/feynman_rule_trigluon_subsub}&=2 g f^{abc}\left(k_\perp^\mu g_\perp^{\nu \rho } - k_\perp^\nu g_\perp^{\mu \rho } \right)\nn\,, \\[3mm]
\fd{4cm}{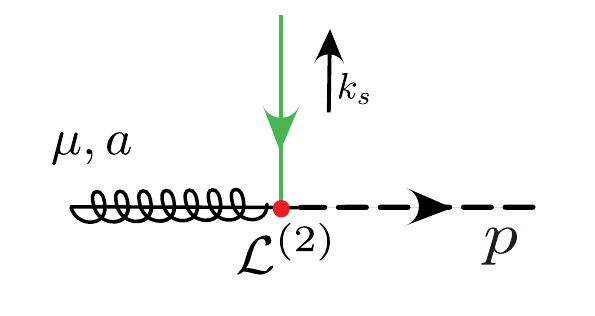}&=-\frac{gT^a}{n \cdot k_s}\left[n^\mu -\frac{n \cdot p }{\bar n \cdot p}\,\bar{n}^\mu  \right]\nn \,,\\
	\fd{5cm}{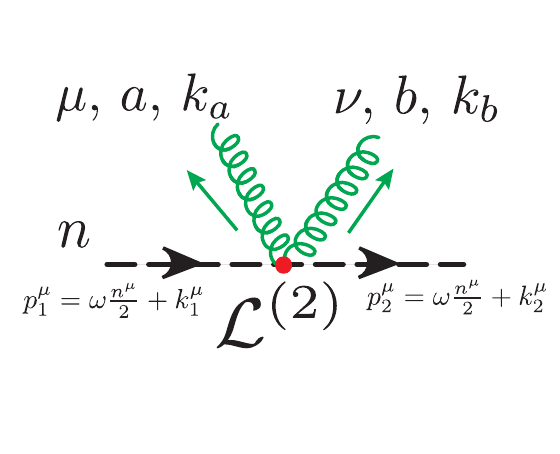}\hspace{-5mm}&=-\frac{g^2 \bnslash}{2\bn\cdot p}\left(\frac{1}{4}[T^a,T^b][\gamma_\perp^\alpha, \gamma_\perp^\beta] + \frac{1}{2}\{T^a,T^b\}g_\perp^{\alpha \beta}\right)\left(g_\perp^{\mu \alpha} g_\perp^{\nu\beta} - \frac{k^\alpha_{a\perp} g_\perp^{\nu\beta}}{n\cdot k_a} n^\mu \right.    \nn \\[-1cm]
	& \left.\quad- \frac{g_\perp^{\mu \alpha}k^\beta_{b\perp}}{n\cdot k_b} n^\nu  +n^\mu n^\nu \frac{k^\alpha_{a\perp}  k^\beta_{b\perp}}{n\cdot k_a\, n\cdot k_b} \right) + \big(a\leftrightarrow b,\mu \leftrightarrow \nu, k_a \leftrightarrow k_b \big)\nn \\
	&=-\frac{g^2 \bnslash}{2\bn\cdot p}T^a T^b\left(\gamma_\perp^\mu \gamma_\perp^\nu - \frac{\Sl{k}_{a\perp} \gamma_\perp^\nu}{n\cdot k_a} n^\mu - \frac{\gamma_\perp^\mu\Sl{k}_{b\perp}}{n\cdot k_b} n^\nu  +n^\mu n^\nu \frac{\Sl{k}_{a\perp}  \Sl{k}_{b\perp}}{n\cdot k_a\, n\cdot k_b} \right) \nn \\
	&\quad+ \big(a\leftrightarrow b,\mu \leftrightarrow \nu, k_a \leftrightarrow k_b \big)\nn\,, \\[2mm]
	\fd{4cm}{figures/feynman_rule_fourgluon_subsub}&=-2g^4 f^{abc}f^{cde} \left[ g^\perp_{\mu \rho} g^\perp_{\nu \sigma} -g^\perp_{\mu \sigma} g^\perp_{\nu \rho} + \frac{n^\mu}{n\cdot k_1}\left(g^\perp_{\nu \rho} k_{1 \sigma}^\perp -g^\perp_{\nu \sigma} k_{1 \rho}^\perp \right)  \nn \right. \\
	&\left.\quad - \frac{n^\nu}{n\cdot k_2}\left(g^\perp_{\mu \rho} k_{2 \sigma}^\perp -g^\perp_{\mu \sigma} k_{2 \rho}^\perp \right) + \frac{n^\mu n^\nu}{n\cdot k_1 n\cdot k_2}\left( k_{1 \rho}^\perp k_{2 \sigma}^\perp - k_{1 \sigma}^\perp k_{2 \rho}^\perp \right) \right]\nn\,.
\end{align}
\end{center}
\vspace{-0.4cm}
\caption{ Feynman rules for the $\cO(\lambda^2)$ Lagrangian when $p_\perp^n=0$ describing the emission of a soft gluon or quark, or the double non-eikonal emission of soft gluons.
 } 
\label{fig:lam2_feynrules}
\end{figure}

In \Eq{eq:lam2_BPS}, terms appear involving $0,1$ and $2$ $\cB_{us}$ fields, as well as the gauge invariant ultrasoft quark field. Since the $\cO(\lambda^2)$ Lagrangians contain various terms we are going to give the Feynman rules under the common assumption of vanishing  label perpendicular momentum of all collinear fields to zero $\cP_\perp =0$. Under this assumption the Feynman rules are given in \Fig{fig:lam2_feynrules}.

It is important to emphasize that since $\cB_{us(n)}^{b\nu}$ has Feynman rules with an infinite number of soft emissions, the terms involving one and two $\cB_{us}$ fields will both contribute to the complete two gluon Feynman rule. In the Feynman rules in \Fig{fig:lam2_feynrules} we have given only the contribution from the Lagrangian insertion involving two $\cB_{us}$ fields. For simplicity we have not given the two soft gluon Feynman rule from $\cL_{\cB}^{(2)}$ which can be straightforwardly derived using the two gluon Feynman rule of the ultrasoft gauge invariant gluon field in \eq{twogluonBus}. These contributions are separately gauge invariant.

The other terms in \Eq{eq:lam2_BPS} involve additional collinear fields. For soft emissions from a collinear line, they can first contribute when there are collinear loops. We will not explicitly compute their loop level contributions, but will further discuss their structure in later sections. Note that at $\cO(\lambda^2)$, we also have contributions from two insertions of the $\cO(\lambda)$ operator. These will be discussed when we consider the complete classification of radiative function for $e^+e^-\to$ dijets in \Sec{sec:RF_thrust}.

\section{Amplitude Level Factorization with Radiative Functions}\label{sec:RadiativeFunction_intro}

In this section we derive factorization formulas in terms of radiative functions for soft emissions at amplitude level. While our goal is to study the factorization of event shapes, and the structure of radiative functions at cross section level, initiating our studies at amplitude level is useful for several reasons. First, it is useful for connecting to the study of the subleading power soft behavior of amplitudes, which in itself is an interesting subject to which the factorization theorems that we derive can be applied. Second, it allows us to connect to other work in the literature on radiative functions, which have typically been formulated at amplitude level. Finally, it also provides a slightly simpler situation to illustrate the general features of factorization involving radiative functions, which will persist at the cross section level. In particular, we will illustrate how radiative functions are defined as integrals along the lightcone of Lagrangian insertions, which dress the leading power Wilson lines, giving rise to a breakdown of eikonalization. 

In this section we will only consider those radiative functions that are relevant for describing tree level soft emissions.  In particular, this eliminates all contributions involving more than two collinear fields. Furthermore, we will use RPI to take each collinear sector in the amplitude to have a total $\cP_\perp=0$, which eliminates $\cO(\lambda)$ contributions to radiative soft gluon emission, as is guaranteed by the LBK theorem \cite{Low:1958sn,Burnett:1967km}. See \cite{Larkoski:2014bxa} for a detailed discussion in the context of SCET.
This leaves us with the following cases of interest
\begin{itemize}
\item Single $\psi_{us}$ emission at $\cO(\lambda)$,
\item Single $\cB_{us}$ emission at $\cO(\lambda^2)$,
\item Double $\cB_{us}$ emission at  $\cO(\lambda^2)$,
\end{itemize}
each of which will be studied in this section. Single $\psi_{us}$ emission could also be studied at $\cO(\lambda^2)$, however, due to fermion number conservation in the soft sector, this can first contribute at cross section level at $\cO(\lambda^4)$, and is therefore not of interest to us here. Terms with additional collinear fields, that do not contribute at tree level, can be treated in an identical manner, and we will briefly comment on loop level contributions at the end of this section.

For convenience a summary of the radiative functions is given in \Tab{tab:rad_func_amp}, which shows the schematic factorization of the amplitude, the tree level Feynman rule for the radiative function, as well as the equation number where its definition can be found. The derivation of the factorizations are given in the text.

After extending the formalism of this section to cross section level factorization in \Sec{sec:fact_RadiativeFunction}, a complete classification of the field content of all radiative jet functions contributing to a physical observable, namely thrust in $e^+e^-\to $ dijets, will be given in \Sec{sec:RF_thrust}. This includes those which first contribute at loop level. By fixing a particular physical process, we will be able to exploit the symmetries of the problem to slightly simplify the structure of the radiative contributions.

{
\renewcommand{\arraystretch}{1.4}
\begin{table}[t!]
\scalebox{0.842}{
\hspace{-1.5cm}\begin{tabular}{| c | c | c |c |c|c| r| }
  \hline                       
  $\begin{matrix} \text{Radiative} \\[-3mm] \text{Function} \end{matrix}$& Factorization & Tree Level Feynman Rule& Equation \\
  \hline
  $\left[\cJ_{\psi {n_j}}^{\balpha i} (k^+)\right]^{\beta_j,s_j}$ & $\fd{3cm}{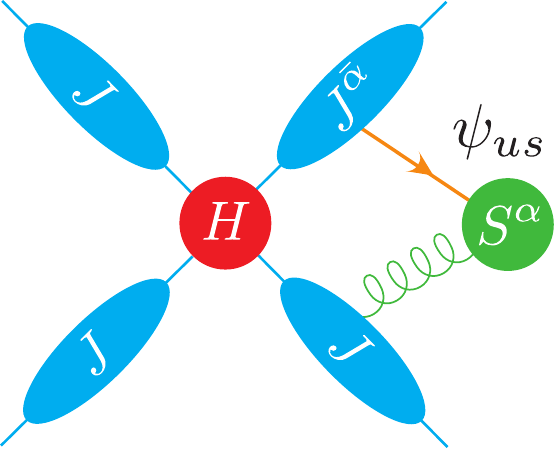}$ &  $\fd{3cm}{figures_b/Kfactor_L1_quark_noWilson_rad_low} = -\left[ \left( \Sl{\epsilon}^A_\perp - \frac{\Sl{p}_{\perp}\, n\cdot \epsilon^A(p)}{\bar n \cdot p} \right)\frac{\Sl{n}}{2 k^+}\right]_{\balpha, s} T^A_{i, \beta_j}$& \Eq{eq:rad_jet_soft_quark_amplitude} \\
  \hline
   $\left[\cJ^{\mu \nu}_{q,A}(k^+)\right]^{i,s}$ & $\fd{3cm}{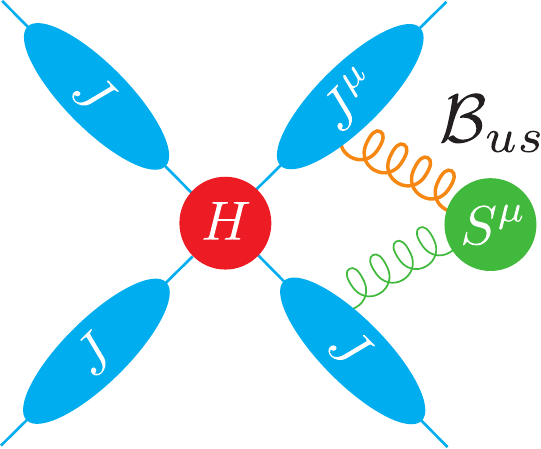}$ & $\fd{3cm}{figures_b/Kfactor_L2_gluon_noWilson_rad_low}= -\left[\bar{u}_n(p)\frac{gT^A  \gamma_\perp^\mu \gamma_\perp^\nu}{\bar n \cdot p\, n\cdot k}\right]^{i,s}$ &  \Eq{eq:rad_jet_soft_gluon_amplitude}\\
  \hline  
   $\left[\cJ_{g, A}^{\mu \nu}(k)\right]^{M \rho} $ & $\fd{3cm}{figures_amp/amp_sub_gluon_fac_low.pdf}$ & $\fd{3cm}{figures_b/Kfactor_L2_gluon_gluon_noWilson_rad_low}= \frac{-igf^{ABM}}{\bar n \cdot p\, n\cdot k} \left( g_\perp^{\mu \rho}\epsilon_\perp^{\nu B} - g_\perp^{\nu \rho}\epsilon_\perp^{\mu B}\right)\,$ &  \Eq{eq:rad_amp_gluon}\\
   \hline
      $\left[\cJ^{\mu \nu }_{q, A B,n}(k)\right]^{i,s} $ & $\fd{3cm}{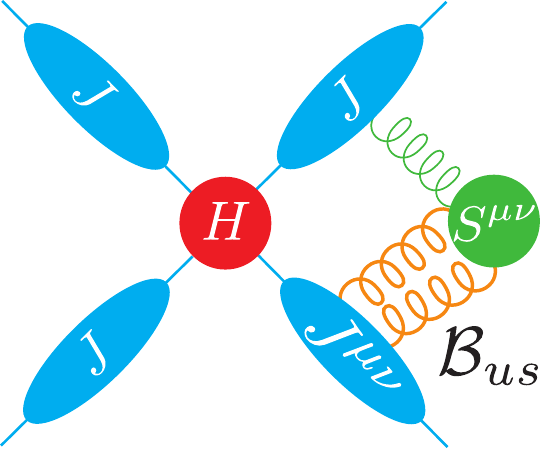}$ & $\fd{3cm}{figures_b/Kfactor_L2_gluon_noWilson_doublerad_low}= \frac{i g^2}{\nbar \cdot p n\cdot k}  \left [\bar{u}_n(p)\left([\gamma^\mu_\perp, \gamma^\nu_\perp] [T^A,T^B]+ g^{\mu \nu}_\perp \{T^A,T^B\} \right) \right]^{i,s}$ &  \Eq{eq:amp_quark_doublegluon}\\
      \hline
         $\left[\cJ_{g, A}^{\mu \nu}(k)\right]^{} $ & $\fd{3cm}{figures_amp/amp_doubleB_low.pdf}$ & $\fd{3cm}{figures_b/Kfactor_L2_gluon_gluon_noWilson_doublerad_low}=\frac{g^2f_{ABC}f_{CDM }}{\nbar \cdot p n\cdot k} \left( g_\perp^{\mu \rho} \epsilon_\perp^{\nu D} - g_\perp^{\nu \rho} \epsilon_\perp^{\mu D} \right) = gf_{ABC}\cJ_{g,C}^{\mu \nu}\,$ &  \Eq{eq:amp_gluon_doublegluon}\\
         \hline
\end{tabular}}
\caption{A summary of the radiative functions with tree level Feynman rules, showing also the schematic factorization of the amplitude, and the equation where the definition of the radiative function can be found. Derivations of the factorizations are given in the text.
}
\label{tab:rad_func_amp}
\end{table}
}

\subsection{Leading Power Amplitude Factorization}\label{sec:LP_amp}

Before considering radiative functions at the amplitude level, we begin by briefly reviewing the well known leading power amplitude level factorization. This will help to establish our notation, as well as to emphasize distinctions when we consider the subleading power case.

To study factorization at the amplitude level, we can proceed as in \Sec{sec:sub-fact}, however, we study only the matrix element 
\begin{align}
\cA_N=\langle X | \cO(0) | 0\rangle\,,
\end{align}
instead of the squared matrix element.
Here $\cO$ is a full theory QCD operator, and $X$ is an $N$-jet state in the full theory. In the soft and collinear limits, we can proceed identically to the factorization at the cross section level, namely we match to the leading power $N$-jet operator in the EFT, which we assume has a single collinear field, $X^{k_i}_{n_i}$, in each collinear sector
\begin{align}
\cO_N^{(0)}=C_N^{(0)}\left( \{ Q_i \} \right) \otimes \prod\limits_{i=1}^N \left[ \delta (\bar n_i Q_i -\bar n\cdot i \partial_n ) X_{n_i}^{\kappa_i}(0)  \right]\,.
\end{align}
Here the $\kappa_i$ labels the parton identity of the $n_i$ collinear field that can either be a quark jet field $\chi_{n_i}$ or a gluon jet field $\cB_{n_i}^\perp$.
Throughout this section, we will assume for simplicity that there is a single such operator, since the structure of the leading power operator will not play a significant role in our discussion. Furthermore, we will suppress explicit contractions of color indices, since they are standard.
The BPS field redefinition factorizes the Hilbert space, and hence the state
\begin{align}
\langle X |=\prod\limits_i  \langle X_{n_i} |  \langle X_{us} | \,,
\end{align}
into collinear states $\langle X_{n_i} | $ and an ultrasoft state $\langle X_{us} |$. With the interactions in the Lagrangian decoupled, the leading power factorization of the matrix element is then simply an algebraic exercise, and we obtain the factorized expression
\begin{align}\label{eq:LP_amp}
\cA^{(0)}_N=\langle X | \cO_N^{(0)} |0\rangle = C^{(0)}_N\left( \{ Q_i \} \right) \prod\limits_i \langle X_{n_i} | \delta (\bar n_i Q_i -\bar n\cdot i \partial_n ) X_{n_i}^{\kappa_i}(0) |0\rangle   \left \langle X_{us} \left | \TO  \prod _{i}Y^{\kappa_i}_{n_i} (0) \right |0 \right\rangle\,.
\end{align}
Here $\kappa_i$ labels both the parton identity of the $n_i$ collinear field, as well as the representation of the Wilson line, as determined by the BPS field redefinition, and $\TO$ denotes time ordering.
This gives rise to the familiar factorization into a hard matching coefficient, coefficient functions describing the collinear radiation along each lightlike direction, and a soft amplitude, 
\begin{align}
\cA^{(0)}_N=C^{(0)}_N\left( \{ Q_i \} \right) \cdot \left( \prod\limits_i \cJ^{\kappa_i}_{n_i} \right ) \cdot \cS_N=\fd{3cm}{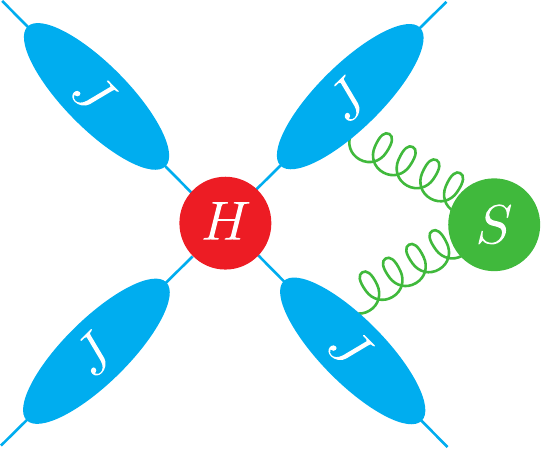}\,.
\end{align}
The leading power collinear function and soft amplitude are defined as
\begin{align}
\cJ^{\kappa_i}_{n_i}=\langle X_{n_i} | \delta (\bar n_i Q_i -\bar n\cdot i \partial_n ) X_{n_i}^{\kappa_i}(0) |0\rangle\,, \qquad \cS_N= \left \langle X_{us} \left | \TO  \prod\limits_{i} Y^{\kappa_i}_{n_i} (0)\right |0 \right \rangle\,.
\end{align}
At leading power the soft function is defined as a matrix element of Wilson lines, which are generated in SCET through the BPS field redefinition. The soft emissions therefore only resolve the color and direction of the collinear legs. To simplify our notation, in \Eq{eq:LP_amp} we have left implicit the contraction of all color indices, and denote it simply by the ``dot" symbol.   No Lorentz or Dirac indices are passed between the jet and soft functions, and therefore the soft degrees of freedom have no sensitivity to the spin of the collinear particles. Furthermore, the factorization is multiplicative, with no convolution in the soft momentum. We will see that when we consider the factorization involving Lagrangian insertions at subleading power, these features no longer hold.

\subsection{Definition of Radiative Functions}\label{sec:RF_amp}
We now  consider  amplitude level factorization at subleading power. Here we will focus solely on contributions from Lagrangian insertions, which will give rise to radiative functions. We have studied the structure of subleading power hard scattering operators extensively in  \cite{Feige:2017zci,Moult:2017rpl}. 
After performing the BPS field redefinition,  the contributions from subleading power Lagrangian insertions to the amplitude take the form
\begin{align}
\cA_N^{(j),\text{rad.}}=C^{(0)}_N \int d^4 x ~ \prod\limits_{n_i} \left \langle X_{n_i} \left |  \left \langle X_{us} \left | \TO\left\{ \cL_{n_i}^{(j)\text{BPS}}(x) \cO_N^{(0)\text{BPS}} \right\} \right |0 \right\rangle \right. \right. \,.
\end{align}
Here $ \cO_N^{(0)\text{BPS}}$ is the leading power BPS redefined $N$ jet operator, and $\cL_{n_i}^{(j)}(x)$ is the $\cO(\lambda^j)$, $j\geq 1$, Lagrangian for the $n_i$ sector, after BPS field redefinition. The ``$\text{rad.}$" superscript on the matrix element emphasizes that this is only the radiative contribution to the amplitude. More generally, one must consider multiple Lagrangian insertions, or Lagrangian insertions with subleading power hard scattering operators, as detailed in  \Eq{eq:xsec_lam2_RF}. These will factorize in a similar manner, and will be discussed in \Sec{sec:fact_RadiativeFunction}.

Unlike the leading power case of \Eq{eq:LP_amp}, where the amplitude factorized into a product of functions, at subleading power this factorization will include  integration variables linking the jet and soft functions. These integration variables will parametrize the position along the light cone direction, and describes the position of the soft emission from the collinear sector. Furthermore, at subleading power  the soft function no longer couples just to the color charge and direction of the jet functions, but can instead couple via Lorentz or Dirac indices in a manner which depends on the spin of the collinear particle.

\subsubsection{Soft Quark Emission}\label{sec:RF_amp_squark}

The simplest case for which to consider the factorization is the $\cL^{(1)}$ emission of a soft quark. Unlike the $\cL^{(1)}$ emission of a soft gluon, this does not vanish when $\cP_\perp$ vanishes, and it has a simpler structure than the $\cL^{(2)}$ insertions. It therefore provides a simple demonstration of the convolution structure which will appear at subleading powers. At the amplitude level, fermionic soft theorems in supersymmetric field theories and supergravity, and their relation to asymptotic symmetries have been considered \cite{Chen:2014xoa,Dumitrescu:2015fej,Avery:2015iix,Lysov:2015jrs}. At the cross section level soft quarks were found to give a leading logarithmic contribution to $B$-physics process \cite{Lee:2004ja} and event shape observables  \cite{Moult:2016fqy,Boughezal:2016zws,Moult:2017jsg}. The cross section level factorization will be discussed in \Sec{sec:fact_RadiativeFunction}.

For radiative soft quark emission at $\cO(\lambda)$, the relevant Lagrangian is 
\begin{align}
{\cal L}_{\chi_n \psi_{us}}^{(1)\text{BPS}} 
&= \bar{\chi}_n g \slashed{\cB}_{n\perp} \psi_{us (n)}+\text{h.c.} \,.
\end{align}
We are therefore interested in the factorization of the matrix element
\begin{align}
\cA_{N,\psi}^{(1),\text{rad.}}= C^{(0)}_N \int d^4 x ~ \prod\limits_{n_i} \left \langle X_{n_i}  \left |  \left \langle X_{us} \left | \TO \left \{ \bar{\chi}_{n_j} g \slashed{\cB}_{n_j\perp} \psi_{us (n_j)}(x) \cO_N^{(0)\text{BPS}} \right \} \right |0 \right \rangle \right. \right. \,.
\end{align}
The subscript $\psi$ labeling the amplitude indicates that a soft fermion is radiated.
Since the Lagrangian insertion appears only in the collinear sector $n_j$, the factorization of the other collinear sectors proceeds exactly as at leading power, giving rise to the leading power jet functions discussed in \Sec{sec:LP_amp}. For concreteness, we assume that the field in the $n_j$ collinear sector is a collinear quark. To simplify the notation in intermediate steps, we will drop the explicit time ordering, and reinstate it only in the final factorized formula. We then have
\begin{align}
\cA_{N,\psi}^{(1),\text{rad.}}=C^{(0)}_N \int d^4 x ~ \langle X_{n_j}|  \Big( \bar{\chi}_{n_j}  g \slashed{\cB}_{n_j\perp} \Big)^\balpha(x) \chi_{n_j}(0) |0\rangle      \langle X_{us} |  \psi^{\alpha}_{us(n_j)} (x)  \prod_i Y^{\kappa_i}_{n_i} (0) |0\rangle \prod\limits_{i\neq j} J^{\kappa_i}_{n_i}\,,
\end{align}
and it remains only to factorize the $n_j$ collinear sector from the soft sector.  To facilitate a comparison with definitions of radiative functions given in the literature, it will be convenient to formulate the convolution between the jet and soft functions in momentum space. Inserting $1=\int d^4y \delta^{(4)}(x-y)=\int d^4y \int \frac{d^4k}{(2\pi)^4} e^{ik\cdot(x-y)}$, we obtain
{
\begin{align}\label{eq:fact_sq_partial}
\cA_{N,\psi}^{(1),\text{rad.}}=&C^{(0)}_N \int \frac{d^4k}{(2\pi)^4}   \left[\int d^4 x ~ e^{ik \cdot x} \left \langle X_{n_j} \left| \ \Big( \bar{\chi}_{n_j}  g \slashed{\cB}_{{n_j}\perp} \Big)^\balpha(x)  \chi_{n_j}(0) \right |0 \right\rangle \right] \nn \\
&\hspace{2cm} \cdot \left[  \int d^4 y ~ e^{-ik \cdot y}  \left  \langle X_{us}  \left |  \psi_{us}^\alpha (y)  \prod_i Y^{\kappa_i}_{n_i} (0)   \right |0 \right \rangle \right] 
\cdot \prod\limits_{i\neq j} J^{\kappa_i}_{n_i}\,.
\end{align}}
In its current form, \Eq{eq:fact_sq_partial} is written as a four dimensional convolution. To regulate this expression in dimensional regularization, one must extend this to a $d=4-2\epsilon$ dimensional convolution. This implies that one cannot separately consider the soft and collinear functions after expansion in dimensional regularization, and therefore that one cannot achieve a renormalized factorization. Furthermore, as written, the factorized expression has not yet made manifest the physical picture of decorating a Wilson line via an insertion of an operator along the light cone. To simplify the convolution structure, we make use of the multipole expansion that has been implemented in the effective theory. Due to the multipole expansion the collinear matrix elements in the effective theory are local in the $n\cdot x$ and $x_\perp$ components, 
\begin{align}\label{eq:local_collinear}
\left \langle X_{n_j} \left | \Big( \bar{\chi}_{n_j}  g \slashed{\cB}_{{n_j}\perp} \Big)^\balpha(x) \chi_{n_j} (0)(0) \right |0 \right\rangle\sim \delta^2(x_\perp) \delta(x^+)\,,
\end{align}
where the $\perp$ and $+$ are in the light cone coordinates with respect to $n_j$.
This can also be seen from the multipole expanded propagator \Eq{eq:prop}. Using this property, we can simplify the \Eq{eq:fact_sq_partial} to a single variable convolution. Focusing just on the $n_j$ and soft sectors, we have
{\scriptsize
\begin{align}
&\int \frac{d^4k}{(2\pi)^4}    \left[  \int d^4 x ~ e^{ik \cdot x} \langle X_{n_j}| \Big( \bar{\chi}_{n_j}  g \slashed{\cB}_{{n_j}\perp} \Big)^\balpha(x) \chi_{n_j}(0) |0\rangle \right]   \left[  \int d^4 y ~ e^{-ik \cdot y}   \langle X_{us} |  \psi_{us({n_j})}^\alpha (y)   \prod_i Y^{\kappa_i}_{n_i} (0)  |0\rangle \right]\nn \\
&=\int \frac{d^4k}{(2\pi)^4}    \left[  \int d x^- ~ e^{ik^+x^-/2} \langle X_{n_j}| \Big( \bar{\chi}_{n_j}  g \slashed{\cB}_{{n_j}\perp} \Big)^\balpha(x) \chi_{n_j}(0) |0\rangle \right]  \left[  \int d^4 y ~ e^{-ik\cdot y}   \langle X_{us} |  \psi_{us({n_j})}^\alpha (y)   \prod_i Y^{\kappa_i}_{n_i} (0)  |0\rangle \right] \nn \\
&=\int \frac{dk^+}{(2\pi)}    \left[  \int d x^- ~ e^{ik^+x^-/2} \langle X_{n_j}| \Big( \bar{\chi}_{n_j}  g \slashed{\cB}_{{n_j}\perp} \Big)^\balpha(x^-) \chi_{n_j}(0) |0\rangle \right]  \left[\frac{dk^-}{2\pi}\frac{d^2 k_\perp}{(2\pi)^2}  \int d^4 y ~ e^{-ik\cdot y}   \langle X_{us} |  \psi_{us({n_j})}^\alpha (y)   \prod_i Y^{\kappa_i}_{n_i} (0)  |0\rangle \right] \nn \\
&=\int \frac{dk^+}{(2\pi)}    \left[  \int d x^- ~ e^{ik^+x^-/2} \langle X_{n_j}| \Big( \bar{\chi}_{n_j}  g \slashed{\cB}_{{n_j}\perp} \Big)^\balpha(x^-) \chi_{n_j}(0) |0\rangle \right]  \left[ \int dy^- ~ e^{-ik^+y^-/2}   \langle X_{us} | \psi_{us({n_j})}^\alpha (y^-)   \prod_i Y^{\kappa_i}_{n_i} (0)  |0\rangle \right] \nn \\
& \normalsize= \int \frac{dk^+}{(2\pi)} \cJ_{\psi {n_j}}^\balpha (k^+)S_{N\psi {n_j}}^\alpha (k^+)\,.
\end{align}
}
Note that here we use the lightcone definition for the $x$ variable as
\begin{align}
x^\mu=x^-\frac{n^\mu}{2}+x^+\frac{\bar n^\mu}{2}+x_\perp\,.
\end{align}
This then gives the momentum space definition of the radiative jet function $\left[\cJ_{\psi {n_j}}^\balpha (k^+)\right]^{i,s}$, and the corresponding soft function $S_{N\psi {n_j}}^\alpha (k^+)$
\be \label{eq:rad_jet_soft_quark_amplitude}
	\left[\cJ_{\psi {n_j}}^\balpha (k^+)\right]^{i,s} = \int dx^- ~ e^{ik^+ x^-/2}\langle X_{n_j}| \Big( \bar{\chi}_{n_j}  g \slashed{\cB}_{{n_j}\perp} \Big)^\balpha(x^-) \chi^{i,s}_{n_j}(0) |0\rangle\,.
\ee
In position space, we have
\begin{align}
&\int \frac{dk^+}{(2\pi)} \cJ_\psi^\balpha (k^+)S_\psi^\alpha (k^+)\nn \\
&=\int d x^-   \left[    \left \langle X_{n_j} \left| \TO\Big( \bar{\chi}_{n_j}  g \slashed{\cB}_{{n_j}\perp} \Big)^\balpha(x^-) \chi_{n_j}(0)  \right |0\right\rangle \right]   \left[   \left \langle X_{us} \left | \TO  \psi_{us({n_j})}^\alpha (x^-)   \prod_i Y^{\kappa_i}_{n_i} (0)  \right |0 \right\rangle \right] \nn \\
&= \int dx^- \cJ_{\psi {n_j}}^\balpha (x^-)S_{N\psi {n_j}}^\balpha (x^-)\,,
\end{align}
where in this final form, we have explicitly reinstated the time ordering.
The factorization for this contribution to the full amplitude is then given by
\begin{align}
\cA_{N,\psi}^{(1),\text{rad.}}= C^{(0)}_N\left( \{ Q_i \} \right)  \prod\limits_{i\neq j} J^{\kappa_i}_{n_i} \int dx^- \cJ_{\psi {n_j}}^\balpha (x^-)S_{N\psi {n_j}}^\alpha (x^-) =\fd{3cm}{figures_amp/amp_sub_quark_fac_low.pdf}\,.
\end{align}
This factorization gives the physical picture of dressing the Wilson line with an operator at a position $x^-$ along the light cone. Due to the multipole expansion, the soft sector still sees the collinear sector as lying exactly on the light cone. Unlike the leading power case, the soft function and jet function both carry a fermionic index. The soft degrees of freedom are therefore aware of the identity of the collinear partons. Other radiative functions will have an analogous structure, but can involve more complicated contractions between the soft and collinear sectors and additional integrals.

The tree level result for the radiative function is given by
\begin{align}\label{eq:1quarklam_feyn_rad}
&\fd{3cm}{figures_b/Kfactor_L1_quark_noWilson_rad_low}= \left[\cJ_{\psi {n_j}}^{\balpha i} (k^+)\right]^{\beta_j,s_j}_{|\text{LO}} = -\left[ \left( \Sl{\epsilon}^A_\perp - \frac{\Sl{p}_{\perp}\, n\cdot \epsilon^A(p)}{\bar n \cdot p} \right)\frac{\Sl{n}}{2 k^+}\right]_{\balpha, s} T^A_{i, \beta_j}\,.
\end{align}
In this initial investigation, we will not consider loop corrections. Although here we have treated an outgoing quark which emits a soft quark and becomes a collinear gluon, the opposite case of a gluon field converting to a quark field can be treated in an identical fashion. 

\subsubsection{Soft Gluon Emission from a Collinear Quark}\label{sec:amp_softgluon}

We now consider the $\cO(\lambda^2)$ emission of a soft gluon insertion. We will consider first the case of the emission from a collinear quark, where we will work through the convolution structure in detail. We will then state the result for the emission from a collinear gluon, which has a similar structure.   For the emission of a soft gluon from a collinear quark, the relevant Lagrangian is
\begin{align}
{\cal L}_{n \xi}^{(2) \text{BPS}} &=  \bar \chi_n  \left(  T^a \gamma^\mu_\perp \frac{1}{\bar \cP}  i \Sl \partial_{us\perp} - i  {\overleftarrow{\Sl \partial}}_{us\perp} \frac{1}{\bar \cP} T^a \gamma^\mu_\perp   \right)   \frac{\Sl \bn}{2} \chi_n  \,   g \cB_{us(n)}^{a\mu} \,.
\end{align}
We therefore must consider the factorization of the matrix element
\begin{align}
\cA_{N_q,\cB_{us}}^{(2),\text{rad.}}=  C^{(0)}_N \int d^4 x ~ \prod\limits_{n_i} \left \langle X_{n_i} \left|  \langle X_{us} |\TO {\cal L}_{n_j \xi}^{(2) \text{BPS}}(x) \cO_N^{(0)\text{BPS}}  \right|0 \right\rangle\,.
\end{align}
As in the previous section, we will drop the explicit time ordering until the final formula.
Due to the presence of the ultrasoft derivative in the Lagrangian, the factorization is slightly more complicated than for the case of a soft quark emission. 

As motivation for the structure that the factorization should take, we can look at the LBK theorem for soft gluon emission at $\cO(\lambda^2)$. For the emission of a soft gluon off of a collinear quark, the LBK theorem can be written as \cite{Larkoski:2014bxa}
\begin{align}
S^{(2)}_{i\psi} \cA_N &=g \frac{2\epsilon_{s\mu} p_{s\nu} }{(\bar n_i \cdot p_i)(n_i\cdot p_s)}  \bar u(p_i) T_i \left \{ n_i^{[\mu} \bar n_i^{\nu]}\frac{\bar n_i \cdot p_i}{2} \frac{\partial}{\partial(\bar n_i \cdot p_i)} \right. \nn \\
&\hspace{5cm}\left. +\gamma_\perp^{[\mu} n_i^{\nu]} \frac{\Sl{\bar n}_i}{4} + p_{s\perp}^{[\mu} \frac{n_i^{\nu]}}{2n_i\cdot p_s} +\frac{1}{4} [\gamma^\mu_\perp,\gamma^\nu_\perp]   \right\} \tilde \cA_N\,,
\end{align}
or in terms of the angular momentum generator $J^{\mu \nu}$, as
\begin{align}
S^{(2)}_{i\psi} \cA_N &=T^i \frac{\epsilon_{s\mu} p_{s\nu} J_i^{\mu \nu}}{p_i \cdot p_s} \cA_N\,.
\end{align}
This expression holds only for as on-shell emission, which cannot be assumed of  the group momentum flowing into the $\cB_{us(n)}$ field, and furthermore it is the complete result for the amplitude, not just the contribution from the radiative functions. Nevertheless, we would like our radiative functions to have a structure which mimics this as closely as possible. In particular, we would like that the ultrasoft derivatives appearing in the subleading power Lagrangian act only on the soft sector.  Furthermore, it suggests that the radiative jet function should carry two Lorentz indices, which are contracted with a $\cB_{us(n)}$ field, and  an ultrasoft derivative. 

To perform the factorization, we assume for concreteness that there is a collinear quark field in the $n_j$ collinear sector. Since the Lagrangian insertion is in the $n_j$ sector we can immediately factorize the other collinear sectors, giving the leading power jet functions, and we obtain
\begin{align}
\cA_{N_q,\cB_{us}}^{(2),\text{rad.}}&= C^{(0)}_N \prod\limits_{i\neq j} J^{\kappa_i}_{n_i} \int d^4 x ~ e^{ik \cdot x}  \\
&\hspace{-1.5cm}\cdot \left \langle X_{n_j} \left | \left \langle X_{us} \left| \bar \chi_{n_j} \left(  T^a \Sl{\cB}_{us({n_j}) \perp} \frac{1}{\bar \cP} i\Sl{\partial}_{us\perp}-  i{\overleftarrow{\Sl{\partial}}}_{us\perp}  \frac{1}{\bar \cP}  T^a  \Sl{\cB}_{us({n_j}) \perp}    \right) \frac{\Sl{\bar n}}{2}\chi_{n_j}(x) \chi_{n_j}^{(0)}(0) \prod_i Y^{\kappa_i}_{n_i} (0)  \right |0 \right \rangle \right. \right. \,. \nn
\end{align}
To achieve a factorization of the ultrasoft derivatives, we can choose the external states to have no $\perp$ residual momentum. After BPS field redefinition,  there are no soft collinear interactions, other than through the single Lagrangian insertions. Therefore, momentum is only passed between the soft and collinear sectors at this vertex. The ultrasoft derivative operator can either act on the group momentum that flows out through the $\cB_{us(n)}$ field, or on a residual momentum component coming from a collinear loop. In dimensional regularization, we have \cite{iain_notes}
\begin{align}
	\sum\limits_{q_l} \int d^d q_r (q_r)^j F(q_l^-, q_l^\perp, q_r^+)=0\,,
\end{align}
where $(q_r)^j$ with $j>0$ denotes positive powers of the $q_r^-$ and $q_r^\perp$ momenta, which are the only residual momenta which appear in the subleading power Lagrangians. Any residual momentum from a collinear loop momentum picked up by the derivative is therefore set to zero. Therefore the derivative picks up just the group momentum of the $\cB_{us(n)}$ field.  Ultimately, this is due to the fact that the ultrasoft momentum of the $\cB_{us(n)}$ field is the only physical ultrasoft momentum flowing in the graphs. At loop level, this statement is slightly non-trivial since the subleading power Lagrangian can couple to closed fermion loops. However, given a $\bar \chi_n$ which produces a fermion in the hard scatter in the $n$ collinear sector, it is possible to choose the momentum routing so that the soft momentum is routed only along the direction of fermion number flow of the collinear operator.  Therefore, the ultrasoft derivative acts just on the soft gluon field, and only one of the tensor structures appears.

For the particular matrix elements of interest, we can therefore perform the following simplifications
{\begin{small}
\begin{align}
& \int d^4 x ~ e^{ik \cdot x} \left \langle X_{n_j}\left| \left\langle X_{us}\left| \bar \chi_n \left(  T^a \Sl{\cB}^a_{us({n_j}) \perp} \frac{1}{\bar \cP} i\Sl{\partial}_{us\perp}-  i{\overleftarrow{\Sl{\partial}}}_{us\perp}  \frac{1}{\bar \cP}  T^a  \Sl{\cB}^a_{us({n_j}) \perp}    \right) \frac{\Sl{\bar n}}{2} \chi_{n_j}(x) \chi_{n_j}^{(0)}(0) \prod_i Y^{\kappa_i}_{n_i} (0) \right |0 \right \rangle \right. \right. \nn \\
&= \int d^4 x ~ e^{ik \cdot x} \left \langle X_{n_j} \left | \left \langle X_{us} \left| \bar \chi_n \left( -  i   \frac{1}{\bar \cP}  T^a  \left[ \Sl{\partial}_{us\perp}    \Sl{\cB}^a_{us({n_j}) \perp} \right]    \right) \frac{\Sl{\bar n}}{2}  \chi_{n_j}(x) \chi_{n_j}^{(0)}(0) \prod_i Y^{\kappa_i}_{n_i} (0)  \right|0 \right \rangle \right. \right. \,.
\end{align}
\end{small}}
Here the square brackets indicate that the derivatives act only within those brackets.
It is now straightforward to factorize this contribution to the amplitude, following the steps in \Sec{sec:RF_amp_squark}, so we will not repeat them explicitly. After simplifying the convolution structure, we find
\begin{align}
\cA_{N,\cB_{us}}^{(2),\text{rad.}}= C^{(0)}_N \int \frac{dk^+}{(2\pi)}    &\left[  \int dx^- ~ e^{ik^+ x^-/2} \left \langle X_{n_j} \left | \bar \chi_n \left(  -  \gamma^\perp_\nu \frac{1}{\bar \cP}  T^A  \gamma^\perp_\mu    \right)\frac{\Sl{\bar n}}{2} \chi_{n_j}(x)\chi_{n_j}(0) \right |0 \right\rangle \right] \nn \\
 &\hspace{-0.2cm}\cdot \left[  \int dy^- ~ e^{-ik^+ y^-/2}  \left \langle X_{us} \left |  \left[i\partial_{\perp}^\nu\cB_{us({n_j})}^{\mu A} (y) \right]   \prod_i Y^{\kappa_i}_{n_i} (0)  \right |0 \right \rangle \right]  \prod\limits_{i\neq j} J^{\kappa_i}_{n_i}  \nn \\
  &= C^{(0)}_N \int \frac{dk^+}{(2\pi)} \cJ_{q,{n_j}}^{\mu \nu A} (k)S_{q,{n_j}}^{\mu \nu A}(k) \prod\limits_{i\neq j} J^{\kappa_i}_{n_i}\,.
\end{align}
Here, we have defined the radiative function $\left[\cJ^{\mu \nu A}_{q, n_j}(k^+)\right]^{i,s}$ as
\be \label{eq:rad_jet_soft_gluon_amplitude}
	\left[\cJ^{\mu \nu A }_{q,n_j}(k^+)\right]^{i,s} = \int dx^- ~ e^{ik^+ x^-/2}  \left \langle X_{n_j} \left | \TO \bar \chi_{n_j} \left(  -  \gamma^\perp_\nu \frac{1}{\bar \cP}  T^A  \gamma^\perp_\mu    \right)\frac{\Sl{\bar n}}{2}\chi_{n_j}(x) \chi^{i,s}_{n_j}(0)  \right |0 \right \rangle\,.
\ee
We note that the subscript $q$ denotes that this describes the radiative emission from a quark, with the fact that it is the radiative emission of a single gluon being specified by the single free adjoint color index.

This radiative function exhibits an explicit coupling of $i\partial_{\perp}^\nu\cB_{us}^\mu$ to something that is reminiscent of the spin orbital angular momentum, inserted at a position along the light cone.  In position space, we have
\begin{align}
\cA_{N,\cB_{us}}^{(2),\text{rad.}}= & C^{(0)}_N\int dx^-    \left[  \left \langle X_{n_j} \left | \TO \bar \chi_{n_j} \left(  -  \gamma^\perp_\nu \frac{1}{\bar \cP}  T^A  \gamma^\perp_\mu    \right)\frac{\Sl{\bar n}}{2}\chi_{n_j}(x) \chi^{i,s}_{n_j}(0)  \right |0 \right \rangle \right] \nn \\
&\hspace{1.65cm}\cdot \left[     \left \langle X_{us} \left | \TO  \left[i\partial_{\perp}^\nu\cB_{us({n_j})}^{\mu A} (x) \right]   \prod_i Y^{\kappa_i}_{n_i} (0)  \right |0 \right\rangle \right] \prod\limits_{i\neq j} J^{\kappa_i}_{n_i}   \nn \\
  &=  C^{(0)}_N \int dx^- \cJ_{q A}^{\mu \nu } (x^-)S_{A}^{\mu \nu}(x^-) \prod\limits_{i\neq j} J^{\kappa_i}_{n_i} = \fd{3cm}{figures_amp/amp_sub_gluon_fac_low.pdf}\,,
\end{align}
where we have reinstated the explicit time ordering.

The tree level result for the radiative function is
\begin{align}\label{eq:1Blam2_fromquark_feyn_rad}
	\fd{3cm}{figures_b/Kfactor_L2_gluon_noWilson_rad_low}~= \left[\cJ^{\mu \nu}_{q,A}(k^+)\right]^{i,s}_{|LO} &= -\left[\bar{u}_n(p)\frac{gT^A  \gamma_\perp^\mu \gamma_\perp^\nu}{\bar n \cdot p\, n\cdot k}\right]^{i,s}\nn \\
	&= -\left[\bar{u}_n(p)\frac{gT^A }{\bar n \cdot p\, n \cdot k}\left( g_\perp^{\mu \nu} + i \sigma_\perp^{\mu \nu} \right)\right]^{i,s}\,.
\end{align}
In \Sec{sec:comp_laenen} we will compare this with results in the literature.

We have focused on the factorization for the particular insertion of the subleading power quark Lagrangian onto a collinear line, which contributes at tree level, and have shown how this gives rise to a radiative function.  At loop level, one can also consider contributions from the other terms in the Lagrangian. For example, one can insert the gluon component of the $\cL^{(2)}$ Lagrangian into an outgoing quark leg. This will contribute, for example through the diagram
\begin{align}
 \fd{3cm}{figures_amp/one_loop_quark_gluon_insertion_low}\,.
\end{align}
The calculation of contributions of this form were performed in the threshold limit in \cite{Bonocore:2016awd}.
The factorization for this is identical to the case of the gluon, which will be discussed in the next section, \sec{amp_softgluon_from_gluon}. As a matter of fact, since the coupling of the Lagrangian is the same, one must just exchange the field at the origin $\cB_{n\perp}\to \chi_n$. Since the goal of this section is to show in detail for several examples how radiative functions at amplitude level arise in SCET, we will not perform a complete classification of the radiative functions at loop level. In \Sec{sec:RF_thrust}, we will perform this classification for the case of the thrust observable in $e^+e^-\to$ dijets, using symmetries specific to the problem to simplify the number of distinct contributions.

\subsubsection{Soft Gluon Emission from a Collinear Gluon}\label{sec:amp_softgluon_from_gluon}

We can also consider the emission of a gauge invariant soft gluon field from a collinear gluon. Since the derivation of the factorization is similar, here we skip most of the steps to quickly get to the final result. However, since the gluon fields carry both color and Lorentz indices and the resulting radiative functions definition will depend on them, in this paragraph we treat these indices explicitly.\\
Even though the $\cO(\lambda^2)$ Lagrangian in the gluon sector is somewhat complicated, it strongly simplifies under the assumption of no label perpendicular momentum flowing through the collinear fields, as explained in \Sec{sec:lam2_Lagrangian}. Therefore, for the emission of a soft gluon from a collinear gluon with $\cP_\perp =0$, the relevant Lagrangian is simply
\begin{align}
	{\cal L}_{n g}^{(2) \text{BPS}} &=  g \text{Tr} \left( \partial_\perp^{[\mu} \cB_{us \perp}^{\nu]}[ \cB^\perp_{n\mu}  , \cB^\perp_{n\nu}   ] \right)\,.
\end{align}
Since we want to study the factorization involving a gluon lagrangian insertion, our hard scattering operator $\cO_{N}^{(0)}$ must contain at least one gluon field. For concreteness let's take $\cO_{N}^{(0)}$ to have a collinear gluon field in the $n_j$ direction. We therefore must consider the factorization of the matrix element
\begin{align}
\cA_{\alpha_1,\dots,\alpha_N}^{(2),\text{rad.}}=  C^{(0)}_{N} \int d^4 x ~ \prod\limits_{n_i} \left \langle X_{n_i} \left|    \langle X_{us} | \TO {\cal L}_{n_j g}^{(2) }(x) \cO_{N \, \alpha_1,\dots,\alpha_N}^{(0)}(0)  \right|0 \right\rangle\,.
\end{align}
where $\alpha_i$ are color indices of the external legs of the amplitude and they belong to the representation determined by the parton identity.\footnote{We use the labels $k_i = q,\bar{q},g$ for the fundamental, anti-fundamental and adjoint representation respectively.}
As usual, after BPS field redefinition we have 
\be
	\cO_{N \alpha_1,\dots,\alpha_N}^{(0)} \to \cO_{N \beta_1, \dots,\beta_N}^{(0) \text{BPS}} \prod_{i=1}^N \left(Y_{n_i }^{k_i}\right)_{\beta_i \alpha_i}\,,\qquad {\cal L}_{n_j g}^{(2) } \to {\cal L}_{n_j g}^{(2) \text{BPS}}\,.
\ee
It is convenient to isolate the gluon leg on which we want to insert the subleading lagrangian by singling out the gluon field $\cB_{n_j \perp}^{\rho \beta_j}$, where $\rho$ is a Lorentz index and $\beta_j$ a color one, from the hard scattering operator, as follow
\be
	\cO_{N \beta_1, \dots,\beta_N}^{(0) \text{BPS}} \prod_{i=1}^N \left(Y_{n_i }^{k_i}\right)_{\beta_i \alpha_i} = \cO_{N \, \beta_{i\neq j}}^{(0)\text{BPS}}\cB_{n_j \perp}^{\rho \beta_j} (Y^g_{n_j})_{\beta_j \alpha_j} \prod_{i\neq j} \left(Y_{n_i }^{k_i}\right)_{\beta_i \alpha_i} \, ,
\ee
where $(Y^g_{n_j})_{\beta_j \alpha_j}$ is the Wilson line\footnote{Note that this is an adjoint Wilson line, so that $(Y^g_{n_j})_{\beta_j \alpha_j} \equiv \left(\cY_{n_j}\right)_{\beta_j \alpha_j}$. In our notation this information is carried by the parton label $g$.} resulting from the BPS field redefinition of the gluon field $\cB_{n_j}^{\mu \beta_j}$.
Since all the gauge invariant gluon fields appearing in this section are perpendicular, we will drop the $\perp$ label on $\cB_{n_j}^\perp$ and $\cB_{us}^\perp$ to lighten the notation. Therefore, after BPS field redefinition, the amplitude reads
{\small
\begin{align}
	\cA_{\alpha_1,\dots,\alpha_N}^{(2),\text{rad.}}=  C^{(0)}_{N} \int d^4 x ~ \left(\prod\limits_{n_i} \Big \langle X_{n_i} \Big|\right) \Big\langle X_{us} \Big|  \TO {\cal L}_{n_j g}^{(2) \text{BPS}}(x) \cO_{N \, \beta_{i\neq j}}^{(0) \text{BPS} }(0) \cB_{n_j}^{\rho \beta_j} (Y^g_{n_j})_{\beta_j \alpha_j} \prod_{i\neq j} \left(Y_{n_i }^{k_i}\right)_{\beta_i \alpha_i}  \Big|0 \Big\rangle\,.
\end{align}}%
Having made the color indices explicit, we can start by factorizing all the $n_i$ collinear jets with $i\neq j$, where $n_j$ is the direction of the collinear jet on which we insert the subleading Lagrangian
{\footnotesize
\begin{align}
\cA_{N_g,\cB_{us}}^{(2),\text{rad.}}&= C^{(0)}_N \prod\limits_{i\neq j} J^{\kappa_i \, \beta_i}_{n_i} \int d^4 x \, e^{ik \cdot x} \left \langle X_{n_j} \left | \left \langle X_{us} \left| g f^{ABC} \left[\partial_\perp^{[\mu} \cB_{us}^{\nu]A}\right] \cB^{B}_{n_j\mu}  \cB^{C}_{n_j\nu}  (x) \cB_{n_j  \rho}^{a}(0) \prod_{i} \left(Y_{n_i }^{k_i} (0)\right)_{\beta_i \alpha_i}   \right |0 \right \rangle \right. \right. \,. \nn
\end{align}}%
After simplifying the convolution structure, we find
\begin{align}
\cA_{N_g,\cB_{us}}^{(2),\text{rad.}}&= C^{(0)}_N  \int \frac{dk^+}{(2\pi)}    \left[  \int dx^- ~ e^{ik^+ x^-/2} \left \langle X_{n_j} \left | g f^{ABC}  \cB^{B}_{n_j\mu} (x) \cB^{C}_{n_j\nu}  (x) \cB_{n_j  \rho}^{M}(0) \right |0 \right\rangle \right] \nn \\
 &\hspace{1.85cm}\cdot \left[  \int dy^- ~ e^{-ik^+ y^-/2}  \left \langle X_{us} \left |  \partial_\perp^{[\mu} \cB_{us}^{\nu]A}(y)  \prod_i \cY^{\kappa_i}_{n_i} (0)  \right |0 \right \rangle \right]  \prod\limits_{i\neq j} J^{\kappa_i \beta_i}_{n_i}  \nn \\
  &= C^{(0)}_N \int \frac{dk^+}{(2\pi)} \cJ_{g, A,n_j}^{\mu \nu } (k^+) S_{g, A,n_j}^{\mu \nu }(k^+) \prod\limits_{i\neq j} J^{\kappa_i}_{n_i}\,.
\end{align}
where we have defined 
\be\label{eq:rad_amp_gluon}
	\left[\cJ_{g, A ,n_j}^{\mu \nu}(k^+)\right]^{M \rho} \equiv  \int dx^- ~ e^{ik^+ x^-/2} \langle X_{n_j}| g f_{ABC} \cB_{n_j}^{\mu B}(x^-)\cB_{n_j}^{\nu C}(x^-)\cB_{n_j}^{M \rho}(0)  |0\rangle \,.
\ee
As before, the subscript $g$ indicates that this describes emission from a gluon, while the adjoint color index indicates that it is describing the emission of a single gluon field.

The tree level Feynman rule for this radiative function is
\begin{align}\label{eq:1Blam_fromgluon_feyn_rad}
&\fd{3cm}{figures_b/Kfactor_L2_gluon_gluon_noWilson_rad_low}= \left[\cJ_{g, A}^{\mu \nu}(k^+)\right]^{M \rho}_{|{LO}} = \frac{-igf^{ABM}}{\bar n \cdot p\, n\cdot k} \left( g_\perp^{\mu \rho}\epsilon_\perp^{\nu B} - g_\perp^{\nu \rho}\epsilon_\perp^{\mu B}\right)\,.
\end{align}

\subsubsection{Double Non-Eikonal Soft Gluon Emission}

In addition to the radiative functions which couple a single $\cB_{us(n)}$ field to the collinear line, the $\cL^{(2)}$ Lagrangian also contains a term that couples to a product $\cB_{us(n)}\cB_{us(n)}$ describing a double emission. It is important to emphasize that the radiative functions of \Sec{sec:amp_softgluon} with a single ultrasoft insertion also have two gluon Feynman rules arising from the Wilson lines present in the gauge invariant definition of the ultrasoft gluon field. However, such additional emissions are effectively eikonal. The double soft radiative function to be studied in this sections describes a genuinely double non-eikonal emission. 

Double soft theorems have attracted some attention in the literature \cite{Low:2015ogb,DiVecchia:2015bfa,Georgiou:2015jfa,McLoughlin:2016uwa}. We note that as compared to some treatments in the literature, the SCET power counting is such that we always take a limit such that the soft gluons are becoming simultaneously soft. This is in contrast to consecutive limits, where the soft limit for two gluons is taken consecutively. See \cite{Klose:2015xoa} for a discussion of the difference between these limits. As emphasized in \cite{ArkaniHamed:2008gz}, double soft limits are interesting as they allow one to probe the group structure of the theory, and are proportional to the structure constants of the theory. We will see that this is indeed true for the double radiative function.

	\paragraph{Emissions from collinear quarks.}

We start with the emission from a fermionic leg. Therefore, we are interested in the factorization of the term in the $\cL_{\chi_n}^{(2)}$ Lagrangian involving two $\cB_{us(n)}$ fields
\begin{align}
	\cL_{\chi_n ,\cB_{us} \cB_{us}}^{(2) } &= \bar \chi_n  \left[ T^a \gamma^\mu_\perp \frac{1}{\bar \cP} T^b \gamma^\nu_\perp   \right]   \frac{\Sl \bn}{2} \chi_n  g \cB_{us(n)}^{a\mu} g \cB_{us(n)}^{b\nu} \nn \\
	& =\bar \chi_n  \left[  \frac{1}{\bar \cP}  [\gamma^\mu_\perp, \gamma^\nu_\perp] [T^a,T^b]+ g^{\mu \nu}_\perp \{T^a,T^b\} \right]   \frac{\Sl \bn}{2} \chi_n  g \cB_{us(n)}^{a\mu} g \cB_{us(n)}^{b\nu} \,,
\end{align}
and in particular, the factorization of the matrix element
\begin{align}
\cA_{N,\cB_{us}\cB_{us}}^{(2),\text{rad.}}= C^{(0)}_N\int d^4 x ~ \prod\limits_{n_i} \left \langle X_{n_i} \left |  \left \langle X_{us} \left| \TO  \cL_{\chi_n ,\cB_{us} \cB_{us}}^{(2) } (x) \cO_N^{(0)\text{BPS}} \right |0 \right\rangle \right. \right.\,.
\end{align}

Since this matrix element contains no derivatives, the derivation of the convolution structure is identical to that for the soft quark emission in \Sec{sec:RF_amp_squark}. We therefore simply give the final result, skipping intermediate steps. We find 
{\normalsize
\begin{align}
\cA_{N,\cB_{us}\cB_{us}}^{(2),\text{rad.}} &=C^{(0)}_N \int \frac{dk^+}{(2\pi)} \nn \\
&\cdot    \left[  \int dx^- ~ e^{ik^+ x^-/2} \left \langle X_{n_j} \left| \bar \chi_{n_j}  \left[  \frac{1}{\bar \cP}  [\gamma^\mu_\perp, \gamma^\nu_\perp] [T^A,T^B]+ g^{\mu \nu}_\perp \{T^A,T^B\} \right]   \frac{\Sl \bn}{2} \chi_{n_j}(x^-) \chi^{i, s}_{n_j}(0) \right |0 \right \rangle \right] \nn \\
 & \cdot \left[  \int dy^- ~ e^{-ik^+ y^-/2}   \left \langle X_{us} \left |  \left[\cB_{us({n_j})\perp}^{\mu A}  \cB_{us({n_j})\perp}^{\nu B} (y^-) \right]   \prod_i Y^{\kappa_i}_{n_i} (0)  \right |0 \right\rangle \right] \prod\limits_{i\neq j} J^{\kappa_i}_{n_i}  \nn \\
  &=C^{(0)}_N \int \frac{dk^+}{(2\pi)} \cJ^{\mu \nu}_{q, AB,n_j}(k^+)\, S_{q, AB,n_j}^{\mu \nu} (k^+)\prod\limits_{i\neq j} J^{\kappa_i}_{n_i}\,.
\end{align}}%
We have defined the radiative function as
{\small
\be \label{eq:amp_quark_doublegluon}
	\left[\cJ^{\mu \nu }_{q ,A B,n_j}(k^+)\right]^{i,s} = \int dx^- ~ e^{ik^+ x^-/2} \left \langle X_{n_j} \left| \bar \chi_{n_j}  \left[  \frac{1}{\bar \cP}  [\gamma^\mu_\perp, \gamma^\nu_\perp] [T_A,T_B]+ g^{\mu \nu}_\perp \{T_A,T_B\} \right]   \frac{\Sl \bn}{2} \chi_{n_j}(x) \chi^{i, s}_{n_j}(0) \right |0 \right \rangle\,.
\ee}%
The subscript $q$ and the two free adjoint indices are meant to indicate that this radiative function describes the double emission of soft gluons from a quark field.

In position space, we have
\begin{align}
\cA_{N,\cB_{us}\cB_{us}}^{(2),\text{rad.}}=C^{(0)}_N\int dx^-    &\left[   \langle X_{n_j}| \TO  \bar \chi_{n_j}  \left[  \frac{1}{\bar \cP}  [\gamma^\mu_\perp, \gamma^\nu_\perp] [T^a,T^b]+ g^{\mu \nu}_\perp \{T^a,T^b\} \right]   \frac{\Sl \bn}{2} \chi_{n_j}(x^-) \chi_{n_j}(0)  |0\rangle \right] \nn \\
 & \left[   \langle X_{us} | \TO  \cB_{us({n_j})\perp}^{\mu a} \cB_{us({n_j})\perp}^{\nu b} (x^-)    \prod_i Y^{\kappa_i}_{n_i} (0)  |0\rangle \right]  \prod\limits_{i\neq j} J^{\kappa_i}_{n_i}  \\
  &=C^{(0)}_N \int dx^-  \cJ^{\mu \nu}_{q, AB,n_j}(x^-)\, S_{q, AB,n_j}^{\mu \nu} (x^-)\prod\limits_{i\neq j} J^{\kappa_i}_{n_i} =\fd{3cm}{figures_amp/amp_doubleB_low.pdf}\,. \nn
\end{align}
The form of the $\cJ_{\cB\cB}$ current is quite interesting. In particular, it involves both the symmetric and anti-symmetric structure constants, each coupling to different Lorentz structures. The antisymmetric color structure constants are associated with the perp components of the orbital momentum generator. 

Evaluating the radiative function at tree level, we find
\begin{align}\label{eq:1Blam_fromquark_feyn_doublerad}
&\fd{3cm}{figures_b/Kfactor_L2_gluon_noWilson_doublerad_low}=\left[\cJ^{\mu \nu }_{q, A B,n}(k^+)\right]^{i,s}_{|LO} = \frac{i g^2}{\nbar \cdot p n\cdot k}  \left [\bar{u}_n(p)\left([\gamma^\mu_\perp, \gamma^\nu_\perp] [T^A,T^B]+ g^{\mu \nu}_\perp \{T^A,T^B\} \right) \right]^{i,s}\,.
\end{align}
It is important to emphasize that this radiative function does not completely describe double soft gluon emission at $\cO(\lambda^2)$. One must also consider the two gluon Feynman rule of the radiative function of \Sec{sec:amp_softgluon} involving a single $\cB_{us(n)}$. Interestingly, since the $\cL^{(1)}$ are directly proportional to $\cP_\perp$, at tree level, there is no contribution to the double soft limit from a product $\cL^{(1)} \cdot \cL^{(1)}$, or from the combination of a $\cO(\lambda)$ hard scattering operator and a $\cL^{(1)}$ insertion.

	\paragraph{Emissions from collinear gluons.}

The final tree level contribution is the emission of two gauge invariant gluon fields from a collinear gluon field. Since the derivation is similar, here we present only the final result. The relevant term in the $\cO(\lambda^2)$ Lagrangian is
\begin{align}
	\cL^{(2)\text{BPS}}&\supset  \text{Tr} \left(g^2 [ \cB_{us}^{\perp \mu}  ,\cB_{us}^{\perp \nu}   ]    [ \cB^\perp_{n\mu}  , \cB^\perp_{n\nu}   ]  \right)  \,.
\end{align}
From this we can derive the factorization of the amplitude using the same procedure as in \Sec{sec:amp_softgluon_from_gluon} to get
\begin{align}
\cA_{N,\cB_{us}\cB_{us}}^{(2),\text{rad.}}&=C^{(0)}_N\int dx^-  \left[   \langle X_{n_j}| f_{ABC}f_{CDE} \cB_{n_j}^{\mu D}\cB_{n_j}^{\nu E}(x)\cB_{n_j}^{M \rho}(0)  |0\rangle \right] \nn \\
 &\hspace{2.1cm}\cdot \left[   \langle X_{us} | \TO  \cB_{us({n_j})\perp}^{\mu A} \cB_{us({n_j})\perp}^{\nu B} (x)    \prod_i Y^{\kappa_i}_{n_i} (0)  |0\rangle \right]  \prod\limits_{i\neq j} J^{\kappa_i}_{n_i}  \\
  &=C^{(0)}_N \int dx^- \cJ_{\cB\cB{n_j}}^{\mu \nu AB}(x^-) S_{\cB\cB{n_j}}^{\mu \nu AB}(x^-)\prod\limits_{i\neq j} J^{\kappa_i}_{n_i}\,. \nn
\end{align}
where we have defined 
\be\label{eq:amp_gluon_doublegluon}
	\left[\cJ_{g, AB,n_j}^{\mu \nu}\right]^{M \rho} \equiv \langle X_{n_j}| g^2 f_{ABC}f_{CDE} \cB_{n_j}^{\mu D}\cB_{n_j}^{\nu E}(x)\cB_{n_j}^{M \rho}(0)  |0\rangle \,.
\ee
Note that 
\be 
	\left[\cJ_{g, AB,n_j}^{\mu \nu}\right]^{M \rho} = gf^{ABC}\left[\cJ_{g, C,n_j }^{\mu \nu}\right]^{M \rho}\,.
\ee
Therefore we have just shown  that the double emission radiative function is completely determined by the single emission one. Since we have proven this relation at the operator level, it is true at all orders in perturbation theory.

Evaluating the radiative function at tree level, we have
\begin{align}\label{eq:1Blam_fromgluon_feyn_doublerad}
&\fd{3cm}{figures_b/Kfactor_L2_gluon_gluon_noWilson_doublerad_low}=\frac{g^2f_{ABC}f_{CDM }}{\nbar \cdot p n\cdot k} \left( g_\perp^{\mu \rho} \epsilon_\perp^{\nu D} - g_\perp^{\nu \rho} \epsilon_\perp^{\mu D} \right) = gf_{ABC}\cJ_{g,C}^{\mu \nu}\,.
\end{align}

\subsection{Comparison to the Literature}\label{sec:comp_laenen}

In this section we make contact with other definitions of radiative functions given in the literature. Radiative functions have been studied in the context of threshold resummation \cite{Laenen:2008gt,Laenen:2008ux,Laenen:2010uz,Bonocore:2014wua,White:2014qia,Bonocore:2015esa,Bonocore:2016awd}, where they have been defined for the case of the emission of a single soft gluon. If partonic initial states are used, then the radiative jet functions defined in the threshold limit are most similar to what we have referred to as amplitude level radiative functions in this section, but with incoming instead of outgoing conventions. This is due to the fact that the kinematics of the threshold limit imply that collinear emissions cannot cross the cut, and therefore the factorization is distinct from the cross section level factorizations we study in \Sec{sec:fact_RadiativeFunction}.

In \Refs{DelDuca:1990gz,Bonocore:2015esa,Bonocore:2016awd} a radiative function\footnote{Note that in~\cite{Bonocore:2015esa} this is called \emph{radiative jet function}. As explained in \Sec{sec:introRadiative} we have preferred to keep the term \emph{radiative jet function} only for objects entering the factorization at the cross section level, in analogy with the leading power \emph{jet functions}. The amplitude level functions are identified as \emph{radiative functions}. Since the comparison in this section is done at the amplitude level we will use the term \emph{radiative function} throughout the section.} was defined as  
\begin{align}\label{eq:Bonocore_function}
J_{\mu,a}(p,n,k) u(p)= \int d^d y e^{-i(p-k)\cdot y} \langle 0 | \Phi_n(\infty, y) \psi(y) j_{\mu,a}(0) |p \rangle\,,
\end{align}
with the non-abelian current, $j_{\mu,a}$, given by \cite{Bonocore:2016awd}
\begin{align}\label{eq:Bonocore_current}
j^\mu_a (x) =g \left( -\bar \psi(x) \gamma^\mu T_a \psi(x) + f_a^{bc} \left[  F^{\mu \nu}_c (x) A_{\nu v}(x) +\partial_\nu (A^\mu_b(x) A^\nu_c(x)) \right]    \right)\,.
\end{align}
Here $\Phi_n(\infty, y)$ is a Wilson line, so that the combination $\Phi_n(\infty, y) \psi(y)$ is equivalent to the gauge invariant field $\chi_n$ in SCET.
This structure is clearly recognized to be of the same general form as the amplitude level radiative functions defined in the previous sections. Following~\cite{Bonocore:2015esa}, the radiative function defined in \eq{Bonocore_function} can be expanded in powers of $\alpha_s$ and at tree level its Feynman rule reads 
\be\label{eq:Bonocore_tree_level}
	J^{\nu(0)} (p,k) = \frac{\Sl{k}\gamma^\nu}{2 p\cdot k} - \frac{p^\nu}{p\cdot k}\,.
\ee
Using this radiative function, the next-to-leading power corrections in the threshold limit were computed to NNLO. We would now like to show that the Feynman rule of the radiative function we have derived in \eq{1Blam2_fromquark_feyn} matches \eq{Bonocore_tree_level} after expanding homogeneously in the SCET power counting and taking the ultrasoft emission to be on-shell and the external quark to be purely collinear.

These two assumptions translate into the following kinematics
\be\label{eq:kinematic_comparison}
	k^\mu = n\cdot k \frac{\bn^\mu}{2} + \bn\cdot k \frac{n^\mu}{2} + k_{\perp}^\mu\,,\qquad k^2 = 0 \,,\qquad p^\mu = \omega \frac{n^\mu}{2} + k^\mu\,.
\ee
We can then kinematically expand the current of \eq{Bonocore_tree_level} as
\begin{align}\label{eq:Bonocore_comparison}
	J^{\nu(0)} (p,k) &= \frac{\left(\Sl{k}_\perp  + \frac{\bnslash}{2} n \cdot k + \frac{\nslash}{2} \bn \cdot k \right) \left(\gamma^\nu_\perp  + \frac{\bnslash}{2} n^\nu + \frac{\nslash}{2} \bn^\nu \right)}{ \omega  n\cdot k +\dots} - 2\frac{n\cdot k \frac{\bn^\mu}{2} + (\cancel{\omega} + \bn\cdot k) \frac{n^\mu}{2} + k_{\perp}^\mu}{ \omega\, n\cdot k +\dots} \nn \\
	&= \frac{1}{\omega\, n\cdot k}\left[\left(\Sl{k}_\perp \gamma_\perp^\nu + P_n \bn \cdot k n^\nu + P_\bn n\cdot k \bn^\nu\right) - (n\cdot k \bn^\mu + \bn\cdot k n^\mu + 2k_{\perp}^\mu) \right] \nn \\
	&= \frac{1}{\omega\, n\cdot k}\left(\Sl{k}_\perp \gamma_\perp^\nu - P_\bn \bn \cdot k n^\nu - \cancel{P_n n\cdot k \bn^\nu} - 2k_{\perp}^\mu \right) \nn \\
	&= \frac{1}{\omega\, n\cdot k}\left(\Sl{k}_\perp \gamma_\perp^\nu + \frac{k_\perp^2}{n\cdot k} n^\nu - 2k_{\perp}^\mu \right)\,.
\end{align}
In the first line we neglected $\omega \frac{n^\mu}{2}$ since we are focusing on the $\cO(\lambda^2)$ expansion of this object. In the third and fourth line we used the properties of the Dirac projectors $P_n, \, P_\bn$ acting on outgoing collinear spinors
\be 
	P_n = \frac{\nslash\bnslash}{4} \,,\qquad P_\bn = \frac{\bnslash\nslash}{4}\,,\qquad \id =P_n + P_\bn \,,\qquad \bar{u}_n P_\bn = \bar{u}_n \,,\qquad \bar{u}_n P_n = 0 \,.
\ee

We can now show how this result is reproduced in our framework. We first note, that in our definition of the radiative function for the emission of a soft gluon from a quark, as discussed in \Sec{sec:amp_softgluon}, we have factorized the ultrasoft derivative and ultrasoft gluon field into a soft function. Therefore, we do not have a correspondence at the level of the radiative function itself with \Eq{eq:Bonocore_comparison}. Instead, we must contract the radiative jet function with the corresponding soft function. Evaluating this at tree level, we have
\be\label{eq:1Blam2_fromquark_feyn}
	\fd{3cm}{figures_b/Kfactor_L2_gluon_noWilson_low}~=~\frac{1}{\omega\, n \cdot k}\left[\Sl k_\perp \gamma_\perp^\nu+ \frac{k_\perp^2}{n\cdot k} n^\nu -2 k_\perp^{\nu}  \right] \,.
\ee
We see that this indeed matches \eq{1Blam2_fromquark_feyn}. This shows, that despite the slightly different organization, the radiative functions in the two approaches are describing the same physics.

While the general form of the radiative functions in~\cite{Bonocore:2016awd} and those defined in this chapter are similar, and indeed we have shown they give the same tree level result, the radiative functions defined in this chapter differ in many aspects from the ones of~\cite{Bonocore:2016awd}. In addition to several trivial differences that are conventional, namely  \cite{Bonocore:2016awd} uses an incoming, instead of outgoing collinear state, and that the Lagrangian insertion occurs at $y=0$, there are some more major differences, which we now elaborate on.

The definitions of \Refs{DelDuca:1990gz,Bonocore:2015esa,Bonocore:2016awd} involve  a four dimensional convolution (which in dimensional regularization must be extended to a $d=4-2\epsilon$ dimensional convolution) instead of the one dimensional convolution along the light cone direction  derived in the effective theory. The ability to simplify the definition to involve a single variable convolution along the light cone relied on the multipole expansion in SCET, which renders the collinear matrix elements local in certain directions. The multipole expansion also implies that no expansions need to be performed after performing any integrals, and that all results are automatically homogeneous in the power counting. We believe that this is essential for achieving a true factorization. In particular, to claim a factorization, it must be that no divergences are generated by the integral in the final convolution variable, and that the convolution variable must be gauge invariant. This seems difficult to achieve if the convolution variable is a $d=4-2\epsilon$ dimensional momenta in dimensional regularization. While we are not yet able to prove that the convolutions in the single scalar lightcone variables in our formulation converge, we believe that the reduction to single variable convolutions that are not dimensionally regularized is an essential first step.

A second major difference in the definitions relates to gauge invariance. The current of \Eq{eq:Bonocore_current} is not by itself a gauge invariant object, and hence neither is the jet function of \Eq{eq:Bonocore_function}. However, as shown in \Refs{DelDuca:1990gz,Bonocore:2015esa,Bonocore:2016awd} it will give a gauge invariant result for the cross section with a single soft emission. This lack of gauge invariance in a radiative function defined in the full theory is perhaps not surprising, as one is attempting to factorize a gluon emission from a quark-antiquark current, but a full theory gluon is not gauge invariant. In the context of QED, where the radiative function was originally defined \cite{DelDuca:1990gz}, this issue was not present, since the photon is not charged, and therefore the radiative function is itself gauge invariant. It is ultimately this difference which makes the extension to the QCD case more difficult. To achieve a true factorization into objects which can be separately renormalized, it seems desirable that each of the objects be separately gauge invariant.

In the effective theory approach to defining the radiative functions the radiative jet and soft functions are each separately gauge invariant, since they are constructed from the gauge invariant gluon, $\cB_{us(n)}$ and quark, $\psi_{us(n)}$ fields which couple to the radiative functions. This leads to a fairly intricate structure of Wilson lines in the definition of the currents, which ensures their gauge invariance. Without the presence of these soft Wilson lines, it is also not clear how the current of \Refs{DelDuca:1990gz,Bonocore:2015esa,Bonocore:2016awd} can describe multiple soft gluon emissions.

Finally, another difference between the two approaches is that due to the manifest power counting in the effective theory, there are a large number of distinct field structures present in the SCET $\cO(\lambda)$ and $\cO(\lambda^2)$ Lagrangians, as compared with only the two terms present in the current of \Eq{eq:Bonocore_current}. These additional terms in the SCET Lagrangian have more than two collinear fields, or have an additional $\cP_\perp$ insertion, so that they first contribute when their is an additional collinear loop. For example, with a collinear loop, we have the distinct diagrams
\begin{align}
\fd{3.7cm}{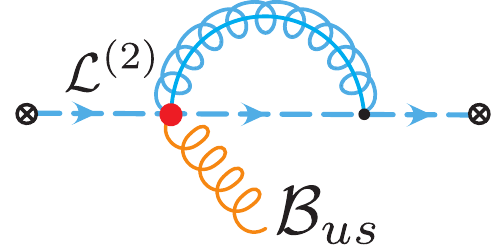}\,,\qquad
\fd{3.7cm}{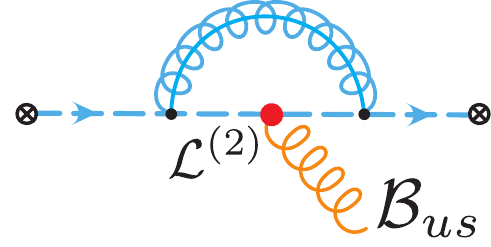}\,,\qquad
\fd{3.7cm}{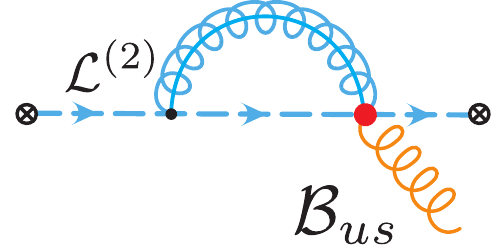}\,.
\end{align}
This is forced upon us in the effective theory by manifest power counting, which does not hold for the current of \Eq{eq:Bonocore_current}, which must be expanded after performing loop integrals. While this admittedly requires more distinct terms to be considered, we believe that they will have a simpler structure, and will facilitate an all orders understanding. Furthermore, it is hoped that only a subset will contribute at a given logarithmic accuracy. The additional terms involving multiple collinear fields will be discussed in more detail in \Sec{sec:RF_thrust} where we will consider the field structure of the complete set of radiative functions which contribute to the thrust event shape.

\section{Cross Section Level Factorization with Radiative Jet Functions}\label{sec:fact_RadiativeFunction}

In this section we show in detail how to perform subleading power factorization at the cross section level for radiative contributions in a gauge invariant manner to all orders in $\alpha_s$ in a non-abelian gauge theory. We derive the explicit structure of the factorization for the insertion of: 
\begin{itemize}
\item A single insertion of a $\cB_{us}$ field from the $\cL^{(2)}$ Lagrangian (\Sec{sec:sub_cross_quark}),
\item A double insertion of the $\psi_{us}$ field from two $\cL^{(1)}$ insertions (\Sec{sec:sub_cross_gluon}).
\end{itemize}
Here we have given the sections where the factorization is given to aid the reader.
Furthermore, we consider when this insertion is on both a collinear quark, or collinear gluon leg.  These are the two contributions to a factorized cross section that are non-vanishing at lowest order in $\alpha_s$, and allow a factorized description for the contributions that have been computed to fixed order in \cite{Moult:2016fqy,Moult:2017jsg}. Since they describe the lowest order contributions to the cross section, their renormalization should capture the LL series. The key result of this section is a factorized expression for these contributions to the cross section as a convolution between separately gauge invariant soft and collinear factors. The convolution is in terms of either one or two one-dimensional variables, which represent the position of operator insertions along the lightcone, similar to the convolution over the observable variable in the leading power factorization review in \Sec{sec:fact_LP}. 

After having worked out these two examples in detail, and illustrating how a gauge invariant factorization can be achieved, in \Sec{sec:RF_thrust}, we then provide an enumeration of the different radiative jet functions which contribute to the event shape thrust in $e^+e^-\to$ dijets to $\cO(\lambda^2)$, including those which first contribute at loop level. This will illustrate the complexity of subleading power factorization in an example of interest.

\subsection{Convolution Structure at Leading Power}\label{sec:fact_LP}

We begin by reviewing the well known leading power factorization for an SCET$_\text{I}$ event shape, $\tau$, in $e^+e^-\to$ dijets. Here the unique leading power operator is
\begin{align}\label{eq:LP_op_fact}
\cO^{(0)\mu}=\bar \chi_n  \gamma^\mu_\perp \chi _{\bar n}\,.
\end{align}
Our focus is on the convolution structure in the observable and momentum, as it is this aspect which will change at subleading power for the radiative jet functions. Therefore we will often suppress explicit Dirac or color indices.

Following the discussion of \Sec{sec:sub-fact}, after performing the BPS field redefinition, factorizing the observable, and Fierzing the Lorentz and Dirac structure (which we will suppress, as it is not important to our current discussion), the leading power cross section can be written as
\begin{align}
	\frac{1}{\sigma_0}\frac{d\sigma^{(0)}}{d\tau}&=H(Q^2) \int d^4x  \int d \tau_n d\tau_\bn d\tau_{us} \delta(\tau -\tau_n -\tau_\bn -\tau_{us}) \cdot\frac{1}{Q N_c} \tr \langle 0 | \bar \chi_n(x)_\alpha \delta(\tau_n-\hat \tau^{(0)}_n)   \chi_n(0)_\delta |0 \rangle \nn \\
&\hspace{-0.4cm}\cdot\frac{1}{Q N_c} \tr \langle 0 |   \chi_\bn(x)_\beta \delta(\tau_{\bar n}-\hat \tau^{(0)}_{\bar n})  \bar \chi_\bn(0)_\gamma   |0 \rangle \cdot \tr \langle 0 |  Y_\bn (x) Y_n^\dagger(x) \delta(\tau_{us}-\hat \tau^{(0)}_{us})  Y_n(0)  Y^\dagger_\bn(0) |0 \rangle\,.
\end{align}
This is derived by considering the factorization of the squared matrix element of \Eq{eq:LP_op_fact}, as was described in \Sec{sec:sum_fact}.
We would now like to simplify this expression, and remove the convolution over the variable $x$, which couples the different functions. To achieve this, we first define Fourier transforms of each of the functions,
\begin{align}
	\frac{1}{QN_c}\tr \langle 0 |  Y_\bn (x) Y_n^\dagger(x) \delta(\tau_{us}-\hat \tau^{(0)}_{us}) Y_n(0)  Y^\dagger_\bn(0) |0 \rangle &= \int \frac{d^4 r}{(2\pi)^4} e^{-ir \cdot x}  \cS(Q\tau_{us}, r)\,, \\
\frac{1}{QN_c} \tr \langle 0 | \bar \chi_n(x)_\alpha \delta(\tau_n-\hat \tau^{(0)}_n)  \chi_n(0)_\delta |0 \rangle&=\int \frac{d^4 l}{(2\pi)^4}   e^{-il\cdot x}  \cJ_n(\tau_n,l,Q) \left(\frac{\Sl{n}}{2} \right)_{\delta \alpha}\,, \nn\\
\frac{1}{QN_c} \tr \langle 0 |   \chi_\bn(x)_\beta \delta(\tau_{\bar n}-\hat \tau^{(0)}_{\bar n})  \bar \chi_\bn(0)_\gamma   |0 \rangle &=\int \frac{d^4 k}{(2\pi)^4} e^{-ik\cdot x} \cJ_\bn(\tau_\bn,k,Q)   \left(\frac{\Sl{\bar n}}{2} \right)_{\beta \gamma}  \,, \nn
\end{align}
and writing the result in terms of light cone coordinates, we can express the cross section as
\begin{align}
	\frac{1}{\sigma_0}\frac{d\sigma^{(0)}}{d\tau}&= H(Q^2) \int d^4x \int d \tau_n d\tau_\bn d\tau_{us} \delta(\tau -\tau_n -\tau_\bn -\tau_{us}) \cdot   Q\int \frac{d^4 r}{(2\pi)^4}  e^{-ir_1 \cdot x}  \nn \\
&\cdot \int \frac{dk^+ dk^- d^2 k_\perp}{2 (2\pi)^4} e^{-i(k^+x^-/2+ k^-x^+/2 -k_\perp \cdot x_\perp)} \cdot \int \frac{dl^+ dl^- d^2 l_{\perp}}{2 (2\pi)^4} e^{-i(l^+x^-/2+ l^-x^+/2 -l_{\perp} \cdot x_\perp)} \nn \\ 
&\cdot   \cS(\tau_{us}, r)     \cJ_n(\tau_n, l,Q)      \cJ_\bn(\tau_\bn, k,Q)\,.
\end{align}
We now use that  $\cJ_\bn$ depends only on $k^-$, and $\cJ_n$ depends only on $k^+$, which follows from the multipole expansion in the effective theory.
Performing the integrals in $k^+, k^\perp$, $l^-, l^\perp$, we then find
\begin{align}
	\frac{1}{\sigma_0}\frac{d\sigma^{(0)}}{d\tau}&=H(Q^2) \int d \tau_n d\tau_\bn d\tau_{us} \delta(\tau -\tau_n -\tau_\bn -\tau_{us})    \nn \\
&\cdot \left[  \int \frac{d^4 r}{(2\pi)^4}    \cS(Q\tau_{us}, r)  \right] \cdot \left[  \int \frac{dk^-}{2\pi Q} \cJ_\bn(\tau_\bn, k^-,Q)  \right] \cdot   \left[   \int \frac{dl^+}{2\pi Q} \cJ_n(\tau_n, l^+,Q)  \right]\,,
\end{align}
which defines  the standard leading power factorization
\begin{align}\label{eq:fact_tau_LP}
	\frac{1}{\sigma_0}\frac{d\sigma^{(0)}}{d\tau}&= Q^5 H(Q^2) \int d \tau_n d\tau_\bn d\tau_{us} \delta(\tau -\tau_n -\tau_\bn -\tau_{us})    S(Q\tau_{us}) J_{\bar n}(Q^2\tau_\bn) J_n(Q^2\tau_n)\,.
\end{align}
Each of $S$, $J_n$ and $J_{\bar n}$ are infrared finite\footnote{We assume that all functions are defined with their appropriate zero-bin subtractions \cite{Manohar:2006nz}.} and gauge invariant. The only coupling between the soft and collinear degrees of freedom is the convolution in the physical observable $\tau$. Diagramatically, we have
\begin{align}
\fd{3cm}{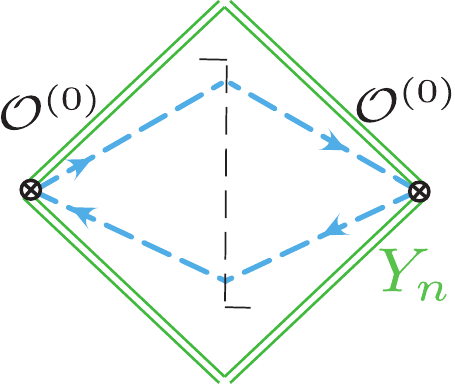}= \fd{3cm}{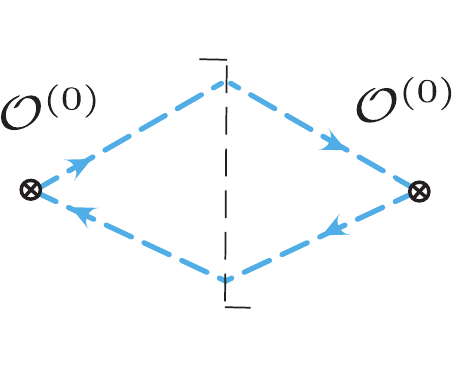} \otimes  \fd{3cm}{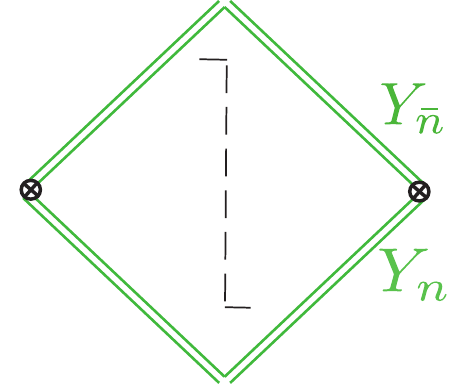}\,,
\end{align}
where the blue lines indicated collinear fields, the double green lines indicate Wilson lines, and the $\otimes$ indicates the convolution in $\tau$.
We have not drawn explicitly the loop corrections, as it is understood that this holds to all orders in $\alpha_s$.
Here, for simplicity, we have only illustrated the factorization of the soft and collinear pieces of the diagrams, since it is this aspect which is relevant for the radiative jet functions. 
This picture will be contrasted with that for the radiative functions derived in the next section.

\subsection{Convolution Structure at Subleading Power}\label{sec:fact_NLP}

We now consider the subleading factorization at cross section level for the contributions coming from Lagrangian insertions and show how they lead to radiative jet functions. We discuss in detail the convolution structure for how these radiative jet functions enter the factorization.

\subsubsection{Soft Quark Emission}\label{sec:sub_cross_quark}

 At the cross section level, two insertions of the $\cL^{(1)}$ soft quark Lagrangian are required to give a non-vanishing contribution. At tree level, such contributions will describe single soft quark emission. Subleading power shape functions arising from two insertions of the $\cL^{(1)}$ soft quark Lagrangian were considered in \cite{Beneke:2002ph,Bosch:2004cb,Lee:2004ja}. They have a similar structure to the matrix elements considered in this section, although the shape functions are defined as matrix elements of $B$ meson states, instead of vacuum matrix elements.

The Lagrangian of interest for soft quark emission is 
\begin{align}
\cL^{(1)\text{BPS}}_{\chi \psi}=\bar \chi_n g \Sl{\cB}_{n\perp} \psi_{us(n)} +\bar \psi_{us(n)} g\Sl{\cB}_{n\perp} \chi_n\,. 
\end{align}
We first consider the insertion of this operator onto a collinear quark line, which at lowest perturbative order corresponds to the emission of a soft quark from a collinear quark. To illustrate this, we use as an example $e^+e^-\to$ dijets with the hard scattering operator of \Eq{eq:LP_op_fact}. The derivation of the factorization requires both a factorization of the spin and color structures, as well as of the momentum convolutions. The factorization of the spin and color structures is in principle a straightforward excerise, which requires, in the present case, the repeated application of Fierz identities to the matrix element
\begin{align}
\int d^4x \int d^4y \int d^4z\, \bigl\langle 0 \bigr|\cO^{(0)\mu \dagger}(x)  \cL^{(1)\text{BPS}}_{\chi \psi}(y) \cL^{(1)\text{BPS}}_{\chi \psi}(z) \cO^{(0)\mu}(0) \bigl|0\bigr\rangle\,.
\end{align}
The details of this are presented in \App{app:Fierzing}, and here we focus on the convolution structure. We note that in general one must consider the presence of evanescent operators, however, since in this chapter we do not consider the renormalization of the matrix elements appearing in the factorization formula, we leave this to future work. After factorizing the spin and color structure, we arrive at the following expression for the radiative contribution to the cross section
\begin{align}
&\frac{1}{\sigma_0}\frac{d\sigma^{(2),\text{rad}}_{\psi, n,q} }{d\tau}= H(Q^2) \int d^4x \int d^4y \int d^4z\int d \tau_n d\tau_\bn d\tau_{us} \delta(\tau -\tau_n -\tau_\bn -\tau_{us}) \nn \\
&\cdot \frac{1}{N_c}\tr \langle 0 |  Y_\bn^\dagger (x) Y_n^\dagger(x) \psi_{us(n)}(y)_\beta \delta(\tau_{us}-\hat \tau^{(0)}_{us}) \bar \psi_{us(n)}(z)_\gamma Y_n(0)  Y_\bn(0) |0 \rangle  \left(\frac{\Sl{n}}{2} \right)_{\beta \gamma} \nn \\
&\cdot \frac{1}{Q N_c} \tr \langle 0 | \chi_n(x)_\beta \bar \chi_n(y)_\gamma \cB^\mu_{n\perp} (y) \delta(\tau_n-\hat \tau^{(0)}_n)\cB^\nu_{n\perp}(z) \chi_n(z)_\alpha \bar \chi_n(0)_\delta |0 \rangle  g^{\mu\nu}_\perp  \left(\frac{\Sl{ n}}{2} \right)_{\beta \gamma}  \left(\frac{\Sl{ n}}{2} \right)_{\alpha\delta}  \nn \\
&\cdot \frac{1}{Q N_c} \tr \langle 0 |   \chi_\bn(x)_\beta \delta(\tau_{\bar n}-\hat \tau^{(0)}_{\bar n})  \bar \chi_\bn(0)_\gamma   |0 \rangle \left(\frac{\Sl{\bar n}}{2} \right)_{\beta \gamma}\,,
\end{align}
which we would like to factorize into hard, jet and soft functions. The superscripts and subscripts labeling the cross section indicate that this is the $\cO(\lambda^2)$ contribution arising from the radiative emission of a soft quark from the $n$ collinear sector. Fourier transforming each of the matrix elements, we have
\begin{align}
&\frac{1}{QN_c}\tr \langle 0 |  Y_\bn^\dagger (x) Y_n^\dagger(x) \psi_{us(n)}(y)_\beta \delta(\tau_{us}-\hat \tau^{(0)}_{us}) \bar \psi_{us(n)}(z)_\gamma Y_n(0)  Y_\bn(0) |0 \rangle \nn \\
&\hspace{1cm}= \int \frac{d^4 r_1}{(2\pi)^4}  \frac{d^4 r_2}{(2\pi)^4}  \frac{d^4 r_3}{(2\pi)^4} e^{-ir_1 \cdot x} e^{-ir_2 \cdot y} e^{-ir_3 \cdot z} S_{\chi n\psi}^{(2)}(\tau_{us}, r_1, r_2, r_3,Q) \left(\frac{\Sl{\bar n}}{2} \right)_{\beta \gamma}\,,\nn \\
&\frac{1}{Q N_c} \tr \langle 0 | \chi_n(x)_\beta \bar \chi_n(y)_\gamma \cB^\mu_{n\perp} (y) \delta(\tau_n-\hat \tau^{(0)}_n)\cB^\nu_{n\perp}(z) \chi_n(z)_\alpha \bar \chi_n(0)_\delta |0 \rangle \nn \\
&\hspace{1cm}=\int \frac{d^4 l_1}{(2\pi)^4}  \frac{d^4 l_2}{(2\pi)^4}  \frac{d^4 l_3}{(2\pi)^4}  e^{-il_1\cdot x} e^{-il_2\cdot y} e^{-il_3\cdot y} \cJ_{\chi n\psi}^{(2)}(\tau_n, l_1, l_2,l_3,Q)g^{\mu\nu}_\perp  \left(\frac{\Sl{\bar n}}{2} \right)_{\beta \gamma}  \left(\frac{\Sl{\bar n}}{2} \right)_{\alpha\delta}\,,\nn\\
&\frac{1}{QN_c} \tr \langle 0 |   \chi_\bn(x)_\beta \delta(\tau_{\bar n}-\hat \tau^{(0)}_{\bar n})  \bar \chi_\bn(0)_\gamma   |0 \rangle=\int \frac{d^4 k}{(2\pi)^4} e^{-ik\cdot x} \cJ_\bn(\tau_\bn, k,Q) \left(\frac{\Sl{n}}{2} \right)_{\beta \gamma}  \,.
\end{align}
Using the locality of the collinear matrix elements, as discussed in our derivation of the amplitude level radiative functions (see \Eq{eq:local_collinear} and the surrounding discussion), we find
{\small
\begin{align}
	&\hspace{-0.25cm}\frac{1}{\sigma_0}\frac{d\sigma^{(2),\text{rad.}}_{\psi, n,q} }{d\tau}=Q^5 H(Q^2) \int d \tau_n d\tau_\bn d\tau_{us} \delta(\tau -\tau_n -\tau_\bn -\tau_{us})  \left[  \int \frac{dk^-}{2\pi Q} \cJ_\bn(\tau_\bn, k^-,Q)  \right] \! \int \frac{dr_2^+}{2\pi Q} \frac{dr_3^+}{2\pi Q} \\
	&\quad\cdot \left[  Q\int \frac{d^4 r_1}{(2\pi)^4}   \int \frac{dr_2^-}{2\pi}  \frac{d^2r_2^\perp}{(2\pi)^2}   \int \frac{dr_3^-}{2\pi}  \frac{d^2r_3^\perp}{(2\pi)^2}       \frac{ S_{\chi n\psi}^{(2)}(\tau_{us}, r_1, r_2,r_3,Q)}  {r_2^+ r_3^+}  \right]   \cdot   \left[  Q^2 \int \frac{dl_1^+}{2\pi} \cJ_{\chi n \psi}(\tau_n, l_1^+, r_2^+,r_3^+) \cdot r_2^+ r_3^+ \right]\nn\\
	&\equiv Q^5 H(Q^2)  \int d \tau_n d\tau_\bn d\tau_{us} \delta(\tau -\tau_n -\tau_\bn -\tau_{us}) J_\bn(Q^2\tau_\bn) \int \frac{dr_2^+}{2\pi Q} \frac{dr_3^+}{2\pi Q}  S_{\chi n\psi}^{(2)}(Q\tau_{us}, r_2^+,r_3^+) J_{\chi n\psi}^{(2)}(\tau_n, r_2^+,r_3^+,Q)\,.\nn
\end{align}}%
Here $J_{n\psi}^{(2)}(\tau_n, r_2^+,r_3^+)$ defines the radiative jet function at the cross section level. Note that it is defined as a collinear matrix element, so all loop corrections are collinear in nature. Diagramatically, we have
\begin{align}
\fd{3cm}{figures_b/soft_quark_diagram_BPS_low.pdf} =  \int dr_2^+ dr_3^+ \fd{3cm}{figures_b/soft_quark_collinear_piece_low.pdf} \otimes  \fd{3cm}{figures_b/soft_quark_diagram_wilsonframe_low.pdf}\,.
\end{align}
At lowest order in $\alpha_s$, the soft function is proportional to $\delta(r_2^+-r_3^+)$, which simplifies the structure of the convolution to a single variable. In the presence of radiative corrections, the full convolution structure is required.
As expected at subleading power, this contribution is first non-vanishing crossing the cut, namely the soft quark. Purely virtual contributions are proportional to $\delta(\tau)$, and hence leading power.  In the case that the soft quark is radiated from the $\bar n$ collinear sector, an identical factorization applies.

The ability to formulate this factorization in a gauge invariant way relies on the use of non-local gauge invariant collinear quark and gluon and soft quark and gluon fields. These operators have a highly intricate Wilson line structure, involving both soft and collinear Wilson lines situated at a variety of positions along different light cones. However, this structure is completely dictated by the symmetries of the effective theory. 

As compared with the leading power factorization of \Eq{eq:fact_tau_LP}, there is a convolution structure in the $+$ component of the soft momentum, in addition to the convolution in the observable $\tau$. This couples the collinear and soft sectors in a more non-trivial way, describing the ``radiation" of a soft parton from the collinear jet at a position along the light cone, as was the case at the amplitude level. Note that while we can perform the factorization of the jet and soft functions into this convolution structure, it is not a priori guaranteed that such convolutions converge. This would indicate a naive breakdown of the factorized expression, or at least that a reorganization is required. This has been explicitly observed in subleading power factorization formulae for $B$-meson decays \cite{Beneke:2003pa}. In defining the jet and soft functions, we have inserted factors of the convolution variables $r_2^+$ and $r_3^+$. This is done so that the lowest order divergence appears entirely in the soft function, allowing it to be extracted using standard plus distributions. We can therefore make sense of the convolutions appearing in the factorization at lowest order in $\alpha_s$.  An understanding of the convolutions appearing in the factorization formulas in this chapter is left to future work. In this chapter we will not consider radiative corrections to the soft and jet functions $S_{\chi n \psi}^{(2)}(\tau_{us}, r_2^+,r_3^+)$, $J_{\chi n\psi}^{(2)}(\tau_n, r_2^+,r_3^+)$, leaving this to future work.  The renormalization of these operators, and the corresponding renormalization group evolution will resum subleading power logarithms in the thrust variable.

We can also perform an identical factorization when the soft quark Lagrangian is inserted on a collinear gluon leg. Here we take as a concrete example the case of thrust in $H\to gg$ with the leading power operator 
\begin{align}
\cO_\cB^{(0)}=-2\omega_1 \omega_2 \delta^{ab} \cB_{\perp \bar n, \omega_2}^a \cdot \cB_{\perp \bar n, \omega_1}^b H\,.
\end{align}
Going through an identical procedure as for the previous case, after performing the Fierzing we arrive at the following expression
\begin{align}
&\frac{1}{\sigma_0}\frac{d\sigma^{(2),\text{rad}}_{\psi, n,g} }{d\tau}= H(Q^2) \int d^4x \int d^4y \int d^4z\int d \tau_n d\tau_\bn d\tau_{us} \delta(\tau -\tau_n -\tau_\bn -\tau_{us}) \nn \\
&\cdot  \langle 0 |  \cY_\bn^{ac} (x) \cY_n^{ad}(x) \bar \psi_{us(n)}(y)_\beta \delta(\tau_{us}-\hat \tau^{(0)}_{us})  \psi_{us(n)}(z)_\gamma \cY_\bn^{\dagger bc}(0)  \cY_n^{\dagger bd}(0) |0 \rangle  \left(\frac{\Sl{n}}{2} \right)_{\beta \gamma} \nn \\
&\cdot \frac{1}{N_c} \tr \langle 0 |  \cB^{d\mu}_{n\perp} (x)   (\bar \chi_n(z)  \Sl{\cB}_{n\perp} (z)  )_\alpha  \delta(\tau_n-\hat \tau^{(0)}_n)   (g\Sl{\cB}_{n\perp} (y) \chi_n(y)   )_\delta \cB^{d\nu}_{n\perp}(0) |0 \rangle  g^{\mu\nu}_\perp  \left(\frac{\Sl{ n}}{2} \right)_{\alpha\delta}  \nn \\
&\cdot \frac{1}{N_c} \tr \langle 0 |   \cB^{c\mu}_{\bn \perp}(x) \delta(\tau_{\bar n}-\hat \tau^{(0)}_{\bar n})   \cB^{c\nu}_{\bn \perp}(0)   |0 \rangle g^{\mu \nu}_\perp\,.
\end{align}
Fourier transforming each of the matrix elements, we have
\begin{align}
&\frac{1}{Q}\langle 0 |  \cY_\bn^{ac} (x) \cY_n^{ad}(x) \bar \psi_{us(n)}(y)_\beta \delta(\tau_{us}-\hat \tau^{(0)}_{us})  \psi_{us(n)}(z)_\gamma \cY_\bn^{\dagger bc}(0)  \cY_n^{\dagger bd}(0) |0 \rangle  \nn \\
&\hspace{1cm}= \int \frac{d^4 r_1}{(2\pi)^4}  \frac{d^4 r_2}{(2\pi)^4}  \frac{d^4 r_3}{(2\pi)^4} e^{-ir_1 \cdot x} e^{-ir_2 \cdot y} e^{-ir_3 \cdot z} S_{\cB n\psi}^{(2)}(\tau_{us}, r_1, r_2, r_3,Q) \left(\frac{\Sl{\bar n}}{2} \right)_{\beta \gamma}\,,\nn \\
&\frac{1}{Q^2 N_c} \tr \langle 0 |  \cB^{d\mu}_{n\perp} (x)   (\bar \chi_n(z)  \Sl{\cB}_{n\perp} (z)  )_\alpha  \delta(\tau_n-\hat \tau^{(0)}_n)   (g\Sl{\cB}_{n\perp} (y) \chi_n(y)   )_\delta \cB^{d\nu}_{n\perp}(0) |0 \rangle \nn \\
&\hspace{1cm}=\int \frac{d^4 l_1}{(2\pi)^4}  \frac{d^4 l_2}{(2\pi)^4}  \frac{d^4 l_3}{(2\pi)^4}  e^{-il_1\cdot x} e^{-il_2\cdot y} e^{-il_3\cdot y} \cJ_{\cB n\psi}^{(2)}(\tau_n, l_1, l_2,l_3,Q)g^{\mu\nu}_\perp    \left(\frac{\Sl{\bar n}}{2} \right)_{\alpha\delta}\,,\nn\\
&\frac{1}{N_c} \tr \langle 0 |   \cB^{c\mu}_{\bn \perp}(x) \delta(\tau_{\bar n}-\hat \tau^{(0)}_{\bar n})   \cB^{c\nu}_{\bn \perp}(0)   |0 \rangle=\int \frac{d^4 k}{(2\pi)^4} e^{-ik\cdot x} \cJ_\bn(\tau_\bn, k,Q)g^{\mu \nu}_\perp  \,.
\end{align}
As before, this can be simplified to convolutions involving just the light cone positions
{\footnotesize
\begin{align}
&\frac{1}{\sigma_0}\frac{d\sigma^{(2),\text{rad.}}_{\psi, n,g} }{d\tau}= Q^5 H(Q^2) \int d \tau_n d\tau_\bn d\tau_{us} \delta(\tau -\tau_n -\tau_\bn -\tau_{us})   \left[  \int \frac{dk^-}{2\pi Q} \cJ_\bn(\tau_\bn, k^-,Q)  \right]  \int \frac{dr_2^+}{2\pi Q} \frac{dr_3^+}{2\pi Q}   \\
&\cdot \left[ Q \int \frac{d^4 r_1}{(2\pi)^4}   \int \frac{dr_2^-}{2\pi}  \frac{d^2r_2^\perp}{(2\pi)^2}   \int \frac{dr_3^-}{2\pi}  \frac{d^2r_3^\perp}{(2\pi)^2}       \frac{ S_{\cB n\psi}^{(2)}(\tau_{us}, r_1, r_2,r_3,Q)}  {r_2^+ r_3^+}  \right]   \cdot   \left[  Q^2 \int \frac{dl_1^+}{2\pi} \cJ_{\cB n \psi}(\tau_n, l_1^+, r_2^+,r_3^+,Q) \cdot r_2^+ r_3^+ \right]\nn\\
	&\equiv Q^5 H(Q^2)  \int d \tau_n d\tau_\bn d\tau_{us} \delta(\tau -\tau_n -\tau_\bn -\tau_{us}) J_\bn(Q^2\tau_\bn) 
	\cdot  \int \frac{dr_2^+}{2\pi Q} \frac{dr_3^+}{2\pi Q}  S_{\cB n\psi}^{(2)}(\tau_{us}, r_2^+,r_3^+,Q) J_{\cB n\psi}^{(2)}(\tau_n, r_2^+,r_3^+,Q)\nn\,.
\end{align}}%
Schematically, this can be illustrated as
\begin{align}
\fd{3cm}{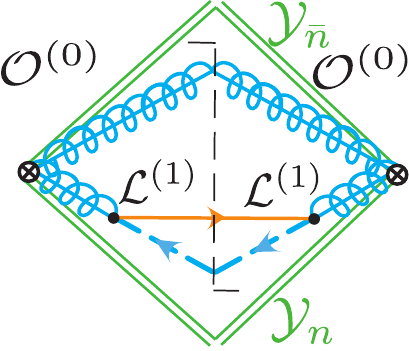} =  \int dr_2^+ dr_3^+ \fd{3cm}{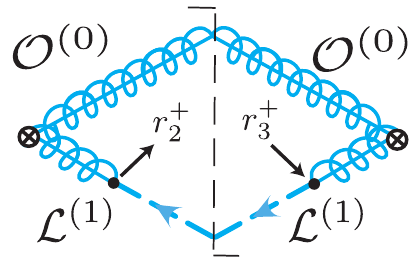} \otimes  \fd{3cm}{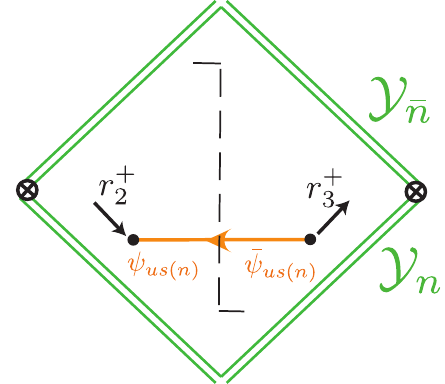}\,.
\end{align}
This takes an identical form to the case of the emission from a collinear quark, with the exception of the detailed structure of the Wilson lines and fields in the jet function.

\subsubsection{Soft Gluon Emission}\label{sec:sub_cross_gluon}

At lowest order in $\alpha_s$, we also have contributions from soft gluon emission. Due to the LBK theorem, these first arise through a single insertion of the $\cL^{(2)}$ Lagrangian. Here we will show how these can be factorized to all orders at the cross section level, for the insertion of the $\cL^{(2)}$ Lagrangian on either a collinear quark or gluon leg.

We begin by considering the case of the insertion on a collinear quark leg, corresponding to soft gluon emission from a collinear quark. We again take as a concrete process thrust for $e^+e^-\to$ dijets, as mediated by the hard scattering operator of \Eq{eq:LP_op_fact}. For convenience, we recall the form of the Lagrangian
\begin{align}
{\cal L}_{n \xi}^{(2) \text{BPS}} &=  \bar \chi_n  \left(  T^a \gamma^\mu_\perp \frac{1}{\bar \cP}  i \Sl \partial_{us\perp} - i  {\overleftarrow{\Sl \partial}}_{us\perp} \frac{1}{\bar \cP} T^a \gamma^\mu_\perp   \right)   \frac{\Sl \bn}{2} \chi_n  \,   g \cB_{us(n)}^{a\mu} \,.
\end{align}
This contributes to the cross section through the interference with the leading power hard scattering operator. As was the case at amplitude level, the factorization for the ultrasoft gluon emission is more complicated due to the presence of the ultrasoft derivative. We would like to arrange the ultrasoft derivative such that it acts only on the ultrasoft fields in the soft function.  This can be done using identical arguments as given in \Sec{sec:RF_amp_squark}. Flipping the derivative to act in the soft sector, and performing the Fierzing of the matrix element to obtain a factorized form, we find
\begin{align}
\frac{1}{\sigma_0}\frac{d\sigma^{(2),\text{rad}}_{\cB_{us}, n} }{d\tau}&= H(Q^2) \int d^4x d^4y \int d \tau_n d\tau_\bn d\tau_{us} \delta(\tau -\tau_n -\tau_\bn -\tau_{us}) \nn \\
&\cdot\frac{1}{QN_c} \tr \langle 0 | \chi_n(x)_\kappa \delta(\tau_n-\hat \tau_n)  \frac{1}{\bar \cP}\bar \chi_n(y)_\rho \chi_n(y)_\alpha \bar \chi_n(0)_\delta |0 \rangle  \left(\frac{\Sl{\bar n}}{2} \right)_{\kappa \rho}  \left(\frac{\Sl{\bar n}}{2} \right)_{\alpha \delta} \nn \\
& \cdot\frac{1}{QN_c} \tr \langle 0 |   \chi_\bn(x)_\beta \delta(\tau_n-\hat \tau_n)  \bar \chi_\bn(0)_\gamma   |0 \rangle \left(\frac{\Sl{ n}}{2} \right)_{\beta \gamma}  \nn \\
&\cdot \frac{1}{N_c} \tr \langle 0 |  Y_\bn^\dagger (x) Y_n^\dagger(x) \delta(\tau_{us}-\hat \tau_{us}) \partial_\perp \cdot \cB_{ us (n) \perp}(y) Y_n(0)  Y_\bn(0) |0 \rangle   \,.
\end{align}
The derivation of the Lorentz, Dirac, and color structure is left to \App{app:Fierzing}. 
Fourier transforming each of the functions
\begin{align}
&\frac{1}{Q N_c}\tr \langle 0 |  Y_\bn^\dagger (x) Y_n^\dagger(x) \delta(\tau_{us}-\hat \tau_{us}) \partial_\perp \cdot \cB_{us(n) \perp }(y) Y_n(0)  Y_\bn(0) |0 \rangle \nn \\
&\hspace{4cm}=\int \frac{d^4 r_1}{(2\pi)^4}  \frac{d^4 r_2}{(2\pi)^4} e^{-ir_1 \cdot x} e^{-ir_2 \cdot y} S_{n \cB_{us}}^{(2)}(\tau_{us}, r_1, r_2,Q)\,,\nn \\
&\frac{1}{Q N_c} \tr \langle 0 | \chi_n(x)_\beta \delta(\tau_n-\hat \tau_n) \frac{1}{\bar \cP}\bar \chi_n(y)_\gamma \chi_n(y)_\alpha \bar \chi_n(0)_\delta |0 \rangle \nn \\
&\hspace{3cm}=\int \frac{d^4 l_1}{(2\pi)^4}  \frac{d^4 l_2}{(2\pi)^4}  e^{-il_1\cdot x} e^{-il_2\cdot y} \cJ_{n \cB_{us}}^{(2)}(\tau_n, l_1, l_2,Q) \left(\frac{\Sl{\bar n}}{2} \right)_{\beta \gamma}  \left(\frac{\Sl{\bar n}}{2} \right)_{\alpha \delta} \,,\nn\\
&\frac{1}{Q N_c} \tr \langle 0 |   \chi_\bn(x)_\beta \delta(\tau_n-\hat \tau_n)  \bar \chi_\bn(0)_\gamma   |0 \rangle=\int \frac{d^4 k}{(2\pi)^4} e^{-ik\cdot x} \cJ_\bn(\tau_\bn, k,Q) \left(\frac{\Sl{ n}}{2} \right)_{\beta \gamma} \,,
\end{align}
and using the locality of the collinear functions to simplify the structure of the convolutions, we find
\begin{align}
	\frac{1}{\sigma_0}\frac{d\sigma^{(2),\text{rad}}_{\cB_{us}, n} }{d\tau}&= Q^5 H(Q^2)  \int d \tau_n d\tau_\bn d\tau_{us} \delta(\tau -\tau_n -\tau_\bn -\tau_{us})\cdot \left[  \int \frac{dk^-}{2\pi} \cJ_\bn(\tau_\bn, k^-,Q)  \right] \nn \\
	&\qquad\cdot   \int \frac{dr_2^+}{2\pi Q}  \cdot \left[  \int \frac{d^4 r_1}{(2\pi)^4}   \int \frac{dr_2^-}{2\pi}  \frac{d^2r_2^\perp}{(2\pi)^2}    \frac{ S_{n\cB_{us}}^{(2)}(\tau_{us}, r_1, r_2,Q)}  {r_2^+}  \right] \nn \\
	&\qquad \cdot   \left[   \int \frac{dl_1^+}{2\pi} \cJ_{n\cB_{us}}(\tau_n, l_1^+, r_2^+,Q) \cdot r_2^+ \right]
	\nn\\&\equiv
	Q^5 H(Q^2)  \int d \tau_n d\tau_\bn d\tau_{us} \delta(\tau -\tau_n -\tau_\bn -\tau_{us})  J_{\bar n}(Q^2\tau_\bn) \nn\\
	&\qquad\cdot \int \frac{dr_2^+}{2\pi Q}  S_{n\cB_{us}}^{(2)}(Q\tau_{us}, r_2^+)  J_{n\cB_{us}}^{(2)}(\tau_n, r_2^+,Q)\,.
\end{align}
Diagramatically, we have
\begin{align}
\fd{3cm}{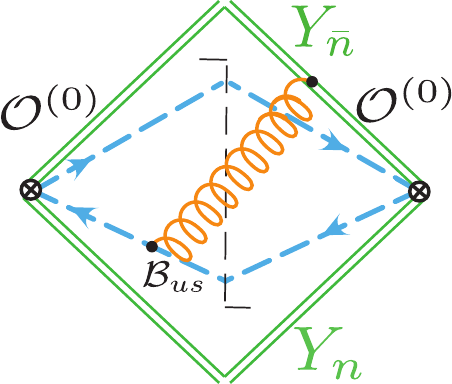}=\int d r_2^+ \fd{3cm}{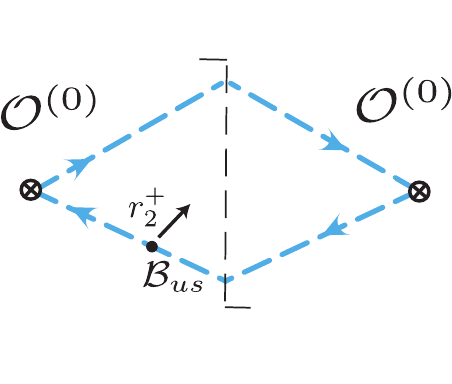} \otimes  \fd{3cm}{figures_b/wilson_framing_radiative_low.pdf}\,.
\end{align}
Here the green lines indicate Wilson lines, and the orange spring indicates the $\cB_{us(n)}$ field. As with the soft quark radiative function, this soft gluon radiative function is first non-vanishing with a single emission crossing the cut.
We have done some rearrangement, multiplying and dividing by a factor of $l_2^+$. This choice is to make the tree level expression for the jet function independent of $l_2^+$. In particular, we find a radiative jet function
\begin{align}
J_{n\cB_{us}}^{(2)}(\tau_n, r_2^+)=\left[   \int \frac{dl_1^+}{2\pi} \cJ_{n\cB_{us}}(\tau_n, l_1^+, r_2^+) \cdot r_2^+ \right]\,,
\end{align}
which incorporates all the collinear dynamics in the $n$ sector from which the $\cB_{us(n)}$ field is emitted. Since it is defined as a collinear matrix element, all loop corrections are collinear in nature, and are factorized from the soft loop corrections, which are described by the soft function.

We can also consider the insertion of an $\cL^{(2)}$ on a collinear gluon line, which corresponds at lowest order to the emission of a soft gluon from a collinear gluon leg. However, here we will find something interesting. If the collinear gluon field is taken to have no label perpendicular momentum, the $\cO(\lambda^1)$ Lagrangian insertion vanishes. However, at $\cO(\lambda^2)$, we have a contribution from the subsubleading gluon lagrangian $\cL_{ng}^{(2)}$, defined in \eq{subsubgluonlagr}. By dropping the terms proportional to the label perpendicular momentum, the relevant insertion comes from the three field operator
\be 
	\cL_{ng}^{(2)} \supset g \text{Tr} \left( \partial_\perp^{[\mu} \cB_{us \perp}^{\nu]}[ \cB^\perp_{n\mu}  , \cB^\perp_{n\nu}   ]\right)\,.
\ee
However, we note that
\begin{align}
	\partial_\perp^{[\mu} \cB_{us \perp}^{\nu]} &\equiv \partial_\perp^{[\mu} \big( S_n^\dagger \partial_\perp^{\nu]}S_n] + S_n^\dagger gA_{us \perp}^{\nu]}S_n \big) \nn \\
	&=[\partial_\perp^{[\mu} S_n^\dagger] [\partial_\perp^{\nu]}S_n] +  S_n^\dagger \partial_\perp^{[\mu} \partial_\perp^{\nu]} S_n+ \partial_\perp^{[\mu} S_n^\dagger gA_{us \perp}^{\nu]}S_n \nn \\
	&= \partial_\perp^{[\mu} gA_{us \perp}^{\nu]} + \cO(g^2)\,.
\end{align}
For the case of dijet production, where the two jets are back to back, all Wilson lines in the soft function are along the light cone directions $n$ or $\bar n$. Therefore, we see that the emission of a soft gluon from $\cL_{ng}^{(2)}$ gives a vanishing contribution to the cross section at lowest order in $\alpha_s$. This fact was used for the calculation of the leading log series in \Ref{Moult:2018jjd}. We will therefore not consider its factorization explicitly, since in this section we have only considered the factorization for contributions at lowest order in $\alpha_s$. Beyond $\cO(\alpha_s)$, there are other contributions that will be discussed in more detail in the next section.

\subsection{Discussion}\label{sec:discuss_squared}

In the section we have derived a factorized expression in terms of a convolution of gauge invariant soft and collinear matrix elements for the emission of a gauge invariant soft quark or gluon field at the cross section level. The main new feature at subleading power, are additional convolutions in the momentum passed between the soft and collinear sectors. Due to the multipole expansion in SCET this can be simplified to a single component $k^+_i$, which represents the position along the light cone. This also follows directly from the SCET power counting since only the $+$-momenta are the same size. To be able to separately renormalize the soft and collinear matrix elements, allowing for a resummation of subleading power logarithms, it is essential that both matrix elements are gauge invariant, and that no additional divergences appear in the single variable convolution. In a non-abelian gauge theory, the ability to formulate a soft emission in a gauge invariant manner is highly non-trivial since soft partons carry non-abelian charge, and it relies on our use of non-local gauge invariant soft quark and gluon fields, and an understanding of the all orders Lagrangian describing their interactions.

Some of the factorized expressions presented in this section are similar to those which have appeared in the $B$-physics literature \cite{Lee:2004ja}. In this case, instead of being vacuum matrix elements, the soft function is a matrix element of $B$ meson states. These were studied for the case of heavy to light decays where the heavy quark is treated using HQET. Nevertheless, their structure as non-local operators with insertions along the light cone is similar. This can be viewed as another advantage of using the operator based approach, namely since the SCET Lagrangians are universal, the same structures will appear in a variety of distinct physical processes allowing for a universal framework. As an example, although we have used for concreteness here the case of dijet production, since the insertions appear in a single collinear sector, this could be extended to N-jet production in a straightforward manner.

In this section we have explicitly worked out the structure, including the Lorentz and Dirac structure for  two examples, namely single soft quark and soft gluon emission, with the goal of showing in detail how the factorization with Lagrangian insertions can be performed, and how it gives rise to radiative functions at the cross section level. These two examples were chosen, since they contribute at lowest order. It should be clear that a similar factorization can be performed for an insertion of any term in the subleading power Lagrangians of \Eqs{eq:L1_fields}{eq:L4_fields}, or for multiple Lagrangian insertions. In \Sec{sec:RF_thrust} we will perform a systematic classification for the case of thrust in $e^+e^-\to$ dijets of the field structure of all possible radiative functions.

\section{Classification of Radiative Contributions for Dijet Event Shapes}\label{sec:RF_thrust}

Having shown how we can achieve a factorization of radiative contributions in terms of a convolution of gauge invariant soft and collinear matrix elements,  in this section we extend this to a complete classification of the radiative contributions for an SCET$_\text{I}$ event shape observable. For concreteness, we will consider thrust in $e^+e^-\to $ dijets, however thrust for $H\to gg$ would have a very similar form. Unlike in the previous section, where we worked out in detail the factorization for those contributions that contribute at lowest order in $\alpha_s$, here we will not explicitly perform the Fierzing and factorization. Instead, we will content ourselves with an understanding of the field structure, leaving the details of the factorization and resummation to future work. Nevertheless, a study of the field content provides valuable insight into the structure and complexity of the factorization at subleading power. Despite the large number of different field structures in the Lagrangians of \Eqs{eq:L1_fields}{eq:L4_fields}, we will see that due to the symmetries of the problem many terms can be shown not to contribute to all orders in $\alpha_s$. For example, we will show that there are no $\cO(\lambda)\sim \cO(\sqrt{\tau})$ power corrections, as expected from perturbative calculations. A much larger reduction in the number of jet and soft functions is achieved if one works to a fixed order in $\alpha_s$, for example $\cO(\alpha_s^2)$, and a study of the field content is sufficient to understand these aspects of the factorization.

In \Sec{sec:vanish}, we show that $\cO(\lambda)$ radiative contributions to the cross section vanish to all order in $\alpha_s$, and in \Sec{sec:novanish} we classify the $\cO(\lambda^2)$ contributions. For those operators that can contribute, we will explicitly give the field structure and contributing diagrams to $\cO(\alpha_s^2)$.

\subsection{Vanishing at $\cO(\lambda)$}\label{sec:vanish}

At $\cO(\lambda)$, the only possible radiative contribution is an $\cL^{(1)}$ insertion into the leading power hard scattering operators. If this $\cL^{(1)}$ insertion involves the conversion of a collinear quark to an ultrasoft quark, it will vanish, since it will give rise to a vacuum matrix element involving an odd fermion number. To show that the remaining possible $\cL^{(1)}$ insertions also vanish, we note that by power counting they all involve either an ultrasoft gluon field, or an ultrasoft derivative operator, but not both. 

First consider the case of an ultrasoft derivative operator. With a single Lagrangian insertion into the leading power operators,  after performing the BPS field redefinition, the ultrasoft derivative operator must act on collinear fields, and is therefore factorized into the jet function. We now use the dimensional regularization rule for residual momenta 
\begin{align}
\sum\limits_{q_l} \int d^d q_r (q_r)^j F(q_l^-, q_l^\perp, q_r^+)=0\,,
\end{align}
where $(q_r)^j$ denotes positive powers of the $q_r^-$ and $q_r^\perp$ momenta, which are the only residual momenta which appear in the subleading power Lagrangians, to show that this vanishes. After eliminating these terms, we can simplify the structure of the  $\cL^{(1)}$ Lagrangian to
\begin{align}
\cL_n^{(1)\text{BPS}}&\sim    \cB_{n\perp} \cB_{n\perp} \cB_{us(n) \perp} \cP_\perp +\bar \chi_n \chi_n \cB_{us(n) \perp} \cP_\perp+  \cB_{n\perp} \cB_{n\perp} \cB_{n\perp} \cB_{us(n) \perp} +\bar \chi_n \chi_n \cB_{n\perp} \cB_{us(n) \perp}\,,
\end{align}
each term of which involves a $\cB_{us(n)}$ field. After performing the factorization, the $\cB_{us(n)}$ field can be factorized into the soft function. However, the soft function must be rotationally invariant about the $n-\bar n$ axes. Since the only other objects which appear in the soft functions at this power are Wilson lines, this implies that it is only possible to form a rotationally invariant soft function from the $\bar n\cdot \cB_{us(n)}$, or $n\cdot \cB_{us(n)}$ components of the field. However, only the  $\cB_{us(n)\perp}$ component appears at this power. Therefore, there are no $\cO(\lambda)$ power corrections from Lagrangian insertions. In  \cite{Feige:2017zci,Moult:2017rpl}, we have shown that $\cO(\lambda)$ contributions from hard scattering operators vanish for both a $q\bar q$ and $gg$ current, as do subleading power corrections to the measurement operator. Therefore, the present analysis provides a complete proof that there are no $\cO(\lambda)$ power corrections to all orders in $\alpha_s$.

\subsection{Contributions at $\cO(\lambda^2)$}\label{sec:novanish}

The $\cO(\lambda^2)$ radiative contributions to the cross section do not vanish. As discussed in \Sec{sec:sub-fact}, in addition to an insertion of the $\cO(\lambda^2)$ Lagrangian, one must also consider two insertions of the $\cO(\lambda)$ Lagrangian, as well as an insertion of the $\cO(\lambda)$ Lagrangian into an $\cO(\lambda)$ hard scattering operator.\footnote{One must also consider subleading corrections to the measurement function. However, as discussed in \Sec{sec:obs_fact}, the subleading power corrections to the measurement function first enter at $\cO(\lambda^2)$, and therefore if we are working only to $\cO(\lambda^2)$, we do not need to consider the interference of the subleading power measurement function with Lagrangian insertions. At higher powers, this contribution could be treated in a similar way to the contributions discussed in this chapter. } The complete expression for the radiative contribution is therefore
{\begin{small}
\begin{align}\label{eq:xsec_lam2_rad}
&\frac{d\sigma}{d\tau}^{(2),\text{rad.}} =  N \sum_{X,i}  \tilde \delta^{(4)}_q   \int d^4x  \bra{0} C_i^{(1)*} \tO_i^{(1)\dagger}(0) \ket{X}\bra{X} \TO( i \cL^{(1)}(x)) C^{(0)} \tO^{(0)}(0) \ket{0}   \delta\big( \tau - \tau^{(0)}(X) \big)+\text{h.c.}    \nn\\
&+ N \sum_{X,i}  \tilde \delta^{(4)}_q   \int d^4x  \bra{0}  C^{(0)*} \tO^{(0)\dagger}(0)\ket{X}\bra{X}\TO ( i \cL^{(1)}(x)) C_i^{(1)} \tO_i^{(1)}(0)  \ket{0}   \delta\big( \tau - \tau^{(0)}(X) \big)+\text{h.c.}    \nn\\
&+ N \sum_X  \tilde \delta^{(4)}_q \int d^4x  \bra{0}  C^{(0)*} \tO^{(0)\dagger}(0) \ket{X}\bra{X} \TO ( i \cL^{(2)}(x)) C^{(0)} \tO^{(0)}(0) \ket{0}   \delta\big( \tau - \tau^{(0)}(X) \big)  +\text{h.c.} \nn\\
&-\frac{N}{2} \sum_X    \tilde \delta^{(4)}_q  \int d^4x \int d^4y \bra{0}\ATO \cL^{(1)}(x) \cL^{(1)}(y)  C^{(0)*} \tO^{(0)\dagger}(0) \ket{X}\bra{X}  C^{(0)} \tO^{(0)}0) \ket{0}   \delta\big( \tau - \tau^{(0)}(X) \big)  +\text{h.c.} \nn\\
&+\frac{N}{2} \sum_X   \tilde \delta^{(4)}_q  \int d^4x  \int d^4y \bra{0}\ATO\cL^{(1)}(x)  C^{(0)*} \tO^{(0)\dagger}(0) \ket{X}\bra{X} \cL^{(1)}(y) C^{(0)} \tO^{(0)}0) \ket{0}  \delta\big( \tau - \tau^{(0)}(X) \big) +\text{h.c.} \,,
\end{align}
\end{small}}%
which is a subset of the complete set of $\cO(\lambda^2)$ contributions given in \Eq{eq:xsec_lam2_RF}.
We will consider the different contributions from each of these cases in turn. For simplicity, we will not separately discuss the different possible positions of the final state cut, since the operator structure is the same in all cases. 

\subsubsection{Contributions from $\cO(\lambda)$ Hard Scattering Operators }\label{sec:lam_hard}
We first consider term the first two lines of \eq{xsec_lam2_rad} involving a single $\cO(\lambda)$ hard scattering operators, $\cO^{(1)}$, and a single $\cO(\lambda)$ Lagrangian insertion, $\cL^{(1)}$.
A detailed derivation of the $\cO(\lambda)$ hard scattering operators for $e^+e^-\to$ dijets was given in \cite{Feige:2017zci}. The operators are of two types, and either involve a collinear quark and a collinear gluon field in the same collinear sector, or two collinear quark fields in the same collinear sector. We can use arguments identical to those presented in \Sec{sec:vanish} to show that these terms do not contribute to all orders in $\alpha_s$. In particular, since the hard scattering operators at $\cO(\lambda)$ do not involve additional ultrasoft fields, the rotational invariance of the soft function, combined with the structure of the $\cO(\lambda)$ Lagrangian, which involves only $\cB_{us(n)\perp}$ fields  guarantees that all contributions involving ultrasoft field insertions vanish. Similarly, all contributions involving insertions of an ultrasoft derivative from the Lagrangian vanish for the same reasons as described in \Sec{sec:vanish}.

\subsubsection{Contributions from $\cO(\lambda^2)$ Lagrangian Insertions}\label{sec:thrust_single_insert}

Next we consider the third line of \eq{xsec_lam2_rad}, involving the T-products of leading power hard scattering operators and $\cL^{(2)}$.
When considering contributions to the cross section,  the structure of the  $\cL^{(2)}$ Lagrangian can be simplified, since one cannot have a single insertion involving an ultrasoft fermion, or involving just ultrasoft derivatives acting in the collinear sector. This reduces the structure down to $10$ distinct field structures
\begin{align}
\cL_n^{(2)\text{BPS}}&\sim  \cB_{n\perp} \cB_{n\perp} \cB_{us(n)} \cB_{us(n)} + \cB_{n\perp} \cB_{n\perp} \partial_{us} \cB_{us(n)}  + \cB_{n\perp} \cB_{n\perp} \cB_{n\perp} \cB_{n\perp} \cB_{us(n)} \nn \\
&+ \cB_{n\perp} \cB_{n\perp} \cB_{us(n)} \cP_\perp^2 +\cB_{n\perp} \cB_{n\perp} \cP_\perp \cB_{n} \cB_{us(n)}+\bar \chi_n \chi_n \cB_{us(n)} \cB_{us(n)} \nn \\
& +\bar \chi_n \chi_n \partial_{us} \cB_{us(n)}  +\bar \chi_n \chi_n \cB_{n\perp} \cB_{n\perp} \cB_{us(n)}+\bar \chi \chi \cB_{us(n)} \cP_\perp^2 +\bar \chi_n \chi_n \cP_\perp \cB_{n} \cB_{us(n)}\,,
\end{align}
the explicit form of which was given in \Eq{eq:lam2_BPS}.
Unlike at $\cO(\lambda)$, these contributions do not vanish.

For each of these different contributions, it is easy to understand at which order they can first contribute. Those with a $\cP_\perp$ insertion will first contribute at one-higher order, as they require an additional collinear emission to provide a non-zero $\perp$ momentum. In particular, we see that while there are a large number of different radiative functions that are required for an all orders descriptions, many first contribute at high loops. Considering only those terms which contribute to $\cO(\alpha_s^2)$, we have only $6$ terms
\begin{align}
\cL_n^{(2)\text{BPS}}&\sim \bar \chi_n \chi_n \partial_{us} \cB_{us(n)}+\bar \chi_n \chi_n \cB_{us(n)} \cB_{us(n)}  + \cB_{n\perp} \cB_{n\perp} \partial_{us} \cB_{us(n)}  +  \cB_{n\perp} \cB_{n\perp} \cB_{us(n)} \cP_\perp^2 \nn \\
& +\bar \chi \chi \cB_{us(n)} \cP_\perp^2 +\bar \chi_n \chi_n \cP_\perp \cB_{n} \cB_{us(n)}\,.
\end{align}
The first term contributes at $\cO(\alpha_s)$. The second term contributes at $\cO(\alpha_s^2)$, and  has been analyzed in detail in \Sec{sec:sub_cross_gluon} where we worked out the complete structure of the factorization, and its tree level diagram is shown in \Tab{tab:LO}. The remaining terms first contribute at $\cO(\alpha_s^2)$, and representative diagrams, along with the field structure of the factorization are given in \Tab{tab:NLO_1} for those with a single $\cB_{us(n)}$ field, and in \Tab{tab:NLO_2} for those with two $\cB_{us(n)}$ fields.

\subsubsection{Contributions from $[\cO(\lambda)]^2$ Lagrangian Insertions }\label{sec:thrust_double_insert}

At $\cO(\lambda^2)$ one receives contributions from two $\cL^{(1)}$ insertions as given by the last two lines of \eq{xsec_lam2_rad}. The non-vanishing contributions can have either one, or two soft fields. Since the $\cL^{(1)}$ has the field structure 
\begin{align}\label{eq:L1_doubleinsert}
\cL_n^{(1)\text{BPS}}&\sim \cB_{n\perp} \cB_{n\perp} \partial_{us} \cP_\perp+\bar \chi_n \chi_n \partial_{us} \cP_\perp  +  \cB_{n\perp} \cB_{n\perp} \cB_{us(n)} \cP_\perp\nn \\
& +\bar \chi_n \chi_n \cB_{us(n)} \cP_\perp+  \cB_{n\perp} \cB_{n\perp} \cB_{n\perp} \cB_{us(n)}  +\bar \chi_n \chi_n \cB_{n\perp} \cB_{us(n)} +    \cB_{n\perp} \cB_{n\perp} \cB_{n\perp} \partial_{us} \nn \\
&+\bar \chi_n \chi_n \cB_{n\perp} \partial_{us} +\bar \chi_n \cB_{n\perp} \psi_{us(n)}\,,
\end{align}
this gives rise to a large number of distinct possibilities. While it is possible to write the radiative functions for all the different possible cross terms of \Eq{eq:L1_doubleinsert}, here we focus only on those that contribute up to $\cO(\alpha_s^2)$. Note that the two insertions do not need to be in the same collinear sector. However, those in distinct sectors first contribute at 3 loops. We therefore do not draw them explicitly.

The terms in the $\cL^{(1)}$ Lagrangian contain at most one ultrasoft field. Since radiative contributions where both $\cL^{(1)}$ insertions involve no collinear fields vanish,  the double $\cL^{(1)}$ insertions can contain either one or two ultrasoft fields. Except for those terms involving the ultrasoft quarks, all terms in the $\cL^{(1)}$ Lagrangian involve either $>2$ collinear fields, or a $\cP_\perp$ operator. In both cases, these first appear at $\cO(\alpha_s^2)$. The only tree level contribution from $\cL^{(1)}$ insertions is the that of the soft quarks, which was discussed in detail in \Sec{sec:sub_cross_quark} where the complete factorization structure was worked out, and whose leading order diagram is shown in \Tab{tab:LO}. In \Tab{tab:NLO_1} we summarize the field structures of the different radiative terms with a single emission arising from two  $\cL^{(1)}$ insertions, as well as a representative diagram at $\cO(\alpha_s^2)$. In all cases, these can be viewed as a propagator correction, marked with a cross, and a $\cB_{us(n)}$ emission from  $\cL^{(1)}$. The contributions with two $\cB_{us(n)}$ fields are shown in \Tab{tab:NLO_2}, along with representative diagrams at $\cO(\alpha_s^2)$. Note that here, and throughout this section, we will explicitly draw soft fields in the radiative jet functions to illustrate where they are attached. These fields will of course be factorized into the soft functions.

It is interesting to briefly consider the simplification of the $\cL^{(1)} \cdot \cL^{(1)}$ terms in the threshold limit. The threshold limit has received much attention in the power corrections literature \cite{Dokshitzer:2005bf,Grunberg:2007nc,Laenen:2008gt,Laenen:2008ux,Grunberg:2009yi,Laenen:2010uz,Almasy:2010wn,Bonocore:2014wua,White:2014qia,deFlorian:2014vta,Bonocore:2015esa,Bonocore:2016awd} due to the large amount of available perturbative data  \cite{Matsuura:1987wt,Matsuura:1988sm,Hamberg:1990np,Anastasiou:2014lda,Anastasiou:2014vaa,Anastasiou:2015ema,Anastasiou:2015yha,Dulat:2017prg}. Up to crossing, the production of a color singlet Drell-Yan pair in a proton proton collision is identical to the case of $e^+e^-\to$ dijets considered here, and therefore power corrections in $(1-z)$ can be considered using the same formalism and operators. In the case of threshold production, collinear particles are kinematically forbidden from crossing the cut into the final state. This greatly reduces the number of possible contributions at $\cO(\alpha_s^2)$. In particular, since for the $\cL^{(1)}\cdot \cL^{(1)}$ to contribute at $\cO(\alpha_s^2)$, the collinear gluon must cross the cut, these terms will not contribute in the threshold limit to $\cO(\alpha_s^3)$.

{
\renewcommand{\arraystretch}{1.4}
\begin{table}[t!]
\scalebox{0.842}{
\hspace{0.2cm}\begin{tabular}{| l | c | c |c |c|c| r| }
  \hline                       
  $T$-Product & Example Diagram & Soft Function& Jet Function \\
  \hline
  $\cL^{(2)}$ & $\fd{3cm}{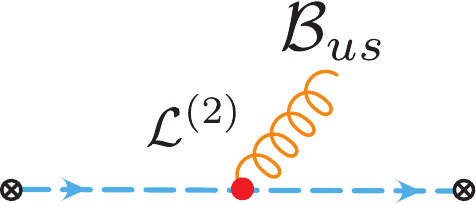}$ &  $\langle0| Y_n  Y_{\bar n}(0) \partial \cB_{us(n)}(x)Y_{\bar n} Y_n(0)   |0\rangle$& $ \langle0| \bar \chi_n(y) \bar \chi_n \chi_n(x)  \chi_n (0)|0\rangle$ \\
  \hline
   $\cL^{(1)}\cdot \cL^{(1)}$ & $\fd{3cm}{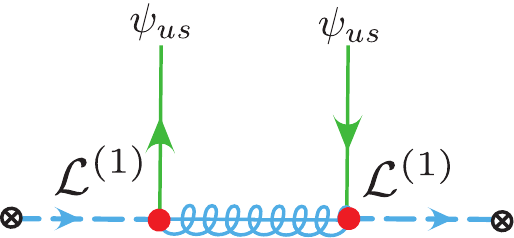}$ & {\begin{small}$ \langle0|Y_n  Y_{\bar n}(0) \bar \psi_{us}(x) \psi_{us}(z)Y_{\bar n} Y_n(0)  |0\rangle$\end{small}} &  {\begin{small}$ \langle0| \bar \chi_n(y) \bar \chi_n \cB_{n\perp} (z)\cdot  \bar \chi_n \cB_{n\perp} (x) \chi_n (0)|0\rangle$  \end{small}} \\
  \hline  
\end{tabular}}
\caption{
Leading order contributions to the $\cO(\lambda^2)$ thrust cross section from radiative functions. The soft gluon emission arises from a single $\cL^{(2)}$ insertions, in accord with the LBK theorem, while the soft quark contribution arises from two $\cL^{(1)}$ insertions. We do not explicitly write the corresponding functions when the soft gluon or quark is emitted from the $\bar n$ collinear sector.
}
\label{tab:LO}
\end{table}
}


{
\renewcommand{\arraystretch}{1.4}
\begin{table}[t!]
\scalebox{0.842}{
\hspace{-0.5cm}\begin{tabular}{| l | c | c |c |c|c| r| }
  \hline                       
  $T$-Product & Example Diagram & Soft Function & Jet Function \\
  \hline
  $\cL^{(2)}$ & $\fd{2.5cm}{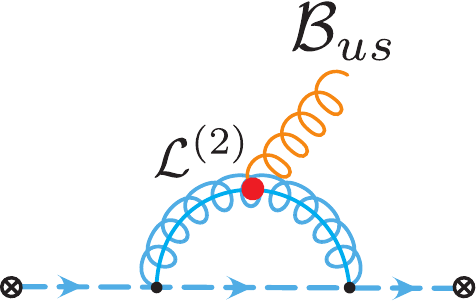}$ &  $\langle0| Y_n  Y_{\bar n}(0)  \partial \cB_{us(n)} (x) Y_{\bar n} Y_n(0)  |0\rangle$& $  \langle0| \bar \chi_n(y) \cB_{n\perp} \cB_{n\perp}(x) \chi_n (0)|0\rangle $ \\
  \hline 
    $\cL^{(2)}$ & $\fd{2.5cm}{figures_classify_all/one_loop_quark_gluon_insertion_low.pdf}$ &  $\langle0| Y_n  Y_{\bar n}(0)\cB_{us(n)}(x)Y_{\bar n} Y_n(0)  |0\rangle$& $ \langle0| \bar \chi_n(y) \cB_{n\perp} \cB_{n\perp}  \cP_\perp^2 (x) \chi_n (0)|0\rangle$ \\
  \hline 
   $\cL^{(2)}$  & $\fd{2.5cm}{figures_classify_all/1collinear_1soft_L2_low.pdf} $ &  $\langle0| Y_n  Y_{\bar n}(0)\partial \cB_{us(n)}(x) Y_{\bar n} Y_n(0) |0\rangle$& $ \langle0| \bar \chi_n(y) \bar \chi \chi  \cP_\perp^2 (x) \chi_n(0)|0\rangle$ \\
  \hline  
     $\cL^{(2)}$  & $\fd{2.5cm}{figures_classify_all/1collinear_1soft_low.pdf}$ &  $\langle0|Y_n  Y_{\bar n}(0) \cB_{us(n)}(x) Y_{\bar n} Y_n(0) |0\rangle$& $ \langle0| \bar \chi_n(y) \bar \chi_n \chi_n \cP_\perp \cB_{n}  (x) \chi_n(0)|0\rangle$ \\
    \hline  
     $\cL^{(1)}\cdot \cL^{(1)}$  & $\fd{2.5cm}{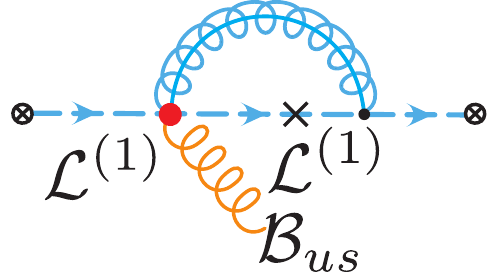}$ &  $\langle0|Y_n  Y_{\bar n}(0) \partial \cB_{us(n)}(z)Y_{\bar n} Y_n(0) |0\rangle$& $\langle0| \bar \chi_n(y) \bar \chi_n \chi_n \cB_{n\perp}(z)\cdot \bar \chi_n \chi_n  \cP_\perp (x) \chi_n (0)|0\rangle$ \\
      \hline  
     $\cL^{(1)}\cdot \cL^{(1)}$  & $\fd{2.5cm}{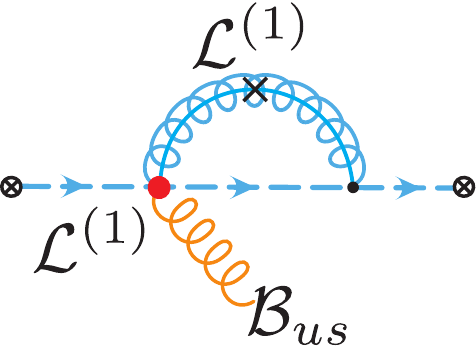}$ &  $\langle0|Y_n  Y_{\bar n}(0) \partial \cB_{us(n)}(z)Y_{\bar n} Y_n(0) |0\rangle$& $ \langle0| \bar \chi_n(y) \bar \chi_n \chi_n \cB_{n\perp} (z)\cdot \cB_{n\perp} \cB_{n\perp}  \cP_\perp (x) \chi_n(0)|0\rangle$ \\
       \hline  
     $\cL^{(1)}\cdot \cL^{(1)}$  & $\fd{2.5cm}{figures_classify_all/double_L1_propagator_quark_c_low}$ &  $\langle0|Y_n  Y_{\bar n}(0) \partial \cB_{us(n)}(z)Y_{\bar n} Y_n(0) |0\rangle$& $ \langle0| \bar \chi_n(y) \bar \chi_n \chi_n \cP_\perp(z)\cdot \cB_{n\perp} \cB_{n\perp}  \cP_\perp (x) \chi_n (0)|0\rangle$ \\
       \hline  
     $\cL^{(1)}\cdot \cL^{(1)}$  & $\fd{2.5cm}{figures_classify_all/double_L1_propagator_quark_d_low}$ &  $\langle0|Y_n  Y_{\bar n}(0) \partial \cB_{us(n)}(z)Y_{\bar n} Y_n(0) |0\rangle$& $ \langle0| \bar \chi_n(y) \bar \chi_n \chi_n \cP_\perp(z)\cdot \bar \chi_n \chi_n  \cP_\perp (x) \chi_n (0)|0\rangle$ \\
       \hline  
     $\cL^{(1)}\cdot \cL^{(1)}$  & $\fd{2.5cm}{figures_classify_all/double_L1_propagator_gluon_a_low}$ &  $\langle0|Y_n  Y_{\bar n}(0)  \partial \cB_{us(n)}(z)Y_{\bar n} Y_n(0) |0\rangle$& $ \langle0|\bar \chi_n(y)  \cB_{n\perp} \cB_{n\perp}  \cP_\perp (z)\cdot \bar \chi_n \chi_n  \cP_\perp (x) \chi_n (0)|0\rangle$ \\
       \hline  
     $\cL^{(1)}\cdot \cL^{(1)}$  & $\fd{2.5cm}{figures_classify_all/double_L1_propagator_gluon_b_low}$ &  $\langle0|Y_n  Y_{\bar n}(0) \partial \cB_{us(n)}(z)Y_{\bar n} Y_n(0) |0\rangle$& $ \langle0|\bar \chi_n(y)  \cB_{n\perp} \cB_{n\perp} \cP_\perp (z)\cdot \cB_{n\perp} \cB_{n\perp}  \cP_\perp (x) \chi_n (0)|0\rangle$ \\    
  \hline 
\end{tabular}}
\caption{
Radiative functions which contribute at $\cO(\lambda^2)$ to the thrust cross section starting at 2 loops with a single soft emission. This includes both terms arising from a single $\cL^{(2)}$ insertion, and from two $\cL^{(1)}$ insertions. We have not explicitly written the functions where the emission occurs in the $\bar n$ collinear sector, since they can easily be obtained from those given. We have also not explicitly written the measurement function.
}
\label{tab:NLO_1}
\end{table}
}
%
%
%
%
{
\renewcommand{\arraystretch}{1.4}
\begin{table}[t!]
\scalebox{0.842}{
\hspace{-0.9cm}\begin{tabular}{| l | c | c |c |c|c| r| }
  \hline                       
  $T$-Product & Example Diagram & Soft Function&Jet Function \\
  \hline
  $\cL^{(2)}$ & $\fd{3cm}{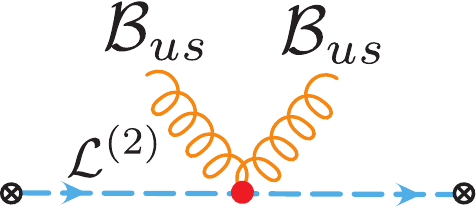}$ &  $\langle0|Y_n  Y_{\bar n}(0) \cB_{us(n)}  \cB_{us(n)} (x) Y_{\bar n} Y_n(0) |0\rangle$& $ \langle0| \bar \chi_n(y) \bar \chi_n \chi_n  (x) \chi_n (0)|0\rangle$ \\
  \hline 
   $\cL^{(1)} \cdot \cL^{(1)} $ & $\fd{3cm}{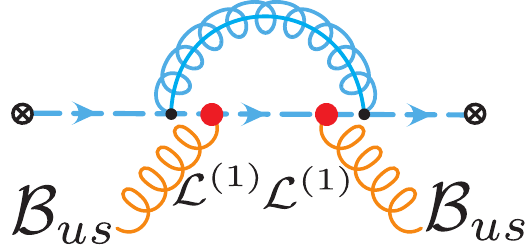}$ &  $\langle0|Y_n  Y_{\bar n}(0) \cB_{us(n)} (x) \cB_{us(n)} (z)Y_{\bar n} Y_n(0)  |0\rangle$& $\langle0| \bar \chi_n(y) \bar \chi_n \chi_n \cP_\perp (z)\cdot  \bar \chi_n \chi_n  \cP_\perp  (x) \chi_n(0)|0\rangle$ \\
  \hline  
     $\cL^{(1)} \cdot \cL^{(1)} $ & $\fd{3cm}{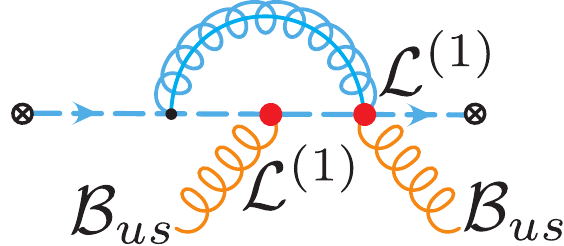}$ &  $\langle0|Y_n  Y_{\bar n}(0) \cB_{us(n)} (x) \cB_{us(n)}(z) Y_{\bar n} Y_n(0) |0\rangle$& $ \langle0| \bar \chi_n(y) \bar \chi_n \chi_n  \cP_\perp(z) \cdot  \bar \chi_n \chi_n \cB_{n\perp}  (x) \chi_n (0)|0\rangle$ \\
  \hline  
       $\cL^{(1)} \cdot \cL^{(1)} $ & $\fd{3cm}{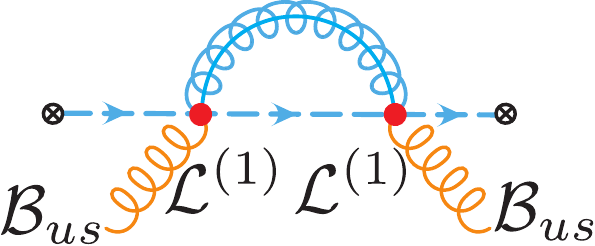}$ &  $\langle0|Y_n  Y_{\bar n}(0) \cB_{us(n)}(x)  \cB_{us(n)}(z) Y_{\bar n} Y_n(0) |0\rangle$& $ \langle0| \bar \chi_n(y) \bar \chi_n \chi_n \cB_{n\perp} (z) \cdot \bar \chi_n \chi_n \cB_{n\perp}  (x) \chi_n (0)|0\rangle$ \\
  \hline  
\end{tabular}}
\caption{
Radiative functions which contribute at $\cO(\lambda^2)$ to the thrust cross section starting at 2 loops with two soft emissions. In the restriction to the threshold limit, only the first $T$-product contributes, since collinear partons cannot cross the cut, making the other terms first contribute at $\cO(\alpha_s^3)$. We have not explicitly written the functions where the emission occurs in the $\bar n$ collinear sector, since they can easily be obtained from those given.
}
\label{tab:NLO_2}
\end{table}
}

\subsection{Discussion}\label{sec:discuss_RF}

The analysis of this section illustrates that the general structure of the Lagrangian contributions is quite complicated. In particular, while there are only several contributions at lowest order in $\alpha_s$, there are a large number of contributions which first appear at higher orders, coming from Lagrangian insertions which involve multiple collinear fields. Indeed, there are terms whose first non-vanishing contribution is at $4$-loops. These contributions do not seem in general to be related to the renormalization group evolution of those operators that contribute at lowest order, and thus appear to complicate the general structure.

 We believe that the leading logarithmic contributions are captured by the renormalization group evolution of the two radiative functions that contribute at lowest order, which are given in \Tab{tab:LO}, and that the Lagrangian insertions involving higher numbers of collinear fields are only required to reproduce subleading logarithms. This has been explicitly verified in the subleading power leading logarithmic resummation \cite{Moult:2018jjd} for thrust in $H\to gg$ in pure Yang-Mills theory (i.e. without quarks), and here a proof of this statement was given under the assumption that the convolutions appearing in the factorization formula converge. However, we do not currently have a more general proof of this statement. This expectation comes from the fact that collinear loops from these operators involving multiple collinear fields are less singular than the identical collinear loops involving collinear Wilson line emissions. If this is true, then the leading logarithmic resummation may take a simple form, but beyond leading logarithmic order an increasing number of operators will be required. We plan to develop the leading logarithmic renormalization group evolution of these leading power operators in a future chapter, and by comparing with explicit calculations it should be possible to understand the role of the operators involving additional collinear fields.

\section{Conclusions}\label{sec:conclusionsRadiative}

In this chapter we have derived an all orders factorization for the emission of soft partons from a jet, expressed as a convolution of gauge invariant soft and collinear matrix elements. Unlike for leading power soft emissions, which are only sensitive to the color charge and direction of the collinear particles, radiative functions couple the momentum of the soft and collinear sectors in a non-trivial manner, and are also sensitive to the spin structure of the collinear particles.  Using SCET, we have shown that the radiative functions are given by matrix elements of universal subleading power Lagrangians describing the interactions of non-local gauge invariant quark and gluon fields. The use of non-local gauge invariant fields is crucial to achieve a gauge invariant factorization for soft parton emission, since in a non-abelian gauge theory the emitted parton carries a charge. The multipole expansion in the effective theory allows the factorization to be expressed as a single variable convolution describing the position along the light cone of the operator insertion, which dresses the leading power Wilson lines, and describes the breakdown of eikonalization. 

A key advantage of our approach is that the radiative functions are derived from factorized matrix elements of the SCET Lagrangians, which describe the interactions of the gauge invariant quark and gluon fields to all orders. This allows us to provide a complete classification of all radiative functions that will contribute at a given power in the expansion. The number of such operators at each given power is finite, and soft and collinear emissions to all orders in $\alpha_s$ are described by the intricate Wilson line structure of the operators, dictated by the symmetries of the effective theory. As one particular example, we introduced a radiative function describing the emission of a soft quark. As was shown in the calculation of subleading power thrust \cite{Moult:2016fqy,Boughezal:2016zws,Moult:2017jsg} contributions from soft quarks give a leading logarithmic contribution at subleading power. Soft quarks are also required to compute subleading power corrections to the threshold limit for Drell-Yan production in the $qg$ channel. We performed a classification of all the radiative functions which contribute for the event shape thrust in $e^+e^-\to$ dijets, and derived in detail the structure of the radiative functions which contribute at tree level. This provides the final missing ingredient in achieving factorization and resummation of event shapes at subleading power. 

Combined with the subleading power operator bases for $\bar q \Gamma q$ \cite{Feige:2017zci,Chang:2017atu} and $gg$ \cite{Moult:2017rpl}, and the analysis of the subleading power measurement function \cite{Feige:2017zci}, all the sources of subleading power corrections in SCET (as were summarized in \Sec{sec:sum_fact}) have now been characterized in detail. It will now be of significant interest to derive a complete subleading power factorization theorem for an event shape observable, and to study the renormalization group evolution of the different components, a direction which we intend to pursue in future work. The renormalization of higher twist matrix elements has been studied  \cite{Balitsky:1987bk,Ratcliffe:1985mp,Ji:1990br,Ali:1991em,Kodaira:1996md,Balitsky:1996uh,Mueller:1997yk,Belitsky:1997zw,Belitsky:1999ru,Belitsky:1999bf,Vogelsang:2009pj,Braun:2009mi}, as has that of subleading power operators in $B$ physics \cite{Hill:2004if,Beneke:2005gs}, and of several power suppressed operators for $e^+e^-\to$ dijets \cite{Freedman:2014uta,Goerke:2017lei} and $N$-jet production \cite{Beneke:2017ztn,Beneke:2018rbh}.  We anticipate that combining all these ingredients will allow for the resummation of subleading power corrections for event shape observables.

\chapter{First Subleading Power Resummation for Event Shapes}\label{sec:subRGE}

\section{Introduction}

Due to the complexity of interacting gauge theories in four dimensions, simplifying limits such as the soft, collinear, or Regge limits play a central role.  These limits are important both phenomenologically, where they often capture dominant contributions to processes of interest, as well as theoretically, where they place important constraints on the structure of amplitudes and cross sections. While well understood at leading power, less is known about the all orders perturbative structure of the subleading power corrections to these limits. These subleading power corrections have recently been attracting a growing level of interest, see for example \cite{Manohar:2002fd,Beneke:2002ph,Pirjol:2002km,Beneke:2002ni,Bauer:2003mga,Hill:2004if,Lee:2004ja,Dokshitzer:2005bf,Trott:2005vw,Laenen:2008ux,Laenen:2008gt,Paz:2009ut,Benzke:2010js,Laenen:2010uz,Freedman:2013vya,Freedman:2014uta,Bonocore:2014wua,Larkoski:2014bxa,Bonocore:2015esa,Bonocore:2016awd,Moult:2016fqy,Boughezal:2016zws,DelDuca:2017twk,Balitsky:2017flc,Moult:2017jsg,Goerke:2017lei,Balitsky:2017gis,Beneke:2017ztn,Feige:2017zci,Moult:2017rpl,Chang:2017atu}. A subset of these analyses consider power corrections to the threshold limit of Drell Yan and related processes, where there are no contributions from power corrections due to real collinear radiation. 

In this chapter we will study the all orders structure of subleading power corrections to both the soft and collinear limits. This requires corrections beyond the type that can be studied from the threshold limit.  Using soft collinear effective theory (SCET) \cite{Bauer:2000ew, Bauer:2000yr, Bauer:2001ct, Bauer:2001yt}, which allows for a systematic power expansion using operator and Lagrangian based techniques, we will show for the first time how subleading power logarithms can be resummed to all orders in $\alpha_s$ for an event shape, which for concreteness we take to be thrust, $T=1-\tau$ \cite{Farhi:1977sg}, with $\tau\ll1$ in the simplified example of pure glue QCD for the process $H\to gg$ mediated by the effective operator $H G_{\mu\nu}^a G^{\mu\nu a}$ obtained by integrating out the top quark. In particular, we will show that at subleading power higher order corrections in $\alpha_s$ exponentiate at leading logarithmic (LL) accuracy into a single logarithmic term multiplying the same type of Sudakov form factor \cite{Sudakov:1954sw} as at leading power. 
Our approach is general, allowing other observables to be considered, and making clear what ingredients are needed to achieve higher logarithmic accuracy, as well as higher orders in the power expansion.

The all orders cross section for the thrust observable can be expanded in powers of $\tau$ (here $\tau$ is taken to be dimensionless), keeping all orders in $\alpha_s$ at each power
\begin{align}\label{eq:intro_expansion}
\frac{\df\sigma}{\df\tau} &=\frac{\df\sigma^{(0)}}{\df\tau} +\frac{\df\sigma^{(1)}}{\df\tau} +\frac{\df\sigma^{(2)}}{\df\tau}+\frac{\df\sigma^{(3)}}{\df\tau} +{\cal O}(\tau)\,.
\end{align}
Here $\df\sigma^{(n)}/\df\tau$ captures to all orders in $\alpha_s$ terms that scale like $\tau^{n/2-1}$, and for thrust the odd powers $d\sigma^{(2\ell+1)}/d\tau$ vanish. The leading power (LP) terms scale as $1/\tau$ (including $\delta(\tau)$) modulo logarithms. Explicitly, we have
\begin{align}
\frac{1}{\sigma_0}\frac{d\sigma^{(0)}}{d\tau} = \sum\limits_{n=0}^\infty \sum\limits_{m=-1}^{2n-1} \left(  \frac{\alpha_s(\mu)}{4\pi} \right)^n  c^{(0)}_{n,m}  {\cal L}_m(\tau)\,,
\end{align}
where ${\cal L}_{m\ge 0}(\tau) = [\theta(\tau)\log^m(\tau)/\tau]_+$ is a standard plus-function which integrates to zero over the interval $\tau\in [0,1]$, and ${\cal L}_{-1}(\tau) = \delta(\tau)$. Here the $c^{(0)}_{n,m}$ coefficients include $\log(\mu/Q)$ dependence, where $Q=m_H$ is the mass of the Higgs boson setting the scale of the hard scattering.
 All orders factorization theorems~\cite{Collins:1981uk,Collins:1985ue,Collins:1988ig,Collins:1989gx} can be proven at leading power for a number of event shape like observables~\cite{Sterman:1986aj,Bauer:2001yt,Bauer:2002nz,Fleming:2007qr,Schwartz:2007ib}. For the particular case of thrust in $H\to gg$, we have~\cite{Korchemsky:1999kt,Fleming:2007qr,Schwartz:2007ib}
\begin{align} \label{eq:fact0}
  \frac{1}{\sigma_0} \frac{\df\sigma^{(0)}}{\df\tau} 
   &=H^{(0)}(Q,\mu)\, \!\!\int\!\! ds_n ds_\bn dk \, \hat \delta_\tau\, 
     J^{(0)}_{g}(s_n,\mu ) ~ J^{(0)}_{g}(s_\bn,\mu) ~  S_{g}^{(0)}(k,\mu)  \,,
\end{align}
where 
\begin{align}
\hat \delta_\tau =\delta\left(\tau- \frac{s_n}{Q^2} -\frac{s_\bn}{Q^2} -\frac{k}{Q}\right)\,,
\end{align}
is the thrust measurement function.
Here $H^{(0)}(Q,\mu)$ is a hard function, $J^{(0)}_{g}(s,\mu )$ are gluon jet functions, and $S_{g}^{(0)}(k,\mu)$ is the adjoint soft function, whose precise definitions will be given in \Eqs{eq:jet_func}{eq:s_func} respectively. We normalize such that at lowest order $H^{(0)}$ is $1$, and the jet and soft functions are $\delta$-functions. The jet and soft functions are gauge invariant infrared finite matrix elements, which obey  simple renormalization group (RG) evolution equations that predict infinite towers of higher order logarithmically enhanced terms. The number of logarithms that are predicted is dictated by the logarithmic accuracy, denoted by N$^k$LL. Explicitly, for the first few orders, a resummation at N$^k$LL can be used to predict all the terms $c^{(0)}_{n,m}$, satisfying
\begin{align} \label{eq:whichms}
\text{LL predicts}: m=2n-1\,, \\
\text{NLL predicts}: m\geq 2n-2\,, \nn \\
\text{NNLL predicts}: m\geq 2n-4\,, \nn \\
\text{N$^3$LL predicts}: m\geq 2n-6\,, \nn 
\end{align}
for any $n$. Technically, for these resummations this counting is applied for $\log(d\sigma^{(0)}/dy)$ where $y$ is Fourier conjugate to $\tau$.\footnote{The standard counting which defines the resummation orders in position space is given by identifying the terms as $\log(d\sigma^{(0)}/dy)\simeq \sum_k  (\alpha_s \log)^k \log |_{\rm LL} + (\alpha_s\log)^k|_{\rm NLL} +  (\alpha_s\log)^k \alpha_s|_{\rm NNLL}+(\alpha_s\log)^k\alpha_s^2|_{\rm N^3LL}+\ldots$. This means that the resummation yields terms beyond those indicated in \Eq{eq:whichms} when expanded at the cross section level.} For thrust, these logarithms were first resummed to NLL in \cite{Catani:1991kz,Catani:1992ua}. Factorization and renormalization has been used to resum large logarithmic contributions to a number of $e^+e^-$ event shapes at leading power at N$^3$LL order \cite{Becher:2008cf,Abbate:2010xh,Chien:2010kc,Hoang:2014wka,Moult:2018jzp}.

Additional terms in \Eq{eq:intro_expansion} are suppressed by powers of $\lambda \sim \sqrt{\tau}$, with odd powers, $d\sigma^{(2\ell+1)}/d\tau$ vanishing, so that the series involves only integer powers of $\tau$ \cite{Beneke:2003pa,Lee:2004ja,Freedman:2013vya,Feige:2017zci}. These power suppressed terms do not involve distributions, and at power $\tau^{\ell-1}$ for $\ell \geq 1$ can be written as
\begin{align}\label{eq:subl_expansion}
\frac{1}{\sigma_0}\frac{d\sigma^{(2\ell)}}{d\tau} 
  =\sum\limits_{n=1}^\infty \sum\limits_{m=0}^{2n-1} \left(  \frac{\alpha_s(\mu)}{4\pi} \right)^n c^{(2\ell)}_{n,m}\, \tau^{\ell-1}\,\log^{m}(\tau)\,.
\end{align}
The structure of the subleading power terms is much less well understood, despite considerable effort. The first non-trivial power corrections are  described by $\df\sigma^{(2)}/\df\tau$, i.e. at  $\cO(\lambda^2)\sim \cO(\tau)$, which we will refer to as next-to-leading power (NLP). The subleading power terms at $\cO(\lambda^2)$ have recently been analytically computed in fixed order to $\cO(\alpha_s^2 \log^3)$ for thrust \cite{Freedman:2013vya,Moult:2016fqy,Boughezal:2016zws} and $N$-jettiness \cite{Moult:2016fqy,Boughezal:2016zws,Moult:2017jsg} for the first time, and the next-to-leading logarithms for $N$-jettiness at $\cO(\alpha_s)$ have been examined in \cite{Boughezal:2018mvf}. There has also been recent work on calculations of power corrections for $p_T$ in Drell-Yan \cite{Balitsky:2017flc,Balitsky:2017gis}, in the Regge limit \cite{Moult:2017xpp,Bruser:2018jnc}, and for subleading power quark mass effects \cite{Liu:2017vkm}. All these calculations have hinted at a simple structure for the power corrections, motivating an all orders understanding.

In a series of papers, we have developed within SCET all the ingredients relevant for the factorization and all orders description at $\cO(\lambda^2)$ for the case of dijet production from a color singlet current. This includes the bases of hard scattering operators \cite{Feige:2017zci,Moult:2017rpl,Chang:2017atu}, the factorization of the measurement function \cite{Feige:2017zci}, and the factorization of `radiative' contributions arising from subleading power Lagrangian insertions \cite{Moult:2019mog}. In this chapter we combine these ingredients, and carry out the resummation of the leading logarithmic (LL) contributions to all orders in $\alpha_s$ for NLP corrections to thrust. In particular, this determines all terms $c^{(2)}_{n,2n-1}$ for any $n$ in \Eq{eq:subl_expansion}, giving all the terms in the series
\begin{align}
\frac{1}{\sigma_0}\frac{d\sigma^{(2)}}{d\tau}&=\left(  \frac{\alpha_s}{4\pi} \right)\, c_{1,1}^{(2)} \log \tau +\left(  \frac{\alpha_s}{4\pi} \right)^2\, c_{2,3}^{(2)} \log^3 \tau+ \left(  \frac{\alpha_s}{4\pi} \right)^3\, c_{3,5}^{(2)} \log^5 \tau +\cdots \,, \\
&=\left( \frac{\alpha_s}{4\pi} \right) 8 C_A \log\tau -
  \left(\frac{\alpha_s}{4\pi}\right)^2
32 C_A^2 \log^3 \tau + \left(\frac{\alpha_s}{4\pi}\right)^3 64 C_A^3
  \log^5 \tau + \ldots \,, \nn
\end{align}
where in the second line we have given the first few terms of the result that we will derive for thrust in pure glue $H\to gg$.
Note that this series starts at $\alpha_s \log \tau$, which has interesting consequences for the resummation. We will show that this necessitates the introduction of new jet and soft functions which arise through mixing, and which we term $\theta$-jet and $\theta$-soft functions. We will analytically solve the corresponding subleading power RG equation involving the mixing, and including the running coupling. We consider for simplicity the case of thrust in $H\to gg$ without fermions, i.e. in a pure SU$(3)$ Yang-Mills theory without matter. This will allow us to illustrate the conceptual complexities of renormalization at the cross section level in the simplest possible setting with a smaller set of operators. The addition of operators relevant for including fermions will be considered in future work.

An outline of this chapter is as follows. In \Sec{sec:renorm} we show in the context of an illustrative example how one can renormalize subleading power jet and soft functions. The illustrative example allows for an understanding of the renormalization to all orders in $\alpha_s$, and allows us to provide complete field theoretical definitions for all functions involved in the RG flow. This involves a new class of jet and soft functions which arise at cross section level through mixing, which we demonstrate is a generic feature at subleading power that is needed to predict the series that starts at $\alpha_s \log \tau$. At $\cO(\lambda^2)$, this gives rise to a $2\times 2$ mixing structure for the RG equations. We study in detail the consistency equations for this type of RG evolution, allowing us to derive powerful and general constraints on the structure of operators that can be mixed into at subleading powers. In \Sec{sec:solution} we solve the general form of the subleading power mixing equation, including the running coupling as is relevant for subleading power resummation in QCD. In \Sec{sec:fact} we apply this to resum the leading logs at subleading power for thrust in pure glue $H\to gg$, deriving the structure of the Sudakov exponent for the subleading power corrections. In \Sec{sec:split} we perform a fixed order check of our result. We explicitly calculate to $\cO(\alpha_s^3)$ the $\cO(\lambda^2)$ leading logarithms, confirming the result predicted by the RG. Furthemore, we interpret the fixed order expansion in terms of information about the $\cO(\alpha_s^n)$ corrections to subleading power splitting functions.  We conclude in \Sec{sec:conc_subRGE}.

\section{Renormalization at Subleading Power}\label{sec:renorm}

In this section we study the structure and completeness of jet and soft functions for renormalization group equations at subleading power.   In \Sec{sec:illexample} we introduce a simple illustrative example which can be studied to all orders from known factorization properties at leading power, and from which many interesting lessons about the structure of subleading power resummation can be deduced.  This example also appears explicitly for thrust in $H\to gg$ from contributions from subleading power kinematic corrections.  In \Sec{sec:renorm_a}, we show that the renormalization of the subleading power jet and soft functions in our illustrative example leads to mixing into jet and soft functions involving $\theta$-functions of the measurement operator, which we term $\theta$-jet and $\theta$-soft functions, and we derive the structure of the RG to all orders in $\alpha_s$.  In \Sec{sec:consistency} we study RG consistency in a setup that is a generalization of our illustrative example in order to derive general constraints at subleading power on the structure of anomalous dimensions and on the appearance of $\theta$-function operators.

\subsection{An Illustrative Example at Subleading Power}\label{sec:illexample}

Our illustrative example of a subleading power factorization is obtained by multiplying the leading power factorization by $\tau$ and using
\begin{align}
  \tau \hat \delta_\tau = \tau \delta(\tau - \tau_n -\tau_\bn - \tau_s) =  (\tau_n+\tau_\bn+\tau_s)    \hat \delta_\tau \,,
\end{align}
which gives a subleading power cross section whose factorized structure follows immediately from the leading power factorization of \Eq{eq:fact0}:
\begin{align} \label{eq:fact_NLP_multTau}
 \frac{1}{\sigma_0} \frac{\df\sigma^{(2)}}{\df\tau} 
   &= H^{(0)}(Q,\mu) \int \frac{ds_n ds_\bn dk}{Q^2}\, \hat \delta_{\tau}\, \left[ s_n  J^{(0)}_{g}(s_n,\mu ) \right]  J^{(0)}_{g}(s_\bn,\mu) S_{g}^{(0)}(k,\mu)    \\
   &+ H^{(0)}(Q,\mu) \int \frac{ds_n ds_\bn dk}{Q^2} \,\hat \delta_{\tau} \, J^{(0)}_{g}(s_n,\mu )  \left[ s_\bn  J^{(0)}_{g}(s_\bn,\mu) \right ]   S_{g}^{(0)}(k,\mu)  \nn \\
   &+ H^{(0)}(Q,\mu)  \int \frac{ds_n ds_\bn dk}{Q} \,  \hat \delta_{\tau}\, J^{(0)}_{g}(s_n,\mu )  J^{(0)}_{g}(s_\bn,\mu)    \left[ k  S_{g}^{(0)}(k,\mu) \right]   \,.\nn
\end{align}
This can be written in terms of subleading power jet and soft functions as
\begin{align}  \label{eq:fact_NLP_multTau_rewrite}
  \frac{1}{\sigma_0} \frac{\df\sigma^{(2)}}{\df\tau}   
      &= H^{(0)}(Q,\mu)\int \frac{ds_n ds_\bn dk}{Q^2}\, \hat \delta_\tau\,  J^{(2)}_{g}(s_n,\mu )  J^{(0)}_{g}(s_\bn,\mu) S_{g}^{(0)}(k,\mu)    \\
   &+ H^{(0)}(Q,\mu) \int\frac{ds_n ds_\bn dk}{Q^2}\, \hat \delta_\tau\,  J^{(0)}_{g}(s_n,\mu )   J^{(2)}_{g}(s_\bn,\mu)   S_{g}^{(0)}(k,\mu)  \nn \\
   &+ H^{(0)}(Q,\mu) \int \frac{ds_n ds_\bn dk}{Q} \,  \hat \delta_\tau\, J^{(0)}_{g}(s_n,\mu )  J^{(0)}_{g}(s_\bn,\mu)     S_{g}^{(2)}(k,\mu)\,. \nn
\end{align}
The superscripts indicate the power of the function, namely those with superscript $(0)$ are LP in the $\tau$ expansion, while  those with superscript $(2)$ are power suppressed by $\lambda^2 \sim \tau$.
In this factorization, $H^{(0)}(Q,\mu)$ is the leading power hard function, which is process dependent, and will not play an important role in the current discussion.
The leading power jet function, which for $H\to gg$ is a gluon jet function, is defined as a matrix element of collinear fields
\begin{align}\label{eq:jet_func}
J^{(0)}_{g}(s,\mu)
  &= \frac{(2\pi)^3}{(N_c^2-1)}\Big\langle 0 \Big|\, \cB^{a}_{n\perp\mu} (0)\,\delta(Q+\bar \cP) \delta^2(\cP_\perp)\, \delta\left(\frac{s}{Q}-\hat \Tau\right)\, \cB^{\mu a}_{n\perp}(0) \,\Big|0\Big\rangle
\,, 
\end{align}
where $\cB^{a\mu}_{n\perp}$, is a gauge invariant gluon field (see \Eq{eq:softgluondef_sRGE} for an explicit definition), and the leading power adjoint soft function is given by
\begin{align}\label{eq:s_func}
S_{g}^{(0)}(k,\mu)=\frac{1}{(N_c^2-1)}  \tr \big\langle 0 \big| \cY^T_{\bar n}(0) \cY_n(0) \delta(k-\hat \Tau)  \cY_n^T(0) \cY_{\bar n}(0) \big|0 \big\rangle\,,
\end{align}
where $\cY_n$, $\cY_\bn$  are adjoint Wilson lines along the given lightlike directions. Explicitly,
\be\label{eq:Wilson_def}
\cY^{bc}_n(x)=\mathbf{P} \exp \left [ g \int\limits^{\infty}_0 ds\, n\cdot A^a_{us}(x+sn) f^{abc}\right]\,.
\ee
In both cases, $\hat \Tau$ is an operator that returns the value of $\Tau$ measured on a given state, where the dimensionless thrust $\tau=\Tau/Q$. In general it can be written in terms of the energy momentum tensor of the effective theory \cite{Lee:2006nr,Sveshnikov:1995vi,Korchemsky:1997sy,Bauer:2008dt,Belitsky:2001ij,Mateu:2012nk}. At tree level, $J^{(0)}_{g}(s,\mu)=\delta(s)+\cO(\alpha_s)$ and $S_{g}^{(0)}(k,\mu)=\delta(k)+\cO(\alpha_s)$.

After multiplying by $\tau$, the operator definitions for the subleading power jet and soft functions appearing in \Eq{eq:fact_NLP_multTau} are simply
\begin{align}\label{eq:tau_funcs}
J^{(2)}_{g,  \delta}(s,\mu)&=\frac{(2\pi)^3}{(N_c^2-1)}   \Big\langle 0 \Big|\, \cB^{\mu a}_{n\perp} (0)\,\delta(Q+\bar \cP) \delta^2(\cP_\perp)\, s\,\delta\left(\frac{s}{Q}-\hat \Tau\right)\, \cB^{\mu a}_{n\perp,\omega}(0) \,\Big|0\Big\rangle\,, \\
S^{(2)}_{g, \delta}(k,\mu)&= \frac{1}{(N_c^2-1)}\tr \langle 0 |  \cY^T_{\bar n}(0) \cY_n(0)\, k ~\delta(k-\hat \Tau) \cY_n^T(0) \cY_{\bar n}(0) |0\rangle\,.\nn
\end{align}
 The subscript $\delta$ is meant to indicate that the measurement function that appears is the same as the leading power measurement. The mass dimension of both functions in \Eq{eq:tau_funcs} is zero. Although this example may appear too trivial, it turns out to become quite interesting when we consider the RG evolution of these subleading power jet and soft functions, which we do next.

\subsection{$\theta$-jet and $\theta$-soft Functions and RG Equations}\label{sec:renorm_a}

The RG for the subleading power jet and soft functions in \Eq{eq:tau_funcs} is easily deduced from the RG evolution of the leading power jet and soft functions. The leading power jet and soft functions satisfy the RG equations
\begin{align} \label{eq:RGES0J0}
\mu \frac{dS^{(0)}_{g}(k,\mu) }{d\mu} 
&=  \int\!\! dk' \: \gamma^S_g(k-k',\mu)\, S^{(0)}_{g}(k',\mu)\,,
\\
\mu \frac{dJ^{(0)}_{g}(s,\mu) }{d\mu}  &=  \int ds'  \gamma^J_{g}(s- s',\mu)\: J^{(0)}_{g}(s',\mu)\,, \nn
\end{align}
where the form of the anomalous dimensions to all orders in $\alpha_s$ is
\begin{align}\label{eq:LP_anom_dim}
 \gamma^S_g(k,\mu) &= 4 \Gamma_{\text{cusp}}^g[\alpha_s] \, \frac{1}{\mu} \biggl[  \frac{\mu\,\theta(k)}{k}  \biggr]_+  + \gamma_g^S[\alpha_s] \: \delta(k) \,,
  \\
  \gamma^J_g(s,\mu) &= -2 \Gamma_{\text{cusp}}^g[\alpha_s] \, \frac{1}{\mu^2} \biggl[  \frac{\mu^2\,\theta(s)}{s}  \biggr]_+  + \gamma_g^J[\alpha_s] \: \delta(s) \,,
   \nn
\end{align}
with $\Gamma^{g}_{\cusp}[\alpha_s]$ the gluon cusp anomalous dimension \cite{Korchemsky:1987wg,Korchemskaya:1992je}.

We can now derive the all orders result for the RG evolution of the subleading power jet and soft functions. Multiplying the leading power soft function by $k$, we find for the soft function
\begin{align}
\mu \frac{d}{d\mu} & k S^{(0)}_{g}(k,\mu) 
=  \int\!\! dk' \: \big((k-k')+k'\big)\gamma^S_g(k-k',\mu)\, S^{(0)}_{g}(k',\mu)\,,  \\
&=\int\!\! dk' \: \big((k-k')+k'\big)  \left\{  4 \Gamma_{\text{cusp}}^g[\alpha_s] \, \frac{1}{\mu} \biggl[  \frac{\mu\,\theta(k-k')}{k-k'}  \biggr]_+  + \gamma_g^S[\alpha_s] \: \delta(k-k')  \right\} S^{(0)}_{g}(k',\mu) \nn \\
&=\int\!\! dk'  4 \Gamma_{\text{cusp}}^g[\alpha_s] \theta(k-k') S^{(0)}_{g}(k',\mu)  + \int\!\! dk' \: \gamma^S_g(k-k',\mu)\, k' S^{(0)}_{g}(k',\mu)\,.\nn 
\end{align}
This implies
\begin{align}\label{eq:tau_mix}
\mu \frac{d}{d\mu}  S^{(2)}_{g,\delta}(k,\mu) &=4\Gamma^g_{\cusp}[\alpha_s] ~   S^{(2)}_{g,\theta}(k,\mu)    + \int\!\! dk' \: \gamma^S_g(k-k',\mu)\, S^{(2)}_{g,\delta}(k',\mu)\,.
\end{align}
Here we have defined the new power suppressed soft function
\begin{align}\label{eq:theta_soft_first}
S^{(2)}_{g,\theta}(k,\mu)&= \frac{1}{(N_c^2-1)} \tr \langle 0 | \cY^T_{\bar n} (0)\cY_n(0) \theta(k-\hat \Tau) \cY_n^T(0) \cY_{\bar n}(0) |0\rangle\,.
\end{align}
We refer to this as a $\theta$-soft function. Its tree level value is $S^{(2)}_{g,\theta}(k,\mu)=\theta(k) +\cO(\alpha_s)$.
This function receives its power suppression from its measurement function, $\theta(k-\hat \Tau)$. In particular, $\theta(\tau)\sim \cO(\tau^0)$, while $\delta(\tau) \sim \cO(1/\tau)$. 

Performing an identical exercise for the jet function, we obtain
\begin{align}\label{eq:tau_jet_mix}
\mu \frac{d}{d\mu}  J^{(2)}_{g,\delta}(s,\mu) 
&=-2\Gamma^g_{\cusp}[\alpha_s] ~   J^{(2)}_{g,\theta}(s,\mu)    + \int\!\! ds' \: \gamma^J_g(s-s',\mu)\, J^{(2)}_{g,\delta}(s',\mu)\,.
\end{align}
Here we have defined the subleading power jet function
\begin{align}\label{eq:theta_op}
J^{(2)}_{g,\theta}(s,\mu)&=\frac{(2\pi)^3}{(N_c^2-1)} \Big\langle 0 \Big|\, \cB^{\mu a}_{n\perp} (0)\,\delta(Q+\bar \cP) \delta^2(\cP_\perp)\, \theta\left(\frac{s}{Q}-\hat \Tau\right)\, \cB^{\mu a}_{n\perp,\omega}(0) \,\Big|0\Big\rangle\,,
\end{align}
which we will refer to as a $\theta$-jet function.  Its tree level value is $J^{(2)}_{g,\theta}(s,\mu)=\theta(s) +\cO(\alpha_s)$. In \cite{Paz:2009ut} it was also found that additional subleading power jet functions whose tree level values were $\theta$-functions were required due to the non-closure of the RG evolution, and it was conjectured that they took the form of \Eq{eq:theta_op}.   Our illustrative example has allowed us to derive the necessity of such operators in a straightforward manner, and prove that here this new function suffices to all orders in $\alpha_s$. More general constraints on the functions that can appear through mixing at subleading power will be derived from the consistency of the RG equations in \Sec{sec:consistency}.

Interestingly, we see that the evolution equation for the power suppressed jet and soft functions are no longer homogeneous evolution equations. In particular, they mix into the $\theta$-jet and $\theta$-soft functions. 
This clearly shows that a new class of subleading power operators, namely the $\theta$-jet and $\theta$-soft operators, are required to renormalize consistently at subleading power in SCET.
These operators do not appear  at amplitude level, but instead arise from mixing at cross section level. It is clear that they have all the correct symmetry properties, as well as the correct power counting, and therefore it is not unexpected that they can be generated by RG evolution.

The renormalization group evolution of the $\theta$-function operators can also be derived by integration of the leading power RG equation. Considering explicitly the soft function, we have
\begin{align}
\mu \frac{d}{d\mu}  S^{(2)}_{g,\theta}(k,\mu) &= \int dk' \theta(k-k') \int dk'' \gamma_g^{S}(k'-k'',\mu) S_g^{(0)}(k'',\mu)  \\
&= \int dk' \gamma_g^{S}(k-k',\mu) \int dk'' \theta(k'-k'')S_g^{(0)}(k'',\mu) \nn \\
&= \int dk' \gamma_g^{S}(k-k',\mu) S_{g,\theta}^{(2)}(k',\mu)\,.\nn
\end{align}
We therefore find that to all orders in $\alpha_s$, the RG for the $\theta$-jet and $\theta$-soft operators is identical to that of the leading power jet and soft functions
\begin{align}\label{eq:theta_RG}
\mu \frac{d}{d\mu}  S^{(2)}_{g,\theta}(k,\mu)& = \int dk' \gamma^S_g(k-k')S^{(2)}_{g,\theta}(k',\mu)\,, \\
\mu \frac{d}{d\mu}  J^{(2)}_{g,\theta}(s,\mu) &= \int ds' \gamma^J_{g}(s- s')J^{(2)}_{g,\theta}(s',\mu)\,.\nn
\end{align}
Together \Eqs{eq:tau_mix}{eq:tau_jet_mix} combined with \Eq{eq:theta_RG} give a simple, closed $2 \times 2$ matrix RG structure for the subleading power jet and soft functions
\begin{align}\label{eq:2by2mix}
\mu \frac{d}{d\mu} \Biggl(\begin{array}{c} J^{(2)}_{g, \delta}(s,\mu) \\[5pt]
  J^{(2)}_{g,\theta}(s,\mu) \end{array} \Biggr) &= \int ds' 
  \Biggl( \begin{array}{cc} 
\gamma^J_{g,\delta\delta}(s-s') \qquad &  \gamma^J_{g,\delta\theta}\, \delta(s-s') \\[5pt]
  0 &  \gamma^J_{g, \theta \theta}(s-s')    
\end{array} \Biggr) 
\Biggl(\begin{array}{c} J^{(2)}_{g,\delta}(s',\mu) \\[5pt] 
J^{(2)}_{g, \theta}(s',\mu) \end{array} \Biggr) \,, \\
\mu \frac{d}{d\mu}\left(\begin{array}{c} S^{(2)}_{g,\delta}(k,\mu) \\ S^{(2)}_{g,\theta}(k,\mu) \end{array} \right) &= \int dk' \left( \begin{array}{cc} \gamma^S_{g,\delta \delta}(k-k',\mu) & \gamma^S_{g,\delta \theta}\, \delta(k-k') \\  0 &  \gamma^S_{g, \theta \theta}(k-k',\mu)   \end{array} \right) \left(\begin{array}{c} S^{(2)}_{g,\tau\delta}(k',\mu) \\ S^{(2)}_{g, \theta}(k',\mu) \end{array} \right) \,. \nn
\end{align}
Fourier transforming to position space 
\begin{align} \label{eq:FT_JS}
\tilde J_x^{(2)}(y) &= \int\!\! ds~ e^{-is y}~  J_x^{(2)}(s) \,,
& \tilde S_x^{(2)}(z) &= \int\!\! dk~ e^{-ik z}~  S_x^{(2)}(k) \,,
\end{align}
(where here the mass dimensions are $[y]=-2$ and $[z]=-1$)
these RG equations become multiplicative
\begin{align}\label{eq:RGyz}
   \mdm \columnspinor{\tilde J^{(2)}_{g,\delta} (y,\mu)}{\tilde J^{(2)}_{g,\theta}(y,\mu)} 
   &= \smallmatrix{\tilde \gamma^J_{g,\delta\delta}(y,\mu)}{ \gamma^J_{g,\delta\theta}[\alpha_s]}{0}{\tilde \gamma^J_{g,\theta\theta}(y,\mu)} \columnspinor{\tilde J^{(2)}_{g,\delta}(y,\mu)}{\tilde J^{(2)}_{g,\theta}(y,\mu)} 
  \,, \\
   \mdm \columnspinor{\tilde S^{(2)}_{g,\delta} (z,\mu)}{\tilde S^{(2)}_{g,\theta}(z,\mu)} 
   &= \smallmatrix{\tilde \gamma^S_{g,\delta\delta}(z,\mu)}{ \gamma^S_{g,\delta\theta}[\alpha_s]}{0}{\tilde \gamma^S_{g,\theta\theta}(z,\mu)} \columnspinor{\tilde S^{(2)}_{g,\delta}(z,\mu)}{\tilde S^{(2)}_{g,\theta}(z,\mu)}
 \,. \nn
\end{align}

For our illustrative example, the RG equations in \Eq{eq:2by2mix} or \Eq{eq:RGyz} are  valid to all orders in $\alpha_s$, and we can identify that
\begin{align}\label{eq:anom_dim_mix_diag}
\gamma^S_{g,\delta \delta}(k, \mu)=  \gamma^S_{g, \theta\theta}(k,\mu)=\gamma_g^S(k,\mu)\,, \\
\gamma^J_{g,\delta \delta}(s, \mu)=  \gamma^J_{g, \theta \theta}(s,\mu)=\gamma_{g}^J(s,\mu) \,, \nn
\end{align}
where $\gamma^S_g(k,\mu)$ and $\gamma_{g}^J(s,\mu)$ are the LP anomalous dimensions in \Eq{eq:LP_anom_dim}. They include the cusp anomalous dimensions, and hence drive double logarithmic evolution. On the other hand, in our illustrative example the off diagonal terms in \Eq{eq:2by2mix} are
\begin{align}\label{eq:anom_dim_mix}
\gamma^S_{g,\delta \theta}&=  4\Gamma^g_{\cusp}[\alpha_s]\,,\\
\gamma^J_{g,\delta \theta}&= -2\Gamma^g_{\cusp}[\alpha_s]\,, \nn
\end{align}
which generate single logarithmic terms.

The particular relations for the anomalous dimensions of \Eqs{eq:anom_dim_mix_diag}{eq:anom_dim_mix}, and in particular the fact that the mixing anomalous dimension is proportional to the cusp anomalous dimension,  is a feature of this specific illustrative example, and will not in general be true. However, the general features of this example will be true at subleading power. In particular, subleading power jet and soft functions will exhibit single logarithmic mixing with $\theta$-function operators, and diagonal anomalous dimensions corresponding to operator self mixing will give rise to double logarithmic evolution.  In \Sec{sec:consistency} we will discuss more general constraints on the subleading power anomalous dimensions and the types of functions which can arise through mixing, from RG consistency constraints in SCET.

From this example we have shown how subleading power jet and soft functions involving $\theta$-function measurement operators arise in a straightforward manner,  we have derived their field structure to all orders in $\alpha_s$, and we have shown that their RG closes in a $2 \times 2$ form.  Before solving this subleading power RG equation, it is also useful to see how this mixing appears from the perspective of a fixed order calculation for the subleading power soft function. This will illustrate that this phenomenon of mixing is  generic at subleading power, due to the fact that subleading power corrections first contribute with a real emission without virtual corrections, and is not simply a feature of the specific example considered here.

At lowest order, the power suppressed soft function vanishes
\begin{align}
\left.S^{(2)}_{g,\delta} (k,\mu)\right|_{\cO(\alpha_s^0)}=\fd{2cm}{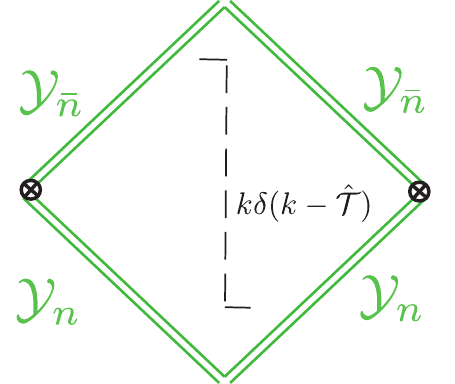}=k\delta(k)=0\,.
\end{align}
With a single emission, we have
\begin{align}
\left.S^{(2)}_{g,\delta} (k,\mu)\right|_{\cO(\alpha_s)}
  =2\fd{2cm}{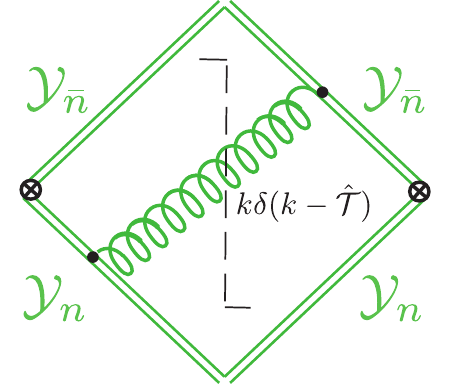}
  =4g^2  \left(  \frac{\mu^2 e^{\gamma_E}}{4\pi} \right)^\epsilon \! C_A\! \int \frac{d^d\ell}{(2\pi)^d} \frac{1}{\ell^+\ell^-} 2\pi\delta(\ell^2) \theta(\ell^0) k \delta(k-Q\hat \tau)\,,
\end{align}
where the measurement function on a single particle state is given by
\begin{align}
k \delta (k -Q\hat \tau)& =k \delta \left(  k -\ell^+\right) \theta(\ell^- -\ell^+) +k \delta \left(  k -\ell^-\right) \theta(\ell^+ -\ell^-)\nn \\
&=2k \delta \left( k -\ell^+ \right) \theta(\ell^- -\ell^+)\,,
\end{align}
using the $\ell^+\leftrightarrow \ell^-$ symmetry of this particular integrand.
Using the delta functions to perform the integrals of the $l_\perp$ and $l^+$, we find
\begin{align}\label{eq:NLP_soft_toy}
S^{(2)}_{g,\delta} (k,\mu)&=\frac{8\alpha_s C_A k^{-\epsilon}}{\Gamma(1-\epsilon)(4\pi)^{1-\epsilon}} \left(  \frac{\mu^2 e^{\gamma_E}}{4\pi} \right)^\epsilon   \int\limits_{k}^\infty \frac{d\ell^-}{2\pi} \frac{1}{(\ell^-)^{1+\epsilon}} =\frac{8\alpha_s C_A e^{\epsilon \gamma_E}}{\Gamma(1-\epsilon)(4\pi)} \left(  \frac{\mu^2 }{k^2} \right)^\epsilon \frac{1}{\epsilon}\nn \\
&=8C_A\frac{\alpha_s(\mu)}{4\pi}  \theta(k) \left(   \frac{1}{\epsilon} +  \log   \frac{\mu^2}{k^2} +\cO(\epsilon)\right)\,.
\end{align}
Here we clearly see that an SCET UV divergence from $\ell^-\to \infty$ appears at the first order at which this power suppressed soft function is non-vanishing.

Although we are considering a specific subleading power example, these two calculations illustrate a general phenomenon at subleading power: subleading power jet and soft functions vanish at lowest order since purely virtual corrections are leading power, scaling like $\delta(\tau)$, and they in general have a UV divergence in SCET at the first perturbative order at which they appear. Without the knowledge of the $\theta$-soft and $\theta$-jet operators, this behavior is confusing, since it is not clear what renormalizes this divergence. However, with an understanding of the presence of these $\theta$-function operators, we can now straightforwardly interpret the fixed order calculation of the subleading power soft function in \Eq{eq:NLP_soft_toy} as operator mixing, and immediately read off the anomalous dimension from the $1/\epsilon$ pole in the standard way.
The operator $S^{(2)}_{g,\theta}$ is non-zero at tree level, and simply gives
\begin{align}
\left. S^{(2)}_{g,\theta}(k,\mu)\right|_{\cO(\alpha_s^0)}=\fd{2cm}{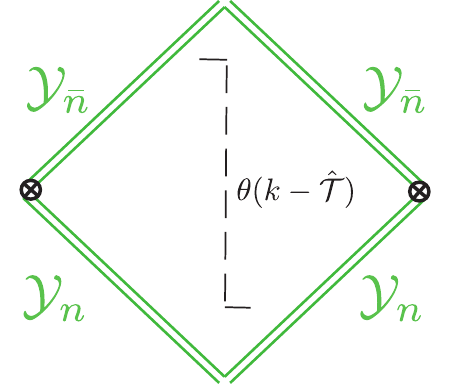} =\theta(k)\,.
\end{align} 
The renormalization of this operator provides the needed counterterm,
and from \Eq{eq:NLP_soft_toy} we find
\begin{align}
  \gamma^S_{g,\delta  \theta}&= 16 \frac{\alpha_s}{4\pi}C_A=  4\frac{\alpha_s}{4\pi}\Gamma^{g,0}_{\cusp}\,,
\end{align}
where $\Gamma^{g,0}_{\cusp}=4C_A$ is the one-loop gluon cusp anomalous dimension. This result is in agreement with our derivation from the known structure of the RG equations in \Eq{eq:tau_mix}. This example clearly resolves any confusion arising in the renormalization of the subleading power operators, which with the addition of subleading power $\theta$-jet and $\theta$-soft functions becomes a standard operator mixing problem.

\subsection{Renormalization Group Consistency}\label{sec:consistency}

Motivated by the structure of the RG equations in our illustrative example, we consider a somewhat more general factorization theorem where the soft and jet sectors have an analogous $2\times2$ mixing structure with some unknown functions that do not appear in the matching, without working under the assumption that these functions take the form of the $\theta$-jet or $\theta$-soft functions of the previous section. The fact that the cross section is $\mu$-independent implies RG consistency equations in SCET that yield relations between the anomalous dimensions of hard, jet, and soft functions, and will allow us to prove on general grounds that the functions appearing through mixing at subleading power must be integrals of the leading power functions in the factorization theorem. This shows that the $\theta$-jet or $\theta$-soft functions appear much more generally than in our illustrative example. It will also allow us to demonstrate that there will always be at least pairs of subleading power $\theta$-soft and $\theta$-collinear functions. 

We consider terms in a subleading power factorization theorem where the power corrections occur in either a jet or soft function with the form
\begin{align}\label{eq:eg_RG_consistency}
 \frac{1}{\sigma_0} \frac{\df\sigma^{(2)}}{\df\tau} 
   &= 2H_1(Q,\mu)\int\!  \frac{ds_n ds_\bn dk}{Q^2}\, \hat \delta_\tau\, J^{(2)}_{\delta}(s_n,\mu )  J^{(0)}(s_\bn,\mu) S^{(0)}(k,\mu)    \\
   &\ \
  + H_2(Q,\mu) \int\! \frac{ds_n ds_\bn dk}{Q} \,  \hat \delta_\tau\, J^{(0)}(s_n,\mu )  J^{(0)}(s_\bn,\mu)   S_{\delta}^{(2)}(k,\mu)  \,,\nn
\end{align}
where we have used the $n\leftrightarrow \bar n$ symmetry to write corrections to the two jet functions into a single expression. Here $H_{1}=1 +{\cal O}(\alpha_s)$ and $H_2=1+{\cal O}(\alpha_s)$ are taken to be dimensionless hard functions. We will assume that these $H_i$ do not mix, so $\mu\frac{d}{d\mu} H_i(Q,\mu)=\gamma_{H_i}(Q,\mu) H_i(Q,\mu)$.  We will also assume that $J^{(2)}_{\delta}$ and $S^{(2)}_{\delta}$ start at ${\cal O}(\alpha_s)$, and obey $2\times2$ mixing equations of the form in \Eq{eq:2by2mix} which has them mix with operators starting at ${\cal O}(\alpha_s^0)$. Importantly, here we do not assume that $J^{(2)}_{\delta}$ and $S^{(2)}_{\delta}$ are related to the functions defined in \Eq{eq:tau_funcs}. We also assume that the terms in \Eq{eq:eg_RG_consistency} close in the renormalization group flow (at least up to some order in the N$^k$LL expansion, though we will shortly focus on LL order). From \Eq{eq:fact_NLP_multTau_rewrite} we see that the expression for the cross section in our illustrative example satisfies all the above assumptions and is a special case of the assumed form. With the above assumptions, our goal is to derive RG consistency equations by demanding the RG invariance of this cross section, $\mu d/d\mu\, d\sigma^{(2)}/d\tau = 0$.

For the analysis of RG consistency it is most convenient to Fourier transform $\tau$ to position space, so that \Eq{eq:eg_RG_consistency} becomes
\begin{align}\label{eq:eg_RG_consistency_y}
 \frac{1}{\sigma_0} \frac{\df\sigma^{(2)}}{\df y}
   &\equiv \int\! d\tau \: e^{-i\tau y}   \frac{1}{\sigma_0} \frac{\df\sigma^{(2)}}{\df\tau}  \\
   &= \frac{2}{Q^2} H_1(Q,\mu) \tilde J^{(2)}_{\delta}\Bigl(\frac{y}{Q^2},\mu \Bigr) \tilde J^{(0)}\Bigl(\frac{y}{Q^2},\mu \Bigr)
   \tilde S^{(0)}\Bigl(\frac{y}{Q},\mu\Bigr)  
   \nn  \\
   &\ \
  + \frac{1}{Q} H_2(Q,\mu) \tilde J^{(0)}\Bigl(\frac{y}{Q^2},\mu \Bigr) \tilde J^{(0)}\Bigl(\frac{y}{Q^2},\mu\Bigr)  \tilde S_{\delta}^{(2)}\Bigl(\frac{y}{Q},\mu\Bigr) 
   \,.\nn
\end{align}
Here $y$ is dimensionless and the Fourier transforms of jet and soft functions are defined as in \Eq{eq:FT_JS}. Differentiating each of the terms in \Eq{eq:eg_RG_consistency_y} and using \Eq{eq:RGES0J0} and the analog of \Eq{eq:RGyz} gives terms involving anomalous dimensions times the same functions back again, plus the terms involving mixing into additional functions. For notational convenience we will refer to these as $\theta$-jet and $\theta$-soft functions, although we will not assume that they take the functional form of the illustrative example result in \Eqs{eq:theta_soft_first}{eq:theta_op}. We therefore arrive at the following consistency equation (here for brevity we suppress the $\mu$ arguments in all functions and anomalous dimensions), 
\begin{align}
0 &= \mu \frac{d}{d\mu} \left[ \frac{d\sigma^{(2)}}{\df y}  \right] 
  \\
 &= \frac{2}{Q^2} \biggl[ \gamma_{H_1}(Q) + \tilde\gamma_{\delta\delta}^J\Bigl(\frac{y}{Q^2}\Bigr) + 
 \tilde\gamma_{J^{(0)}}\Bigl(\frac{y}{Q^2}\Bigr)  +
 \tilde\gamma_{S^{(0)}}\Bigl(\frac{y}{Q}\Bigr) \biggr] 
  H_1(Q) \tilde J^{(2)}_{\delta}\Bigl(\frac{y}{Q^2}\Bigr) \tilde J^{(0)}\Bigl(\frac{y}{Q^2} \Bigr) \tilde S^{(0)}\Bigl(\frac{y}{Q}\Bigr)  
  \nn\\
 &\quad + \frac{2}{Q^2} \gamma_{\delta\theta}^J[\alpha_s] 
  H_1(Q) \tilde J^{(2)}_{\theta}\Bigl(\frac{y}{Q^2}\Bigr) \tilde J^{(0)}\Bigl(\frac{y}{Q^2} \Bigr) \tilde S^{(0)}\Bigl(\frac{y}{Q}\Bigr)  
  \nn\\ 
 & \quad + \frac{1}{Q} \biggl[ \gamma_{H_2}(Q) + 
  2 \tilde\gamma_{J^{(0)}}\Bigl(\frac{y}{Q^2}\Bigr) +
 \tilde\gamma_{\delta\delta}^S\Bigl(\frac{y}{Q}\Bigr) \biggr] 
  H_2(Q) \tilde J^{(0)}\Bigl(\frac{y}{Q^2} \Bigr) \tilde J^{(0)}\Bigl(\frac{y}{Q^2}\Bigr)  \tilde S_{\delta}^{(2)}\Bigl(\frac{y}{Q}\Bigr) 
 \nn\\
 &\quad + \frac{1}{Q} \gamma_{\delta\theta}^{S}[\alpha_s] 
  H_2(Q) \tilde J^{(0)}\Bigl(\frac{y}{Q^2} \Bigr) \tilde J^{(0)}\Bigl(\frac{y}{Q^2}\Bigr)  \tilde S_{\theta}^{(2)}\Bigl(\frac{y}{Q}\Bigr) 
  \,. \nn
\end{align}
Using the relation between anomalous dimensions that follows from the leading power consistency relation, $\gamma_{H^{(0)}}(Q)+2 \tilde \gamma_{J^{(0)}}(y/Q^2)+\tilde \gamma_{S^{(0)}}(y/Q)=0$, dividing by $\bigl[ \tilde J^{(0)}\bigl(\frac{y}{Q^2} \bigr)\bigr]^2  \tilde S^{(0)}\bigl(\frac{y}{Q}\bigr)$, and multiplying by $iy$ simplifies this result to
\begin{align}
 0 &= 2 H_1(Q) \biggl[ \gamma_{H_1}(Q) -\gamma_{H^{(0)}}(Q) + \tilde\gamma_{\delta\delta}^J\Bigl(\frac{y}{Q^2}\Bigr) -
 \tilde\gamma_{J^{(0)}}\Bigl(\frac{y}{Q^2}\Bigr)  \biggr] 
 \,\Biggl[
 \frac{ \frac{iy}{Q^2} \, \tilde J^{(2)}_{\delta}\!\bigl(\frac{y}{Q^2}\bigr) }
 {\tilde J^{(0)}\bigl(\frac{y}{Q^2} \bigr) } \Biggr]
  \nn \\
 & \quad +   H_2(Q)   \biggl[ \gamma_{H_2}(Q) -\gamma_{H^{(0)}}(Q) + 
 \gamma_{\delta\delta}^S\Bigl(\frac{y}{Q}\Bigr) -
 \tilde\gamma_{S^{(0)}}\Bigl(\frac{y}{Q}\Bigr) \biggr] 
 \,\Biggl[ 
  \frac{\frac{iy}{Q}\,  \tilde S_{\delta}^{(2)}\!\bigl(\frac{y}{Q}\bigr) }
  { \tilde S^{(0)}\bigl(\frac{y}{Q}\bigr) } \Biggr] 
 \nn\\
 &\quad + 2 H_1(Q) \, \gamma_{\delta\theta}^J[\alpha_s] \Biggl[
 \frac{\frac{iy}{Q^2}\,\tilde J^{(2)}_{\theta}\!\bigl(\frac{y}{Q^2}\bigr)}
  { \tilde J^{(0)}\bigl(\frac{y}{Q^2} \bigr) } \Biggr]
  + H_2(Q)\,  \gamma_{\delta\theta}^{S}[\alpha_s] \Biggl[
 \frac{\frac{iy}{Q} \, \tilde S_{\theta}^{(2)}\!\Bigl(\frac{y}{Q}\Bigr)}
 { \tilde S^{(0)}\bigl(\frac{y}{Q}\bigr) } \Biggr] 
  \,. 
\end{align}
This consistency equation is quite non-trivial since it involves separate functions of each of $Q$, $y/Q^2$, and $y/Q$. Specializing to LL order we include only the logarithmic terms from the anomalous dimensions in the first two lines, and only the ${\cal O}(\alpha_s)$ terms for the anomalous dimensions in the last line. This gives
\begin{align} \label{eq:LLconsistency}
 0 &= \Biggl[
 \frac{ \frac{iy}{Q^2} \, \tilde J^{(2)}_{\delta}\!\bigl(\frac{y}{Q^2},\mu\bigr) }
 {\frac{\alpha_s(\mu)}{4\pi}\, \tilde J^{(0)}\bigl(\frac{y}{Q^2},\mu \bigr) } \Biggr]^{\rm LL}
  \frac{\alpha_s^2(\mu)}{(4\pi)^2} 
 \biggl\{ 2 \bigl(\Gamma_{H_1}^0 -\Gamma_{H^{(0)}}^0 \bigr) \log\frac{\mu^2}{Q^2} + 
 2 \bigl( \Gamma_{\delta\delta}^{J0} -  \Gamma^0_{J^{(0)}} \bigr) \log\frac{iy\mu^2}{Q^2}  \biggr\}
  \nn \\
 & +  \biggl[ \frac{ H_2(Q,\mu) }{ H_1(Q,\mu) }\biggr]^{\rm LL} \Biggl[ 
  \frac{\frac{iy}{Q}\,  \tilde S_{\delta}^{(2)}\!\bigl(\frac{y}{Q},\mu\bigr) }
  {\frac{\alpha_s(\mu)}{4\pi}\, \tilde S^{(0)}\bigl(\frac{y}{Q},\mu\bigr) } \Biggr]^{\rm LL} 
 \frac{\alpha_s^2(\mu)}{(4\pi)^2} \biggl\{ \bigl(\Gamma_{H_2}^0 -\Gamma^0_{H^{(0)}} \bigr) \log\frac{\mu^2}{Q^2} + 
 \bigl( \Gamma_{\delta\delta}^{S0} - \Gamma^0_{S^{(0)}} \bigr) 
   \log\frac{i y \mu}{Q} \biggr\}
 \nn\\
 &+   \Biggl[
 \frac{\frac{iy}{Q^2}\,\tilde J^{(2)}_{\theta}\!\bigl(\frac{y}{Q^2},\mu\bigr)}
  { \tilde J^{(0)}\bigl(\frac{y}{Q^2},\mu\bigr) } \Biggr]^{\rm LL} 
  \, 2 \frac{\alpha_s(\mu)}{4\pi}\, \gamma_{\delta\theta}^{J0} 
  +  \biggl[\frac{ H_2(Q,\mu) }{ H_1(Q,\mu) }\biggr]^{\rm LL}  \Biggl[
 \frac{\frac{iy}{Q} \, \tilde S_{\theta}^{(2)}\!\Bigl(\frac{y}{Q},\mu\Bigr)}
 { \tilde S^{(0)}\bigl(\frac{y}{Q},\mu\bigr) } \Biggr]^{\rm LL} 
 \, \frac{\alpha_s(\mu)}{4\pi}\, \gamma_{\delta\theta}^{S0} 
  \,, 
\end{align}
where we have restored the $\mu$ arguments. The $0$ superscripts on the anomalous dimensions here indicate that these are the lowest order term in these anomalous dimensions (which are simple numbers). In the first two lines we have included a $1/\alpha_s(\mu)$ since $\tilde J_\delta^{(2)}$ and $\tilde S_\delta^{(2)}$ themselves start at ${\cal O}(\alpha_s)$. This way all terms in square brackets in \Eq{eq:LLconsistency} start at ${\cal O}(\alpha_s^0)$.  Since $\mu$ is  arbitrary, all ratios of hard, jet, and soft functions in square brackets in \Eq{eq:LLconsistency} can each be thought of as a LL series, $\big[\cdots\big]^{\rm LL} =  \sum_{k=0}^\infty a_k [\alpha_s(\mu) \log^2(X)]^k$, where $X=\mu^2/Q^2$, $X=y\mu^2/Q^2$, or $X=y\mu/Q$ for ratios of hard, jet, or soft functions respectively (or the analogs with running coupling effects which does not change the arguments below). The coefficients $a_k$ in these series are numbers that depend on powers of the corresponding anomalous dimensions for the objects in that square bracket. 

To see what \Eq{eq:LLconsistency} implies, first consider the ratio of jet functions in the first line. In the case of our illustrative example from \Sec{sec:illexample} we have $\tilde J_\delta^{(2)}/J^{(0)} \propto d/d(y/Q^2) \, \log \tilde J^{(0)}$, so it is safe to assume that this ratio of jet functions is a non-trivial function of $y/Q^2$. The first line of \Eq{eq:LLconsistency} can then not cancel against the terms in the second line since they have different functional dependence on $y$ and $\mu/Q$. Nor can it cancel against the terms on the third line, since they start at different orders in $\alpha_s$.  This implies that the curly bracket on the first line of \Eq{eq:LLconsistency} vanishes. Due to the presence of two independent types of logarithms in this bracket this immediately implies relations between the cusp anomalous dimension coefficients for these functions at LL order:
\begin{align} \label{eq:consistencyJ}
  \Gamma_{H_1}^0 &= \Gamma_{H^{(0)}} \,, 
 & \Gamma_{\delta\delta}^{J0} &=  \Gamma_{J^{(0)}} \,.
\end{align}
For the same reason the curly bracket on the second line of \Eq{eq:LLconsistency} must also vanish, which then implies the following LL anomalous dimension relations:
\begin{align} \label{eq:consistencyS}
  \Gamma_{H_2}^0 &= \Gamma_{H^{(0)}} \,, 
 & \Gamma_{\delta\delta}^{S0} &=  \Gamma_{S^{(0)}} \,.
\end{align}
Together these imply that $\Gamma_{H_1}^0=\Gamma_{H_2}^0$, which gives $[H_2(Q,\mu)/H_1(Q,\mu)]^{\rm LL}=1$. 

In \Eq{eq:LLconsistency} this then leaves only the LL mixing terms, where the remaining constraint now takes the form
\begin{align}
 0=  \Biggl[
 \frac{\frac{iy}{Q^2}\,\tilde J^{(2)}_{\theta}\!\bigl(\frac{y}{Q^2},\mu\bigr)}
  { \tilde J^{(0)}\bigl(\frac{y}{Q^2},\mu\bigr) } \Biggr]^{\rm LL} 
  \, 2 \, \gamma_{\delta\theta}^{J0} 
  +   \Biggl[
 \frac{\frac{iy}{Q} \, \tilde S_{\theta}^{(2)}\!\Bigl(\frac{y}{Q},\mu\Bigr)}
 { \tilde S^{(0)}\bigl(\frac{y}{Q},\mu\bigr) } \Biggr]^{\rm LL} 
 \,  \gamma_{\delta\theta}^{S0} 
 \,.
\end{align}
In our illustrative example the two square brackets here are both equal to $1$.  The RG consistency implies that this is actually a much more general result, true for any operators satisfying the assumptions set out at the beginning of this section. In particular, since the two square brackets have different functional dependence, $y/Q^2$ and $y/Q$ respectively, they must both be independent of these variables. This gives:\footnote{More generally the RHS of the results in \Eq{eq:thetaconsistency} could be constants, but we choose to normalize $\tilde J^{(2)}_{\theta}$ and $\tilde S_{\theta}^{(2)}$ so these constants are both $1$.}
\begin{align} \label{eq:thetaconsistency}
 & \Biggl[
 \frac{\frac{iy}{Q^2}\,\tilde J^{(2)}_{\theta}\!\bigl(\frac{y}{Q^2},\mu\bigr)}
  { \tilde J^{(0)}\bigl(\frac{y}{Q^2},\mu\bigr) } \Biggr]^{\rm LL} = 1 \,,
 & \Biggl[
 \frac{\frac{iy}{Q} \, \tilde S_{\theta}^{(2)}\!\Bigl(\frac{y}{Q},\mu\Bigr)}
 { \tilde S^{(0)}\bigl(\frac{y}{Q},\mu\bigr) } \Biggr]^{\rm LL} & =1 \,.
\end{align}
This then leaves a simple relation between the mixing anomalous dimensions
\begin{align} \label{eq:gamthetaconsitency}
  2\gamma_{\delta \theta}^{J0} + \gamma_{\delta\theta}^{S0} = 0 \,,
\end{align}
which we also found in our illustrative example. 
In momentum space \Eq{eq:thetaconsistency} implies that
\begin{align} \label{eq:JSthetaconstraint}
  J_\theta^{(2)}(s,\mu)^{\rm LL} &= \int_0^s \! ds'\, J^{(0)}(s',\mu)^{\rm LL} 
   \,,
 & S_\theta^{(2)}(k,\mu)^{\rm LL} &= \int_0^k \! dk'\, S^{(0)}(k',\mu)^{\rm LL} 
   \,.
\end{align}
While true in our illustrative example, viewed as a more general constraint  this result is quite interesting. For more general operators defining $J_\delta^{(2)}$ and $S_\delta^{(2)}$ it might not be a priori clear (without performing higher order loop and gluon emission calculations) how to define the operators giving the $J_\theta^{(2)}$ and $S_\theta^{(2)}$ that one mixes into. The RG consistency result in \Eq{eq:JSthetaconstraint}  implies that the required $J_\theta^{(2)}$ and $S_\theta^{(2)}$ functions agree with those defined from the cumulative of the leading power operators, at least at LL order.   The RG consistency results in \Eqs{eq:consistencyJ}{eq:consistencyS} furthermore imply that the LL cusp anomalous dimensions of $J_\delta^{(2)}$ and $S_\delta^{(2)}$ are the same as those for the jet and soft functions at leading power. Note that although $\gamma_{\theta\theta}^J$ or $\gamma_{\theta\theta}^S$ do not appear explicitly in the RG consistency equation, they are present in the LL expressions for $\tilde J^{(2)}_{\theta}$ and $\tilde S^{(2)}_{\theta}$ and hence are constrained by \Eq{eq:thetaconsistency}.

This example also illustrates another important point. There must always be (at least) a pair of functions at subleading power whose renormalization group evolution is tied by consistency. This is also clear from the fact that when evaluated at their natural scales, the subleading power $J_\delta^{(2)}$ and $S_\delta^{(2)}$ functions are $0+\cO(\alpha_s)$, and not $\delta(\tau)+\cO(\alpha_s)$ as at leading power. Thus if one chooses to run all functions to the canonical scale of  either of the subleading power functions, this function will simply not contribute at LL accuracy. To see this explicitly, we can use the evolution equations to run all functions in the position space factorization theorem from their canonical scales $\mu_H^2\sim Q^2$, $\mu_J^2\sim Q^2/iy$, or $\mu_S^2\sim Q^2/(iy)^2$ to a common scale $\mu^2$.  This gives
\begin{align}
  \frac{1}{\sigma_0} \frac{\df\sigma^{(2)}}{\df y}
  &= \frac{2}{Q^2} H_1(Q,\mu_H) U_{H_1}(Q,\mu_H,\mu) 
  U_{J}^{(0)}\Bigl(\frac{y}{Q^2},\mu_J,\mu\Bigr) 
  \tilde J^{(0)}\Bigl(\frac{y}{Q^2},\mu_J \Bigr)
  U_{S}^{(0)}\Bigl(\frac{y}{Q},\mu_S,\mu\Bigr) 
 \nn\\
 &\quad\ \times
 \tilde S^{(0)}\Bigl(\frac{y}{Q},\mu_S\Bigr)  
 \biggl[ U_{\delta\delta}^J\Bigl(\frac{y}{Q^2},\mu_J,\mu\Bigr) \tilde J^{(2)}_{\delta}\Bigl(\frac{y}{Q^2},\mu_J \Bigr) +
  U_{\delta\theta}^J\Bigl(\frac{y}{Q^2},\mu_J,\mu\Bigr) \tilde J^{(2)}_{\theta}\Bigl(\frac{y}{Q^2},\mu_J \Bigr) \biggr]
   \nn  \\
   &\ \
  + \frac{1}{Q} H_2(Q,\mu) U_{H_2}(Q,\mu_H,\mu) 
  \Bigl[ U_J^{(0)}\Bigl(\frac{y}{Q^2},\mu_J,\mu \Bigr) \tilde J^{(0)}\Bigl(\frac{y}{Q^2},\mu_J \Bigr) \Bigr]^2 
  \nn\\
  & \quad\ \times 
  \biggl[ U_{\delta\delta}^{S}\Bigl(\frac{y}{Q},\mu_S,\mu\Bigr) 
   \tilde S_{\delta}^{(2)}\Bigl(\frac{y}{Q},\mu_S\Bigr)  + 
 U_{\delta\theta}^{S}\Bigl(\frac{y}{Q},\mu_S,\mu\Bigr) 
   \tilde S_{\theta}^{(2)}\Bigl(\frac{y}{Q},\mu_S\Bigr)  \biggr]
  \,.
\end{align}
Here the $U_H$, $U_S$ and $U_J$ factors are evolution kernels for the various hard, jet, and soft functions. For our analysis of $H\to gg$ in pure glue QCD  their explicit form will be given later in the text.
At LL order we can then use that
\begin{align}
 \tilde J^{(2)}_{\delta}(y/Q^2,\mu_J )  &=0+\cO(\alpha_s)\,,
 & \tilde S^{(2)}_{\delta}(y/Q,\mu_S ) &=0+\cO(\alpha_s)\,,
\end{align}
which implies that the terms with the $U_{\delta\delta}^J$ and $U_{\delta\delta}^S$ kernels are not needed at this order. We can also simplify the LL result by using $\tilde S^{(0)} = 1$ and $\tilde J^{(0)}=1$ (we allow here a non-trivial overall numeric factor from $H_1$ and $H_2$ at tree level). 
The LL resummed result then simplifies to 
\begin{align} \label{eq:LLresum_arb_mu}
  \frac{1}{\sigma_0} \frac{\df\sigma^{(2)\,{\rm LL}}}{\df y}
  &= \frac{2 H_1}{Q^2} U_{H_1}(Q,\mu_H,\mu) 
  U_{J}^{(0)}\Bigl(\frac{y}{Q^2},\mu_J,\mu\Bigr) 
  U_{S}^{(0)}\Bigl(\frac{y}{Q},\mu_S,\mu\Bigr) 
 U_{\delta\theta}^J\Bigl(\frac{y}{Q^2},\mu_J,\mu \Bigr) \tilde J^{(2)}_{\theta}\Bigl(\frac{y}{Q^2},\mu_J \Bigr) 
   \nn  \\
   &\ \
  + \frac{H_2}{Q}  U_{H_2}(Q,\mu_H,\mu) 
  \Bigl[ U_J^{(0)}\Bigl(\frac{y}{Q^2},\mu_J,\mu \Bigr)  \Bigr]^2   
  U_{\delta\theta}^{S}\Bigl(\frac{y}{Q},\mu_S,\mu\Bigr) 
   \tilde S_{\theta}^{(2)}\Bigl(\frac{y}{Q},\mu_S\Bigr)  
  \,.
\end{align}
Finally we can use the RG consistency freedom that says the same result is obtained no matter what value we pick for $\mu$. For example, taking $\mu=\mu_J$ we have $U_{\delta \theta}^J(y/Q^2,\mu_J,\mu_J)=0$ which removes the first term, and $U_{J}^{(0)}(y/Q^2,\mu_J,\mu_J)=1$ which simplifies the second, leaving
\begin{align}  \label{eq:LLresum_pick_muJ}
  \frac{1}{\sigma_0} \frac{\df\sigma^{(2)\,{\rm LL}}}{\df y}
  &=  \frac{H_2}{Q}  U_{H_2}(Q,\mu_H,\mu_J)  
  U_{\delta\theta}^{S}\Bigl(\frac{y}{Q},\mu_S,\mu_J\Bigr) 
   \tilde S_{\theta}^{(2)}\Bigl(\frac{y}{Q},\mu_S\Bigr)  
  \,.
\end{align}
In this form the  LL resummed result is obtained completely from the subleading power soft function. If instead we had chosen $\mu=\mu_S$, then $U_{\delta\theta}^S(y/Q,\mu_S,\mu_S)=0$ would have removed the second term in \Eq{eq:LLresum_arb_mu}, and the result would have been expressed entirely from the first term that involves the subleading power jet functions, which can be simplified using $U_S^{(0)}(y/Q,\mu_S,\mu_S)=1$.  This equivalence between different resummed formula is an expression of the LL consistency result in \Eq{eq:gamthetaconsitency} at the level of the cross section. We will use \Eq{eq:LLresum_pick_muJ} to simplify the resummation for thrust at next-to-leading power in \Sec{sec:fact}.

\section{Solution to the Subleading Power RG Mixing Equation}\label{sec:solution}

Having illustrated that the renormalization of subleading power jet and soft functions generically involves mixing with $\theta$-jet and $\theta$-soft operators, in this section we solve a general form of the subleading power RG equations involving mixing, including the running coupling $\alpha_s(\mu)$. This solution will be sufficient for all cases required in this chapter, and we believe that it will be of general utility for subleading power resummation.

We consider a function, $F$, which obeys an RG equation of the form of \Eq{eq:2by2mix}.  To remove the convolution structure, we work in Fourier (or Laplace) space, with a variable $y$ conjugate to a momentum variable $k$ of dimension $p$. Defining
\begin{align}
\tilde F(y)= \int dk~ e^{-ik y}~ F(k),
\end{align}
the RG equation for $\tilde F$ is then multiplicative
\be\label{eq:RG_to_solve}
	 \mdm \columnspinor{\tilde F^{(2)}_{\delta} (y,\mu)}{\tilde F^{(2)}_{\theta}(y,\mu)} = \smallmatrix{\tilde \gamma_{11}(y,\mu)}{ \gamma_{12}[\alpha_s]}{0}{\tilde \gamma_{22}(y,\mu)} \columnspinor{\tilde F^{(2)}_{\delta}(y,\mu)}{\tilde F^{(2)}_{\theta}(y,\mu)} \,.
\ee
Here, to simplify notation, we have defined
\begin{align}
\tilde \gamma_{11} (y,\mu) &=\Gamma_{11}[\alpha_s] \log \bigl(ie^{\gamma_E} (y-i0)\mu^p  \bigr)+\gamma_{11}[\alpha_s]\,, \\
\tilde \gamma_{22} (y,\mu) &=\Gamma_{22}[\alpha_s] \log \bigl(ie^{\gamma_E} (y-i0)\mu^p \bigr) +\gamma_{22}[\alpha_s]\,.\nn
\end{align}
To shorten the equations, we will not explicitly write the branch cut prescription in the following. The off-diagonal mixing term, $\gamma_{12}[\alpha_s]$, does not contain logarithms.

\subsection{General Solution}

We will solve the subleading power mixing equation without the constraint that $\tilde \gamma_{11}=\tilde\gamma_{22}$, as occurred in the example of \Sec{sec:renorm_a}. We do this both because we believe that this solution will be relevant for the renormalization of more general functions at subleading power, as well as to illustrate how the standard leading power Sudakov exponential arises as a special limit when $\tilde \gamma_{11}= \tilde \gamma_{22}$, but not more generally.

We can write the all orders solution to the differential equation of \Eq{eq:RG_to_solve} as
\begin{align}\label{eq:solution_gen}
\tilde F^{(2)}_{ \delta} (y,\mu)
 = U_{\delta\delta}(y,\mu,\mu_0)\,  \tilde F^{(2)}_{\delta}(y,\mu_0) 
  + U_{\delta\theta}(y,\mu,\mu_0)\, \tilde F^{(2)}_{\theta}(y,\mu_0)  \,,
\end{align}
with
\begin{align}\label{eq:solution_gen_b}
 U_{\delta\delta}(y,\mu,\mu_0) &= \exp\biggl[\: \int\limits_{\mu_0}^\mu \dmmp \,\tilde \gamma_{11}(y,\mu') \biggr] \,,
 & U_{\delta\theta}(y,\mu,\mu_0) &= U_{\delta\delta}(y,\mu,\mu_0) X(y,\mu, \mu_0) \,,
\end{align}
where $X$ satisfies
\begin{align}
\mdm X(y,\mu, \mu_0) =e^{-\int\limits_{\mu_0}^\mu \dmmp \tilde \gamma_{11}(y,\mu') }  \gamma_{12}[\alpha_s(\mu)] \,   e^{~\int\limits_{\mu_0}^\mu \dmmp \tilde \gamma_{22}(y,\mu')} \,,
\end{align}
and the boundary condition $X(y,\mu_0,\mu_0)=0$. 
Solving for $X$, we have
\begin{align}\label{eq:solution_X_first}
X(y,\mu, \mu_0) &=\int\limits_{\mu_0}^\mu \frac{d\mu^{''}}{\mu^{''}} e^{-\int\limits_{\mu_0}^{\mu''} \dmmp \tilde \gamma_{11}(y,\mu') }  \gamma_{12}[\alpha_s(\mu '')] \,   e^{~\int\limits_{\mu_0}^{\mu''} \dmmp \tilde \gamma_{22}(y,\mu')} \\
&=  \int\limits_{\mu_0}^\mu \frac{d\mu^{''}}{\mu^{''}} \: \gamma_{12}  [\alpha_s(\mu'')]\: \exp\Biggl( {-\int\limits_{\mu_0}^{\mu''} \dmmp [ \tilde \gamma_{11}(y,\mu') -\tilde \gamma_{22}(y,\mu') ] }  \Biggr)
 \,.\nn
\end{align}
We can derive a closed analytic form for $X$ order by order in the anomalous dimensions, including the running coupling. For the remainder of this section we consider the solution at LL order, where the anomalous dimensions take the form 
\be\label{eq:LLapproxconditions}
\tilde \gamma_{ii}(y,\mu) 
  =\Gamma_{ii}^0 \, \frac{\alpha_s(\mu)}{4\pi}\, \log \left (\frac{\mu^p}{\mu_y^p} \right ) 
 \,,\qquad \gamma_{12}[\alpha_s]
  =\gamma_{12}^0\, \frac{\alpha_s(\mu)}{4\pi} \,,
\ee
where $\Gamma^0_{11}$, $\Gamma^0_{22}$, $\gamma^0_{12}$ are numbers, and  we have defined the mass dimension $1$ variable $\mu_y$ by
\be\label{eq:muydef}
	\frac{1}{\mu^p_y} \equiv e^{\gamma_E}i(y-i0) \,.
\ee
Note that at LL order we need only the logarithmic term for the diagonal anomalous dimensions $\tilde\gamma_{11}(y,\mu)$ and $\tilde\gamma_{22}(y,\mu)$. The non-logarithmic term is needed for the off-diagonal term $\gamma_{12}[\alpha_s]$ because of the fact that the boundary terms in \Eq{eq:solution_gen} start at different orders, $\tilde F^{(2)}_{\delta}(y,\mu_0)\sim {\cal O}(\alpha_s)$ and $\tilde F^{(2)}_{\theta}(y,\mu_0) \sim {\cal O}(\alpha_s^0)$. 

To include the effects of running coupling, we use the standard approach of switching to an integration in $\alpha_s$ instead of $\mu$ through the change of variables
\begin{align}
	\frac{d\mu}{\mu}= \frac{d\alpha_s}{\beta[\alpha_s]}\,.
\end{align}
At LL-order, we can use the LL $\beta$ function which gives
\begin{align}\label{eq:changeofvariable}
	\frac{d\mu}{\mu}= -\frac{2\pi}{\beta_0}\frac{d\alpha_s}{\alpha_s^2}\,, \qquad \beta_0 = \frac{11}{3} C_A - \frac{4}{3}T_F n_f\,,
\end{align}
We also rewrite the logarithm appearing in the anomalous dimension as
\be\label{eq:logmuasalpha}
	\log\Bigl( \frac{\mu}{\mu_y}\Bigr) = -\frac{2\pi}{\beta_0}\int_{\alpha_s(\mu_y)}^{\alpha_s(\mu)} \daap = \frac{2\pi}{\beta_0}\left(\frac{1}{\alpha_s(\mu)}-\frac{1}{\alpha_s(\mu_y)}\right) = \frac{2\pi}{\beta_0 \alpha_s(\mu_y)}\left(\frac{\alpha_s(\mu_y)}{\alpha_s(\mu)}-1\right)\,.
\ee
We then have
\begin{align}  \label{eq:Uddresult}
	U_{\delta\delta}(y,\mu,\mu_0) &= \exp\Biggl\{~\Gamma_{11}^0\int\limits_{\mu_0}^\mu \dmmp \left(\frac{\alpha_s(\mu')}{4\pi}\right)\log \left (\frac{\mu^{'p}}{\mu^p_y} \right )\Biggr\}  \\
	&=\exp\left[\frac{p\pi\Gamma^0_{11}}{\beta_0^2\alpha_s(\mu_0)} \left(\frac{1}{r} - 1 + \log r \right) \right]\left(\frac{\mu^p_y}{\mu^p_0}\right)^{\frac{\Gamma^0_{11}}{2\beta_0}\log(r) }\,, \nn
\end{align}
where 
\be\label{eq:rdefinition}
	r\equiv\frac{\alpha_s(\mu)}{\alpha_s(\mu_0)} \,,
\ee 
and at this order we take the boundary conditions
 \begin{align}
 \tilde F^{(2)}_{\delta}(y,\mu_0)=0\,, \qquad \tilde F^{(2)}_{\theta}(y,\mu_0)=\frac{1}{i(y-i0)}\,.
 \end{align}
Recall that $1/i(y-i0)$ is the Fourier transform of $\theta(k)$.
Thus at LL the solution becomes
\begin{align}\label{eq:solution_gen_LL}
	\tilde F^{(2) \LL}_{ \delta} (y,\mu) = U_{\delta\theta}(y,\mu,\mu_0)  \, \frac{1}{i(y-i0)} \,,
\end{align}
with the evolution kernel given by 
\begin{align} \label{eq:ULLrunning}
	U_{\delta\theta}^{\LL}(y,\mu,\mu_0) &=\exp\left[\frac{p\pi\Gamma^0_{11}}{\beta_0^2\alpha_s(\mu_0)} \left(\frac{1}{r} - 1 + \log r \right) \right]\left(\frac{\mu^p_y}{\mu^p_0}\right)^{\frac{\Gamma^0_{11}}{2\beta_0}\log(r) }X^{\LL}(y,\mu, \mu_0) \,.
\end{align}
Using \Eqs{eq:changeofvariable}{eq:logmuasalpha} we can compute $X(y,\mu, \mu_0)$ in terms of the running coupling as
\begin{align}\label{eq:Xmostgeneral}
	\!X(y,\mu, \mu_0) &= -\frac{\gamma^0_{12}}{2\beta_0} \int_{\alpha_s(\mu_0)}^{\alpha_s(\mu)} \frac{\df \alpha_s'}{\alpha_s'} \exp\left\{\!\frac{p \pi }{\beta_0^2}\Bigl(\Gamma^0_{11} -\Gamma^0_{22}\Bigr)\! \int_{\alpha_s(\mu_0)}^{\alpha_s'} \frac{\df \alpha_s''}{\alpha_s''}\left(\frac{1}{\alpha_s''}-\frac{1}{\alpha_s(\mu_y)}\right) \!\right\} 
     \nn \\
 &= -\frac{\gamma^0_{12}}{2\beta_0} \int_{\alpha_s(\mu_0)}^{\alpha_s(\mu)} \frac{\df \alpha_s'}{\alpha_s'} \exp\left\{ \frac{p \pi}{\beta_0^2}\left(\Gamma^0_{11} -\Gamma^0_{22}\right) \left[\frac{1}{\alpha_s(\mu_0)} - \frac{1}{\alpha_s'} - \frac{1}{\alpha_s(\mu_y)} \log\frac{\alpha_s'}{\alpha_s(\mu_0)} \right]\right\}
     \nn \\
 &= -\frac{\gamma^0_{12}}{2\beta_0} \int_{\phi(\mu)}^{\phi(\mu_0)} 
  \frac{\df \phi'}{\phi'} \exp\left\{ \phi(\mu_0) - \phi' -\phi(\mu_y) \log\frac{\phi(\mu_0)}{\phi'} \right\}
  \,,
\end{align}
where in the last line we used the definition
\begin{align}
	\phi(\mu)\equiv \frac{p\pi(\Gamma^0_{11}-\Gamma^0_{22})}{\beta_0^2\,\alpha_s(\mu)}\,.
\end{align}
The final integral gives the LL solution
\begin{align} \label{eq:XLLresult}
	X^\LL(y,\mu, \mu_0)
	&=-\frac{\gamma^0_{12}}{2\beta_0} e^{\phi(\mu_0)}\left[ r^{-\phi(\mu_y)} E\Bigl(1-\phi(\mu_y),\phi(\mu)\Bigr)-E\Bigl(1-\phi(\mu_y),\phi(\mu_0)\Bigr) \right] ,
\end{align}
where $E(n,z)$ is the exponential integral function 
\begin{align}
 E(n,z)=\int_1^{\infty }  \! \frac{dt}{t^n} \: e^{-zt}\,.
\end{align}
Plugging these results into \Eq{eq:solution_gen} we obtain the general solution to the subleading RG at LL order in terms of the results in \Eqs{eq:Uddresult}{eq:XLLresult}:
\begin{align} \label{eq:solution_LL}
\tilde F^{(2)}_{ \delta} (y,\mu)^{\rm LL}
 =  U_{\delta\delta}^{\rm LL}(y,\mu,\mu_0)\, X^{\rm LL}(y,\mu,\mu_0)\,
     \frac{1}{i(y-i0)}  \,.
\end{align}

For illustration we can take the limit 
without the running coupling, set $\mu_0 = \mu_y$, and assume\footnote{Note that we made no assumption on the signs of the $\Gamma_{11}^0$ and $\Gamma_{22}^0$ which can be negative. If $\Gamma_{11}^0<\Gamma_{22}^0$, the result involves an imaginary error function ($\Erfi$) instead of the error function ($\Erf$).} $\Gamma_{11}^0>\Gamma_{22}^0$ which gives  
\begin{align}\label{eq:ULLfixedalphas}
	U_{\delta\theta}^{LL}(y,\mu,\mu_y)\biggr|_{\alpha_s(\mu) = \alpha_s}
   &
      =\gamma^0_{12} \,\frac{\alpha_s}{8\pi} \,
    \sqrt{\frac{\pi}{\Delta_\Gamma}}\,\Erf\biggl[ 
    \sqrt{\Delta_{\Gamma}} \,  \log\frac{\mu}{\mu_y} \biggr]\,
    \exp\left[ p \, \Gamma^0_{11}\, \frac{\alpha_s}{8\pi}\, \log^2\frac{\mu}{\mu_y}  \right]  ,
\end{align}
where $\Delta_\Gamma \equiv\left(\frac{\alpha_s}{8\pi}\right)p\left(\Gamma^0_{11}-\Gamma^0_{22}\right)$ and Erf is the error function, $\Erf(x)=(2/\sqrt{\pi}) \int_0^x e^{-t^2} dt$ which expanded around $x=0$ reads $\Erf(x)=2x/\sqrt{\pi}-2x^3/(3 \sqrt{\pi })+\cO(x^5)$. The kernel in \Eq{eq:ULLfixedalphas} is easily interpreted as the standard Sudakov factor with fixed coupling multiplied by the error function arising from the integral over the difference of Sudakov exponentials in \Eq{eq:solution_X_first}.  The solutions in \Eqs{eq:ULLrunning}{eq:ULLfixedalphas} emphasize that there is a closed form solution in terms of elementary functions, and that in the most general case we will not necessarily get a simple Sudakov exponential at subleading power. We also emphasize that in all the LL results $\gamma^0_{12}$ appears only as an overall factor.

\subsection{Solution With Equal Diagonal Entries}
To gain further insight into the form of the LL solution to the subleading power RG it is instructive to restrict our attention to the case $\Gamma^0_{11}=\Gamma^0_{22}$ which is the relevant one for the subleading soft and jet functions considered in \Sec{sec:renorm}. With $\Gamma^0_{11}=\Gamma^0_{22}$, we have $\phi=0$ so that $X$ simplifies to 
\begin{align}
X^\LL(y,\mu,\mu_0)\big|_{\Gamma^0_{11}
 =\Gamma^0_{22}}&=-\frac{\gamma^0_{12}}{2\beta_0} \int_{\alpha_s(\mu_0)}^{\alpha_s(\mu)} \frac{\df \alpha_s'}{\alpha_s'} = -\frac{\gamma^0_{12}}{2\beta_0}\log r \,,
\end{align}
where $r$ was defined in \Eq{eq:rdefinition} and the evolution kernel simplifies to
\begin{align} \label{eq:ULLrunningsameGamma}
	U_{\delta\theta}^{\LL}(y,\mu,\mu_0)\big|_{\Gamma^0_{11}=\Gamma^0_{22}} &=-\frac{\gamma^0_{12}}{2\beta_0}\, 
   \log r\,\exp\left[\frac{p\pi\Gamma^0_{11}}{\beta_0^2\alpha_s(\mu_0)} \left(\frac{1}{r} - 1 + \log r \right) \right]\left(\frac{\mu^p_y}{\mu^p_0}\right)^{\frac{\Gamma^0_{11}}{2\beta_0}\log(r) } \,.
\end{align}
Therefore with $\Gamma_{11}^0=\Gamma_{22}^0$ we recover a simple Sudakov evolution at LL. For this case the final expression for the LL resummed function in position space is
\be\label{eq:FLL}
	\tilde F^{(2)\LL}_{\delta}(y,\mu)
  = -\frac{\gamma^0_{12}}{2\beta_0}\log r\exp\left[\frac{p\pi\Gamma^0_{11}}{\beta_0^2\alpha_s(\mu_0)} \left(\frac{1}{r} - 1 + \log r \right) \right]\left(\frac{\mu^p_y}{\mu^p_0}\right)^{\frac{\Gamma^0_{11}}{2\beta_0}\log(r) } \frac{1}{i(y-i0)}\,.
\ee

To obtain the expression for $F^{(2)\LL}_{\delta}(k,\mu)$ we  transform \Eq{eq:FLL} back to momentum space  which gives
\begin{align}\label{eq:Fmomspacetext}
	F^{(2)\LL}_{\delta}(k,\mu)
  &= 	U^{\LL}_{\delta\theta}(k,\mu,\mu_0) \, \theta(k)  \,, 
\end{align}
where the evolution kernel is obtained with the simple replacement $\mu^p_y\to k$,
\begin{align}\label{eq:Umomspace}
U^{\LL}_{\delta\theta}(k,\mu,\mu_0) &= - \frac{\gamma^0_{12}}{2\beta_0}\log r\exp\left[\frac{p\pi\Gamma^0_{11}}{\beta_0^2\alpha_s(\mu_0)} \left(\frac{1}{r} - 1 + \log r \right) \right] \left(\frac{k}{\mu^p_0}\right)^{\frac{\Gamma^0_{11}}{2\beta_0}\log(r) }  \,.
\end{align}
Further details about why this simple replacement suffices at LL are given in \App{sec:inversefourier}.

For concreteness, let us now consider the case where the subleading function $F^{(2)}_\delta(k,\mu)$ is the subleading power soft function of \Eq{eq:tau_funcs}. The soft function depends on a momentum variable of dimension $p=1$ and from \Eqs{eq:anom_dim_mix_diag}{eq:anom_dim_mix} we have that for $S^{(2)}_{g,\delta}(k,\mu)$ the anomalous dimensions are\footnote{The minus sign for $\Gamma^0_{11}$ comes from the fact that Laplace transforming \Eq{eq:LP_anom_dim} we have 
\be 
	\frac{1}{\mu} \biggl[  \frac{\mu\,\theta(k)}{k}  \biggr]_+ \to -\log(ye^{\gamma_E} \mu)\,,  \nn
\ee 
therefore giving 
\be
	\underbrace{-4 \Gamma_{\text{cusp}}^{g,0}}_{\Gamma^0_{11}} \frac{\alpha_s}{4\pi}\log(ye^{\gamma_E} \mu)\,. \nn
\ee }
\begin{align}
	\tilde \gamma_{11} (k,\mu) = \tilde \gamma_{22} (k,\mu) &= \gamma^{S}_{g} (k,\mu) &&\implies&& \Gamma^0_{11} = \Gamma^0_{22}= -4 \Gamma^{g,0}_\cusp = -16 C_A\,, \\
	\gamma_{12} [\alpha_s] &= 4\Gamma^g_\cusp [\alpha_s] &&\implies && \gamma_{12}^0 = 4 \Gamma^{g,0}_\cusp = 16 C_A\,. \nn
\end{align}
Using these results in \Eq{eq:Umomspace} we obtain
\begin{align}\label{eq:Smomspacetext}
	S^{(2)\LL}_{g,\delta}(k,\mu) &= -\theta(k) \frac{2 \Gamma^{g,0}_\cusp}{\beta_0}\log r\exp\left[-\frac{4\pi \Gamma^{g,0}_\cusp}{\beta_0^2\alpha_s(\mu_0)} \left(\frac{1}{r} - 1 + \log r \right) \right] \left(\frac{k}{\mu_0}\right)^{\frac{-2 \Gamma^{g,0}_\cusp}{\beta_0}\log(r) }  \,.
\end{align}
We can resum logarithms in the subleading power soft function by running from the canonical scale of the soft function $\mu_0=\mu_S=Q \tau$, to an arbitrary scale $\mu$. Hence,
\begin{align}\label{eq:SresummedQt}
	S^{(2)\LL}_{g,\delta}(Q\tau,\mu) &= -\theta(\tau) \frac{2 \Gamma^{g,0}_\cusp}{\beta_0}\log \left(\frac{\alpha_s(\mu)}{\alpha_s(Q\tau)}\right)   \\
	&\ \ \times\exp\left[-\frac{4\pi \Gamma^{g,0}_\cusp}{\beta_0^2 \alpha_s(Q\tau)} \left(\frac{\alpha_s(Q\tau)}{\alpha_s(\mu)} - 1 + \log  \frac{\alpha_s(\mu)}{\alpha_s(Q\tau)}\right) \right]
	 . \nn
\end{align}
If we ignore the running of the coupling, this simplifies to
\be\label{eq:SLLfixedcoupling}
	 S^{(2)\LL}_{g,\delta}(Q\tau,\mu)\bigg|_{\alpha_s(\mu)=\alpha_s} \!\!\!\!\!\! = \theta(\tau)4 \Gamma^{g,0}_\cusp\left(\frac{ \alpha_s}{4\pi}\right)\log \left(\frac{\mu}{Q\tau}\right)\exp\left[-2 \Gamma^{g,0}_\cusp\left(\frac{\alpha_s}{4\pi}\right)\log^2 \left(\frac{\mu}{Q\tau}\right)\right] ,
\ee
where the physical interpretation is quite clear. Expanding this structure perturbatively in $\alpha_s$, we have
\begin{align}
S^{(2)}_{g, \delta} (Q\tau,\mu)\bigg|_{\alpha_s(\mu)=\alpha_s} \!\!\!\!\!\!= \theta(\tau) \left[ \left(\frac{ \alpha_s}{4\pi}\right)\gamma^0_{12} \log\left( \frac{\mu}{Q \tau} \right) +  \frac{1}{2} \left(\frac{ \alpha_s}{4\pi}\right)^2 \gamma^0_{12} \Gamma^0_{11} \log^3\left( \frac{\mu}{Q\tau} \right) +\cdots    \right] .
\end{align}
We see that the first single logarithm is generated by the mixing into the $\theta$-function operators, and then this is dressed by a double logarithmic Sudakov that is driven by the diagonal entries in the mixing matrix, namely the cusp anomalous dimensions. This shows again how the single log appearing in the fixed order expansion is generated through RG evolution, namely through operator mixing. Therefore, as desired, all large logarithms are generated through RG evolution, and they are resummed to all orders by solving the subleading power RG equation with mixing. We also see that the operator mixing is absolutely crucial, since the entire LL result comes from the mixing which starts the evolution. 

For completeness, we present also the result for the subleading jet function after LL evolution. The anomalous dimensions are derived in \Eqs{eq:anom_dim_mix_diag}{eq:anom_dim_mix} and are related to the soft function ones via RG consistency. 
\begin{align}
	\tilde \gamma_{11} (k,\mu) = \tilde \gamma_{22} (k,\mu) &= \gamma^{J}_{g} (k,\mu) &&\implies&& \Gamma^0_{11} = \Gamma^0_{22}= 2 \Gamma^{g,0}_\cusp = 8 C_A\,,
\nn\\
    \gamma_{12}[\alpha_s] &= -\frac{1}{2} \gamma_{\delta\theta}^{S0} 
    &&\implies&& \gamma^0_{12} = -2 \Gamma^{g,0}_\cusp = -8 C_A \,.
\end{align}
The canonical scales for $J^{(2)}_{g, \delta} (s,\mu)$ are given by
\begin{align}
s = \mu_J^2 = \mu_0^2 = Q^2\tau \,\quad\implies p = 2\,.
\end{align}
Therefore, we find 
\begin{align}
	J^{(2)}_{g, \delta} (Q^2\tau,\mu) &= \theta(\tau) \frac{ \Gamma^{g,0}_\cusp}{\beta_0}\log \left(\frac{\alpha_s(\mu)}{\alpha_s(Q\sqrt{\tau})}\right)   \\
	&\ \ \times\exp\left[\frac{4\pi \Gamma^{g,0}_\cusp}{\beta_0^2 \alpha_s(Q\sqrt{\tau})} \left(\frac{\alpha_s(Q\sqrt{\tau})}{\alpha_s(\mu)} - 1 + 
	\log\frac{\alpha_s(\mu)}{\alpha_s(Q\sqrt{\tau})} \right) \right] .\nn
\end{align}
Therefore, as with the case of the soft function, our analytic solution of the subleading power mixing equation resums the logarithms at subleading power.

\section{Leading Logarithmic Resummation at Next-to-Leading Power}\label{sec:fact}

In this section we will apply the formalism for the resummation of subleading power jet and soft functions developed in the previous sections to resum the leading logarithms for thrust in pure glue $H\to gg$. This is a standard example used to study gluon jets.  We have chosen to restrict ourselves to the case of pure glue to demonstrate in the simplest setting the resummation of subleading power logarithms for a physical process and to highlight the role of the $\theta$-jet and $\theta$-soft operators and operator mixing. The inclusion of fermion operators and the extension to other processes is interesting, and will be considered in future work.

The complete structure of power corrections for dijet event shapes in SCET has been described in detail in the literature, where all relevant ingredients have been studied. In the effective theory, there are three sources of power corrections\footnote{The decomposition into these different classes of power corrections depends on the particular organization of the effective theory being used, but the final result does not.}
\begin{itemize}
\item Subleading power hard scattering operators \cite{Kolodrubetz:2016uim,Moult:2017rpl,Feige:2017zci,Chang:2017atu,Beneke:2017ztn}
\item Subleading power expansion of measurement operators and kinematics \cite{Freedman:2013vya,Feige:2017zci,Moult:2016fqy}
\item Subleading power Lagrangian insertions  \cite{Beneke:2002ni,Chay:2002vy,Manohar:2002fd,Pirjol:2002km,Beneke:2002ph,Bauer:2003mga,Moult:2019mog}
\end{itemize}
It was shown in \cite{Moult:2019mog} that there are no radiative contributions for pure glue $H\to gg$ at NLP at LL order. Therefore we need only consider the first two categories, namely hard scattering operators, and kinematic and measurement expansions, to derive the leading logarithms. We therefore write the cross section as
\begin{align}\label{eq:split}
\frac{1}{\sigma_0}\frac{\df\sigma^{(2)}_{\text{LL}} }{\df\tau} &= \frac{1}{\sigma_0}\frac{\df\sigma^{(2)}_{\kin,\text{LL}} }{\df\tau}+ \frac{1}{\sigma_0}\frac{\df\sigma^{(2)}_{\hard,\text{LL}} }{\df\tau} \,,
\end{align}
where we have put the subscript `LL` to emphasize that we will only give LL expressions for the factorization of the components, and will not include operators that first contribute at higher logarithmic order. In the next two sections we will explicitly work out the factorization and resummation for these two contributions. In both cases the resummation reduces to the mixing equation solved in \Sec{sec:solution}, allowing us to immediately derive the resummed result for thrust at subleading power. 

It is important to emphasize before continuing that the exact split between the terms in \Eq{eq:split} depends on the choice of momentum routing used to setup the factorization, although the final result for the factorization does not. For example, terms involving ultrasoft derivatives in $T$-products or hard scattering operators can in certain cases be eliminated from the hard term through a choice of momentum routing, and will then appear as kinematic corrections. However,  subleading power corrections from operators with additional ultrasoft fields are unambiguously in the hard component. We will define a convenient split in \Sec{sec:kin_corrections}.

\subsection{Kinematic and Observable Corrections}\label{sec:kin_corrections}

We begin by considering corrections from the expansion of the phase space (kinematics) and the thrust observable definition.
These were also considered in the fixed order calculations of \cite{Moult:2016fqy,Moult:2017jsg}, but here we will show how they can be treated to all orders as is required for factorization and resummation. In \cite{Feige:2017zci} it was shown through explicit calculation that the contributions from the thrust measurement function in our formalism do not contribute at LL order. We therefore only need to consider corrections to the phase space here.

\subsubsection{Factorization}\label{sec:kin_corrections_fact}

At subleading power, in addition to considering the expansion of the matrix elements which enter into the cross section, one must also consider power corrections arising from kinematic constraints on the phase space which can be neglected at leading power. To understand this issue we begin by writing the $N$ particle phase space
\begin{align}
\sigma = L_H \int \prod\limits_{i=1}^{N} \dbar^d p_i C(p_i) (2\pi)^4 \delta^4 \Bigl(q-\sum p_i\Bigr) |\cM|^2\,.
\end{align}
Here $q^2=Q^2$ is the momentum of the scattering, $\dbar^d p = d^d p/(2 \pi)^d$, $C(p)=2\pi \delta(p^2) \theta(p^0)$ is the on-shell particle constraint, and $L_H$ is the leptonic tensor. We now consider a final state consisting of $n$-collinear particles with total sector label mometum $\bar n \cdot k_n$, $\bar n$-collinear particles  with total sector label mometum $ n \cdot k_\bn$, and soft particles with total sector momentum $k_s$. Since $n\cdot k_s \sim \bar n \cdot k_s \sim \lambda^2$, at leading power, we can expand the momentum conserving delta function, and the incoming momentum $q$ fixes the large momentum of the collinear sector, namely
\begin{align}
\delta\Bigl(n\cdot q-\sum n\cdot p_i\Bigr)\, \delta\Bigl(\bn \cdot q-\sum \bn\cdot p_i\Bigr)
  =\delta(n\cdot q- n\cdot k_\bn)\, \delta(\bn \cdot q- \bn\cdot k_n)\,.
\end{align}
However, when working at subleading powers, we need to consider the power corrections to this formula, which we refer to as kinematic corrections. These can be organized in a number of different ways. Here we describe a way which seems particularly convenient for the process we are considering.

\begin{figure}
\begin{center}
\subfloat[]{\label{fig:routings_a}
  \includegraphics[width=0.37\textwidth]{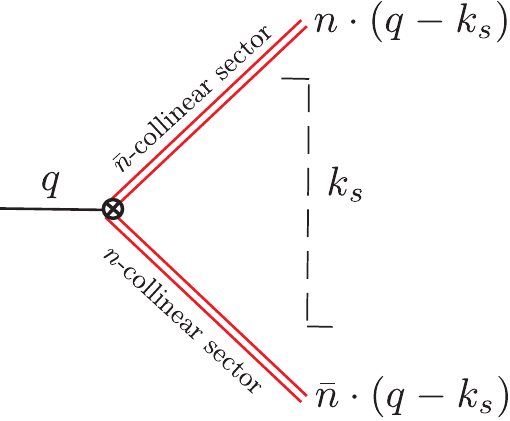}
  }
  \subfloat[]{\label{fig:routings_b}
  \includegraphics[width=0.35\textwidth]{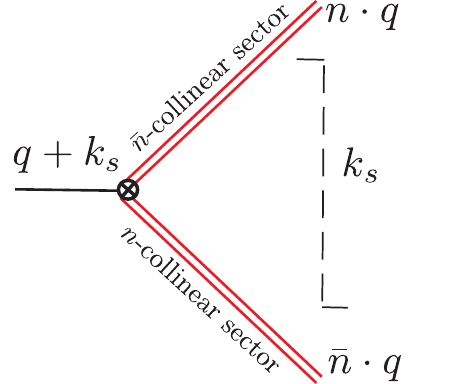}
  }
  \end{center}
  \caption{Two different routings for the soft momentum. In a) the additional soft momentum is routed into the collinear sectors. In b) the additional momentum is routed in through the hard scattering vertex, simplifying the large momentum routed into the collinear sectors.}
  \label{fig:routings}
\end{figure}

In SCET, exact momentum conservation for both label and residual components is implemented in all diagrams. 
Residual momenta must then be routed in the diagram, and unlike at leading power, their effects on the kinematics must be kept to the required power. This routing can be chosen arbitrarily, as long as it is done consistently for all contributions.\footnote{In particular, as mentioned above, this routing determines whether some contributions enter as kinematic or hard power corrections in the decomposition of \Eq{eq:split}.} As an example, consider the routing of the residual momentum from the soft sector. The most naive routing is shown in \Fig{fig:routings_a}. Here we imagine that the soft sector has a total momentum  $k_s$. This momentum must be extracted from the collinear sectors. The residual $n\cdot k_s\sim \lambda^2$ and $\bar n \cdot k_s \sim \lambda^2$ must be kept in the calculations of the collinear sector when working at $\cO(\lambda^2)$, complicating the calculations by requiring us to include $\partial_{us}$ acting on collinear lines.  Here we can still neglect the residual perp momentum of the soft sector, since this enters first as $k_\perp^2\sim \lambda^4$, which is beyond the order to which we work.

A more convenient routing is shown in \Fig{fig:routings_b}. Here, we instead route $q+k_s$ into the hard scattering vertex. The collinear sectors then have exactly $n\cdot q$ and $\bar n \cdot q$ as their large momentum contributions, and all kinematics in the final state is exact.  All kinematic corrections for this routing can be obtained by expanding the phase space factor in the leptonic tensor, which takes the form
\begin{align}\label{eq:lep_tensor_expand}
\frac{1}{(Q+ k_s)^4}=\frac{1}{Q^4}-2\frac{n\cdot k_s}{Q^5}-2\frac{\bar n\cdot k_s}{Q^5} +\cO(\tau^2)\,.
\end{align}
We therefore introduce the measurement functions
\begin{align}
n\cdot \hat k_s = \sum\limits_{i\in S} n\cdot k_s^i\,, \qquad  \bar n\cdot \hat k_s = \sum\limits_{i\in S} \bar n\cdot k_s^i\,,
\end{align}
where the sum is over all soft particles.
To LL accuracy we can make the replacement $n\cdot \hat k_s \to n\cdot \hat k_s \theta( \bn\cdot \hat k_s -n\cdot \hat k_s)$ and $\bn\cdot \hat k_s \to \bn\cdot \hat k_s \theta( n\cdot \hat k_s -\bn\cdot \hat k_s)$, since after multiplying the eikonal integrand $1/(l^+l^-)$ by $l^+$ (or $l^-$), the divergence responsible for the anomalous dimension comes only from the region of phase space where $l^-$ (or $l^+$) is unconstrained by the measurement. These kinematic corrections therefore combine to give the full thrust measurement function
\begin{align}
 n\cdot \hat k_s \theta( \bn\cdot \hat k_s -n\cdot \hat k_s)  + \bn\cdot \hat k_s \theta( n\cdot \hat k_s -\bn\cdot \hat k_s) = Q\hat \tau_s.
\end{align}
The $n\cdot k_n$  and $\bar n \cdot k_\bn$ residual momentum of each of the two collinear sectors can also be routed into the current in the exact same manner, leading to power correction given by $Q\hat \tau_n$ and  $Q\hat \tau_\bn$ respectively.

 We therefore find that the kinematic corrections arising from the phase space expansion give exactly the power suppressed jet and soft functions considered in \Sec{sec:renorm}, namely
\begin{align}
J^{(2)}_{g,\delta}(s, \mu)&=\frac{(2\pi)^3}{(N_c^2-1)}\langle 0 | \cB^{\mu a}_{n\perp}(0)\,\delta(Q+\bar \cP) \delta^2(\cP_\perp)\,  s~\delta\left(\frac{s}{Q}-\hat \Tau\right) \cB^{\mu a}_{n\perp}(0) |0 \rangle\,, \\
S^{(2)}_{g,\delta}(k, \mu)&=\frac{1}{(N_c^2-1)} \tr \langle 0 |  \cY^T_{\bar n}(0) \cY_n(0) k ~\delta(k-\hat \Tau) \cY_n^T(0) \cY_{\bar n}(0)|0\rangle\,. \nn
\end{align}
Indeed, this is one of the reasons why these particular subleading power jet and soft functions were used as an example in \Sec{sec:renorm}.

We can now write down an all orders factorization for the full contribution from kinematic corrections to the cross section at $\cO(\tau)$
\begin{align}\label{eq:fact_kin}
\frac{d\sigma_{\kin,\text{LL}}^{(2)}}{d\tau}&= n_\kin \int \frac{ds_n ds_\bn dk}{Q^2} \hat \delta_\tau    H^{(0)} (Q,\mu) J^{(2)}_{g,\delta} (s_n, \mu)  J_{g}^{(0)} (s_\bn, \mu)  S_{g}^{(0)}(k, \mu)   \\
   &+n_\kin \int \frac{ds_n ds_\bn dk}{Q^2} \hat \delta_\tau H^{(0)} (Q,\mu) J_{g}^{(0)} (s_n, \mu) J^{(2)}_{g,\delta}(s_\bn, \mu)   S_{g}^{(0)} (k, \mu) \nn  \\
& +n_\kin \int \frac{ds_n ds_\bn dk}{Q} \hat \delta_\tau H^{(0)}  (Q,\mu) J_{g}^{(0)} (s_n, \mu)  J_{g}^{(0)}(s_\bn, \mu)   S^{(2)}_{g,\delta}(k, \mu)  \,. \nn
\end{align}
The factorization for the kinematic corrections is therefore exactly the form considered in \Eq{eq:fact_NLP_multTau_rewrite}. 
We have explicitly put the subscript LL, to emphasize that beyond LL there would be additional contributions.
Here the integer constant
\begin{align}\label{eq:nkin_def}
n_\kin= -2\,,
\end{align}
is a normalization factor, effectively the number of times this contribution enters, which is obtained from \Eq{eq:lep_tensor_expand}. We have extracted it as a constant so as to be able to clearly track it, and distinguish it from other integer factors that will appear.

\subsubsection{Resummation}\label{sec:kin_corrections_renorm}

Since the kinematic contributions give exactly the illustrative example considered in \Sec{sec:renorm}, we can immediately perform the resummation of logarithms for this contribution using the solution to the mixing RG equation given in \Sec{sec:solution}. For concreteness, we can run both the soft and hard functions to the jet scale, $\mu_J =Q \sqrt{\tau}$ from their natural scales, $\mu_H=Q$ and $\mu_S=Q\tau$. At leading log order we can set $H^{(0)}(Q,Q)=1$ and $S^{(2)}_{g, \theta} (Q\tau,Q\tau) = \theta(\tau)$. We therefore have
\begin{align}\label{eq:kin_in_kernels}
\frac{1}{\sigma_0}\frac{\df\sigma^{(2)}_{\kin,\text{LL}} }{\df\tau}&=-2 \,U_H(Q, Q\sqrt{\tau})U^S_{g,\delta\theta}(Q\tau,Q \sqrt{\tau})\, \theta(\tau)\,, 
\end{align}
Here the hard evolution kernel is that of the leading power hard function.
\begin{align}\label{eq:LPHkernel}
U_H(Q,Q\sqrt{\tau}) = \exp\left\{-\frac{4\pi \Gamma^{g,0}_\cusp}{\beta_0^2\alpha_s(Q)} \left[ \frac{\alpha_s(Q)}{\alpha_s(Q\sqrt{\tau})} - 1 + \log \left( \frac{\alpha_s(Q\sqrt{\tau})}{\alpha_s(Q)}\right) \right] \right\} \,.
\end{align}
where $ \Gamma^{g,0}_\cusp =4C_A$ is the one-loop gluon cusp anomalous dimension.
The resummed soft function is given by the combination
\begin{align}
	S^{(2)}_{g, \delta}(Q\tau, \mu=Q \sqrt{\tau}) 
=  U^S_{g,\delta\theta}(Q\tau,Q \sqrt{\tau})\, 
  S^{(2)}_{g, \theta} (Q\tau,\mu_0=Q\tau)\,,
\end{align}
and by taking the result of \Eq{eq:SresummedQt} with $\mu = Q\sqrt{\tau}$, we have that the evolution kernel for the soft function at LL reads
\begin{align}\label{eq:SkinLLrunning}
	U^{S\,\LL}_{g,\delta\theta}(Q\tau,Q\sqrt{\tau}) &= - \frac{2 \Gamma^{g,0}_\cusp}{\beta_0}\log \left(\frac{\alpha_s(Q\sqrt{\tau})}{\alpha_s(Q\tau)}\right) \\
	&\ \ \times \exp\left\{-\frac{4\pi \Gamma^{g,0}_\cusp}{\beta_0^2 \alpha_s(Q\tau)} \left[\frac{\alpha_s(Q\tau)}{\alpha_s(Q\sqrt{\tau})} - 1 + \log  \left(\frac{\alpha_s(Q\sqrt{\tau})}{\alpha_s(Q\tau)}\right)\right] \right\} \nn\,.
\end{align}
Plugging these expressions for the evolution kernels into \Eq{eq:kin_in_kernels}, we find that the resummed result for the kinematic contributions is given by
\begin{align}\label{eq:SkinLLrunning_b}
	\frac{1}{\sigma_0}\frac{\df\sigma^{(2)}_{\kin,\text{LL}} }{\df\tau}&= \theta(\tau) \frac{4 \Gamma^{g,0}_\cusp}{\beta_0}\log \left(\frac{\alpha_s(Q\sqrt{\tau})}{\alpha_s(Q\tau)}\right) \exp\biggl\{-\frac{4\pi \Gamma^{g,0}_\cusp}{\beta_0^2 } \biggl[\frac{2}{\alpha_s(Q\sqrt{\tau})} - \frac{1}{\alpha_s(Q\tau)} - \frac{1}{\alpha_s(Q)} \nn\\
	&\quad +  \frac{1}{\alpha_s(Q\tau)}\log  \left(\frac{\alpha_s(Q\sqrt{\tau})}{\alpha_s(Q\tau)}\right) + \frac{1}{\alpha_s(Q)}\log  \left(\frac{\alpha_s(Q\sqrt{\tau})}{\alpha_s(Q)}\right)\biggr] \biggr\}\,.
\end{align}
Simplifying to the case of a fixed coupling and plugging in $\Gamma^{g,0}_\cusp=4C_A$, the kinematic contribution at leading log reads
\begin{align}
\frac{1}{\sigma_0}\frac{\df\sigma^{(2)}_{\kin,\text{LL}} }{\df\tau}&=\left(  \frac{\alpha_s}{4\pi} \right)16 C_A \theta(\tau)\log(\tau) e^{- \frac{\alpha_s}{4\pi} \Gamma^{g,0}_{\text{cusp}} \log^2(\tau)} \,. 
\end{align}
This is a remarkably simple result, involving double logarithmic asymptotics governed by the cusp anomalous dimension. However, this is not surprising since these corrections arise from a multiplication of the leading power result by $\tau$.

\subsection{Hard Scattering Operators}\label{sec:hard_operators}

The second class of contributions that are required for the LL description at NLP arise from corrections to the scattering amplitudes themselves, which in this case are described by subleading power hard scattering operators in the EFT.  A complete basis of hard scattering operators at $\cO(\lambda^2)$ for $H\to gg$ was derived in \cite{Moult:2017rpl}. 

At subleading powers, it becomes important to work in terms of gauge invariant fields, even at the ultrasoft scale.  Leading power interactions between soft and collinear particles in the effective theory can be decoupled to all orders using the BPS field redefinition \cite{Bauer:2002nz}, which for the gluon operator reads
\be \label{eq:B_BPSfieldredefinition}
\cB^{a\mu}_{n\perp}\to \cY_n^{ab} \cB^{b\mu}_{n\perp}\,.
\ee
This factorizes the Hilbert space into separate soft and collinear sectors. After performing the BPS field redefinition, operators in the effective theory can be written in terms of gauge invariant soft and collinear gluon fields
\begin{align} \label{eq:softgluondef_sRGE}
g \cB^{a\mu}_{us(i)}&= \left [   \frac{1}{in_i\cdot \partial_{us}} n_{i\nu} i G_{us}^{b\nu \mu} \cY^{ba}_{n_i}  \right] \,, \qquad g\cB_{n_i\perp}^{A\mu} =\left [ \frac{1}{\bar \cP}    \bar n_{i\nu} i G_{n_i}^{B\nu \mu \perp} \cW^{BA}_{n_i}         \right]\,,
\end{align}
where $\cY$ and $\cW$ are adjoint soft and collinear Wilson lines (see \Eq{eq:Wilson_def}). Due to the presence of the Wilson lines, these gauge invariant fields have Feynman rules at every order in $\alpha_s$. An identical construction exists for collinear and soft fermions, although they will not be needed here since we focus on pure Yang-Mills theory.

The subleading power operators that contribute to the LL cross section involve either an insertion of the $\cB_{n\perp}$, or $\cB_{us}$ operators. The relevant operators, along with their tree level matching coefficients which are required for LL resummation, are given in \Tab{tab:ops}.  The leading power operator is also given for convenience. An important simplification which occurs for the soft operators is that their Wilson coefficients are fixed by reparametrization invariance (RPI) \cite{Larkoski:2014bxa}. In particular, we have the all orders relation
\begin{align} \label{eq:usRPIrelation_sRGE}
C^{(2)}_{\cB \bn(us)}&=-\frac{\partial C^{(0)} }{\partial \omega_1} 
\,, 
\end{align}
and similarly for $n\leftrightarrow \bar n$. As we will see, this will provide a significant simplification, since it fixes the anomalous dimensions of these soft operators. This relationship can be viewed as a manifestation of the Low-Burnett-Kroll theorem \cite{Low:1958sn,Burnett:1967km}, where the connection with our SCET based approach has been explained in detail in \cite{Larkoski:2014bxa}.

{
\renewcommand{\arraystretch}{1.4}
\begin{table}[t!]
\begin{center}
\scalebox{0.842}{
\begin{tabular}{| l | c | c |c |c|c| r| }
  \hline                       
   Operator & Tree Level Matching Coefficient \\
  \hline
    $\cO_\cB^{(0)}=C^{(0)} \delta^{ab} \cB_{\perp \bar n, \omega_2}^a \cdot \cB_{\perp \bar n, \omega_1}^b H$& $C^{(0)}=-2\omega_1 \omega_2\,.$ \\
  \hline
   $\cO^{(2)}_{\cP \cB1}=C^{(2)}_{\cP \cB1}i f^{abc} \cB^a_{n\perp,\omega_1}\cdot \left[  \cP_\perp \cB^b_{\bar n \perp,\omega_2}\cdot  \right] \cB_{\bar n \perp,\omega_3}^c    H$ &  $C^{(2)}_{\cP \cB1}=-\left( \frac{1}{2}\right)4g \left(  2+\frac{\omega_3}{\omega_2}+ \frac{\omega_2}{\omega_3}  \right)$ \\
   \hline
   $\cO^{(2)}_{\cP \cB2}=C^{(2)}_{\cP \cB2} if^{abc}\left[ \cP_\perp \cdot \cB_{\bar n \perp,\omega_3}^a \right] \cB^b_{n\perp,\omega_1} \cdot \cB_{\perp \bar n, \omega_2}^c    H$ & $C^{(2)}_{\cP \cB2}=4g\left( 2+\frac{\omega_3}{\omega_2}  + \frac{\omega_2}{\omega_3}\right)$\\
  \hline  
  $\cO^{(2)}_{\cB(us(n))}=C^{(2)}_{\cB \bn(us)} \left(i  f^{abd}\, \big({\cal Y}_n^T {\cal Y}_{\bar n}\big)^{dc}\right)  \left (  \cB^a_{n\perp, \omega_1} \cdot \cB^b_{\bar n \perp, \omega_2} \bar n \cdot g\cB^c_{us(n)} \right)$ & $C^{(2)}_{\cB \bn(us)}=-2 \omega_2$ \\
  \hline
  $\cO^{(2)}_{\cB(us(\bar n))}= C^{(2)}_{\cB n(us)} \left(i  f^{abd}\, \big({\cal Y}_{\bar n}^T {\cal Y}_{n}\big)^{dc}\right)  \left ( \cB^a_{n\perp, \omega_1} \cdot \cB^b_{\bar n \perp, \omega_2} n\cdot g\cB^c_{us(\bar n)} \right)$ & $C^{(2)}_{\cB n(us)}=-2\omega_1$ \\
  \hline
\end{tabular}}
\end{center}
\caption{
Hard scattering operators that contribute to the LL cross section to $\cO(\lambda^2)$, along with their tree level matching coefficients. These operators and matching coefficients were derived in \cite{Moult:2017rpl}.
}
\label{tab:ops}
\end{table}
}

The operators which contribute to the fixed order leading logarithms were identified in the calculation of  \cite{Moult:2017jsg} as those which contribute a logarithm at the lowest order in perturbation theory. The leading logarithms to all orders are then obtained by the renormalization of these contributions, which dresses them with an all orders resummation of double logarithms. To prove that this is indeed the case, we can assume that there exists a jet or soft function that first contributes at some higher order, for concreteness $\alpha_s^2$,  and that this contribution is leading logarithmic, and hence contributes as $\alpha_s^2 \log^3(\tau)$. With our understanding of the renormalization of subleading jet and soft functions, we know that this implies that this function must be renormalized by a subleading power $\theta$-function type operator, since it can't be a self renormalization. Taking $\mu d/d\mu$, the anomalous dimension of such a LL mixing contribution would have to be of the form $\gamma \sim \log^2(\mu/\mu_0)$. However, it is know that anomalous dimensions in SCET can be at most linear in logarithms, which is required by RG consistency. This argument was first presented in \cite{Manohar:2003vb} in the context of leading power RG consistency. Since this argument relies only on the additive properties of the logarithm, it applies also here. This implies that the operators appearing in \Tab{tab:ops} are sufficient to derive the LL resummation.

\subsubsection{Factorization}\label{sec:hard_operators_factorization}

With an understanding of the operators that contribute, it is now straightforward to write down a factorization for their contributions, which is sufficient for the LL resummation. Detailed accounts of the factorization of matrix elements at subleading power have been given in \cite{Lee:2004ja,Beneke:2004in,Hill:2004if,Freedman:2013vya,Moult:2019mog}. Since the focus of this chapter is on the LL resummation through the mixing with the $\theta$-jet and $\theta$-soft operators, here we simply present the final result for the factorization. Since there are only a small number of operators that appear due to our restriction to a pure glue final state we find a simple LL factorization formula
\begin{align}\label{eq:fact_hard}
\frac{1}{\sigma_0}\frac{d\sigma_{\hard,\text{LL}}^{(2)}}{d\tau}&=n_\hard \int  \frac{ds_n ds_\bn dk}{Q}  \hat \delta_\tau H_{n \cdot \cB}(Q,\mu)     S_{\bn\cB_{us}}^{(2)}(k, \mu)J^{(0)}_{g}(s_n, \mu)\ J^{(0)}_{g}(s_\bn,\mu) \\
&+  n_\hard \int \frac{ds_n ds_\bn dk}{Q^2} \hat \delta_\tau \int d\omega~ H_{\cB\cP}(\omega, Q, \mu) S_{g}^{(0)}(k, \mu) J_{\cB\cP}^{(2)}(s_\bn,\omega, \mu) J_{g}^{(0)}(s_n, \mu)\,. \nn
\end{align}
Here 
\begin{align}\label{eq:nhardvalue}
n_\hard=2\,,
\end{align}
is a combinatorial factor from the equality of $S_{\bn \cB_{us}}^{(2)}$ and $S_{n \cB_{us}}^{(2)}$ in the first line, and from correcting both jet functions and taking the symmetric combination in the second.
This factorization involves a power suppressed soft function
\begin{align}
S_{\bn \cB_{us}}^{(2)}(k, \mu)=&\frac{if^{abd}}{N_c^2-1}\tr \langle 0 |  (\cY_n^T(0) \cY_\bn(0))^{dc} \bar n \cdot g\cB^c_{ us (n)}(0) \delta(k-\hat \Tau) (\cY_n(0) \cY^T_{\bar n}(0))^{ab} |0 \rangle \,,
\end{align}
which arises from the insertion of the $\cB_{us}$ field into the standard leading power soft function. Here we have absorbed the $g$ from the matching coefficient into the soft function. As with the previous subleading power soft functions we have defined in \Eqs{eq:tau_funcs}{eq:theta_soft_first}, this subleading power soft function has mass dimension zero. This factorization also involves a  subleading power jet function
\begin{align}
 \cJ^{(2)}_{\cB\cP}(s, \omega,\mu)& =\\
 &\hspace{-1cm}\frac{(2\pi)^3}{(N_c^2-1)}\frac{Q^2}{\omega (Q-\omega)}\langle 0| [\cB_{\perp \bar n,\omega}(0) [g\cB_{\perp \bar n}(0) \cdot \cP_\perp^\dagger] \delta(Q+\bar \cP) \delta^2(\cP_\perp)\, \delta\left(\frac{s}{Q}-\hat \Tau\right)  \cB_{\perp \bar n}(0) |0\rangle\,,\nn
\end{align}
which arises from the hard scattering operators involving an additional $\cB_{\perp}$ field, and $\cP_\perp$ operator. We have again absorbed the $g$ from the matching coefficient into the definition of the jet function, and as with the subleading power jet functions of \Eqs{eq:tau_funcs}{eq:theta_op} we have defined this jet function to have mass dimension 0.  This jet function involves a convolution in an additional label variable, which is the label momentum of one of the $\cB_\perp$ fields. However, at LL this does not play a role in its renormalization.

\subsubsection{Resummation}\label{sec:hard_operators_renormalization}

Using the factorized expression for the hard scattering operators, we can resum their contribution to the cross section to LL accuracy. To simplify the LL analysis as much as possible, we can exploit consistency relations in the RG equations. As mentioned in \Sec{sec:consistency}, since the subleading power jet and soft functions start as $\cO(\alpha_s)$, we can always choose to eliminate one of them. In the present case, it is convenient to choose to run to the jet scale, where 
\begin{align}
J_{\cB\cP}^{(2)}(s,\omega,\mu)&= 0+\cO(\alpha_s)\,.
\end{align}
With this choice, we do not need to consider the power suppressed jet functions. 

We do, however, have to consider the renormalization of the subleading power soft functions, and the hard function $H_{\bn \cdot \cB}$. However, as described in \Sec{sec:hard_operators}, the anomalous dimension of this hard function is fixed by RPI due to the relation of \Eq{eq:usRPIrelation_sRGE}. This can be seen by differentiating the RG equation for the leading power Wilson coefficient, whose all orders structure is
\begin{align}
\mu \frac{d}{d\mu}C^{(0)} (\omega_1, \omega_2, \mu)=\gamma_C (\omega_1, \omega_2, \mu) C^{(0)} (\omega_1, \omega_2, \mu)\,.
\end{align}
Taking the derivative with respect to $\omega_1$, and switching the order of differentiation, we find 
\begin{align}
\mu \frac{d}{d\mu} \left[  \frac{\partial}{\partial\omega_1} C^{(0)} (\omega_1, \omega_2, \mu)   \right]& = \frac{\partial}{\partial\omega_1} [\gamma_C (\omega_1, \omega_2, \mu)]  C^{(0)} (\omega_1, \omega_2, \mu)  \\
&+ \gamma_C(\omega_1, \omega_2, \mu) \frac{\partial}{\partial\omega_1}   C^{(0)}  (\omega_1, \omega_2, \mu)\,.\nn
\end{align}
The all orders form of the anomalous dimension for the leading power matching coefficient is given by
\begin{align}\label{eq:hard_RG_LP}
\gamma_C (\omega_1,\omega_2,\mu) = \Gamma^g_\cusp[\alpha_s(\mu)] \log \left( \frac{-\omega_1 \omega_2}{\mu^2}  \right) +\gamma_C[\alpha_s(\mu)]\,,
\end{align}
where the second term $\gamma_C[\alpha_s(\mu)]$ is the non-cusp anomalous dimension, which contains no logarithms, and drives the single logarithmic evolution.
The leading double logarithmic evolution is governed by the cusp component. The differentiation in the first component removes the double log component, and therefore we have that to LL accuracy
\begin{align}
\mu \frac{d}{d\mu} \left[  \frac{\partial}{\partial\omega_1} C^{(0)} (\omega_1, \omega_2, \mu)  \right] = \gamma_C(\omega_1, \omega_2, \mu) \left[ \frac{\partial}{\partial\omega_1}  C^{(0)}  (\omega_1, \omega_2, \mu) \right]\,.
\end{align}
This shows that the  LL RG evolution for the subleading power hard scattering operators involving a $\cB_{{us}}$ is identical to that for the leading power hard function, and in particular, is driven by the cusp anomalous dimension.

Finally, the self mixing anomalous dimension of the subleading power soft function is also fixed by RG consistency. In particular, the jet functions appearing in the factorization of \Eq{eq:fact_hard} are the leading power jet functions, and their anomalous dimensions are given in \Eq{eq:LP_anom_dim}. Combining this with the known anomalous dimension for the hard function, it implies by RG consistency relations of \Sec{sec:consistency} that the self mixing anomalous dimension of the subleading power soft function is equal to that of the leading power soft function to LL.

We therefore only need to compute the mixing anomalous dimensions into the $\theta$ function operators for the soft functions involving the $\cB_{{us}}$ operators.  Computing the one loop matrix element of the power suppressed soft function, we find
\begin{align}
\left. S_{\bn \cB_{us}}^{(2)}(k, \mu)\right|_{\cO(\alpha_s)}&= \fd{2cm}{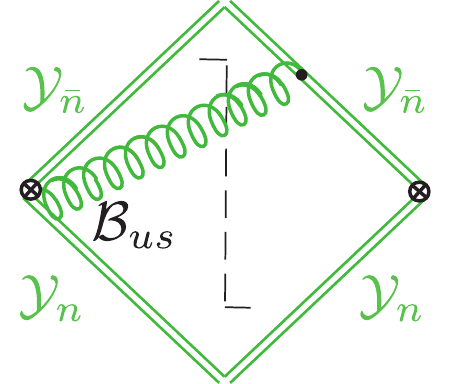}+\fd{2cm}{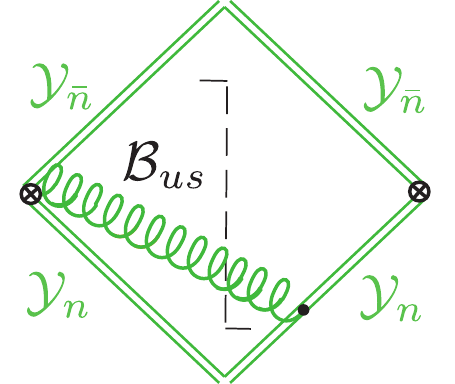} \\
&=g^2  \left(  \frac{\mu^2 e^{\gamma_E}}{4\pi} \right)^\epsilon  C_A \int \frac{d^dl}{(2\pi)^d} \left( \frac{2}{l^+}+ \frac{2}{l^-} \right )2\pi\delta(l^2) \theta(l^0) k \delta(k-Q\hat \tau) \nn\\
&=4C_A\frac{\alpha_s(\mu)}{4\pi}\theta(k) \left(   \frac{1}{\epsilon} +  \log \left(  \frac{\mu^2}{k^2}\right)  +\cO(\epsilon)\right)\,. \nn
\end{align}
As with the illustrative example of \Eq{sec:renorm}, we see that this soft function mixes with a $\theta$-function operator. The RG consistency  relations of \Sec{sec:consistency} imply that the all orders structure of the function being mixed into is that of the adjoint soft function $\theta$-function operator of \Eq{eq:theta_soft_first}. We note that this is a highly non-trivial statement, which would be difficult to prove in perturbation theory, but is dictated by the RG consistency equations of the EFT. We therefore find a $2\times 2$ mixing structure
\begin{align}\label{eq:2by2mix_S_mod}
\mu \frac{d}{d\mu}\left(\begin{array}{c} S_{\bn\cB_{us}}(k,\mu) \\ S_{g,\theta}(k,\mu) \end{array} \right) &= \int dk' \left( \begin{array}{cc} \gamma^S_{\bn \cdot \cB_{us}}(k-k',\mu) & \gamma_{\bn\cdot \cB_{us} \theta}\, \delta(k-k') \\  0 &  \gamma^S_{g,\theta \theta}(k-k',\mu)   \end{array} \right) \left(\begin{array}{c} S_{\bn\cB_{us}}(k',\mu) \\ S_{g, \theta}(k',\mu) \end{array} \right) \,,
\end{align}
where to LL accuracy,
\begin{align}\label{eq:anom_dim_Bus}
\gamma^S_{\bn \cdot \cB_{us}}(k,\mu)&= 4 \Gamma_{\text{cusp}}^{g0} \frac{\alpha_s(\mu)}{4\pi} \, \frac{1}{\mu} \biggl[  \frac{\mu\,\theta(k)}{k}  \biggr]_+\,, \\
\gamma_{\bn\cdot\cB_{us} \theta}&= 8C_A \frac{\alpha_s(\mu)}{4\pi}  \,.\nn
\end{align}
This therefore determines all the anomalous dimensions that are required for LL resummation at NLP. Since the RG equation takes exactly the form already solved in \Sec{sec:solution}, we can immediately use those results to perform the resummation.

Just as for the kinematic contribution, here we run all the functions to the jet scale, $\mu_J^2 =Q^2 \tau$. At their natural scales, $\mu_H=Q$ and $\mu_S=Q\tau$, the hard and the soft function are respectively\footnote{$H_{\bar n \cdot \cB}$ is related to the Wilson coefficient $C^{(2)}_{\cB \bn(us)}$ of the hard scattering operator. From Table \ref{tab:ops} we see that at LP we have $|C^{(0)}(Q,Q)|^2 = 4 Q^4$, and these factors are contained in the normalization factor $\sigma_0$. At subleading power this factor is coming from the interference of $O^{(2)}_{\cB \bn(us)}$ with $O^{(0)}$, which gives $C^{(2)}_{\cB \bn(us)}(Q,Q)C^{(0)}(Q,Q) = 4 Q^3$. In \Eq{eq:fact_hard} one can see the extra ${1}/{Q}$ in the prefactor of the factorization theorem which is precisely the ratio of the tree level subleading Wilson coefficient by the LP one. Thus our $H_{\bar n \cdot \cB}(Q,Q)$ is normalized so that it is dimensionless and equal to $1$ at tree level.} $H_{\bar n \cdot \cB}(Q,Q) = 1$ and  $S^{(2)}_{g, \theta} (Q\tau,Q\tau) = \theta(\tau)$. Using $n_\hard = 2$ from \Eq{eq:nhardvalue}, the hard scattering operator contribution is
\begin{align}
\frac{1}{\sigma_0}\frac{\df\sigma^{(2)}_{\hard,\text{LL}} }{\df\tau}
  &=2\, U_{H_{\bar n \cdot \cB}}(Q, Q \sqrt{\tau}) \,
    U^S_{\bn  B_{us}}(Q\tau,Q \sqrt{\tau}) \, \theta(\tau) \,.
\end{align}
As was shown above, the hard evolution kernel $U_{H_{\bar n \cdot \cB}}(Q, Q \sqrt{\tau})$ is identical to that for the leading power operator, which is quoted in \Eq{eq:LPHkernel}. The soft function takes an identical form to that given in \Eq{eq:Fmomspacetext}, but with $k=\mu_0= Q\tau$ and the anomalous dimensions from \Eq{eq:anom_dim_Bus}. Hence, we get
\begin{align}
S^{(2)\LL}_{g,\delta}(Q\tau,Q\sqrt{\tau}) &= -\theta(\tau) \frac{8 C_A}{\beta_0}\log (r) \,\exp\left[-\frac{4\pi \Gamma^{g,0}_\cusp}{\beta_0^2 \alpha_s(Q\tau)} \left(\frac{1}{r} - 1 + \log(r)\right) \right] \,, 
\end{align} 
where here we have
\begin{align}
 r = \frac{\alpha_s(Q\sqrt{\tau})}{\alpha_s(Q\tau)}\,.
\end{align}
Combining these pieces together, we have 
\begin{align}
	\frac{1}{\sigma_0}\frac{\df\sigma^{(2)}_{\hard,\text{LL}} }{\df\tau}&= -\theta(\tau) \frac{2 \Gamma^{g,0}_\cusp}{\beta_0}\log \left(\frac{\alpha_s(Q\sqrt{\tau})}{\alpha_s(Q\tau)}\right) \exp\biggl\{-\frac{4\pi \Gamma^{g,0}_\cusp}{\beta_0^2 } \biggl[\frac{2}{\alpha_s(Q\sqrt{\tau})} - \frac{1}{\alpha_s(Q\tau)} - \frac{1}{\alpha_s(Q)} \nn\\
	&\quad +  \frac{1}{\alpha_s(Q\tau)}\log  \left(\frac{\alpha_s(Q\sqrt{\tau})}{\alpha_s(Q\tau)}\right) + \frac{1}{\alpha_s(Q)}\log  \left(\frac{\alpha_s(Q\sqrt{\tau})}{\alpha_s(Q)}\right)\biggr] \biggr\} \,.
\end{align}
As with the kinematic contribution to the cross section, we find that the contribution from hard scattering operators resums at LL accuracy into a Sudakov exponential governed by the cusp anomalous dimension.

It is important to emphasize that the simplicity of this result is largely due to the restriction to LL. At LL accuracy the anomalous dimensions do not involve additional convolution variables in the subleading power jet and soft functions, and are purely multiplicative in these variables. This significantly simplifies the structure, with the primary ingredient to achieve renormalization and resummation being the mixing with the $\theta$-jet and $\theta$-soft functions. Beyond LL, the $\theta$-jet and $\theta$-soft will continue to play an important role, but the convolution structure will become more complicated. 

\subsection{Resummed Result for Thrust in $H\to gg$ at Next-to-Leading Power}\label{sec:resum}

\begin{figure}
\begin{center}
  \vspace{-0.2cm}
  \label{fig:sudakovs_a}
  \raisebox{-0.5cm}{a)}\hspace{-1cm}
  \includegraphics[width=0.49\textwidth]{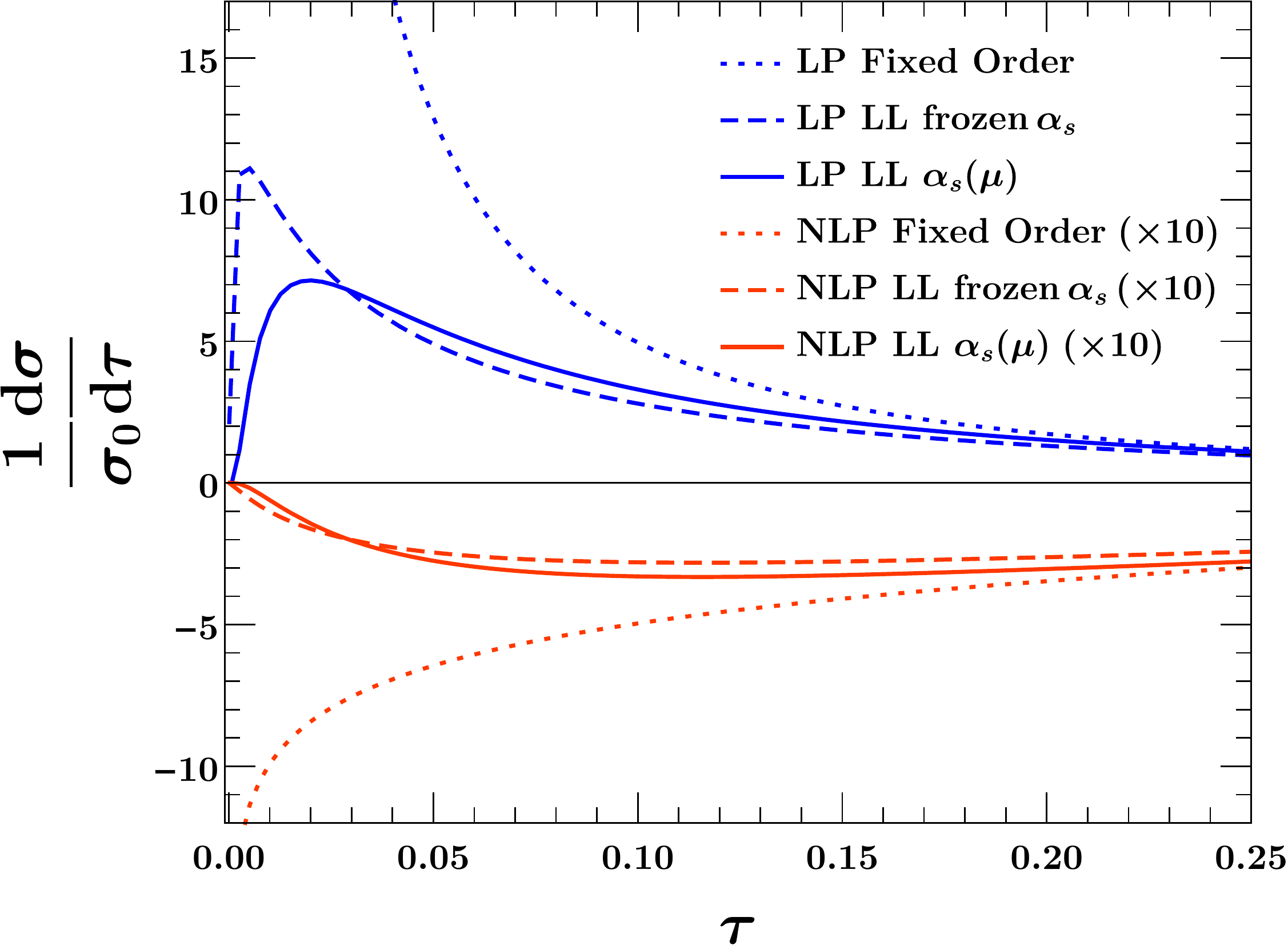}
  \hspace{1cm}\raisebox{-0.5cm}{b)}\hspace{-1.2cm} 
  \includegraphics[width=0.505\textwidth]{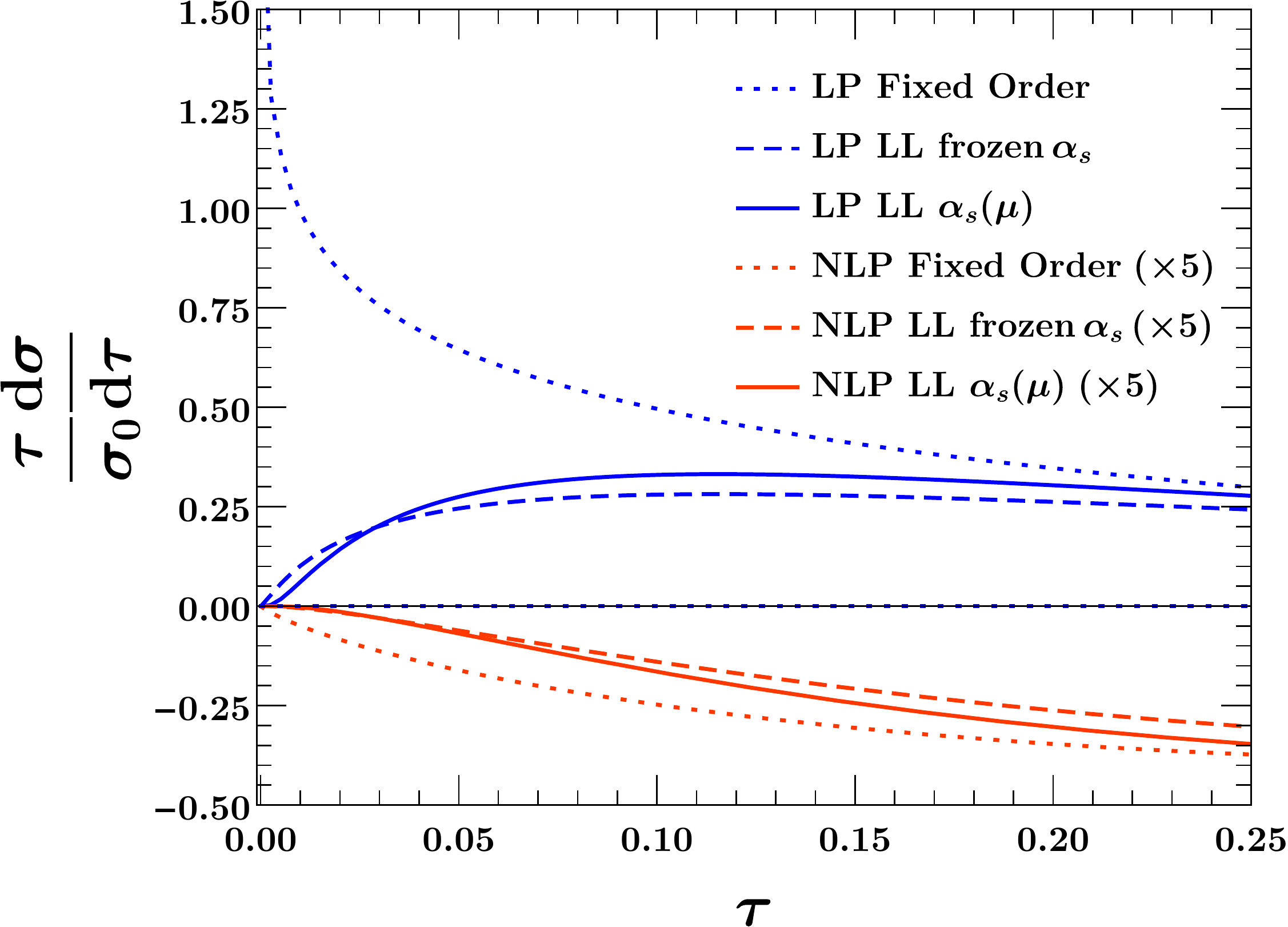}
  \end{center}
\vspace{-0.2cm}
  \caption{Plots of the LP and NLP fixed order and resummed predictions for thrust in pure glue $H\to gg$, with and without running coupling. In a) we show $d\sigma/d\tau$ and in b) we show $\tau d\sigma/ d\tau$. Resummation at LP cures a $1/\tau$ divergence, while resummation at NLP overturns a much weaker logarithmic divergence, leading to a broader shoulder.}
  \label{fig:sudakovs}
\end{figure}

Having resummed the two different contributions to the cross section in \Eq{eq:split}, we can now give a resummed result for thrust in pure glue $H\to gg$.
 Adding together the different contributions, each of which is dressed by the same Sudakov exponential, we find
\begin{align}\label{eq:resum_nlp}
\frac{1}{\sigma_0}\frac{\df\sigma^{(2)}_{\text{LL}} }{\df\tau} &= \frac{1}{\sigma_0}\frac{\df\sigma^{(2)}_{\kin,\text{LL}} }{\df\tau}+ \frac{1}{\sigma_0}\frac{\df\sigma^{(2)}_{\hard,\text{LL}} }{\df\tau} \nn\\
&= \theta(\tau)\frac{8C_A}{\beta_0}\log \left(\frac{\alpha_s(Q\sqrt{\tau})}{\alpha_s(Q\tau)}\right) \exp\biggl\{-\frac{4\pi \Gamma^{g,0}_\cusp}{\beta_0^2 } \biggl[\frac{2}{\alpha_s(Q\sqrt{\tau})} - \frac{1}{\alpha_s(Q\tau)} - \frac{1}{\alpha_s(Q)} \nn\\
	&\quad +  \frac{1}{\alpha_s(Q\tau)}\log  \left(\frac{\alpha_s(Q\sqrt{\tau})}{\alpha_s(Q\tau)}\right) + \frac{1}{\alpha_s(Q)}\log  \left(\frac{\alpha_s(Q\sqrt{\tau})}{\alpha_s(Q)}\right)\biggr] \biggr\}\,.
\end{align}
With a fixed coupling, \Eq{eq:resum_nlp} simplifies to
\begin{align}\label{eq:resum_nlp_fixed}
\frac{1}{\sigma_0}\frac{\df\sigma^{(2)}_{\text{LL}} }{\df\tau}\Big|_{\alpha_s(\mu)=\alpha_s} &= \left(  \frac{\alpha_s}{4\pi} \right) 8C_A \theta(\tau) \log(\tau) e^{-4C_A \frac{\alpha_s}{4\pi} \log^2(\tau)}   \,. 
\end{align}
This shows the exponentiation of the subleading power logarithms into a Sudakov form factor governed by the cusp anomalous dimension, and is one of the main results of this chapter. We note that this result is simply $-\tau$ multiplying the LP result with LL resummation. This simplicity is in part related to the fact that we have chosen a simple event shape example, and is not expected to hold in general at LL, nor beyond LL.
In \Sec{sec:split} we will check this result to $\cO(\alpha_s^3)$ by expanding known results for the amplitudes \cite{Garland:2001tf,Garland:2002ak,Gehrmann:2011aa}, and find complete agreement.

This resummation tames the (integrable) singularity in the subleading power cross section as $\tau \to 0$. A plot of the LL NLP resummed cross section is shown in \Fig{fig:sudakovs}, along with the NLP fixed order results, and the LP results. Results with and without running coupling are shown. We use $\alpha_s(m_Z)=0.118$ for the running coupling $\alpha_s(\mu)$, and when we freeze the coupling, we use $\alpha_s=\alpha_s(m_H)=0.113$.  The NLP results are multiplied by a factor of 10 in \Fig{fig:sudakovs} a) and a factor of 5 in \Fig{fig:sudakovs} b) to make them visible. Due to the fact that the NLP result is not enhanced by a factor of $1/\tau$ it leads to a much broader result, peaked at large values of $\tau$. This has interesting consequences for the effect of the running coupling. In particular, at subleading power the running coupling has a much smaller effect, since the distribution is more suppressed at smaller values of $\tau$.  At higher powers, resummation is not required for the cross section to go to zero as $\tau \to 0$, since the corrections behave as $\tau^n \log^m(\tau)$, with $n>0$. Nevertheless,  RG equations are still useful for predicting higher order terms in the perturbative expansion.

\section{Subleading Power Collinear Limit and Fixed Order Check}\label{sec:split}

In this section we check our resummed result for thrust to $\cO(\alpha_s^3)$ by explicitly calculating the power corrections to this order. This is achieved by exploiting a relation between the LL result and the subleading power collinear limit of the involved amplitudes. We also discuss flipping around this logic, and using the resummed results to constrain corrections in the collinear limit at $n$th-loop order. In particular, for $H\to ggg$ we will show that the same loop corrections dress terms that appear at leading and next-to-leading order in the power expansion. 

The $N$-loop fixed order result at NLP can be written as \cite{Moult:2016fqy,Moult:2017jsg}
\begin{align}\label{eq:constraint_setup}
\frac{1}{\sigma_0}\frac{\df\sigma^{(2,N)}}{\df\tau}
  = & \sum_{\kappa}\sum_{i=0}^{2N-1} \frac{c_{\kappa,i}}{\epsilon^i} \left( \frac{\mu^{2N}}{Q^{2N} \tau^{m(\kappa)}}   \right)^\epsilon
 + \dots
\,,\end{align}
where the dots involve terms that are first relevant beyond LL order. Our superscript $(j,N)$ notation denotes the subleading power at order $j$ and loop order $N$. Here the sum over $\kappa$ is over different possible combinations of soft, collinear, or hard particles entering the $N$-loop result, and the power $m(\kappa)$ appearing in \Eq{eq:constraint_setup} depends on this combination. For example, a single emission at NLP can be either soft, or collinear, and we have
\begin{align} \label{eq:classes1}
\text{soft:} \qquad &\kappa=s\,, \qquad m(\kappa) =2\,, \\
\text{collinear:} \qquad &\kappa=c\,, \qquad m(\kappa)=1 \,.\nn
\end{align}
For a more detailed discussion see \cite{Moult:2016fqy,Moult:2017jsg}. By demanding cancellation of poles in $1/\epsilon$, as is required for an infrared and collinear safe observable, one can derive relations between contributions involving different numbers of hard, collinear and soft particles, which were used in \cite{Moult:2016fqy,Moult:2017jsg} to simplify the NNLO fixed order calculation of the NLP leading logarithms. In particular, in \cite{Moult:2016fqy,Moult:2017jsg}, it was shown that the complete result for the leading logarithms for thrust can be written at any order purely in terms of the $N$-loop hard-collinear coefficient describing a single collinear splitting
\begin{align}\label{eq:constraints_final}
\frac{1}{\sigma_0}\frac{\df\sigma^{(2,N)}}{\df\tau}
&= c_{hc,2N-1} \log^{2N-1} \tau +\cdots
\,.\end{align}
Here the dots denote subleading logarithms.
More precisely, here $c_{hc,2N-1}$ is the result for the leading $1/\epsilon$ divergence (as in \Eq{eq:constraint_setup}) with $N-1$ hard loops correcting a single collinear splitting. One class of diagram that contributes is 
\begin{align}
\fd{5cm}{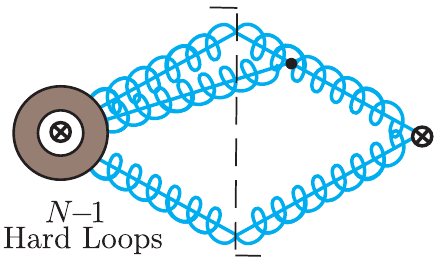}\,,\nn
\end{align}
but there will also be hard loop corrections to the amplitudes on both sides of the cut.
This relation will allow us to check our result obtained from renormalization group evolution to $\cO(\alpha_s^3)$ by expanding known results for $H\to ggg$ at two loops \cite{Gehrmann:2011aa}. In addition, it will also allow us to use our result for the all orders logarithms in thrust derived from RG evolution to understand the subleading power collinear limit at higher orders.

\subsection{General Structure}\label{sec:split_gen_structure}

Before presenting our result for the expanded amplitude squared in the collinear limit, we begin by reviewing the known IR structure of amplitudes, which we will use to organize our result. The IR structure of amplitudes is summarized by the dipole formula~\cite{Catani:1998bh} and its generalization~\cite{Dixon:2008gr,Becher:2009qa,Gardi:2009qi,Almelid:2015jia}, which provides a prediction for all the IR $1/\epsilon$ poles of scattering amplitudes at $n$ loops~(recall that we use $\alpha_s/(4 \pi)$ as the loop expansion parameter). Here we only need the full QCD amplitude for $H\to$ three partons at $n$-loops
\begin{align}
M^{(n)}=M^{(n)}_{\text{dipole}} + M^{(n)}_{R}\,.
\end{align}
Here $M^{(n)}_{\text{dipole}} $ contains all $1/\epsilon$ poles, while the remainder part $M^{(n)}_{R}$ is finite but still carries functional dependence on the kinematics that can become singular in certain limits (it is typically called the finite term but we will not use this naming scheme here). When integrating over these regions of phase space, $M^{(n)}_{R}$ must be known to all orders in $\epsilon$, and does contribute to the LL result.  More explicitly, at one-loop, we have
\begin{align}
M^{(1)} = I^{(1)}(\epsilon) M^{(0)} + M^{(1)}_{R} \,.
\end{align}
Here $I^{(1)}(\epsilon)$ is an operator in color space that can be predicted from the infrared structure of the scattering process. Using the color-charge operator notation, $I^{(1)}(\epsilon)$ can be written as~\cite{Catani:1998bh}
\begin{align}
  \label{eq:Ione}
  I^{(1)}(\epsilon) = \frac{\alpha_s}{4\pi} \frac{e^{-\epsilon \gamma_E}}{\Gamma(1 - \epsilon)} \sum_{i} \frac{1}{\mathbf{T}_i^2} \left(\mathbf{T}_i^2 \frac{1}{\epsilon^2} + \gamma_i \frac{1}{\epsilon} \right) \sum_{j\neq i} \mathbf{T}_i \cdot \mathbf{T}_j \left(\frac{\mu^2 e^{-i \pi}}{2 p_i \cdot p_j} \right)^\epsilon \,,
\end{align}
where $\mathbf{T}_i$ is the color-charge operator of massless parton $i$, $\gamma_i$ is the associated quark/gluon anomalous dimension, and we assume all QCD partons are outgoing for simplicity. 
In this chapter, we have focused only on deriving a leading logarithmic result for thrust at subleading power. One obvious source of leading logarithmic contributions comes from the leading divergent terms in the amplitudes~\cite{Moult:2016fqy,Moult:2017jsg}, which exponentiate trivially. For $H \to g(p_1) g(p_2) g(p_3)$ in pure glue QCD, we have
\begin{align}\label{eq:Catani_exp}
 M_\text{dipole,LL} = \exp\left[ - \frac{\alpha_s}{4 \pi} \frac{C_A}{\epsilon^2} \left(\left(-\frac{\mu^2}{s_{12}}\right)^\epsilon 
+ \left(-\frac{\mu^2}{s_{13}}\right)^\epsilon + \left(-\frac{\mu^2}{s_{23}}\right)^\epsilon \right)  \right] M^{(0)}\,,
\end{align}
where $s_{ij} = (p_i + p_j)^2$. The subscript LL denotes that only terms contributing to thrust at LL are kept. Note that Eq.~\eqref{eq:Catani_exp} contains not only divergent terms, but also finite terms through the expansion of $\epsilon$.
After squaring the amplitudes and integrating over the phase space, the leading divergences at ${\cal O}(\alpha_s^{n+1})$ become $\alpha_s^{n+1}/\epsilon^{2 n + 1}$ at NLP, and give rise to leading logarithms for the thrust cross section. In general, the remainder part $M_R$ are not known to exhibit an iterative structure to all orders.

Typically, LL resummation at LP is carried out either by using the coherent branching formalism~\cite{Webber:1983if,Marchesini:1983bm,Marchesini:1987cf} which makes use of strongly ordered real radiation, or by computing anomalous dimensions from virtual ultraviolet divergences to hard, jet, and soft functions in SCET. However, by consistency this LL resummation also provides interesting information about higher order virtual loop corrections to a single collinear splitting.  In the next section we discuss this at both LP and NLP.  Further details for the leading power case can be found in \App{sec:LLfromCollinear}.  For this analysis both the dipole and remainder terms contribute.  Although the remainder terms do not have explicit poles in $\epsilon$, they do not necessarily vanish in the soft or colllinear limits, and in particular contain logarithms in these limits.  We will use our all orders understanding of the leading logarithms for thrust derived in Sec.~\ref{sec:fact} to show that the remainder terms also exhibit interesting exponentiation patterns.

\subsection{Subleading Power Collinear Splitting}\label{sec:split_oneloop}

To perform the expansion of the squared amplitudes in the collinear limits, we use the results of \cite{Gehrmann:2011aa}. These are in a particularly convenient form for our purposes, namely they are already expressed in a decomposition into the dipole and remainder terms.

For $H \to g(p_1)g(p_2)g(p_3)$, the collinear power expansion at amplitude level is controlled by $s = P^2 = (p_1 + p_2)^2$, the invariant mass of a pair of gluons. 
At tree level, the leading power result is given by
\begin{align}
  \label{eq:33}
  |M^{(0,0)}|^2 = 2 \tilde{\lambda}^2 \frac{(1-z+z^2)^2}{z
  (1-z)} \frac{Q^2}{s} \,,
\end{align}
where $\tilde{\lambda}^2 = 128 N_c \lambda^2 \pi^2$, $\lambda$ is the effective coupling of dimension $5$ Higgs-gluon-gluon operator, and $z$ is the longitudinal momentum fraction of $p_1$ with respective to $P$ in the collinear limit.
The next-to-leading power collinear expansion is
\begin{align}
  \label{eq:34}
  |M^{(2,0)}|^2 = 2 \tilde{\lambda}^2 \frac{1 + 2 z - 3 z^2 + 2 z^3 - z^4}{z (1-z)} \,.
\end{align}
Here we have used a double superscript notation where the first superscript indicates the power in $s/Q^2$, and the second indicates the order in $\alpha_s$. Eq.~\eqref{eq:34} contains end-point singularity in the momentum fraction, which is regularized by the $d$ dimension phase space measure. 
For the purpose of extracting the leading logarithms, it is only
necessary to consider the $z\to 0$ or $z\to 1$ limit. In the current
case the two limits are identical, and we find
\begin{align}
  |M^{(2,0)}|_\text{LL}^2   =  \tilde{\lambda}^2 \frac{2}{z
   (1-z)}  \,,
\end{align}
where we use subscript LL to denote that only the end-point singular term in $z$ is retained.
We can use these to define the tree level LP and NLP splitting functions, valid at LL level
\begin{align}\label{eq:split_def}
P_{gg,\text{LL}}^{(0,0)}\ & = \frac{Q^2}{s} \frac{2}{z(1-z)}\,, \quad
P_{gg,\text{LL}}^{(2,0)}\ = \frac{2}{z(1-z)}\,.
\end{align}
Here we see the explicit suppression in $s/Q^2$ of the NLP result.
We then have
\begin{align}
  \label{eq:35}
  |M^{(0,0)}|_\text{LL}^2 & = \tilde{\lambda}^2  P_{gg,\text{LL}}^{(0,0)}  \,, \qquad
  |M^{(2,0)}|_\text{LL}^2 =  \tilde{\lambda}^2  P_{gg,\text{LL}}^{(2,0)}\,.
\end{align}

Using \Eq{eq:Catani_exp} it is trivial to give the all loop result for squared amplitude for the terms predicted by dipole formula. We find
\begin{align}
|M|^2_\text{dipole,LP,LL}&\!\! = \tilde{\lambda}^2  P_{gg,\text{LL}}^{(0,0)}\exp\left( {F_{\text{dipole}}} \right)\,, \qquad
|M|^2_\text{dipole,NLP,LL} \!\!= \tilde{\lambda}^2  P_{gg,\text{LL}}^{(2,0)}\exp\left( {F_{\text{dipole}}} \right)\,,
\end{align}
where
\begin{align}
\label{eq:dipole}
F_{\text{dipole}}= \frac{\alpha_s \mu^{2 \e}}{4 \pi}  \frac{(-2
  C_A)}{\e^2} \left( [(1-z)Q^2]^{-\e} + s^{-\e} + [z Q^2]^{-\e}\right)\,.
\end{align}
Interestingly, the form of the dipole term guarantees that its leading logarithmic loop corrections are independent of the power expansion. The power expansion arises only in the expansion of the tree level amplitude squared.

Much more interesting are the remainder terms of the amplitude, whose all order form is not predicted. We can begin by looking at their form at one-loop.
By inspecting the higher order in $\e$ terms in the
remainder term of the amplitude, we can write down an all-order-in-$\e$ expression
for the leading transcendental piece of the remainder terms (i.e. the piece required to give the LL for thrust). We find
\begin{align}
  \label{eq:47}
       2 \mathrm{Re}\big[
&\,  M^{(0)*}M_R^{(1)}\big]\Big|_\text{LP,LL} = 
 - 2 C_A \tilde  \lambda^2  P_{gg,\text{LL}}^{(0,0)}  
\nn\\ &\,
\times \frac{\alpha_s\,\mu^{2 \epsilon} }{4\pi}  \Bigg[\left(\frac{[Q^2]^{-\e}}{\e^2} - \frac{[z (1-z) Q^2]^{-\e}}{\e^2}
  \right) - \left(\frac{[s]^{-\e}}{\e^2} - \frac{[z(1-z)
  s]^{-\e}}{\e^2}  \right)\Bigg]\,.
\end{align}
The structure of this leading transcendental component of the remainder term is quite interesting. Expanding it, we see that both the $1/\epsilon^2$ and $1/\epsilon$ poles cancel, giving a finite result
\begin{align}
&\Bigg[\left(\frac{[Q^2]^{-\e}}{\e^2} - \frac{[z (1-z) Q^2]^{-\e}}{\e^2}
  \right) - \left(\frac{[s]^{-\e}}{\e^2} - \frac{[z(1-z)
  s]^{-\e}}{\e^2}  \right)\Bigg] \nn \\
&= \frac{[Q^2]^{-\epsilon}  }{2} \left(  -\log^2 \left( \frac{s}{Q^2} \right)  -\log^2(z(1-z)) +\log^2 \left( \frac{s(1-z)z}{Q^2}   \right) +\cO(\epsilon)  \right)\,.
\end{align}
However, we see that this term secretly contains leading poles in $1/\epsilon$ when written in the form of \Eq{eq:constraint_setup} and therefore will contribute to the LL result at LP. The reason is that when integrating over the momentum fraction $z$ using $d$ dimension phase space measure, there is a mismatch in the exponent of $z$ between different terms. Since this is a non-traditional way to obtain the leading logarithms for the thrust distribution, we provide a more detailed explanation in Appendix~\ref{sec:LLfromCollinear}. For the NLP terms, we find the exact same structure, with only a different prefactor
\begin{align}
  \label{eq:48}
         2 \mathrm{Re}\big[
&\,  M^{(0)*}M_R^{(1)}\big]\Big|_\text{NLP,LL} = - 2 C_A
\tilde   \lambda^2  P_{gg,\text{LL}}^{(2,0)}  
\nn\\ &\,
\times \frac{\alpha_s\,\mu^{2 \epsilon} }{4\pi}  \Bigg[\left(\frac{[Q^2]^{-\e}}{\e^2} - \frac{[z (1-z) Q^2]^{-\e}}{\e^2}
  \right) - \left(\frac{[s]^{-\e}}{\e^2} - \frac{[z(1-z)
  s]^{-\e}}{\e^2}  \right)\Bigg]\,.
\end{align}
Interestingly, as was the case for the dipole terms, we again see that the transcendental structure is the same at LP and NLP, and it just multiplies the tree level splitting function.

Going to two loops, quite interestingly, we find that the remainder term is
\begin{align}
  \label{eq:49}
& \,  2 \mathrm{Re}\big[ M^{(0)*}M_R^{(2)}\big] +
  M_R^{(1)*}M_R^{(1)}\Big|_\text{LP,LL} =  
 2 C_A^2 \tilde  \lambda^2  P_{gg,\text{LL}}^{(0,0)}   
\nn\\ &\qquad \times \frac{\alpha_s^2\,\mu^{4 \epsilon} }{(4\pi)^2} 
 \Bigg[\left(\frac{[Q^2]^{-\e}}{\e^2} - \frac{[z (1-z) Q^2]^{-\e}}{\e^2}
  \right) - \left(\frac{[s]^{-\e}}{\e^2} - \frac{[z(1-z)
  s]^{-\e}}{\e^2}  \right)\Bigg]^2\,.
\end{align}
Note that only the $\Ord{\e^0}$ terms in Eq.~\eqref{eq:49} are
explicitly verified using the two-loop amplitudes. To verify to higher
order in $\e$, one needs to know the two-loop amplitudes also to higher order in $\e$, which are currently not available in the
literature. However, since this contribution is related to the hard-collinear contribution in the effective theory, the renormalizability of the effective theory guarantees that the all loop result can be obtained by RG evolution of the lowest order result. To LL, the power expansion in the amplitudes acts only on the kinematic factors giving rise to the lowest order splitting functions in \Eq{eq:split_def}, but not on the transcendental function. This therefore fixes the all order in $\e$ form of Eq.~\eqref{eq:49}. Compared with Eq.~\eqref{eq:47}, we have the relation
\begin{align}
  \label{eq:50}
  & \, \frac{2 \mathrm{Re}\big[ M^{(0)*}M_R^{(2)}\big] +
  M_R^{(1)*}M_R^{(1)} \Big|_\text{LP,LL}}{|M^{(0,0)}|_\text{LL}^2} =  
\frac{1}{2!} \left( \frac{2 \mathrm{Re}\big[
    M^{(0)*}M_R^{(1)}\big]\Big|_\text{LP,LL}}{|M^{(0,0)}|_\text{LL}^2}\right)^2\,,
\end{align}
that is, the remainder term also exponentiates. Similarly, for the
NLP piece, we have
\begin{align}
  \label{eq:51}
    & \, \frac{2 \mathrm{Re}\big[ M^{(0)*}M_R^{(2)}\big] +
  M_R^{(1)*}M_R^{(1)} \Big|_\text{NLP,LL}}{|M^{(2,0)}|_\text{LL}^2} =  
\frac{1}{2!} \left( \frac{2 \mathrm{Re}\big[
      M^{(0)*}M_R^{(1)}\big]\Big|_\text{NLP,LL}}{|M^{(2,0)}|_\text{LL}^2} \right)^2\,.
\end{align}
Here we observe exponentiation of the remainder term at LP and NLP, and furthermore, we again see that the transcendental structure at both LP and NLP is identical. 

With the expanded result for the squared amplitude, we can simply integrate it over the collinear phase space to obtain the result for thrust. We find
\begin{align}
  \label{eq:52}
  \frac{1}{\sigma_0} \frac{\df\sigma^{(2)}}{\df \tau} =
 \left( \frac{\alpha_s}{4\pi} \right) 8 C_A \log\tau -
  \left(\frac{\alpha_s}{4\pi}\right)^2
32 C_A^2 \log^3 \tau + \left(\frac{\alpha_s}{4\pi}\right)^3 64 C_A^3
  \log^5 \tau +\cO(\alpha_s^4) \,.
\end{align}
This agrees with the result derived from the RG in \Eq{eq:resum_nlp}, and provides an explicit check at $\cO(\alpha_s^3)$ of the result from the RG. The terms  to $\cO(\alpha_s^2)$ were also computed in \cite{Moult:2017jsg} using this technique. The $\cO(\alpha_s^3)$ term has not previously appeared in the literature.

We can now use the higher order terms predicted by the RG to study the collinear limit at higher loop orders. In particular, since we have derived using the RG that the leading logarithms for thrust exponentiate into a Sudakov, given in \Eq{eq:resum_nlp}, the all-loop expansion of the amplitudes in the collinear limit must agree with this exponentiation.

We have already shown that at least to two loops, the leading logarithmic contribution of the remainder terms exponentiate.
Combined with the exponentiation of the dipole terms, we conjecture that to all orders, amplitudes in the collinear limit through to NLP exponentiate, namely
\begin{align}\label{eq:split_allorders}
\left. \left[ M^* M \right]\right |_{\text{LP,LL}}&=  \tilde \lambda^2 P_{gg,\text{LL}}^{(0,0)} e^{F_\text{dipole} + F_R}  \,,
\quad
\left. \left[ M^* M \right]\right |_{\text{NLP,LL}}=  \tilde \lambda^2 P_{gg,\text{LL}}^{(2,0)} e^{F_\text{dipole} + F_R}  \,,
\end{align}
where 
\begin{align}
\label{eq:remainder}
F_R=\frac{\alpha_s \mu^{2 \epsilon}}{4 \pi}  (-2 C_A)  \left[ \left(\frac{[Q^2]^{-\e}}{\e^2} - \frac{[z (1-z) Q^2]^{-\e}}{\e^2}
  \right) - \left(\frac{[s]^{-\e}}{\e^2} - \frac{[z(1-z)
  s]^{-\e}}{\e^2}  \right) \right]\,.
\end{align}
In particular, this result reproduces the leading logarithms in thrust obtained through RG evolution to all loop order in \Eq{eq:resum_nlp}. Note that this is an amplitude level statement, and while we have explicitly checked it to two loops, and when integrated over $z$ it agrees with our result obtained from the RG for thrust, which provides a strong check, we phrase it only as a conjecture, since it is possible $z$ dependent terms that do not give rise to leading logarithms for the thrust observable could be present.
This seems to imply an interesting iterative structure for the remainder terms of the amplitude, which is relevant for leading logarithmic resummation, and goes beyond the dipole formula.  This would be interesting to investigate further, and we hope that the study of subleading power limits will lead to a further understanding.

Here we have only considered the case of $H\to ggg$, but it is important to understand the universality of the above subleading power splitting functions, and in particular of their loop corrections, even at a given logarithmic accuracy. The universality of subleading power collinear factorization has been studied at tree level in \cite{Nandan:2016ohb}, but it would be interesting to try to extend it to all loop order using the techniques in this chapter. A perhaps related question is the definition of an infrared finite remainder function in planar ${\cal N}=4$ SYM, where a clever definition of exponentiated terms can lead to a better behaved remainder function~\cite{Caron-Huot:2016owq}. 

\section{Conclusions}\label{sec:conc_subRGE}

In this chapter we have, for the first time, resummed to all orders in $\alpha_s$  subleading power logarithms for the thrust observable to LL accuracy for pure glue $H\to gg$. We have shown that the subleading power logarithms exponentiate to all orders into a Sudakov exponential controlled by the cusp anomalous dimension multiplying a  logarithm, see \Eq{eq:resum_nlp}. Resummation is achieved by RG evolution of gauge invariant non-local Wilson line operators and its accuracy is systematically improvable. 

The renormalization of subleading power jet and soft functions requires the introduction of a new class of universal soft and collinear functions, which we termed $\theta$-jet and $\theta$-soft functions. These functions, which involve $\theta$-functions of the measurement, appear through operator mixing, and we argued that they will play a general role in renormalization and resummation at subleading powers. We introduced a simple example which allowed us to understand the structure of these functions to all orders in $\alpha_s$, as well as to derive their renormalization group evolution, which we proved closes into a $2\times2$ mixing equation. We analytically solved this subleading power RG mixing equation, including the effects of running coupling.

We checked our result derived from RG evolution to $\cO(\alpha_s^3)$ by direct calculation of the power corrections. Using consistency relations from the cancellation of IR poles, the leading logarithms can be derived entirely from the collinear limit, allowing us to use our all orders result derived from the RG equations to understand higher order loop corrections to the subleading power collinear limit. We showed explicitly that to two-loops all leading transcendental pieces in the collinear and subleading power collinear limit exponentiate. We conjectured that this exponentiation holds to all loop order, and showed that this results in agreement with the results for the thrust observable derived from RG evolution. This seems to indicate an interesting structure for the IR finite terms in the subleading power collinear limits, beyond what is predicted by the dipole formula, and it would be interesting to investigate this further.  

Since this represents the first all orders resummation of NLP logarithms for an event shape, there are many interesting directions in which it can be extended. In particular,  it will be important to extend our results to higher logarithmic accuracy to understand what universal structures persist. The simplicity of the leading logarithmic structure to all powers suggests the possibility of a simple structure. It will also be interesting to study subleading power corrections for other observables, such as $q_T$ or in the threshold limit, as well as to extend the calculation to the $N$-jet case, for example for the $N$-jettiness observable \cite{Stewart:2010tn}. The renormalization of amplitude level hard scattering operators for the $N$-jet case was recently considered \cite{Beneke:2017ztn}, which provides an important ingredient in this direction. Our work provides a path for the systematic resummation of subleading power logarithms for event shapes, and we hope that this will lead to an improved understanding of the all orders structure of the subleading power soft and collinear limits.

\chapter{\boldmath Subleading Power Rapidity Divergences and\\ Power Corrections for $q_T$}\label{sec:ptNLL}

\section{Introduction}\label{sec:introPtNLL}

Observables in quantum field theory that are sensitive to soft and collinear emissions
suffer from potentially large logarithms in their perturbative predictions.
The structure of these logarithms depends on the observable in question.
For a large class of phenomenologically relevant observables,
these logarithms arise from emissions that are widely separated in rapidity, as opposed to, or in addition to, the more standard case of logarithms from a hierarchy of virtualities.
At leading order in the associated power expansion, these rapidity logarithms can be resummed to all orders in $\alpha_s$ using rapidity evolution equations. Historically these include the well-known massive Sudakov form factor \cite{Collins:1989bt},  Collins-Soper \cite{Collins:1981uk,Collins:1981va,Collins:1984kg}, BFKL  \cite{Kuraev:1977fs,Balitsky:1978ic,Lipatov:1985uk},  and rapidity renormalization group \cite{Chiu:2011qc,Chiu:2012ir} equations.

The resummation of such rapidity logarithms is necessary for a number of applications, including the $q_T$ spectrum for small $q_T$ in color-singlet processes (see e.g.~\refscite{Ji:2004wu, Bozzi:2005wk, Becher:2010tm, GarciaEchevarria:2011rb, Chiu:2012ir, Wang:2012xs, Neill:2015roa, Ebert:2016gcn, Bizon:2017rah, Chen:2018pzu, Bizon:2018foh}), double parton scattering (see e.g.~\refscite{Diehl:2011yj, Manohar:2012jr, Buffing:2017mqm}), jet-veto resummation (see e.g.~\refscite{Banfi:2012jm, Becher:2013xia, Stewart:2013faa}), recoil sensitive event-shape observables (see e.g.~\refscite{Dokshitzer:1998kz, Becher:2012qc, Larkoski:2014uqa, Moult:2018jzp}), multi-differential observables (see e.g.~\refscite{Laenen:2000ij,Procura:2014cba, Marzani:2015oyb, Lustermans:2016nvk, Muselli:2017bad, Hornig:2017pud, Kang:2018agv, Michel:2018hui}), processes involving massive quarks or gauge bosons (see e.g.~\refscite{Ciafaloni:1998xg, Fadin:1999bq, Kuhn:1999nn, Chiu:2007yn,Chiu:2009mg, Gritschacher:2013pha, Hoang:2015vua, Pietrulewicz:2017gxc}), and small-$x$ resummations that go beyond the simplest applications of BFKL (see e.g.~\refscite{Catani:1990eg,Balitsky:1995ub,Kovchegov:1999yj,JalilianMarian:1996xn,JalilianMarian:1997gr,Iancu:2001ad}).
In all these cases, the resummation was performed at leading power (LP), and at present very little is known about the structure of rapidity logarithms and their associated evolution equations at subleading power.

There has been significant interest and progress in studying power corrections \cite{Manohar:2002fd, Beneke:2002ph, Pirjol:2002km, Beneke:2002ni, Bauer:2003mga,Laenen:2008gt,Laenen:2010uz,Larkoski:2014bxa} both in the context of $B$-physics (see e.g.~\refscite{Mantry:2003uz,Hill:2004if, Mannel:2004as, Lee:2004ja, Bosch:2004cb, Beneke:2004in, Tackmann:2005ub, Trott:2005vw,Paz:2009ut,Benzke:2010js}) and for collider-physics
cross sections (see e.g.~\refscite{Dokshitzer:2005bf,Laenen:2008ux,Freedman:2013vya,Freedman:2014uta,Bonocore:2014wua,Bonocore:2015esa,Kolodrubetz:2016uim,Bonocore:2016awd,Moult:2016fqy,Boughezal:2016zws,DelDuca:2017twk,Balitsky:2017flc,Moult:2017jsg,Goerke:2017lei,Balitsky:2017gis,Beneke:2017ztn,Feige:2017zci,Moult:2017rpl,Chang:2017atu,Boughezal:2018mvf,Ebert:2018lzn,Bahjat-Abbas:2018hpv}). Recently, progress has been made also in understanding the behaviour of matrix elements in the subleading soft and collinear limit \cite{Bhattacharya:2018vph} in the presence of multiple collinear directions using spinor-helicity formalism.
In \refcite{Moult:2018jjd} the first all-order resummation at subleading power for collider observables was achieved for a class of power-suppressed kinematic logarithms in thrust including both soft and collinear radiation. More recently in~\refcite{Beneke:2018gvs} subleading power logarithms for a class of corrections in the threshold limit have also been resummed. In both cases the subleading power logarithms arise from widely separated virtuality scales, and their resummation make use of effective field theory techniques.
Given the importance of observables involving nontrivial rapidity scales, it is essential to extend these recent subleading-power results to such observables, and more generally, to understand the structure of rapidity logarithms and their evolution equations at subleading power.

In this chapter, we initiate the study of rapidity logarithms at subleading power, focusing on their structure in fixed-order perturbation theory. We show how to consistently regularize subleading-power rapidity divergences, and highlight several interesting features regarding their structure. In particular, power-law divergences appear at subleading power, which give nontrivial contributions and must be handled properly. We introduce a new ``pure rapidity'' regulator and an associated ``pure rapidity'' $\overline{\rm MS}$-like renormalization scheme. This procedure is homogeneous in the power expansion, meaning that it does not mix different orders in the power expansion, which significantly simplifies the analysis of subleading power corrections. We envision that it will benefit many applications.

As an application of our formalism, we compute the complete $\cO(\alpha_s)$ power-suppressed contributions for $q_T$ for color-singlet production, which provides a strong check on our regularization procedure.
We find the interesting feature that the appearing power-law rapidity divergences yield derivatives of PDFs in the final cross section. Our results provide an important ingredient for improving the understanding of $q_T$ distributions at next-to-leading power (NLP). They also have immediate practical applications for understanding and improving the performance of fixed-order subtraction schemes based on the $q_T$ observable \cite{Catani:2007vq}.

To systematically organize the power expansion, we use the soft collinear effective theory (SCET) \cite{Bauer:2000ew, Bauer:2000yr, Bauer:2001ct, Bauer:2001yt}, which provides operator and Lagrangian based techniques for studying the power expansion in the soft and collinear limits. The appropriate effective field theory for observables with rapidity divergences is \scetii~\cite{Bauer:2002aj}. In this theory, rapidity logarithms can be systematically resummed using the rapidity renormalization group (RRG) \cite{Chiu:2011qc,Chiu:2012ir} in a similar manner to virtuality logarithms. The results derived here extend the rapidity renormalization procedure to subleading power, and we anticipate that they will enable the resummation of rapidity logarithms at subleading power.

The outline of this chapter is as follows. In \sec{rapidity_general}, we give a general discussion of the structure and regularization of rapidity divergences at subleading power.
We highlight the issues appearing for rapidity regulators that are not homogeneous in the power-counting parameter, focusing on the $\eta$ regulator as an explicit example. We then introduce and discuss the pure rapidity regulator, which is homogeneous.
In \sec{qT}, we derive a master formula for the power corrections to the color-singlet $q_T$ spectrum at $\cO(\as)$, highlighting several interesting features of the calculation.
We also give explicit results for Higgs and Drell-Yan production, and
perform a numerical cross check to validate our results. We conclude in \sec{conc_pt}.

\section{Rapidity Divergences and Regularization at Subleading Power}
\label{sec:rapidity_general}

Rapidity divergences naturally arise in the calculation of observables sensitive to the transverse momentum of soft emissions. In a situation where we have a hard interaction scale $Q$ and the relevant transverse momentum $k_T$ of the fields is small compared to that scale, $\lambda \sim k_T / Q \ll 1$, the appropriate effective field theory (EFT) is \scetii~\cite{Bauer:2002aj}, which contains modes with the following momentum scalings
\begin{align} \label{eq:modes_introPtNLL}
 &&&n{-}\text{collinear}: \quad k_n \sim Q \, (\lambda^2, 1, \lambda) &&\implies &&k^-/Q \sim 1
\,,\\ \nn
&&&\bn{-}\text{collinear}: \quad k_{\bn} \sim Q \, (1, \lambda^2, \lambda) &&\implies &&k^-/Q \sim \lambda^2
\,,\\\nn
&&&\text{soft}: \hspace{1.8cm} k_s \sim Q \, (\lambda, \lambda, \lambda) &&\implies &&k^-/Q \sim \lambda\,.
\end{align}
Here we have used lightcone coordinates $(n\cdot k, \bn \cdot k, k_\perp) \equiv (k^+, k^-, k_\perp)$, defined with respect to two lightlike reference vectors $n^\mu$ and $\bn^\mu$. For concreteness, we take them to be $n^\mu =(1,0,0,1)$ and $\bn^\mu=(1,0,0,-1)$.
Unlike \sceti\ where the modes are separated in virtuality, in \scetii~the modes in the EFT have the same virtuality, but are distinguished by their longitudinal momentum ($k^+$ or $k^-$), or equivalently, their rapidity $e^{2y_k} = k^-/k^+$. This separation into modes at hierarchical rapidities introduces divergences, which arise when $k^+/k^-\to \infty$ or $k^+/k^-\to 0$  \cite{Collins:1992tv,Manohar:2006nz,Collins:2008ht,Chiu:2012ir,Vladimirov:2017ksc}. These so-called rapidity divergences are not regulated by dimensional regularization, which is boost invariant and therefore cannot distinguish modes that are only separated in rapidity.

Rapidity divergences can be regulated by introducing a rapidity regulator that breaks boost invariance, allowing the modes to be distinguished, and logarithms associated with the different rapidity scales to be resummed. The rapidity divergences cancel between the different sectors of the effective theory, since they are not present in the full theory. They should not be thought of as UV, or IR, but as arising from the factorization in the EFT. By demanding invariance with respect to the regulator, one can derive renormalization group evolution equations (RGEs) in rapidity.
In SCET, a generic approach to rapidity evolution was introduced in \refscite{Chiu:2011qc, Chiu:2012ir}. These rapidity RGEs allow for the resummation of large logarithms associated with hierarchical rapidity scales.

At leading power in the EFT expansion, the structure of rapidity divergences and the associated rapidity renormalization group are well understood by now, and they have been studied to high perturbative orders (see e.g.\ \refcite{Li:2016ctv} at three-loop order). Indeed, in certain specific physical situations involving two lightlike directions, rapidity divergences can be conformally mapped to UV divergences \cite{Hatta:2008st,Caron-Huot:2015bja,Caron-Huot:2016tzz,Vladimirov:2016dll,Vladimirov:2017ksc}, giving a relation between rapidity anomalous dimensions and standard UV anomalous dimensions. However, little is known about the structure of rapidity divergences or their renormalization beyond the leading power.\footnote{For some interesting recent progress for the particular case of the subleading power Regge behavior for massive scattering amplitudes in $\cN=4$ super Yang-Mills theory, see \refcite{Bruser:2018jnc}.}

In this section, we discuss several interesting features of rapidity divergences at subleading power, focusing on the perturbative behavior at next-to-leading order (NLO). At subleading power there are no purely virtual corrections at NLO, and so we will focus on the case of the rapidity regularization of a single real emission, which allow us to identify and resolve a number of subtleties.
After a brief review of the structure of rapidity-divergent integrals at leading power in \sec{rapDivLP}, we discuss additional issues that arise at subleading power in \sec{rapDivNLP}. We discuss in detail the behavior of the $\eta$ regulator at subleading power, highlighting effects that are caused by the fact that it is not homogeneous in the power expansion. In \sec{upsilonreg}, we introduce the pure rapidity regularization, which regulates rapidity instead of longitudinal momentum and which we find to significantly simplify the calculation at subleading power.
Finally, in \sec{distribution}, we discuss the distributional treatment of power-law divergences,
which arise at subleading power.

\subsection{Review of Rapidity Divergences at Leading Power}
\label{sec:rapDivLP}

We begin by reviewing the structure of rapidity divergent integrals at leading power.  As mentioned above, we restrict ourselves to the case of a single on-shell real emission, which suffices at NLO. Defining $\delta_+(k^2)=\theta(k^0) \delta(k^2)$,
its contribution to a cross section sensitive to the transverse momentum $\kt$ of the emission is schematically given by
\begin{align} \label{eq:sigma_schematic}
\df\sigma(\kt) &\sim \frac{2}{k_T^2} \int\df k^0 \df k^z \, \delta_+(k^2)\, g(k)
\nn\\
&= \frac{1}{k_T^2} \int_0^\infty \frac{\df k^-}{k^-}\, g(k)\Big|_{k^+ = k_T^2/k^-}
 = \frac{1}{k_T^2} \int_0^\infty \frac{\df k^+}{k^+}\, g(k)\Big|_{k^- = k_T^2/k^+}
\,.\end{align}
Here, we have extracted the overall $1/k_T^2$ behaviour, and $g(k)$ is an observable and process dependent function, containing the remaining phase-space factors and amplitudes.
The precise form of $g(k)$ is unimportant, except for the fact that it includes
kinematic constraints on the integration range of $k^\pm$,
\begin{equation} \label{eq:kin_constr_toy}
 g(k) \sim \theta(k^\pm - k_\text{min}^\pm)\, \theta( k_\text{max}^\pm - k^\pm)
\,.\end{equation}
For our discussion we take $k_T > 0$ such that we can work in $d=4$ dimensions.
In the full theory, \eq{sigma_schematic} is finite, with the apparent singularities for $k^\pm \to 0$ or $k^\pm \to \infty$ being cut off by the kinematic constraints in \eq{kin_constr_toy}.
In the effective theory, one expands \eq{sigma_schematic} in the soft and collinear limits specified in \eq{modes_introPtNLL}. This expansion also removes the kinematic constraints,
\begin{equation}
	\underbrace{\,k_\text{min}^\pm \to 0\,}_{\text{soft and collinear limits}}\,,\qquad \underbrace{\,k_\text{max}^\pm \to +\infty\,}_{\text{soft limit}}\,,
\end{equation}
such that individual soft and collinear contributions acquire explicit divergences as $k^\pm \to 0$ or $k^\pm \to \infty$. This is actually advantageous, since the associated logarithms can now be tracked by these divergences. To regulate them, we introduce a regulator $R(k,\eta)$, where $\eta$ is a parameter such that $\lim_{\eta\to0}R(k,\eta)=1$. By construction, inserting $R(k,\eta)$ under the integral in \eq{sigma_schematic} does not affect the value of $\df\sigma(\kt)$ when taking $\eta\to0$ in the full calculation.
To describe the limit $k_T \ll Q$, we expand \eq{sigma_schematic} in the soft and collinear limits described by the modes in \eq{modes_introPtNLL}.
To be specific, the soft limit of \eq{sigma_schematic} is obtained by evaluating the integrand together with the regulator $R(k,\eta)$ using the soft scaling $k_s$ of \eq{modes_introPtNLL}, and expanding in $\lambda$,
\begin{align} \label{eq:sigma_schematic_soft}
\df\sigma_s(\kt)
&\sim \frac{1}{k_T^2} \int_0^\infty\! \frac{\df k_s^-}{k_s^-}\, g(k_s)\Big|_{k_s^+ = k_T^2/k_s^-}\, R(k_s,\eta)
\nn \\
&= \frac{1}{k_T^2} \int_0^\infty\! \frac{\df k^-}{k^-}\, g_s(0)\, R(k, \eta)
\times \bigl[ 1 + \cO(\lambda) \bigr]
\,.\end{align}
Since the leading-power result must scale like $1/k_T^2$, the
LP soft limit $g_s(k^\mu=0)$ must be a pure constant, which implies that the kinematic constraints in \eq{kin_constr_toy} are removed.
This introduces the aforementioned divergences as $k^-\to0$ or $k^-\to\infty$, which are now regulated by $R(k,\eta)$.

The analogous expansion in the collinear sectors is obtained by inserting the $k_n$ or $k_\bn$ scalings of \eq{modes_introPtNLL} into \eq{sigma_schematic}, and expanding in $\lambda$,
\begin{align} \label{eq:sigma_schematic_collinear}
\df\sigma_n(\kt)
&\sim \frac{1}{k_T^2} \int_0^\infty\! \frac{\df k_n^-}{k_n^-}\, g(k_n)\Big|_{k_n^+ = k_T^2/k_n^-}\, R(k_n,\eta)
\nn\\
&= \frac{1}{k_T^2} \int_0^Q\! \frac{\df k^-}{k^-}\, g_n\biggl(\frac{k^-}{Q}\biggr)\, R(k,\eta)
\times \bigl[ 1 + \cO(\lambda) \bigr]
\,,\nn\\
\df\sigma_\bn(\kt)
&\sim \frac{1}{k_T^2} \int_0^\infty\! \frac{\df k_\bn^+}{k_\bn^+}\, g(k_\bn)\Big|_{k_\bn^- = k_T^2/k_\bn^+}\, R(k_\bn,\eta)
\nn\\
&= \frac{1}{k_T^2} \int_0^Q\! \frac{\df k^+}{k^+}\, g_\bn\biggl(\frac{k^+}{Q}\biggr)\, R(k,\eta)
\times \bigl[ 1 + \cO(\lambda) \bigr]
\,.\end{align}
In this case, only the lower bound on $k^\pm$ is removed by the power expansion, while the upper limit is given by the relevant hard scale $Q$.
The expansion of $g(k_n)$ in the collinear limit can still depend on the momentum $k^-/Q \sim \cO(\lambda^0)$, as indicated by the functional form of $g_n(k^-/Q)$, and likewise for the $\bn$-collinear limit.

Without the rapidity regulator, the integrals in \eqs{sigma_schematic_soft}{sigma_schematic_collinear} exhibit a logarithmic divergence as $k^\pm \to 0$ or $k^\pm \to \infty$, which is not regulated by dimensional regularization or any other invariant-mass regulator. Since $k^+ k^- = k_T^2$ is fixed by the measurement, this corresponds to a divergence as the rapidity $y_k = (1/2) \ln(k^-/k^+) \to \pm\infty$.
The rapidity regulator $R(k,\eta)$ regulates these divergence by distinguishing the soft and collinear modes. To ensure a cancellation of rapidity divergences in the effective theory, it should be defined  as a function valid on a full-theory momentum $k$, which can then be expanded in the soft or collinear limits. Since there are no divergences in the full theory, this guarantees the cancellation of divergences in the EFT expansion.

At leading power a variety of regulators have been proposed. Since the divergences are only logarithmic, and the focus has not been on higher orders in the power expansion, there are not many constraints from maintaining the power counting of the EFT. Therefore, a variety of regulators have been used, including hard cutoffs \cite{Balitsky:1995ub,JalilianMarian:1997gr,Kovchegov:1999yj,Manohar:2006nz}, tilting Wilson lines off the lightcone \cite{Collins:1350496},  the delta regulator \cite{Chiu:2009yx}, the $\eta$ regulator \cite{Chiu:2011qc,Chiu:2012ir}, the analytic regulator \cite{Beneke:2003pa,Chiu:2007yn,Becher:2011dz}, and the exponential regulator \cite{Li:2016axz}.

At subleading power, we will discuss in more detail the application of the $\eta$ regulator, which can be formulated at the operator level by modifying the Wilson lines appearing in the SCET fields as \cite{Chiu:2011qc,Chiu:2012ir}
\begin{align} \label{eq:reg_soft}
S_n(x)&= \sum\limits_{\text{perms}} \exp \biggl[ -\frac{g}{n\cdot \cP} \frac{w\, |2\,\cP^z|^{-\eta/2}}{\nu^{-\eta/2}}\, n \cdot A_s \biggr] \,, \\
\label{eq:reg_coll}
W_{n}(x) &= \sum\limits_{\text{perms}} \exp \biggl[  -\frac{g}{ \bar n\cdot \cP} \frac{w^2 \,|2\,\cP^z|^{-\eta}}{\nu^{-\eta}}\,  \bar n \cdot A_{n} \biggr] \,,
\end{align}
where $S_n$ and $W_n$ are soft and collinear Wilson lines.
The operator $\cP$ picks out the large (label) momentum flowing into the Wilson line, $\nu$ is a rapidity regularization scale, $\eta$ a parameter exposing the rapidity divergences as $1/\eta$ poles, and $w$ a bookkeeping parameter obeying
\begin{align}\label{eq:omegadef}
\nu\frac{\partial w(\nu)}{\partial\nu} = -\frac{\eta}{2} w(\nu)\,,\qquad\lim_{\eta \to 0} w(\nu)=1
\,.\end{align}
Note that at leading power, one can replace $|2\cP^z| \to |\bn \cdot \cP|$ in \eq{reg_coll},
as employed in \refscite{Chiu:2011qc,Chiu:2012ir},
while at subleading power we will show that this distinction is actually important.
The $\eta$ regulator was extended in \refcite{Rothstein:2016bsq} to also regulate Glauber exchanges in forward scattering, where regulating Wilson lines alone does not suffice.

\subsection{Rapidity Regularization at Subleading Power}
\label{sec:rapDivNLP}

We now extend our discussion to subleading power, where we will find several new features.
First, while at leading power, rapidity divergences arise only from gluons,  at subleading power rapidity divergences can arise also from soft quarks.  Soft quarks have also been rapidity-regulated to derive the quark Regge trajectory \cite{Moult:2017xpp}.
Here, since we consider only the case of a single real emission crossing the cut, this simply means that we must regulate both quarks and gluons. More generally, one would have to apply a rapidity regulator to all operators in the EFT, as has been done for the case of forward scattering in \refcite{Rothstein:2016bsq}. It would be interesting to understand if these subleading rapidity divergences can also be conformally mapped to UV divergences of matrix elements, as was done for the rapidity divergences in the leading power $q_T$ soft function in \refscite{Vladimirov:2016dll,Vladimirov:2017ksc}.

Second, the structure of rapidity divergences becomes much richer at subleading power, placing additional constraints on the form of the rapidity regulator to maintain a simple power expansion. This more interesting divergence structure follows directly from power counting.
For example, the subleading corrections to the soft limit can be obtained by expanding the integrand in \eq{sigma_schematic_soft} to higher orders in $\lambda$.
The power counting for soft modes in \eq{modes_introPtNLL} implies that the first $\cO(\lambda)$ power suppression can only be given by additional factors of $k^-/Q$ or $k^+/Q$ in \eq{sigma_schematic_soft}.
At the next order, $\cO(\lambda^2)$, one can encounter additional factors $(k^+/Q)^2, (k^-/Q)^2$.
The possible structure of rapidity-divergent integrals in the soft limit up to $\cO(\lambda^2)$ is thus given by%
\footnote{We can also have integrals with an additional factor of $k_T/Q$ or $k^2_T/Q^2$, which however do not change the structure of the integrand and can thus be treated with the same techniques as at leading power.}
\begin{align}\label{eq:sublintegrals}
 \cO(\lambda^0): \qquad
 & \int_0^\infty \frac{\df k^-}{k^-} \softregulator
\,,\\\nn
 \cO(\lambda^1): \qquad
 & \int_0^\infty \frac{\df k^-}{k^-} \biggl(\frac{k^-}{Q}\biggr) \softregulator
 \,,\quad
 \int_0^\infty \frac{\df k^-}{k^-} \biggl(\frac{k^+}{Q}\biggr) \softregulator
\,,\\\nn
 \cO(\lambda^2): \qquad
 & \int_0^\infty \frac{\df k^-}{k^-} \biggl(\frac{k^-}{Q}\biggr)^2 \softregulator
 \,, \quad
 \int_0^\infty \frac{\df k^-}{k^-} \biggl(\frac{k^+}{Q}\biggr)^2 \softregulator
\,,\end{align}
where it is understood that $k^+ = k_T^2 / k^-$.
We can see that the $\cO(\lambda^0)$ limit only produces logarithmic divergences,
while the power-suppressed corrections give rise to power-law divergences.
The prototypical rapidity-divergent integral encountered in the soft limit is thus given by
\begin{align}\label{eq:toy_integrals_soft}
\Isoftgen{\alpha}
&= \int_0^\infty \frac{\df k^-}{k^-} \biggl(\frac{k^-}{Q}\biggr)^\alpha \softregulator
\,,\end{align}
where $\alpha$ counts the additional powers of $k^-$.

A similar situation occurs in the collinear sectors.
In the $n$-collinear limit, $k \sim Q(\lambda^2, 1, \lambda)$, the large momentum $k^-$ is not suppressed with respect to $Q$, such that the power suppression can only arise from explicit factors of $k_T^2$.
(Of course, $k^+ \sim \cO(\lambda^2)$ can also give a suppression, but it can always be reduced back to $k^+ = k_T^2 / k^-$.)
Similarly, in the $\bn$-collinear limit $k^+$ is unsuppressed, and power suppressions only arise from $k_T^2$.
However, the structure of the collinear expansion of $g(k)$ is richer than in the soft case, because there is always a nontrivial dependence on the respective unsuppressed ratio $k^\mp/Q$.
To understand this intuitively, consider the splitting of a $n$-collinear particle into two on-shell $n$-collinear particles with momenta
\begin{align}
 p_1^\mu&=(Q-k^-)\frac{n^\mu}{2}+k_\perp^\mu+\frac{k_T^2}{Q-k^-} \frac{\bar n^\mu}{2}
\,,\qquad
 p_2^\mu=k^-\frac{n^\mu}{2}-k_\perp^\mu+\frac{k_T^2}{k^-} \frac{\bar n^\mu}{2}
\,.\end{align}
The associated Lorentz-invariant kinematic variable is given by
\begin{align}
s_{12} = (p_1 + p_2)^2 = \frac{k_T^2 Q^2}{k^-(Q-k^-)}\,.
\end{align}
Expanding any function of $s_{12}$ in $k_T$ thus gives rise to additional factors of the large momentum $k^-$.
Thus, in general, expanding $g(k_n)$ in the collinear limit can give rise to both positive and negative powers of $k^-$ that accompany the power-suppression in $k_T^2$.
These factors are of course not completely independent, as the sum of all soft and collinear contributions must be rapidity finite, i.e., any rapidity divergences induced by these additional powers of $k^-$ must in the end cancel against corresponding divergences in the soft and/or other collinear contributions.
In summary, the generic form of integrals in the collinear expansion is given by
\begin{align}\label{eq:toy_integrals_collinear}
\Ingen{\alpha}
&= \int_0^Q \frac{\df k^-}{k^-} \biggl(\frac{k^-}{Q}\biggr)^\alpha g_n \biggl(\frac{k^-}{Q}\biggr) \nregulator
\,, \\
\Ibngen{\alpha}
&=\int_0^Q \frac{\df k^+}{k^+} \biggl(\frac{k^+}{Q}\biggr)^\alpha g_\bn \biggl(\frac{k^+}{Q} \biggr)\bnregulator
\,.\end{align}
Here, $g_n(x)$ and $g_\bn(x)$ are regular functions as $x \to 0$.
At LP, only $\alpha=0$ contributes, which gives rise to logarithmic divergences,
while at subleading power for $\alpha\neq 0$ we again encounter power-law divergences.
As we will see in \sec{distribution}, these power-law divergences have a nontrivial effect,
namely they lead to derivatives of PDFs in the perturbative expansion for hadron collider processes.

The presence of power-law divergences at subleading power also implies that more care must be taken to ensure that the regulator does not unnecessarily complicate the power counting of the EFT. For example, with the exponential regulator \cite{Li:2016axz}, or with a hard cutoff, power-law divergences lead to the appearance of powers of the regulator scale, and hence break the homogeneity of the power expansion of the theory.

Furthermore, at leading power one also has the freedom to introduce and then drop subleading terms to simplify any stage of the calculation. While this may seem a general feature and not appear very related to the regularization of rapidity divergences, we will see in a moment that this freedom, explicitly or not, is actually used in most of the rapidity regulators in the literature.

In summary, having a convenient-to-use regulator at subleading power imposes stronger constraints than at leading power.
In particular, we find that the regulator
\begin{itemize}
 \item must be able to regulate not only Wilson lines, but all operators, including those generating soft quark emissions,
 \item must be able to deal not only with logarithmic divergences, but also with power-law divergences without violating the power counting of the EFT by inducing power-law mixing,
 \item and should be homogeneous in the power-counting parameter $\lambda$ to minimize mixing between different powers.
\end{itemize}
The first requirement means one cannot use regulators acting only on Wilson lines, such as taking Wilson lines off the light-cone as in~\refcite{Collins:1350496}, the $\delta$ regulator as used in \refscite{Chiu:2009yx,GarciaEchevarria:2011rb}, and the $\eta$ regulator as used in \refscite{Chiu:2011qc,Chiu:2012ir}, while the $\eta$ regulator as modified and employed in \refscite{Rothstein:2016bsq,Moult:2017xpp} and the analytic regulator of \refcite{Becher:2011dz} can be used.
The second requirement is satisfied by all dimensional regularization type regulators, such as the $\eta$ regulator or analytic regulator, but not by those that are more like a hard cutoff, including the exponential regulator \cite{Li:2016axz}.
To highlight the last point, in the following we discuss in more detail the properties of the $\eta$ regulator at subleading power.

\subsubsection[The \texorpdfstring{$\eta$}{eta} Regulator at Subleading Power]
{\boldmath The $\eta$ Regulator at Subleading Power}
\label{sec:etareg}

In the $\eta$ regulator, one regulates the $k^z$ momentum of emissions through the regulator function
(see \eq{reg_soft})
\begin{equation} \label{eq:Rz}
R_z(k,\eta) = w^2 \biggl|\frac{2 k^z}{\nu}\biggr|^{-\eta} = w^2 \nu^\eta |k^- - k^+|^{-\eta}
\,.\end{equation}
For a single massless emission this corresponds to regulating its phase-space integral as
\begin{equation}
 \int \df^d k\, \delta_+(k^2) \quad\to\quad
 \int\df^d k\, \delta_+(k^2)\, R_z(k,\eta)
 = w^2\nu^\eta \int\df^d k \,\delta_+(k^2)\, |k^- - k^+|^{-\eta}
\,.\end{equation}
In the soft limit $k^+ \sim k^- \sim \lambda Q$, the regulator is homogeneous in $\lambda$ and therefore does not need to be expanded. The prototypical soft integral in \eq{toy_integrals_soft} evaluates to
\begin{align}\label{eq:soft_integrals_a}
\Isoftz{\alpha}
&= w^2 \nu^\eta \int_0^\infty \frac{\df k^-}{k^-} \biggl(\frac{k^-}{Q}\biggr)^\alpha \biggl|k^- - \frac{k_T^2}{k^-}\biggr|^{-\eta}
 \nn\\&
 = w^2 \biggl(\frac{\nu}{k_T}\biggr)^{\eta} \biggl(\frac{k_T}{Q}\biggr)^\alpha \cos\biggl(\frac{\alpha \pi}{2}\biggr)
   \, \sin\biggl(\frac{\eta \pi}{2}\biggr) \,
   \frac{1}{\pi} \Gamma(1-\eta) \Gamma\Bigl(\frac{\eta}{2} - \frac{\alpha}{2}\Bigr)
   \Gamma\Bigl(\frac{\eta}{2} + \frac{\alpha}{2}\Bigr)
\,.\end{align}
Symmetry  under $\alpha \leftrightarrow -\alpha$ implies that
\begin{align}
\Isoftz{-\alpha} =  \biggl(\frac{k_T^2}{Q^2}\biggr)^{-\alpha} \Isoftz{\alpha}
\,.\end{align}
This reflects the symmetry under exchanging $k^- \leftrightarrow k^+$,
which is not broken by the $\eta$ regulator.
One can easily deduce the behavior as $\eta\to0$ from \eq{soft_integrals_a}.
Since $\sin(\eta) \sim \eta$, a pole in $\eta$ can only arise if both $\Gamma$ functions have poles,
which requires $\alpha = 0$.
A finite result is obtained if exactly one $\Gamma$ function yields a pole, which requires $\alpha$ to be even.
For odd $\alpha$, the expression vanishes at $\eta = 0$.
Hence, the exact behavior for $\eta\to0$ is given by
\begin{alignat}{2} \label{eq:soft_integrals}
\Isoftz{0} &= \frac{2}{\eta} + \ln\frac{\nu^2}{k_T^2} + \cO(\eta)
\,, \nn\\
\Isoftz{\alpha} &=0 \qquad &&(\alpha~\mathrm{odd})
\,, \nn\\
\Isoftz{\alpha} &= \frac{2}{|\alpha|}\biggl(\frac{k_T}{Q}\biggr)^\alpha + \cO(\eta)
\qquad &&(\alpha~\mathrm{even})
\,.\end{alignat}
In particular, since the $\eta$ regulator behaves like dimensional regularization,
it is well-behaved for power-law divergences and the soft integrals only give rise
to poles from the logarithmic divergences.

In the collinear sector, the behavior is more complicated at subleading power,
because the regulator factor $2k^z = k^- - k^+$ is not homogeneous in $\lambda$.
At leading power~\cite{Chiu:2011qc,Chiu:2012ir,Rothstein:2016bsq}, one takes advantage of the fact that $2 k^z \to k^-$ in the $n$-collinear limit and $2 k^z \to k^+$ in the $\bn$-collinear limit, so that the expanded result correctly regulates the collinear cases, and makes it symmetric under the exchange $n \leftrightarrow \bn$. A fact that will be important for our analysis is that this power expansion induces higher order terms.  These terms have never been considered in the literature since they are not important at leading power.
However, at subleading power one can no longer neglect the subleading component of the regulator.
Implementing the $\eta$ regulator at subleading power in the collinear limits
thus requires to expand the regulator \eq{Rz} itself,
\begin{align} \label{eq:Rz_expanded}
R_z(k_n,\eta) &= w^2 \nu^{\eta} \biggl| k_n^- - \frac{k_T^2}{k_n^-} \biggl|^{-{\eta}}
 = w^2 \, \biggl|\frac{k_n^-}{\nu}\biggl|^{-\eta} \, \biggl[1 + {\eta}\, \frac{k_T^2}{(k_n^-)^2} +\cO(\lambda^4) \biggr]
\,, \nn\\
 R_z(k_\bn,\eta) &= w^2 \biggl| \frac{k_\bn^+}{\nu} \biggr|^{-\eta} \biggl[1 + {\eta}\, \frac{k_T^2}{(k_\bn^+)^2} +\cO(\lambda^4) \biggr]
\,.\end{align}
Applying this to the general LP integral in the $n$-collinear sector, \eq{toy_integrals_collinear} with $\alpha=0$, we obtain
\begin{align}\label{eq:218}
\Inz{0}
&= w^2 \int_0^Q \frac{\df k^-}{k^-} \biggl|\frac{k^-}{\nu}\biggl|^{-\eta} g_n \biggl(\frac{k^-}{Q}\biggr)
\nn\\ & \quad
+ \eta\, w^2\, \frac{k_T^2}{Q^2} \int_0^Q \frac{\df k^-}{k^-}  \biggl|\frac{k^-}{\nu}\biggl|^{-\eta} \biggl(\frac{k^-}{Q}\biggr)^{-2} g_n \biggl(\frac{k^-}{Q}\biggr)
+ \cO(\lambda^4)
\,,\end{align}
and analogously for $\Ibnz{0}$.
Here, the first line is the standard LP integral, while the second line arises from expanding the regulator and is suppressed by $k_T^2/Q^2 \sim \lambda^2$.
While it is also proportional to $\eta$, the remaining integral can produce a $1/\eta$ rapidity divergence to yield an overall finite contribution.

In \sec{qT}, we will see explicitly that these terms from expanding the regulator are crucial to obtain the correct final result at subleading power.
However, in practice they are cumbersome to track in the calculation and yield complicated structures.
To establish an all-orders factorization theorem, the mixing of different orders in the power expansion due to the regulator becomes a serious complication. Hence, it is desirable to employ a rapidity regulator that is homogeneous in $\lambda$. We will present such a regulator in the following \sec{upsilonreg}.

\subsection{Pure Rapidity Regularization}
\label{sec:upsilonreg}

We wish to establish a rapidity regulator that is homogeneous at leading power such that it does not mix LP and NLP integrals, as observed in \sec{etareg} for the $\eta$ regulator.
This can be achieved by implementing the regulator similar to the $\eta$ regulator of \refscite{Chiu:2011qc,Chiu:2012ir,Rothstein:2016bsq}, but instead of regulating the momentum $k^z$ with factors of $w |2k^z/\nu|^{-\eta/2}$, one regulates the rapidity $y_k$ of the momentum $k^\mu$, where
\begin{align}
y_k \equiv \frac{1}{2}\ln\frac{\bn\cdot k}{n\cdot k }
\,.\end{align}
To implement a regulator involving rapidity we use\footnote{
	Note that we can implement the pure rapidity regulator in terms of label and residual momentum operators for example as
	\begin{equation}
	w^2\, \upsilon^{\eta}\,
	\biggl|\frac{\bn\cdot(\cP + \partial)}{n\cdot(\cP + \partial)}\biggr|^{-\eta/2}
	\,.\end{equation}
	where the label momentum operator $\cP$ picks out the large $\cO(\lambda^0)$ momentum component
	of the operator it acts on, while $\partial$ picks out the $\cO(\lambda)$ or $\cO(\lambda^2)$ components. 
	In this case, the operator
	\begin{align}
	\hat Y = \frac{1}{2}\ln\frac{\bn\cdot(\cP + \partial)}{n\cdot(\cP + \partial)}
	\end{align}
	picks out the rapidity of the operator it acts on.
}
factors of
\begin{equation}\label{eq:def_of_upsilon}
w^2\upsilon^{\eta}\left|\frac{\bn\cdot k}{n\cdot k}\right|^{-\eta/2} 
 = w^2 \upsilon^{\eta} e^{-y_k\eta}
\,.\end{equation}
Here we have defined a rapidity scale $\upsilon$ (\verb|\upsilon|)
which is the analog of the scale $\nu$ (\verb|\nu|) in the $\eta$ regulator.
Although $\upsilon$ is dimensionless, in contrast to the dimensionful $\nu$, it still shares the same properties as pure dimensional regularization. In particular, it will give rise to poles in $\eta$ that can be absorbed in $\MS$-like rapidity counterterms.
To ensure $\upsilon$ independence of \eq{def_of_upsilon}, we introduced a bookkeeping parameter $w=w(\upsilon)$ in analogy to the bookkeeping parameter $w(\nu)$ in the $\eta$ regulator, see \eq{omegadef} and \refcite{Chiu:2012ir}.
Also note that this regulator does not affect UV renormalization, which in \scetii~arises from transverse momenta going to infinity and thus is orthogonal to regulating rapidity.

We call \eq{def_of_upsilon} the \emph{pure rapidity regulator}, and pure rapidity regularization the procedure of regulating rapidity divergences using \eq{def_of_upsilon}. When only the $1/\eta$ poles are subtracted we then refer to the renormalized result as being in the \emph{pure rapidity renormalization scheme}.

If we want to make the rapidity scale $\upsilon$ into a true rapidity scale $\Upsilon$, then we can change variables as
\begin{align}
\upsilon \equiv e^\Upsilon  \,.
\end{align}
With this definition \eq{def_of_upsilon} becomes
\begin{align}  \label{eq:def_of_CapUpsilon}
w^2 \upsilon^{\eta} e^{-y_k\eta}
\equiv w^2 e^{\eta(\Upsilon - y_k)} 
 \,,
\end{align}
and the factor regulating divergences depends on a rapidity difference between the scale parameter $\Upsilon$ and $y_k$.

It is interesting to consider the behavior of amplitudes regulated with \eq{def_of_CapUpsilon} under a reparameterization transformation known as RPI-III~\cite{Manohar:2002fd}, which takes $n^\mu \to e^{-\beta} n^\mu$ and $\bn^\mu \to e^{\beta} \bn^\mu$ for some, not necessarily infinitesimal, constant $\beta$.
For a single collinear sector, this can be interpreted as a boost transformation.
Since RPI transformations can be applied independently for each set of collinear basis vectors $\{n_i,\bar n_i\}$ they in general constitute a broader class of symmetry transformations in SCET. 
Prior to including a regulator for rapidity divergences all complete SCET amplitudes are invariant under such transformations. 
All previous rapidity regulators violate this symmetry.
For the pure rapidity regulator in \eq{def_of_CapUpsilon} we have $y_k\to y_k+\beta$, so the transformation is quite simple.%
\footnote{
	Any operators that are defined such that they transform under RPI-III, will do so by a factor $e^{k\beta}$, where $k$ is their RPI-III charge.
	The pure rapidity regulator therefore has an RPI-III charge of $-\eta$. 
	This leads to rapidity-renormalized collinear and soft functions in SCET which carry this charge.
	When considering any observable like a cross section, the combined charge of the renormalized functions describing this observable is zero. 
} 
It can be compensated by defining the rapidity scale to transform like a rapidity,
$\Upsilon \to \Upsilon+\beta$.
Therefore, the $\upsilon^\eta$ factor in the regulator does for RPI-III what the usual
$\mu^\epsilon$ factor does for the mass-dimensionality in dimensional regularization.

As an example of the application of this new regulator, we consider again a real emission with momentum $k^\mu$. The regulator function $R(k,\eta)$ that follows from \eq{def_of_upsilon} is given by
\begin{align} \label{eq:vita_regulatorrapidity}
R_Y(k,\eta) &= w^2\, \upsilon^{\eta}\, \biggl|\frac{k^-}{k^+} \biggr|^{-\eta/2}
 = w^2\, \upsilon^{\eta}\, e^{-\eta\, y_k}
\,.\end{align}
The real-emission phase space is then regulated as
\begin{align}
\int \df^d k\, \delta_+(k^2) &\quad\to\quad  \int \df^d k \,\delta_+(k^2)\,  R_Y(k,\eta)
 =  \int \df^d k\, \delta_+(k^2) \, w^2\, \upsilon^{\eta}\, e^{-\eta\, y_k}
\,.\end{align}
A peculiar feature of the pure rapidity regulator is that it renders the prototypical soft integrals scaleless such that they vanish. That is, using \eq{vita_regulatorrapidity} in
\eq{toy_integrals_soft}, we obtain
\begin{align}\label{eq:vanishingsoft}
\IsoftY{\alpha}
&= \int_0^\infty \frac{\df k^-}{k^-} \biggl(\frac{k^-}{Q}\biggr)^\alpha R_Y(k,\eta)
= w^2\, \upsilon^\eta\, k_T^\eta\, Q^{-\alpha} \int_0^\infty \df k^-\, (k^-)^{\alpha - \eta - 1} = 0
\,.\end{align}
The final integrals are scaleless and vanish for all integer values of $\alpha$, just like scaleless integrals vanish in dimensional regularization.\footnote{Technically one can find terms of the form $1/\eta - 1/\eta$, which can be set to zero via analytic continuation in the standard manner.}

Considering the collinear sectors, the prototypical collinear integrals in \eq{toy_integrals_collinear} with $R_Y(k,\eta)$ become
\begin{align} \label{eq:RYcollinear}
\InY{\alpha}
&= w^2\, \upsilon^\eta\, k_T^{+\eta}\,Q^{-\alpha}
\int_0^Q \df k^-\, (k^-)^{\alpha-\eta-1} \, g_n \biggl(\frac{k^-}{Q}\biggr)
\,,\nn\\
\IbnY{\alpha}
&= w^2\, \upsilon^\eta\, k_T^{-\eta}\,Q^{-\alpha}
\int_0^Q \df k^+\, (k^+)^{\alpha+\eta-1} \, g_\bn \biggl(\frac{k^+}{Q} \biggr)
\,.\end{align}
Although the regulator does not act symmetrically in the $n$-collinear and $\bn$-collinear sectors, the asymmetry is easy to track by taking $\eta \leftrightarrow -\eta$ and $\upsilon\leftrightarrow 1/\upsilon$ when swapping $n\leftrightarrow \bn$ and $k^+\leftrightarrow k^-$.
Since $R_Y(k,\eta)$ is homogeneous in $\lambda$, it does not generate any subleading power terms, in contrast to \eq{218} for the $\eta$ regulator. In particular, the LP integral becomes
\begin{align}
\InY{0}
&= w^2\, \upsilon^\eta\, k_T^{+\eta}
\int_0^Q \frac{\df k^-}{(k^-)^{1+\eta}} \, g_n \biggl(\frac{k^-}{Q}\biggr)
\nn \\*
&= w^2\, \Bigl(\upsilon \frac{k_T}{Q}\Bigr)^\eta
\int_0^Q \df k^- \biggl[-\frac{1}{\eta}\, \delta(k^-) + \frac{1}{Q} \cL_0\biggl(\frac{k^-}{Q}\biggr)\,
 + \ord{\eta} \biggr]g_n \biggl(\frac{k^-}{Q}\biggr)
\,,\end{align}
where we used the standard distributional identity $1/x^{1+\eta} = -\delta(x)/\eta + \cL_0(x) + \ord{\eta}$ to extract the $1/\eta$ divergence. (See \sec{distribution} below for a more general
discussion.)
Taking $\eta \to -\eta$, the analogous $1/\eta$ pole in the $\bn$-collinear sector
has the opposite sign, such that the $1/\eta$ poles cancel when adding the $n$-collinear and $\bn$-collinear contributions.
This is a general feature in all cases where the soft contribution vanishes as in \eq{vanishingsoft}.

Some comments about the features of the pure rapidity regulator are in order:
\begin{itemize}
  \item It involves the rapidity
  \begin{equation}
   e^{2y_k} \equiv \frac{\bn\cdot k}{n\cdot k}\,,
   \end{equation}
  and therefore breaks boost invariance as required to regulate rapidity divergences. The boost invariance is restored by the dimensionless $\upsilon$ rapidity scale, analogous to how the dimensionful mass scale $\mu$ in dimensional regularization restores the dimensionality.
  \item Rapidity divergences appear as $1/\eta$ poles, allowing the definition of the pure rapidity renormalization scheme as a dimensional regularization-like scheme.
  \item At each order in perturbation theory, the poles in $\eta$ and the $\upsilon$-dependent pieces cancel when combining the results for the $n$-collinear, $\bn$-collinear, and soft sectors.
  \item The pure rapidity regulator is homogeneous\footnote{In cases where it is possible to combine label and residual momenta in the phase space integral that needs to be rapidity regulated.} in the SCET power counting parameter $\lambda$. Therefore it does not need to be power expanded, and hence does not mix contributions at different orders in the power expansion.
  \item For the case of a single real emission considered here:
  \begin{itemize}
     \item Soft integrals and zero-bin \cite{Manohar:2006nz} integrals are scaleless and vanish.
     \item It follows that the $\eta$ poles and the $\upsilon$ dependent pieces cancel between the $n$-collinear and $\bn$-collinear sectors.
     \item The results for the $n$-collinear and $\bn$-collinear sectors are not identical but are trivially related by taking $\eta \leftrightarrow -\eta$ and $\upsilon\leftrightarrow 1/\upsilon$ when swapping $n\leftrightarrow \bn$.
  \end{itemize}
\end{itemize}
The introduction of this new pure rapidity regulator allows us to regulate rapidity divergences at any order in the EFT power expansion, while maintaining the power counting of the EFT independently at each order.

Although in this chapter we will only use pure rapidity regularization for a single real emission at fixed order, we note that one can derive a rapidity renormalization group for the pure rapidity regulator by imposing that the cross section must be independent of $\upsilon$.
Similar to the $\eta$ regulator, this regulator is not analytical and can also be used to properly regulate virtual and massive loops. This will be discussed in detail elsewhere.

To conclude this section we note that the pure rapidity regulator can be seen as a particular case of a broader class of homogeneous rapidity regulators given by
\begin{align} \label{eq:vita_regulator}
R_c(k,\eta)
= w^2\, \upsilon^{(1-c)\eta/2}\,
\biggl|\frac{k^-}{\nu}\biggr|^{-\eta/2} \biggl|\frac{k^+}{\nu}\biggr|^{-c \eta/2}
\,,\end{align}
where $c\ne1$ is an arbitrary parameter governing the antisymmetry between the $n$-collinear and $\bn$-collinear sectors.
As for the pure rapidity regulator, this regulator is homogeneous in $\lambda$ and renders the same class of soft integrals scaleless.
However, it requires an explicit dimensionful scale $\nu$ to have the correct mass dimension.
Note that for $c=1$, \eq{vita_regulator} only depends on the boost invariant product $k^+k^-$ and therefore does not regulate rapidity divergences.
For $c=-1$, it recovers the pure rapidity regulator and the dependence on $\nu$ cancels.
Lastly, for $c=0$ and massless real emissions, \eq{vita_regulator} essentially reduces to the regulator of \refcite{Becher:2011dz}.
We choose $c=-1$ because it yields the same finite terms in the $n$-collinear and $\bn$-collinear functions, and thus has enhanced symmetry.
Choosing a different value of $c$ shifts terms between the two sectors, see also \app{master_formula_c}. The combined result is always independent of $c$.

\subsection{Distributional Treatment of Power Law Divergences}
\label{sec:distribution}

To complete our treatment of rapidity divergences at subleading power, we show how their distributional structure can be consistently treated when expanded against a general test function.
In particular, we will see that the power-law rapidity divergences lead to derivatives of PDFs. 

In the collinear limit at NLP, we obtain divergent integrals of the form
\begin{equation} \label{eq:toy_dist}
 \int_0^Q \frac{\df k^-}{Q} \frac{g_n(k^-/Q)}{(k^-/Q)^{a + \eta}}
\,,\end{equation}
which appear for both the $\eta$ regulator (with $a=1-\alpha=1,2,3$ at NLO) and the pure rapidity regulator (with $a=1-\alpha=1,2$ at NLO).

The function $g_n(k^-/Q)$ is defined to be regular for $k^-/Q\to 0$.
If it is known analytically, we can in principle evaluate the integral in \eq{toy_dist} analytically and expand the result for $\eta \to 0$ to obtain the regularized expression.
However, $g_n(k^-/Q)$ is typically not given in analytic form.
In particular, for $pp$ collisions it contains the parton distribution functions (PDFs) $f(x)$.
Therefore, to extract the rapidity divergence, we need to expand $1/{(k^-)^{a + \eta}}$ in $\eta$ in a distributional sense.
To do so, we first change the integration variable from $k^-$ to the dimensionless variable $z$ defined through $k^- = Q (1-z)$,
such that \eq{toy_dist} becomes
\begin{align} \label{eq:toy_dist2}
\int_0^1 \df z\,\frac{\tilde g(z)}{(1-z)^{a+\eta}} \,,\qquad \tilde g(z) = g_n(1-z)
\,.\end{align}
In \eq{toy_dist2}, the rapidity divergence arises as $z\to1$.
For $a=1$, it can be extracted using the standard distributional identity
\begin{align} \label{eq:plus_dist_1}
 \frac{1}{(1-z)^{1+\eta}} = - \frac{\delta(1-z)}{\eta} + \cL_0(1-z) + \cO(\eta)
\,,\end{align}
where $\cL_0(y) = [\theta(y)/y]_+$ is the standard plus distribution and we remind the reader that its convolution against a test function $\tilde g(z)$ is given by
\begin{align}
 \int_x^1 \df z\, \tilde g(z) \cL_0(1-z) = \int_x^1 \df z \, \frac{\tilde g(z) - \tilde g(1)}{1-z} + \tilde g(1) \underbrace{\int_x^1 \df z \, \cL_0(1-z)}_{\ln(1-x)}
 \,,\quad x \in [0,1]
\,.\end{align}
For $a>1$, these distributions need to be generalized to higher-order plus distributions subtracting higher derivatives as well.
For example, for $a=2$ one obtains
\begin{align} \label{eq:double_plus}
  \frac{1}{(1-z)^{2+\eta}} &
 = \frac{\delta'(1-z)}{\eta} - \delta(1-z) + \cL_0^{++}(1-z) + \cO(\eta)
\,,\end{align}
where the second-order plus function $\cL_0^{++}(1-z)$ regulates the quadratic divergence $1/(1-z)^2$.
Its action on a test function $\tilde g(z)$ is given by a double subtraction,
\begin{align}
 \int_x^1 \df z \, \tilde g(z) \cL_0^{++}(1-z) &
 = \int_x^1 \df z \, \frac{\tilde g(z) - [\tilde g(1) + \tilde g'(1)(z-1)]}{(1-z)^2}
 \nn\\*&\quad
 + \tilde g(1) \underbrace{\int_x^1 \df z\, \cL_0^{++}(1-z)}_{-x/(1-x)}
 +\, \tilde g'(1) \underbrace{\int_x^1 \df z\, (z-1)  \cL_0^{++}(1-z)}_{-\ln(1-x)}
\,.\end{align}
In \app{plus_distr}, we give more details on these distributions, generalizing to arbitrary $a \ge 1$.
Note that the second-order plus function has also appeared for example in \refcite{Mateu:2012nk}.

Eq.~\eqref{eq:double_plus} implies the appearance of derivatives of delta functions, $\delta'(1-z)$, which will induce derivatives of the PDFs that are contained in $\tilde g(z)$. The appearance of such derivatives in subleading power calculations was first shown in \refcite{Moult:2016fqy} in the context of \sceti-like observables. However, in such cases they arose simply from a Taylor expansion of the momentum being extracted from the PDF.
Here, they also arise from power-law divergences, a new mechanism to induce derivatives of PDFs.
Recently, power-law divergences inducing derivatives of PDFs have appeared also in the study of \sceti-like observables involving multiple collinear directions at subleading power~\cite{Bhattacharya:2018vph}. We believe they are a general feature of calculations beyond leading power.

In practice, the higher-order distributions can be cumbersome to work with.
Instead, we find it more convenient to use integration-by-parts relations to reduce the divergence in \eq{toy_dist2} to the linear divergence $1/(1-z)$, which yields explicit derivatives of the test function.
For the cases $a=2$ and $a=3$ we encounter in \sec{qT}, this gives
\begin{align} \label{eq:plus_dist_2}
 \int_x^1 \df z \frac{\tilde g(z)}{(1-z)^{2+\eta}} &
 = \tilde g'(1) \biggl(\frac{1}{\eta} - 1\biggr) - \frac{\tilde g(x)}{1-x}
   - \int_x^1 \df z \, \tilde g'(z) \cL_0(1-z) + \cO(\eta)
\,,\\ \label{eq:plus_dist_3}
 \int_x^1 \df z \frac{\tilde g(z)}{(1-z)^{3+\eta}} &
 = \tilde g''(1) \biggl(- \frac{1}{2\eta} + \frac34\biggr) - \frac{\tilde g(x)+(x-1) \tilde g'(x)}{2(1-x)^{2}}
 \nn\\*&\quad + \frac{1}{2} \int_x^1 \df z \, \tilde g''(z) \cL_0(1-z) + \cO(\eta)
\,.\end{align}
\Equations{plus_dist_2}{plus_dist_3} can be used to write the kernels fully in terms of a standard $\cL_0$, but they must be applied within the integral to directly yield derivatives of the test function $\tilde g(z)$.

In our application in \sec{qT}, $\tilde g(z)$ will always involve the PDF $f(x/z)$ and vanish at $z=x$.
We can thus also write \eqs{plus_dist_2}{plus_dist_3} as operator equations,
\begin{align} \label{eq:plus_dist_2_operator}
 \frac{1}{(1-z)^{2+\eta}} &\quad\to\quad
 \biggl[ \biggl(\frac{1}{\eta}-1\biggr) \delta(1-z) - \cL_0(1-z)\biggr] \frac{\df}{\df z}
\,, \\  \label{eq:plus_dist_3_operator}
 \frac{1}{(1-z)^{3+\eta}} &\quad\to\quad
 \biggl[ \delta(1-z) \biggl(- \frac{1}{2 \eta} + \frac{3}{4} \biggr) + \frac{1}{2} \cL_0(1-z) \biggr] \frac{\df^2}{\df^2 z}
 + \frac{\tilde g'(x)}{2(1-x)} \delta(1-z)
\,.\end{align}
Note that the second relation is quite peculiar, as we have to add the boundary term proportional to $g'(x)$,
and thus cannot be interpreted as a distributional relation.
In our calculation in \sec{qT}, this term will not contribute due to an overall suppression by $\eta$, such that only the divergent term in \eq{plus_dist_3_operator} needs to be kept.

\section{Power Corrections for Color-Singlet \texorpdfstring{\boldmath $q_T$}{qT} Spectra}
\label{sec:qT}

In this section we use our understanding of rapidity regularization at subleading power
to compute the perturbative power corrections to the transverse momentum $q_T$ in color-singlet production
at invariant mass $Q$, which is one of the most well studied observables in QCD.
Schematically, the cross section differential in $q_T$ can be expanded as
\begin{align}
 \frac{\df\sigma}{\df q_T^2} = \frac{\df\sigma^{(0)}}{\df q_T^2} + \frac{\df\sigma^{(2)}}{\df q_T^2} + \cdots
\,,\end{align}
where $\sigma^{(0)}$ is the leading-power cross section and $\sigma^{(2n)}$ the N$^n$LP cross section.
In general, in this section we will denote power suppression in $\cO(\lambda^n)$ with $\lambda \sim q_T/Q$ relative to the leading-power result through superscripts $^{(n)}$.
The $\sigma^{(2n)}$ terms scale like
\begin{align}
 \frac{\df\sigma^{(2n)}}{\df q_T^2} \sim \frac{1}{q_T^2} \biggl(\frac{q_T^2}{Q^2}\biggl)^{n}
\,,\end{align}
and hence only the LP cross section is singular as $q_T \to 0$.
In particular, $\sigma^{(0)}$ contains Sudakov double logarithms $\log^2(Q/q_T)$.

The factorization of $\sigma^{(0)}$ in terms of transverse-momentum dependent PDFs (TMDPDFs) was first shown by Collins, Soper, and Sterman in \refscite{Collins:1981uk,Collins:1981va,Collins:1984kg} and later elaborated on by Collins in \refcite{Collins:1350496}.
Its structure was also studied in \refscite{Catani:2000vq,deFlorian:2001zd,Catani:2010pd}.
The factorization was also studied in the framework of SCET by various groups, see e.g.\ \refscite{Becher:2010tm, GarciaEchevarria:2011rb, Chiu:2012ir}.
Using the notation of \refcite{Chiu:2012ir}, the factorized LP cross section for the production of a color-singlet final state $L$ with invariant mass $Q$ and total rapidity $Y$ in a proton-proton collision can be written as%
\footnote{We suppress that for gluon-gluon fusion, $H$ and $B$ carry polarization indices.}
\begin{align} \label{eq:sigma0}
 \frac{\df \sigma^{(0)}}{\df Q^2 \df Y \df^2 \qt} &
 = \sigma_0 \!\sum_{i,j} H_{ij}(Q, \mu)
   \!\int\!\! \df^2\bt \, e^{\img \qt \cdot \bt}
   \tilde B_i\bigl(x_a, \bt, \mu, \nu\bigr) \tilde B_j\bigl(x_b, \bt, \mu, \nu\bigr)
   \tilde S(b_T, \mu, \nu)
\,,\!\end{align}
where $x_{a,b} = Q e^{\pm Y}/\Ecm$ are the momentum fractions carried by the incoming partons.
In \eq{sigma0}, $H_{ij}$ is the hard function describing virtual corrections to the underlying hard process $ij \to L$,
the $\tilde B_i$ are TMD beam functions in Fourier space and $\tilde S$ is the TMD soft function in Fourier space.
While $H_{ij}$ only depends on the $\MS$ renormalization scale $\mu$, the beam and soft functions also depend on the rapidity renormalization scale $\nu$.

For nonperturbative $q_T \sim b_T^{-1} \sim \LQCD$, the $\tilde B_i$ become genuinely nonperturbative functions,
while for perturbative $q_T \sim b_T^{-1} \gg \LQCD$ they can be matched perturbatively onto PDFs,
\begin{align}\label{eq:q_match}
 \tilde B_i(x,\bt,\mu,\nu) &= \sum_j \int_x^1 \frac{\df z}{z} \, \tilde \cI_{ij}\Bigl(\frac{x}{z},\bt,\mu,\nu\Bigr) f_j(z,\mu)
\,.\end{align}
The perturbative kernels $\cI_{ij}$ are known to two loops \cite{Catani:2011kr,Catani:2012qa,Gehrmann:2012ze,Gehrmann:2014yya},
and the soft function $\tilde S$ is known to three loops \cite{Echevarria:2015byo,Luebbert:2016itl,Li:2016ctv}.
This has allowed resummation to next-to-next-to-next-to leading logarithmic accuracy \cite{Bizon:2017rah,Chen:2018pzu,Bizon:2018foh}.

Recently, there has been some progress towards a nonperturbative factorization of the NLP cross section $\df\sigma^{(2)}/\df q_T^2$, which involves higher twist PDFs \cite{Balitsky:2017gis,Balitsky:2017flc}.
Here, we are interested in studying the perturbative power corrections to the NLP terms,
where one can perform an OPE to match onto standard PDFs.
At subleading power, the perturbative kernels also involve (higher) derivatives of distributions,
which can always be reduced to standard distributions acting on derivatives of PDFs.
The NLP cross section at $\cO(\as)$ thus takes the form
\begin{align} \label{eq:sigma_NLP_introPtNLL}
 &\frac{\df\sigma^{(2,1)}}{\df Q^2 \df Y \df q_T^2}
= \hat\sigma^\LO(Q,Y) \, \frac{\as}{4\pi}
    \int_{x_a}^1 \frac{\df z_a}{z_a} \int_{x_b}^1 \frac{\df z_b}{z_b}
\nn\\&\quad\times \biggl[
   f_i\biggl(\frac{x_a}{z_a}\biggr) f_j\biggl(\frac{x_b}{z_b}\biggr)
   C_{f_i f_j}^{(2,1)}(z_a, z_b, q_T)
  + \frac{x_a}{z_a} f'_i\biggl(\frac{x_a}{z_a}\biggr) \frac{x_b}{z_b} f'_j\biggl(\frac{x_b}{z_b}\biggr)
  C_{f'_i f_j'}^{(2,1)}(z_a, z_b, q_T)
  \nn\\&\qquad
   + \frac{x_a}{z_a} f'_i\biggl(\frac{x_a}{z_a}\biggr) f_j\biggl(\frac{x_b}{z_b}\biggr)
   C_{f_i' f_j}^{(2,1)}(z_a, z_b, q_T)
  + f_i\biggl(\frac{x_a}{z_a}\biggr) \frac{x_b}{z_b} f'_j\biggl(\frac{x_b}{z_b}\biggr)
  C_{f_i f_j'}^{(2,1)}(z_a, z_b, q_T)
\biggr]
\,,\end{align}
where $\hat \sigma^\LO$ is the LO partonic cross section which serves as an overall normalization.
The $C^{(2,1)}_{a b}$ are perturbative coefficients, expressed in terms of distributions, and we suppress the explicit $Q$ and $Y$ dependence in the kernels $C^{(2,1)}_{a b}$.
In general, at order $\as^n$ their logarithmic structure is
\begin{align}
 C^{(2,n)}_{ab}(z_a, z_b, q_T) = \sum_{m=0}^{2n-1} C^{(2,n)}_{ab,m}(z_a, z_b) \ln^m\frac{Q^2}{q_T^2}
\,.\end{align}
More explicitly, at NLO they have the form
\begin{align}
 C^{(2,1)}_{ab}(z_a, z_b, q_T) = C^{(2,1)}_{ab,1}(z_a, z_b) \ln\frac{Q^2}{q_T^2} + C^{(2,1)}_{ab,0}(z_a, z_b)
\,,\end{align}
i.e.\ they only contain a single logarithm $\ln(Q^2/q_T^2)$ and a $q_T$-independent piece.
(Note that due to the dependence on $z_{a,b}$, it will yield a $Q^2$ and $Y$ dependence.)
We emphasize that in the form given here, all logarithms have been extracted,
and the $q_T$ distribution is directly expressed in terms of PDFs and their derivatives.

In the following, we will derive a master formula to obtain the NLO NLP kernels $C^{(2,1)}_{ab}$
for arbitrary color-singlet processes, as well as the explicit results for Higgs and Drell-Yan production.
The study of higher perturbative orders, and the derivation of a factorization and resummation is left to future work.
However, we do wish to comment on one complication which occurs for $q_T$ at higher orders, that we have not addressed. Unlike for beam thrust, at NNLO and beyond, one can have power-suppressed contributions at small $q_T$ from two hard partons in the final state that are nearly back-to-back such that their transverse momenta balance to give a small total $q_T$. At NNLO, this is at most a constant power correction, since it is not logarithmically enhanced. but at higher orders it can have a logarithmic contribution. These power corrections are of a different nature than those discussed here, and are not captured as an expansion in the soft and collinear limits about the Born process.

The remainder of this section is organized as follows. In \sec{master_formula}, we derive the
master formula for the NLP corrections using the $\eta$ regulator, showing in particular that the terms from expanding the regulator contribute. In \sec{master_formula_2}, we rederive this master formula in  pure rapidity regularization, which will be simpler due to the fact that one does not have additional terms from the expansion of the regulator, and due to the fact that the soft sector is scaleless.
In \sec{results_singlet}, we then apply the master formula to derive explicit results for Drell-Yan and gluon-fusion Higgs production. In \sec{discuss}, we discuss our results and compare them with
the known NLP results for beam thrust. Finally in \sec{numerics}, we provide a numerical
validation of our results.

\subsection{Master Formula for Power Corrections to Next-to-Leading Power}
\label{sec:master_formula}

We consider the production of a color-singlet final state $L$ at fixed invariant mass $Q$ and rapidity $Y$,
measuring the magnitude of its transverse momentum $q_T^2 = |\qt|^2$.
The underlying partonic process is
\begin{align}
 a(p_a) + b(p_b) \to L(p_1, \cdots) + X(k_1, \cdots)
\,,\end{align}
where $a,b$ are the incoming partons and $X$ denotes additional QCD radiation.
Following the notation of \refcite{Ebert:2018lzn}, we express the cross section as
\begin{align} \label{eq:sigma1}
 \frac{\df\sigma}{\df Q^2 \df Y \df q_T^2} &
 = \int_{0}^{1}\!\! \df \zeta_a \df \zeta_b\, \frac{f_a(\zeta_a)\, f_b(\zeta_b)}{2 \zeta_a \zeta_b \Ecm^2}
   \int\!\biggl(\prod_i \frac{\df^d k_i}{(2\pi)^d} (2\pi) \delta_+(k_i^2) \biggr)
   \int\!\!\frac{\df^d q}{(2\pi)^d} \, |\cM(p_a, p_b; \{k_i\}, q)|^2
   \nn\\* &\quad\times
   (2\pi)^d \delta^{(d)}(p_a + p_b - k - q) \, \delta(Q^2 - q^2)
   \, \delta\biggl(Y - \frac{1}{2}\ln\frac{q^-}{q^+}\biggr)
   \, \delta\bigl(q_T^2 - |\kt|^2\bigr)
\,.\end{align}
Here, the incoming momenta are given by
\begin{align} \label{eq:p_ab}
 p_a^\mu = \zeta_a \Ecm \frac{n^\mu}{2}
\,,\qquad
 p_b^\mu = \zeta_b \Ecm\frac{\bn^\mu}{2}
\,,\end{align}
$k = \sum_i k_i$ is the total outgoing hadronic momentum, and $q$ is the total leptonic momentum.
In particular, $\kt = \sum_i \vec k_{i,T}$ is the vectorial sum of the transverse momenta of all emissions.
Since the measurements are not affected by the details of the leptonic final state,
the leptonic phase-space integral has been absorbed into the matrix element,
\begin{align} \label{eq:M2_1}
 |\cM(p_a, p_b; \{k_i\}, q)|^2 &= \int\df\Phi_L(q) \, |\cM(p_a, p_b; \{k_i\}, \{p_j\})|^2
\,,\nn\\
  \df\Phi_L(q) &= \prod_j \frac{\df^d p_{j}}{(2\pi)^d} (2\pi) \delta_+(p_{j}^2 - m_j^2)
\, (2\pi)^d \delta^{(d)}\Bigl(q - \sum_j p_j\Bigr)
\,.\end{align}
The matrix element $\cM$ also contains the renormalization scale $\mu^{2\eps}$,
as usual associated with the renormalized coupling $\alpha_s(\mu)$, and may also contain virtual corrections.

There is an important subtlety when measuring the transverse momentum $q_T$ using dimensional regularization,
as the individual transverse momenta $\vec k_{i,T}$ are continued to $2-2\eps$ dimensions.
The measurement function $\delta(q_T^2 - |\kt|^2)$ in \eq{sigma1} can thus be interpreted
either as measuring the magnitude in $2-2\eps$ dimensions or the projection onto $2$ dimensions.
This scheme dependence cancels in the final result, but can lead to different intermediate results.
At the order we are working, both choices give identical results, so for simplicity of the following manipulations
we specify to measuring the magnitude in $2-2\eps$ dimension.
For detailed discussions, see e.g.\ \refscite{Jain:2011iu,Luebbert:2016itl}.

The $\delta$ functions measuring the invariant mass $Q$ and rapidity $Y$ fix the incoming momenta to be
\begin{align} \label{eq:zeta_ab}
 \zeta_a(k) &= \frac{1}{\Ecm} \Bigl(k^- +  e^{+Y} \sqrt{Q^2 + k_T^2} \Bigr)
\,,\quad
 \zeta_b(k) = \frac{1}{\Ecm} \Bigl(k^+ +  e^{-Y} \sqrt{Q^2 + k_T^2} \Bigr)
\,.\end{align}
Equation~(\ref{eq:sigma1}) can now be simplified to
\begin{align} \label{eq:sigma2}
 \frac{\df\sigma}{\df Q^2 \df Y \df q_T^2} &
 = \int\!\biggl(\prod_i \frac{\df^d k_i}{(2\pi)^d} (2\pi) \delta_+(k_i^2) \biggr)
   \frac{f_a(\zeta_a)\, f_b(\zeta_b)}{2 \zeta_a \zeta_b \Ecm^4}
   \Msquared(Q, Y; \{k_i\}) \, \delta\bigl(q_T^2 - |\kt|^2\bigr)
\,,\end{align}
where we introduced the abbreviation
\begin{align} \label{eq:Msquared}
 \Msquared(Q, Y; \{k_i\}) \equiv |\cM(p_a, p_b, \{k_i\}, q=p_a+p_b - k)|^2
\,.\end{align}
This emphasizes that the squared matrix element depends only on the Born measurements $Q$ and $Y$,
which fix the incoming momenta through \eqs{p_ab}{zeta_ab}, and the emission momenta $k_i$.
The restriction that $\zeta_{a,b} \in [0,1]$ is kept implicit in \eq{sigma2} through the support of the proton PDFs.

\subsubsection{General Setup at NLO}
\label{sec:generalsetup}

For reference, we start with the LO cross section following from \eq{sigma2},
\begin{align} \label{eq:sigmaLO}
 \frac{\df\sigma^\LO}{\df Q^2 \df Y \df q_T^2} &
 = \frac{f_a(x_a)\, f_b(x_b)}{2 x_a x_b \Ecm^4}
   \Msquared^\LO(Q, Y)\, \delta\bigl(q_T^2\bigr)
\,,\end{align}
where
\begin{align} \label{eq:xab}
 x_a = \frac{Q e^Y}{\Ecm} \,,\quad x_b = \frac{Q e^{-Y}}{\Ecm}
\,,\end{align}
and $\Msquared^\LO$ is the squared matrix element in the Born kinematics, see \eq{Msquared}.
For future reference, we also define the LO partonic cross section, $\hat \sigma^{\rm LO}(Q,Y)$, by
\begin{align} \label{eq:sigmaLO_2}
  \frac{\df\sigma^\LO}{\df Q^2 \df Y \df q_T^2} &
 = \hat\sigma^\LO(Q,Y) \, f_a(x_a)\, f_b(x_b) \, \delta(q_T^2)
\,,\qquad
  \hat\sigma^\LO(Q,Y) = \frac{\Msquared^\LO(Q, Y)}{2 x_a x_b \Ecm^4}
\,.\end{align}
At NLO, the virtual correction only contributes at leading power and is proportional to $\delta(q_T^2)$.
At subleading power, it suffices to consider the real correction, given from \eq{sigma2} by
\begin{align} \label{eq:sigmaNLO_1}
 \frac{\df\sigma}{\df Q^2 \df Y \df q_T^2} &
 = \int\! \frac{\df^d k}{(2\pi)^d}\, (2\pi) \delta_+(k^2)\,
   \frac{f_a(\zeta_a)\, f_b(\zeta_b)}{2 \zeta_a \zeta_b \Ecm^4}\,
   \Msquared(Q, Y; \{k\}) \, \delta\bigl(q_T^2 - |\kt|^2\bigr)
\nn\\&
 = \frac{q_T^{-2\eps}}{(4\pi)^{2-\eps} \Gamma(1-\eps)} \int_0^\infty \frac{\df k^-}{k^-}
   \frac{f_a(\zeta_a)\, f_b(\zeta_b)}{2 \zeta_a \zeta_b \Ecm^4}\,
   \Msquared(Q, Y; \{k\})\bigg|_{k^+ = q_T^2/k^-}
\,.\end{align}
In the following, we will mostly keep the symbol $k^+$ often leaving the use of the relation $k^+ = k_T^2/k^- = q_T^2/k^-$ to the end,
since this makes the symmetry under $k^+ \leftrightarrow k^-$ manifest.
The integral in \eq{sigmaNLO_1} is finite as the physical support of the PDFs, $0 \le \zeta_{a,b} \le 1$, cuts off the integral in $k^-$.
As discussed in \sec{rapDivLP}, these constraints will be expanded for small $q_T \ll Q$, after which the integral becomes rapidity divergent.
To regulate the integral, we use the $\eta$ regulator where one inserts a factor of $w^2 |2 k^z/\nu|^{-\eta}$ into the integral,
\begin{align} \label{eq:sigmaNLO_2}
 \frac{\df\sigma}{\df Q^2 \df Y \df q_T^2} &
 = \frac{q_T^{-2\eps}}{(4\pi)^{2-\eps} \Gamma(1-\eps)} \int_0^\infty \frac{\df k^-}{k^-}
   w^2 \nu^\eta \biggl|k^- - \frac{q_T^2}{k^-}\biggr|^{-\eta}
   \frac{f_a(\zeta_a)\, f_b(\zeta_b)}{2 \zeta_a \zeta_b \Ecm^4}\,
   \Msquared(Q, Y; \{k\})
\,.\end{align}
We now wish to expand \eq{sigmaNLO_2} in the limit of small $\lambda \sim q_T / Q \ll 1$.
Using the knowledge from the EFT, this can be systematically achieved by employing the scaling of \eq{modes_introPtNLL},
\begin{align} \label{eq:modes_pt}
 &n{-}\text{collinear}: \quad k_n \sim Q \, (\lambda^2, 1, \lambda)
\,,\\
 \nn&\bn{-}\text{collinear}: \quad k_{\bn} \sim Q \, (1, \lambda^2, \lambda)
\,,\\
 \nn&\text{soft}: \hspace{1.8cm} k_s \sim Q \, (\lambda, \lambda, \lambda)
\,,\end{align}
for the momentum $k$.
By inserting each of these scalings into \eq{sigmaNLO_2} and expanding the resulting expression
to first order in $\lambda$, one precisely obtains the soft and beam functions as defined
in the $\eta$ regulator. This illustrative exercise is shown explicitly in Appendix A of \cite{Ebert:2018gsn}.
Here, we are interested in the first nonvanishing power correction, which occurs at $\cO(\lambda^2) \sim \cO(q_T^2/Q^2)$. We will explicitly show that the $\cO(\lambda)$ linear power correction vanishes.
To compute the $\cO(\lambda^2)$ result, we will consider the soft and collinear cases separately, deriving master formulas for all scalings
applicable to any color-singlet production.
The power-suppressed operators and Lagrangian insertions required to calculate these directly will be presented in \refcite{Chang:NLP}.

\subsubsection[Soft Master Formula for \texorpdfstring{$q_T$}{qT}]
{\boldmath Soft Master Formula for $q_T$}
\label{sec:soft_master}

We first consider the case of a soft emission $k \sim Q(\lambda,\lambda,\lambda)$.
In this limit, the incoming momenta from \eq{zeta_ab} are expanded as
\begin{align}
 \zeta_a(k) &
 = x_a \biggl[ 1 + \frac{k^- e^{-Y}}{Q} + \frac{k_T^2}{2 Q^2} + \cO(\lambda^3) \biggr]
 \equiv x_a \biggl[1 + \Delta_a^{(1)} + \Delta_a^{(2)} + \cO(\lambda^3) \biggr]
\,,\nn\\
 \zeta_b(k) &
 = x_b \biggl[ 1 + \frac{k^+ e^{+Y}}{Q} + \frac{k_T^2}{2 Q^2} + \cO(\lambda^3) \biggr]
 \equiv x_b \biggl[1 + \Delta_b^{(1)} + \Delta_b^{(2)} + \cO(\lambda^3) \biggr]
\,,\end{align}
where as usual $k^+ = k_T^2/k^- = q_T^2/k^-$, $x_{a,b} = Q e^{\pm Y}/\Ecm$ as in \eq{xab},
and the terms in square brackets correspond to $\cO(\lambda^0)$,
$\cO(\lambda^1)$, and $\cO(\lambda^2)$, respectively.
It follows that the PDFs and flux factor are expanded as
\begin{align} \label{eq:phi_soft}
 \Phi \equiv \frac{f_a(\zeta_a) f_b(\zeta_b)}{\zeta_a \zeta_b} &
 = \frac{f_a(x_a) f_b(x_b)}{x_a x_b}
 + \frac{1}{x_a x_b} \biggl\{ \frac{k^- e^{-Y}}{Q} \bigl[ x_a f'_a(x_a) \, f_b(x_b) - f_a(x_a) f_b(x_b) \bigr]
   + (\rm{sym.}) \biggr\}
 \nn\\&\quad
 + \frac{1}{x_a x_b} \biggl\{
   \frac{(k^- e^{-Y})^2}{Q^2} \Bigl[ f_a(x_a) f_b(x_b) - x_a f'_a(x_a) \, f_b(x_b) + \frac{1}{2} x_a^2 f''_a(x_a) \, f_b(x_b) \Bigr]
\nn\\&\qquad\qquad\quad
   + \frac{k_T^2}{2Q^2} \bigl[ x_a f'_a(x_a) \, x_b f'_b(x_b) - x_a f'_a(x_a) \, f_b(x_b) \bigr]
   + (\rm{sym.})
 \biggr\} + \cO(\lambda^3)
\nn\\&
 \equiv \frac{1}{x_a x_b} \bigl[ \Phi^{(0)} + \Phi^{(1)} + \Phi^{(2)} \bigr] + \cO(\lambda^3)
\,.\end{align}
Here, $(\rm{sym.})$ denotes simultaneously flipping  $a \leftrightarrow b$ and letting $k^- \to k^+, Y \to -Y$.
For brevity, we introduced the abbreviation $\Phi^{(n)}$ for the $\cO(\lambda^n)$ pieces.
Note that we expanded to the second order in $\lambda$, as the $\cO(\lambda^1)$ piece will vanish
and the first nonvanishing correction in fact arises at $\cO(\lambda^2)$.

The expansion of the matrix element is process dependent, and we define the expansion in the soft limit through
\begin{align} \label{eq:M_soft}
 \Msquared_s(Q, Y; \{k\}) = \Msquared_s^{(0)}(Q, Y; \{k\}) + \Msquared_s^{(1)}(Q, Y; \{k\}) + \Msquared_s^{(2)}(Q, Y; \{k\}) + \cO(\lambda)
\,.\end{align}
The LP matrix element scales as $\Msquared_s^{(0)} \sim \lambda^{-2}$,
such that $\int \df k_T^2 \, \Msquared_s^{(0)} \sim \lambda^0$.
The next two matrix elements are each suppressed by an additional order in $\lambda$ relative to the one before.

Plugging the expansions \eqs{phi_soft}{M_soft} back into \eq{sigmaNLO_2} and collecting terms in $\lambda$,
the soft limit through $\cO(\lambda^2)$ is obtained as
\begin{align} \label{eq:sigma_soft_2}
 \frac{\df\sigma_s}{\df Q^2 \df Y \df q_T^2} &
 = \frac{q_T^{-2\eps}}{(4\pi)^{2-\eps} \Gamma(1-\eps)} \frac{1}{2 x_a x_b \Ecm^4}
   \int_0^\infty \frac{\df k^-}{k^-} w^2 \nu^{\eta} \biggl|k^- - \frac{q_T^2}{k^-}\biggr|^{-\eta}
   \\*\nn&\quad
   \times \biggl\{
      \Phi^{(0)} \Msquared_s^{(0)}(Q, Y; \{k\})
    + \Bigl[ \Phi^{(0)} \Msquared_s^{(1)}(Q, Y; \{k\}) + \Phi^{(1)} \Msquared_s^{(0)}(Q, Y; \{k\}) \Bigr]
   \\*\nn&\qquad
    + \Bigl[ \Phi^{(0)} \Msquared_s^{(2)}(Q, Y; \{k\}) + \Phi^{(1)} \Msquared_s^{(1)}(Q, Y; \{k\}) + \Phi^{(2)} \Msquared_s^{(0)}(Q, Y; \{k\}) \Bigr]
   \biggr\}
\,.\end{align}
The first term in curly brackets is the leading-power result, the second term the $\cO(\lambda)$ contribution, and the last line contains the $\cO(\lambda^2)$ contribution.
Since each of these terms has a homogeneous scaling in $\lambda$,
they can only contribute integer powers of $k^-$, yielding integrals of
the form $\Isoftz{\alpha}$ given in \eq{soft_integrals_a}.

\paragraph{\boldmath Leading Power  [$\cO(\lambda^0)$]}

The leading soft limit of the squared amplitude $\Msquared$ is universal and given by
\begin{align} \label{eq:Msquared_soft_LP}
 \Msquared_s^{(0)}(Q,Y; \{k\}) &
 = \frac{16 \pi \as \muMS^{2\eps} \mathbf{C}}{k_T^2} \times \Msquared^\LO(Q,Y)
\,,\end{align}
where $\muMS$ is the renormalization scale in the MS scheme and $\mathbf{C} = C_F, C_A$ is the Casimir constant for the $q\bar{q}$ and $gg$ channel,
and the limit vanishes for any other channel.
The cross section at LP thus becomes
\begin{align} \label{eq:sigma_soft_LP}
 \frac{\df\sigma^{(0)}_s}{\df Q^2 \df Y \df q_T^2} &
 = \Isoftz{0}\, \frac{q_T^{-2\eps}}{(4\pi)^{2-\eps} \Gamma(1-\eps)}
   \frac{\Phi^{(0)} \Msquared_s^{(0)}(Q, Y; \{k\})}{2 x_a x_b \Ecm^4}
\,.\end{align}

\paragraph{\boldmath $\cO(\lambda)$}

Here, we show that power corrections at $\cO(\lambda) \sim \cO(q_T/Q)$ vanish at NLO.
At this order, we can let $\eps\to0$ to obtain the cross section from \eq{sigma_soft_2} as
\begin{align} \label{eq:sigma_soft_lambda1}
 \frac{\df\sigma_s^{(1)}}{\df Q^2 \df Y \df q_T^2} &
 = \frac{1}{2 (4\pi)^2 x_a x_b \Ecm^4}
   \int_0^\infty \frac{\df k^-}{k^-} w^2 \nu^{\eta} \biggl|k^- - \frac{q_T^2}{k^-}\biggr|^{-\eta}
   \\*\nn&\quad\times
   \Bigl[ \Phi^{(0)} \Msquared_s^{(1)}(Q, Y; \{k\}) + \Phi^{(1)} \Msquared_s^{(0)}(Q, Y; \{k\}) \Bigr]
\,.\end{align}
From \eq{phi_soft}, the expansion of the phase space is given by
\begin{align}
 \Phi^{(1)} &
 = \frac{k^- e^{-Y}}{Q} \bigl[ x_a f'_a(x_a) \, f_b(x_b) - f_a(x_a) f_b(x_b) \bigr]
 + (\rm{sym.})
\,.\end{align}
From \eq{Msquared_soft_LP}, we know that $\Msquared^{(0)}_s \sim \Msquared^\LO / k_T^2$,
so $\Phi^{(1)} \Msquared^{(0)} \sim k^-, k_T^2/k^-$.
Hence, this contribution to \eq{sigma_soft_lambda1} is proportional to $\Isoftz{\pm1} = 0$, see \eq{soft_integrals} for odd $\alpha$, and therefore vanishes.
The NLP expansion $\Msquared_s^{(1)}$ of the matrix element is suppressed by $\cO(\lambda)$ relative to $\Msquared^\LO$,
which from power counting can only be given by either $k^-$ or $k^+ = k_T^2/k^-$.
Hence, the $\Phi^{(0)} \Msquared^{(1)}$ term is also proportional to $\Isoftz{\pm1} = 0$ and vanishes as well.

More generally, power counting combined with the behavior of the integrals in \eq{soft_integrals} shows that at NLO, the power expansion is in $q_T^2/Q^2$. It would be interesting to extend this proof to higher perturbative orders.
We also remark that the collinear limit will not have a $\cO(\lambda)$ expansion at all,
and thus the consistency condition that rapidity divergences cancel between soft and collinear sectors
already implies that the soft NLP result cannot contribute to the leading logarithm.

\paragraph{\boldmath Next-to-Leading Power [$\cO(\lambda^2)$]}

The first nonvanishing power correction thus arises at $\cO(\lambda^2)\sim \cO(q_T^2/Q^2)$.
To derive a general master formula at this order, we decompose the expansion of the matrix element
according to the possible dependence on $k^\pm$, which follows from power counting and mass dimension,
\begin{align} \label{eq:altM}
 \Msquared_s^{(0)}(Q,Y; \{k\}) &= \frac{1}{k_T^2} \,\altM{0}(Q,Y)
\,,\nn\\
 \Msquared_s^{(1)}(Q,Y; \{k\}) &= \frac{1}{k_T^2} \biggl[ \frac{k^+}{Q} \,\altM{1}_+(Q,Y) + \frac{k^-}{Q} \,\altM{1}_-(Q,Y) \biggr]
\,,\nn\\
 \Msquared_s^{(2)}(Q,Y; \{k\}) &= \frac{1}{k_T^2} \biggl[ \frac{(k^+)^2}{Q^2} \,\altM{2}_{++}(Q,Y) + \frac{k_T^2}{Q^2} \,\altM{2}_{00} + \frac{(k^-)^2}{Q^2}  \,\altM{2}_{--}(Q,Y) \biggr]
\,.\end{align}
The expansion is defined such that all $\altM{i}$ have the same mass dimension.
We now only need to plug \eq{altM} back into \eq{sigma_soft_2}, collect the powers of $k^-$
(using that $k^+ = k_T^2 / k^-$) and apply \eq{soft_integrals_a}.
Only terms proportional to $\Isoftz{0}$ will yield a divergence in $\eta$,
and thus constitute the LL correction at NLP, while all other terms contribute at NLL.
We find
\begin{align} \label{eq:sigma_soft_NLP_LL}
 \frac{\df\sigma^{(2),\text{LL}}_s}{\df Q^2 \df Y \df q_T^2} &
 = \frac{1}{2 (4\pi)^2 x_a x_b \Ecm^4} \frac{1}{Q^2}
   w^2 \biggl(\frac{2}{\eta} + \ln\frac{\nu^2}{q_T^2} \biggr) \times \biggl\{
   \nn\\*&\hspace{1.cm}
   f_a(x_a) f_b(x_b) \biggl[ \altM{2}_{00}
      - e^{-Y} \altM{1}_+(Q,Y) - e^{Y} \altM{1}_-(Q,Y)
   \biggr]
   \nn\\*&\hspace{1.cm}
   + x_a f'_a(x_a) \, f_b(x_b) \biggl[ e^{-Y} \altM{1}_+(Q,Y) - \frac{1}{2} \altM{0}(Q,Y) \biggr]
   \nn\\*&\hspace{1.cm}
   + f_a(x_a) \, x_b f'_b(x_b) \biggl[ e^{+Y} \altM{1}_-(Q,Y) - \frac{1}{2} \altM{0}(Q,Y)  \biggr]
   \nn\\*&\hspace{1.cm}
   + x_a f'_a(x_a) \, x_b f'_b(x_b)\, \altM{0}(Q,Y)
   \biggr\}
\,,\end{align}
and
\begin{align} \label{eq:sigma_soft_NLP_NLL}
 \frac{\df\sigma^{(2),\text{NLL}}_s}{\df Q^2 \df Y \df q_T^2} &
 = \frac{1}{2 (4\pi)^2 x_a x_b \Ecm^4} \frac{1}{Q^2} \times \biggl\{
   \nn\\*&\hspace{1.cm}
   f_a(x_a) f_b(x_b) \biggl[
      \altM{0}(Q,Y) \Bigl( e^{-2Y} + e^{+2Y} \Bigr)
      + \altM{2}_{++}(Q,Y) + \altM{2}_{--}(Q,Y)
      \nn\\*&\hspace{3.cm}
      - e^{-Y} \altM{1}_-(Q,Y) - e^{Y} \altM{1}_+(Q,Y)
   \biggr]
   \nn\\*&\hspace{1.cm}
   + x_a f'_a(x_a) \, f_b(x_b) \biggl[ e^{-Y} \altM{1}_-(Q,Y) - e^{-2Y} \altM{0}(Q,Y) \biggr]
   \nn\\*&\hspace{1.cm}
   + f_a(x_a) \, x_b f'_b(x_b) \biggl[ e^{+Y} \altM{1}_+(Q,Y) - e^{+2Y} \altM{0}(Q,Y) \biggr]
   \nn\\*&\hspace{1.cm}
   + x_a^2 f''_a(x_a) \, f_b(x_b) \, \frac{e^{-2Y}}{2} \altM{0}(Q,Y)
   \nn\\*&\hspace{1.cm}
   + f_a(x_a) \, x_b^2 f''_b(x_b) \, \frac{e^{+2Y}}{2} \altM{0}(Q,Y)
   \biggr\}
\,.\end{align}
An interesting feature of \eq{sigma_soft_NLP_NLL} is the appearance of double derivatives of the PDFs,
arising from the expansion of $f[\zeta(k)]$ through $\cO(\lambda^2)$.
Most terms in \eqs{sigma_soft_NLP_LL}{sigma_soft_NLP_NLL} also exhibit an explicit
rapidity dependence, which is surprising for the boost-invariant observable $q_T$.
In fact, we will see explicitly that the full soft expansion exactly cancels against
rapidity-dependent terms in the collinear expansions, yielding a rapidity-independent final result. This behavior is expected since the rapidity dependence arises from the rapidity-dependent regulator, and therefore we expect that they should cancel in the final regulator independent result.

\subsubsection[Collinear Master Formula for \texorpdfstring{$q_T$}{qT}]
{\boldmath Collinear Master Formula for $q_T$}
\label{sec:collinear_master}

We next consider the case of a $n$-collinear emission $k \sim Q (\lambda^2,1,\lambda)$,
from which one can easily obtain the $\bn$-collinear case from symmetry.
Here, it is important to consistently expand the rapidity regulator in \eq{sigmaNLO_2} in the $n$-collinear limit,
\begin{equation} \label{eq:regulator_NLP}
  w^2 \nu^\eta \biggl|k^- - \frac{q_T^2}{k^-} \biggr|^{-\eta}
= w^2 \biggl|\frac{k^-}{\nu} \biggr|^{-\eta} \biggl[1 + \eta\,\frac{q_T^2}{(k^-)^2} + \cO(\lambda^4) \biggr]
\,.\end{equation}
Applying this to \eq{sigmaNLO_2} yields
\begin{align} \label{eq:sigma_coll_1}
 \frac{\df\sigma}{\df Q^2 \df Y \df q_T^2} &
 = \frac{q_T^{-2\eps}}{(4\pi)^{2-\eps} \Gamma(1-\eps)} \int_0^\infty \frac{\df k^-}{k^-}
   w^2 \biggl|\frac{k^-}{\nu} \biggr|^{-\eta} \biggl(1 + \eta\,\frac{q_T^2}{(k^-)^2} \biggr)
   \frac{f_a(\zeta_a)\, f_b(\zeta_b)}{2 \zeta_a \zeta_b \Ecm^4}\,
   \Msquared(Q, Y; \{k\})
\,.\end{align}
We now expand all pieces in $\lambda$.
The incoming momenta from \eq{zeta_ab} are expanded as
 \begin{align}
 \zeta_a(k) &
 = x_a \biggl[ \biggl(1 + \frac{k^- e^{-Y}}{Q}\biggr) + \frac{q_T^2}{2 Q^2} \biggr] + \cO(\lambda^4)
 \equiv x_a \biggl[ \frac{1}{z_a} + \Delta_a^{(2)} \biggr]  + \cO(\lambda^4)
\,,\nn\\
 \zeta_b(k) &
 = x_b \biggl[ 1 + \biggl(\frac{k^+ e^{+Y}}{Q} + \frac{q_T^2}{2 Q^2} \biggr) \biggr] + \cO(\lambda^4)
 \equiv x_b \biggl[1 + \Delta_b^{(2)} \biggr] + \cO(\lambda^4)
\,,\end{align}
where we grouped the terms of common scaling together
and defined $k^- = Q e^Y (1 - z_a)/z_a$.
(Recall that the superscript $^{(2)}$ denotes the suppression by $\lambda^2$.)
Expanding the PDFs and flux factors in $\lambda$, we obtain
\begin{align} \label{eq:flux_ncoll}
 \frac{f_a(\zeta_a)\, f_b(\zeta_b)}{\zeta_a \zeta_b} &
 = \frac{z_a}{x_a x_b} f_a\Bigl(\frac{x_a}{z_a}\Bigr) f_b(x_b)
 \nn\\* & \quad
 + \frac{z_a}{x_a x_b} \frac{q_T^2}{2Q^2} \biggl[
    \frac{(1-z_a)^2 - 2}{1-z_a} f_a\Bigl(\frac{x_a}{z_a}\Bigr) f_b(x_b)
   + x_a f'_a\Bigl(\frac{x_a}{z_a}\Bigr) f_b(x_b)
   \nn\\*&\qquad\qquad\qquad\quad
   + \frac{1+z_a}{1-z_a} f_a\Bigl(\frac{x_a}{z_a}\Bigr) \, x_b f'_b(x_b)
 \biggr] + \cO(\lambda^4)
\,.\end{align}
The expansion of the matrix element is process dependent, and we define it by
\begin{align} \label{eq:M_coll}
 \Msquared(Q, Y; \{k\}) = \Msquared_n^{(0)}(Q, Y; \{k\}) + \Msquared_n^{(2)}(Q, Y; \{k\}) + \cO(\lambda^4)
\,.\end{align}
Note that in contrast to the soft limit, there is no $\cO(\lambda)$ suppressed term here.

Next, we switch the integration variable in \eq{sigma_coll_1} via
\begin{align} \label{eq:km_to_za}
 k^- = Q e^Y \frac{1 - z_a}{z_a}
\,,\qquad
 \int_0^\infty \frac{\df k^-}{k^-} = \int_{x_a}^1 \frac{\df z_a}{z_a (1-z_a)}
\,,\end{align}
where the lower bound on the $z_a$ integral follows from the physical support of the PDF $f_a(x_a/z_a)$.
Inserting eqs.\ \eqref{eq:flux_ncoll} -- \eqref{eq:km_to_za} into \eq{sigma_coll_1}
and collecting the $\cO(\lambda^0)$ and $\cO(\lambda^2)$ pieces,
we obtain the leading $n$-collinear limit as
\begin{align} \label{eq:sigma_coll_LP}
 \frac{\df\sigma_n^{(0)}}{\df Q^2 \df Y \df q_T^2} &
 = \frac{q_T^{-2\eps}}{(4\pi)^{2-\eps} \Gamma(1-\eps)}
   w^2 \biggl|\frac{Q e^Y}{\nu}\biggr|^{-\eta}
   \int_{x_a}^1 \frac{\df z_a}{z_a}  \frac{z_a^{1+\eta}}{(1-z_a)^{1+\eta}}
   \frac{f_a(x_a/z_a) f_b(x_b)}{2 x_a x_b \Ecm^4}
   \nn\\*&\hspace{6.5cm}\times
   \Msquared_n^{(0)}(Q, Y; \{k\})
\,,\end{align}
which can be evaluated to obtain the known LP beam function.
For the NLP correction, we can let $\eps\to0$ to obtain
\begin{align} \label{eq:sigma_coll_NLP}
 \frac{\df\sigma^{(2)}_n}{\df Q^2 \df Y \df q_T^2} &
 = \frac{w^2}{(4\pi)^2} \biggl|\frac{Q e^Y}{\nu}\biggr|^{-\eta}
   \int_{x_a}^1 \frac{\df z_a}{z_a}  \frac{z_a^{1+\eta}}{(1-z_a)^{1+\eta}}
   \frac{1}{2 x_a x_b \Ecm^4} \, \biggl\{
    f_a\Bigl(\frac{x_a}{z_a}\Bigr) f_b(x_b) \Msquared_n^{(2)}(Q, Y; \{k\})
   \nn\\*&\quad
   + \frac{q_T^2}{2 Q^2} \Msquared_n^{(0)}(Q, Y; \{k\}) \biggl[ \frac{(1-z_a)^2 - 2}{1-z_a} f_a\Bigl(\frac{x_a}{z_a}\Bigr) f_b(x_b)
   + x_a f'_a\Bigl(\frac{x_a}{z_a}\Bigr) \, f_b(x_b)
   \nn\\*&\qquad
   + \frac{1+z_a}{1-z_a} \,f_a\Bigl(\frac{x_a}{z_a}\Bigr) \, x_b f'_b(x_b)
   + \frac{2 \eta}{e^{2Y}} \frac{z_a^2}{(1-z_a)^2} \, f_a\Bigl(\frac{x_a}{z_a}\Bigr) f_b(x_b)
   \biggr]
   \biggr\}
\,.\end{align}
The corresponding result in the $\bn$-collinear case reads
\begin{align} \label{eq:sigma_bncoll_NLP}
 \frac{\df\sigma^{(2)}_{\bn}}{\df Q^2 \df Y \df q_T^2} &
 = \frac{w^2}{(4\pi)^2} \biggl|\frac{Q e^{-Y}}{\nu}\biggr|^{-\eta}
   \int_{x_b}^1 \frac{\df z_b}{z_b}  \frac{z_b^{1+\eta}}{(1-z_b)^{1+\eta}}
   \frac{1}{2 x_a x_b \Ecm^4} \, \biggl\{
    f_a(x_a) f_b\Bigl(\frac{x_b}{z_b}\Bigr) \Msquared_\bn^{(2)}(Q, Y; \{k\})
   \nn\\*&\quad
   + \frac{q_T^2}{2 Q^2} \Msquared_\bn^{(0)}(Q, Y; \{k\}) \biggl[ \frac{(1-z_b)^2 - 2}{1-z_b} f_a(x_a) f_b\Bigl(\frac{x_b}{z_b}\Bigr)
   + f_a(x_a) \, x_b f'_b\Bigl(\frac{x_b}{z_b}\Bigr)
   \nn\\*&\qquad
   + \frac{1+z_b}{1-z_b} \,x_a f'_a(x_a) \, f_b\Bigl(\frac{x_b}{z_b}\Bigr)
   + \frac{2 \eta}{e^{-2Y}} \frac{z_b^2}{(1-z_b)^2} \, f_a(x_a) f_b\Bigl(\frac{x_b}{z_b}\Bigr)
   \biggr]
   \biggr\}
\,.\end{align}
As discussed in \sec{distribution}, a striking feature of \eqs{sigma_coll_NLP}{sigma_bncoll_NLP} is the appearance of power divergences $1/(1-z)^{2+\eta}$ and even $1/(1-z)^{3+\eta}$, which can be regulated using higher-order plus distributions, see also \app{plus_distr}.
Here, we find it more convenient to employ the integration-by-parts relations in \eqs{plus_dist_2}{plus_dist_3} to write the kernels fully in terms of standard plus distributions, at the cost of inducing explicit derivatives of the PDFs.
In order to apply these relations, we need to identify all divergences in $1/(1-z)^2$ and $1/(1-z)^3$.
To do so, first note that the LP matrix element scales as
\begin{equation}
 \Msquared_n^{(0)} \sim \frac{k^-}{k_T^2} P(z,\eps) \sim (1-z) P(z,\eps)
\,,\end{equation}
where $P$ is the appropriate splitting function in $d=4-2\eps$ dimensions, which itself scales like $P(z,\epsilon)\sim 1/(1-z)$.
Due to the overall prefactor of $k^- \sim (1-z)$, the LP matrix element is finite as $z\to1$. Power counting implies that the subleading matrix element can at most yield one additional pole $1/(1-z)$.
Motivated by these two observations, we write the expanded squared amplitude as
\begin{align} \label{eq:altM_coll}
 \Msquared_n^{(0)}(Q,Y;\{k\}) &= \altM{0}_n(z_a)
\,,\nn\\
 \Msquared_n^{(2)}(Q, Y; \{k\}) &= \frac{k_T^2}{2Q^2} \frac{\altM{2}_n(z_a)}{1-z_a}
\,,\end{align}
and likewise for $\Msquared_\bn$ in the $\bn$-collinear limit.
The power suppression of $\Msquared_n^{(2)}$ is made manifest by extracting the factor $k_T^2/Q^2$.
For brevity, we suppress any dependence of $\altM{0}_n$ and $\altM{2}_n$ on $Q$ and $Y$.

Inserting \eq{altM_coll} into \eq{sigma_coll_NLP}, collecting powers of $(1-z_a)$,
and applying the distribution identities eqs.\ \eqref{eq:plus_dist_1}, \eqref{eq:plus_dist_2} and \eqref{eq:plus_dist_3},
the LL contribution at NLP is obtained as
\begin{align} \label{eq:sigma_coll_NLP_LL}
 \frac{\df\sigma^{(2),\text{LL}}_n}{\df Q^2 \df Y \df\cO} &
 = \frac{1}{(4\pi)^2} \frac{q_T^2}{2Q^2}
   \frac{1}{2 x_a x_b \Ecm^4}\,w^2 \biggl(\frac{1}{\eta} - \ln\frac{Q e^Y}{\nu} \biggr)
   \nn\\&\quad \times  \biggl\{
   f_a(x_a) f_b(x_b) \Bigl[ \altMp{2}_n(1) - 2 \altMp{0}_n(1) \Bigr]
   + f_a(x_a) \, x_b f'_b(x_b) \Bigl[ \altM{0}_n(1) + 2 {\altMp{0}_n}(1) \Bigr]
   \nn\\*&\qquad
   + x_a f'_a(x_a) f_b(x_b) \Bigl[ \altM{0}_n(1) - \altM{2}_n(1) \Bigr]
   - 2 x_a f'_a(x_a) \, x_b f'_b(x_b)  \altM{0}_n(1)
   \biggr\}
\,.\end{align}
Here, we used that the LL result is proportional to $\delta(1-z_a)$ to cancel the $z_a$ integral in \eq{sigma_coll_LP},
and the $\altMp{i}_n(1)$ are the derivative of $\altM{i}_n(z_a)$ at $z_a=1$.
Similarly, we obtain the NLL contribution as
\begin{align} \label{eq:sigma_coll_NLP_NLL}
 \frac{\df\sigma^{(2),\text{NLL}}_n}{\df Q^2 \df Y \df q_T} &
 = \frac{1}{(4\pi)^2} \frac{q_T^2}{2Q^2}
   \frac{1}{2 x_a x_b \Ecm^4}
   \int_{x_a}^1 \frac{\df z_a}{z_a}
   \nn\\*&\quad \times  \biggl\{
   f_a\Bigl(\frac{x_a}{z_a}\Bigr) f_b(x_b) \Bigl\{
      \delta(1-z_a) \Bigl[ \altM{2}_n(1) - \altMp{2}_n(1) - 2 \altM{0}_n(1) + 2 \altMp{0}_n(1) \Bigr]
      \nn\\*&\hspace{3.5cm}
      {- e^{-2Y} \delta(1-z_a) \Bigl[ 2 \altM{0}_n(1) + 4 \altMp{0}_n(1) + \altMpp{0}_n(1) \Bigr] }
      \nn\\*&\hspace{3.5cm}
      + z_a \cL_0(1-z_a) \Bigl[ 2 \altMp{0}_n(z_a) - \altMp{2}_n(z_a)\Bigr]
      + z_a \altM{0}_n(z_a)
   \Bigr\}
   \nn\\*&\qquad
   + \frac{x_a}{z_a} f'_a\Bigl(\frac{x_a}{z_a}\Bigr) f_b(x_b) \biggl\{
      \delta(1-z_a) \Bigl[ \altM{2}_n(1) - 2 \altM{0}_n(1) \Bigr]
      \nn\\*&\hspace{4cm}
      {+ 2e^{-2Y} \delta(1-z_a) \Bigl[ \altM{0}_n(1) + \altMp{0}_n(1) \Bigr] }
      \nn\\*&\hspace{4cm}
      + \cL_0(1-z_a) \Bigl[ \altM{2}_n(z_a) + (z_a^2-2)\altM{0}_n(z_a) \Bigr]
   \biggr\}
   \nn\\*&\qquad
   + f_a\Bigl(\frac{x_a}{z_a}\Bigr) \, x_b f'_b(x_b) \biggl\{
      \delta(1-z_a) \Bigl[ \altM{0}_n(1) - 2 \altMp{0}_n(1) \Bigr]
      \nn\\*&\hspace{4cm}
     - z_a \cL_0(1-z_a) \Bigl[ \altM{0}_n(z_a) + (1+z_a) \altMp{0}_n(z_a) \Bigr]
   \biggr\}
   \nn\\*&\qquad
   + \frac{x_a}{z_a} f'_a\Bigl(\frac{x_a}{z_a}\Bigr) \, x_b f'_b(x_b) \Bigl[
      2 \delta(1-z_a) \altM{0}_n(1) + (1+z_a) \altM{0}_n(z_a) \cL_0(1-z_a)
   \Bigr]
   \nn\\*&\qquad
   {
   - \Bigl(\frac{x_a}{z_a}\Bigr)^2 f''_a\Bigl(\frac{x_a}{z_a}\Bigr) f_b(x_b)
     \, \delta (1-z_a) e^{-2Y} \altM{0}_n(1)
   }
   \biggr\}
\,.\end{align}
Here, all terms with an explicit rapidity dependence arise from the expansion of the regulator itself,
see \eq{regulator_NLP}.
In practice, they will exactly cancel against the soft NLL result \eq{sigma_soft_NLP_NLL}.

\subsection{Derivation of the Master Formula in Pure Rapidity Regularization}
\label{sec:master_formula_2}

In \sec{master_formula}, we used the $\eta$ regulator of the form $|2 k^z / \nu|^{-\eta}$ to derive
the master formula.
In this section, we repeat the derivation of the master formula using the pure rapidity regulator introduced in \sec{upsilonreg}.
As discussed there, this regulator has the advantage that it is homogeneous in the power expansion, which reduces the number of terms at subleading power.
Furthermore, it renders the soft sector scaleless. 
The result using the generalization of the pure rapidity regulator, \eq{vita_regulator}, is shown in \app{master_formula_c} for completeness.

The derivation of the $n$-collinear expansion proceeds similar to the calculation shown in \sec{collinear_master}.
In \eq{sigma_coll_NLP}, one has to replace the regulator factor by
\begin{equation}
 \biggl|\frac{k^-}{\nu}\biggr|^{-\eta} = \biggl|\frac{Q e^Y}{\nu}\biggr|^{-\eta} \biggl|\frac{1-z_a}{z_a}\biggr|^{-\eta}
 ~\to~
 \upsilon^{\eta}\biggl|\frac{k^-}{k^+}\biggr|^{-\eta/2} = \upsilon^{\eta} \biggl|\frac{k^-}{q_T}\biggr|^{-\eta} = \upsilon^{\eta} \biggl|\frac{Q e^Y}{q_T}\biggr|^{-\eta}\biggl|\frac{1-z_a}{z_a}\biggr|^{-\eta}
\end{equation}
and drop the terms in $\smallupsilon/e^{2Y}$, as they are fully induced by the expansion of the regulator.
The NLP LL result is then easily obtained from \eq{sigma_coll_NLP_LL} by replacing $\nu \to q_T \upsilon$,
\begin{align} \label{eq:sigma_ncoll_NLP_LL_vita}
 \frac{\df\sigma^{(2),\text{LL}}_n}{\df Q^2 \df Y \df q_T^2} &
 = \frac{1}{(4\pi)^2} \frac{q_T^2}{2Q^2}
   \frac{1}{2 x_a x_b \Ecm^4}\,w^2\biggl(\frac{1}{\smallupsilon} - \ln\frac{Q e^Y}{q_T} + \ln(\upsilon) \biggr)
   \nn\\*&\quad \times  \biggl\{
   f_a(x_a) f_b(x_b) \Bigl[ \altMp{2}_n(1) - 2 \altMp{0}_n(1) \Bigr]
   + f_a(x_a) \, x_b f'_b(x_b) \Bigl[ \altM{0}_n(1) + 2 \altMp{0}_n(1) \Bigr]
   \nn\\*&\qquad
   + x_a f'_a(x_a) f_b(x_b) \Bigl[ \altM{0}_n(1) - \altM{2}_n(1) \Bigr]
   - 2 x_a f'_a(x_a) \, x_b f'_b(x_b)  \altM{0}_n(1)
   \biggr\}
\,.\end{align}
In the $\bn$-collinear limit, one has to replace the regulator factor
\begin{equation}
 \biggl|\frac{k^+}{\nu}\biggr|^{-\eta} = \biggl|\frac{Q e^{-Y}}{\nu}\biggr|^{-\eta} \biggl|\frac{1-z_b}{z_b}\biggr|^{-\eta}
 ~\to~
 \upsilon^{\eta}\biggl|\frac{k^-}{k^+}\biggr|^{-\eta/2} = \upsilon^{\eta} \biggl|\frac{k^+}{q_T}\biggr|^{\eta} = \upsilon^{\eta} \biggl|\frac{Q e^{-Y}}{q_T}\biggr|^{\eta} \biggl|\frac{1-z_b}{z_b}\biggr|^{\eta}
\end{equation}
and drop terms in $\eta/e^{-2Y}$ in \eq{sigma_bncoll_NLP}.
The NLP LL result is then obtained from \eq{sigma_coll_NLP_LL} by replacing $\eta \to -\eta, \nu \to q_T / \upsilon$ and exchanging $a\leftrightarrow b$ as
\begin{align} \label{eq:sigma_nbarcoll_NLP_LL_vita}
 \frac{\df\sigma^{(2),\text{LL}}_\bn}{\df Q^2 \df Y \df q_T^2} &
 = \frac{1}{(4\pi)^2} \frac{q_T^2}{2Q^2}
   \frac{1}{2 x_a x_b \Ecm^4}\,w^2\biggl(-\frac{1}{\smallupsilon} - \ln\frac{Q e^{-Y}}{q_T} - \ln(\upsilon) \biggr)
   \\&\quad \times  \biggl\{
   f_a(x_a) f_b(x_b) \Bigl[ \altMp{2}_\bn(1) - 2 \altMp{0}_\bn(1) \Bigr]
   + f_a(x_a) \, x_b f'_b(x_b) \Bigl[ \altM{0}_\bn(1) - \altM{2}_\bn(1) \Bigr]
   \nn\\*&\qquad
   + x_a f'_a(x_a) f_b(x_b)  \Bigl[ \altM{0}_\bn(1) + 2 {\altMp{0}_\bn}(1) \Bigr]
   - 2 x_a f'_a(x_a) \, x_b f'_b(x_b)  \altM{0}_\bn(1)
   \biggr\}\nn
\,.\end{align}
Summing \eqs{sigma_ncoll_NLP_LL_vita}{sigma_nbarcoll_NLP_LL_vita}, the poles in $\smallupsilon$ precisely cancel,
and the dependence on $e^Y$ and $\bigupsilon$ cancel as well to yield a pure logarithm in $\ln(Q/q_T)$.
This cancellation has to occur between the two collinear sectors,
as there are no contributions from the soft sector.

The NLP NLL result for the pure rapidity regulator is identical to that in \eq{sigma_coll_NLP_NLL} upon dropping all rapidity-dependent pieces,
which we have explicitly verified by repeating the derivation in \sec{collinear_master} using the pure rapidity regulator.
This provides a highly nontrivial check of our regularization procedure, and our understanding of subleading-power rapidity divergences.

\subsection{Next-to-leading Power Corrections at NLO}
\label{sec:results_singlet}

In this section, we give explicit results for the full NLP correction at NLO
for gluon-fusion Higgs and Drell-Yan production in all partonic channels.
Since both are $s$-channel processes, their power corrections
are always proportional to their Born cross sections, and we express the NLP result at $\cO(\as)$ as
\begin{align} \label{eq:sigma_NLP}
 &\frac{\df\sigma^{(2,1)}}{\df Q^2 \df Y \df q_T^2}
= \hat\sigma^\LO(Q) \, \frac{\as}{4\pi}
    \int_{x_a}^1 \frac{\df z_a}{z_a} \int_{x_b}^1 \frac{\df z_b}{z_b}
\nn\\*&\quad\times \biggl[
   f_i\biggl(\frac{x_a}{z_a}\biggr) f_j\biggl(\frac{x_b}{z_b}\biggr)
   C_{f_i f_j}^{(2,1)}(z_a, z_b, q_T)
  + \frac{x_a}{z_a} f'_i\biggl(\frac{x_a}{z_a}\biggr) \frac{x_b}{z_b} f'_j\biggl(\frac{x_b}{z_b}\biggr)
  C_{f'_i f_j'}^{(2,1)}(z_a, z_b, q_T)
\nn\\*&\qquad
   + \frac{x_a}{z_a} f'_i\biggl(\frac{x_a}{z_a}\biggr) f_j\biggl(\frac{x_b}{z_b}\biggr)
   C_{f_i' f_j}^{(2,1)}(z_a, z_b, q_T)
  + f_i\biggl(\frac{x_a}{z_a}\biggr) \frac{x_b}{z_b} f'_j\biggl(\frac{x_b}{z_b}\biggr)
  C_{f_i f_j'}^{(2,1)}(z_a, z_b, q_T)
\biggr]
\,.\end{align}
Here, we suppress the explicit $Q$ and $Y$ dependence in the kernels $C^{(2,1)}_{a b}$.

The required $H+j$ and $Z+j$ amplitudes are conveniently expressed in terms of the Mandelstam variables
\begin{align} \label{eq:mandelstam}
 s_{ab} &= 2 p_a \cdot p_b = Q^2 + 2 q_T^2 + \Bigl( k^+ e^Y + k^- e^{-Y}\Bigr) \sqrt{Q^2 + q_T^2}
\,,\nn \\
 s_{ak} &= -2 p_a \cdot k = - q_T^2 - k^+ e^{+Y} \sqrt{Q^2 + q_T^2}
\,,\nn \\
 s_{bk} &= -2 p_b \cdot k = - q_T^2 - k^- e^{-Y} \sqrt{Q^2 + q_T^2}
\,,\end{align}
which allows us to straightforwardly obtain the LP and NLP expansions in both the soft and collinear limits,
as required by the collinear and soft master formulas.
In the following, we only give the final results after combining soft,
$n$-collinear, and $\bn$-collinear power corrections. The results were computed
separately using both regulators, which provides a highly nontrivial check of our calculation.

\subsubsection{Gluon-Fusion Higgs Production}
\label{sec:ggH_results}

We first consider on-shell Higgs production in gluon fusion in the $m_t \to \infty$ limit,
for which the LO partonic cross section is given by
\begin{align}
 \hat\sigma^\LO(Q)
 = \frac{\Msquared^\LO(Q,Y)}{2 x_a x_b \Ecm^4}
 = 2\pi \delta(Q^2 - m_H^2) \frac{|\cM_{gg\to H}^\LO(Q)|^2}{2 Q^2 \Ecm^2}
\,.\end{align}
The LO matrix element in $d=4-2\epsilon$ dimensions is given by
\cite{Dawson:1990zj,Djouadi:1991tka}
\begin{align}\label{eq:M2_gg_H}
 |\cM^\LO_{gg\to H}(Q)|^2 &= \frac{\as^2 Q^4}{576 \pi^2 v^2} \biggl(\frac{4\pi\muMS^2}{m_t^2}\biggr)^{2\eps} \frac{\Gamma^2(1+\eps)}{1-\eps}
\,.\end{align}
At NLO, there are three distinct partonic channels, $gg\to Hg$, $q\bar q \to H g$, and $gq \to H q$,
which we consider separately.
Here, we calculate the full LL and NLL kernels for all channels.
The LL results will be summarized in \sec{discuss}.

\paragraph{\boldmath $gg \to H g$}

The spin- and color-averaged squared amplitude for $g(p_a) + g(p_b) \to H(q) + g(k)$ is given by \cite{Dawson:1990zj}
\begin{align} \label{eq:M2_ggHg}
 \Msquared_{gg \to H g}(Q, Y, \{k\}) &
 = \Msquared^\LO_{gg\to H}(Q) \times \frac{8 \pi \as C_A \muMS^{2\eps}}{ Q^4 (1-\eps)}
\nn\\* & \quad\times
   \biggl[ (1-2\eps) \frac{Q^8 + s_{ab}^4 + s_{ak}^4 + s_{bk}^4}{s_{ab} s_{ak} s_{bk}}
   + \frac{\eps}{2} \frac{(Q^4 + s_{ab}^2 + s_{ak}^2 + s_{bk}^2)^2}{s_{ab} s_{ak} s_{bk}} \biggr]
\,.\end{align}
The full result from combining the soft, $n$-collinear, and $\bn$-collinear contributions is given by
\begin{align} \label{eq:C_ggHg}
  C_{f_g f_g}^{(2,1)}(z_a, z_b, q_T) &
  = 2 C_A \frac{1}{Q^2} \biggl\{
      \biggl[8 \ln\frac{Q^2}{q_T^2} + 12 \biggr] \delta(1-z_a) \delta(1-z_b)
      \nn\\*&\hspace{2cm}
      + \delta(1-z_a)\biggl[ -8 + \frac{3}{z_b} + z_b - 12 z_b^2 + 9 z_b^3 + 8 \cL_0(1-z_b) \biggr]
      \nn\\*&\hspace{2cm}
      + \biggl[ -8 + \frac{3}{z_a} + z_a - 12 z_a^2 + 9 z_a^3 + 8 \cL_0(1-z_a) \biggr] \delta(1-z_b)
    \biggr\}
\,,\nn\\
  C_{f'_g f_g}^{(2,1)}(z_a, z_b, q_T) &
  = 2 C_A \frac{1}{Q^2} \biggl\{
    \biggl[- \ln\frac{Q^2}{q_T^2} - 1 \biggr] \delta(1-z_a) \delta(1-z_b)
      \nn\\*&\hspace{2cm}
      + \delta(1-z_a)\biggl[ 2 + \frac{2}{z_b^2} + \frac{1}{z_b} + z_b +  3 z_b^2 - \cL_0(1-z_b) \biggr]
      \nn\\*&\hspace{2cm}
      + \biggl[ 4 + \frac{2}{z_a} - 2 z_a + 5 z_a^2 - 3 z_a^3 - \cL_0(1-z_a) \biggr] \delta(1-z_b)
    \biggr\}
\,,\nn\\
  C_{f_g f'_g}^{(2,1)}(z_a, z_b, q_T) &
  = 2 C_A \frac{1}{Q^2} \biggl\{
      \biggl[- \ln\frac{Q^2}{q_T^2} - 1 \biggr] \delta(1-z_a) \delta(1-z_b)
      \nn\\*&\hspace{2cm}
      + \delta(1-z_a)\biggl[ 4 + \frac{2}{z_b} - 2 z_b + 5 z_b^2 - 3 z_b^3 - \cL_0(1-z_b)\biggr]
      \nn\\*&\hspace{2cm}
      + \biggl[ 2 + \frac{2}{z_a^2} + \frac{1}{z_a} + z_a + 3 z_a^2 - \cL_0(1-z_a) \biggr] \delta(1-z_b)
    \biggr\}
\,,\nn\\
  C_{f'_g f'_g}^{(2,1)}(z_a, z_b, q_T) &
  = 2 C_A \frac{1}{Q^2} \biggl\{
       \biggl[2\ln\frac{Q^2}{q_T^2} + 4 \biggr] \delta(1-z_a) \delta(1-z_b)
      \nn\\*&\hspace{2cm}
      + \delta(1-z_a)\biggl[ -1 + \frac{1}{z_b^2} - z_b^2 + 2 \cL_0(1-z_b)\biggr]
      \nn\\*&\hspace{2cm}
      + \biggl[ -1 + \frac{1}{z_a^2} - z_a^2 + 2 \cL_0(1-z_a)\biggr] \delta(1-z_b)
    \biggr\}
\,.\end{align}
Substituting these results into \eq{sigma_NLP} yields the NLP cross section for $gg\to Hg$ at NLO.

\paragraph{\boldmath $gq \to Hq$}

The $gq\to Hq$ channel has power corrections at both LL and NLL.
The spin- and color-averaged squared amplitude for $g(p_a) + q(p_b) \to H(q) + q(k)$ is given by \cite{Dawson:1990zj}
\begin{align} \label{eq:M2_gqHq}
 \Msquared_{gq \to Hq}(Q, Y, \{k\})
 = - \Msquared^\LO_{gg\to H}(Q) \times 8 \pi \as C_F \muMS^{2\eps} \frac{1}{Q^4 s_{bk}}
  \Bigl[ s_{ab}^2 + s_{ak}^2 - \eps (s_{ab}+s_{ak})^2 \Bigr]
\,.\end{align}
The full result from combining the soft, $n$-collinear, and $\bn$-collinear contributions is given by
\begin{align} \label{eq:C_gqHq}
  C_{f_g f_q}^{(2,1)}(z_a, z_b, q_T) &
  = 2 C_F \frac{1}{Q^2} \biggl\{
      \biggl[ \ln\frac{Q^2}{q_T^2} + 3 \biggr] \delta(1-z_a) \delta(1-z_b)
      \nn\\*&\hspace{2cm}
      + \delta(1-z_a) \biggl[ -1 + \frac{3}{z_b} + 2 z_b - 2 z_b^2 + \cL_0(1-z_b) \biggr]
      \nn\\*&\hspace{2cm}
      + \biggl[ \frac{1}{z_a} + \cL_0(1-z_a) \biggr] \delta(1-z_b)
    \biggr\}
\,,\nn\\
  C_{f'_g f_q}^{(2,1)}(z_a, z_b, q_T) &
  = 2 C_F \frac{1}{Q^2} \biggl\{
      \biggl[ \ln\frac{Q^2}{q_T^2} + 2 \biggr] \delta(1-z_a) \delta(1-z_b)
      \nn\\*&\hspace{2cm}
      + \delta(1-z_a) \biggl[ \frac{2+z_b - z_b^3}{z_b^2} + \cL_0(1-z_b) \biggr]
      \nn\\*&\hspace{2cm}
      + \biggl[ \frac{1}{z_a} + \cL_0(1-z_a) \biggr] \delta(1-z_b)
    \biggr\}
\,,\nn\\
  C_{f_g f'_q}^{(2,1)}(z_a, z_b, q_T) &
  = 2 C_F \frac{1}{Q^2} \delta(1-z_a) \biggl(\frac{2}{z_b} - \frac{3}{2} z_b + z_b^2 \biggr)
\,,\nn\\
  C_{f'_g f'_q}^{(2,1)}(z_a, z_b, q_T) &
  = 2 C_F \frac{1}{Q^2} \delta(1-z_a) \biggl(-\frac12 + \frac{1}{z_b^2} + \frac{z_b}{2} \biggr)
\,.\end{align}
Substituting these results into \eq{sigma_NLP} yields the NLP cross section for $gq\to Hq$ at NLO.

\paragraph{\boldmath $qg \to Hq$}

The result for $qg \to Hq$ can be obtained from \eq{C_gqHq}
by exchanging $f_q \leftrightarrow f_g$ and $a \leftrightarrow b$,
\begin{align} \label{eq:C_qgHq}
  C_{f_q f_g}^{(2,1)}(z_a, z_b, q_T) &
  = 2 C_F \frac{1}{Q^2} \biggl\{
      \biggl[ \ln\frac{Q^2}{q_T^2} + 3 \biggr] \delta(1-z_a) \delta(1-z_b)
      \nn\\*&\hspace{2cm}
      + \biggl[ -1 + \frac{3}{z_a} + 2 z_a - 2 z_a^2 + \cL_0(1-z_a) \biggr] \delta(1-z_b)
      \nn\\*&\hspace{2cm}
      + \delta(1-z_a) \biggl[ \frac{1}{z_b} + \cL_0(1-z_b) \biggr]
    \biggr\}
\,,\nn\\
  C_{f'_q f_g}^{(2,1)}(z_a, z_b, q_T) &
  = 2 C_F \frac{1}{Q^2} \biggl(\frac{2}{z_a} - \frac{3}{2} z_a + z_a^2 \biggr) \delta(1-z_b)
\,,\nn\\
  C_{f_q f'_g}^{(2,1)}(z_a, z_b, q_T) &
  = 2 C_F \frac{1}{Q^2} \biggl\{
      \biggl[ \ln\frac{Q^2}{q_T^2} + 2 \biggr] \delta(1-z_a) \delta(1-z_b)
      \nn\\*&\hspace{2cm}
      + \biggl[ \frac{2+z_a - z_a^3}{z_a^2} + \cL_0(1-z_a) \biggr] \delta(1-z_b)
      \nn\\*&\hspace{2cm}
      + \delta(1-z_a) \biggl[ \frac{1}{z_b} + \cL_0(1-z_b) \biggr]
    \biggr\}
\,,\nn\\
  C_{f'_q f'_g}^{(2,1)}(z_a, z_b, q_T) &
  = 2 C_F \frac{1}{Q^2} \biggl(-\frac12 + \frac{1}{z_a^2} + \frac{z_a}{2} \biggr) \delta(1-z_b)
\,.\end{align}
Substituting these results into \eq{sigma_NLP} yields the NLP cross section for $qg\to Hq$ at NLO.

\paragraph{\boldmath $q\bar{q} \to Hg$}

The $q\bar q \to Hg$ channel has no leading logarithms and thus only contributes at NLL.
The spin- and color-averaged squared amplitude is given by \cite{Dawson:1990zj}
\begin{align}
 \Msquared_{q\bar q \to Hg}(Q, Y, \{k\}) &
 = \Msquared^\LO_{gg\to H}(Q) \times \frac{64\pi}{3} \as C_F \muMS^{2\eps} \frac{1-\eps}{Q^4 s_{ab}}
   \bigl[ s_{ak}^2 + s_{bk}^2 - \eps (s_{ak}+s_{bk})^2 \bigr]
\,.\end{align}
The results for the kernels are given by
\begin{align} \label{eq:C_qqHg}
  C_{f_q f_\bq}^{(2,1)}(z_a, z_b, q_T) &
  = \frac{16 C_F}{3} \frac{1}{Q^2} \biggl[ \delta(1-z_a) \biggl(1 + \frac{1}{z_b} - 2 z_b\biggr)
    + \biggl(1 + \frac{1}{z_a} - 2 z_a\biggr) \delta(1-z_b) \biggr]\,,
\nn\\
  C_{f'_q f_\bq}^{(2,1)}(z_a, z_b, q_T) &
  = \frac{16 C_F}{3} \frac{1}{Q^2} \frac{(1-z_a)^2}{z_a} \delta(1-z_b)\,,
\nn\\
  C_{f_q f'_\bq}^{(2,1)}(z_a, z_b, q_T) &
  = \frac{16 C_F}{3} \frac{1}{Q^2} \delta(1-z_a) \frac{(1-z_b)^2}{z_b}
\,,\nn\\
  C_{f'_q f'_\bq}^{(2,1)}(z_a, z_b, q_T) &= 0
\,.\end{align}
Substituting these results into \eq{sigma_NLP} yields the NLP cross section for $q\bar{q} \to Hg$ at NLO.

\subsubsection{Drell-Yan Production}
\label{sec:DY_results}

We next consider the Drell-Yan process $p p \to Z/\gamma^* \to \ell^+ \ell^-$, and for brevity denote it as $p p \to V$.
In contrast to on-shell Higgs production, it is important to be able to include off-shell effects.
The LO partonic cross section is given by
\begin{align}
 \hat\sigma^\LO(Q) = \frac{4\pi \alpha_{em}^2}{3 N_c Q^2 \Ecm^2}
 \biggl[ Q_q^2 + \frac{(v_q^2 + a_q^2)(v_\ell^2 + a_\ell^2) - 2 Q_q v_q v_\ell (1 - m_Z^2/Q^2)}{(1-m_Z^2/Q^2)^2 + m_Z^2 \Gamma_Z^2 / Q^4} \biggr]
\,,\end{align}
where $Q$ is the dilepton invariant mass, $v_{\ell,q}$ and $a_{\ell,q}$ are the standard
vector and axial couplings of the leptons and quarks to the $Z$ boson,
and the $\ell^+ \ell^-$ phase space has already been integrated over.
At NLO , there are two distinct partonic channels, $q\bar q\to Vg$ and $qg \to Vq$,
which we consider separately.
Here, we calculate the full LL and NLL kernels for all channels.
The LL results will be summarized in \sec{discuss}.

\paragraph{\boldmath $q\bar q\to Vg$}

We first consider the $q\bar q\to Vg$ channel, for which the spin- and color-averaged squared amplitude is given by \cite{Gonsalves:1989ar}
\begin{align} \label{eq:M2_qqbar_Zg}
 |\cM_{q\bar q\to Vg}|^2 &
 = |\cM_{q\bar q\to V}|^2 \times \frac{8 \pi \as C_F \muMS^{2\eps}}{Q^2}
  \left[ (1-\epsilon) \left(\frac{s_{ak}}{s_{bk}}+\frac{s_{bk}}{s_{ak}}\right)+\frac{2 s_{ab} Q^2}{s_{ak} s_{bk}} -2\epsilon \right]
\,.\end{align}
The full result from combining the soft, $n$-collinear, and $\bn$-collinear contributions is given by
\begin{align} \label{eq:C_qqZg}
  C_{f_q f_{\bar q}}^{(2,1)}(z_a, z_b, q_T) &
  = 2 C_F \frac{1}{Q^2} \biggl[
      -4 \delta(1-z_a)\delta(1-z_b)
      \nn\\*&\hspace{2cm}
      - \delta(1-z_a) \frac{1+z_b^2-4z_b^3}{2z_b}
      - \frac{1+z_a^2-4z_a^3}{2z_a} \delta(1-z_b)
    \biggr]
\,,\nn\\
  C_{f'_q f_{\bar q}}^{(2,1)}(z_a, z_b, q_T) &
  = 2 C_F \frac{1}{Q^2} \biggl\{
      \biggl[- \ln\frac{Q^2}{q_T^2} - 1\biggr] \delta(1-z_a)\delta(1-z_b)
      \nn\\*&\hspace{2cm}
      + \delta(1-z_a) \biggl[ \frac32 + \frac{1}{2z_b} + z_b - \cL_0(1-z_b) \biggr]
      \nn\\*&\hspace{2cm}
      - \biggl[ \frac{1+z_a+2z_a^3}{2z_a} + \cL_0(1-z_a) \biggr] \delta(1-z_b)
    \biggr\}
\,,\nn\\
  C_{f_q f'_{\bar q}}^{(2,1)}(z_a, z_b, q_T) &
  = 2 C_F \frac{1}{Q^2} \biggl\{
      \biggl[- \ln\frac{Q^2}{q_T^2} - 1\biggr] \delta(1-z_a)\delta(1-z_b)
      \nn\\*&\hspace{2cm}
      - \delta(1-z_a) \biggl[ \frac{1+z_b+2z_b^3}{2z_b} + \cL_0(1-z_b) \biggr]
      \nn\\*&\hspace{2cm}
      + \biggl[ \frac32 + \frac{1}{2z_a} + z_a - \cL_0(1-z_a) \biggr] \delta(1-z_b)
    \biggr\}
\,,\nn\\
  C_{f'_q f'_{\bar q}}^{(2,1)}(z_a, z_b, q_T) &
  = 2 C_F \frac{1}{Q^2} \biggl\{
      \biggl[ 2 \ln\frac{Q^2}{q_T^2} + 4\biggr] \delta(1-z_a)\delta(1-z_b)
      \nn\\*&\hspace{2cm}
      + \delta(1-z_a) \biggl[ \frac{1-2z_b-z_b^2}{2z_b} + 2 \cL_0(1-z_b) \biggr]
      \nn\\*&\hspace{2cm}
      + \biggl[ \frac{1-2z_a-z_a^2}{2z_a} + 2 \cL_0(1-z_a) \biggr] \delta(1-z_b)
    \biggr\}
\,.\end{align}
Substituting these results into \eq{sigma_NLP} yields the NLP cross section for $q\bar{q} \to Vg$ at NLO.

\paragraph{\boldmath $qg\to Vq$}

The spin- and color-averaged squared amplitude for the $qg\to Vq$ channel is given by \cite{Gonsalves:1989ar}
\begin{align} \label{eq:M2_qg_Zq}
 \Msquared_{qg \to Vq}(Q,Y,\{k\})
 = - \Msquared^\LO_{q \bar q \to V}(Q) \times \frac{8 \pi \as T_F \muMS^{2\eps}}{Q^2 (1-\eps)}
  \left[ (1-\epsilon) \left(\frac{s_{ab}}{s_{bk}}+\frac{s_{bk}}{s_{ab}}\right)+\frac{2 s_{ak} Q^2}{s_{ab} s_{bk}} -2\epsilon \right]
\,.\end{align}
The full result from combining the soft, $n$-collinear, and $\bn$-collinear contributions is given by
\begin{align} \label{eq:C_qgZq}
  C_{f_q f_g}^{(2,1)}(z_a, z_b, q_T) &
  = 2 T_F \frac{1}{Q^2} \biggl\{
    \delta(1-z_a) \delta(1-z_b)
    \nn\\*&\hspace{2cm}
    + \delta(1-z_a) \frac{-1 + z_b^2 + 24 z_b^3 - 18 z_b^4}{2z_b}
    + 2 z_a \delta(1-z_b)
    \biggr\}
\,,\nn\\
  C_{f'_q f_g}^{(2,1)}(z_a, z_b, q_T) &
  = 2 T_F \frac{1}{Q^2} \biggl\{
    \biggl[ \ln\frac{Q^2}{q_T^2} + 2\biggr] \delta(1-z_a) \delta(1-z_b)
    \nn\\*&\hspace{2cm}
    + \delta(1-z_a) \biggl[ \frac{1-z_b-2z_b^2-6z_b^3}{2z_b} + \cL_0(1-z_b) \biggr]
    \nn\\*&\hspace{2cm}
    + \bigl[ 1 - z_a + \cL_0(1-z_a) \bigr] \delta(1-z_b)
    \biggr\}
\,,\nn\\
  C_{f_q f'_g}^{(2,1)}(z_a, z_b, q_T) &
  = 2 T_F \frac{1}{Q^2} \delta(1-z_a) \frac{-1 + 5 z_b + z_b^2 - 10 z_b^3 + 6 z_b^4}{2z_b}
\,,\nn\\
  C_{f'_q f'_g}^{(2,1)}(z_a, z_b, q_T) &
  = 2 T_F \frac{1}{Q^2} \delta(1-z_a) \frac{1-z_b + 2 z_b^3}{2z_b}
\,.\end{align}
Substituting these results into \eq{sigma_NLP} yields the NLP cross section for $q g \to V q$ at NLO.

\paragraph{\boldmath $g q\to Vq$}

The result for $gq \to Vq$ can be obtained from \eq{M2_qg_Zq}
by exchanging $a \leftrightarrow b$ and $f_q \leftrightarrow f_g$,
\begin{align} \label{eq:C_gqZq}
  C_{f_g f_q}^{(2,1)}(z_a, z_b, q_T) &
  = 2 T_F \frac{1}{Q^2} \biggl\{
    \delta(1-z_a) \delta(1-z_b)
    \nn\\*&\hspace{2cm}
    + 2 \delta(1-z_a) z_b
    + \frac{-1 + z_a^2 + 24 z_a^3 - 18 z_a^4}{2z_a} \delta(1-z_b)
    \biggr\}
\,,\nn\\
  C_{f'_g f_q}^{(2,1)}(z_a, z_b, q_T) &
  = 2 T_F \frac{1}{Q^2} \frac{-1+5z_a+z_a^2-10z_a^3 + 6z_a^4}{2z_a} \delta(1-z_b)
\,,\nn\\
  C_{f_g f'_q}^{(2,1)}(z_a, z_b, q_T) &
  = 2 T_F \frac{1}{Q^2} \biggl\{
    \biggl[ \ln\frac{Q^2}{q_T^2} + 2\biggr] \delta(1-z_a) \delta(1-z_b)
    \nn\\*&\hspace{2cm}
    + \delta(1-z_a) \bigl[ 1 - z_b + \cL_0(1-z_b) \biggr]
    \nn\\*&\hspace{2cm}
    + \biggl[ \frac{1-z_a-2z_a^2-6z_a^3}{2z_a} + \cL_0(1-z_a) \biggr] \delta(1-z_b)
    \biggr\}
\,,\nn\\
  C_{f'_g f'_q}^{(2,1)}(z_a, z_b, q_T) &
  = 2 T_F \frac{1}{Q^2} \frac{1-z_a+2z_a^3}{2z_a} \delta(1-z_b)
\,.\end{align}
Substituting these results into \eq{sigma_NLP} yields the NLP cross section for $g q \to V q$ at NLO.

\subsection{Discussion}
\label{sec:discuss}

Since the full calculation of the power corrections is rather involved, and contains a number of moving pieces, here we highlight several interesting features of the calculation, and compare them to
the perturbative power corrections for beam thrust. For the purposes of this discussion, it is convenient to recall the form of the LL power corrections for the Born partonic configurations
\begin{align} \label{eq:sigma_Higgs_LL_introPtNLL}
 \frac{\df\sigma^{(2),\text{LL}}_{g g \to H g}}{\df Q^2 \df Y\df q_T^2} &
 = \hat\sigma^\LO_{gg\to H}(Q) \times \frac{\as C_A}{4\pi} \frac{2}{Q^2} \ln\frac{Q^2}{q_T^2}
   \Bigl[ 8 f_g(x_a) f_g(x_b) +f^{gg}_{\text{uni}}(x_a,x_b) \Bigr]\,,\\
 \frac{\df\sigma^{(2),\text{LL}}_{q \bq \to V g}}{\df Q^2 \df Y\df q_T^2} &
 = \hat\sigma^\LO_{q\bar q\to V}(Q) \times \frac{\as C_F}{4\pi} \frac{2}{Q^2} \ln\frac{Q^2}{q_T^2}
   \Bigl[ f^{q\bar q}_{\text{uni}}(x_a,x_b) \Bigr]
\,,\nn\end{align}
where
\begin{align}
f^{gg}_{\text{uni}}(x_a,x_b)&=- x_a f'_g(x_a) f_g(x_b) - f_g(x_a) \, x_b f'_g(x_b)+ 2 x_a f'_g(x_a) \, x_b f'_g(x_b)\,, \\
f^{q\bq}_{\text{uni}}(x_a,x_b)&=- x_a f'_q(x_a) f_\bq(x_b) - f_q(x_a) \, x_b f'_\bq(x_b)+ 2 x_a f'_q(x_a) \, x_b f'_\bq(x_b)\,,
\end{align}
are identical up to switching of the labels on the PDFs. For the channels with a quark emission, we have
\begin{align}
\frac{\df\sigma^{(2),\text{LL}}_{g q \to H q}}{\df Q^2 \df Y\df q_T^2} &
 = \hat\sigma^\LO_{gg\to H}(Q) \times \frac{\as C_F}{4\pi} \frac{2}{Q^2} \ln\frac{Q^2}{q_T^2}
   \Bigl[ f_g(x_a) f_q(x_b) + f^{gq}_{\text{uni}}(x_a,x_b) \Bigr]
\,, \\
 \frac{\df\sigma^{(2),\text{LL}}_{g q \to V q}}{\df Q^2 \df Y\df q_T^2} &
 = \hat\sigma^\LO_{q\bar q\to Z}(Q) \times \frac{\as T_F}{4\pi} \frac{2}{Q^2} \ln\frac{Q^2}{q_T^2}
   \, \Bigl[ f^{qg}_{\text{uni}}(x_b,x_a) \Bigr]\,,
\end{align}
where
\begin{align}
f^{gq}_{\text{uni}}(x_a,x_b)&=x_a f'_g(x_a) f_q(x_b)\,, \\
f^{qg}_{\text{uni}}(x_b,x_a)&=f_g(x_a) \, x_b f'_q(x_b)\,,
\end{align}
are again identical up to the switching of the labels on the PDFs.

First, we note that these results involve a more complicated structure of derivatives than the power corrections to the \sceti~beam thrust observable, where at most a single derivative appeared in a given term \cite{Moult:2016fqy, Boughezal:2016zws, Moult:2017jsg}. Furthermore, for beam thrust, at LL there are no derivatives for the channels involving quark emission. Interestingly, the explanation for this arises from very different reasons in the soft and collinear sectors. In the soft sector, it is a simple consequence of the modified power counting of the soft modes, which implies that they must be expanded to two orders in the power counting. In the collinear sector, where the power counting is the same for $q_T$ and beam thrust, it arises from the presence of the power law singularities, which must be expanded against the PDFs. The cancellation of rapidity divergences between the soft and collinear sectors therefore exhibits a much more nontrivial relationship.

Another feature of the LL power corrections is the independence from explicit factors of the color-singlet rapidity $Y$, suggesting that the expansion parameter is indeed $q_T^2/Q^2$, as is expected from the fact that $q_T$ is boost invariant. In fact, the rapidity dependence is induced purely by the PDFs and their derivatives. This is particularly interesting for the case of Drell-Yan, where the only terms that contribute arise from derivatives acting on the PDFs, which leads to a more nontrivial rapidity dependence, and in particular, a rapidity dependence that is different from that at leading power. This has potentially interesting implications for power corrections for $q_T$ subtractions, and we will show this rapidity dependence numerically in \sec{numerics}.

It is also interesting to discuss the universality of these results between Higgs and Drell-Yan production. For the case of beam thrust, the LL results are related by a Casimir scaling, $C_A \leftrightarrow C_F$. Here we see explicitly that this is not the case for $q_T$. However, we see that all terms involving the derivatives of the PDFs are universal up to exchanges of the partonic indices, and it is only the coefficients of the $ff$ PDF structure that are non-universal.  One way of understanding this difference in universality between beam thrust, which is an \sceti~observable, and $q_T$, which is an \scetii~observable, is the different power counting of the soft sector. Since soft momenta in \scetii~scale as ${\cal O}(\lambda)$ rather than ${\cal O}(\lambda^2)$ this requires that for $q_T$ the soft matrix element must be expanded to one higher power, at which point there is a breaking of their universality.
However, the terms involving derivatives of the PDF get part of their power suppression from expanding the momenta entering the PDFs, and therefore are effectively expanded to the same power as for an \sceti~observable such as beam thrust. It would be interesting to understand this universality structure in more detail, in particular how it extends to other processes, and to higher orders.

\subsection{Numerical Results}
\label{sec:numerics}

In this section, we validate our results by numerically comparing the NLP spectrum to the full $q_T$ spectrum,
which we obtain by numerically integrating \eq{sigmaNLO_1}.
For Drell-Yan production, we fix $Q = m_Z = 91.1876~{\rm GeV}$ and use $\as(m_Z) = 0.118$.
For Higgs production, we work in the on-shell limit with $Q = m_H = 125~{\rm GeV}$
and $\as(m_H) = 0.1126428$ corresponding to a three-loop running from $\as(m_Z)$.
In both cases, we use $\Ecm=13~{\rm TeV}$ and the NNPDF31 NNLO PDFs~\cite{Ball:2017nwa} with
fixed factorization and renormalization scales $\mu_f = \mu_r = Q$.
We also fix the rapidity to $Y=2$ to have a nontrivial test of the rapidity dependence of our results
and to break the degeneracy between the $qg$ and $gq$ channels.

We compare the nonsingular cross section at NLO$_0$,%
\footnote{From the point of view of the $q_T$ factorization theorem, the leading-order Born process is $p p \to X$, and hence $\sigma^\LO \sim \delta(q_T^2)$. A nonvanishing transverse momentum is first obtained for $p p \to X+j$, which is the real part of the NLO correction to $p p \to X$, but the LO contribution for $q_T>0$. For clarity, we denote this order as NLO$_0$ to stress that it is counted with respect to the Born process $p p \to X$.}
which is obtained by subtracting all singular terms which diverge as $1/q_T^2$ from the full $q_T$ spectrum,
against our predictions for the NLP cross section.
The dependence of the nonsingular cross section on $q_T$ is given by
\begin{align} \label{eq:xs_nonsing}
 \frac{\df\sigma^{\rm nonsing}_{{\rm NLO}_0}}{\df Q^2 \df Y \df q_T^2} =
 c_1(Q,Y) \ln\frac{Q^2}{q_T^2} + c_0(Q,Y) + \cO\biggl(\frac{q_T^2}{Q^2}\biggr)
\,,\end{align}
where $c_1$ is predicted by the LL term at NLP and $c_0$ is predicted by the NLL term at NLP.
Note that $c_0$ is independent of $q_T$, but has a nontrivial dependence on $Q$ and $Y$.
The $\cO(q_T^2)$ corrections arise at subsubleading power.

\begin{figure*}[t!]
 \centering
 \includegraphics[height=5.6cm]{plots/qt_spectrum_ggH_Y2_loglog.pdf}
   \hfill
 \includegraphics[height=5.6cm]{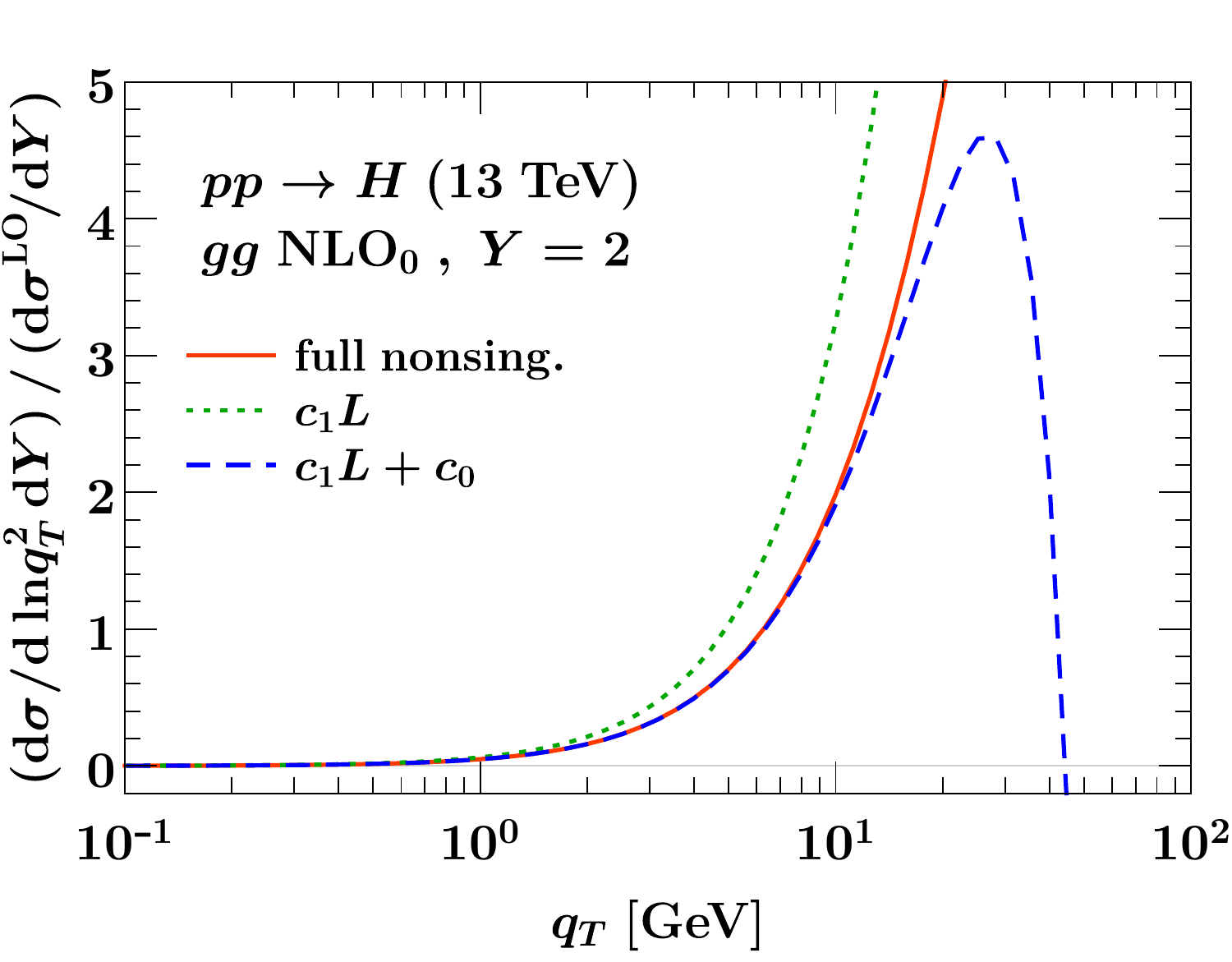}
 \includegraphics[height=5.6cm]{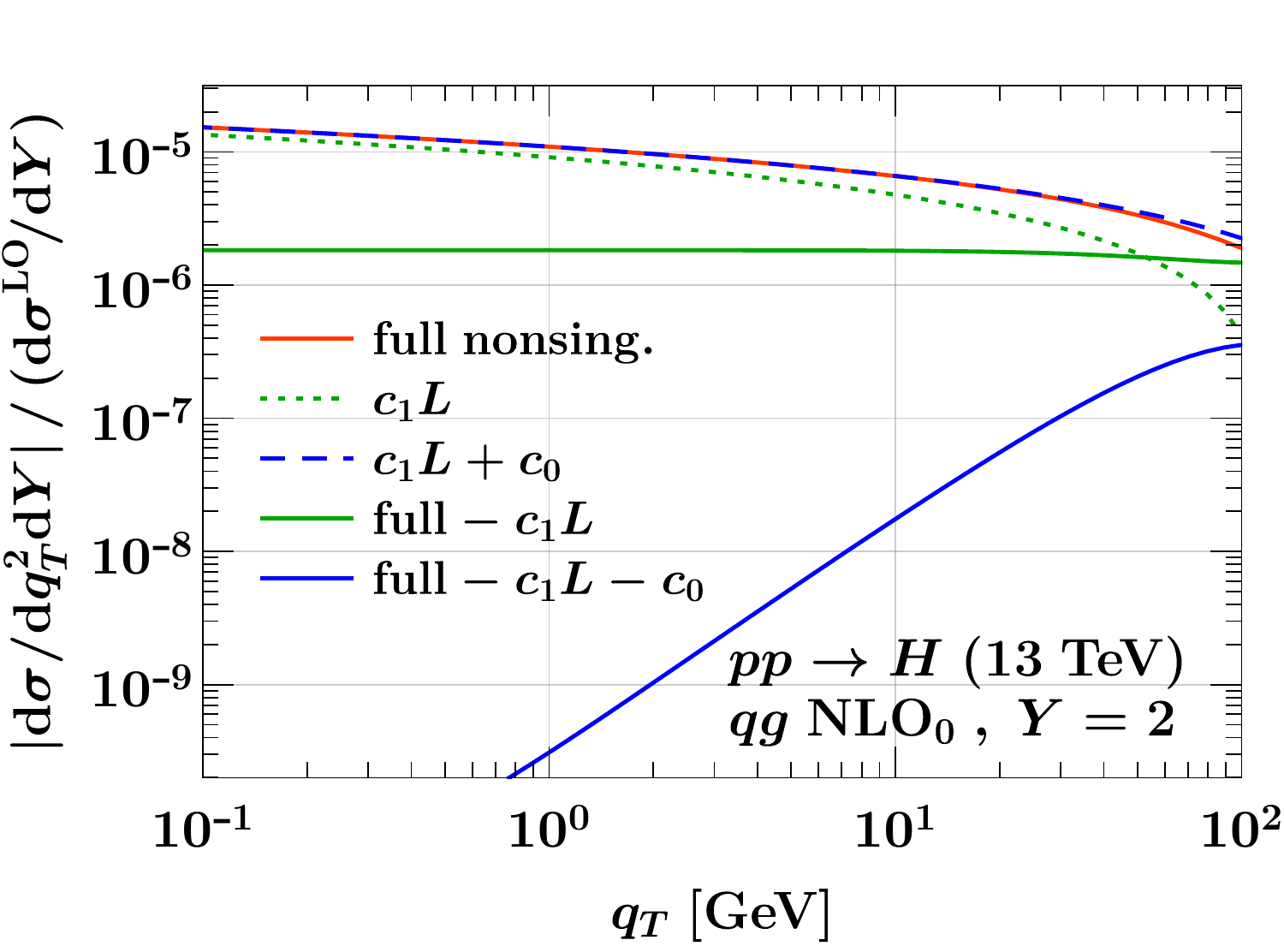}%
   \hfill
 \includegraphics[height=5.6cm]{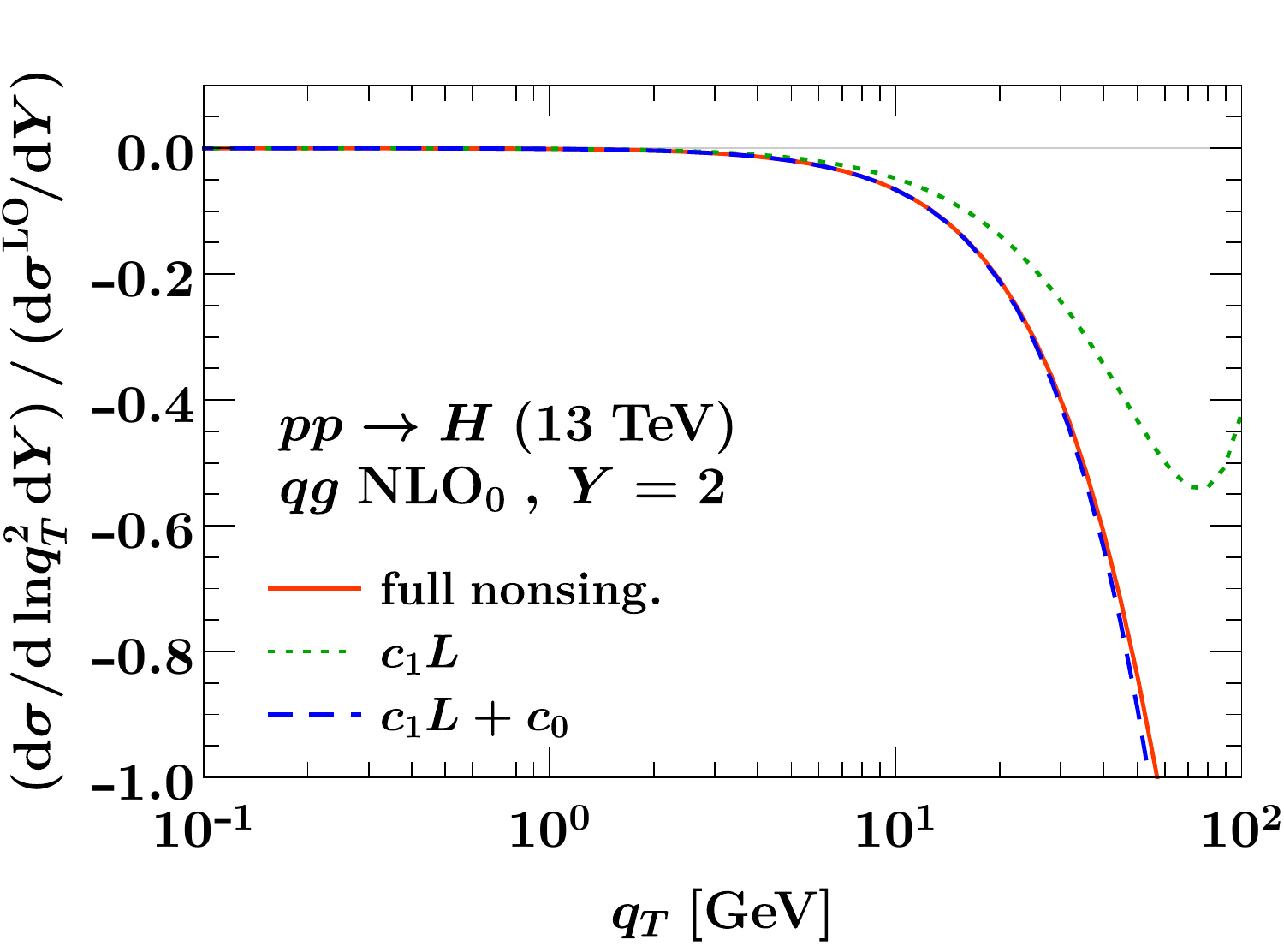}
 \includegraphics[height=5.6cm]{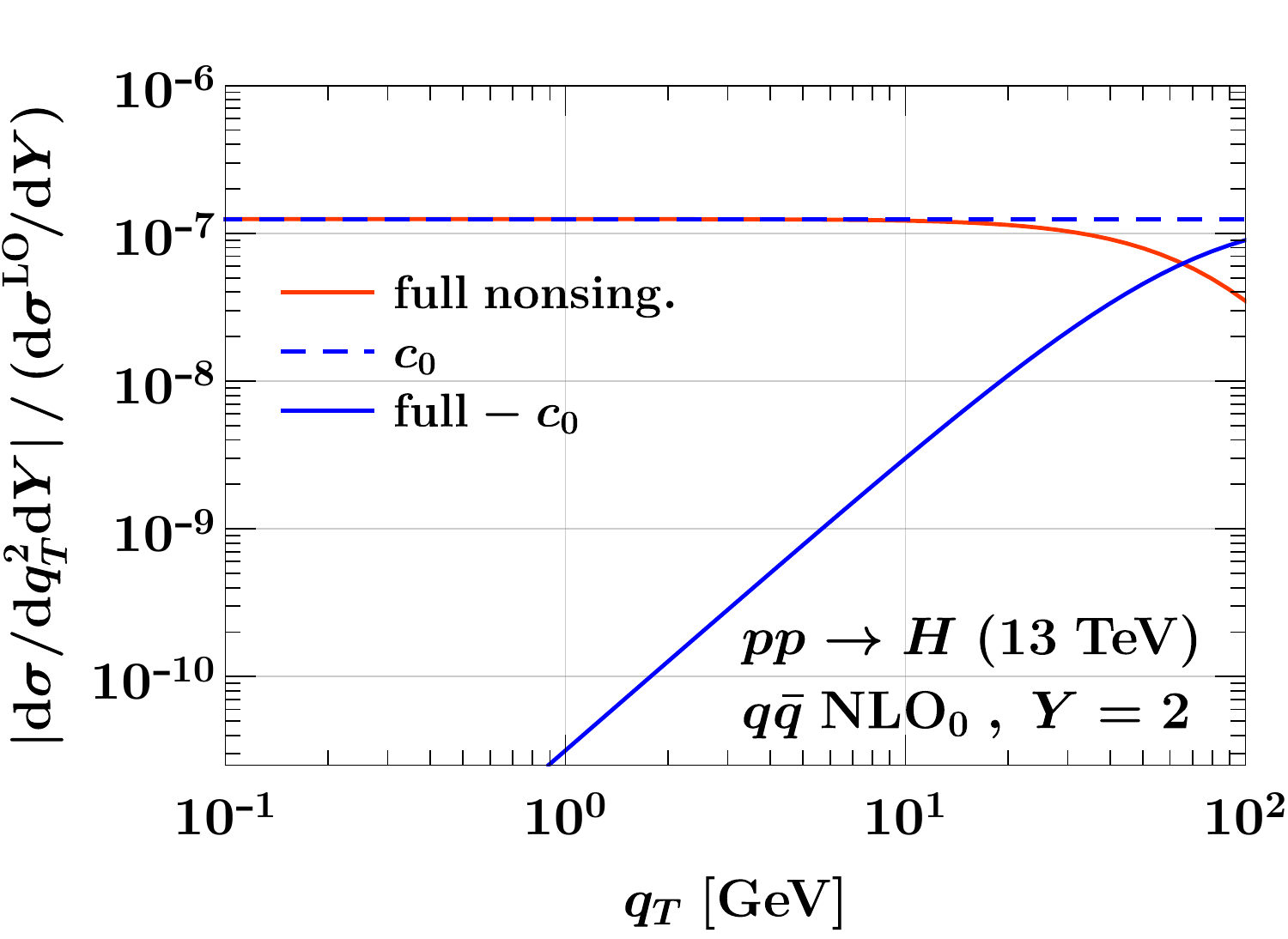}%
   \hfill
 \includegraphics[height=5.6cm]{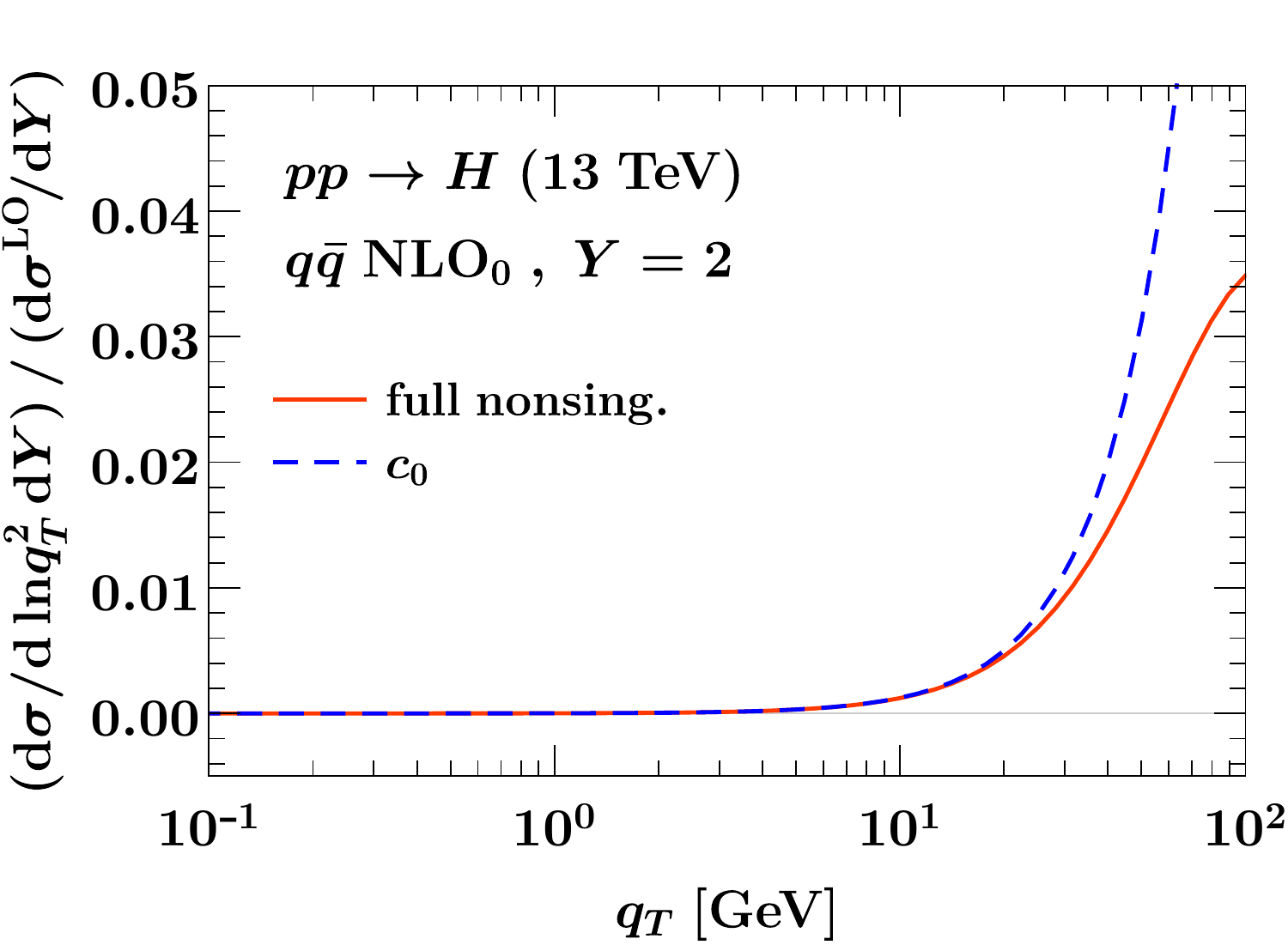}
\caption{Comparison of the LL and NLL corrections at subleading power with the full
nonsingular $q_T$ spectrum for all partonic channels contributing to Higgs production at NLO$_0$.}
 \label{fig:Higgs}
\end{figure*}

\begin{figure*}[t!]
 \centering
 \includegraphics[height=5.6cm]{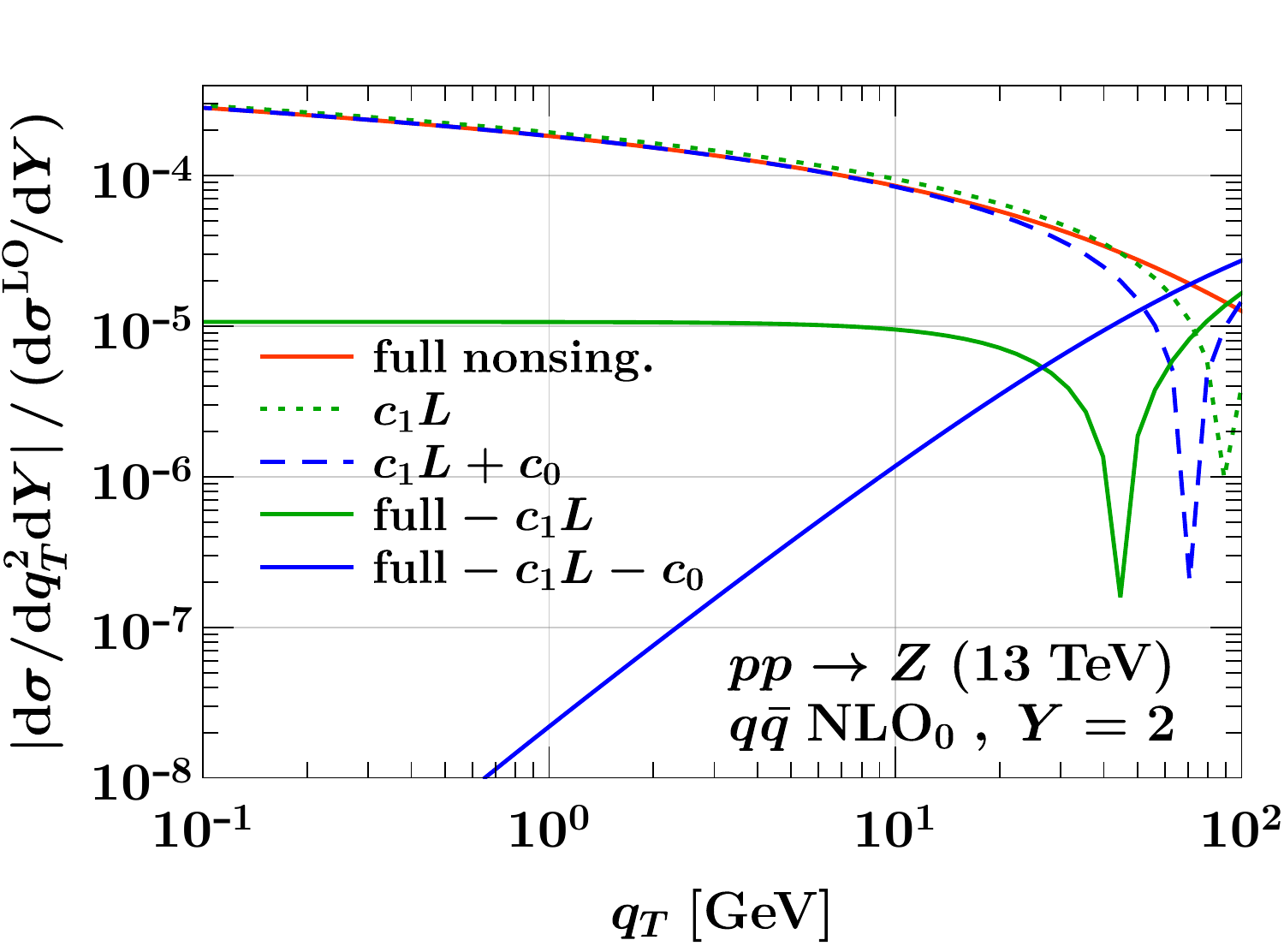}%
   \hfill
 \includegraphics[height=5.6cm]{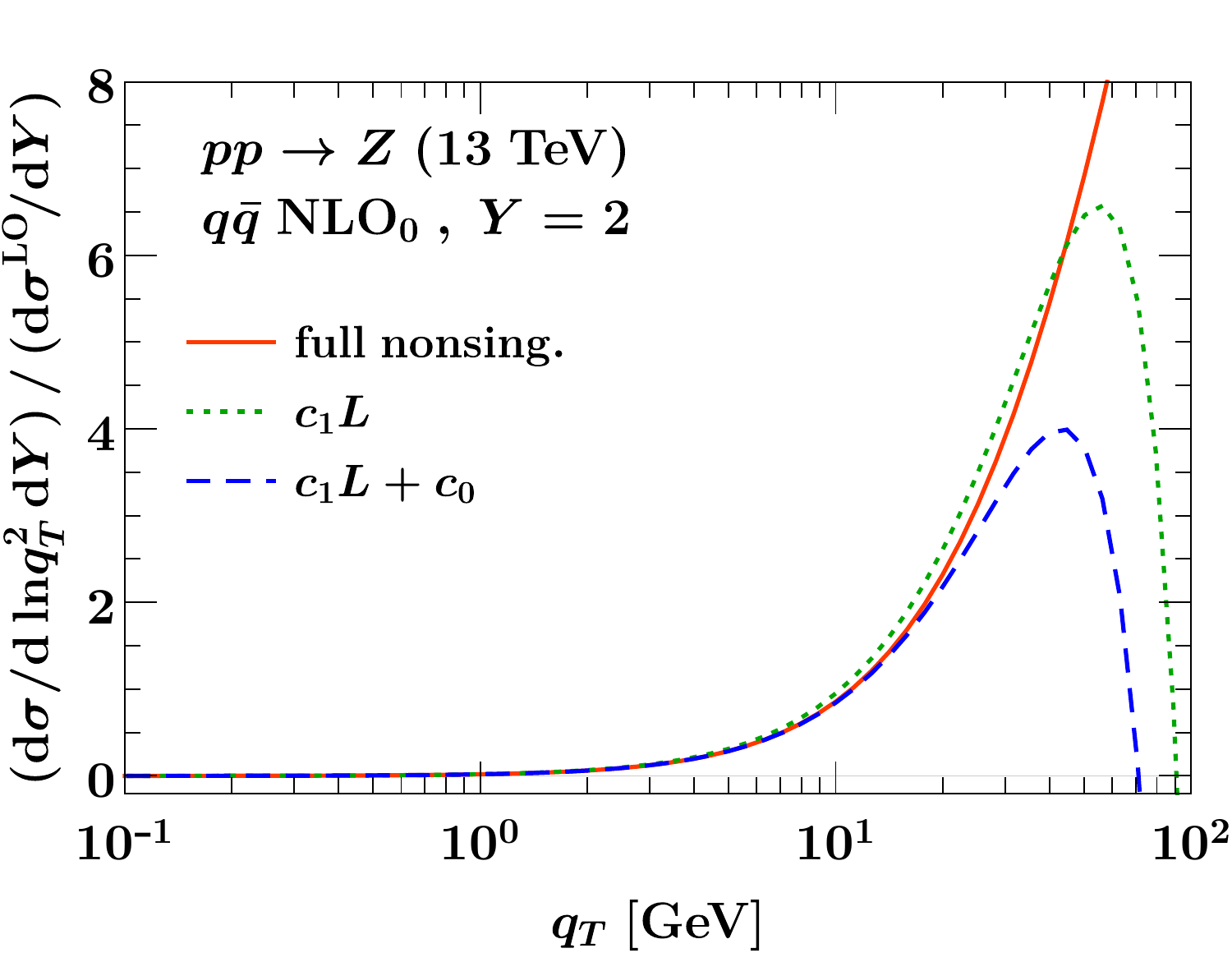}
 \includegraphics[height=5.6cm]{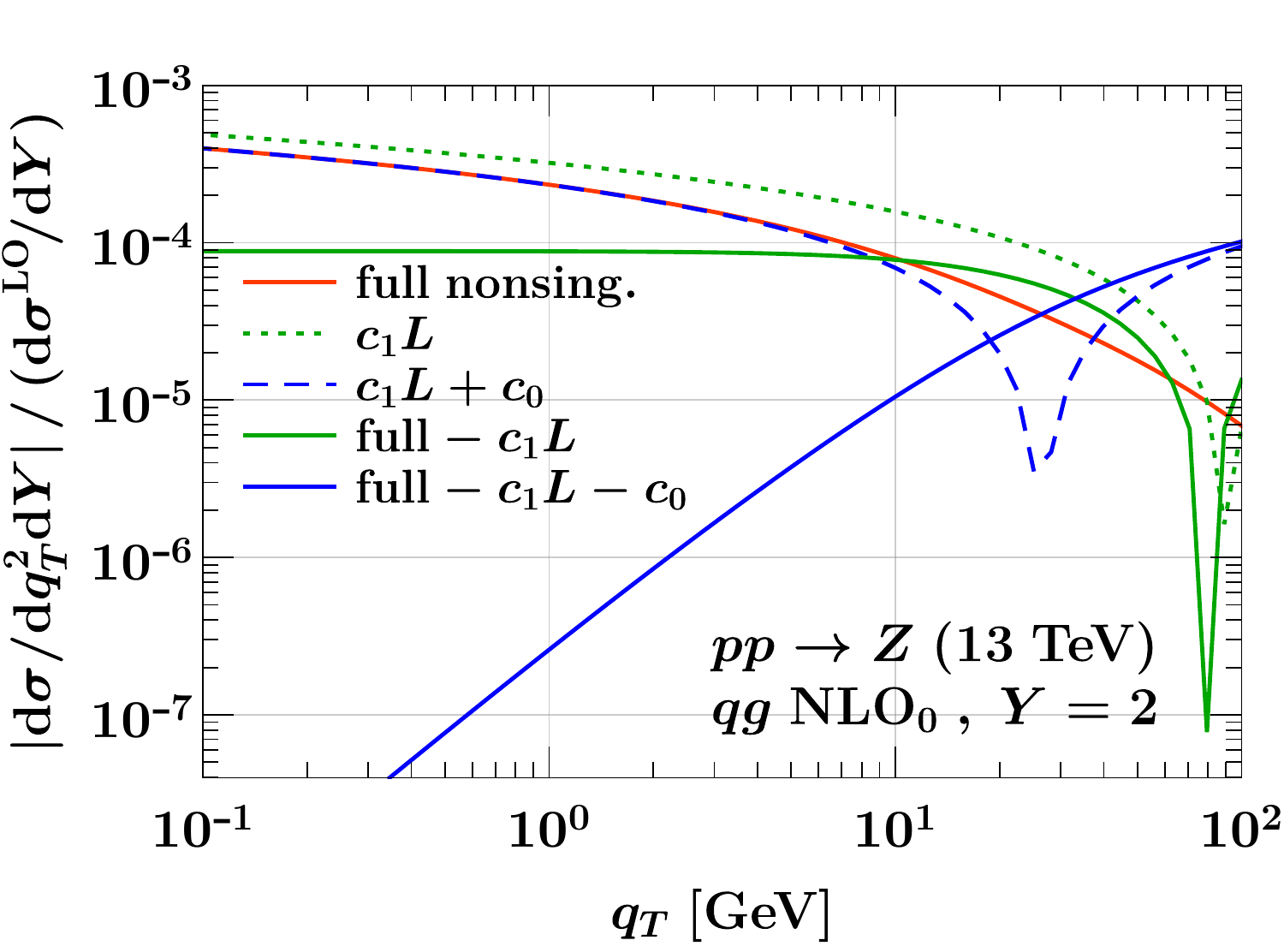}%
   \hfill
 \includegraphics[height=5.6cm]{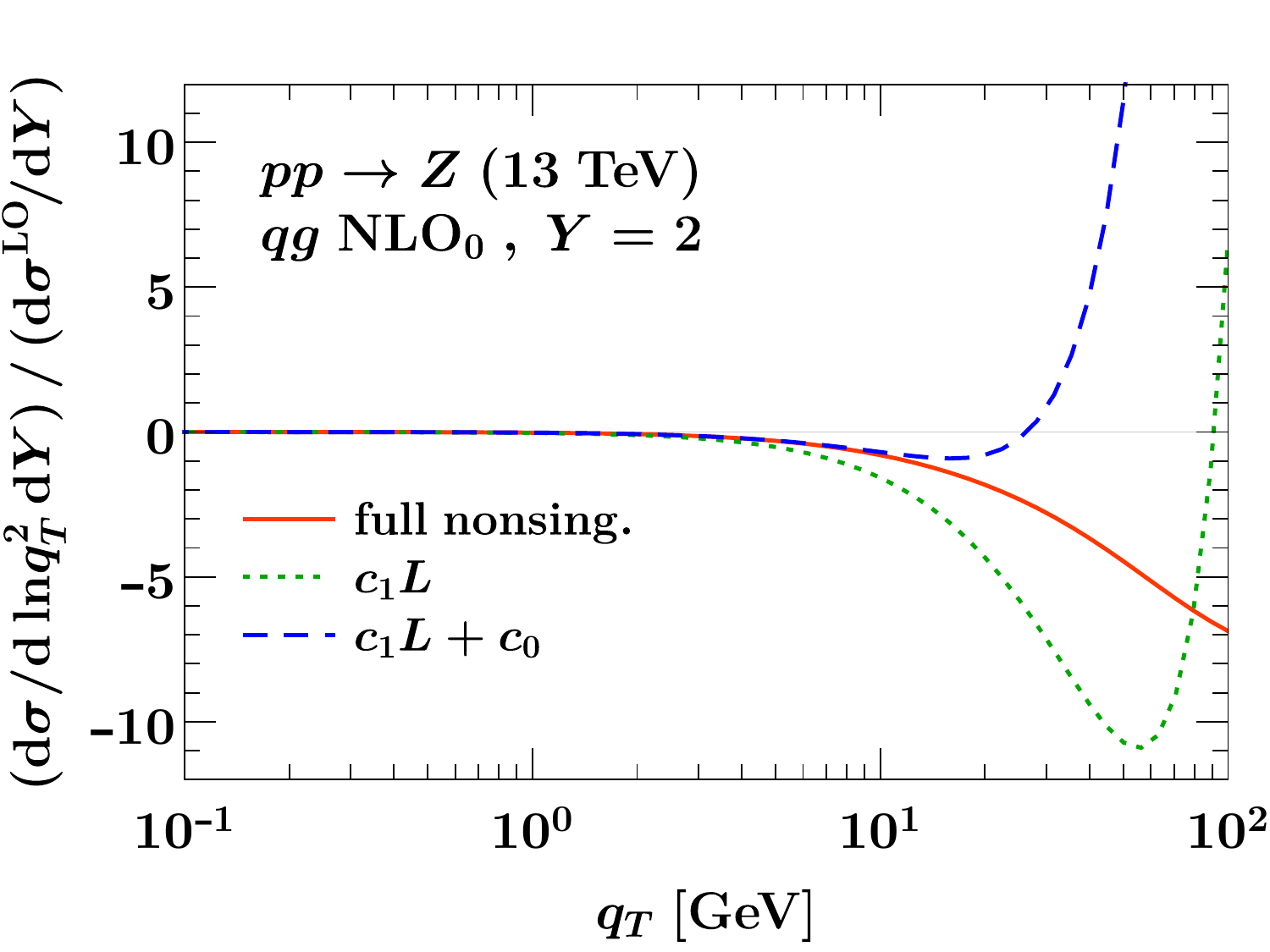}
\caption{Comparison of the LL and NLL corrections at subleading power with the full
nonsingular $q_T$ spectrum for all partonic channels contributing to Drell-Yan production at NLO$_0$.}
 \label{fig:DY}
\end{figure*}

\begin{figure*}[t!]
 \centering
 \includegraphics[height=5.6cm]{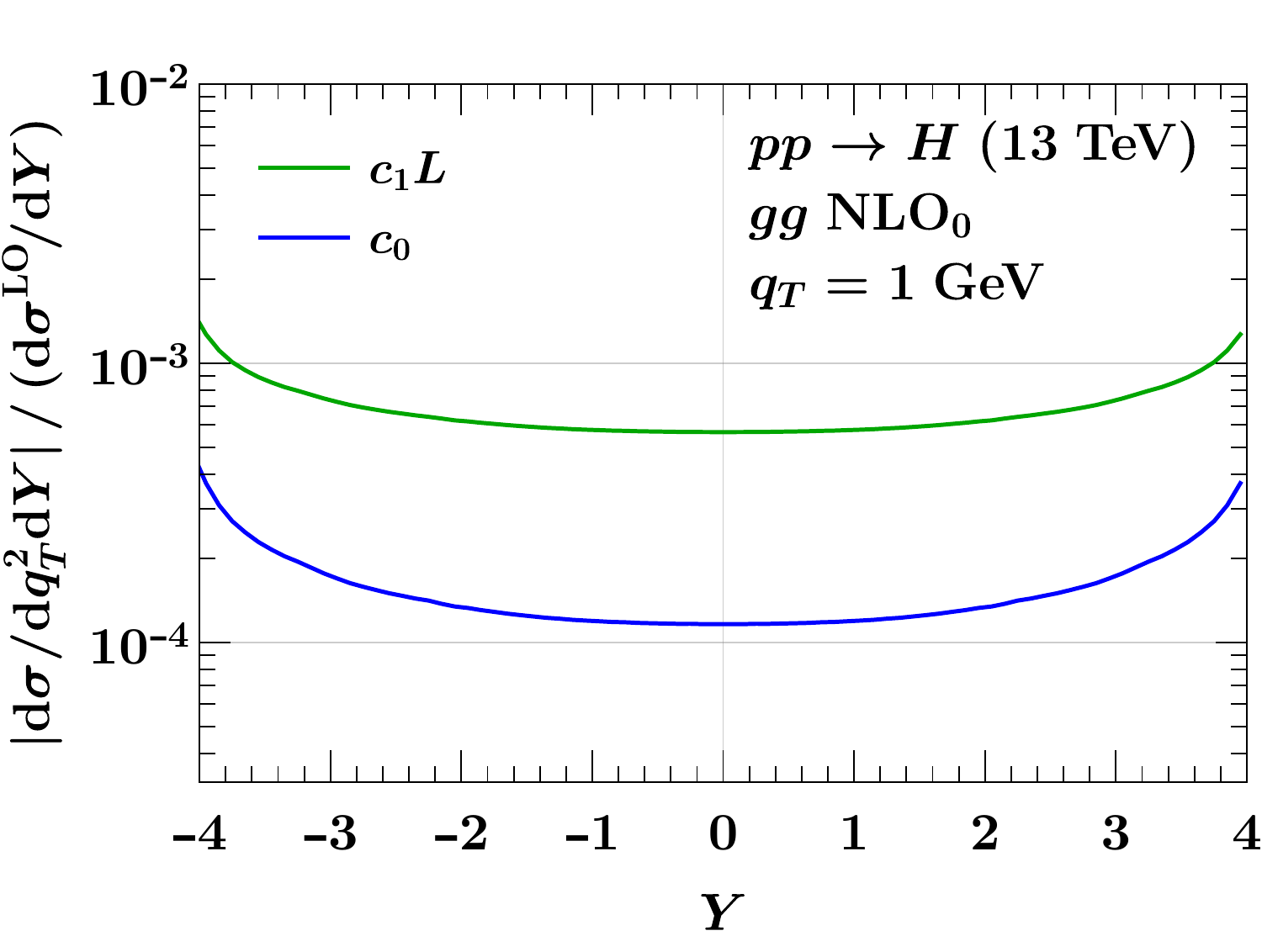}%
   \hfill
 \includegraphics[height=5.6cm]{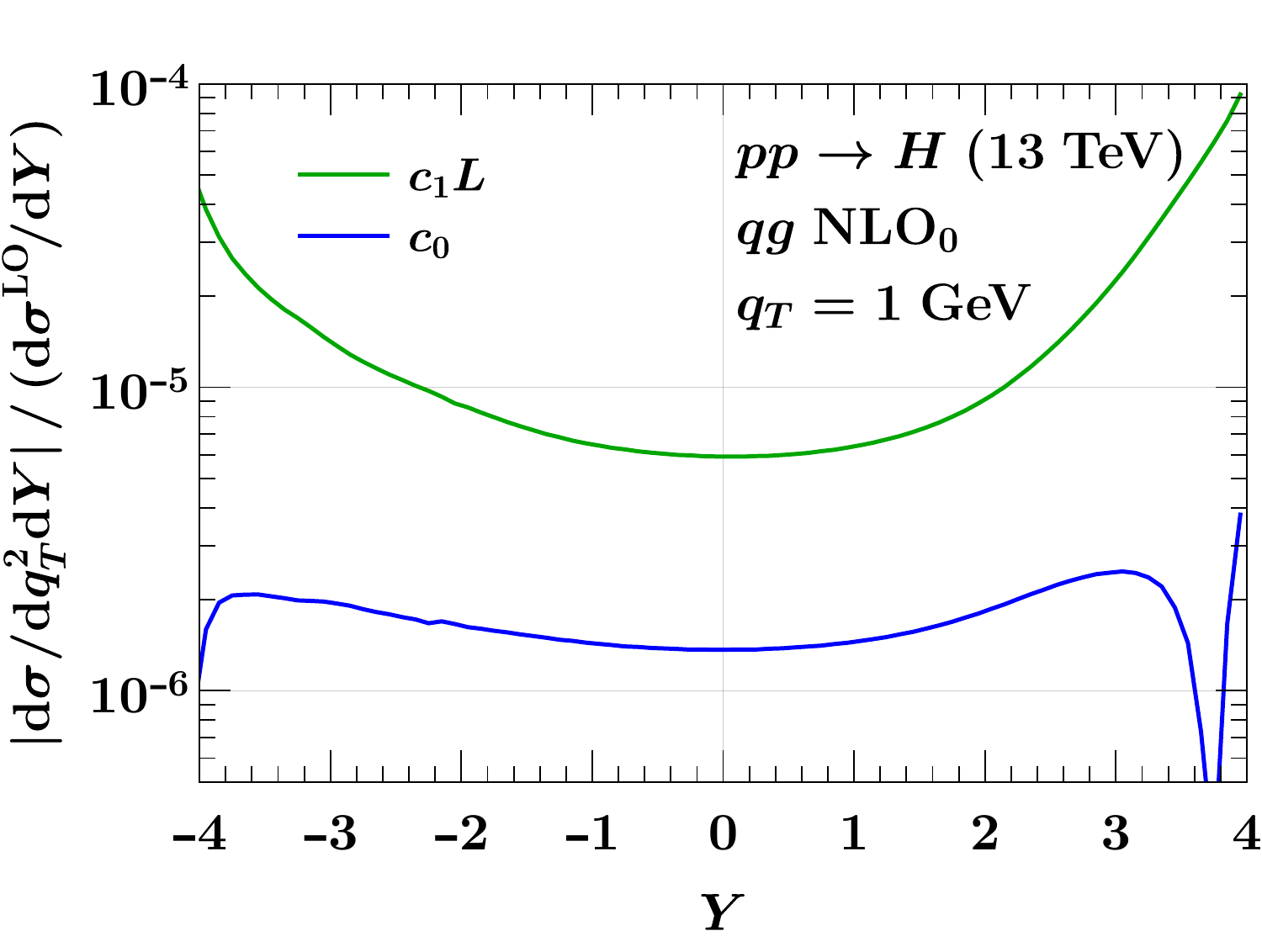}\\
 \includegraphics[height=5.6cm]{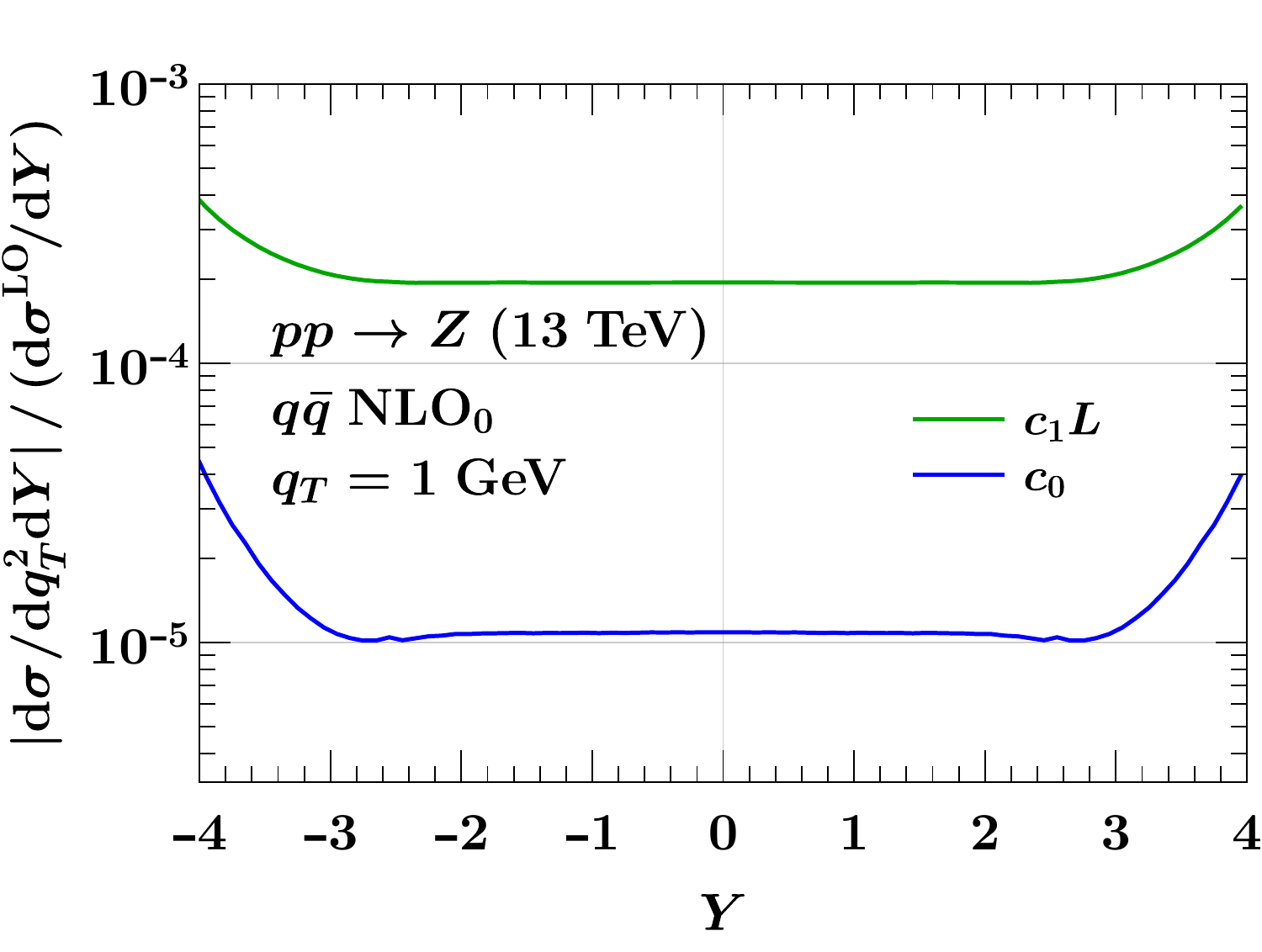}%
  \hfill
 \includegraphics[height=5.6cm]{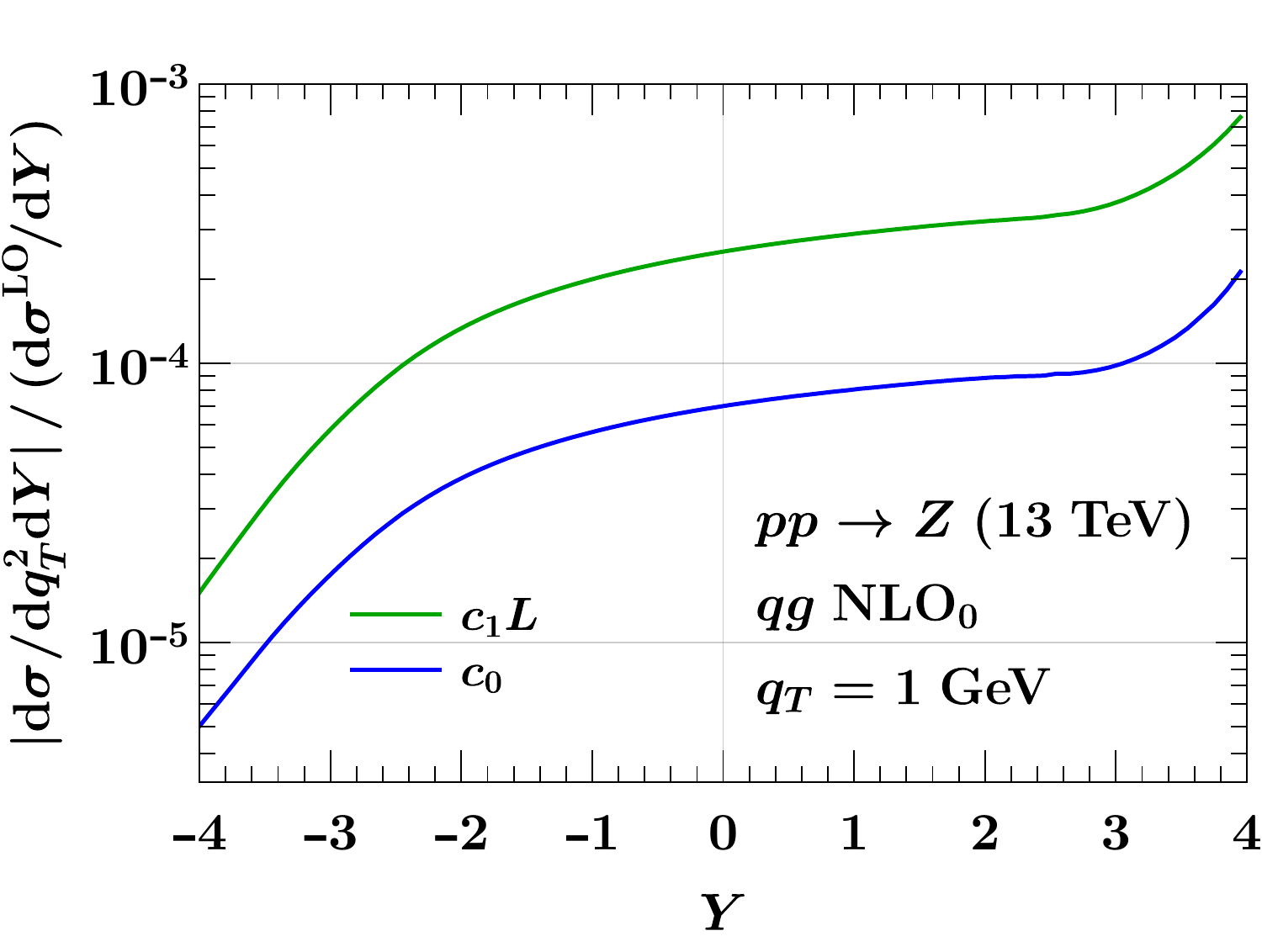}%
 \caption{Rapidity dependence of the LL (green) and NLL (blue) power corrections for Higgs and Drell-Yan production at NLO, relative to the LO rapidity dependence.
  The $q\bar{q}$ channel for Higgs production is not shown, as its LL power corrections vanish.
 }
 \label{fig:rapidity}
\end{figure*}

In \fig{Higgs}, we show the $q_T$ spectrum for all channels contributing to Higgs production.
The corresponding results for Drell-Yan production are shown in \fig{DY}.
In the left panel, we compare the nonsingular $q_T$ spectrum (solid red) against the NLP LL (green dashed) and full NLP (blue dashed) predictions.
For all channels, the NLP NLL result is an excellent approximation of the nonsingular spectrum up to $q_T \sim 10~\GeV$.
The solid green line shows the nonsingular spectrum minus the NLP LL correction, which in all cases is almost perfectly constant up to $q_T \sim 10~\GeV$, as expected from the structure of \eq{xs_nonsing}.
The solid blue line shows the nonsingular spectrum minus the full NLP correction, which vanishes as $q_T^2$ for small $q_T$ as expected from \eq{xs_nonsing}. This provides a strong numerical check
of our analytic results of the NLP contributions.
The right panels of \figs{Higgs}{DY} compare the nonsingular spectrum $q_T^2 \, \df\sigma/\df q_T^2$ with the NLP LL and NLP NLL approximations.
Again, we find excellent agreement up to $q_T \sim 10~\GeV$.

In \fig{rapidity}, we show the rapidity dependence of the power corrections for the $gg$ and $qg$ channels for Higgs production and for the $q\bar{q}$ and $qg$ channels for Drell-Yan production.
We show the individual NLP terms as given in \eq{xs_nonsing}, with the LL term proportional to $c_1$ shown in green and the NLL term proportional to $c_0$ shown in blue.
Since their $q_T$ dependence is trivial, we fix $q_T = 1~\GeV$, which only affects the overall size of the LL term, and we normalize the results to the LO rapidity spectrum.
Despite the fact that the kernels have no explicit rapidity dependence, we observe a nontrivial rapidity dependence due to the PDF derivatives, and in the case of the $qg$ channels also because they involve different PDFs than the Born process.
This is different than the case of beam thrust, which for certain definitions has an explicit rapidity dependence through factors of $e^{\pm Y}$ in both the LL and the NLL kernels \cite{Moult:2016fqy, Moult:2017jsg, Ebert:2018lzn}.
The rapidity dependence is particularly interesting for Drell-Yan production, where the term proportional to the PDFs themselves vanishes, see \eq{sigma_Higgs_LL_introPtNLL}, and so the power corrections are determined solely by the structure of the PDF derivatives.
At large values of $|Y|$, this leads to a relatively large dependence of the power corrections on the rapidity.
For Higgs production this effect is more moderate due to the appearance of a term proportional to PDFs as present at LO, which dominates the rapidity dependence.
This observation, which we believe is likely to persist at higher perturbative orders, could have important implications in the context of $q_T$ subtractions \cite{Catani:2007vq},
where it is important to understand the rapidity dependence of the power corrections. Our results suggest that the rapidity dependence may be well behaved for the case of Higgs production
but could be more problematic for Drell-Yan production. We leave the investigation of the structure at higher perturbative orders to future work.

\section{Conclusions}
\label{sec:conc_pt}

In this chapter, we have studied in detail the structure and consistent regularization of rapidity divergences at subleading order in the power expansion. We have discussed several
new features appearing at subleading power that put additional requirements on the rapidity regulator.
As a result, most of the rapidity regulators that have been used in the literature at leading power
become either unsuitable or inconvenient at subleading power.
In particular, we have shown that the $\eta$ regulator, which in principle can be applied at subleading power, is not homogeneous in the power expansion, which leads to undesirable complications at subleading power.
We have introduced a new pure rapidity regulator, which is homogeneous in the power counting. It allows us to regulate rapidity divergences appearing in $q_T$ distributions at any order in the power expansion, while respecting the power counting of the EFT. This significantly simplified the analysis of rapidity divergences and the associated logarithms at subleading power. It would be interesting to study its application to other physical problems of interest and to further study its properties.

We have also found a rich structure of power-law divergences at subleading power, which can
have a nontrivial effect on the final NLP result. Furthermore, at subleading power, rapidity divergences arise not only from gluons, but also from quarks. It would be interesting to further understand their formal properties.

As an explicit application of our formalism to a physical observable, we considered the $q_T$ spectrum for color-singlet production, for which we computed the complete NLP corrections, i.e., including both the logarithmic and nonlogarithmic contributions, at fixed $\ord{\alpha_s}$.
This provides a highly nontrivial test of our regulator.
In this case, the power-law rapidity divergences have the effect of inducing derivatives of the PDFs in the final NLP result for the $q_T$ spectrum. We also find that unlike for the case of beam thrust, where the LL power corrections for Higgs and Drell-Yan production are related by $C_A\leftrightarrow C_F$, this is not the case for the LL power corrections for $q_T$, which have a different structure for these two processes.

Our results represent a first important step in systematically studying subleading power corrections for observables with rapidity divergences. It opens the door for addressing a number of interesting questions. It will be important to extend our results and to better understand the structure of subleading-power rapidity divergences at higher perturbative orders. As a particularly interesting application, the power corrections for the $q_T$ spectrum can be used to improve the numerical performance and to better understand the systematic uncertainties of $q_T$ subtractions, whose feasibility at next-to-next-to-next-to-leading order has recently been demonstrated in \refcite{Cieri:2018oms} for Higgs production. We also hope that recent advances in the renormalization at subleading power, which has enabled the all-orders resummation of subleading-power logarithms, can also be extended to enable the resummation of subleading-power rapidity logarithms, with possible applications in a variety of contexts.

\chapter{Collinear Expansion for Color Singlet Cross Sections}\label{sec:collinear_expansion_chapter}
\section{Introduction}
Our knowledge of the structure of quantum field theory (QFT) is rapidly advancing. 
On the one hand this steady progress allows us to answer fundamental questions about the interactions of nature by  deriving precise predictions for the outcome of scattering experiments that can be compared with experimental observation.
On the other hand we learn about the mathematical structures that underly this description.

Progress in QCD perturbation theory has allowed us to venture to predictions at next-to-next-to-next-to leading order (N$^3$LO) in the strong coupling constant for select inclusive and differential cross sections at the Large Hadron Collider (LHC)~\cite{Anastasiou:2015ema,Dreyer:2016oyx,Mistlberger:2018etf,Cieri:2018oms,Dulat:2018bfe,Dreyer:2018qbw,Duhr:2019kwi,Duhr:2020seh,Duhr:2020kzd}.
Resummation of kinematic limits of cross sections has reached the similarly astounding precision for a multitude of observables~\cite{Becher:2008cf, Abbate:2010xh, Hoang:2015hka, Bonvini:2014joa, Schmidt:2015cea, H:2019dcl, Ebert:2017uel, Chen:2018pzu, Bizon:2018foh}.
Nevertheless, the difficulty of describing the scattering of fundamental particles is ever rising with increasing demand for precision and for more complex observables. 
To overcome seemingly insurmountable complexity, parametric or kinematic expansions have proven highly effective.
For example, expanding the gluon fusion Higgs boson production cross section around the production threshold of the Higgs boson allowed for the computation of the first hadron collider cross section at N$^3$LO in QCD perturbation theory~\cite{Anastasiou:2015ema}.

Kinematic expansions in hadron collisions have been studied since a long time. For example, such expansions provide the bases of factorization theorems for inclusive processes in hadron collisions~\cite{Bodwin:1984hc,Collins:1984kg,Collins:1985ue,Sterman:1986aj,Collins:1988ig,Collins:1989gx,Catani:1989ne,Lustermans:2019cau}.
They have also been used to derive universal quantities like emission currents or splitting amplitudes (see for example \refscite{Altarelli:1977zs,DelDuca:1989jt,Mangano:1990by,Bern:1994zx,Campbell:1997hg,Catani:1998nv,Catani:2000vq, deFlorian:2001zd,Becher:2009cu,Dixon:2010zz,Catani:2010pd,Duhr:2013msa,Duhr:2014nda,Li:2014afw,Anastasiou:2018fjr,DelDuca:2019ggv,Dixon:2019lnw}), for studying the high energy behavior of amplitudes and cross sections (see for example \refscite{Kuraev:1977fs,Balitsky:1978ic,Lipatov:1985uk,Catani:1990eg,Mueller:1994jq,Korchemskaya:1996je,DelDuca:2001gu,Caron-Huot:2017zfo,Caron-Huot:2020grv}) as well as in the calculation of counterterms for subtraction algorithms (see for example~\refscite{Gehrmann_De_Ridder_2012,DelDuca:2013kw,Czakon:2014oma,Herzog:2018ily,Magnea:2018hab,Delto:2019asp}).
In the method of regions, one expands Feynman integrals in all relevant kinematic limits to simplify their evaluation~\cite{Beneke:1997zp}.
More generally they can be used to study divergence structures of Feynman integrals~\cite{Anastasiou:2018rib} or to approximate hadronic cross sections~\cite{Dittmaier:2014qza,Bonetti:2017ovy,Lindert:2018iug,Anastasiou:2018adr,Liu:2019tuy,Dreyer:2020urf}.
Soft-Collinear Effective Theory (SCET) is based on the kinematic expansion of scattering amplitudes and the realisation that such limits can be described by effective field theories~\cite{Bauer:2000ew, Bauer:2000yr, Bauer:2001ct, Bauer:2001yt, Bauer:2002nz}.
These techniques have also been used to derive the factorization of several infrared observables for color-singlet processes at hadron colliders, see for example \refscite{Stewart:2009yx,Becher:2010tm,GarciaEchevarria:2011rb,Chiu:2012ir,Monni:2016ktx,Buffing:2017mqm,Tackmann:2012bt,Banfi:2012jm,Becher:2012qa,Liu:2012sz,Procura:2014cba,Lustermans:2019plv,Monni:2019yyr}.

In this article we detail a technique for the efficient expansion of differential partonic cross sections for the production of a color singlet final state $h$ in hadron-hadron collisions in the kinematic limit that all radiation produced alongside $h$ is collinear to one of the collision axis of our scattering process.
The method outlined here is based on the work mentioned before and extends existing technology.
It also shares many similarities with the method developed in \refscite{Anastasiou:2015yha,Anastasiou:2013srw,Anastasiou:2014lda,Anastasiou:2014vaa,Dulat:2017prg} to expand cross sections around the limit of all radiation being soft.
Our expansion is carried out at the integrand level, i.e., before loop or phase space integrals are carried out. 
The resulting expressions can be interpreted diagrammatically.
This in turn greatly simplifies the analytic computation of matrix elements by employing powerful loop integration techniques like the reverse unitarity framework~\cite{Anastasiou2003,Anastasiou:2002qz,Anastasiou:2003yy,Anastasiou2005,Anastasiou2004a} or integration-by-part (IBP) identities~\cite{Chetyrkin:1981qh,Tkachov:1981wb}.
Our expansion is systematically improvable as we can compute to arbitrarily high power in our expansion parameter. 
The mathematical functions that appear in each term of the expansion are determined by the first few expansion coefficients. 

The collinear expansion of cross sections can find many applications in the computation of higher order corrections to scattering processes.
Cross sections for the production of hard probes $h$ can be approximated by performing a systematic collinear expansion.
Recently, an all-order factorization theorem was derived for the first order in this collinear expansion~\cite{Lustermans:2019cau}.
While the usefulness of our expansion technique depends on the specific observable in question,
it is obvious that key observables like the rapidity or transverse momentum of a hard probe are amenable to such an expansion.
We demonstrate the applicability of our collinear expansion to the rapidity distribution of the Higgs boson produced via gluon fusion. By calculating the collinear expansion to its second order, we demonstrate the excellent convergence of our series towards the full result at NNLO in perturbation theory.

Kinematic limits of cross sections can also be used to identify universal structures of quantum field theories.
Our expansion technique allows to gain access to splitting functions or integrated counter terms that may find application in subtraction algorithms used for the computation of fully differential cross sections. 
Universal building blocks that find their application in the resummation of perturbative cross sections can be accessed efficiently using this expansion technique. 

One example of such universal building blocks are so-called beam functions~\cite{Stewart:2009yx,Stewart:2010qs}
which arise in SCET and play a crucial role in factorization theorems of hadronic observables.
We demonstrate how to relate beam functions to the kinematic limit of our perturbative cross sections and how they can be extracted efficiently.
Specifically, we investigate the transverse momentum $(q_T)$ dependent beam functions and $N$-jettiness ($\Tau_N$) beam function.
We illustrate our method by computing these quantities through NNLO, up to the second order in the dimensional regularization parameter $\eps$, confirming recent results in the literature~\cite{Luo:2019hmp,Luo:2019bmw,Baranowski:2020xlp}.
These results are necessary input for the calculation of aforementioned beam functions at N$^3$LO in QCD,
where much progress has been already made for the quark $\Tau_N$ beam function~\cite{Melnikov:2018jxb,Melnikov:2019pdm,Behring:2019quf}, and which has already been achieved for the quark $q_T$ beam function and TMDPDF~\cite{Luo:2019szz}.
In our companion papers~\cite{Ebert:2020yqt,Ebert:2020unb}, we complete this task by computing the $q_T$ and $\Tau_0$ beam functions in all channels at N$^3$LO based on the methods outlined in this article.

In recent years the universal structure of cross sections beyond leading power in kinematic expansions within SCET have been explored \cite{Manohar:2002fd,Beneke:2002ph,Pirjol:2002km,Beneke:2002ni,Bauer:2003mga,Hill:2004if,Lee:2004ja,Benzke:2010js,Freedman:2014uta,Kolodrubetz:2016uim,Moult:2016fqy,Moult:2017jsg,Beneke:2017vpq,Feige:2017zci,Moult:2017rpl,Chang:2017atu,Moult:2017xpp,Alte:2018nbn,Beneke:2018gvs,Beneke:2017ztn,Beneke:2018rbh,Moult:2018jjd,Ebert:2018lzn,Ebert:2018gsn,Bhattacharya:2018vph,Beneke:2019kgv,Moult:2019mog,Beneke:2019mua,Moult:2019uhz,Moult:2019vou,Liu:2019oav,Liu:2020ydl,Liu:2020eqe}.
As this avenue of research is still growing rapidly, our expansion techniques may provide analytic information towards the structure of cross sections at higher power.  
In fact, the method developed in this chapter is inspired by the calculation of power corrections in fixed order SCET for $\Tau_0$ \cite{Moult:2016fqy,Moult:2017jsg,Ebert:2018lzn} and $q_T$~\cite{Ebert:2018gsn}.
It will be interesting to extend these studies to higher order in $\as$ and the power expansion.
We hope that our techniques will provide readily accessible tools for the computation of yet unknown universal building blocks.

This article is structured as follows: In \sec{setup} we setup a parameterization for differential cross sections for color singlet production at hadron colliders.
This will mainly serve to develop a notation and to identify the objects that we aim to expand. 
In \sec{collinear_expansion_xs} we introduce the general strategy of expanding differential hadronic cross sections around the collinear limit, identifying the relevant kinematic regions and formally defining what we intend by collinear expansion.
We then continue the discussion about collinear expansions in \sec{collinear_expansion} by showing in practice how to perform the collinear expansion for squared matrix elements.
We will show explicit examples of the expansion of two loop cut diagrams at leading and beyond leading power, both for real radiation as well as for loop corrections.
In \sec{SCET} we explain how our collinear expansion of cross section is related to the effective field theory framework of SCET and in particular to the factorization of hadronic differential cross sections.
In \sec{beamfunctions} we review the role of SCET beam functions in the factorization of hadronic differential cross sections and we show that they are naturally connected to the leading term of our collinear expansion of cross sections. We discuss in detail how to obtain beam functions both in the case of $q_T$ and $\Tau_N$.
In \sec{rapidity} we apply our formalism to compute the rapidity spectrum of the Higgs in gluon fusion at NNLO in QCD via the collinear expansion of the partonic cross section.
We conclude in \sec{conclusions_collexp}.

\section{Setup for differential cross sections}
\label{sec:setup}

In this section, we develop the notation for differential cross sections at hadron colliders.
In \sec{setup_general}, we introduce our generic notation for the production of a colorless hard probe $h$ in a proton-proton collision.
In \sec{setup_phasespace} we provide a detailed derivation of the required differential phase space.

\subsection{General setup and notation}
\label{sec:setup_general}

We consider the production of a colorless hard probe $h$ and an additional hadronic state $X$ in a proton-proton collision.
Examples of such processes are the gluon fusion production cross section of a Higgs boson or the hadronic production of a $Z$ boson or virtual photon (Drell-Yan).
\begin{align} \label{eq:process_hadr}
 P(P_1) + P(P_2) \quad\to\quad h(-p_h) + X(-k)
\,.\end{align}
Here, $P_{1,2}$ are the momenta of the incoming protons, which in the hadronic center-of-mass frame are given by
\begin{align} \label{eq:P1P2}
 P_1^\mu = \sqrt{S} \frac{n^\mu}{2} \,,\qquad P_2^\mu = \sqrt{S} \frac{\bn^\mu}{2}
\,,\end{align}
where $S = (P_1 + P_2)^2$ is the hadronic center-of-mass energy and the protons are aligned along the directions
\begin{align} \label{eq:nnb}
 n^\mu = (1,0,0,1) \,,\qquad \bn^\mu = (1,0,0,-1)
\,.\end{align}
In \eq{process_hadr}, $p_h$ is the momentum of the hard probe $h$, and $k$ is the total momentum of the hadronic state $X$,
and as indicated both momenta are taken to be incoming.

The hadronic process in \eq{process_hadr} receives contributions from the partonic processes
\begin{align} \label{eq:process_part}
 i(p_1) + j(p_2) \quad\to\quad h(-p_h) + X_n(-p_3, \dots, -p_{n+2})
\,,\end{align}
where $i$ and $j$ are the flavors of the incoming partons, and their momenta are given by
\begin{align} \label{eq:p1p2}
 p_1^\mu = x_1 P_1^\mu \,,\qquad p_2^\mu = x_2 P_2^\mu
\,,\end{align}
such that the partonic center of mass energy is given by
\begin{align}
 s = (p_1 + p_2)^2 = x_1 x_2 S
\,.\end{align}
In \eq{process_part}, $X_n$ is a hadronic final state consisting of $n\ge0$ partons with momenta $\{p_3, \cdots, p_{n+2}\}$ and total momentum $k^\mu \equiv \sum_{i>2} p_i^\mu$. 

We are interested in describing processes that are differential in the four momentum $p_h^\mu$,
which we parameterize in terms of its rapidity $Y$ and virtuality $Q$,
\beq
\label{eq:hadrvardef}
Y=\frac{1}{2}\log\left(\frac{\bar n\cdot p_h}{n\cdot p_h}\right),\hspace{1cm} Q^2=p_h^2\,,
\eeq
and by momentum conservation its transverse momentum $p_{h \perp}^\mu$ is fixed to be $p_{h \perp}^\mu = -k_\perp^\mu$.
The momentum $k^\mu$ is parameterized in terms of the variables
\bea
\label{eq:vardef}
\wa=-\frac{\bar n\cdot k}{\bar n \cdot p_1},\hspace{1cm}
\wb=-\frac{ n\cdot k}{ n \cdot p_2},\hspace{1cm}
x=\frac{k^2}{(\bar n\cdot k)(n\cdot k)} = 1- \frac{\vec{k}_\perp^2}{(\bar n\cdot k)(n\cdot k)} \,.
\eea

We refer to the hadronic cross section differential in the above variables as the general differential cross section,
\beq
\label{eq:sigma_differential}
 \frac{\df\sigma}{\df Q^2 \df Y \df \wa \df \wb \df  x}  =  \frac{\sigma_0}{\tau} \sum_{i,j} x_1 f_i\left(x_1\right) x_2 f_j\left(x_2\right) \frac{\df \eta_{ij}}{ \df Q^2  \df \wa \df \wb \df  x}  \,.
\eeq
Here, the sum runs over all possible initial state configurations $i,j$, the $f_i(x)$ denote the parton distribution functions, and $\df \eta_{ij} / (\df Q^2  \df \wa \df \wb \df  x)$ is the general partonic coefficient function.
\eq{sigma_differential} is normalized by $\sigma_0$, which contains all constant factors appearing in the Born level cross section.
The Bjorken momentum fractions $x_{1,2}$ can be expressed in terms of the variables introduced above.
\bea
\label{eq:xidef}
x_1&=& \frac{x_1^B} {z_1}    =x_1^B  \biggl[\sqrt{1+(k_T/Q)^2} - \frac{\bn \cdot k}{Q} e^{-Y} \biggr]
\,,\nn\\
x_2&=&  \frac{x_2^B} {z_2}  = x_2^B \biggl[\sqrt{1+(k_T/Q)^2} - \frac{n \cdot k}{Q} e^{+Y} \biggr]
\,,\eea
where the momentum fractions appearing at Born level are given by
\beq \label{eq:xiBdef}
x_1^B =\sqrt{\tau} e^Y,\hspace{1cm}x_2^B =\sqrt{\tau} e^{-Y}
\,,\eeq
where $\tau=Q^2 / S$
and we use the functions 
\beq
\label{eq:zdef}
z_1=\sqrt{\frac{1-\wa}{1-\wb}} \sqrt{1-\wa-\wb+\wa\wb x},\hspace{0.5cm}z_2=\sqrt{\frac{1-\wb}{1-\wa}} \sqrt{1-\wa-\wb+\wa\wb x}.
\eeq
At Born level, $k^\mu=0$, such that the momentum fractions $x_{1,2}$ reduce to $x_{1,2}^B$,
while in the presence of real radiation the kinematic constraint $k^\mu < 0$ dictates that $x_{1,2} \ge x_{1,2}^B$.

The general partonic coefficient function in \eq{sigma_differential} is given by
\begin{align} 
\label{eq:sigma_part}
  \frac{\df \eta_{ij}}{ \df Q^2  \df \wa \df \wb \df  x}  &
 = \frac{\tau}{\sigma_0} \frac{\cN_{ij}}{2 Q^2} \sum_{X_n}\int  \frac{\df\Phi_{h+n} }{ \df \wa \df \wb \df  x}\, |\cM_{ij\to h+X_n}|^2
\,.\end{align}
Here, the sum runs over all hadronic final states $X_n$ consisting of $n$ partons,
and $\df\Phi_{h+n}$ is the phase space measure of the $h+X_n$ final state which will be discussed in more detail in \sec{setup_phasespace}.
$|\cM_{ij\to h+X_n}|^2$ is the associated squared matrix element summed over final and initial state colors and helicities.
We have also pulled out the overall normalization factor $\cN_{ij}$, related to the spins and polarizations of the incoming partons. 
Depending on the initial state, it is given by
\begin{align}
 \cN_{gg} &= \frac{1}{4(n_c^2-1)^2(1-\epsilon)^2}
\,,\nn\\
 \cN_{qg} = \cN_{gq} &= \frac{1}{4(n_c^2-1) n_c(1-\epsilon)}
\,,\nn\\
 \cN_{qq} = \cN_{q\bq} = \cN_{qq'} = \cN_{q\bq'} &= \frac{1}{4n_c^2}
\,.\end{align}
Here, $g$, $q$ ($\bq)$ and $q'$ ($\bq'$) indicate a gluon, (anti-)quark, and (anti-)quark of different flavor than $q$, respectively.

We expand the general partonic coefficient function in $\as$ as
\begin{align} \label{eq:eta_ij_1}
 &\frac{\df\eta_{ij} }{\df Q^2 \df w_1 \df w_2 \df x}
 = \sum_{\ell=0}^\infty \left(\frac{\as}{\pi}\right)^{\ell}
   \frac{\df\eta_{ij}^{(\ell)}}{\df Q^2 \df w_1 \df w_2 \df x}
\\\nn&
 = \eta_{ij}^V \delta(w_1)\delta(w_2)\delta(x)
 \,+\, \sum_{\ell=1}^\infty \left(\frac{\as}{\pi}\right)^{\ell}
   \sum_{n,m=1}^\ell w_1^{-1-m\eps} w_2^{-1-n\eps}
   \frac{\df\eta_{ij}^{(\ell,m,n)}(w_1,w_2,x,Q^2)}{\df Q^2 \df w_1 \df w_2 \df x}
\,.\end{align}
Here, $\eta_{ij}^V$ contains the Born cross section and purely virtual corrections,
and can itself be expanded in $\as/\pi$ with the first term $\eta_{ij}^{V\,(0)}=\delta_{\bar{i}j}$ for flavour diagonal processes like Drell-Yan or Higgs production.
The $\eta_{ij}^{(\ell,m,n)}$ are separately holomorphic in the vicinity of $w_1 = 0$ or $w_2 = 0$.

The differential cross section for a specific observable $\Obs$ that only depends on $p_h^\mu$ and $k^\mu$ is obtained from our general differential cross section given in \eq{sigma_differential} as
\beq
\label{eq:sigma_hadr}
\frac{\df \sigma}{\df Q^2 \df Y \df\Obs} =
 \sigma_0 \sum_{i,j} f_i(x_1^B) \otimes_{x_1^B} \frac{\df \eta_{ij}(x_1^B,x_2^B)}{\df Q^2 \df Y \df\Obs}   \otimes_{x_2^B} f_j(x_2^B).
\eeq
Here, the convolution integral is defined as
\beq
\label{eq:convdef}
f(x) \otimes_x g(x)  = \int_x^1 \frac{\df z}{z} \,f(z) g\left(\frac{x}{z}\right)
\,.\eeq
The corresponding partonic coefficient function differential in $\Obs$ is given by
\bea
\label{eq:partcoef_special}
\frac{\df \eta_{ij}(y_1,y_2)}{\df Q^2 \df Y \df\Obs} &=& \int_0^1 \df x \int_0^\infty \df \wa\df \wb \,
\delta\left(y_1-z_1\right)  \delta\left(y_2-z_2\right)
\nn\\&&\times\,
\delta\bigl[\Obs-\Obs(Q,Y,\wa,\wb,x)\bigr] \, \frac{\df\eta_{ij}}{ \df Q^2  \df \wa \df \wb \df  x}
\,,\eea
where $\Obs(Q,Y,\wa,\wb,x)$ picks out the value of the observable at a given phase space point.
Note that in the above equation the variables $z_i$ are still functions of $\wa$, $\wb$ and $x$ as specified in \eq{zdef}.

The partonic coefficient function in~\eqref{eq:sigma_hadr} contains ultraviolet (UV) and infrared (IR) divergences.
We regulate such divergences using conventional dimensional regularisation by extending the space time dimension by an infinitesimal amount to be $d=4-2\epsilon$.
UV divergences are removed by renormalization in the $\MSbar$ scheme.
IR singularities are removed by the standard mass factorization redefinition of the PDFs.
Specifically, the unsubtracted PDF $f_i(x)$ is given in terms of the finite PDF in the $\MSbar$ scheme $f^R_i(x)$ as
\begin{align} \label{eq:pdf_renorm}
 f_i(x) &= \sum_j \Gamma_{ij}(z) \otimes_z f_j^R(z)
\,,\end{align}
where the sum runs over all parton flavors $j$, $\Gamma_{ij}$ is the PDF counterterm that is known through three loops~\cite{Moch:2004pa, Vogt:2004mw}, and we suppress the associated factorization scale $\mu$.
This allows us to write the hadronic differential cross section of \eq{sigma_hadr} in terms of finite quantities,
\beq
\label{eq:sigma_hadr_finite}
\frac{\df \sigma}{\df Q^2 \df Y \df\Obs} =
\sigma_0 \sum_{i,j} f_i^R(x_1^B) \otimes_{x_1^B} \frac{\df \eta_{ij}^R(x_1^B,x_2^B)}{\df Q^2 \df Y \df\Obs}   \otimes_{x_2^B} f_j^R(x_2^B)\,,
\eeq
with
\beq
 \frac{\df \eta_{ij}^R(z_1,z_2)}{\df Q^2 \df Y \df\Obs}=\sum_{k,\ell} \Gamma_{ki}(z_1) \otimes_{z_1}  \frac{\df \eta_{k\ell}(z_1,z_2)}{\df Q^2 \df Y \df\Obs}\otimes_{z_2} \Gamma_{\ell j}(z_2)\,.
\eeq

\subsection{Differential phase space}
\label{sec:setup_phasespace}

To derive the phase space differential in the variables defined in \eq{vardef}, we start from the generic expression for the phasespace of the $h+X_n$ system,
\begin{align} 
\label{eq:measure}
 \df\Phi_{h+n} &=
  \frac{\df^dp_h}{(2\pi)^d} (2\pi)\delta_+(p_h^2-Q^2) \,
  \left[\prod\limits_{i=3}^{n+2}\frac{\df^dp_i}{(2\pi)^d} (2\pi)\delta_+(p_i^2)\right]
   (2\pi)^d \delta^d(p_1+p_2+p_h+k)
\,,\end{align}
where
\beq
 \delta_+(p^2-m^2)=\theta(-p^0-m)\delta(p^2-m^2)
\,,\eeq
and $k^\mu= \sum\limits_{i=3}^{n+2} p_i^\mu$ is the total momentum of $X_n$.
Next, we separate the integration over $p_h$ and $k$ by inserting the unity
\beq
1=\int \frac{\df^dk}{(2\pi)^d} (2\pi)^d\delta^d(k-p_3-\dots-p_{n+2})\int_0^\infty \frac{\df\mu^2}{2\pi}(2\pi)\delta_+(k^2-\mu^2)\,.
\eeq
This splits the $h{+}X_n$ phase space measure into an integral over the phase space $\Phi_2^\text{m}$ for two massive particles
and the phase space $\Phi_n^0$ for $n$ massless partons of total invariant mass $\mu^2$,
\begin{align} \label{eq:Phi_hn}
 \df\Phi_{h+n} = \int_0^{\infty} \frac{\df\mu^2}{2\pi} \df\Phi_2^\text{m}(\mu^2) \, \df\Phi_n^0(\mu^2)
\,.\end{align}
The two phase space measures are defined as
\begin{align} \label{eq:phi_2_and_0}
 \df\Phi_2^\mathrm{m}(\mu^2) &=
 \frac{\df^d p_h}{(2\pi)^d}\,(2\pi)\delta_+(p_h^2-Q^2) \frac{\df^d k}{(2\pi)^d} (2\pi)\delta_+(k^2-\mu^2)\,(2\pi)^d\delta^d(p_1+p_2+p_h+k)
\,,\nn\\
 \df\Phi^0_n(\mu^2) &=
 \biggl[\prod\limits_{i=3}^{n+2}\frac{\df^dp_i}{(2\pi)^d} (2\pi)\delta_+(p_i^2)\biggr]
 (2\pi)^d \delta^d\biggl(k-\sum\limits_{i=3}^{n+2} p_i\biggr)
\,.\end{align}
The on-shell constraint for $p_h$ is used together with the definition of the rapidity of \eq{hadrvardef} to define the born momentum fractions $x_{1,2}^B$.
Transforming from $k^\mu$ to the variables introduced in \eq{vardef}, we obtain the desired result for the differential phase space,
\begin{align} \label{eq:dPhi_wa_wb_x}
 \frac{\df\Phi_{h+n}}{\df \wa\df \wb \df x} &
 = \frac{1}{2 (4\pi)^{2-\epsilon}\Gamma(1-\epsilon)} \left(\wa\wb s\right)^{1-\epsilon} (1-x)^{-\epsilon}
   \theta[x(1-x)] \theta(\wa)\theta(\wb) \df\Phi_n^0\left(s \wa\wb x\right)
\,.\end{align}
In the special case of having zero or one final state parton, \eq{dPhi_wa_wb_x} becomes
\begin{align}
 \frac{\df\Phi_{h+0}}{\df \wa\df \wb \df x} &= \frac{(2\pi)}{s} \delta(x) \delta(\wa)\delta(\wb)
\,,\nn\\
 \frac{\df\Phi_{h+1}}{\df \wa\df \wb \df x} &= \frac{(\wa\wb s)^{-\eps}}{2 (4\pi)^{1-\epsilon}\Gamma(1-\epsilon)} \delta(x) \theta(\wa)\theta(\wb)
\,.\end{align}
The inclusive phase space volume is obtained by integrating over the differential phase space volume.
\beq
\Phi_{h+n}=\int dw_1 dw_2 dx \delta\left(1-w_1-w_2+w_1 w_2 x -Q^2/s\right)  \df\Phi_{h+n} .
\eeq
\section{Collinear expansion of color-singlet cross sections}
\label{sec:collinear_expansion_xs}

In this section, we introduce the general strategy of expanding cross sections around the collinear limit.
We begin by identifying the key kinematic regions in which we want to expand cross sections in \sec{expansions_intro}.
Next, we define the collinear expansion of hadronic cross sections in \sec{expansions_xsection}. 
Finally, we comment on the use of different coordinates in performing a collinear expansion in \sec{expandingvariables}.
We will provide explicit examples on how to implement this in practice for matrix elements in \sec{collinear_expansion}.

In this section, it will be very convenient to work with light-cone coordinates%
\footnote{Note that another popular conventions in the literature defines light-cone coordinates through the decomposition $p^\mu = p^- \frac{\bn^\mu}{\sqrt2} + p^+ \frac{n^\mu}{\sqrt2} + p_\perp^\mu$ with $p^\pm = (p^0 \pm p^z)/\sqrt2$.}.
We decompose a momentum $p^\mu$ as
\begin{align}\label{eq:lcdef}
 p^\mu &= p^+ \frac{\bn^\mu}{2} + p^- \frac{n^\mu}{2} + p_\perp^\mu \equiv (p^+, p^-, p_\perp)
\,,\end{align}
where the $p^\pm$ components are explicitly given by
\begin{align}\label{eq:lcdef2}
 p^- &= \bn \cdot p = p^0 + p^z \,,\quad p^+ = n \cdot p = p^0 - p^z
\,,\end{align}
and $p_\perp$ is the remaining transverse component.

\subsection{Expanding cross sections around the collinear limit}
\label{sec:expansions_xsection}
We now discuss the expansion of hadronic cross sections of \eq{sigma_hadr} around the particular limit where all final state radiation becomes collinear to one of the incoming proton momenta.

Let us define our collinear expansion: we want to expand around the limit where all real momenta are treated as $n$-collinear, and thus the total momentum $k^\mu$ of the hadronic final-state is $n$-collinear as well, \emph{i.e.}\ it it scales as
\begin{align} \label{eq:k_ncollinear}
 k^\mu \sim k^- \frac{n^\mu}{2} + \lambda^2 k^+ \frac{\bn^\mu}{2} + \lambda k_\perp^\mu
\,.\end{align}
We then want to expand the hadronic differential cross section in \eq{sigma_hadr} to obtain a power series in $\lambda$,
\beq
\frac{\df \sigma}{\df Q^2 \df Y \df\Obs} = \lambda^{-2} \frac{\df \sigma^{(0)}}{\df Q^2 \df Y \df\Obs}  + \lambda^{-1} \frac{\df \sigma^{(1)}}{\df Q^2 \df Y \df\Obs} +\dots.
\eeq
Here, the leading-power cross section $\sigma^{(0)}$ scales as $\lambda^{-2}$,%
\footnote{This leading-power collinear limit precisely corresponds to the generalized threshold limit of \refcite{Lustermans:2019cau}.}
the next-to-leading power (NLP) cross section%
\footnote{Note that for a large class of observables, as for example $q_T$ and beam thrust, the odd powers in this series vanish. It is therefore common practice to indicate as NLP the first non vanishing contribution beyond leading power, which in those cases would be $\sigma^{(2)}$.} 
$\sigma^{(1)}$ as $\lambda^{-1}$, and so forth.
Depending on the observable $\Obs$, this series may start at higher orders in $\lambda$,
but for the infrared-sensitive observables discussed in this chapter we always encounter a leading $\cO(\lambda^{-2})$ term.

It is desirable that Born quantities like $Q^2$ and $Y$ are unaffected by the expansion we want to carry out, as they set the hard scales of the process.
The importance of this for expansions at subleading power in $\lambda$ was already stressed in \refscite{Ebert:2018gsn,Ebert:2018lzn}.
As a consequence, the Bjorken momentum fractions given in \eq{xidef} need to be expanded.
Expressing them in terms of hard quantities and the momentum $k$ we find
\begin{align} \label{eq:x12_expanded}
 \frac{x_1}{x_1^B} &=  \sqrt{1 + k_T^2 / Q^2} - \frac{k^-}{Q} e^{-Y}
      = 1 - \frac{k^- e^{-Y}}{Q} + \cO(\lambda^2)
\,,\nn\\
 \frac{x_2}{x_2^B} &= \sqrt{1 + k_T^2 / Q^2} - \frac{k^+}{Q} e^{+Y}
      = 1 + \cO(\lambda^2)
\,.\end{align}
Since the momentum fractions enter as arguments of the PDFs, a pure hadronic expansion to higher orders in $\lambda$ will automatically involve derivatives of PDFs, as firstly noted for $\Tau_0$ in~\refcite{Moult:2016fqy}.
Furthermore, the variables $\wa$ and $\wb$ we introduced in \eq{vardef} must also be expanded,
\beq
\label{eq:omexp}
\wa=\frac{-k^-}{x_1 \sqrt{S}}=\frac{-k^-}{x_1^B \sqrt{S}-k^-}+\mathcal{O}(\lambda^2)\hspace{1cm}\wb=\frac{-k^+}{x_2 \sqrt{S}}=\frac{-k^+}{x_2^B \sqrt{S}}+\mathcal{O}(\lambda^4)
\,,\eeq
where $x_{1,2}^B$ are the momentum fractions at Born level, see \eq{xiBdef}.
As a consequence we find that the $n$-collinear limit of \eq{partcoef_special} becomes
\bea
\label{eq:partoniccoefexp}
&&\nlim\frac{\df \eta_{ij}(y_1,y_2)}{\df Q^2 \df Y \df \Obs} =\delta\left(1-y_2\right)  \int_0^1 \df x \int_0^\infty \df \wa\df \wb \,   \delta\left[y_1-(1-\wa)\right]  \\
&&\hspace{0.5cm}\times \nlim \left\{\delta\left[\Obs-\Obs(Q,Y,\wa,\wb,x)\right] \frac{\df\eta_{ij}}{ \df Q^2  \df \wa \df \wb \df  x}\right\}\nn
\,,\eea
where $w_{1,2}$ are evaluated according to \eq{omexp}.
The definition of our observable $\Obs$ itself may not be invariant under rescaling according to our power counting.
In order to achieve a pure expansion of the hadronic cross section we may either expand the observable constraint or solve the constraint using one of the remaining integration variables and expand subsequently.
We address how the general partonic coefficient function $ \frac{\df\eta_{ij}}{ \df Q^2  \df \wa \df \wb \df  x}$ can be expanded in \sec{collinear_expansion}. 

Constructing a collinear expansion can be done with different objectives in mind. 
One objective can be to obtain a pure series expansion of the hadronic cross section as discussed above.
Another objective can be to simplify the computation of the partonic coefficient function which does not require a pure expansion of the hadronic cross section.  
In the latter scenario one would only expand the partonic coefficient function $\eta_{ij}$ on the right-hand side of \eq{partcoef_special}, but not expand the $w_{1,2}$ and the momentum fractions $x_{1,2}$ as presented above.
This approach can also serve as a suitable proxy to a collinear expansion, where parts of the cross section are kept exact.

\subsection{Expansions using different coordinates}
\label{sec:expandingvariables}

So far, we defined our power counting such that the invariant mass $Q^2$ and rapidity $Y$ of the produced hard probe $h$ scale homogeneously as $\cO(\lambda^0)$ and the lightcone components of the total momentum $k^\mu$ of the hadronic final state have a homogeneous power counting in $\lambda$. This is reasonable, since one can only measure directly the final-state particles in the hadronic collision, which are then used to define the power counting. In particular, $Q^2$ and $Y$ are the only hard scales in the considered hadronic cross section $\df\sigma/(\df Q^2 \df Y \df \Obs)$.

This setup immediately implies that the momenta $p_1$ and $p_2$ of the incoming partons do not have a homogeneous power counting, as it is evident from their explicit expressions in terms of $Q^2$, $Y$ and $k^\mu$,
\begin{align} \label{eq:p1p2_2}
 p_1^-(Q^2,Y,k^+,k^-,x)  &= -k^- + e^{+Y}\sqrt{Q^2 + k^+ k^- (1-x)}
\,,\nn\\
 p_2^+(Q^2,Y,k^+,k^-,x) &= -k^+ + e^{-Y}\sqrt{Q^2 + k^+ k^- (1-x)}
\,,\end{align}
see \eq{xidef}. Thus, $p_1$ and $p_2$ give rise to an infinite tower of power corrections in $\lambda$, which in turn requires an expansion of $\wa$ and $\wb$ used to define the general partonic coefficient function, as shown in \eq{omexp}.

Since one has access to all incoming and outgoing momenta in the calculation of the partonic coefficient function, the collinear expansion can also be defined by assigning a homogeneous power counting to $p_1$, $p_2$ and $k$. Since this assignment is only meaningful for the partonic process, we refer to it as \emph{partonic collinear expansion}. In this approach, the rescaling appropriate for the collinear limit is given by
\begin{align} \label{eq:partonic_expansion}
 p_1^- \to p_1^- \,,\quad p_2^+ \to p_2^+ \,,\quad \wa = -\frac{k^-}{p_1^-} \to \wa \,,\quad \wb = -\frac{k^+}{p_2^+} \to \lambda^2 \wb \,,\quad x \to x
\,.\end{align}
The key advantage of this assignment is that $\wa$ and $\wb$, which are the natural variables to express the partonic partonic coefficient function, now have homogeneous power counting and do not need to be re-expanded themselves. 
Thus, the collinear expansion has been reduced to an expansion in $\wb$.
The drawback of the partonic collinear expansion is that the rapidity $Y$ of hard probe $h$ no longer uniformly scales as $\cO(\lambda^0)$, which is evident from the expression
\begin{align}
 Y(p_1^-,p_2^+,k^+,k^-,x) &
 = \frac{1}{2}\log\left(\frac{p_1^- + k^-}{p_2^+ + k^+}\right)
 = \frac{1}{2}\log\left(\frac{p_1^- + k^-}{p_2^+}\right) + \cO(\lambda^2)
\,,\end{align}
which now is a quantity derived from $p_1^-$ and $p_2^+$, rather than fixing these as in \eq{p1p2_2}.

Comparing the rescalings in \eqs{omexp}{partonic_expansion}, we see that both approaches agree at leading power, but differ at subleading power. Since there is a well-defined relation between the two approaches, one can easily obtain one expansion from the other, but care has to be taken to consistently apply the power expansion.

In practice, each choice of defining the expansion has its advantages and disadvantages. We can discuss these by classifying the expansions according to the choice of independent variables used to express the partonic coefficient function, which by Lorentz invariance only requires four independent invariables.
It is useful to summarize the above observations for the following possibilities:
\begin{itemize}
 \item $(Q^2,Y,k^+,k^-,x)$: This parameterization has the advantage that is entirely expressed in terms of information about the final state momenta, including the Born measurements $Q$ and $Y$ of the hard probe $h$.
 As properties of the final state, $Q$ and $Y$ need not be expanded, and the collinear expansion is a strict expansion in $k^\mu$.
 Since partonic matrix elements are typically more concise when expressed in terms of the incoming momenta and the final-state radiation, the main drawback of this expansion is that it leads to lengthier expressions for the expanded matrix element.
 Furthermore, measuring the rapidity $Y$ fixes a reference frame for all momenta, such that boost invariance is not manifest anymore.
 \item $(Q^2, \wa, \wb, x)$: Since $\wa$ and $\wb$ are defined as ratios of Lorentz scalars, boost invariance is manifest in this parameterization. Its disadvantage is that $\wa$ and $\wb$ do not have manifest power counting in terms of the observables $Q^2$ and $Y$, and instead must be expanded in $k^\mu$ according to their definitions in \eq{vardef}. Alternatively, one can assign homogeneous power counting to $\wa$ and $\wb$ using \eq{partonic_expansion}, which then requires to expand the rapidity $Y$ in $\lambda$.
 \item $(p_1^-,p_2^+,k^+,k^-,x)$: Here, we trade $Q^2$ and $Y$ for the lightcone momenta $(p_1^-,p_2^+)$ of the incoming partons. This parameterization has the advantage of expressing everything in terms of the momenta of massless particles, i.e.\ the incoming momenta and the hadronic radiation.
 A disadvantage of this parameterization is that $p_1^-$ and $p_2^+$ do not have manifest power counting in terms of hadronic variables $Q^2$ and $Y$, and thus must be expanded in $\lambda$.
\end{itemize}
These parameterizations are of course equivalent, and in practice the preferred parameterization depends on the intended application. While the general illustration of the power expansion is made most manifest using $(Q^2,Y,k^+,k^-,x)$, expanding the partonic cross sections is simplified using ($Q^2, \wa, \wb, x)$.

Finally, we give the explicit relation between the different parameterizations.
We can change variables from $(\wa,\wb)$ to $(k^+,k^-)$ using
\begin{align} \label{eq:sigma_jacob_2}
 \frac{\df\eta_{ij}}{\df Q^2 \df Y \df k^+ \df k^- \df x} &
 = \frac{z(\wa,\wb)}{Q^2} \frac{\df\eta_{ij}}{ \df Q^2 \df \wa \df \wb \df x}
   \bigg|_{\substack{\wa = \wa(Q,Y,k) \\ \wb = \wb(Q,Y,k)}}
\,,\end{align}
where the required variable transformations are given by
\begin{align}
 \wa(Q,Y,k) &= \frac{-k^-}{p_1^-(Q,Y,k)}
 \,,\quad
 p_1^-(Q,Y,k) = -k^- + e^{+Y}\sqrt{Q^2 + k^+ k^- (1-x)}
\,,\nn\\
 \wb(Q,Y,k) &= \frac{-k^+}{p_2^+(Q,Y,k)}
 \,,\quad
 p_2^+(Q,Y,k) = -k^+ + e^{-Y}\sqrt{Q^2 + k^+ k^- (1-x)}
\,,\nn\\
 z(\wa,\wb) &= 1 - \wa - \wb + x \wa \wb
\,,\end{align}
and for brevity we keep implicit that $k^\mu$ is parameterized in terms of $(k^+,k^-,x)$.
Note that since $k^+$ and $k^-$ are defined in the hadronic center-of-mass frame, manifest boost-invariance is lost, and thus \eq{sigma_jacob_2} becomes explicitly $Y$-dependent.
\eq{sigma_jacob_2} makes it clear that defining the power counting in terms of $Q^2$ and $k$ requires a expansion of $\wa$ and $\wb$ on the right hand side.

We can further change variables from $(Q^2,Y)$ to $(p_1^-,p_2^+)$,
\begin{align} \label{eq:sigma_jacob_3}
 \frac{\df\eta_{ij}}{\df p_1^- \df p_2^+ \df k^+ \df k^- \df x} &
 = \frac{\df\eta_{ij}}{\df Q^2 \df Y \df k^+ \df k^- \df x}
   \bigg|_{\substack{Q^2 = Q^2(p_1,p_2,k) \\ Y = Y(p_1,p_2,k)}}
\,,\end{align}
where the required variable transformations are given by
\begin{align}
 Q^2 = (p_1^- + k^-) (p_2^+ + k^+) - (1-x) k^+ k^-
\,,\quad
 Y = \frac12 \ln\frac{p_1^- + k^-}{p_2^+ + k^+}
\,.\end{align}
Here, fixing the power counting of $p_1^-$, $p_2^+$ and $k$ requires to expand $Q^2$ and $Y$ accordingly.

\section{Collinear expansion of matrix elements}
\label{sec:collinear_expansion}

In this section we show how the technique of collinear expansions developed in the previous section is applied in practice.
To setup our conventions for this section, we first discuss the phase space volume for producing the hard probe $h$ with additional emissions in \sec{expansions_ps}, before illustrating the collinear expansion of matrix elements explicitly for both real radiation in \sec{expansions_real} and for loop integrals in \sec{expansions_loop}.

Throughout this section, we will consider the scenario where $k^\mu$ is collinear to the incoming parton with momentum $p_1^\mu = p_1^- n^\mu/2$. According to \eq{modes}, this implies that we assign the following scaling to $k$:
\begin{align}
 k^\mu = (k^+,k^-,k_\perp) \sim ( \lambda^2, 1, \lambda)
\,.\end{align}
In order to obtain a strict power series expansion of the hadronic cross section it is necessary to expand the partonic momentum components $p_1^-$ and $p_2^+$ around the collinear limit.
For the purpose of this section we instead perform a \emph{partonic} collinear expansion (see \sec{expandingvariables}), treating $p_{1,2}$ as external variables and thus as $\cO(\lambda^0)$ quantities.
All final results are functions of $k^-/p_1^-$ and $k^+/p_2^+$ and one can straightforwardly recover a pure expansion in terms of hadronic observables following \sec{expandingvariables}.

\subsection{Collinear phase space}
\label{sec:expansions_ps}

The phase space volume for producing the hard probe $h$ with two emissions, as defined in \eq{Phi_hn}, is given by
\begin{align} \label{eq:PS2}
 \Phi_{h+2}&
 = \int\! \frac{ \df\Phi_{h+2} }{\df \wa\df \wb \df x} 
 = \includegraphics[valign=c,width=4cm]{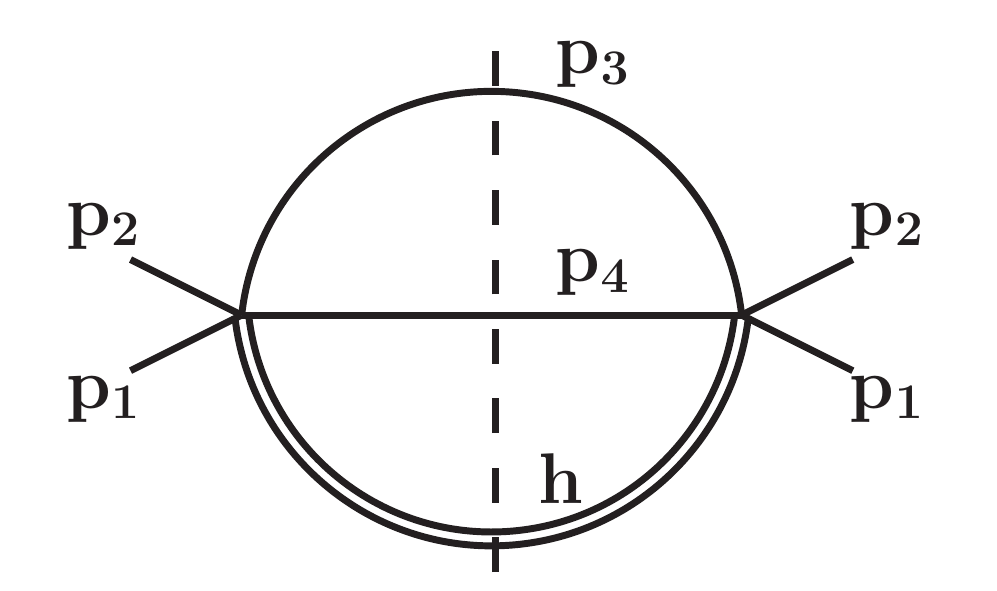} \nn\\
& = \frac{1}{s^2}\frac{ (k^+k^-)^{1-2 \eps } (1-x)^{-\eps } x^{-\eps }}{128 \pi ^3 (1-2 \eps ) \Gamma (1-2 \eps )}
\,.\end{align}
It follows immediately that the phase space volume in \eq{PS2} scales as $\Phi_{h+2} \sim \lambda^{2-4\eps}$.

As usual, we take all momenta to be incoming, and denote the total momentum of all outgoing partons by $k = p_3 + p_4$.
In the above diagram and those below, the dashed line indicates the on-shell constraints of the final state particles, with the solid lines representing massless partons and the double line representing the heavy color-singlet state $h$.

The scaling of  $\Phi_{h+2} \sim \lambda^{2-4\eps}$ can be easily deduced without calculating the actual phase space integrals.
Since $k$ is treated collinear to $p_1$, both final-state momenta $p_3$ and $p_4$ must be collinear to $p_1$ as well.
The associated integration measures and $\delta$ functions entering \eqs{Phi_hn}{phi_2_and_0} transform as
\begin{align}
 \int\df^d p_i \, \delta_+(p_i^2) ~\rightarrow~ \lambda^{2-2\eps} \int\df^d p_i \,\delta_+(p_i^2)
\,,\qquad
 \delta(k^2) ~\rightarrow~ \lambda^{-2} \delta(k^2)
\,.\end{align}
As a consequence, the double-real phase space measure scales as
\begin{align} \label{eq:phi20_coll}
 \int\!\frac{ \df\Phi_{h+2} }{\df \wa\df \wb \df x} ~\rightarrow~ \lambda^{2 - 4\eps} \int\!\frac{ \df\Phi_{h+2} }{\df \wa\df \wb \df x} 
\,,\end{align}
which is precisely the scaling observed in \eq{PS2}.
Similarly, it follows that the more general case of the $h$ + $n$ real emission phase space has the scaling
\begin{align} \label{eq:phin0_coll}
 \int\!\frac{ \df\Phi_{h+n} }{\df \wa\df \wb \df x} ~\rightarrow~ \lambda^{n(2-2\eps) - 2}  \int\!\frac{ \df\Phi_{h+n} }{\df \wa\df \wb \df x} 
\,.\end{align}

\subsection{Collinear limit of real radiation}
\label{sec:expansions_real}

We consider the following example of a more complicated, purely real Feynman integral,
\begin{align} \label{eq:RR}
 I_{\rm RR}
 = \includegraphics[valign=c,width=4cm]{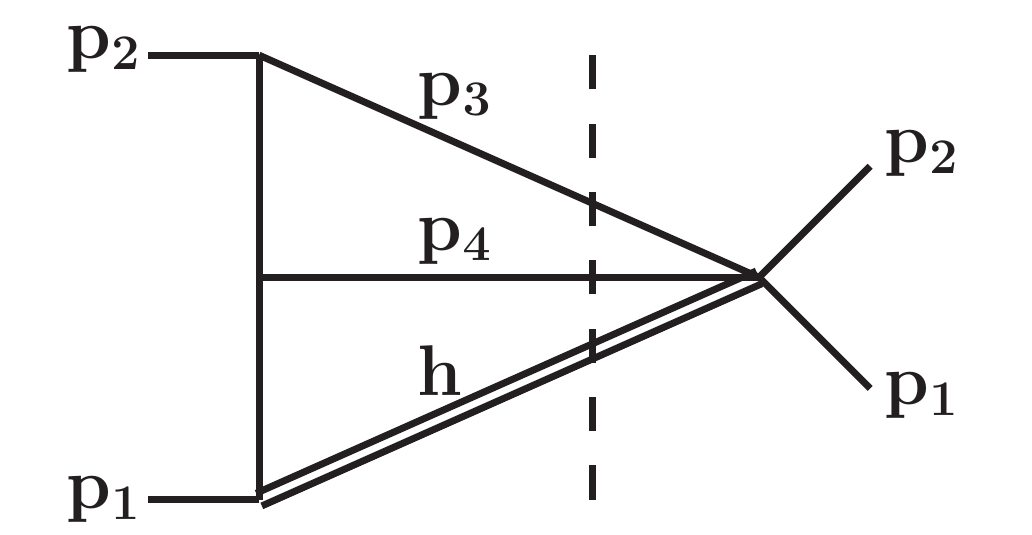}
 = \int\frac{ \df\Phi_{h+2} }{\df \wa\df \wb \df x}  \frac{1}{(p_2+p_3)^2(p_2+p_3+p_4)^2}
\,.\end{align}
Let us first consider the case where both $p_3$ and $p_4$ are collinear to $p_2$.
In this scenario, since  both propagators in \eq{RR} only involve collinear momenta, and thus scale homogeneously as $\lambda^2$ under the $\bn$-collinear rescaling of \eq{modes} and no expansion of \eq{RR} in $\lambda$ is needed.

In contrast, if we consider $p_3$ and $p_4$ to be collinear to $p_1$, then the second propagator is not homogeneous in $\lambda$ anymore, as it contains both $n$-collinear and $\bn$-collinear momenta.
To expand the propagator in this limit, we apply the $n$-collinear rescaling of \eq{modes} to $p_3$ and $p_4$,
\begin{align} \label{eq:p34_nbcoll}
 p_{3,4}^\mu ~\to~  p_{3,4}^- \frac{n^\mu}{2} + \lambda^2\,p_{3,4}^+ \frac{\bn^\mu}{2} + \lambda\,\, p_{3,4\,\perp}^\mu
\,.\end{align}
With these rescalings, it is now straightforward to expand the second propagator in \eq{RR} in $\lambda$,
\begin{align} \label{eq:RR_expansion}
 \frac{1}{(p_2+p_3+p_4)^2} &
 = \frac{1}{p_2^+ (p_3^- + p_4^-) + 2 p_3 \cdot p_4}
 ~\stackrel{p_1-\rm coll}{\longrightarrow}~
 \frac{1}{p_2^+ (p_3^- + p_4^-) + \lambda^2 \, 2 p_3 \cdot p_4}
\nn\\&
 = \sum_{n=0}^\infty (\lambda^2)^n \frac{(-2 p_3 \cdot p_4)^n}{\bigl[p_2^+ (p_3^- + p_4^-) \bigr]^{n+1}}
\,.\end{align}
For $n=0$, this propagator is linear in the real momenta $p_3$ and $p_4$, and thus corresponds to an eikonal propagator.
Higher orders in $\lambda$ only involve pure powers of the eikonal propagator, thus yielding a relatively simple structure of the expansion.
Together with \eq{phi20_coll}, the integral in \eq{RR} can thus be expanded as
\begin{align} \label{eq:RR2}
 I_{\rm RR} &
 ~\stackrel{p_1-\rm coll}{\longrightarrow}~
 \sum_{n=0}^\infty (\lambda^2)^{n+1-2\eps}
   \int\frac{ \df\Phi_{h+2} }{\df \wa\df \wb \df x} \frac{(-2 p_3 \cdot p_4)^i}{(p_2+p_3)^2 \bigl[p_2^+ (p_3^- + p_4^-) \bigr]^{n+1}}
\,.\end{align}
This expansion can be represented diagrammatically as
\begin{align}
 \includegraphics[valign=c,width=3cm]{Diagrams/RR.pdf} ~\rightarrow~
 \lambda^{2-4\eps} \Biggl[ \includegraphics[valign=c,width=3cm]{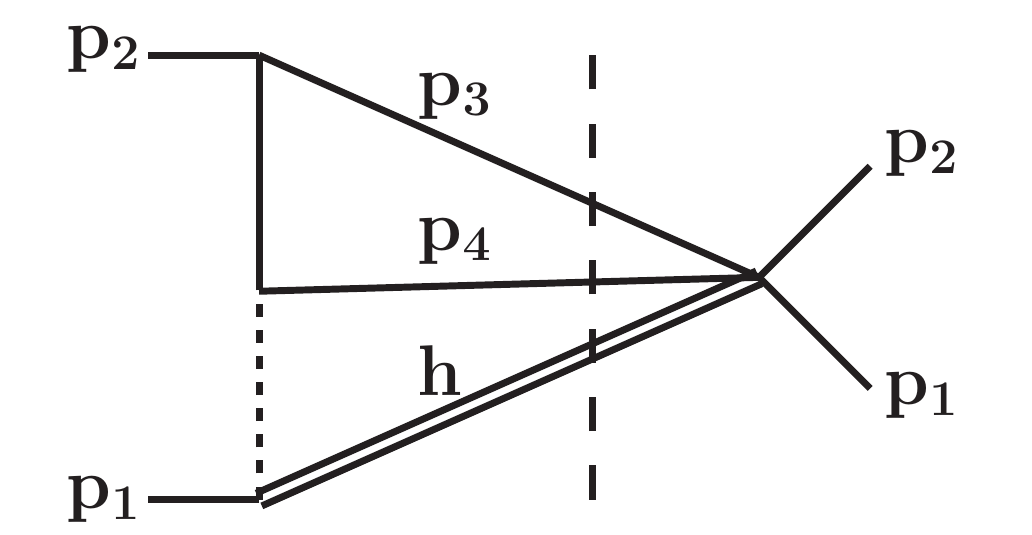}
   - \lambda^2 \includegraphics[valign=c,width=3cm]{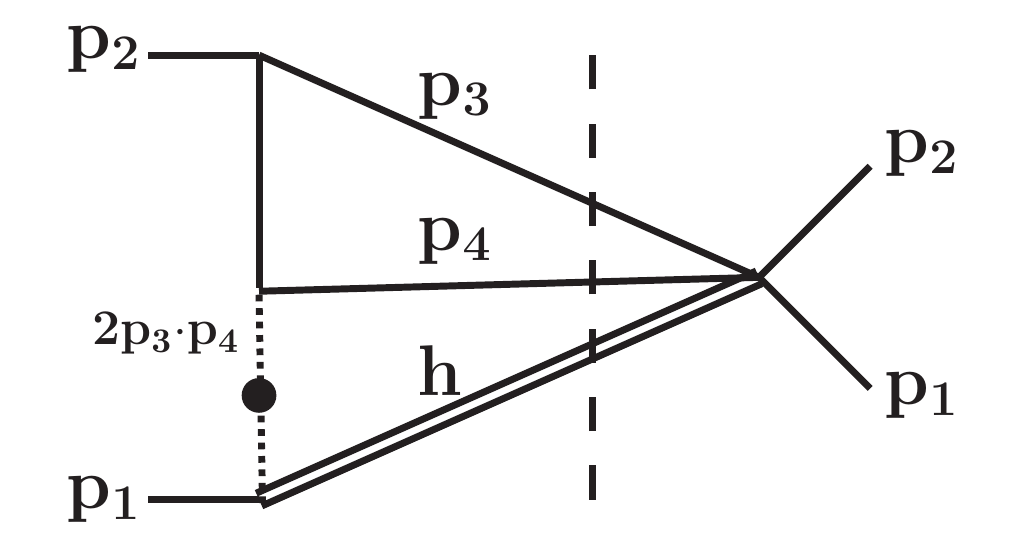}
   + \cO(\lambda^3)
 \Biggr]
\,.\end{align}
Here, the dotted line indicates the expanded (eikonal) propagator and the dot on the line represents higher powers of this propagator. The label denotes the additional kinematic factor arising from \eq{RR_expansion}.

The expansion in \eq{RR2} results in several advantages.
First, we observe that each term in the expansion is homogeneous under the $n$-collinear rescaling transformation in \eq{p34_nbcoll}.
As a consequence, we may directly determine the functional dependence of each term in the expansion on $k^+$ similarly to the case of the phase space volume.
In other words, the resulting functions will be simpler since they only depend on $k^+$ via a multiplicative pre-factor.
Second, the structure of expanded Feynman integrals is amenable to IBP reduction techniques via the framework of reverse unitarity~\cite{Anastasiou2003,Anastasiou:2002qz,Anastasiou:2003yy,Anastasiou2005,Anastasiou2004a}.
The benefit is that the appearing integrals can be related to so called master integrals. 
In our example we find the IBP relations
\begin{alignat}{3} \label{eq:coefeqs}
 \includegraphics[valign=c, width=4cm]{Diagrams/RRColl0.pdf} &= -\frac{1-2\eps}{\eps (p_2^+ k^-)^2 } &&\times  \includegraphics[valign=c, width=4cm]{Diagrams/RRPS.pdf}
\,,\\ \label{eq:coefeqs2}
 \includegraphics[valign=c, width=4cm]{Diagrams/RRColl1.pdf} &= -\frac{k^+ x}{p_2^+}\frac{1-2\eps}{\eps(p_2^+ k^-)^2 }  &&\times  \includegraphics[valign=c, width=4cm]{Diagrams/RRPS.pdf}
\,.\end{alignat}
Clearly, it is very advantageous that any higher order term in our expansion is related to the same master integrals as the first, which in our example is just the phase space volume.
The unexpanded integral of our example in \eq{RR} is itself related to the phase space volume by an IBP identity,
\begin{align}
 \includegraphics[valign=c, width=4cm]{Diagrams/RR.pdf} &= -\frac{(1-2 \eps) }{\eps (p_2^+ k^-)^2 }\left(1+\frac{k^+ x}{p_2^+}\right)^{-1}  \times  \includegraphics[valign=c, width=4cm]{Diagrams/RRPS.pdf}
\,.\end{align}
From this we can easily see that the coefficients obtained in \eq{coefeqs} and \eq{coefeqs2} correspond exactly to the coefficients of the expansion of the exact result.

In summary, we outlined a procedure that allows us to perform an expansion of real radiation integrals around the limit of all final state partons becoming collinear to an initial state momentum.
This expansion is carried out by simply performing the appropriate collinear rescaling transformation of \eq{modes} on all final state parton momenta and subsequently expanding the integrand of our real radiation integral in the artificial parameter $\lambda$, prior to actually evaluating the integral.
Each term in the expansion in $\lambda$ then corresponds to exactly one term in the expansion of the integral in $k^+$. 
The computation of the terms in the expansion is greatly facilitated by applying techniques like IBP identities via the reverse unitarity framework.
In particular, any term appearing at higher orders in the expansion will be expressible in terms of master integrals that appear already in the first few terms.

\subsection{Expansion of loop integrals}
\label{sec:expansions_loop}

In contrast to the phase space integral over real momenta considered in \sec{expansions_real}, where the requirement of $k$ being collinear to $p_1$ restricted $p_{3,4}$ to be collinear to $p_1$ as well, such a restriction does not appear for loop momenta.
Despite this, it is still useful to expand loop integrals in a similar fashion around the hard, collinear and soft regions.
As discussed in \sec{expansions_intro}, for factorization proofs this is crucial to separate these different regions into distinct matrix element,
while in the method-of-regions approach of \refcite{Beneke:1997zp} it used to simplify loop integrals by expanding the integrand in all relevant limits and combining their individual results.

Here, we will show for a simple example how one can easily approximate and expand loop integrals in the discussed regimes, and that the sum of all regions indeed reproduces the full result. This will be illustrated using the following real-virtual diagram,
\begin{align} \label{eq:RV}
 I_{\rm RV} &
 = \includegraphics[valign=c,width=4cm]{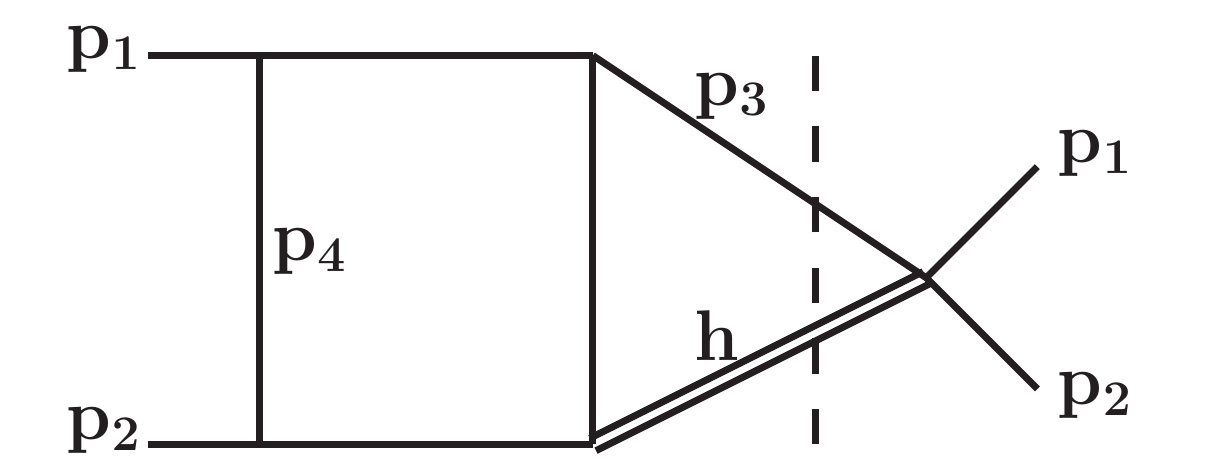}
 = \int\frac{ \df\Phi_{h+1} }{\df \wa\df \wb \df x} \frac{\df^d p_4}{(2\pi)^d}
   \frac{1}{p_4^2 \, (p_1 + p_4)^2 \, (p_1 + p_3 + p_4)^2 \, (p_2 - p_4)^2}
\nn\\&
 = \frac{\img c_\Gamma}{128 \pi^4 \eps^2}\delta(x) \biggl[
   \frac{(k^+ k^-)^{-\eps}}{s^4 (-s)^{\eps}} \sum_{n,m=0}^\infty\frac{(\eps+1)_n \, (\eps+2)_{m+n}}{(n+m+1) \, m!\,n!\, (\eps+2)_n}  \left(\frac{k^+}{p_2^+}\right)^{m} \left(\frac{k^-}{p_1^-}\right)^{n}
 \nn\\&\hspace{2.7cm}
 - \frac{(k^+ k^-)^{-\eps}}{s^3 (p_1^- k^+)^{1+\eps}} (4 \pi)^{-2\eps}\, _2F_1\left(1,-\eps; 1-\eps; \frac{k^-}{p_1^-}\right)
   \biggr]
\,.\end{align}
Here, the total final-state hadronic momentum is $k = p_3$, and thus the $p_3$ integral is actually fixed.
In \eq{RV}, $(a)_n = \Gamma(a+n)/\Gamma(a)$ is the (rising) Pochhammer symbol, and we abbreviate common loop factors by
\begin{align}
 c_\Gamma  = \frac{\Gamma(1+\eps) \Gamma(1-\eps)}{\Gamma(1-2\eps)}
\,.\end{align}
As before, we consider the limit where $k$ is collinear to $p_1$, such that $k^+ \sim \cO(\lambda^2)$ and $k^- \sim \cO(\lambda^0)$.
This immediately implies that \eq{RV} scales as
\begin{align} \label{eq:RV_expanded}
 I_{\rm RV} &\stackrel{\rm coll}{\longrightarrow}
\delta(x)  \frac{\img c_\Gamma }{128 \pi^4 \eps^2} \biggl[
   \lambda^{-2\eps} \frac{(k^+ k^-)^{-\eps}}{s^4 (-s)^{\eps}} \sum_{n,m=0}^\infty \lambda^{2m} \frac{(\eps +1)_n \, (\eps +2)_{m+n}}{(n+m+1) \, m!\,n!\, (\eps +2)_n}  \left(\frac{k^+}{p_2^+}\right)^{m} \left(\frac{k^-}{p_1^-}\right)^{n}
 \nn\\&\hspace{3cm}
 - \lambda^{-2-4\eps} \frac{(k^+ k^-)^{-\eps}}{s^3 (p_1^- k^+)^{1+\eps}} (4 \pi)^{-2\eps}\, _2F_1\left(1,-\eps; 1-\eps;  \frac{k^-}{p_1^-}\right)
   \biggr]
\,.\end{align}
The second line has homogeneous scaling in $\lambda^{-2-4\eps}$, and is the dominant contribution in the limit $\lambda\to0$.
We will see below that this result is entirely from the region where the loop momentum is collinear to $p_1$.
In other words, the leading-power limit of $I_{\rm RV}$ arises from the region where both loop \emph{and} real momenta are collinear to $p_1$.
The first line in \eq{RV_expanded} does not scale homogeneously in $\lambda$, but is suppressed at least as $\cO(\lambda^2)$ compared to the leading-power limit. We will see that this line entirely arises from the region where the loop momentum is hard.
In particular, the two contributions have different fractional scalings in $\lambda^{-2\eps}$ and $\lambda^{-4\eps}$, respectively.
These scalings arise entirely from the loop integral measures, and thus can be easily distinguished between the different contributions.

\subsubsection{Collinear limit}

We first consider the loop momentum $p_4$ to be collinear to the incoming parton with momentum $p_1$.
According to \eq{modes}, we hence transform
\begin{align}
 p_4^\mu ~\to~  p_4^- \frac{n^\mu}{2} + \lambda^2\, p_4^+ \frac{\bn^\mu}{2} + \lambda\,\, p_{4\perp}^\mu
\,.\end{align}
The first three propagators in \eq{RV} scale homogeneously as $\cO(\lambda^{-2})$ under this rescaling,
while the last propagator in \eq{RV} is not homogeneous in $\lambda$ and must be expanded,
\begin{align} \label{eq:coll_prop_exp}
 \frac{1}{(p_2 - p_4)^2}  &
 = \frac{1}{-2 p_2 \cdot p_4 + p_4^2}
 ~\stackrel{p_1-\rm coll}{\longrightarrow}~
 \frac{1}{-2 p_2 \cdot p_4 + \lambda^2 p_4^2}
 = \sum_{n=0}^\infty \lambda^{2n} \frac{(-p_4^2)^n}{(-2 p_2 \cdot p_4)^{n+1}}
\,.\end{align}
Together with \eq{phin0_coll}, this allows us to expand the integrand in \eq{RV} as
\begin{align} \label{eq:RV_ncoll}
 I_{\rm RV}
 \stackrel{p_1-\rm coll}{\longrightarrow}&~
 \lambda^{-2-4\eps} \sum_{n=0}^\infty \lambda^{2n} \!\!\int\!\!\df\Phi_{h+1} \frac{\df^d p_4}{(2\pi)^d} \frac{(-p_4^2)^n}{p_4^2 \, (p_1 + p_4)^2 \, (p_1 + p_3 + p_4)^2 \, (-2 p_2 \cdot p_4)^{n+1}}
 \\\nn=&~
 \lambda^{-2-4\eps} \biggl[~
   \int\frac{ \df\tilde\Phi_{h+1} }{\df \wa\df \wb \df x}  \frac{\df^d p_4}{(2\pi)^d} \frac{1}{p_4^2 \, (p_1 + p_4)^2 \, (p_1 + p_3 + p_4)^2 \, (-2 p_2 \cdot p_4)}
   \\\nn&\hspace{1.2cm}
   - \lambda^2 \int\frac{ \df\Phi_{h+1} }{\df \wa\df \wb \df x}  \frac{\df^d p_4}{(2\pi)^d} \frac{1}{(p_1 + p_4)^2 \, (p_1 + p_3 + p_4)^2 \, (-2 p_2 \cdot p_4)^2} + \cO(\lambda^4) \biggr]
\,.\end{align}
The overall scaling in $\lambda^{-4\eps}$ arises from $\df^d p_4 \sim \lambda^{4-2\eps}$ and $\df\Phi_{h+1} \sim \lambda^{-2\eps}$, and thus is independent of the structure of the diagram itself.
Each order of the expanded integrand is now homogeneous in $\lambda$.
The expansion in \eq{RV_ncoll} can be illustrated graphically as
\begin{align} \label{eq:RV_ncoll_diag}
 I_{\rm RV} &
 \stackrel{\rm coll}{\longrightarrow}
 \lambda^{-2-4\eps} \biggl[ \includegraphics[valign=c,width=4cm]{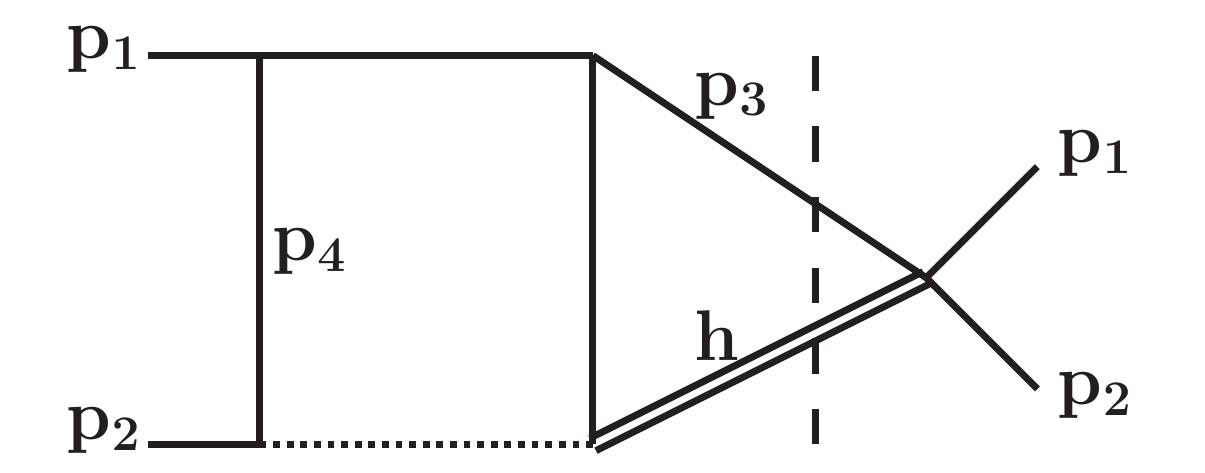}
 - \lambda^2 \includegraphics[valign=c,width=4cm]{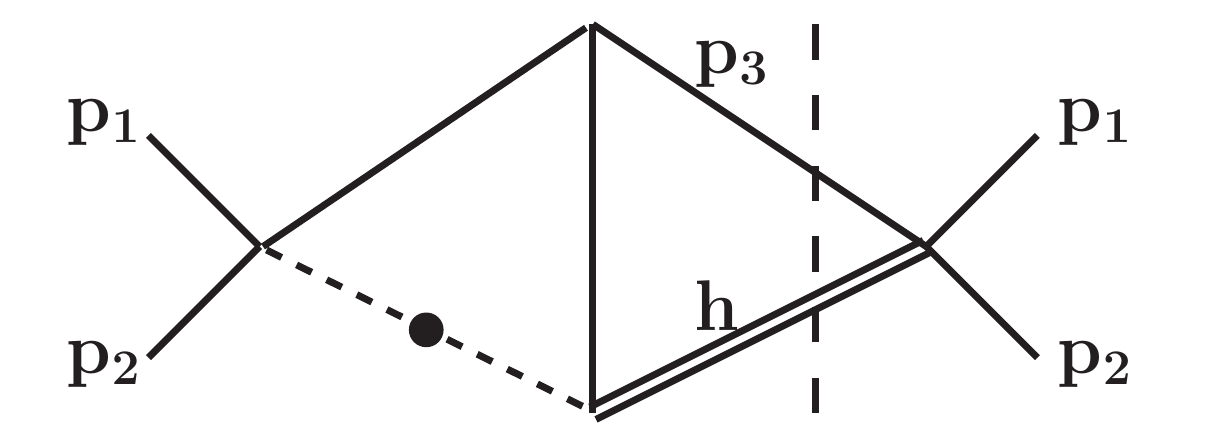} + \cO(\lambda^4) \biggr]
\,.\end{align}
Here, the dotted lines are linear (eikonal) propagators, and the dot on the line denotes that the propagator is raised to one power.
Note that in the second diagram, the explicit $1/p_4^2$ propagator is canceled, indicated by the contracted vertex.
Being able to represent collinear expanded diagrams again in a diagrammatic fashion is extremely useful.
In particular, the structure observed in the collinear loop expansion makes it possible to use IBP techniques for the computation of loop and phase space integrals.

The leading-power integral in \eq{RV_ncoll} can be evaluated as
\begin{align} \label{eq:RV_ncoll_LP}
 I_{\rm RV} & \stackrel{\rm coll}{=}
 - \delta(x) \lambda^{-2-4\eps} \frac{\img c_\Gamma}{128 \pi^4 \eps^2} \frac{(k^+ k^-)^{-\eps}}{s^3 (p_1^- k^+)^{1+\eps}}  (4 \pi)^{-2\eps}  {}_2F_1\left(1,-\eps; 1-\eps; \frac{k^-}{p_1^-}\right) \times \bigl[1 + \cO(\lambda^2) \bigr]
\,,\end{align}
and thus correctly reproduces the last line of \eq{RV_expanded}.
Note that the higher-order terms in $\lambda$, such as the second integral in \eq{RV_ncoll}, can be shown to vanish identically in dimensional regularization.

\subsubsection{Hard limit}

The hard region is characterised by treating the loop momentum as uniformly larger than our expansion parameter $\lambda$, while the final state momentum $p_3$ is still treated as collinear to $p_1$.
Only one propagator in \eq{RV} involves $p_3$, and can be expanded in $\lambda$ as
\begin{align} \label{eq:hardpropexp}
 \frac{1}{(p_1 + p_3 + p_4)^2} =&~
 \frac{1}{\bigl[  p_4^2 + p_4^+(p_1^- + p_3^-)  \bigr] + p_3^+ (p_1^- + p_4^-) + 2 p_{3\perp} \cdot p_{4\perp}}
 \nn\\ \underset{p_3 \, \parallel \, p_1}{\overset{p_4\, \rm hard}{\longrightarrow}}&~
 \frac{1}{\bigl[  p_4^2 + p_4^+(p_1^- + p_3^-)  \bigr] + \lambda^2 p_3^+ (p_1^- + p_4^-) + 2 p_{3\perp} \cdot p_{4\perp}}
 \nn\\ =&~
 \sum_{n=0}^\infty (-\lambda)^n \frac{\bigl[\lambda p_3^+ (p_1^- + p_4^-) + 2 p_{3\perp} \cdot p_{4\perp} \bigr]^n}{\bigl[ p_4^+ (p_1^- + p_3^-) + p_4^2 \bigr]^{n+1}}
\,.\end{align}
All other propagators in \eq{RV} scale as $\cO(\lambda^0)$ and are not expanded.
Together with the rescaling of the phase space measure according to \eq{phin0_coll}, the leading-power hard limit of \eq{RV} becomes
\begin{align} \label{eq:RV_hard}
 I_{\rm RV} &
 ~\stackrel{\rm hard}{\longrightarrow}~
 \lambda^{-2\eps} \int\frac{ \df\Phi_{h+1} }{\df \wa\df \wb \df x}  \frac{\df^d p_4}{(2\pi)^d} \frac{1}{p_4^2 \, (p_1 + p_4)^2 \, (p_2 - p_4)^2 \bigl[ p_4^+ (p_1^- + p_3^-) + p_4^2 \bigr]}
 \times \bigl[1 + \cO(\lambda) \bigr]
\,.\end{align}
The overall scaling in $\lambda^{-2\eps}$ arises entirely from the phase space measure, as the hard loop measure scales as $\df^d p_4 \sim \lambda^0$. This shows that hard and collinear loops never have the same dependence on $\eps$, and thus can be easily distinguished by their overall scalings.

Despite the modified propagator in this integral, it can still be subjected to the usual loop integration techniques like IBPs and differential equations.
The same holds true for all higher order terms in the expansion of the full loop integral.
The explicit example in \eq{RV_hard} is also easily performed using Feynman parameters. We obtain
\begin{align} \label{eq:RV_hard2}
 I_{\rm RV} &~\stackrel{\rm hard}{\longrightarrow}~
 \frac{\img c_\Gamma}{128 \pi^4 \eps (1+\eps)} \delta(x)  \lambda^{-2\eps}\frac{(k^+k^-)^{-\eps }}{p_2^+ k^- s^{3+\eps}}
 \left[1-\left(1+\frac{k^-}{p_1^-}\right)^{-1-\eps}\right]
 \times \bigl[1 + \cO(\lambda) \bigr]
\,.\end{align}
This result exactly agrees with the infinite sum over $n$ in \eq{RV_expanded} evaluated for $m=0$, \emph{i.e.}\ the $\cO(\lambda^{-2\eps})$ to \eq{RV_expanded}.
Furthermore, every higher-order term in the expansion in $k^+$ of the second to last line of \eq{RV} corresponds to exactly one term in the integrand expansion of $I_{\rm RV}$ in $\lambda$.
Terms proportional to odd powers of $\lambda$ drop out identically.
Since higher order terms in the expansion essential just modify the powers of the propagators at the integrand level according to \eq{hardpropexp} it is particularly convenient to use IBP techniques in such a computation.

\subsubsection{Anticollinear limit}

For completeness, we also consider the limit where $p_4$ is collinear to the incoming parton with momentum $p_2$,
in contrast to $k$ which is chosen collinear to $p_1$.
According to \eq{modes}, we hence transform
\begin{align}
 p_4^\mu ~\to~ \lambda^2\, p_4^- \frac{n^\mu}{2} + p_4^+ \frac{\bn^\mu}{2} + \lambda\,\, p_{4\perp}^\mu
\,,\qquad
 \df^d p_4 ~\to~ \lambda^d \df^d p_4
\,.\end{align}
With this rescaling, we need to expand two propagators of the integrand in \eq{RV},
\begin{align} \label{eq:bncoll_prop_exp}
 \frac{1}{(p_1 + p_4)^2} &\stackrel{p_2-\rm coll}{\longrightarrow} \frac{1}{p_1^- p_4^+} \times \bigl[1 + \cO(\lambda^2)\bigr]
\,,\nn\\
 \frac{1}{(p_1 + p_3 + p_4)^2} &\stackrel{p_2-\rm coll}{\longrightarrow} \frac{1}{(p_1^- + p_3^-) p_4^+} \times \bigl[1 + \cO(\lambda^2)\bigr]
\,,\end{align}
For brevity, we only show the two leading terms each.
Together with \eq{phin0_coll}, this allows us to expand the integrand in \eq{RV} as
\begin{align} \label{eq:RV_nbcoll}
 I_{\rm RV} &
 \stackrel{p_2-\rm coll}{\longrightarrow}
 \lambda^{-4\eps} \int\frac{ \df\Phi_{h+1} }{\df \wa\df \wb \df x}  \frac{\df^d p_4}{(2\pi)^d}
   \frac{1}{p_4^2 \, (p_1^- p_4^+) \, [(p_1^- + p_3^-) p_4^+] \, (p_2 - p_4)^2} \times \bigl[1 + \cO(\lambda^2) \bigr]
\,,\end{align}
which is scaleless and thus vanishes in pure dimensional regularization.
Note that the integral from expanding the propagators through $\cO(\lambda^n)$ scales as $\lambda^{n-4\eps}$.
Since the only term with this $\eps$ dependence in \eq{RV_expanded} is fully given by the $n$-collinear limit of \eq{RV_ncoll_LP}, the $p_2$-collinear limit must in fact vanish to all orders in $\lambda$.

\subsubsection{Soft limit}
\label{sec:soft_limit}
We want to compare the result of the collinear expansion to a soft expansion of Feynman diagrams.
To obtain the purely soft region, we rescale the loop momentum $p_4$ in \eq{RV} as
\begin{align} \label{eq:p4_soft}
 p_4^\mu
 ~\stackrel{\rm soft}{\longrightarrow}~ \lambda^2 p_4^- \frac{n^\mu}{2} + \lambda^2 p_4^+ \frac{\bn^\mu}{2} + \lambda^2 p_{4\perp}^\mu
\,.\end{align}
To obtain the soft-collinear overlap, we first rescale $p_4$ as collinear, followed by a subsequent soft rescaling,
\begin{align} \label{eq:p4_zerobin}
 p_4^\mu
 ~\stackrel{\rm coll}{\longrightarrow}~ p_4^- \frac{n^\mu}{2} + \lambda^2 p_4^+ \frac{\bn^\mu}{2} + \lambda p_{4\perp}^\mu
 ~\stackrel{\rm soft}{\longrightarrow}~ \lambda^2 p_4^- \frac{n^\mu}{2} + \lambda^2 p_4^+ \frac{\bn^\mu}{2} + \lambda^2 p_{4\perp}^\mu
\,.\end{align}
Let us explicitly discuss the transformation of two of the propagators in \eq{RV} under \eq{p4_zerobin},
\begin{alignat}{3} \label{eq:blubb_1}
 &\frac{1}{(p_2 - p_4)^2} = \frac{1}{p_4^2 - p_2^+ p_4^-}
 && ~\stackrel{\rm coll}{\longrightarrow}~ \frac{1}{\lambda^2 p_4^2 - p_2^+ p_4^-}
 && ~\stackrel{\rm soft}{\longrightarrow}~ \frac{1}{\lambda^2 (\lambda^2\, p_4^2 - p_2^+ p_4^-)}
\,,\nn\\
 &\frac{1}{(p_1 + p_4)^2} = \frac{1}{p_4^2 + p_1^- p_4^+}
 && ~\stackrel{\rm coll}{\longrightarrow}~ \frac{1}{\lambda^2 (p_4^2 + p_1^- p_4^+)}
 && ~\stackrel{\rm soft}{\longrightarrow}~ \frac{1}{\lambda^2 (\lambda^2 \, p_4^2 + p_1^- p_4^+)}
\,.\end{alignat}
In the first case, rescaling the collinear limit as soft only amounts to an overall rescaling by $\lambda^{-2}$, but does not change the relative scaling of the two terms in the propagator.
In the second case, we observe both that only one term in the denominator gets rescaled in the soft limit, and thus  one will encounter a different kinematic structure when expanding this propagator in $\lambda$ than in the collinear limit.
However, in both cases shown in \eq{blubb_1}, it is easy to see that the soft-collinear limit is identical to taking the soft limit directly.
The same holds for the two other propagators in \eq{RV} that are not explicitly shown here.
In conclusion, we find that at the diagram level, the soft-collinear overlap is equal to soft limit itself.

Finally, we note that the leading-power soft limit of \eq{RV} is given by
\begin{align} \label{eq:RV_soft}
 I_{\rm RV} &~\stackrel{\rm soft}{\longrightarrow}~
 \lambda^{-2-6\eps} \int\frac{ \df\tilde\Phi_{h+1} }{\df \wa\df \wb \df x}  \frac{\df^d p_4}{(2\pi)^d}
   \frac{1}{p_4^2 \, (p_1^- p_4^+) \, [(p_1^- + p_3^-) p_4^+ + p_1^- p_3^+ ] \, (-p_2^+ p_4^-)}
   \times \bigl[1 + \cO(\lambda) \bigr]
\,.\end{align}
This integral is scaleless and vanishes in dimensional regularization.

\subsection{Discussion}
\label{sec:expansions_discussion}

To summarize the key results of this section, we have shown how Feynman diagrams can be systematically expanded in their collinear limit by assigning the appropriate scalings to all loop and real momenta, which allows one to expand the integrand in $\lambda$.
In particular, the expanded integrands allow for a diagrammatic representation and are amenable to standard integral techniques such as IBP~\cite{Chetyrkin:1981qh,Tkachov:1981wb} relations or the method of differential equations~\cite{Kotikov:1990kg,Kotikov:1991hm,Kotikov:1991pm,Henn:2013pwa,Gehrmann:1999as}.
This significantly simplifies evaluating the expanded integrals compared to the exact integral, and thus provides a convenient strategy to approximate Feynman diagrams in the collinear limit.
The illustrated methods are conceptually very simple, and thus easily extend to more complicated diagrams with additional external partons or multi-loop integrals.

In the case of real radiation, the requirement that the total real momentum $k^\mu$ is collinear implies that all real momenta are collinear individually.
This does not apply for loop momenta, which are not confined to be in the collinear region. 
As a consequence we need to follow the method of regions~\cite{Beneke:1997zp} and compute the regions where the loop momenta are hard and where they are collinear.
The sum of both regions yields the correct expansion of our Feynman integrals.
The results of the different regions give rise to different scalings as $\lambda^{-4\eps}$ and $\lambda^{-2\eps}$, respectively. 
This difference is entirely due to the loop measure, and thus hard and collinear contributions can be easily identified by their scaling exponent.
In other words, since the expansion of the loop integrand itself is a simple Laurent series in $\lambda$, the loop measure fully determines the non-integer powers of $\lambda^{-n \eps}$.

We also discussed the soft limit of matrix elements. 
We found in an explicit example that the soft region of a loop integral can be obtained by first computing the collinear region of this integral and subsequently taking the soft limit. 
As a matter of fact this property holds more generally. 
The soft-collinear overlap of a partonic coefficient function can either be computed by first performing the collinear expansion and then the soft expansion, or vice versa.
More precisely, expanding the first $n$ terms in the collinear expansion around the production threshold up to $n$ terms will correctly reproduce the $n^{\text{th}}$ power in the threshold expansion.
This provides a stringent test of the collinear expansion by comparing to existing analytic results for which a threshold expansion was performed.
It also provides a considerable simplification for the calculation of collinear master integrals.
For example, if the method of differential equations is utilized to compute master integrals for the collinear expansion, then the boundary conditions for these differential equations can be chosen to be the threshold-expanded integrals.
For the computation of threshold expanded integrals see for example \refscite{Anastasiou:2013srw,Anastasiou:2015yha,Zhu:2014fma}.

\section{Kinematic expansions and SCET}
\label{sec:SCET}

Differential cross sections may be kinematically enhanced in all different momentum regions shown in \eq{modes}. 
Above we only discussed the expansion cross sections around one particular limit, namely the collinear limit.
However, in order to perform a physically sensible and consistent expansion of a hadronic cross section we need to expand in observable quantities.
A collinear expansion of a hadronic cross section alone typically does not satisfy this requirement.
In order to obtain a physical expansion in an observable all momentum regions where the observable is kinematically enhanced must be considered.
Depending on the observable of interest, the necessary ingredients to achieve this goal may vary. 

Soft-Collinear Effective Theory (SCET)~\cite{Bauer:2000ew, Bauer:2000yr, Bauer:2001ct, Bauer:2001yt, Bauer:2002nz} provides an excellent tool to organize the expansion in such kinematic limits, and we discuss in \sec{facthm} how the tools developed in the previous section connect to factorization theorems derived in SCET.
In such factorization theorems, it is crucial to account for the overlap of regions when combining multiple kinematic expansions, which we address in \sec{zero_bin}.

\subsection{Kinematic expansions and factorisation theorems}
\label{sec:facthm}

The momentum regions shown in \eq{modes} are precisely the basis for the formulation of SCET, which is an effective field theory describing QCD in its collinear and soft limits, \emph{i.e.}~the leading infrared region.
Schematically, the SCET Lagrangian is expanded as
\begin{align} \label{eq:LSCET}
 \cL_{\rm SCET} &= \cL_{\rm SCET}^{(0)} + \sum_{k>0} \cL^{(k)}
\,.\end{align}
Here, the superscript $^{(0)}$ indicates the leading-power (LP) terms in the expansion in $\lambda \ll 1$,
where as before $\lambda$ is an auxiliary power counting parameter.
The $\cL^{(k)}$ indicate subleading power Lagrangians~\cite{Manohar:2002fd,Beneke:2002ph,Pirjol:2002km,Beneke:2002ni,Bauer:2003mga,Feige:2017zci,Moult:2017rpl,Chang:2017atu,Moult:2017xpp,Beneke:2017ztn,Beneke:2018rbh} that are suppressed by $\lambda^k$ w.r.t.\ to the leading power.
The leading-power SCET Lagrangian can be organized as
\begin{align} \label{eq:LSCET_LP}
 \cL_{\rm SCET}^{(0)} &= \cL_h^{(0)} + \cL_n^{(0)}  + \cL_\bn^{(0)} + \cL_s^{(0)} + \cL_\cG^{(0)}
\,.\end{align}
Here, $\cL^{(0)}_h$ contains the hard scattering operators mediating the underlying hard interaction,
and $\cL^{(0)}_{n,\bn,s}$ are the SCET Lagrangians for $n$-collinear, $\bn$-collinear and soft fields as defined by \eq{modes}, respectively.%
\footnote{For soft modes, $m=1$, this is referred to as SCET$_{\rm II}$~\cite{Bauer:2002aj}, otherwise as SCET$_{\rm I}$.}
More generally, in the presence of multiple collinear directions as required e.g.\ for multijet processes, \eq{LSCET_LP} contains a sum over all relevant collinear directions $\{n_i\}$.
SCET also allows for a treatment of Glauber modes, which appear as non-local potentials in $\cL_\cG^{(0)}$, the leading power Glauber Lagrangian~\cite{Rothstein:2016bsq}.

In SCET, factorization is achieved by a field redefinition of soft and collinear fields which decouples the soft and collinear Lagrangians from each other \cite{Bauer:2001yt}. These modes can still interact with each other through the Glauber Lagrangian $\cL_\cG^{(0)}$, which thus can break factorization. In this work we will consider observables where the Glauber contributions from $\cL_\cG^{(0)}$ either cancel identically \cite{Bodwin:1984hc,Collins:1984kg,Collins:1985ue,Collins:1988ig,Collins:1989gx,Collins:1350496,Diehl:2015bca} or start contributing to higher perturbative orders that the one we consider in this work \cite{Gaunt:2014ska,Zeng:2015iba}.

In SCET, the leading kinematic regions are made manifest and decoupled from each other at the Lagrangian level, and this greatly simplifies the derivation of factorization formulas.
For suitable factorizable infrared-sensitive observables $\Obs$, which we take to vanish as $\Obs\to0$ in the Born limit, the hadronic cross section \eq{sigma_hadr_finite} can be factorized as \cite{Collins:1984kg, Stewart:2009yx}
\begin{equation}  \label{eq:fact_thm}
\frac{\df\sigma}{\df Q^2 \df Y \df \Obs}
= \sigma_0 \sum_{i,j} H_{ij}(Q^2)\, \bigl[B_i(x_1^B,\Obs)\otimes B_j(x_2^B,\Obs)\otimes S(\Obs) \bigr]
\times \bigl[1 + \cO(\Obs/Q) \bigr]
\,.\end{equation}
As usual, $Q$ and $Y$ are the invariant mass and rapidity of the colorless final state.
The sum runs over all flavor combinations $(i,j)$ contributing at Born level, $i j \to h$,
and $\sigma_0$ is the corresponding Born partonic cross section.%
\footnote{For ease of notation, we suppress the possibility of $\sigma_0$ depending on the flavors $i,j$.}
The hard function $H_{ij}$ encodes virtual corrections to the Born process $ij\to h$, \emph{i.e.}\ it is given as the corresponding renormalized form factor.
The beam functions $B_i(x,\Tau)$ encode the probability to extract a parton of type $i$ with momentum fraction $x$ from the proton, together with the contribution from collinear radiation to the observable $\Tau$,
while the soft function $S(\Tau)$ encodes the effect of soft exchange between the protons.
Since $S(\Tau)$ only differs between quark- and gluon-induced processes, we suppress an explicit flavor label.
Finally, $\otimes$ denotes a convolution in $\Obs$, whose precise structure depends on the chosen observable $\Obs$, and often can be made multiplicative in a suitable conjugate space.
Note that in \eq{fact_thm} we suppress explicit renormalization scales that are present in all functions.

The factorisation of degrees of freedom at the Lagrangian level makes the ingredients for the various functions in \eq{fact_thm} evident.
The hard function, $n$-collinear and $\bn-$collinear beam functions and the soft function are each defined in terms of only hard, $n$-collinear, $\bn$-collinear  and soft degrees of freedom, respectively.
This implies that the expansion techniques developed in this article are perfectly suited to determine beam functions from a perturbative computation using a pure collinear expansion of cross sections.
Here, it is important that both real and loop momenta are expanded as collinear.
We will provide explicit examples by obtaining the NNLO beam functions for $\Obs=q_T$ and $\Obs=\Tau_N$ ($N$-jettiness) in \sec{beamfunctions}.
We also note that in a similar fashion, one can also obtain the soft function by considering a purely soft expansion.

We also stress that since SCET is an effective field theory, it can be systematically extended by including the power-suppressed Lagrangians $\cL^{(k>0)}$ in \eq{LSCET}. This is the EFT analog of expanding cross sections to subleading order in $\lambda$ about the soft and collinear limits.
However, at subleading powers, collinear and soft interactions do not simply factorize similar to \eq{fact_thm} anymore, and factorization theorems and the resummation of large logarithms become much more involved~\cite{Hill:2004if,Lee:2004ja,Benzke:2010js,Freedman:2014uta,Moult:2016fqy,Moult:2017jsg,Beneke:2017vpq,Beneke:2017ztn,Feige:2017zci,Moult:2017rpl,
Chang:2017atu,Moult:2017xpp,Alte:2018nbn,Beneke:2018gvs,Beneke:2018rbh,Moult:2018jjd,Ebert:2018lzn,Ebert:2018gsn,Bhattacharya:2018vph,Beneke:2019kgv,Moult:2019mog,Beneke:2019mua,Moult:2019uhz,Moult:2019vou,Liu:2019oav}. 
Since our expansion technique allows us to perform collinear expansions of partonic cross sections to arbitrary order in $\lambda$, we hope that it will also provide insights into the structure of factorisation theorems beyond the leading power, and that it can be used to determine universal quantities like generalizations of soft and beam functions at subleading power.

\subsection{Soft-collinear overlap and zero-bin subtractions}
\label{sec:zero_bin}

In order to obtain a full description of a cross section in its infrared limit, we need to combine all collinear and soft regions. Schematically, we expand
\begin{align} \label{eq:zero_bin_1}
 \lim_{\rm IR} \frac{\sigma}{\df Q^2 \df Y \df \Obs} &
 = \frac{\sigma^{(n)}}{\df Q^2 \df Y \df \Obs} + \frac{\sigma^{(\bn)}}{\df Q^2 \df Y \df \Obs} + \frac{\sigma^{(s)}}{\df Q^2 \df Y \df \Obs} + \cdots
\,,\end{align}
where the $\sigma^{(n,\bn,s)}$ correspond to the expansion of the cross sections where all emissions are treated  as $n$-collinear, $\bn$-collinear and soft, respectively.
The ellipses denote mixings of these cases, as well as power-suppressed corrections.
Note that here in the following, we do not consider the hard region. While it is required to obtain an infrared-finite cross section, it corresponds to physics at the hard scale $\mu^2 \sim Q^2$, and does not affect the soft-collinear overlap discussed in the following.

In practice, \eq{zero_bin_1} is often too naive, as there is a nontrivial overlap between the collinear and soft regions. This arises because the soft limit of a squared matrix element is equal to the soft limit of the \emph{collinear limit} of the same matrix element.
As discussed and illustrated in more detail in \sec{soft_limit}, this can be understood since the soft limit can be equivalently obtained by either directly rescaling
\begin{align} \label{eq:soft_1}
 k^\mu = (k^+,k^-,k_\perp)
 \quad\stackrel{\mathrm{soft}}{\xrightarrow{\hspace*{1.5cm}}} \quad
 (\lambda^2, \lambda^2, \lambda^2)
\,,\end{align}
or by first rescaling into the collinear limit with a subsequent soft rescaling,
\begin{align} \label{eq:soft_2}
 k^\mu = (k^+,k^-,k_\perp)
 \quad\stackrel{n-\mathrm{collinear}}{\xrightarrow{\hspace*{1.5cm}}} \quad
 (1, \lambda^2, \lambda)
 \quad\stackrel{\mathrm{soft}}{\xrightarrow{\hspace*{1.5cm}}} \quad
 (\lambda^2, \lambda^2, \lambda^2)
\,.\end{align}
Since the second rescaling only lowers the scaling of each component, no information is lost, and \eqs{soft_1}{soft_2} produce the same expansion of a matrix element.

Consequently, when one integrates over a collinearly-rescaled momentum, the integral will always contain contributions from the soft region. Schematically, if we write
\begin{align}
 \int\df^d p \, f(p^+, p^-, p_\perp)
 \quad\stackrel{n-\mathrm{coll}}{\xrightarrow{\hspace*{1.cm}}}\quad
 \lambda^d \!\int\!\df p^+ \df p^- \df^{d-2} \vec p_\perp \, f^{(n)}(p^+ \sim \lambda^2, p^- \sim 1, p_\perp \sim \lambda)
\end{align}
for the collinear expansion $f^{(n)}$ of an arbitrary integrand $f$, then clearly the integration range extends into a region where the assumed collinear scaling is not justified.
In particular, the $p^-$ integral extends to $p^- \to 0$, which corresponds to a soft region.
This contribution to the soft region can be identified and extracted by further expanding $f^{(n)}$ as indicated in \eq{soft_2}.

In conclusion, the collinear limit of the cross section has an overlap with the soft limit, which can be extracted by an additional reexpansion in the soft limit, which has been demonstrated explicitly for a mixed real-virtual integral in \sec{soft_limit}.
We thus need to modify \eq{zero_bin_1} as
\begin{align} \label{eq:zero_bin_2}
  \lim_{\rm IR} \frac{\sigma}{\df Q^2 \df Y \df \Obs} &
 = \left[ \frac{\sigma^{(n)}}{\df Q^2 \df Y \df \Obs} - \frac{\sigma^{(n \to s)}}{\df Q^2 \df Y \df \Obs}\right]
 + \left[ \frac{\sigma^{(\bn)}}{\df Q^2 \df Y \df \Obs} - \frac{\sigma^{(\bn \to s)}}{\df Q^2 \df Y \df \Obs}\right]
 \nn\\&\quad
 + \frac{\sigma^{(s)}}{\df Q^2 \df Y \df \Obs} + \cdots
\,,\end{align}
where the soft limit of the collinear cross sections are denoted by $\sigma^{(n \to s)}$ and $\sigma^{(\bn \to s)}$, respectively. The terms in brackets hence correspond to the true $n$- and $\bn$-collinear limits of the cross section.
Note that in general, $\sigma^{(n \to s)} \ne \sigma^{(s)}$, because the observable $\Tau$ itself has to be expanded in the collinear and soft limits.

Let us connect these observations to the corresponding treatment in SCET.
As a modal EFT, SCET is built to separately describe soft and collinear modes, and hence as a matter of principle collinear momenta are not allowed to overlap with the soft sector.
In practice, it is not feasible to introduce a cutoff between soft and collinear modes.
Instead, one follows the same strategy outlined above: after calculating a collinear integral, one subtracts its soft limit to obtain the pure collinear result.
This procedure is referred to as zero-bin subtraction \cite{Manohar:2006nz}, and is crucial to a well-defined separation of modes in the EFT.
In practice, the zero-bin subtractions are often absent in dimensional regularization or equal to the soft function itself, and thus can be easily taken into account.

\section{Beam functions from the collinear limit}
\label{sec:beamfunctions}

In this section we show how the collinear expansions can be used to compute beam functions. We briefly introduce the notion of beam functions in \sec{beamdef}, and then show in \sec{beam_funcs_strategy} how they are related to the collinear expansion of cross sections developed before. Our method is briefly contrasted with other methods of calculating beam functions in \sec{beam_func_methods}. We show explicitly how to obtain the $N$-jettiness and the $q_T$ beam functions at NNLO using this method in \sec{Tau_beam_func} and \sec{qT_beam_func}, respectively.

\subsection{Beam functions}
\label{sec:beamdef}

Beam functions are defined as gauge-invariant hadronic matrix element that measure the large lightcone momentum entering the hard interaction, as well as the contribution to the observable $\Tau$ from collinear radiation.
For example, the quark beam function $B_q$ is defined in SCET as~\cite{Stewart:2009yx}
\begin{align} \label{eq:beam_def}
 B_q(x = p^-/P^-, \Tau) = \big< p_n(P) \big| \bar\chi_n(0) \frac{\slashed \bn}{2} \bigl[ \delta(p^- - \bn \cdot \cP) \, \delta(\Tau - \hat \Tau) \, \chi_n(0) \bigr] \big| p_n(P) \bigr>
\,.\end{align}
Here, the $\chi_n = W_n^\dagger q$ are collinear quark fields defined in SCET as quark fields dressed with collinear Wilson lines $W_n$, $p_n(P)$ is a proton moving along the $n$-direction with momentum $P$, and $\bn\cdot\cP$ is the SCET momentum operator that determines the lightcone momentum of all fields to its right.
By boost invariance, the beam function only depends on the momentum fraction $x = p^-/P^-$.
Similarly, $\hat\Tau$ is the appropriate measurement operator determining the observable $\Tau$ in terms of all momenta of the fields to its right.

Beam functions are a natural generalization of PDFs, which in SCET are defined as~\cite{Bauer:2002nz}
\begin{align} \label{eq:pdf_def}
 f_q(x = p^-/P^-) = \bigl<p_n(P) \big| \bar\chi_n(0) \frac{\slashed n}{2} \bigl[ \delta(p^- - \bn \cdot \cP) \, \chi_n(0) \bigr] \big| p_n(P)\bigr>
\,.\end{align}
Comparing \eqs{beam_def}{pdf_def}, it is evident that the beam function extends the PDF by measuring an additional observable $\Tau$ on top of the longitudinal momentum fraction carried by the struck parton.

Both beam functions and PDFs are in general intrinsically nonperturbative matrix elements.
For perturbative $\Tau \gg \lqcd$, one can perform an operator product expansion of the beam function onto the PDF~\cite{Stewart:2009yx},
\begin{align} \label{eq:matching_B}
 B_i(x,\Tau,\mu) = \sum_j \, \cI_{ij}(x,\Tau,\mu)\otimes_x  f_j^R(x,\mu) \times \bigl[1 + \cO(\lqcd/\Tau)\bigr]
\,.\end{align}
Here, the only nonperturbative input is given in terms of the PDFs, while the matching kernel $\cI_{ij}$ are perturbatively calculable.

For completeness, we remark that PDFs and beam functions can also be defined without invoking SCET by expressing the collinear quark fields $\chi_n$ in terms of standard quark fields and collinear Wilson lines $W_n$, which are defined as path-ordered exponentials of the gluon field projected onto the appropriate collinear direction. Beam functions are also often written as the Fourier transform of a position-space correlator, where the separation between the quark fields corresponds to the exchanged momentum and often avoids the need for the momentum operator $\cP$ in \eq{beam_def}. PDFs and TMDPDFs were originally defined in this way~\cite{Collins:1981uw,Collins:1984kg}, and the equivalence of both formulations was also discussed in the context of $\Tau_N$ beam functions in \refscite{Stewart:2009yx,Stewart:2010qs}.
Note that the study of parton distributions from lattice QCD requires the definition in position space, see e.g.~\refscite{Ji:2014hxa,Ji:2018hvs,Ebert:2018gzl,Ebert:2019okf,Ebert:2019tvc,Ji:2019sxk,Ji:2019ewn,Vladimirov:2020ofp,Ebert:2020gxr} for recent progress towards calculating TMDPDFs on lattice, and \refscite{Cichy:2018mum, Ji:2020ect} for a more general overview of parton physics from lattice QCD. For perturbative calculations, both formulations are equivalent.

\subsection{General strategy}
\label{sec:beam_funcs_strategy}

In \sec{facthm}, we discussed that the hard, beam and soft functions in the factorized cross section in \eq{fact_thm} are each defined only in terms of the hard, collinear and soft modes of \eq{modes}, respectively. Hence, in the limit where all loop and final-state momenta are treated as $n$-collinear, the hard function, the $\bn$-collinear beam function, and the soft function only contribute at their respective tree level, where they are normalized to unity and flavor diagonal. Thus, the strict $n$-collinear limit of \eq{fact_thm} is given by
\begin{equation} \label{eq:fact_thm_n}
\strictlim \frac{\df\sigma}{\df Q^2 \df Y \df \Tau}
 = \sigma_0 \sum_{i,j} B_i(x_1^B,\Tau) f_j(x_2^B)
\,,\end{equation}
where we remind the reader that the flavor sum runs over all flavors contributing at Born level, $i j \to h$, and $\sigma_0$ is the associated Born partonic cross section.

We remark that \eq{fact_thm_n} is to be understood at the bare level, as it for example does not encode scale independence. Indeed, as we will see, even after UV renormalization and IR subtraction one encounters leftover poles in $\eps$, which in the full factorized cross section in \eq{fact_thm} cancel with the other ingredients.

In the following, we assume that the Born process is diagonal in flavor, \emph{i.e.} only the $gg$ channel (as in Higgs production in gluon fusion) or the $q\bar q, \bar q q$ channels (as in Drell-Yan or $b\bar{b}$ initiated Higgs production) contribute, where $q$ is an arbitrary quark flavor. With this assumption,  we can fix $j = \bar i$ in \eq{fact_thm_n}, which allows us to easily read off the \emph{bare} beam function by comparing with the $n$-collinear limit of the cross section given in \eqs{sigma_hadr}{partcoef_special},
\begin{align} \label{eq:beam_master}
 B_i(x_1^B,\Tau) &
 = \sum_{j} \int_{x_1^B}^1 \frac{\df z_1}{z_1} f_j\Bigl(\frac{x_1^B}{z_1}\Bigr)
   \times \int_0^1 \df x \int_0^\infty \df \wa\df \wb \, \delta\left[z_1-(1-\wa)\right]
   \nn\\&\qquad\times
   \strictlim \left\{\delta\left[\Obs-\Obs(Q,Y,\wa,\wb,x)\right] \frac{\df\eta_{j \bar i}}{ \df Q^2  \df \wa \df \wb \df  x}\right\}
\,.\end{align}
By fixing the flavor of the $\bn$-collinear parton as $\bar i$, we extract the correct beam function for the flavor $i$ in a flavor-diagonal process.

\eq{beam_master} has precisely the structure of \eq{matching_B}, and we can immediately read off the bare matching kernel as the collinear limit of the partonic coefficient function,
\bea \label{eq:I_bare}
 \cI_{ij}^{\rm bare}(z,\Tau)&=&
   \int_0^1 \df x \int_0^\infty \df \wa\df \wb \, \delta\left[z-(1-\wa)\right]  \nonumber\\
&\times& \strictlim \left\{\delta\left[\Obs-\Obs(Q,Y,\wa,\wb,x)\right] \frac{\df\eta_{j\bar i}}{ \df Q^2  \df \wa \df \wb \df  x}\right\}
\,.\eea
We stress that the partonic coefficient function here is limited to strictly collinear modes only.
In contrast, in the collinear expansion for cross sections discussed before, we also included non-collinear modes when computing loop integrals.
However, we showed that the collinear and non-collinear modes can easily be separated by looking at their respective generalized scaling behaviour.
Extracting the required parts is consequently easy. 
In the strictly collinear limit the general partonic coefficient function of \eq{eta_ij_1} becomes
\begin{align}
 &\strictlim \frac{\df\eta_{j\bar i} }{\df Q^2 \df w_1 \df w_2 \df x}
\\\nn&
 = \delta_{j\bar i} \delta(w_1)\delta(w_2)\delta(x)
 \,+\, \sum_{\ell=1}^\infty \left(\frac{\as}{\pi}\right)^{\ell} w_2^{-1-l\eps}
   \sum_{m=1}^\ell w_1^{-1-m\eps}
   \frac{\df\eta_{j\bar i}^{(\ell,m,n)}(w_1,0,x,Q^2)}{\df Q^2 \df w_1 \df w_2 \df x}
\,.\end{align}
The strict collinear limit for the partonic coefficient function for the observable $\Tau$ in \eqs{beam_master}{I_bare} is then obtained in analogy to \eq{partoniccoefexp}.

A special case of \eq{I_bare} is the bare matching kernel differential in $\wa$, $\wb$ and $x$ itself, from which one can project out all other beam functions we interested in.
In fact, this double-differential beam function can be related to the fully unintegrated parton distribution first formulated in \refscite{Collins:2007ph,Rogers:2008jk} and within SCET in \refscite{Mantry:2009qz,Jain:2011iu}, where it is also known as double-differential beam function (dBF).
Importantly, in general the projection of $(\wa,\wb,x)$ onto the desired observable $\Tau$ only holds at the bare level, not after renormalization of the dBF~\cite{Jain:2011iu}.
The renormalization of the dBF is also significantly more complicated than that of the $\Tau_N$ and $q_T$ beam functions we are interested in, see \refscite{Gaunt:2014xxa,Gaunt:2020xlc} for explicit results at NNLO.

$\cI_{ij}^{\rm bare}$ still contains infared poles that cancel upon PDF renormalization in \eq{beam_master}.
Even after $\as$ renormalization, this still leaves divergences that cancel in the cross section when combining the $n$-collinear limit with the $\bn$-collinear and soft limits, but are remnant in the bare matching kernel.
In the EFT, these divergences are of ultraviolet origin and thus can be absorbed in the standard fashion through a counterterm.
Subtracting both IR and UV poles in this manner, we obtain the renormalized matching kernel as
\begin{equation} \label{eq:I_master}
 \cI_{ij}(x,\Tau,\mu)
 = \sum_{j'} \Gamma_{j j'}(z, \eps) \otimes_z Z_B^i (\Tau,\mu,\eps) \otimes_\Tau \hat Z_{\as}(\mu,\eps) \, \cI_{ij'}^{\rm bare}(x,\Tau,\eps)
\,,\end{equation}
where $\otimes_\Tau$ denotes the appropriate convolution in $\Tau$.
According to \eqs{I_bare}{I_master}, we can obtain the beam function matching kernel as follows:
\begin{enumerate}
 \item Obtain the bare kernel $\cI_{ij}^{\rm bare}$ from the strict collinear limit of the partonic cross section.
 \item Apply $\alpha_s$ renormalization through $\hat Z_{\as}$, which renormalizes the bare coupling constant $\as^b$ in the $\MSbar$ scheme.
 \item Subtract the EFT UV divergences with the beam-function counterterm $Z_B^i$.
       This renormalization does not change the parton flavor $i$, and only differs between quark and gluons, but is independent of the quark flavor.
       In general, this counterterm enters through a convolution in $\Tau$, which can be trivialized by going to suitable conjugate space.
 \item Subtract IR divergences by convolving with the PDF counterterm $\Gamma_{jj'}$, which as usual mixes parton flavors.
\end{enumerate}
Since the $\Gamma_{jj'}$ and $Z_B^i$ commute, one can freely rearrange their order in \eq{I_master}.
Since the beam function counter term $Z_B^i$ gives rise to the renormalization group equation of the beam function, in practice one can either predict $Z_B^i$ from the RGEs and check that this cancels all poles in $\eps$, or determine $Z_B^i$ by absorbing all poles remaining after QCD UV and IR subtraction and verify that it reproduces the RGE dictated by the EFT. For the $\Tau_N$ and $q_T$ beam functions, this is discussed in more detail in our companion papers \cite{Ebert:2020yqt,Ebert:2020unb}.

In \eq{I_bare}, we assumed that the partonic coefficient function is taken in the strict $n$-collinear limit.
As discussed in \sec{zero_bin}, for certain observables there can be overlap with the soft limit, which in the factorized cross section in \eq{fact_thm} is already accounted for by the soft function.
In such instances, one has to subtract off the soft-collinear overlap,
\begin{align} \label{eq:I_bare_0bin}
 \cI_{ij}^{\rm bare}(z,\Tau) &
 = \int_0^1 \df x \int_0^\infty \df \wa\df \wb \, \delta\left[z-(1-\wa)\right]
 \nn\\&\quad\times
 \left[\strictlim\frac{\df \eta_{j\bar{i}}}{\df Q^2 \df Y \df \Tau} - \slim \strictlim\frac{\df \eta_{j\bar{i}}}{\df Q^2 \df Y \df \Tau} \right]
\,.\end{align}
The second term in the above equation denotes that the collinear limit is further re-expanded in the soft limit.

\subsection{Comparison to alternative methods}
\label{sec:beam_func_methods}

Before illustrating our method for the $\Tau_N$ and $q_T$ beam functions in \secs{Tau_beam_func}{qT_beam_func}, we briefly contrast our approach to methods previously used in the literature.
Here, we focus on how to calculate the bare beam function, since the renormalization and subtraction of UV and IR divergences always proceeds in the same fashion.

Most calculations of beam functions explicitly calculate matching coefficients from matrix element of the beam function operator, see e.g.~\refscite{Stewart:2010qs,Gaunt:2014xga,Gaunt:2014cfa,Gaunt:2014xxa,Gehrmann:2012ze,Gehrmann:2014yya,Gangal:2016kuo,Echevarria:2016scs,Gutierrez-Reyes:2019rug,Luo:2019hmp,Luo:2019bmw,Luo:2019szz,Gaunt:2020xlc}. Let us explain some features of this approach for the concrete example of a quark beam function as shown in \eq{beam_def}, whose bare matching kernel $\cI_{qj}$ is obtained by evaluating the matrix element in \eq{beam_def} with an external on-shell parton of flavor $j$. In \refcite{Stewart:2010qs}, the analytic structure of these matrix elements was discussed in detail for the $\Tau_N$ beam function, and it was shown that one can calculate it by taking the discontinuity of matrix elements of the time-ordered operator.

Firstly, this implies that the beam function can be calculated using SCET Feynman rules.
Since a single collinear sector in SCET is equal to a boosted copy of QCD, one can equivalently employ QCD Feynman rules. In this case, eikonal vertices arise from the Feynman rules of the Wilson lines $W_n$ that are part of the collinear quark fields $\chi_n = W_n^\dagger q$. These can be avoided in lightcone gauge, where $\bn \cdot A = 0$ such that $W_n = 1$, but similar terms arise from the gluon propagator in ligthcone gauge. Since the beam function is defined as a gauge-invariant matrix element, both approaches yield equal results.

The discontinuity can be obtained by using the Cutkosky rules~\cite{Cutkosky:1960sp} (see also \refcite{Ellis:1996nn}), which corresponds to taking particles exchanged between the quark fields in \eq{beam_def} on-shell. This is analogous to our approach, where we explicitly consider on-shell radiation into the final state.
Alternatively, one can not apply an on-shell constraint and integrate over all particles, and explicitly take the discontinuity afterwards. Both approaches are discussed in more detail in \refcite{Gaunt:2014cfa}, where they are referred to as on-shell and dispersive method, respectively.

An alternative method that does not directly rely on the definition of the beam function in SCET was pointed out in \refcite{Ritzmann:2014mka}, where it was shown that one can equivalently calculate the beam function from phase-space integrals over QCD splitting functions. This approach was used in \refscite{Baranowski:2020xlp,Melnikov:2018jxb,Melnikov:2019pdm,Behring:2019quf}, where the required splitting function at N$^3$LO was obtained following the method of \refcite{Catani:1999ss}. This approach requires to use a physical gauge where gluons are explicitly transverse, for example the lightcone gauge $\bn \cdot A = 0$.

Similar to \refcite{Ritzmann:2014mka}, our method does not rely on directly calculating SCET matrix elements.
However, our approach is manifestly gauge invariant as it is based on a physical cross section, similar to the direct calculations.
The connection of our approach to these previous methods can be understood as follows: 
Prior to integrating over real radiation, the collinear expansion reproduces precisely the collinear limit of QCD, which in the SCET approach is immediately encoded in the structure of the SCET matrix element, whereas in the approach of \refcite{Ritzmann:2014mka} it is obtained from the QCD splitting function.
In practice, one advantage of our method is that it can be easily integrated with standard methods of generating Feynman diagrams.
One can then use standard methods to evaluate the integrals using IBPs~\cite{Chetyrkin:1981qh,Tkachov:1981wb} and the method of differential equations~\cite{Kotikov:1990kg,Kotikov:1991hm,Kotikov:1991pm,Henn:2013pwa,Gehrmann:1999as} in the reverse unitarity framework~\cite{Anastasiou2003,Anastasiou:2002qz,Anastasiou:2003yy,Anastasiou2005,Anastasiou2004a} over the real radiation phase space, keeping only the total momentum $k$ fixed. This intermediate result, $\df\eta_{ij} / (\df Q^2 \df\wa \df\wb \df x)$, is the bare fully differential beam function, from which one can then project out the desired beam functions.

\subsection[\texorpdfstring{$\Tau_N$}{TauN} beam functions]
           {\boldmath $\Tau_N$ beam functions}
\label{sec:Tau_beam_func}

\subsubsection{Factorization}
\label{sec:tau0_factorization}

$N$-jettiness is an inclusive event shape that yields an $N$-jet resolution variable.
It was first introduced in \refcite{Stewart:2010tn}, and its factorization was derived using SCET in \refscite{Stewart:2009yx, Stewart:2010tn, Jouttenus:2011wh}.
Since the same beam function appears for all $\Tau_N$, we focus only on the simplest case $\Tau_0$, also known as beam thrust, that is relevant to color-singlet processes.
Beam thrust is defined as \cite{Stewart:2010tn, Jouttenus:2011wh}
\begin{align} \label{eq:Tau0}
\Tau_0 = \sum_i {\rm min}\biggl\{ \frac{2 q_1 \cdot (-k_i)}{Q_a},\, \frac{q_2 \cdot (-k_i)}{Q_b} \biggr\}
\,.\end{align}
Here, $q_{1,2}$ are the Born-projected momenta of the incoming partons, given by
\begin{align}
 q_1^\mu = x_1^B \sqrt{S} \frac{n^\mu}{2} = Q e^Y \frac{n^\mu}{2}
\,,\qquad
 q_2^\mu = x_2^B \sqrt{S} \frac{\bn^\mu}{2} = Q e^{-Y} \frac{\bn^\mu}{2}
\,,\end{align}
where as before $Q$ and $Y$ are the invariant mass and rapidity of the color-singlet final state $h$, respectively.
The sum in \eq{Tau0} runs over all final-state particles excluding $h$, and as usual all final-state momenta are taking as incoming.
The $Q_{a,b}$ are measures that determine different definitions of 0-jettiness.
The original definitions are~\cite{Stewart:2009yx, Berger:2010xi}
\begin{alignat}{4} \label{eq:Tau0_2}
 &\text{leptonic:}\quad & Q_a &= Q_b = Q \,,\qquad
  & \Tau_0^{\rm lep} &= -\sum_i \min \Bigl\{ e^Y n \cdot k_i \,,\, e^{-Y} \bn \cdot k_i \Bigr\}
\nn\\
 &\text{hadronic:}\qquad & Q_{a,b} &= Q \, e^{\pm Y} \,,
 & \Tau_0^{\rm cm} &= -\sum_i \min \Bigl\{ n \cdot k_i \,, \bn \cdot k_i \Bigr\}
\,.\end{alignat}
The precise choice does not affect the calculation of the beam function, but it becomes important for the calculation of power corrections~\cite{Ebert:2018lzn}.
We note in passing that at subleading power, the leptonic definition is clearly preferred as it gives rise to smaller power corrections that the hadronic definition \cite{Moult:2016fqy,Ebert:2018lzn}.

At small $\Tau_0 \ll Q$, the cross section can be factorized as~\cite{Stewart:2009yx}
\begin{align} \label{eq:Tau0_fact}
 \frac{\df\sigma}{\df Q^2 \df Y \df \Tau_0} &
= \sigma_0 \sum_{i,j} H_{ab}(Q^2, \mu) \int \! \df t_a \, \df t_b \,
   B_a(t_a, x_1^B, \mu) \, B_b(t_b, x_2^B, \mu) \,
   S\Bigl(\Tau_0 - \frac{t_a}{Q_a} - \frac{t_b}{Q_b}, \mu\Bigr)
\nn \\ &\qquad
\times \Bigl[1 + \cO\Bigl(\frac{\Tau_0}{Q}\Bigr)\Bigr]
\,.\end{align}
As indicated, this factorization holds up to power corrections suppressed by $\Tau_0/Q$ that were studied in~\refscite{Moult:2016fqy,Moult:2017jsg,Boughezal:2018mvf,Ebert:2018lzn,Boughezal:2019ggi} and the relevant SCET operators have been derived in \refscite{Feige:2017zci,Moult:2017rpl,Chang:2017atu}.
In the case of fiducial cuts applied to the decay products of $h$, these corrections can be enhanced as $\cO(\sqrt{\Tau_0/Q})$ \cite{Ebert:2019zkb}.
Furthermore, starting at N$^4$LO it also receives contributions from perturbative Glauber-gluon exchanges that are not captured by \eq{Tau0_fact}~\cite{Gaunt:2014ska, Zeng:2015iba}.

The beam function $B_i(t,x,\mu)$, sometimes also referred to as the virtuality-dependent beam function, appears in the factorization of all $\Tau_N$~\cite{Stewart:2010tn}, deep-inelastic scattering \cite{Kang:2013nha}, and in the factorization of color-singlet processes in the generalized threshold limit \cite{Lustermans:2019cau}.
It is known at NNLO~\cite{Stewart:2010qs,Berger:2010xi,Gaunt:2014xga,Gaunt:2014cfa}, and we compute it at N$^3$LO for all partonic channels in our companion paper~\cite{Ebert:2020unb}. Previous progress towards the calculation of the quark beam function at N$^3$LO was made in \refscite{Melnikov:2018jxb,Melnikov:2019pdm,Behring:2019quf}.

In \eq{Tau0_fact}, the beam functions are defined to measure the $Q_{a,b}$-independent combinations $t_a = -q_1^ -k^+$ and $t_b = - q_2^+ k^-$, while the measurement-dependent normalization factors $Q_{a,b}$ only arise in the convolution in \eq{Tau0_fact}.
This definition naturally arises because $\Tau_0$ simplifies in the $n$-collinear limit to
\begin{align} \label{eq:Tau0_n}
 \nlim \Tau_0 = \sum_i \frac{2 q_1 \cdot (-k_i)}{Q_a} = \frac{q_1^- (-k^+)}{Q_a}
\,,\end{align}
and similarly in the $\bn$-collinear limit.

The soft function in \eq{Tau0_fact} only differs between quark annihilation and gluon fusion, but is independent of quark flavors, and we suppress the explicit color index in \eq{Tau0_fact}.
$S(\Tau,\mu)$ is a hemisphere soft function for two incoming lightlike Wilson lines.
Through NNLO, it is equal to the hemisphere soft function for $e^+ e^- \to \rm{dijets}$ \cite{Stewart:2009yx,Kang:2015moa}, which itself is known at NNLO  \cite{Stewart:2009yx,Stewart:2010qs,Schwartz:2007ib,Fleming:2007xt,Kelley:2011ng,Monni:2011gb,Hornig:2011iu}.

\subsubsection[Calculation of \texorpdfstring{$\Tau_N$}{TauN}-dependent beam functions]
              {\boldmath Calculation of $\Tau_N$-dependent beam functions}
\label{sec:tau0_beam_funcs}

Since the collinear limit of $\Tau_0$ given in \eq{Tau0_n} only depends on the total momentum $k^\mu$ of all real emissions, the $\Tau_N$ beam function can be calculated using the method outlined in \sec{beam_funcs_strategy}. In contrast, the soft limit of \eq{Tau0} requires knowledge of all individual momenta $\{k_i\}$, and thus can not be calculated in this fashion.

Using \eqs{I_bare}{Tau0_n}, we can calculate the bare beam function kernel as
\begin{align} \label{eq:cI_tau_bare}
 \cI_{ij}^{\rm bare}(z,t,\eps) &
 = \int_0^1 \df x \int_0^\infty \df w_1  \df w_2
   \, \delta[z-(1-\wa)] \, \delta\bigl(t - Q^2 w_2 \bigr)
   \strictlim\frac{\df\eta_{j\bar{i}} }{\df Q^2 \df w_1 \df w_2 \df x}
\,.\end{align}
The zero-bin for the $\Tau_N$ beam function is known to be scaleless and thus vanishes in pure dimensional regularization, and hence need not be included explicitly \cite{Stewart:2010tn}.
Note that $w_2 > 0$ implies $t > 0$, which we keep implicit. 

The bare kernel contains UV divergences from the limit $\wa=1-z\to0$ and $\wb=t/Q^2\to0$, which are both regulated using dimensional regularization. The divergences from small $t$ can be made manifest through the standard identity
\begin{align} \label{eq:distr_tau}
 \frac{1}{\mu^2} \left(\frac{\mu^2}{t}\right)^{1 + a \eps} = - \frac{\delta(t)}{a \eps} 
 + \left[\frac{1}{t}\right]_+ + a \eps \left[\frac{\ln(t/\mu^2)}{t}\right]_+  + \cO[(a \eps)^2]
\,,\end{align}
where $\left[\ln^n x / x\right]_+$ is the standard plus distributions.
Following \sec{beam_funcs_strategy}, we obtain the renormalized matching as
\begin{equation} \label{eq:Iij_ren}
 \cI_{ij}(t,z,\mu)
 = \sum_{k} \int\df t' \, Z_B^i(t-t',\eps,\mu) \, \int_z^1\!\frac{\df z'}{z'}
   \, \Gamma_{jk}\Bigl(\frac{z}{z'},\eps\Bigr) \,
   \hat Z_{\as}(\mu,\eps) \, \cI_{i k}(t',z',\eps)
\,,\end{equation}
where the structure of convolution in $t$~\cite{Stewart:2010qs,Berger:2010xi} is made explicit.
In practice, it is more useful to perform the renormalization in Fourier or Laplace space, where the convolution in $t$ turns into a simple product. In particular, the structure of $Z_B^i$ can be easily predicted from the beam function RGE in Fourier space. For details on this, we refer to \refcite{Ebert:2020unb}.

We have implemented the described procedure at one loop through $\cO(\eps^4)$ and at two loops through $\cO(\eps^2)$, as required for the calculation of the three-loop beam function.
We use the collinear limit of the cross sections for Higgs and Drell-Yan production to extract the gluon and quark beam functions, respectively.
As intermediate checks, we verified that the UV and IR counterterms correctly cancel all appearing divergences.
The final renormalized results agrees with the NNLO results reported in \refscite{Gaunt:2014xga,Gaunt:2014cfa},
and the higher-order terms in $\eps$ agree with \refcite{Baranowski:2020xlp}.

\subsection[\texorpdfstring{$q_T$}{qT} beam functions]
           {\boldmath $q_T$ beam functions}
\label{sec:qT_beam_func}

\subsubsection{Factorization}
\label{sec:qT_factorization}

The factorization of the transverse-momentum ($\qt$) distribution of a colorless probe $h$ in the limit $q_T \ll Q$ was first derived by Collins, Soper, and Sterman (CSS) in \refscite{Collins:1981uk,Collins:1981va,Collins:1984kg} and elaborated on in \refscite{Catani:2000vq, deFlorian:2001zd, Catani:2010pd, Collins:1350496}. The factorization was also discussed using SCET in \refscite{Becher:2010tm, GarciaEchevarria:2011rb, Chiu:2012ir, Li:2016axz}.
The factorized cross section is commonly formulated in Fourier (impact parameter) space, with $\bt$ being Fourier-conjugate to $\qt$, as this significantly simplifies the resummation of large logarithms~\cite{Ebert:2016gcn}. We write the factorized $\qt$ spectrum as
\begin{align} \label{eq:qt_fact}
 \frac{\df \sigma}{\df Q^2 \df Y \df^2 \qt} &
 = \sigma_0 \sum_{i,j} H_{ij}(Q^2,\mu) \int\!\df^2\bt \, e^{\img\,\qt \cdot \bt} \,
   \,\tilde B_i\Bigl(x_1^B, b_T, \mu, \frac{\nu}{\omega_a}\Bigr)
   \,\tilde B_j\Bigl(x_2^B, b_T, \mu, \frac{\nu}{\omega_b}\Bigr)
   \, \tilde S(b_T, \mu, \nu)
\nn\\&\quad
 \times \left[1 + \cO\left(q_T^2/Q^2\right) \right]
\,.\end{align}
It receives power corrections suppressed by $q_T^2/Q^2$, which were studied at fixed order in perturbation theory in \refcite{Ebert:2018gsn}. The study of their all-order structure has been initiated using the SCET operator formalism in \refscite{Kolodrubetz:2016uim,Feige:2017zci,Moult:2017rpl,Chang:2017atu,Chang:NLP},
and and their nonperturbative structure has been explored in \refscite{Balitsky:2017gis,Balitsky:2017flc}.
These corrections are enhanced as $\cO(q_T/Q)$ when applying fiducial cuts to $h$ \cite{Ebert:2019zkb},
but for Drell-Yan and Higgs production can be uniquely included in the factorization theorem~\cite{Ebert:fiducial}, and are also linear when one includes radiation from massive final states \cite{Buonocore:2019puv}.

TMD factorization is complicated by the fact that the bare beam and soft functions not only contain IR and UV divergences,
but also so-called rapidity divergences. These must be regularized using a dedicated rapidity regulator,
and after removing the regulator this gives rise to the rapidity renormalization scale $\nu$.
Several such regulators are known in the literature \cite{Collins:1981uk,Collins:1350496,Becher:2010tm,Becher:2011dz,GarciaEchevarria:2011rb,Chiu:2011qc,Chiu:2012ir,Li:2016axz,Rothstein:2016bsq,Ebert:2018gsn}, leading to several equivalent schemes for defining TMD beam and soft functions.
It is also common to combine beam and soft functions into a $\nu$-independent TMDPDF as
\begin{equation}
 \tilde f_i(x,b_T,\mu,\zeta_i) = \tilde B_i(x, b_T, \mu, \nu/\sqrt\zeta_i) \sqrt{\tilde S(b_T,\mu,\nu)}
\,,\end{equation}
where $\zeta_i \propto \omega_i^2$ is known as the Collins-Soper scale \cite{Collins:1981va,Collins:1981uk}.

The TMD beam and soft functions appearing in \eq{qt_fact} are known at NNLO in various regulators \cite{Catani:2011kr,Catani:2012qa,Gehrmann:2012ze,Gehrmann:2014yya,Echevarria:2016scs,Gutierrez-Reyes:2019rug,Luo:2019hmp,Luo:2019bmw,Echevarria:2015byo,Luebbert:2016itl}.
The quark beam function and the soft function are also known at N$^3$LO \cite{Li:2016ctv,Luo:2019szz} using the exponential regulator of \refcite{Li:2016axz}.

An important remark is in order concerning differences between quark- and gluon-induced processes.
In the quark case, \eq{qt_fact} exactly applies, while the gluon beam functions can also depend on the gluon helicity due the vectorial nature of $\bt$.
As first pointed out in \refcite{Catani:2010pd}, the gluon beam function can be decomposed into a polarization-independent piece $B_1$ and a polarization-dependent piece $B_2$ as
\begin{align} \label{eq:Bg_decomposition}
 \tilde B^{\rho\lambda}_g(x,b_T,\mu,\nu)
 = \frac{g_\perp^{\rho\lambda}}{2} \tilde B_1(x,b_T,\mu,\nu)
 + \biggl(\frac{g_\perp^{\rho\lambda}}{2} - \frac{b_\perp^\rho b_\perp^\lambda}{b_\perp^2} \biggr) \tilde B_2(x,b_T,\mu,\nu)
\,.\end{align}
Here, $b_\perp^\mu$ is a Minkowski four vector with $b_\perp^2 = -b_T^2$, and $g_\perp^{\rho\lambda}$ is the transverse component of the metric tensor.
In this case, the hard function in \eq{qt_fact} also depends on the helicities of the colliding gluons.

We will only focus on the production of scalar particles such as a Higgs boson, where the hard function has the trivial helicity structure
\begin{align}
 H_{gg}^{\rho\lambda\rho'\lambda'}(Q,\mu) = H_{gg}(Q,\mu) g_\perp^{\rho\rho'} g_\perp^{\lambda\lambda'}
\,.\end{align}
Thus, the only combination that enters the factorized cross section in this case is
\begin{align} \label{eq:HBB_gg}
 H_{gg\,\rho\lambda\rho'\lambda'} \tilde B^{\rho\lambda}_g \tilde B^{\rho'\lambda'}_g
 = H_{gg} \tilde B^{\rho\lambda}_g \tilde B_{g\,\rho\lambda}
 = \frac12 H_{gg} \bigl[ \tilde B_1 \tilde B_1 + \tilde B_2 \tilde B_2 \bigr]
\,,\end{align}
where we suppress the arguments of all functions for brevity.
Since $\tilde B_2$ describes a spin flip of the incoming gluon, it vanishes at tree level, and thus the $\tilde B_2 \tilde B_2$ term first contributes at $\cO(\as^2)$.
Thus, for a scalar process, the $\tilde B_2 \tilde B_2$ term does not show up in the strict $n$-collinear limit, which hence can be used to calculate $\tilde B_1$ in the same fashion as for the quark case.
Nevertheless, $\tilde B_2$ could be calculated with the same technique for a different process that induces a cross term $\tilde B_1 \tilde B_2$, for example the production of a pseudoscalar probe $h$.
We also note that since $\tilde B_2$ is already known at NNLO \cite{Gutierrez-Reyes:2019rug,Luo:2019bmw}, the $\tilde B_2 \tilde B_2$ term in \eq{HBB_gg} is already known at N$^3$LO.

\subsubsection[Calculation of \texorpdfstring{$q_T$}{qT}-dependent beam functions]
              {\boldmath Calculation of $q_T$-dependent beam functions}
\label{sec:qT_beam_funcs}

In our setup for the differential hadronic cross section, \eq{sigma_differential}, we measured the transverse momentum of $h$ indirectly through
\begin{align} \label{eq:qT_measure}
 x = 1 - \frac{k_T^2}{k^+ k^-} = 1 - \frac{k_T^2}{s \wa \wb}
\,,\end{align}
as by momentum conservation $k_T = q_T$.
Both are defined as the magnitude of a $d-2$-dimensional vector, with the associated solid angle already integrated over in the phase space measure.
The $q_T$ measurement can also be defined in different schemes to account for extending the transverse vector into $d-2$ dimensions, but the scheme dependence must cancel in the renormalized beam functions.
For a more detailed discussion, see e.g.~\refcite{Luebbert:2016itl}.

Using \eqs{I_bare}{qT_measure} together with the leading-power relation $Q^2 = z s$, we obtain the matching kernel of the beam function as
\begin{align} \label{eq:cI_qT_naive}
 \cI_{ij}^{\rm naive}(z,q_T,\eps) &
  = \int_0^1 \df x \, \int_0^\infty \df w_1 \, \df w_2 \, \delta[z-(1-\wa)]
   \delta\biggl(q_T^2 - \frac{1-x}{z} w_1 w_2 Q^2 \biggr)
   \nn\\&\qquad\times
   \strictlim\frac{\df\eta_{j\bar{i}} }{\df Q^2 \df w_1 \df w_2 \df x}
\,.\end{align}
Here, the superscript $^{\rm naive}$ indicates that this is not yet the final result for the bare matching kernel, as it requires further manipulation.
First, we note that \eq{cI_qT_naive} contains divergences as $x\to1$ or $z\to1$ that are not regulated by dimensional regularization, and are a manifestation of the aforementioned rapidity divergences.
In our setup, we must regulate these with a regulator that acts only on the total radiation momentum $k^\mu$, but not on individual emissions. The only such regulator known in the literature is the exponential regulator of \refcite{Li:2016axz}, where one inserts a factor $\exp[2 \tau e^{-\gamma_E} k_i^0]$ into the phase of each real emission $k_i$. Inserting this regulator into \eq{cI_qT_naive} and solving the $\delta$ functions, we obtain
\begin{align} \label{eq:cI_qT_naive_2}
 \cI_{ij}^{\rm naive}(z,q_T,\eps,\tau/\omega) &
 = \lim_{\substack{\tau\to0\\\eps\to0}} \int_0^1 \df x \, \frac{1}{(1-x) (1-z)}
   \exp\biggl[-\tau e^{-\gamma_E} \frac{q_T^2}{\omega} \frac{z}{(1-z)(1-x)} \biggr]
   \nn\\&\hspace{1.2cm}\times
  \strictlim\frac{\df\eta_{j \bar{i}} }{\df Q^2 \df w_1 \df w_2 \df x}
   \bigg|_{ w_2 = \frac{q_T^2}{Q^2} \frac{z}{(1-x)(1-z)},\,w_1=1-z}
\,,\end{align}
where we defined the so-called label momentum of the beam function as $\omega = Q e^Y$.
In \eq{cI_qT_naive_2}, all divergences as $x\to1$ and $z\to1$ are manifestly regulated by the exponential, and any leftover divergences are regulated by dimensional regularization.
As indicated, the limit $\tau\to0$ should be taken before the limit $\eps\to0$.

To proceed, we Fourier transform to the conjugate $\bt$ space, which trades convolutions in $\qt$ for simple products in Fourier space. In $d-2$ dimensions, the Fourier transform reads
\begin{align} \label{eq:cI_qT_bare_3}
 \tilde \cI_{ij}^{\rm naive}(z,b_T,\eps,\tau/\omega)
 = \int\df^{d-2}\qt \, e^{-\img \bt \cdot \qt} \cI_{ij}^{\rm naive}(z,q_T,\eps,\tau/\omega)
\,.\end{align}
We can then apply the zero-bin subtraction to subtract overlap with the soft function, see \sec{zero_bin}, which for the exponential regulator is equivalent to dividing by the soft function in Fourier space~\cite{Luo:2019hmp}.
This in turn completes the manipulations that forced us the introduce the label $^{\rm naive}$ before their execution.
Next, we can apply the usual UV and IR counterterms to obtain the renormalized matching kernel as
\begin{equation} \label{eq:I_ren_qT}
 \tilde \cI_{ij}(x,b_T,\mu,\nu/\omega)
 = \sum_{j'} \Gamma_{jj'}(z, \eps) \otimes_z \tilde Z_B^i(\eps,\mu,\nu/\omega) \hat Z_{\as}(\mu,\eps)
   \frac{\tilde \cI_{ij'}^{\rm naive}(z,b_T,\eps,\tau/\omega)}{\tilde S(b_T,\eps,\tau)}
\,,\end{equation}
where following \refcite{Li:2016ctv} we identify the rapidity renormalization scale as $\nu = 1/\tau$.
The all-order structure of the beam function counter term $\tilde Z_B^i$ can be predicted from the beam function RGE, which we show in detail in \refcite{Ebert:2020yqt}.

We have implemented the described procedure at NLO through $\cO(\eps^4)$ and at NNLO through $\cO(\eps^2)$, as required for the calculation of the three-loop beam function.
We use the collinear limit of the cross sections for Higgs and Drell-Yan production to extract the gluon and quark beam functions, respectively. Since the bare soft function required in \eq{I_ren_qT} has not been published beyond NLO, we have similarly calculated it from the soft limit of the cross section.
Our bare results agree with those of \refscite{Luo:2019hmp,Luo:2019bmw}, and the renormalized beam functions also agree with \refcite{Luebbert:2016itl}.

\section{Collinear expansion of rapidity distributions}
\label{sec:rapidity}
Computing analytic coefficient functions at high orders is a complicated task, and finding suitable approximations can be vital.
Here we demonstrate that our expansion techniques have the potential to approximate the rapidity spectrum of color neutral hard probes.
We perform a computation of the first two terms in the collinear expansion of the rapidity distribution of the Higgs boson produced via gluon fusion at NNLO.
This application also demonstrates that our technique allows one to relatively easily obtain predictions beyond leading power of the kinematic expansion.

The required partonic matrix elements were calculated exactly in \refcite{Gehrmann_De_Ridder_2012}, and the differential distribution was obtained for example in \refcite{Anastasiou2005}.
Currently, this observable is known at N$^3$LO computed via a threshold expansion~\cite{Dulat:2018bfe} and via an approximate differential computation~\cite{Cieri:2018oms}.
The exact computation of the partonic coefficient function is still elusive due to its extreme difficulty, and a collinear expansion of the same could provide a useful ingredient in future phenomenological studies.

We integrate out the degrees of freedom of the top quark and work in an effective theory that couples the Higgs boson directly to gluons~\cite{Inami1983,Shifman1978,Spiridonov:1988md,Wilczek1977,Chetyrkin:1997un,Schroder:2005hy,Chetyrkin:2005ia,Kramer:1996iq,Kniehl:2006bg}. 
We generate all required Feynman diagrams with QGRAF~\cite{Nogueira_1993} and perform their collinear expansion up to second term as illustrated in \sec{collinear_expansion}.
We then employ IBP identities~\cite{Chetyrkin:1981qh,Tkachov:1981wb} to reduce the expanded diagrams to master integrals, which we then compute using the framework of reverse unitarity~\cite{Anastasiou2003,Anastasiou:2002qz,Anastasiou:2003yy,Anastasiou2005,Anastasiou2004a} and the method of differential equations~\cite{Kotikov:1990kg,Kotikov:1991hm,Kotikov:1991pm,Henn:2013pwa,Gehrmann:1999as}.
With this we obtain the bare partonic coefficient function
\beq
\frac{\df\eta_{ij}}{\df Q^2 \df w_1 \df w_2 \df x}\Big|_{w_2 \sim \lambda^2}
\eeq
expanded up to the second term in $\omega_2$. 
Next, we perform a variable transformation from $(\omega_1,\omega_2)\rightarrow (z_1,z_2)$ via \eq{xidef} and \eqref{eq:partcoef_special}, and replace the variable $x$ by $\xi$ via
\beq
x=\frac{\xi (z_1+z_2){}^2}{\bigl[\xi z_1 (1-z_2) + z_2 (1 + z_1)\bigr] \bigl[\xi z_2 (1-z_1) + z_1(1+z_2)\bigr]}.
\eeq
The expansion in  $\omega_2$ is comparable to an expansion in $\bar z_2=1-z_2$, as can be seen by applying the rescaling transformation of \eq{modes}. We find
\beq
\bar z_1=1-z_1=\omega_1 +\mathcal{O}(\lambda^2),\hspace{1cm} \bar z_2=\omega_2 \frac{2-\omega_1(1+x)}{2(1-\omega_1)}+\mathcal{O}(\lambda^4).
\eeq
Introducing the variable $\xi$ has the advantage that its integration domain is independent from $z_1$ and $z_2$ and ranges from 0 to 1.
Next, we expand the partonic coefficient function after this change of variables up the second power in $\bar z_2$ and integrate over $\xi$.
The result is an approximation for the partonic coefficient function of \eq{partcoef_special} with the observable integrated out.
We perform UV renormalisation and combine our partonic matrix elements with collinear counter terms in order to obtain a finite partonic coefficient function through NNLO. 

Obtaining the equivalent expansion in $\bar z_1$ can easily be done by simply relabelling the variables, $\bar z_1\leftrightarrow \bar z_2$.
With this we obtain the following approximation for the full renormalized partonic coefficient function,
\begin{align} \label{eq:approxeta}
 \frac{\df\eta^{R,\,\text{approx.}}_{ij}(z_1,z_2) }{\df Q^2 \df Y} &
=\frac{\df\eta^{R}_{ij}}{\df Q^2 \df Y} \Big|_{\bar z_2 \sim \lambda^2}
+\frac{\df\eta^{R}_{ij}}{\df Q^2 \df Y} \Big|_{\bar z_1 \sim \lambda^2}
-\frac{\df\eta^{R}_{ij}}{\df Q^2 \df Y} \Big|_{\bar z_{1,2} \sim \lambda^2}
\quad
 + \cO(\lambda^2)
\,.\end{align}
The last term in the above equation removes the overlap in the two expansions. 
The hadronic cross section expanded to this order is then obtained by inserting \eq{approxeta} into \eq{sigma_hadr_finite},
\beq \label{eq:approx_sigma}
\frac{\df \sigma(x_1^B,x_2^B)}{ \df Q^2 \df Y} =
\tau \sigma_0 \sum_{i,j} f_i^R(x_1^B) \otimes_{x_1^B} \ \frac{\df\eta^{R,\,\text{approx.}}_{ij}(x_1^B,x_2^B) }{\df Q^2 \df Y}  \otimes_{x_2^B} f_j^R(x_2^B),
\eeq
Note that the leading-power limit of \eqs{approxeta}{approx_sigma} precisely correspond to the leading-power generalized threshold factorization theorem of \refcite{Lustermans:2019cau}, cf.~their eqs.~(17) and (18).

\begin{figure*}
\centering
\includegraphics[width=0.95\textwidth]{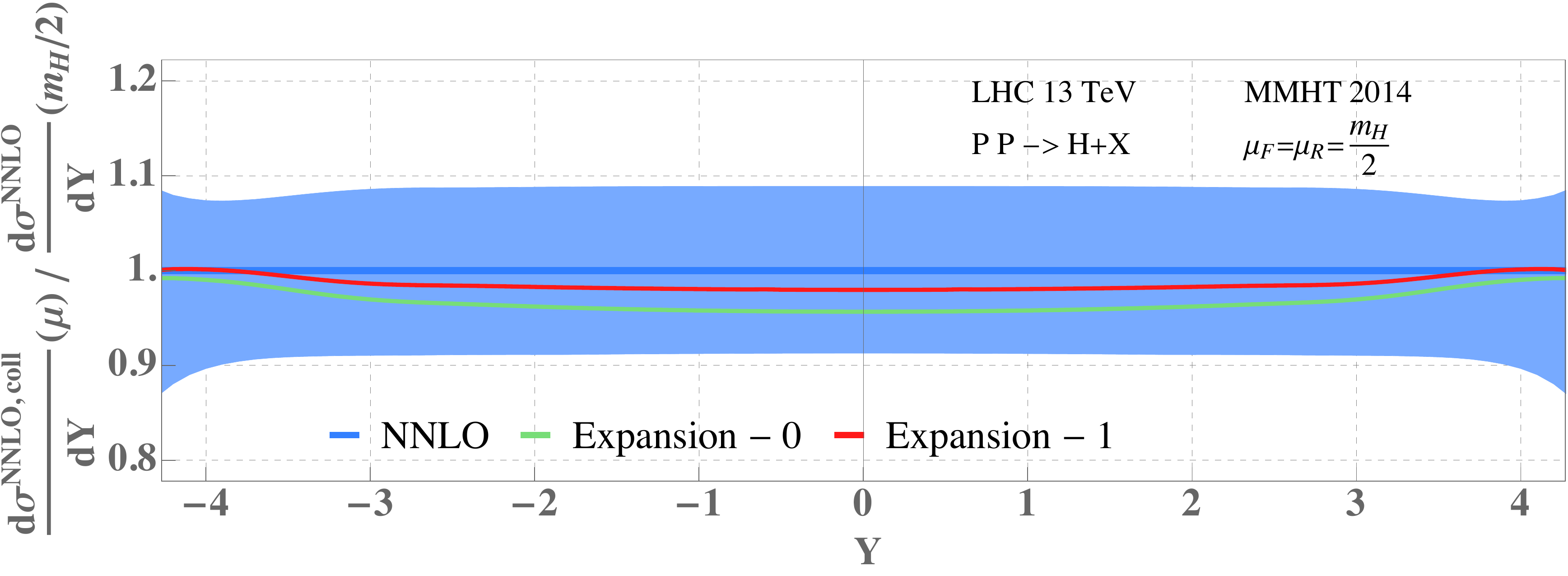}
\caption{Comparison of the Higgs boson rapidity distribution in gluon fusion obtained with a collinear expansion, normalized to the exact results of \refcite{Dulat:2017aa}.}
\label{fig:rap}
\end{figure*}

We have implemented the approximate partonic coefficient function in \eq{sigma_hadr_finite} in a private C++ code. Note that we only expanded the NNLO correction to the partonic coefficient coefficient, but keep the lower orders exact.
To illustrate our results numerically, we evaluate \eq{approx_sigma} for the LHC with a center-of-mass energy of $13$ TeV using the MMHT14 parton distribution functions~\cite{Harland-Lang:2014zoa}.
Figure~\ref{fig:rap} shows the rapidity distribution obtained with this collinear expansion normalized to the exact results obtained from \refcite{Dulat:2017aa}.
The green line shows our result using only the first term in the collinear expansion, while the red line shows the result including also the second term in the collinear expansion.
The blue band in the figure represents the variation of the cross section under a variation of the factorization and renormalisation scale by factor of two around their central values $\mu_F=\mu_R=m_H/2$.
We observe that the collinear expansion approximates the shape of the rapidity spectrum quite well, in particular towards large values of $|Y|$.
This is kinematically expected, as large rapidities enforce all final-state radiation to be collinear to the corresponding incoming parton, such that the collinear expansion is in fact the correct kinematic limit, see also~\refcite{Lustermans:2019cau}.
In addition, including the second-order term in the expansion clearly improves the results, illustrating that the collinear expansion indeed can be used to produce systematically improvable approximations of key collider physics observables.

The computation of the expanded partonic coefficient functions was greatly simplified compared to the computation of the exact result obtained e.g.\ in \refcite{Dulat:2017aa}.
Explicitly, the complexity of the analytic formulae is greatly reduced, and the function space required to express the coefficient function is much simpler.
We expect that a similarly drastic simplification will also occur when applying our method at N$^3$LO, which is a natural application of this research.

\section{Conclusions}
\label{sec:conclusions_collexp}

We have developed a method to efficiently expand differential cross sections for the production of colorless final states in hadron collisions around the particular kinematic limit that all hadronic final-state radiation becomes collinear to one of the colliding hadrons. This yields a generalized power expansion in a power counting parameter $\lambda$ characterizing this limit.

A key feature of our method is that the expansion is systematically improvable, as it allows to compute to arbitrary order in the power counting parameter $\lambda$.
Furthermore, $\lambda$ is treated as a purely symbolic power counting parameter agnostic of the actual observable. This greatly simplifies the expansion, as it can be carried out at the integrand level, i.e. before any phase space or loop integrations are carried out.
Subsequently, carrying out phase space and loop integrals is greatly facilitated as integrands become simpler as a result of the expansion.
Moreover, the expanded integrands have again a diagrammatic nature very much like the original Feynman integrands they were derived from.
This observation makes it manifest that widely used and powerful loop integration techniques like IBP relations and the method of differential equations are applicable to the coefficients of the collinear expansion.
We also stress that the basic functions (the so-called master integrals) required in the computation of higher orders in the expansion are already obtained in the lowest few nontrivial orders of the expansion.

Our method also sheds light on the connection between the collinear limit of hadronic cross sections and factorization theorems derived in SCET. The latter include so-called beam functions, universal quantities defined as hadronic matrix elements of collinear fields in SCET, which can be related to standard light-cone PDFs through convolutions with perturbative matching kernels. We have shown that these kernels are precisely given by the first term in a strict collinear expansion of hadronic cross sections.
As a first application of this, we reproduced the matching kernels for the $N$-jettiness and $q_T$ beam functions at NNLO from a collinear expansion of the NNLO cross sections for the Drell-Yan process and for Higgs boson production in gluon fusion.

As another application of the collinear expansion, we have demonstrated its usefulness to efficiently calculate approximate hadron collider cross sections.
By combining the collinear expansion with the limit where one partonic momentum fraction becomes equal to its Born value, $x_i\to x_i^B$, 
we obtained the first two terms in the collinear expansion of the rapidity distribution of a Higgs boson produced in gluon fusion through NNLO in QCD perturbation theory.
This example illustrates not only that key collider observables can be approximated with high accuracy using our technique, but also that results beyond the leading power can be easily obtained.

In summary, the method of collinear expansions is a great tool to study the infrared limit of QCD. At leading power in the collinear expansion, it provides access to the universal beam functions governing the collinear limit, which we employ to calculate the $\Tau_N$ and $q_T$ beam functions at N$^3$LO in two companion papers~\cite{Ebert:2020unb,Ebert:2020yqt}.
We also believe that the collinear expansions will similarly shed light on the universal structure of hadron collision processes beyond the leading power. Finally, it provides a powerful tool to achieve cutting-edge phenomenological predictions at very high orders in perturbation theory.

\chapter{Fermionic Glauber Operators and Quark Reggeization}\label{sec:qregge}

\section{Introduction} \label{sec:introqregge}

The study of limits of amplitudes and cross sections plays an important role in our understanding of gauge theories by providing constraints on higher order calculations, as well as a glimpse at the all orders structure of the theory. One limit that has been intensely studied since the early days of field theory, both in QED~\cite{Gell-Mann:1964aya,Mandelstam:1965zz,McCoy:1976ff,Grisaru:1974cf} and QCD~\cite{Fadin:1975cb,Kuraev:1976ge,Lipatov:1976zz,Kuraev:1977fs,Balitsky:1978ic,Lipatov:1985uk,Lipatov:1995pn}, is the Regge or forward limit, $|t| \ll s$. The simplicity of this limit lead to the discovery of integrability in QCD~\cite{Lipatov:1993yb,Faddeev:1994zg}, and allows for an understanding at finite coupling in $\cN=4$ super Yang-Mills theory~\cite{Bartels:2014mka,Basso:2014pla,Sprenger:2016jtx}. In this limit large logarithms, $\log(s/|t|)$, appear in the perturbative expansion at weak coupling, and their resummation dresses the $t$-channel propagator, leading to an amplitude that behaves as $(s/|t|)^{\omega}$, where $\omega$ is the Regge trajectory. This behavior is referred to as Reggeization, and directly predicts terms in the higher order perturbative expansion of amplitudes, placing important constraints on their structure (see e.g.~\cite{Dixon:2011pw,Dixon:2014iba,Dixon:2015iva,Caron-Huot:2016owq,Dixon:2016nkn} for applications). The Regge trajectory for the gluon is known to two loops in QCD~\cite{Fadin:1995xg,Fadin:1996tb,Korchemskaya:1996je,Blumlein:1998ib,Fadin:1998py}, and to three loops in non-planar $\cN=4$ \cite{Henn:2016jdu}. Recently there has been progress in understanding the breaking of naive Reggeization, and Regge-cut contributions, leading to a more complete picture of forward scattering at higher loops~\cite{Bret:2011xm,DelDuca:2011ae,Caron-Huot:2013fea,Caron-Huot:2016tzz,Caron-Huot:2017fxr}. At the cross section level the resummation is described by the Balitsky--Fadin--Kuraev--Lipatov (BFKL) equation~\cite{Kuraev:1977fs,Balitsky:1978ic}. 

A powerful approach for studying the limits of gauge theories is the use of effective field theory (EFT) techniques. The framework of soft collinear effective theory (SCET)~\cite{Bauer:2000ew, Bauer:2000yr, Bauer:2001ct, Bauer:2001yt} has been widely used to study the soft and collinear limits of QCD, including power suppressed contributions in these limits (see e.g.~\cite{Larkoski:2014bxa,Moult:2016fqy,Kolodrubetz:2016uim,Moult:2017rpl,Feige:2017zci}). Recently an EFT for forward scattering~\cite{Rothstein:2016bsq} was developed in the framework of SCET, providing a systematic way of analyzing the Regge limit at higher perturbative orders and at higher powers in the expansion in $|t|/s$. In~\cite{Rothstein:2016bsq}, the leading power operators that describe the exchange of $t$-channel Glauber gluons were derived, and it was shown that their rapidity renormalization~\cite{Chiu:2012ir,Chiu:2011qc} gives rise to amplitude level Reggeization and the cross section level BFKL equation. For other approaches to studying the subleading power corrections in the Regge limit see~\cite{Amati:1987wq,Amati:1987uf,Amati:1990xe,Amati:1992zb,Amati:1993tb,Akhoury:2013yua,Luna:2016idw}.

In this paper we apply the EFT for forward scattering to the Reggeization of the quark. This is interesting for a number of reasons.
First, quark exchange in the $t$-channel provides the leading contribution for certain flavor configurations in $2\to 2$ forward scattering in QCD, such as $q \bar{q} \to gg$ and $q \bar{q} \to \gamma \gamma$, and is thus important for understanding the behavior of such amplitudes.
Second, the Reggeization of the quark is power suppressed relative to that of the gluon, and therefore provides a simple case for studying the structure of SCET at subleading power in the Regge limit. Third, the application to quark Reggeization further develops the operator based framework, which together with~\cite{Rothstein:2016bsq} provides a description of the Regge limit for both quark and gluon exchanges which seamlessly interfaces with the standard SCET for the study of hard scattering.

The study of the Reggeization of the quark has a long history. In QED, the photon does not Reggeize due to the abelian nature of the theory, but the electron does, providing the first field theoretic derivation of Regge phenomenon~\cite{Gell-Mann:1964aya,Mandelstam:1965zz,McCoy:1976ff,Grisaru:1974cf}. The BFKL equation for $e^+e^-\to \gamma \gamma$ has also been studied in QED~\cite{Sen:1982xv}. In QCD, the Reggeization of the quark has received less attention since it is at subleading power compared to the Reggeization of the gluon. It was first studied in~\cite{Fadin:1976nw,Fadin:1977jr}, and Reggeization was proven to leading logarithmic (LL) order in~\cite{Bogdan:2006af}. Under the assumption of Reggeization, the two-loop Regge trajectory for the quark was derived in~\cite{Bogdan:2002sr} from the next-to-next-to-leading order $2\to2$ scattering amplitudes in QCD~\cite{Anastasiou:2000kg,Anastasiou:2000ue,Anastasiou:2001sv,Glover:2001af,Bern:2002tk}. Interestingly, to this order it is the same as the Regge trajectory of the gluon, up to so-called Casimir scaling, i.e. replacing $C_A \to C_F$.

The emphasis of this paper is the development of the EFT framework for forward scattering, with the hope of facilitating progress in understanding the structure of the Regge limit of QCD. 
We derive the operators describing the $t$-channel exchange of a Glauber quark in the Regge limit. These operators are fixed by the symmetries of the effective theory, constraints from power and mass dimension counting, and explicit matching calculations. They describe certain soft and collinear gluon radiation to all orders, and have not previously appeared in the literature. For a single emission, they reduce to the vertex of Fadin and Sherman~\cite{Fadin:1976nw,Fadin:1977jr}, which is the analogue of the Lipatov vertex~\cite{Kuraev:1976ge} for the case of a Reggeized quark. 
As a demonstration of our framework, we verify explicitly at one-loop that the rapidity renormalization of our potential operators leads to the Reggeization of the quark at the amplitude level and to the BFKL equation at the cross section level, thus providing another LL proof of these results but in the modern language of renormalization. We also show that it is simple to derive results for amplitudes in the $\bar 6$ and $15$ color channels by considering the simultaneous exchange of a Glauber quark and a Glauber gluon.

An outline of this paper is as follows. In \Sec{sec:review} we briefly review the formulation of SCET with Glauber gluon operators from~\cite{Rothstein:2016bsq}.  In \Sec{sec:ferm_glaub} we derive the structure of the fermionic Glauber operators. We consider Glauber quark exchanges between two collinear particles as well as between a collinear and a soft particle, and discuss their power counting. We also give the relevant Feynman rules.
In \Sec{sec:tree_level} we perform a tree level matching calculation onto the operators, which is sufficient to fix their precise form to all orders in $\alpha_s$. In \Sec{sec:quark_reggeize} we derive the one-loop Reggeization of the quark using the rapidity renormalization of the operators. We also show that rapidity finite contributions arising from box graphs with both a Glauber quark and a Glauber gluon reproduce known results in the $\bar 6$ and $15$ channel. In \Sec{sec:quark_BFKL} we derive the BFKL equation for $q \bar{q} \to \gamma \gamma$, and show that it is equivalent to the standard BFKL equation up to Casimir scaling. We conclude and discuss future directions in \Sec{sec:conc}.

\section{SCET with Glauber Operators}\label{sec:review}
In this section we briefly review the structure of SCET with Glauber operators, following~\cite{Rothstein:2016bsq}. This also allows us to define the notation used throughout the paper. We will gloss over many subtleties in the construction of the effective theory, and refer the interested reader to~\cite{Rothstein:2016bsq} for a more detailed discussion.   

SCET is an effective theory of QCD that describes the interactions of collinear and soft particles~\cite{Bauer:2000ew, Bauer:2000yr, Bauer:2001ct, Bauer:2001yt, Bauer:2002nz}. Let us focus on the single lightlike direction relevant for 2 to 2 forward scattering (multiple lightlike directions are considered in~\cite{Rothstein:2016bsq}). We define two reference vectors $n^\mu$ and $\bn^\mu$ 
such that $n^2 = \bn^2 = 0$ and $n\sdt\bn = 2$. Any momentum $p$ can then be written as
\begin{equation} \label{eq:lightcone_dec_qr}
p^\mu = \bn\sdt p\,\frac{n^\mu}{2} + n\sdt p\,\frac{\bn^\mu}{2} + p^\mu_{\perp}\
\,.\end{equation}
A particle is referred to as ``$n$-collinear'' if it has momentum $p$ close to the $\vec{n}$ direction, or more precisely, if the components of its momentum scale as $(n\!\cdot\! p, \bn \!\cdot\! p, p_{\perp}) \sim 
(\la^2,1,\la)$. Here $\la \ll 1$ is a formal power counting parameter, which is determined by the scales defining the measurement or kinematic limits. We will write the SCET fields for $n$-collinear quarks and gluons, as $\xi_{n}(x)$ and $A_{n}(x)$. In addition to describing collinear particles, SCET also describes soft particles, which have momenta that scale as $(\lambda,\lambda, \lambda)$, and are described in the EFT by separate quark and gluon fields, $q_{s}(x)$ and $A_{s}(x)$. This theory is sometimes called SCET$_\text{II}$~\cite{Bauer:2002aj}.

The SCET Lagrangian is expanded as
\begin{align} \label{eq:SCETLagExpand_qr}
\cL_{\text{SCET}}=\cL_\hard+\cL_\dyn= \cL^{(0)} + \cL_G^{(0)} +\sum_{i\geq0} \cL_\hard^{(i)}+\sum_{i\geq1} \cL^{(i)}\,,
\end{align}
with each term having a definite power counting, ${\cal O}(\lambda^i)$, denoted by the superscript. As written, the SCET Lagrangian is divided into three different contributions. The $ \cL_\hard^{(i)}$ contain hard scattering operators, and are derived by a matching calculation, and are process dependent. The $\cL^{(i)}$ describe the long wavelength dynamics of soft and collinear modes in the effective theory, and are universal. The leading power Glauber Lagrangian $\cL_G^{(0)}$ describes interactions between soft and collinear modes in the form of potentials, which break factorization unless they can be shown to cancel. It is derived in~\cite{Rothstein:2016bsq} and discussed below.

Operators in SCET are formed from gauge invariant building blocks. The gauge invariant $n$-collinear quark and gluon fields are defined as
\begin{align} \label{eq:chiB_qr}
\chi_{{n}}(x) &= \Bigl[W_{n}^\dagger(x)\, \xi_{n}(x) \Bigr]
\,,\qquad 
\cB_{{n}\perp}^\mu(x)
= \frac{1}{g}\Bigl[ W_{n}^\dagger(x)\,i  D_{\perp}^\mu W_{n}(x)\Bigr]
 \,,
\end{align}
with analogous definitions for $\bn$-collinear fields.
The collinear Wilson line is given by
\begin{align}\label{eq:Wilsonline}
W_n =\left[  \sum\limits_{\text{perms}} \exp \left(  -\frac{g}{\bar \cP } \bar n \cdot A_n(x)  \right) \right]\,,
\end{align}
where $\cP$ is the so-called label operator, which picks out the large component of a given momentum. These operators involve non-local Wilson lines, but are local at the scale of the dynamics of the EFT. The gauge invariant soft fields are defined in a similar manner, with
\begin{align}\label{eq:gauge_soft}
\cB_{S\perp}^{\bar n \mu}=\frac{1}{g}[S_\bn^\dagger i D^\mu_{S\perp} S_{\bar n}]\,, \qquad \cB^{n\mu}_{S\perp}=\frac{1}{g} [S_n^\dagger i D^\mu_{S\perp} S_n]\,.
\end{align}
These operators involve Wilson lines of soft gluons, and are non-local at the soft scale.

The leading power Glauber Lagrangian in SCET$_\text{II}$~\cite{Rothstein:2016bsq} is given by
\begin{align}\label{eq:Glauber_Lagrangian}
\cL_G^{\text{II}(0)} 
&= e^{-ix\cdot \cP} \sum\limits_{n,\bar n} \sum\limits_{i,j=q,g}   \cO_n^{iB} \frac{1}{\cP_\perp^2} \cO_s^{BC}   \frac{1}{\cP_\perp^2} \cO_{\bar n}^{jC}   + e^{-ix\cdot \cP} \sum\limits_{n} \sum\limits_{i,j=q,g} \cO_n^{iB}   \frac{1}{\cP_\perp^2} \cO_s^{j_n B}\,,
\end{align}
which gives contributions that scale as $\cO(\lambda^0)$. Glauber modes are not dynamical in the EFT but are incorporated through $\frac{1}{\cP_\perp^2}$ potentials, which are instantaneous in the light cone directions and non-local in the $\perp$ direction. 
In Eq.~(\ref{eq:Glauber_Lagrangian}) the first term describes the scattering of $n$ and $\bar n$ collinear particles, while the second term describes the scattering of collinear particles with soft particles. This Lagrangian is exact and does not receive matching corrections in $\alpha_s$ since no hard interactions are being integrated out~\cite{Rothstein:2016bsq}. Moreover, iterated potentials are reproduced by time ordered products ($T$-products) in the effective theory.

Each term in Eq.~(\ref{eq:Glauber_Lagrangian}) is written in a factorized form with gauge invariant operators that sit at different rapidities. The $n$-collinear operators are given by
\begin{align}
\cO_n^{qB} = \bar \chi_n T^B \frac{\Sl{\bar n}}{2} \chi_n \,, \qquad \cO_n^{gB} = \frac{i}{2} f^{BCD} \cB^C_{n\perp \mu} \frac{\bar n}{2} \sdt (\cP+\cP^\dagger ) \cB^{D\mu}_{n\perp}\,,
\end{align}
with $\bar n$-collinear operators identical under the replacement $n\leftrightarrow \bar n$. The soft operators are given by
\begin{align}\label{eq:iain_ira_soft}
\cO_s^{BC}&=8\pi \alpha_s \bigg \{   \cP^\mu_\perp S_n^\dagger S_{\bar n} \cP_{\perp \mu} -\cP^\perp_\mu g\tilde \cB_{S\perp}^{n\mu} S_n^\dagger S_{\bar n} - S_n^\dagger S_{\bar n} g \tilde \cB_{S\perp}^{\bar n\mu} \cP^\perp _\mu -g \tilde\cB_{S\perp}^{n\mu} S_n^\dagger S_{\bar n} g\tilde\cB_{S\perp \mu}^{\bar n}      \nn \\
&\qquad  -\frac{n^\mu \bar n^\nu}{2}   S_n^\dagger ig\tilde G^{\mu \nu}_s S_{\bar n}    \bigg \} ^{BC}    \,,  \nn \\
\cO_s^{q_n B}&=8\pi \alpha_s \left\{  \bar \psi_S^n T^B \frac{\Sl n}{2}  \psi^n_S  \right \} \,, \nn\\
\cO_s^{g_n B}&= 8\pi \alpha_s \left \{   \frac{i}{2} f^{BCD} \cB^{nC}_{S\perp \mu} \frac{n}{2} \sdt (\cP +\cP^\dagger) \cB_{S\perp}^{nD\mu} \right \}\,.
\end{align}
In \Eq{eq:Glauber_Lagrangian}, the operator $\cO_s^{BC}$ connects two operators of different collinear sectors, and describes an arbitrary number of soft gluon emissions from the forward scattering. For zero emissions, it reduces to $8\pi \alpha_s \cP_\perp^2 \delta^{BC}$, which, together with the factors of $1/\cP_\perp^2$ in Eq.~(\ref{eq:Glauber_Lagrangian}), reproduces the expected $1/\cP_\perp^2$ tree level Glauber potential between two collinear partons.
For a single emission, it reduces to the Lipatov vertex~\cite{Kuraev:1976ge}. The Feynman rules for two soft emissions can be found in~\cite{Rothstein:2016bsq}. 

SCET with Glauber operators provides an operator based formalism for studying Glauber exchanges, and the Regge limit of QCD. For example, amplitude level Reggeization and the BFKL equation can be derived in the EFT through the renormalization group evolution of the operators~\cite{Rothstein:2016bsq}. The role of Glauber exchanges for factorization violation can also be explicitly computed within this framework, as discussed in~\cite{Rothstein:2016bsq}. For example, it was used in~\cite{Schwartz:2017nmr} to give direct computations of the collinear factorization violation in spacelike splitting functions that was first found and computed in~\cite{Catani:2011st}. Higher order leading power calculations in the framework used here were also made in \cite{Zhou:2017his}.

\section{Fermionic Glauber Operators}\label{sec:ferm_glaub}

Having reviewed SCET with Glauber gluon operators, in this section we extend the framework to include Glauber quark operators. In \Sec{sec:nn_ops} we describe the structure of the $n-\bn$ scattering operators, and in \Sec{sec:ns_ops} we describe the structure of the $n$-$s$ scattering operators. In \Sec{sec:regs} we discuss the regulators beyond dimensional regularization that are required for calculating with these operators at loop level. The precise structure of the operators presented in this section are derived from the symmetries of the effective theory, power counting and mass dimension constraints, and matching calculations, and are discussed in detail in \Sec{sec:tree_level}.

\subsection{$n$-$\bar n$ Operator Structure}\label{sec:nn_ops}

In this section we present the structure of the $n$-$\bar n$ scattering 
operators that describe the forward scattering of partons in the $n$ and $\bar n$ collinear sectors through the $t$-channel exchange of a Glauber quark. Analogous to the gluon case, in \Eq{eq:Glauber_Lagrangian}, we write the Lagrangian in the factorized form
\begin{align}\label{eq:Glauber_Lagrangian_cquark}
\cL^{\text{II}(1)} \supset e^{-ix\cdot \cP} \sum\limits_{n,\bar n} \bar{\cO}_{\bar n} \frac{1}{\Sl{\cP}_\perp}  \cO_s   \frac{1}{\Sl{\cP}_\perp} \cO_{n}  \,,
\end{align}
where $\cO_{\bar n}$ and $\cO_n$ describe fields in the collinear sectors, while $\cO_s$ describes fields in the soft sector, which sits at an intermediate rapidity between the two collinear sectors. The superscript $\text{II}$ denotes that we are working in SCET$_\text{II}$, and the superscript $(1)$ denotes that this will give contributions that scale as $\cO(\lambda)$. The factors of $\Sl{\cP}_\perp$ indicate that this is a non-local potential, and reflect the fermionic nature of the Glauber quark. We have kept the color and Dirac indices implicit.  To simplify the notation, we will often refer to the operator as
\begin{align} \label{eq:Glauber_nn}
\cO_{\bar n n}=\bar{\cO}_{\bar n} \frac{1}{\Sl{\cP}_\perp}  \cO_s   \frac{1}{\Sl{\cP}_\perp} \cO_{n}\,.
\end{align}
 
In \Eq{eq:Glauber_Lagrangian_cquark}, we have used the $\supset$ notation to emphasize that this is only the component of the subleading Lagrangian, ${\cal L}^{(1)}$, that describes the $t$-channel exchange of a Glauber quark. In particular, it does not describe $\cO(\lambda)$ power corrections to the $t$-channel exchange of a Glauber gluon, or of compound states. In general, there are other operators consistent with the symmetries of the effective theory as well as with power and mass dimension counting that can be written down. For example, in \Eq{eq:Glauber_Lagrangian_cquark}, one may replace $\frac{1}{\Sl{\cP}_\perp}$ with $\frac{1}{\cP_\perp^2}$, and appropriately modify the numerator with an additional derivative or gluon field to satisfy power and mass dimension counting. In \Sec{sec:tree_level} we will show that $\cO_{\bar n n}$ is sufficient for tree level matching, and therefore any additional operators have vanishing Wilson coefficients at this order. Moreover, we find that the one-loop renormalization of $\cO_{\bar n n}$ does not produce additional operators. Hence, \Eq{eq:Glauber_Lagrangian_cquark} is the complete basis of operators for describing quark Reggeization at LL order.  We have not ruled out the presence of additional fermionic exchange operators from one-loop matching, and we leave the study of the general operator basis to future work.

The exchange of a quark necessarily changes the fermion number in each collinear sector. In particular, there are $8$ scattering configurations: 
\begin{align}\label{eq:8configs}
\fd{2.8cm}{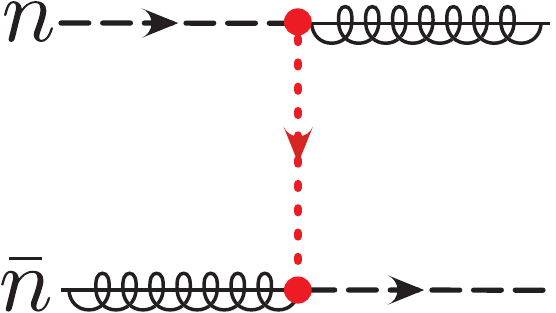} \qquad 
\fd{2.8cm}{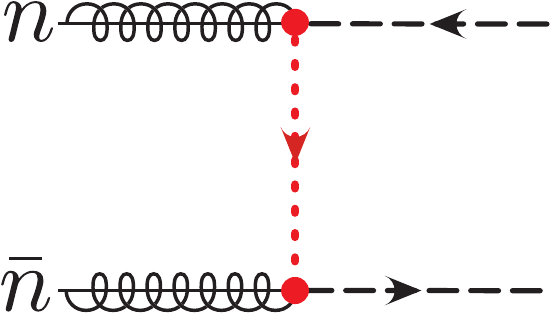}\qquad 
\fd{2.8cm}{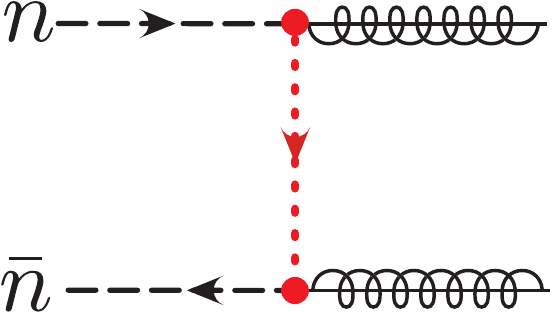} \qquad 
\fd{2.8cm}{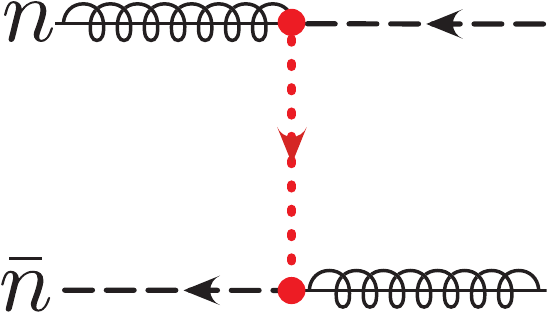} \nn \\[10pt]
\fd{2.8cm}{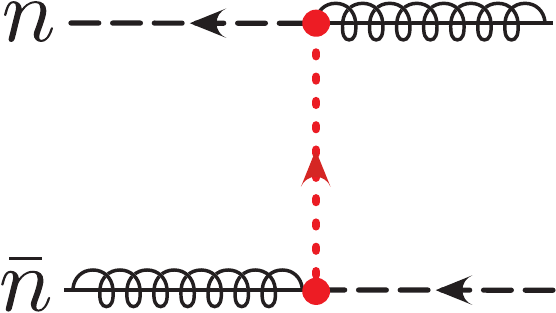} \qquad 
\fd{2.8cm}{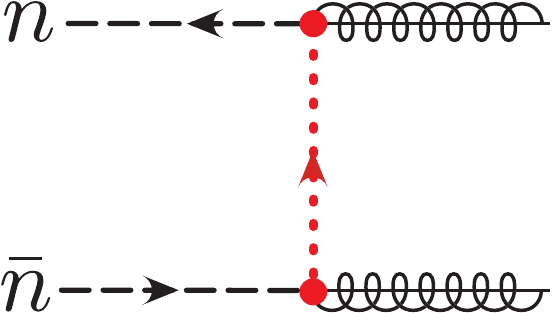}\qquad 
\fd{2.8cm}{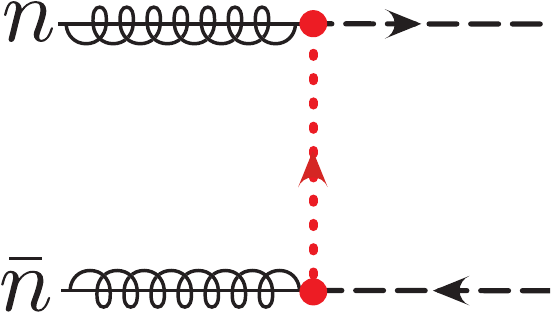} \qquad 
\fd{2.8cm}{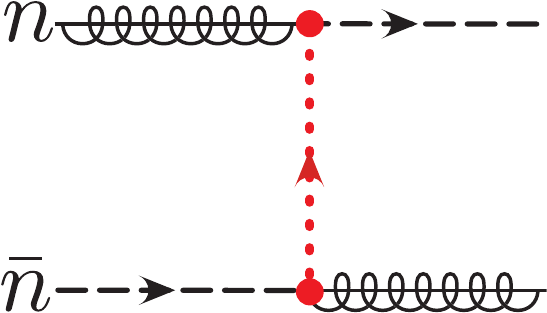}\,,
\end{align}
where the red dotted line denotes the Glauber quark.
Importantly, in \Eq{eq:Glauber_Lagrangian_cquark}, there is a sum over the directions $n$ and $\bar n$, as well as an implicit sum over the label momentum $\cP$. This implies that all $8$ possible collinear-collinear forward scattering configurations are generated from \Eq{eq:Glauber_Lagrangian_cquark}. For scattering configurations that preserve fermion number in each collinear sector, \Eq{eq:Glauber_Lagrangian_cquark} contributes through $T$-products, starting at $\cO(\lambda^2)$ with $T$-products of the above diagrams.

Changing the fermion number in each collinear sector implies that the collinear operators appearing in \Eq{eq:Glauber_nn} must contain a quark and a gluon at tree level. The least suppressed collinear operators with this property in SCET will involve just one quark and one gluon building block, since the collinear fields of \eq{chiB_qr} scale as $\cO(\lambda)$ so that any additional field would bring additional suppression. Moreover the forward scattering condition implies that any momentum structure between the quark and the gluon building block is fixed so that the two building blocks are connected by a simple product. The two objects should combine to spin-$1/2$, and the only non-vanishing object that does this while preserving the desired parity and chirality properties is $\gamma_{\perp}^\mu$. Therefore the collinear operators appearing in \Eq{eq:Glauber_nn} are given by
\begin{align}\label{eq:coll_ops}
\cO_{\bar n}=\Sl{\cB}_{\perp \bar n} \chi_{\bar n}\,, \qquad \cO_{n}=\Sl{\cB}_{\perp n} \chi_n\,.
\end{align}
(Additional factors of $\nslash$ or $\bnslash$ introduced here can be eliminated in the combination in \eq{Glauber_nn} by projection relations.) Having fixed the collinear operators, we can derive the constraints on the soft operator $\cO_s$: 
\begin{itemize}
	\item Counting mass dimensions, the two collinear operators are together dimension-5 while the two Glauber quark potentials subtract two. Since the Lagrangian has mass dimension four, the mid-rapidity soft operator must therefore have mass dimension one.
	\item The Lagrangian, the collinear operators, and the Glauber quark potentials are all RPI III invariant, and hence the soft operator must be RPI III invariant.
	\item The only operators to have mass dimension one that scale as $\cO(\lambda^0)$ are the label momentum operators $\bn \cdot \cP,\, n \cdot \cP$, which are neither RPI III invariant nor soft operators.\footnote{All derivative operators in the soft sector scale as $\cO(\lambda)$.} Therefore the soft operator must scale at least as $\cO(\lambda)$.
	\item Given the quantum numbers of the collinear operators, the soft operator must be a matrix in both color (in the fundamental representation) and in Dirac space.
	\item The soft operator must be soft gauge invariant.
\end{itemize}
These constraints imply that the most general mid-rapidity soft operator can be formed only by the gluon gauge invariant building blocks $\cB^{n}_{S \perp}$ and $\cB^{\bn}_{S \perp}$ of \eq{gauge_soft} (the soft quark operator $\psi_S$ is suppressed), gauge invariant products of soft Wilson lines $S^\dagger_n S_\bn$, and $\cP_\perp$, the only RPI III invariant soft momentum operator. In \Sec{sec:soft_match} we will fix the coefficients of the building blocks via a matching calculation and the resulting mid-rapidiy soft operator for the glauber quark 3 rapidity Lagrangian will be shown to be
\begin{equation}\label{eq:soft_operator_quark}
	\cO_s = -2\pi \alpha_s \left[ S_{\bar n}^\dagger S_{n} \Sl{\cP}_\perp + \Sl{\cP}_\perp S_{\bar n}^\dagger S_{n} - S_{\bar n}^\dagger S_n g\Sl{\cB}^{n}_{S\perp}-  g\Sl{\cB}^{\bar n}_{S\perp} S_{\bar n}^\dagger S_n \right]\,.
\end{equation}
Note the identity
\begin{align}
\cP^\mu_\perp S_\bn^\dagger S_{n} - S_\bn^\dagger S_{n} g \cB_{S\perp}^{n \mu} =S_\bn^\dagger S_{n} \cP^\mu_\perp -g\cB_{S\perp}^{\bn \mu} S_\bn^\dagger S_{n} \,,
\end{align}
which enables rewriting the soft operator in \eq{soft_operator_quark} in a more compact but less symmetric form. The power counting of the operators is $\cO_n \sim \cO_{\bar n} \sim \lambda^2$ and $\cO_s\sim \lambda$. Using the power counting formula of \cite{Rothstein:2016bsq} which subtracts 2 for a mixed $n$-$\bn$-soft operator, we then find that $\cO_{n\bar n}$ contributes at ${\cal O}(\lambda)$ as stated above.

The structure of the soft operator $\cO_s$ in \Eq{eq:soft_operator_quark} is significantly simpler than for the gluon case, $\cO_s^{BC}$ in \Eq{eq:iain_ira_soft}, due to the difference in mass dimension between fermionic and bosonic propagators.  In the gluon case, $\cO_s$ is exact: it is not corrected at higher orders in perturbation theory since Glauber exchange is instantaneous in both time and longitudinal position, and there is no hard contribution that is integrated out~\cite{Rothstein:2016bsq}. While we expect this to be the case here, due to the possibility of the additional operators mentioned below \Eq{eq:Glauber_nn} appearing at higher orders, and the behavior of power suppressed terms from loop diagrams, it is more complicated to show that this is true in this case, and we leave it to future work.

\begin{figure}
\begin{center}
\begin{align}
\fd{2.8cm}{figures/tree_level_1_low.pdf} &=  \bar u_{\bar n}(p_3) \Sl{\epsilon}_{\!\perp} \! (p_2) T^A \Bigg[  -i g^2 \frac{1}{\Sl{q}_\perp}  \Bigg]  \Sl{\epsilon}_{\!\perp} \!(p_4) T^B u_n(p_1) \nn \\[10pt]
\fd{2.8cm}{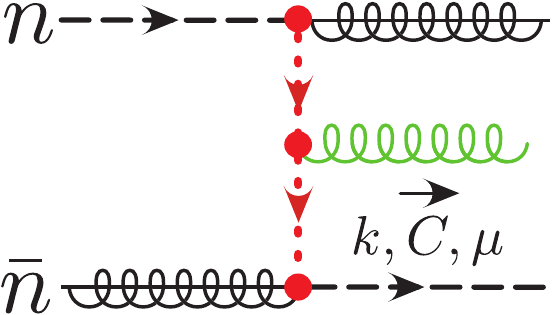} &=\bar u_{\bar n}(p_3) \Sl{\epsilon}_{\!\perp} \! (p_2) T^A \Bigg[   i g^3~ T^C \frac{1}{\Sl{q}_\perp}  \bigg( \gamma^\mu_\perp - \frac{(\Sl{q}_\perp+\Sl{k}_\perp) n^\mu}{n\cdot k}      \nn \\
&\ + \frac{\Sl{q}_\perp  \bar n^\mu}{\bar n \cdot k} \bigg) \frac{1}{\Sl{q}_\perp +  \Sl{k}_\perp}   \Bigg]  \Sl{\epsilon}_{\!\perp} \!(p_4) T^B u_n(p_1)  \nn \\[10pt]
\fd{3.6cm}{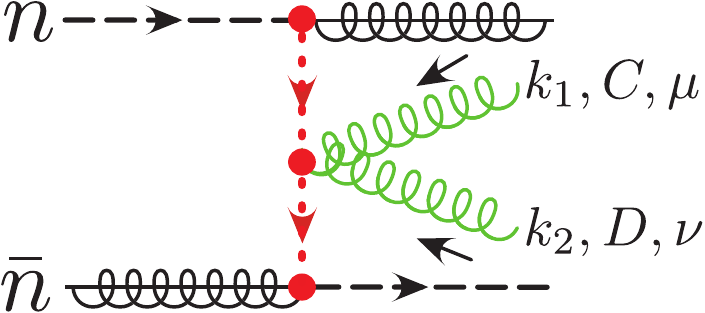}  &= \bar u_{\bar n}(p_3) \Sl{\epsilon}_{\!\perp} \! (p_2) T^A \Bigg[ -i g^4 T^C T^D \frac{1}{\Sl{q}_{\perp} +\Sl{k}_{1\perp} + \Sl{k}_{2\perp}} \bigg( \frac{ n^\nu \gamma^\mu_{\perp} }{ n \cdot k_2} 
-\frac{ \bn^\mu \gamma^\nu_{\perp}}{ \bn \cdot k_1}   \nn \\
&\  + \frac{(\Sl{q}_{\perp} +\Sl{k}_{1\perp} + \Sl{k}_{2\perp})  \ \bn^\mu   \bn^\nu  }{2\bn \cdot( k_1+ k_2) \bn \cdot k_1 } + \frac{\Sl{q}_\perp n^\mu  n^\nu }{2n \cdot( k_1+ k_2) n \cdot k_1}   \nn \\
&\ 
- \frac{( \Sl{k}_{1\perp } +  \Sl{q}_\perp) \bn^\mu n^\nu }{\bn\cdot k_1\,n \cdot k_2} \bigg) \frac{1}{\Sl{q}_\perp} + \Big\{ (C, \mu, k_1) \leftrightarrow (D, \nu, k_2) \Big\}  \Bigg]  \Sl{\epsilon}_{\!\perp} \!(p_4) T^B u_n(p_1) 
 \nn
\end{align}
\end{center}
\caption{Feynman rules for tree level $qg$ forward scattering with zero, one and two soft gluon emissions, generated by the soft operator $\cO_s$.  Soft emissions at higher orders in $\alpha_s$ are also produced by $\cO_s$.
}\label{fig:no_emission_feynrule}
\end{figure}

The soft operator $\cO_s$ describes the emission of soft gluons from the forward scattering to all orders in $\alpha_s$.
The Feynman rules for $qg$ forward scattering with zero, one and two soft gluon emissions are given in \Fig{fig:no_emission_feynrule}. The one emission Feynman rule gives the classic result of Fadin and Sherman \cite{Fadin:1976nw,Fadin:1977jr}, which we will refer to as the Fadin-Sherman vertex. The two emission Feynman rule has not, to our knowledge, appeared in the literature before. It will be required in our derivation of the quark Reggeization through rapidity renormalization (although only a particularly simple projection appears).

\subsection{$n$-$s$ Operator Structure}\label{sec:ns_ops}
In addition to the $n$-$\bar n$ scattering operators, the effective theory also includes operators that describe $n$-$s$ (and $\bn-s$) forward scattering.  We write the Lagrangian for soft collinear forward scattering as
\begin{align}\label{eq:Glauber_Lagrangian_squark}
\cL_G^{\text{II}(1/2)} \supset e^{-ix\cdot \cP} \sum\limits_{n}  \bar{\cO}_n \frac{1}{\Sl{\cP}_\perp}  \cO^{n}_s  + \bar{\cO}^{n}_s \frac{1}{\Sl{\cP}_\perp} \cO_n  \,.
\end{align}
Here the superscript $1/2$ indicates that this Lagrangian contribution scales as $\cO(\lambda^{1/2})$ relative to the leading power contribution. 
These operators play an important role in the rapidity renormalization, contributing through $T$-products in the effective theory. In particular, their contribution scales as $\cO(\lambda^{1/2}) \cdot \cO(\lambda^{1/2})= \cO(\lambda)$, which is at the same order as the $n$-$\bar n$ forward scattering operators. We will use the shorthand
\begin{align}\label{eq:Glauber_ns}
\cO_{ns}=\bar{\cO}_n \frac{1}{\Sl{\cP}_\perp}  \cO^{n}_s  
 \,.
\end{align}

As in \Eq{eq:Glauber_Lagrangian_cquark}, we have used the $\supset$ symbol in \Eq{eq:Glauber_Lagrangian_squark} to emphasize that this is not the complete Lagrangian at $\cO(\lambda^{1/2})$, and includes only the operators required for describing quark Reggeization at LL order

In \Eq{eq:Glauber_Lagrangian_squark}, the sum over the direction $n$, the implicit sum over the label momentum $\cP$, and the presence of both $\cO_{ns}$ and its hermitian conjugate generates all possible scattering configurations, namely:
\begin{align}
\fd{2.8cm}{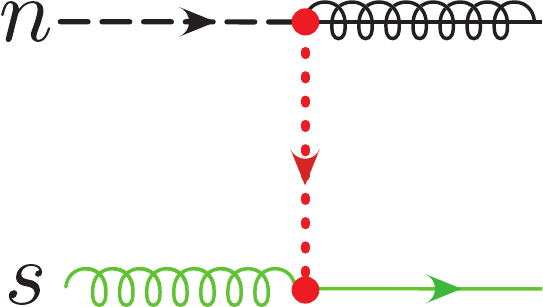} \qquad 
\fd{2.8cm}{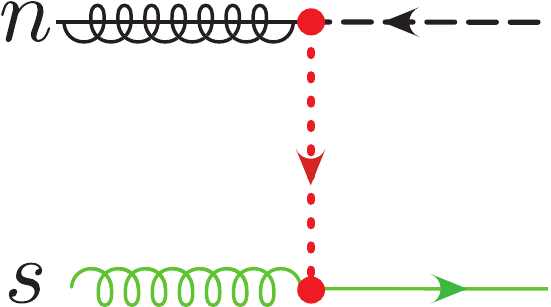}\qquad 
\fd{2.8cm}{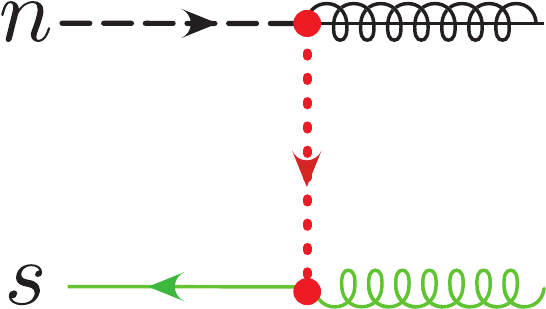} \qquad 
\fd{2.8cm}{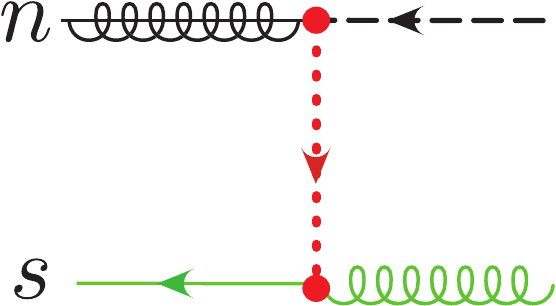}\nn 
\phantom{\,.} \\[10pt]
\fd{2.8cm}{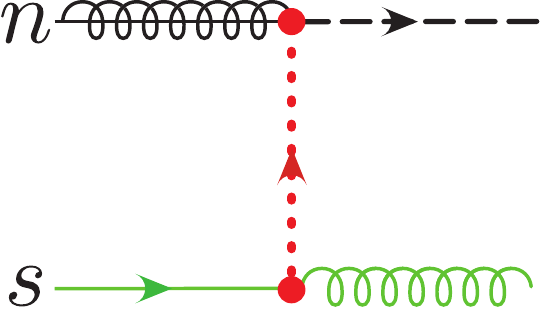} \qquad 
\fd{2.8cm}{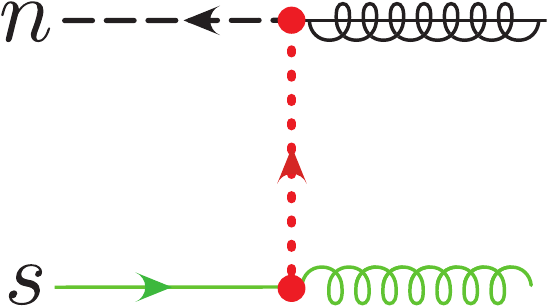}\qquad 
\fd{2.8cm}{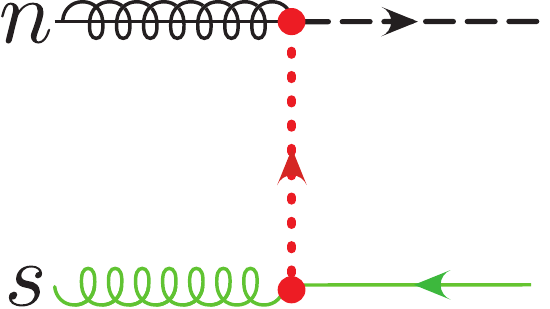} \qquad 
\fd{2.8cm}{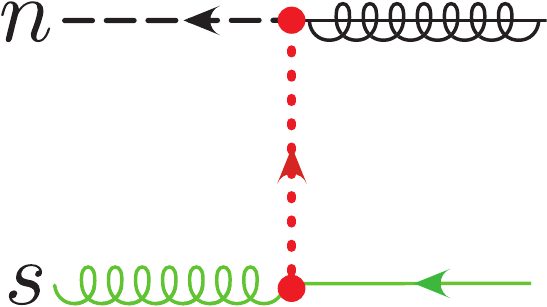}\,.
\end{align}

The $\cO_n$ operators in \Eq{eq:Glauber_Lagrangian_squark} are identical to those in \Eq{eq:coll_ops}. The $\cO_s^{n}$ operators have a similar structure, but we include a prefactor that arises from tree level matching:
\begin{align}\label{eq:n_s_ops}
\cO^{n}_s=-4\pi \alpha_s \Sl{\cB}_{\perp S}^n \psi^n_S\,, 
\qquad \bar{\cO}^n_s=-4\pi \alpha_s \bar \psi^n_S \Sl{\cB}_{\perp S}^n \,.
\end{align}
The power counting of the operators is $\cO_n \sim \cO_{\bar n} \sim \lambda^2$ and $\cO_s^n \sim\cO_s^{\bar n} \sim \lambda^{3/2}$. Using the power counting formula of \cite{Rothstein:2016bsq}, where we subtract 3 for a mixed $n$-soft or $\bn$-soft operator, we then find that $\cO_{ns}$ gives a contribution scaling as ${\cal O}(\lambda^{1/2})$, as stated.

\subsection{Regulators for Rapidity and Glauber Potential Singularities}\label{sec:regs}

As discussed extensively in \cite{Rothstein:2016bsq}, the Glauber Lagrangian requires both the regularization of rapidity divergences, as well as the regularization of divergences associated with Glauber exchanges. Here we use identical regulators to those defined in  \cite{Rothstein:2016bsq}.

Rapidity divergences are regulated using the $\eta$-regulator of \cite{Chiu:2012ir,Chiu:2011qc}. In this regulator the soft and collinear Wilson lines are modified as
\begin{align}
S_n&= \left[  \sum\limits_{\text{perms}} \exp \left(  -\frac{g}{n\cdot \cP} \frac{\omega |2\cP^z|^{-\eta/2}}{\nu^{-\eta/2}} n \cdot A_s(x)  \right) \right]\,, \nn \\
W_{n}&= \left[  \sum\limits_{\text{perms}} \exp \left(  -\frac{g}{ \bar n\cdot \cP} \frac{\omega^2 |\bar n\cdot \cP|^{-\eta}}{\nu^{-\eta}}  \bar n \cdot A_{n}(x)  \right) \right]\,,
\end{align}
with analogous modifications for $S_{\bar n}$ and $W_{\bar n}$. Here $\omega$ is a formal bookkeeping parameter which satisfies
\begin{align}
\nu \frac{\partial}{\partial \nu} \omega^2(\nu) =-\eta~ \omega^2(\nu) \,, \qquad \lim_{\eta \to 0}\, \omega(\nu)=1\,.
\end{align}
For convenience we set $\omega=1$ throughout our calculations since it can be trivially restored. 

Singularities from Glauber exchanges are also regulated using the $\eta$-regulator. In particular, a factor of $\omega |2q^z|^{-\eta} \nu^\eta$ is included for each Glauber exchange, where $q$ is the Glauber momentum. This can be formulated at the level of the Glauber Lagrangian, and can be shown to be routing independent \cite{Rothstein:2016bsq}. We regulate divergences associated with Glauber quarks in an identical manner, and show the consistency of this regulator at one-loop through our calculations of the Reggeization, the BFKL equation, and the box diagrams with simultaneous exchange of a Glauber quark and a Glauber gluon.

\section{Tree Level Matching}\label{sec:tree_level}

In this section we consider tree level matching between QCD and SCET. This, combined with the symmetries of the effective theory as well as constraints from power and mass dimension counting, will allow us to fix the structure of the operators, as given in the previous section. In \Sec{sec:nbarn_match} and \Sec{sec:ns_match} we perform the matching with zero soft emissions. In \Sec{sec:soft_match} we present the most general form of the soft operator $\cO_s$, and fix its structure with tree level matching.

We will use the following alternative notation for Feynman diagrams involving Glauber quark exchange, distinguishing the Glauber quark exchange from a Glauber gluon exchange by including an arrow on the red dotted line:
\begin{align}\label{eq:collinear_potential}
\fd{2.8cm}{figures/tree_level_1_low.pdf}\equiv\fd{2.6cm}{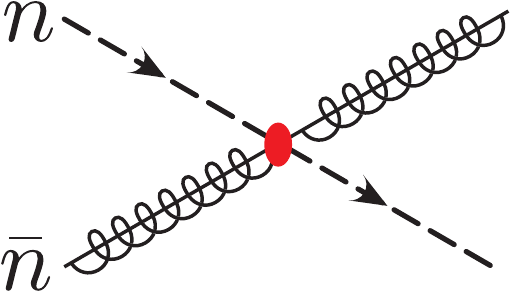}\,, \qquad 
\fd{2.8cm}{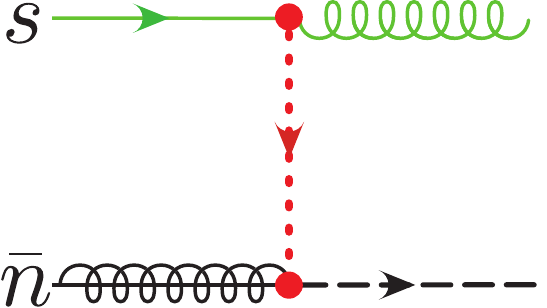}\equiv\fd{2.3cm}{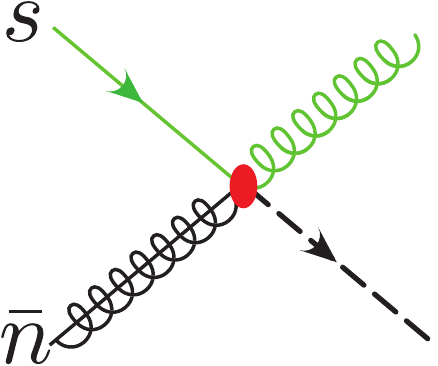}\,,
\end{align}
where we have illustrated with particular configurations of $n$-$\bar n$ and $n$-$s$ scattering. The notation with the red dotted line shows the $t$-channel exchange explicitly, while the notation with the red elliptical blob emphasizes the potential nature of the forward scattering operators.

\subsection{$n$-$\bar n$ Scattering}\label{sec:nbarn_match}

We begin with the matching for the $n$-$\bar n$ scattering operator. For definiteness, we take the configuration $q(p_1^n) + g(p_2^{\bn}) \to  g(p_4^n) + q(p_3^{\bn})$, and choose our momenta as 
\begin{align}
p_{1\perp}&=-p_{4\perp}=q_\perp/2 \,, \qquad p_{2\perp}=-p_{3\perp}=-q_\perp/2 \,.
\end{align}
For this choice, the positive $q_\perp$ is aligned with the fermion number flow.
Expanding the full theory result in the forward limit, we find
\begin{align}\label{eq:full_theory_zero}
\fd{3.5cm}{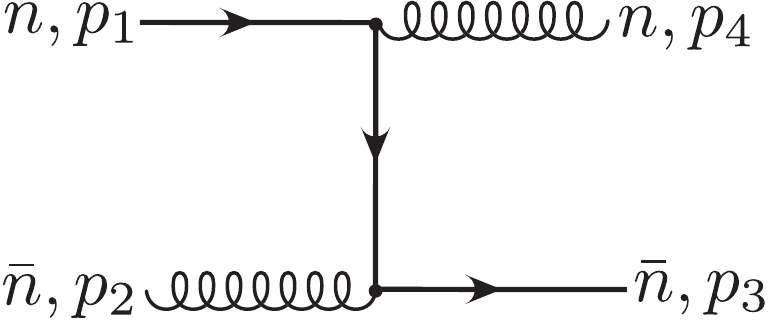}=-4\pi i \alpha_s \bar u_{\bar n}(p_3) \Sl{\epsilon}_\perp (p_2) T^A \frac{\Sl{q}_\perp}{q_\perp^2}  \Sl{\epsilon}_\perp (p_4) T^B u_n(p_1)\,.
\end{align}
This is reproduced in the effective theory by the zero emission Feynman rule of the forward scattering operator $\cO_{\bn n}$:
\begin{align}\label{eq:Onsmatching}
\fd{2.6cm}{figures/fermion_regge_vertex.pdf} =\langle \cO_{\bn n} \rangle=\left \langle \bar \chi_{\bar n} \Sl{\cB}_{\perp \bar n} \frac{1}{\Sl{\cP}_\perp} (-4\pi \alpha_s \Sl{\cP}_\perp) \frac{1}{\Sl{\cP}_\perp} \Sl{\cB}_{\perp n} \chi_n \right \rangle\,.
\end{align}
In particular, this defines the normalization of the soft operator $\cO_s$ with zero emissions, but does not probe the structure of the soft Wilson lines or the soft gluon fields within $\cO_s$.

\subsection{$n$-$s$ Scattering}\label{sec:ns_match}

The expansion of the full theory diagram in \Eq{eq:full_theory_zero} also fixes the structure of the $n$-$s$ operators. In particular, we immediately see that it is reproduced by the zero emission Feynman rule of the forward scattering operator $\cO_{\bn s}$:
\begin{align}
\fd{2.3cm}{figures/soft_scatter_regge_vertex_low.pdf}= \langle \cO_{\bn s} \rangle = \left \langle\bar \chi_{\bar n} \Sl{\cB}_{\perp \bar n} \frac{1}{\Sl{\cP}_\perp}  \left( -4\pi \alpha_s \Sl{\cB}_{\perp S}^\bn \psi^\bn_S \right)\right\rangle\,.
\end{align}
This simple matching, combined with constraints from power counting, mass dimension and the symmetries of the effective theory, therefore fixes the form of the operators $\cO_{ns}$ and $\cO_{\bn s}$. Once again these are the only operators that appear from tree level matching.

\subsection{Matching to the Soft Operator}\label{sec:soft_match}

To derive the precise structure of the soft operator $\cO_s$, we must consider matching with soft gluon emissions. We begin by deriving the most general form of the soft operator consistent with constraints from power counting, mass dimension and the symmetries of the effective theory. We then use matching calculations to fix the free coefficients in the operator.

As discussed in \Sec{sec:nn_ops}, the soft operator must have mass dimension $1$, scale as $\cO(\lambda)$, and be composed of gauge invariant building blocks in the effective theory such as $\cP_\perp$, $\cB^{n}_\perp$, $\cB^{\bar n}_\perp$ and Wilson lines. Since the total $\perp$ momentum of the Lagrangian is zero, we have $\cP_\perp=\cP_\perp^\dagger$, and therefore we can choose to write the operator in terms of $\cP_\perp$. Hermiticity requires that the operator satisfies (up to $\gamma_0$ factors that are absorbed by the collinear operators in $\cO_{\bn n}$)
\begin{align}
\cO_s=\cO^\dagger_s \big|_{n \leftrightarrow \bar n} \,.
\end{align}
The above constraints do not prohibit the appearance of an arbitrary number of soft Wilson lines since these have mass dimension $0$ and scale as $\cO(\lambda^0)$. However, due to the physical picture of these Wilson lines as arising from the emission of gluons off the partons involved in the forward scattering, we will require that each term in the soft operator has two Wilson lines. These soft Wilson lines can appear both explicitly, as well as inside the gauge invariant soft gluon fields, defined in \Eq{eq:gauge_soft}, and both must be counted. The constraint of having two soft Wilson lines leads to the following allowed combinations:
\begin{align}
S_{\bar n}^\dagger S_n \cB^{n\mu}_{S\perp}\,, \qquad \cB^{\bar n \mu}_{S\perp} S_{\bar n}^\dagger S_n\,, \qquad \cB^{n\mu}_{S\perp} S_n^\dagger S_{\bar n} \,, \qquad S_n^\dagger S_{\bar n} \cB^{\bar n \mu}_{S\perp}\,.
\end{align}

Given these constraints, the most general structure of the operator is
\begin{align}\label{eq:gen_soft}
\cO_s &= -4\pi \alpha_s \left[ \frac{C_1}{2} \left(g\Sl{\cB}^{n}_{\perp s}  S_n^\dagger S_{\bar n} + S_n^\dagger S_{\bar n} g\Sl{\cB}_{\perp s}^{\bar n} \right) +\frac{C_2}{2} \left(S_{\bar n}^\dagger S_n g\Sl{\cB}^{n}_{S\perp} +  g\Sl{\cB}^{\bar n}_{S\perp} S_{\bar n}^\dagger S_n \right) \right. \nn \\
&+\frac{C_3}{2} \left( S_n^\dagger S_{\bar n} \Sl{\cP}_\perp + \Sl{\cP}_\perp S_n^\dagger S_{\bar n} \right) \left. +\frac{C_4}{2}\left( S_{\bar n}^\dagger S_{n} \Sl{\cP}_\perp + \Sl{\cP}_\perp S_{\bar n}^\dagger S_{n}\right) \right ] \,.
\end{align}
The tree level matching with zero emission in \Sec{sec:nbarn_match} gives the relation
\begin{align}\label{eq:zeromatchingrelation}
C_3+C_4=1\,.
\end{align}
In the next section, we derive additional coefficient relations by considering soft emissions, which probe the structure of the soft Wilson lines and the soft gluon fields. Note that the general form of the soft operator in \Eq{eq:gen_soft} includes both combinations $S_n^\dagger S_{\bar n}$ and $S_{\bar n}^\dagger S_{n}$. In the Glauber gluon case, the soft operator $\cO_s^{BC}$ in \Eq{eq:iain_ira_soft} has only one of these combinations, corresponding to the ordering of the operators $\cO_n^{iB}$, $\cO_s^{BC}$ and $\cO_\bn^{jC}$ in \Eq{eq:Glauber_Lagrangian}. We will see that this also holds in the Glauber quark case, and in particular we will show that $C_1=C_3=0$ for the ordering of operators in \Eq{eq:Glauber_Lagrangian_cquark}.

\subsubsection{One Soft Emission}\label{sec:one_soft}

\begin{figure}[t!]
	\begin{center}
		\raisebox{2.5cm}{
			\hspace{-14.4cm}
			a)
		} \\[-83pt]
		\hspace{-0.5cm}
		\raisebox{0.32cm}{
			\includegraphics[width=0.16\columnwidth]{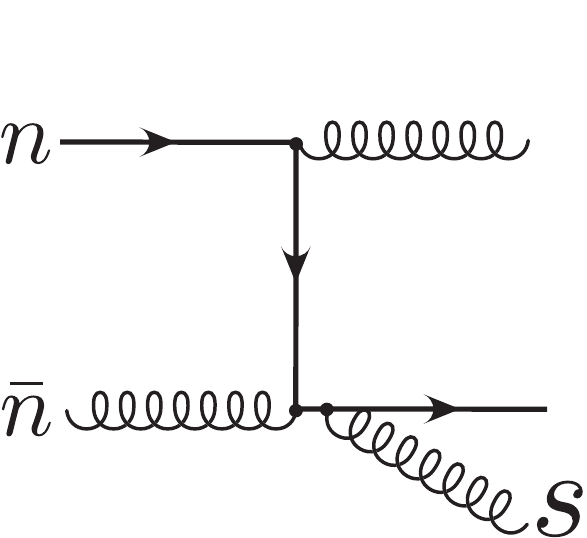}
		}\hspace{0.2cm}
		\raisebox{0.3cm}{
			\includegraphics[width=0.15\columnwidth]{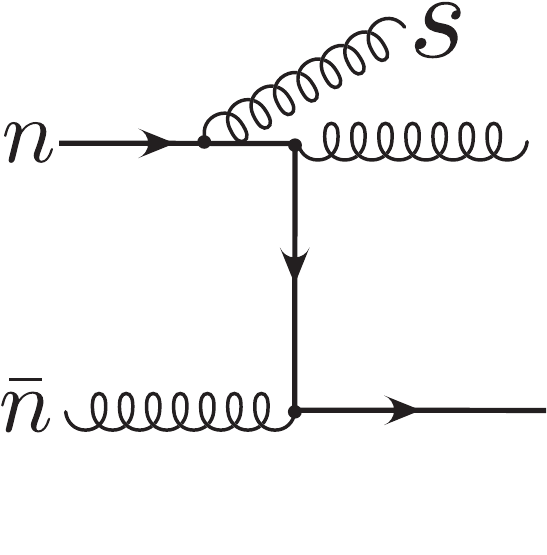} 
		}\hspace{0.2cm}
		\raisebox{0.3cm}{
			\includegraphics[width=0.16\columnwidth]{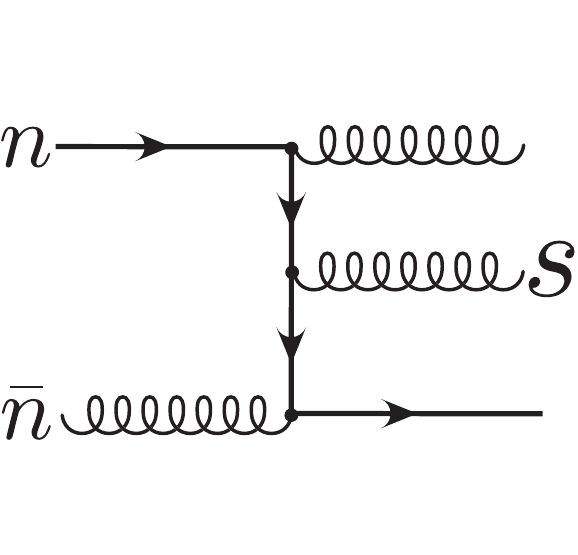}
		}\hspace{0.2cm}
		\raisebox{0.3cm}{
			\includegraphics[width=0.18\columnwidth]{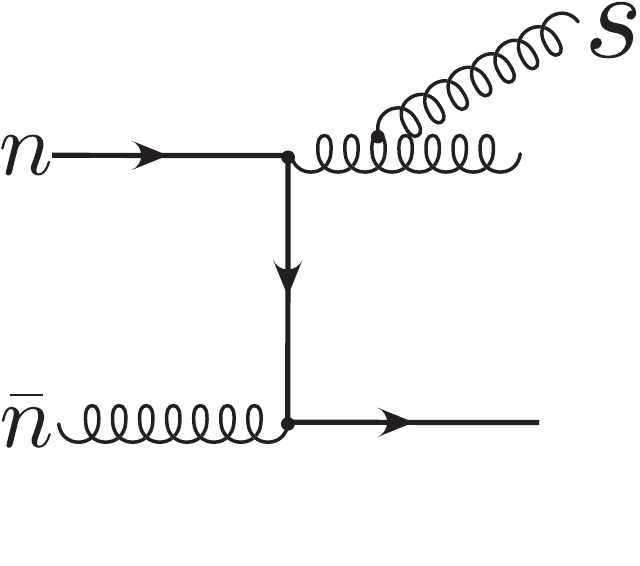}
		}\hspace{0.2cm}		
		\raisebox{0.3cm}{
			\includegraphics[width=0.16\columnwidth]{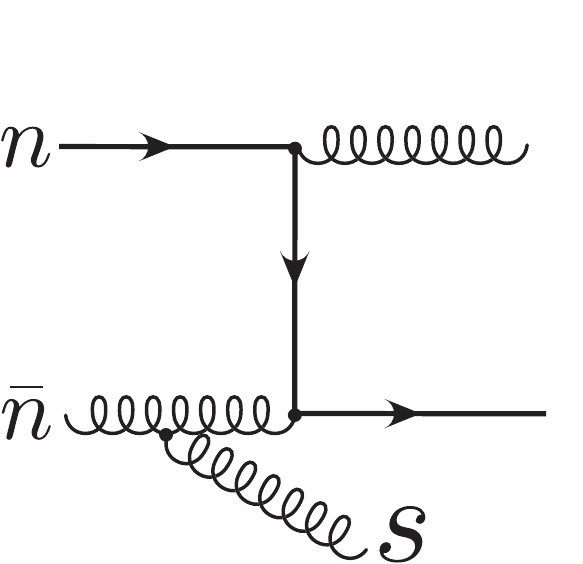}
		}\hspace{0.2cm}
		\\[8pt]
				\raisebox{2cm}{
			\hspace{-1.5cm}
			b)\hspace{3.1cm} 
		} \\[-50pt]	
		\hspace{-0.85cm}
		\includegraphics[width=0.18\columnwidth]{figures/fermion_lipatov_1emission_low.pdf}				
	\end{center}
	\vspace{-0.4cm}
	\caption{\setcaptionskip
	(a) Full theory and  (b) effective theory graphs with a single soft emission. We refer to the effective theory vertex as the Fadin-Sherman vertex since it first appeared in \cite{Fadin:1976nw,Fadin:1977jr}.}
	\label{fig:single_emission}
	\setmainskip
\end{figure}

The single emission diagrams in the full theory and effective theory are shown in \Fig{fig:single_emission}.
Expanded to a single emission with outgoing momentum $k$, the soft operator is given by
\begin{align}\label{eq:soft_op_1emission}
 \cO_s &=-4\pi \alpha_s \left[ \vphantom{\frac{C_3}{2}}   (C_1+C_2)g \Sl{A}_{s\perp} \right. \nn \\
  &-\left(  \frac{C_1}{2}+ \frac{C_2}{2} \right) \left( \frac{gT^A n\cdot A^A_{sk}}{n\cdot k}+\frac{gT^A \bar n\cdot A^A_{sk}}{\bar n \cdot k}  \right) (\Sl{q}_\perp +\Sl{k}_\perp) \nn \\
  &-\left(  \frac{C_3}{2}- \frac{C_4}{2} \right) \left( \frac{gT^A n\cdot A^A_{sk}}{n\cdot k}-\frac{gT^A \bar n\cdot A^A_{sk}}{\bar n \cdot k}  \right) (\Sl{q}_\perp +\Sl{k}_\perp) \nn \\
  &+\left(  \frac{C_1}{2}+ \frac{C_2}{2} \right) \Sl{q}_\perp \left( \frac{gT^A n\cdot A^A_{sk}}{n\cdot k}+\frac{gT^A \bar n\cdot A^A_{sk}}{\bar n \cdot k}  \right) \nn \\
  &\left. -\left(  \frac{C_3}{2}- \frac{C_4}{2} \right)  \Sl{q}_\perp \left( \frac{gT^A n\cdot A^A_{sk}}{n\cdot k}-\frac{gT^A \bar n\cdot A^A_{sk}}{\bar n \cdot k}  \right) \right ]\,.
\end{align}
To fix $C_1+C_2$, we only need the perpendicular polarization, which comes from the full theory diagram
\begin{align}
\fd{2.8cm}{figures/tree_level_full_theory_1emission_3.pdf} \
=i 4\pi \alpha_s  \bar u_{\bar n} \Sl{\epsilon}_{\perp} T^A \frac{\Sl{q}_\perp}{q_\perp^2}  \gamma^\rho_\perp T^c  \frac{(\Sl{q}_\perp  +\Sl{k}_\perp )}{(q_\perp+k_\perp)^2} \Sl{\epsilon}_{\perp} T^B u_n\,.
\end{align}
In the effective theory, we have
\begin{align}
\fd{2.8cm}{figures/fermion_lipatov_1emission_low.pdf}=-i4\pi \alpha_s  (C_1+C_2) \bar u_{\bar n} \Sl{\epsilon}_{\perp} T^A \frac{\Sl{q}_\perp}{q_\perp^2}  \gamma^\rho_\perp T^c  \frac{(\Sl{q}_\perp  +\Sl{k}_\perp )}{(q_\perp+k_\perp)^2} \Sl{\epsilon}_{\perp} T^B u_n\,,
\end{align}
and thus the constraint from matching is
\begin{align}\label{eq:onematchingrelationperp}
C_1+C_2=-1\,.
\end{align}
The Wilson line structure is probed using the $n\cdot A$ and $\bar n \cdot A$ polarizations of the emission. From the remaining four diagrams in the full theory, we find
\begin{align}
&\fd{3cm}{figures/tree_level_full_theory_1emission_1.pdf}+
\fd{2.8cm}{figures/tree_level_full_theory_1emission_2.pdf}+
\fd{3.3cm}{figures/tree_level_full_theory_1emission_4.pdf}+
\fd{3cm}{figures/tree_level_full_theory_1emission_5.pdf}\nn \\[5pt]
&=-i 4\pi  \alpha_s \bar u_{\bar n} \Sl{\epsilon}_{\perp} T^A \left[  \left( \frac{gT^A n\cdot A^A_{sk}}{n\cdot k}  \right) (\Sl{q}_\perp +\Sl{k}_\perp)
  - \Sl{q}_\perp \left( \frac{gT^A \bar n\cdot A^A_{sk}}{\bar n \cdot k}  \right) \right ] \Sl{\epsilon}_{\perp} T^B u_n \,.
\end{align}
Upon comparing with \Eq{eq:soft_op_1emission}, we derive the relation
\begin{align}\label{eq:onematchingrelation}
C_1+C_2&=(C_3-C_4)\,. 
\end{align}

The constraints derived from zero and one emission matching, given in Eqs.~(\ref{eq:zeromatchingrelation}),~(\ref{eq:onematchingrelationperp}) and~(\ref{eq:onematchingrelation}), have the solution $C_1+C_2=-1$, $C_3=0$ and $C_4=1$. The remaining degeneracy between the coefficients $C_1$ and $C_2$ can be broken by matching with two soft emissions.

\subsubsection{Two Soft Emissions}\label{eq:two_soft}

\begin{figure}[t!]
	\begin{center}
		\raisebox{2cm}{
			\hspace{-14.4cm}
			a)
		} \\[-83pt]
		\hspace{-0.5cm}
		\raisebox{0.35cm}{
			\includegraphics[width=0.16\columnwidth]{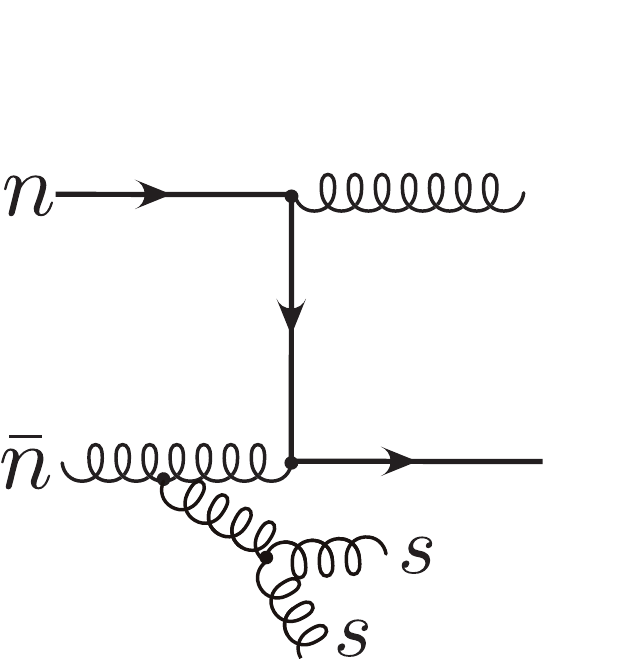}
		}\hspace{0.2cm}
		\raisebox{0.3cm}{
			\includegraphics[width=0.16\columnwidth]{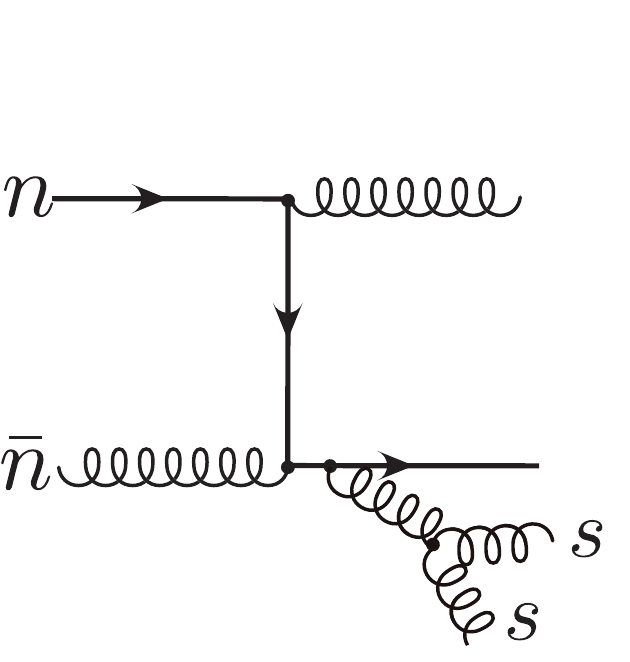} 
		}\hspace{0.2cm}
		\raisebox{0.3cm}{
			\includegraphics[width=0.16\columnwidth]{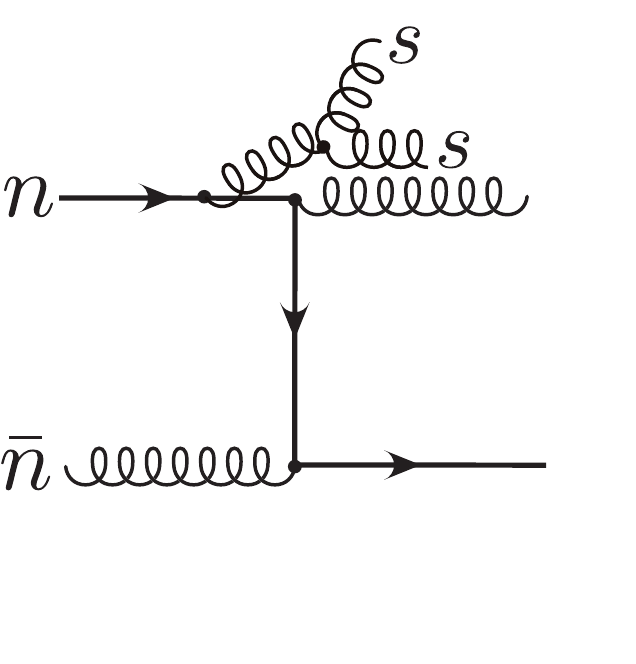}
		}\hspace{0.2cm}
		\raisebox{0.3cm}{
			\includegraphics[width=0.16\columnwidth]{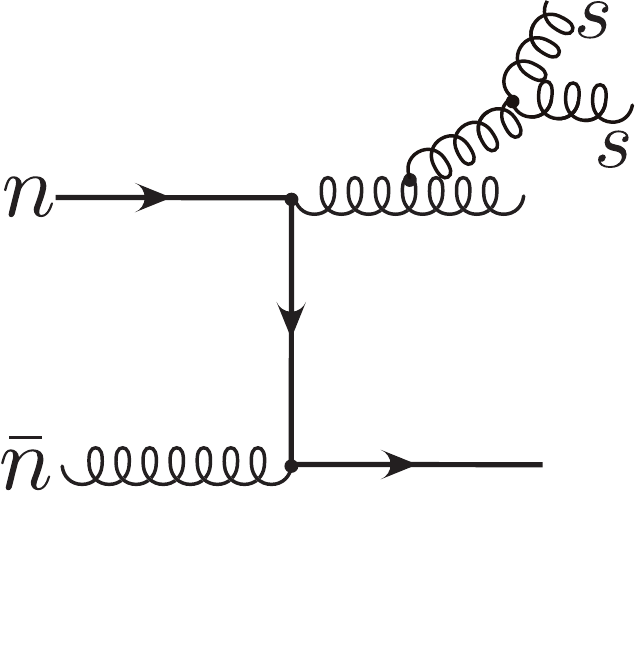}
		}\hspace{0.2cm}
		\\[8pt]
		\hspace{-0.55cm}
		\includegraphics[width=0.16\columnwidth]{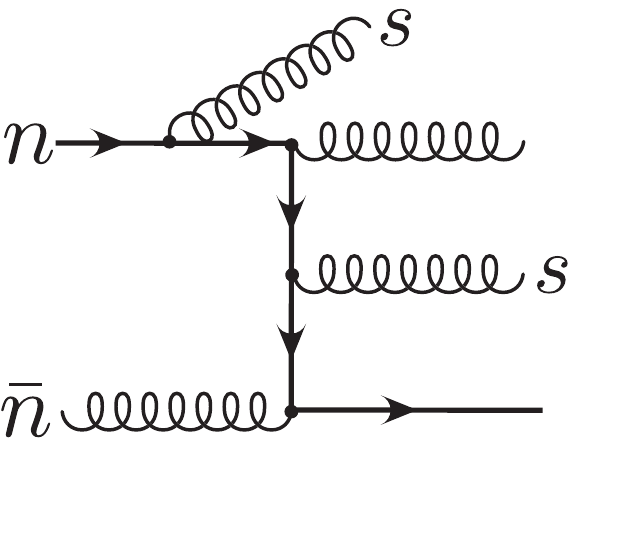}
		\hspace{0.2cm}
		\includegraphics[width=0.16\columnwidth]{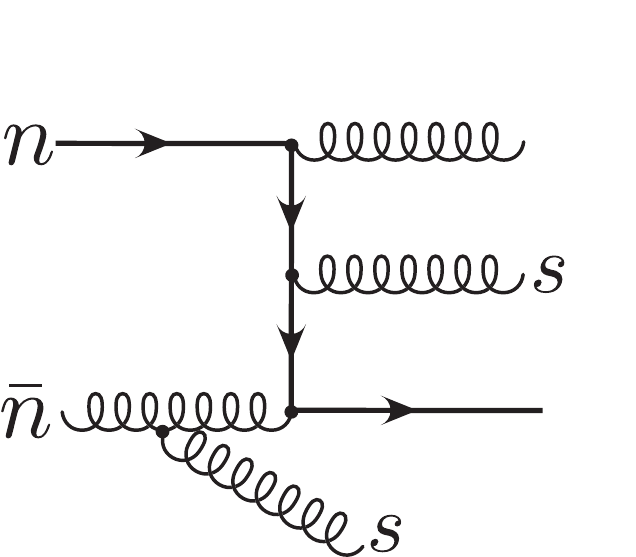}
		\hspace{0.2cm}
		\includegraphics[width=0.16\columnwidth]{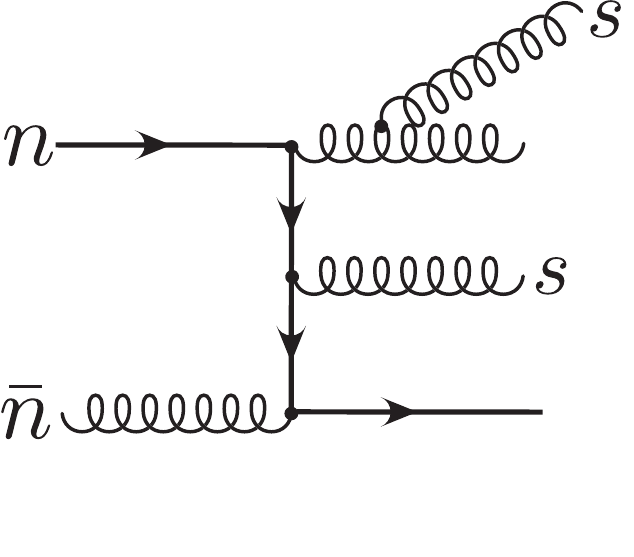}				
		\hspace{0.2cm}
		\includegraphics[width=0.16\columnwidth]{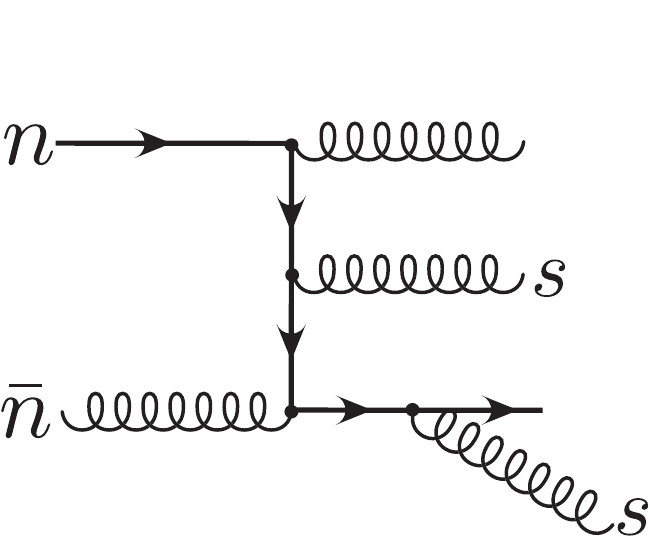}							
		\\[8pt]
		\hspace{-0.55cm}
		\raisebox{-0.1cm}{\includegraphics[width=0.13\columnwidth]{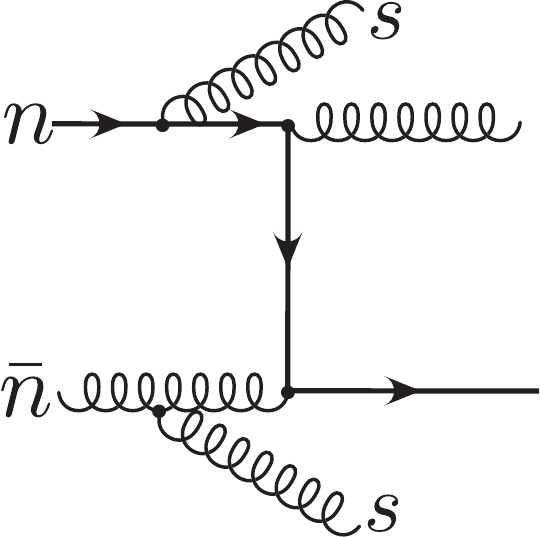}}
		\hspace{0.2cm}
		\raisebox{0.0cm}{\includegraphics[width=0.15\columnwidth]{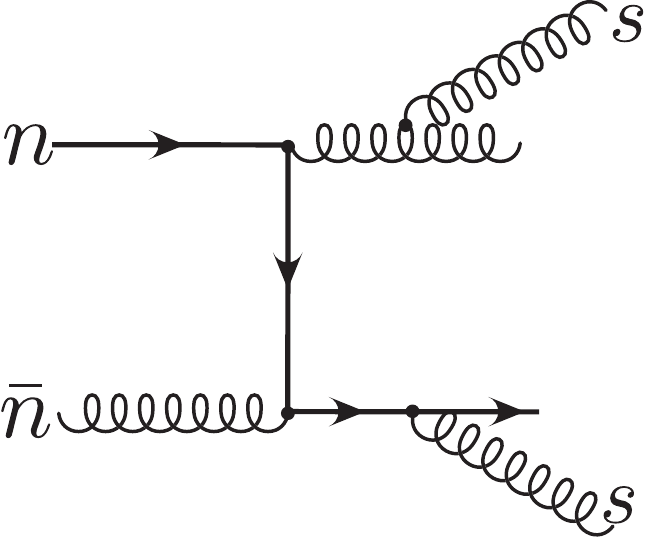}}
		\hspace{0.2cm}
		\raisebox{0.0cm}{\includegraphics[width=0.15\columnwidth]{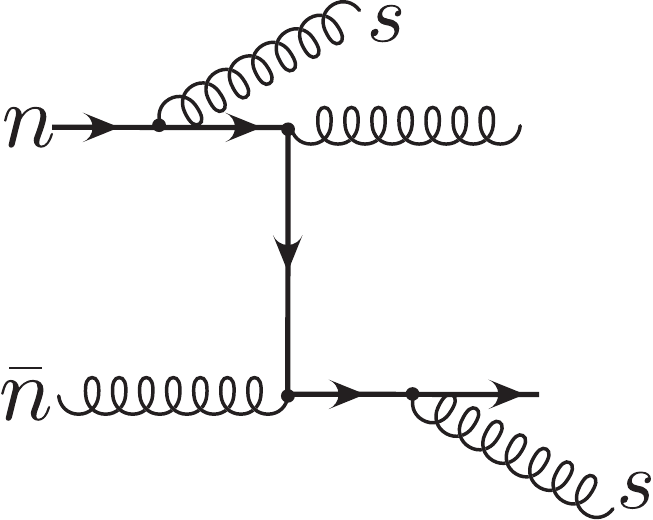}}				
		\hspace{0.2cm}
		\raisebox{-0.1cm}{\includegraphics[width=0.15\columnwidth]{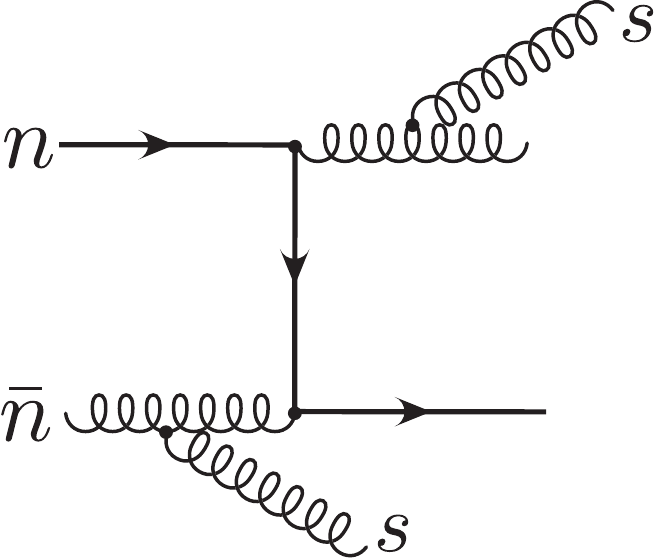}}							
		\\[8pt]
		\hspace{-0.85cm}
		\includegraphics[width=0.15\columnwidth]{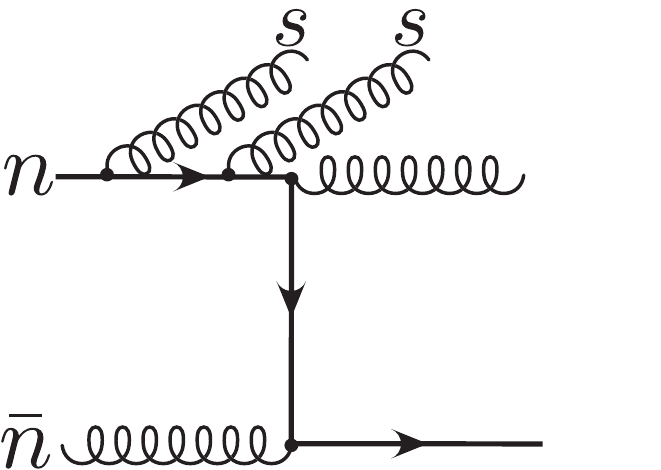}
		\hspace{0.2cm}
		\includegraphics[width=0.15\columnwidth]{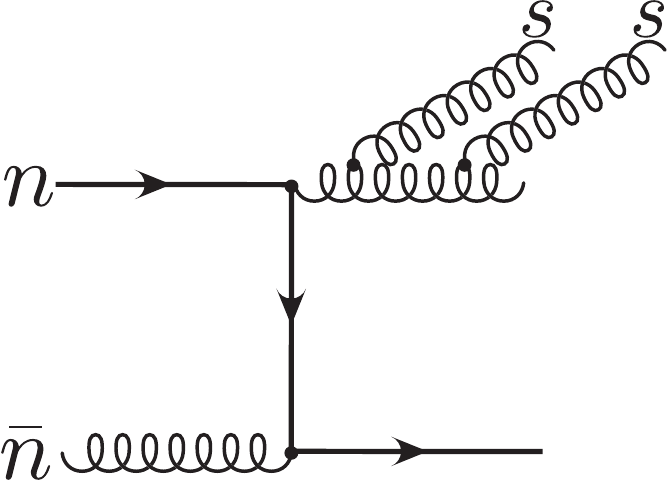}
		\hspace{0.2cm}
		\includegraphics[width=0.15\columnwidth]{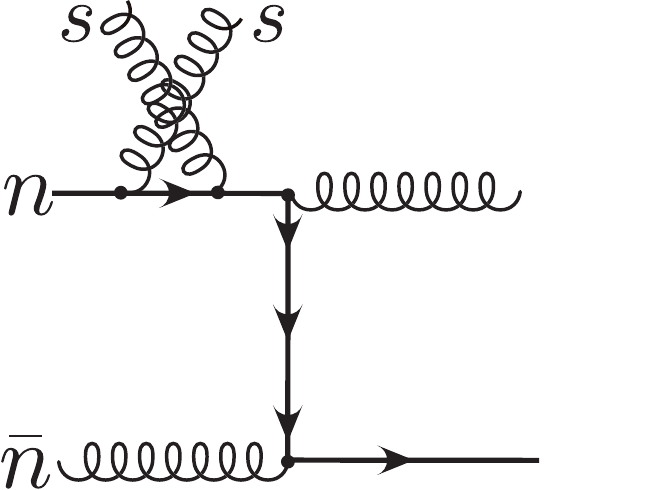}				
		\hspace{0.2cm}
		\includegraphics[width=0.15\columnwidth]{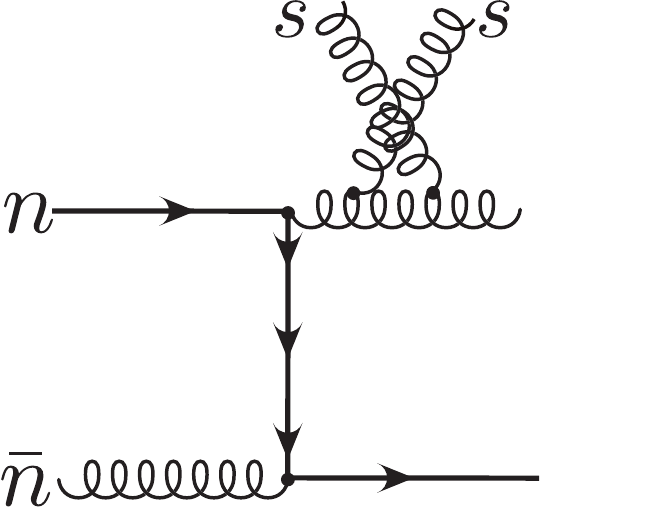}
				\hspace{0.2cm}
		\includegraphics[width=0.15\columnwidth]{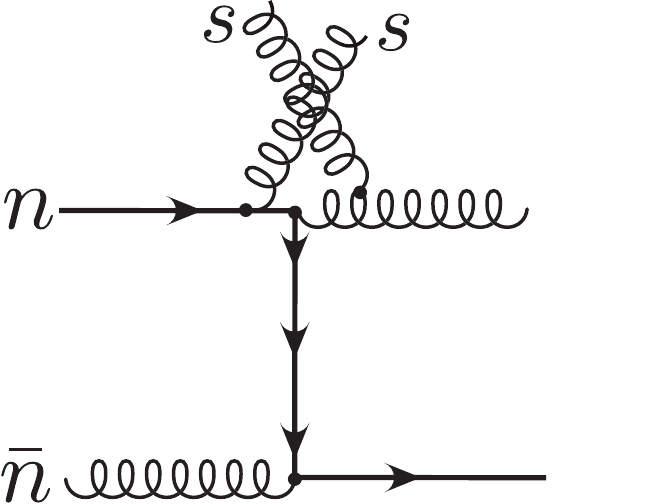}
				\hspace{0.2cm}
		\includegraphics[width=0.15\columnwidth]{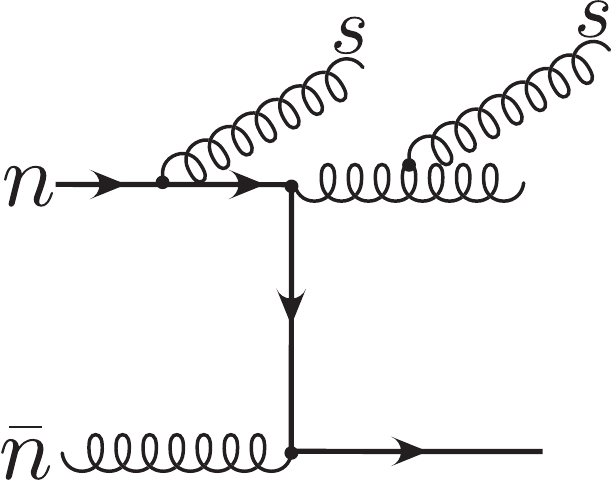}							
				\\[8pt]
		\hspace{-0.85cm}
		\includegraphics[width=0.15\columnwidth]{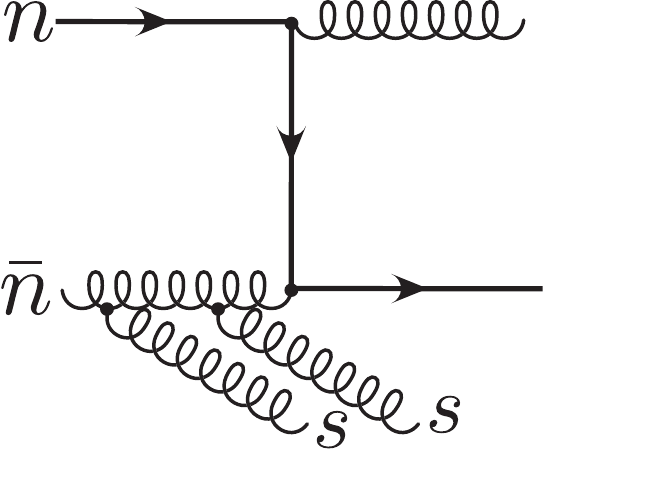}
		\hspace{0.2cm}
		\includegraphics[width=0.15\columnwidth]{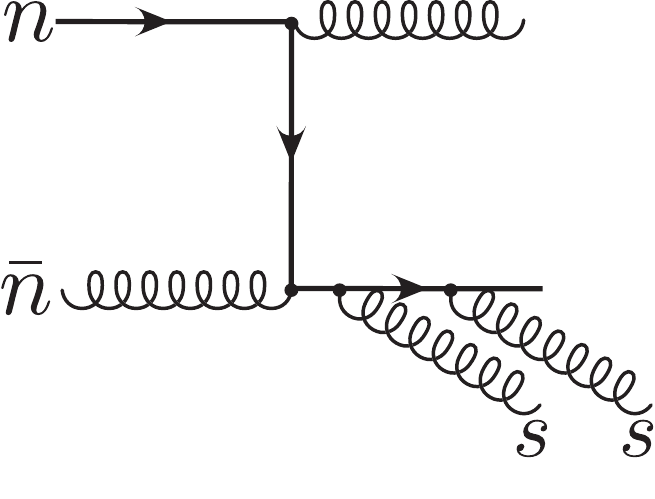}
		\hspace{0.2cm}
		\includegraphics[width=0.15\columnwidth]{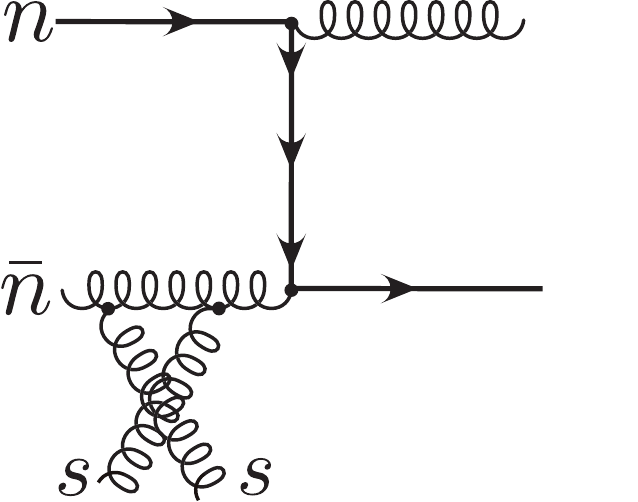}				
		\hspace{0.2cm}
		\includegraphics[width=0.15\columnwidth]{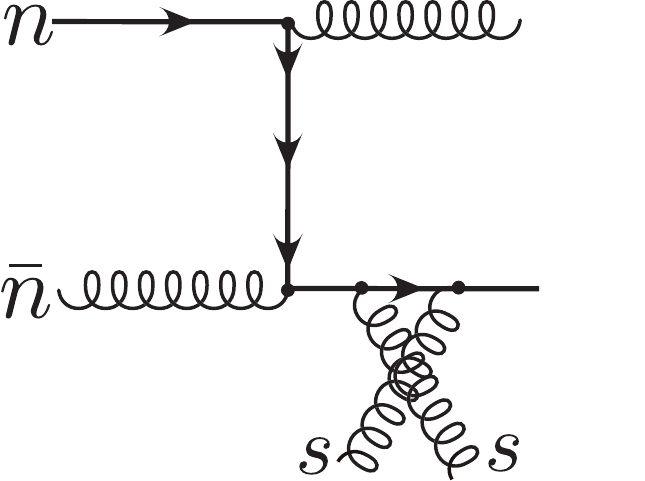}
				\hspace{0.2cm}
		\includegraphics[width=0.15\columnwidth]{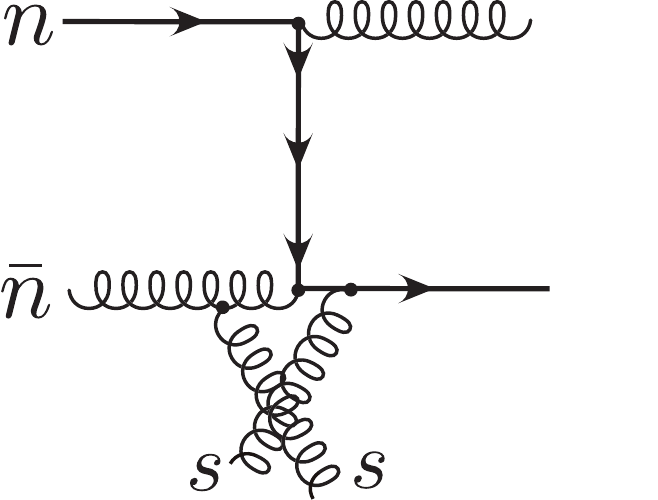}
				\hspace{0.2cm}
		\includegraphics[width=0.15\columnwidth]{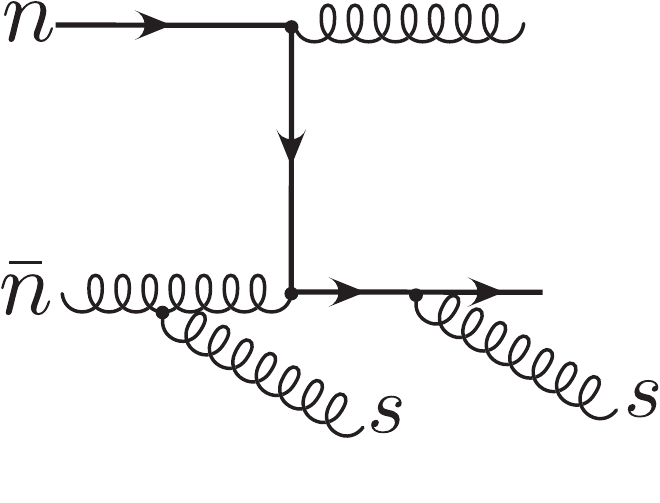}							
				\\[4pt]
		\hspace{-0.85cm}
		\raisebox{0.0cm}{\includegraphics[width=0.15\columnwidth]{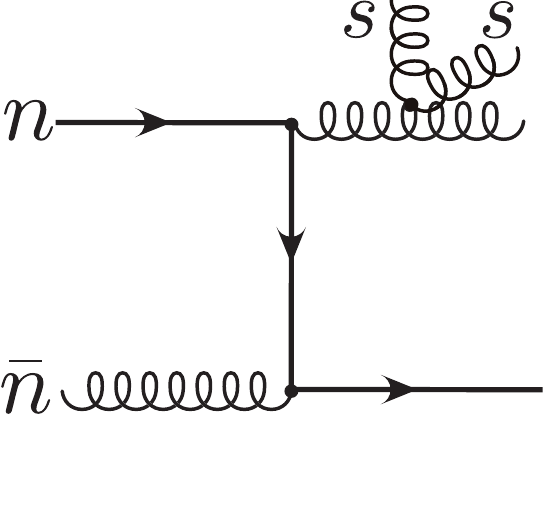}}
		\hspace{0.2cm}
		\raisebox{0.0cm}{\includegraphics[width=0.15\columnwidth]{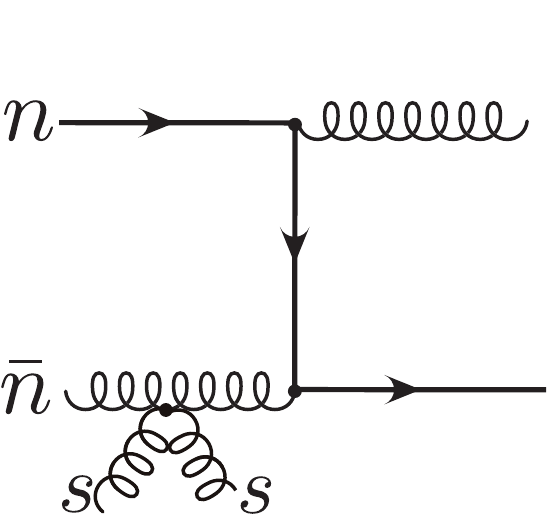}}			
		\hspace{0.2cm}
		\raisebox{0.5cm}{\includegraphics[width=0.15\columnwidth]{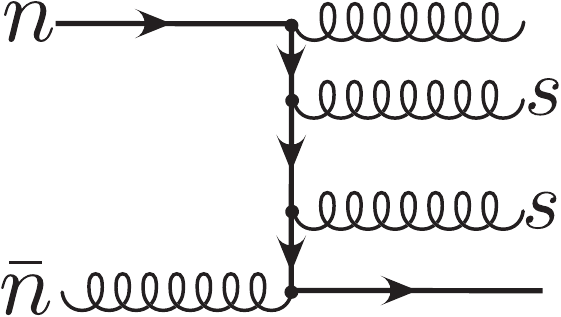}}
				\hspace{0.2cm}
		\raisebox{0.5cm}{\includegraphics[width=0.15\columnwidth]{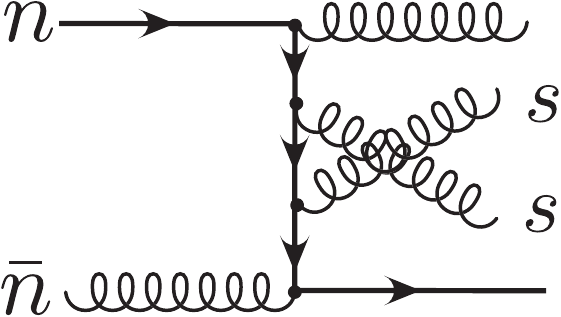}}						
		\\[8pt]
				\raisebox{2cm}{
			\hspace{-14.5cm}
			b)
		} \\[-50pt]	
		\hspace{-0.85cm}
		\includegraphics[width=0.18\columnwidth]{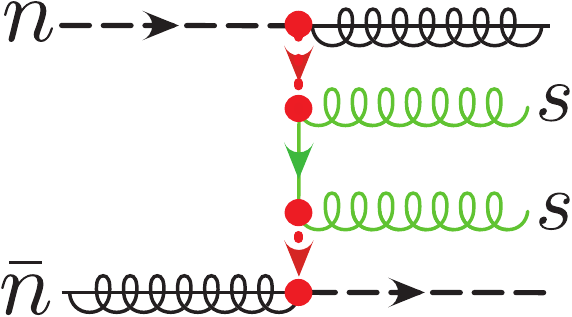}
		\hspace{0.2cm}
		\includegraphics[width=0.18\columnwidth]{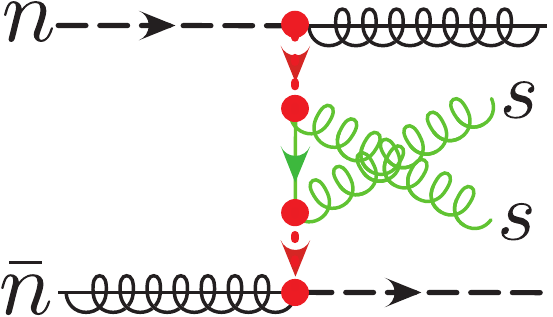}
		\hspace{0.2cm}
		\includegraphics[width=0.18\columnwidth]{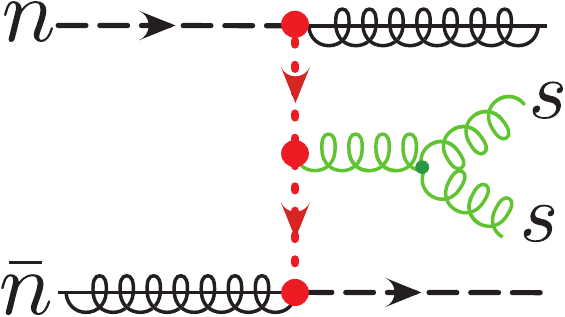}				
		\hspace{0.2cm}
		\includegraphics[width=0.18\columnwidth]{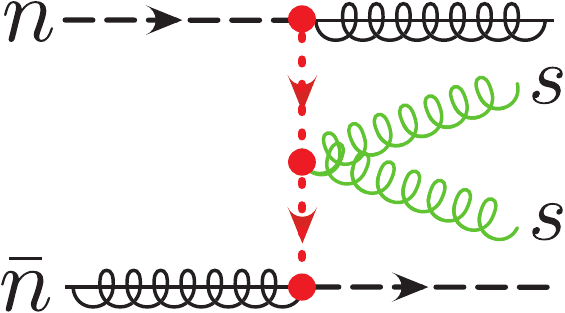}					
	\end{center}
	\vspace{-0.4cm}
	\caption{\setcaptionskip
		(a) Full theory and (b) effective theory graphs with two soft emissions. In the effective theory, the first three graphs are $T$-product contributions, and the fourth graph is the two emission Feynman rule from the Fadin-Sherman vertex.}
	\label{fig:double_emission}
	\setmainskip
\end{figure}

The double emission diagrams in the full theory and effective theory are shown in \Fig{fig:double_emission}. Note that the operators for $n$-$s$ and $\bar n$-$s$ forward scattering enter the matching through $T$-product contributions. 

Instead of performing the complete two emission matching, we will assume that only one ordering of Wilson lines appears, as in the case of the leading power Glauber Lagrangian $\cL_G^{\text{II}(0)}$. This is motivated also by the patterns found in one emission matching as well as the structure of diagrams in \Fig{fig:double_emission} for the two emission matching. We leave a general proof of this statement to future work.
Under this assumption, we have $C_1=0$, which completely fixes the form of our soft operator to the final form given in \Eq{eq:soft_operator_quark}:
\begin{equation}
  \boxed{\cO_s = -2\pi \alpha_s \left[ S_{\bar n}^\dagger S_{n} \Sl{\cP}_\perp + \Sl{\cP}_\perp S_{\bar n}^\dagger S_{n} - S_{\bar n}^\dagger S_n g\Sl{\cB}^{n}_{S\perp}-  g\Sl{\cB}^{\bar n}_{S\perp} S_{\bar n}^\dagger S_n \right]}\,.
\end{equation}

The particular ordering of the Wilson lines, $S_\bn^\dagger S_n$, appearing in $\cO_s$ in \Eq{eq:soft_operator_quark} corresponds to the ordering of the collinear and soft operators in \Eq{eq:Glauber_Lagrangian}, and to the scattering configuration employed in our matching. The soft operator written with the opposite ordering is obtained simply by the replacement $n \leftrightarrow \bn$ in \Eq{eq:soft_operator_quark}.

\section{Quark Reggeization from Rapidity Renormalization}\label{sec:quark_reggeize}
In this section we consider the renormalization of the Glauber operators to derive the Reggeization of the quark. The renormalization should be done at the level of the squared amplitude, including both virtual and real contributions, to obtain IR finite results. Nevertheless, with careful interpretation of the IR divergences, the virtual diagrams can be examined at the amplitude level, and we will see that the solution to the rapidity renormalization group equation (RGE) corresponds to the Reggeization of the quark.

For quark-gluon scattering, we can decompose the color structure of the $t$-channel exchange as $3\, \otimes \, 8 = 3 \oplus  \bar 6 \oplus 15$.
Explicitly, if we decompose the amplitude using the color basis
\begin{align}\label{eq:Colordecomp}
\cM=2\left( T^A T^{B}\right)_{ij} \cA +2 \left( T^{B} T^A \right)_{ij} \cB +\delta^{AB} \delta_{ij} \cC  \,,
\end{align}
then the contributions to the $3$, $\bar 6$ and $15$ color structures are given by \cite{Bogdan:2002sr},
\begin{align}
\cM_3&=2C_F \cA -\frac{1}{N} \cB +\cC\,, \\
\cM_{\bar 6}&=-\cB+\cC\,, \\
\cM_{15}&=\cB+\cC \,.\label{eq:Colordecomp2}
\end{align}
In this section we will focus on the Reggeization of the $3$ channel at LL order, which corresponds to dressing the tree-level $t$-channel quark exchange. In the study of Reggeization, it is conventional to also decompose the amplitude so that it has a definite signature under crossing, i.e.,  $\cM^{\pm}=\frac{1}{2}\left[ \cM \pm \cM(s\leftrightarrow t)   \right]$. Indeed, it is known that it is the positive signature $3$ channel that builds upon the lowest order quark exchange and Reggeizes at LL order. The negative signature channel is suppressed by an $\alpha_s$, and has a series that starts at next-to-leading logarithmic (NLL) order, which is beyond the order we are working.

In \Sec{sec:consistency_qregge} we setup the notation and present the structure for the renormalization of the Glauber quark operators. We also derive consistency relations among the anomalous dimensions of the soft and collinear operators, which provide important checks on our calculation. In Secs.~\ref{sec:one_loop_collinear} and \ref{sec:one_loop_soft} we compute the anomalous dimension of the collinear and soft operators. In \Sec{sec:RG_solve} we solve the RGE and demonstrate the Reggeization of the quark.

The $\bar 6$ and $15$ channels are generated by the simultaneous exchange of both a Glauber quark and a Glauber gluon. These diagrams are rapidity finite at lowest order, and will be considered in \Sec{sec:glauber_box}.

\subsection{RG Structure and Consistency Relations}\label{sec:consistency_qregge}

For the collinear sector, there is no mixing and the renormalization has the structure
\begin{align}
\cO_n^{\, \bare} =V_{\cO_n} \cO_n \,, \qquad V_{\cO_n} =(1+\delta V_n) 
\,, 
\end{align}
with analogous relations for the $\bar{n}$ sector. Following~\cite{Rothstein:2016bsq}, we use the notation ``$V$'' instead of the traditional ``$Z$'' for renormalization factors to remind the reader that these are only virtual contributions and may still depend on IR regulator.

For the soft operator $\cO_s^n$, there is no mixing and we have
\begin{align}
\cO_{s}^{n \, \bare} = V_{\cO_s^n} \cO_s^n \,, \qquad V_{\cO_s^n}  =(1+\delta V_s^n) 
\,, 
\end{align}
with analogous relations for the $\bar{n}$ sector. For the soft operator $\cO_s$, the renormalization group structure is more complicated due to mixing with $T$-products of $\cO_s^n$ and $\cO_s^\bn$. 
This is discussed in detail for the Glauber gluon case in~\cite{Rothstein:2016bsq}.  The structure in our case is given by
\begin{align}
&\vec \cO_{s}^{ \, \bare}=\hat V_{\cO_{s}} \cdot \vec \cO_{s}\,, \nonumber \\
&\vec \cO_{s} =\left(\begin{array}{c} \cO_{s} \\ i\int d^4x~ T~ \cO_s^\bn(x) \bar{\cO}_s^n(0) \end{array} \right)\,, \quad 
\hat V_{\cO_{s}}=\left(\begin{array}{cc} 1+\delta V_s&0 \\ \delta V_s^T& 
\ V_{\cO_s^\bn}  V_{\bar{\cO}_s^n}  \end{array} \right)\,.
\end{align}
Importantly, due to the relative difference in the power counting of $\cO_s$ to that of $\cO_s^\bn$ and $\cO_s^n$, both components in $\vec \cO_{s}$ are the same order in the power counting. 

The renormalization group structure above, for both the collinear and soft sectors, is simpler than for the case of Glauber gluon operators, which involves mixing between quark and gluon operators that leads to the universality of Reggeization~\cite{Rothstein:2016bsq}. In the present case, there is only a non-trivial mixing in the soft sector.

The $\mu$ and $\nu$ anomalous dimensions are derived by demanding the $\mu$ and $\nu$ invariance of the bare operators as usual. Since our operators do not have Wilson coefficients and the soft and collinear fields are at the same $\mu$ scale, we expect their $\mu$ anomalous dimension to vanish, as in the case of $\cL_G^{(0)}$~\cite{Rothstein:2016bsq}. Therefore, we focus here on the $\nu$ anomalous dimensions, which give rise to rapidity renormalization, and the Reggeization. 

We have the standard relations
\begin{align}
\cO^{ \bare}= V_{\cO} \cdot \cO(\nu, \mu) \,, \qquad \nu \frac{\partial }{\partial \nu} \cO(\nu, \mu)=\gamma^\nu_{\cO} \cdot \cO(\nu, \mu) \,, \qquad \gamma^\nu_{\cO}=-V_{\cO}^{-1} \cdot \nu  \frac{\partial }{\partial \nu} V_{\cO} \,,
\end{align}
for $\cO = \cO_n \,, \cO_s^n\,, \cO_s$ and for the operators describing the $\bar{n}$ sector. 
For the soft operator $\cO_s$, which undergoes mixing, the anomalous dimension has the form
\begin{align}
\hat \gamma^\nu_{\cO_{s}}=\left(\begin{array}{cccc} \gamma_{s\nu}^{\text{dir}}&&&0 \\[2pt] 
\gamma_{s\nu}^{T}&&& \gamma^\nu_{\cO_s^\bn}  \gamma^\nu_{\bar{\cO}_s^n} \end{array} \right)\,.
\end{align}

The fact that there is no overall $\nu$ dependence in $n$-$\bar n$ scattering and $n$-$s$ scattering leads to relations among the anomalous dimensions. The consistency for $n$-$\bar n$ scattering is derived at the level of the time evolution operator, and one must consider all possible contributions from $T$-products involving $\cL_G^{\text{II}(0)}$, $\cL^{\text{II}(1/2)}$, and $\cL^{\text{II}(1)}$. At one-loop, this simplifies considerably, and we have
\begin{align}
\nu \frac{\partial }{\partial \nu}  \left( \cO_{\bn n} + i \int d^4x \ T \ \cO_{\bn s}(x)  \cdot \bar{\cO}_{n s}(0)  \right)=0\,.
\end{align}
Note again that this has homogeneous power counting. By differentiating the time evolution of the $n$-$\bar n$ scattering and the $n$-$s$ scattering, we can derive the following relations between anomalous dimensions
\begin{align}\label{eq:ADrelations}
\gamma^\nu_{\cO_n}=\gamma^\nu_{\cO_{\bar n}}\,, \qquad \gamma^{\text{dir}}_{s\nu} + \gamma^{T}_{s\nu}=-\gamma^\nu_{\cO_n}-\gamma^\nu_{\cO_{\bar n}}
\,, \qquad \gamma^\nu_{\cO_s^n}=-\gamma^\nu_{\cO_n}\,.
\end{align}

\subsection{One-Loop Virtual Anomalous Dimension for the Collinear Operator}\label{sec:one_loop_collinear}

\begin{figure}
\begin{center}
\begin{tabular}{cccc}
\fd{3cm}{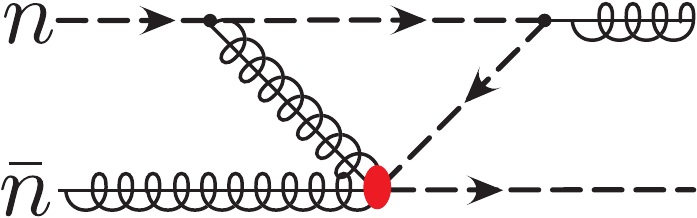} &
\fd{3cm}{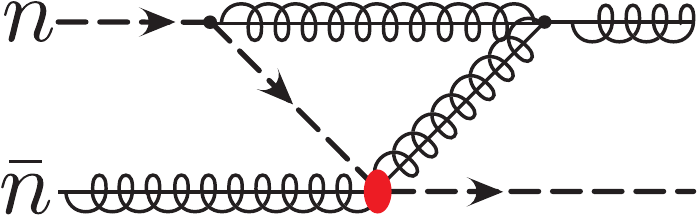} &
\fd{2cm}{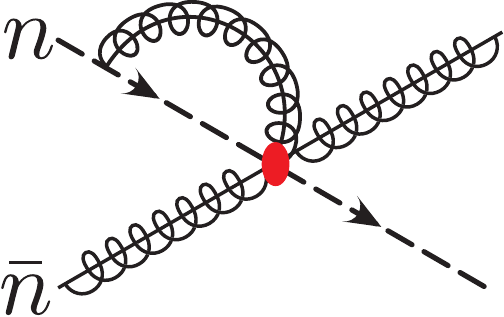} &
\fd{2cm}{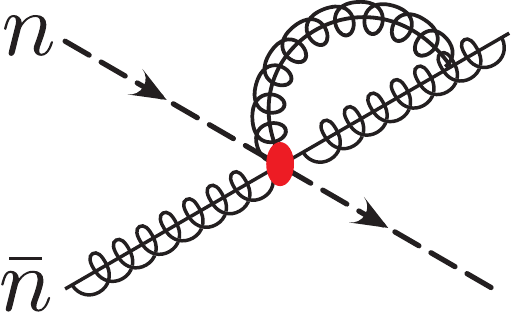} \\
a) & b) & c) & d)
\end{tabular}
\end{center}
\caption{One-loop virtual contributions to the renormalization of the collinear operator $\cO_n$. The V graphs are labeled a) and b), and the Wilson line graphs are labeled c) and d). 
}\label{fig:collinearAD}
\end{figure}

In this section we compute the one-loop virtual contributions to the renormalization of the collinear operator $\cO_n$. The two types of contributions are shown in Fig.~\ref{fig:collinearAD}, which we refer to as V graphs and Wilson line graphs. All the integrals can be evaluated following~\cite{Rothstein:2016bsq}, and we therefore only give the final results. It is sufficient to consider external gluons with perpendicular polarization, which  simplifies the calculation. We employ a gluon mass, $m$, as an IR regulator to ensure that all poles in $\epsilon$ are of UV origin. The IR regulator will explicitly appear in the rapidity anomalous dimension $\gamma^\nu_{\cO_n}$, and in the Regge trajectory.

In the following, we display only contributions to the $1/\eta$ pole (e.g., ignoring coupling and wavefunction renormalization), and denote finite pieces with ellipses. For the V graphs, we find
\begin{align}\label{eq:Vgraphs}
\text{Fig.~\ref{fig:collinearAD}a} &= (4 \pi \alpha_s)^2 (2C_F - C_A)  {\bar u}_{\bar n} \gamma_\perp^\mu T^A { \slashed{q}_\perp \over q_\perp^2} 
\int \dbar^d k  {\iota^\epsilon \mu^{2\epsilon}  |\bar n \cdot k|^{-\eta} \nu^\eta \slashed{k}_\perp (\slashed{k}_\perp + \slashed{q}_\perp) {\bar n} \cdot p_1   \over (k^2-m^2) (k+q)^2 (k+p_1)^2 {\bar n} \cdot k}  \gamma_\perp^\nu T^B u_n + \dots \nn  \\ 
&= -i 4 \pi \alpha_s {\bar u}_{\bar n} \gamma_\perp^\mu T^A { \slashed{q}_\perp \over q_\perp^2} \gamma_\perp^\nu T^B u_n \ {\alpha_s \over 2 \pi} \left( C_F - {C_A \over 2} \right)  {g (\epsilon, \mu^2/t) \over \eta} + \dots \,,
\\
\text{Fig.~\ref{fig:collinearAD}b} &= - (4 \pi \alpha_s)^2 C_A  {\bar u}_{\bar n} \gamma_\perp^\mu T^A { \slashed{q}_\perp \over q_\perp^2} 
 \int \dbar^d k   { \iota^\epsilon \mu^{2\epsilon} |\bar n \cdot k|^{-\eta} \nu^\eta \slashed{k}_\perp (\slashed{k}_\perp + \slashed{q}_\perp) {\bar n} \cdot p_4   \over (k^2-m^2) (k+q)^2 [( k - p_4)^2 - m^2] {\bar n} \cdot k} \gamma_\perp^\nu T^B u_n  + \dots \nn \\ 
&= -i 4 \pi \alpha_s {\bar u}_{\bar n} \gamma_\perp^\mu T^A { \slashed{q}_\perp \over q_\perp^2} \gamma_\perp^\nu T^B u_n   \ {\alpha_s \over 2 \pi} {C_A \over 2}   {g (\epsilon, \mu^2/t) \over \eta} + \dots\,,
\end{align}
where
\begin{align}
g(\epsilon, \mu^2/t) = e^{\epsilon \gamma_E} \left(\mu^2 \over -t \right)^\epsilon \cos(\pi \epsilon) \Gamma(-\epsilon) \Gamma(1+2 \epsilon) \,.
\end{align}
These results are independent of the IR regulator $m$, with $t$ regulating the IR region. For the Wilson line graphs, we find
\begin{align}
\text{Fig.~\ref{fig:collinearAD}c} &= - (4 \pi \alpha_s)^2 (2C_F - C_A)  \int \dbar^d k  { \iota^\epsilon \mu^{2\epsilon} |\bar n \cdot k|^{-\eta} \nu^\eta  {\bar n} \cdot p_1   \over (k^2 - m^2)  (k+p_1)^2 {\bar n} \cdot k}     {\bar u}_{\bar n} \gamma_\perp^\mu T^A { \slashed{q}_\perp \over q_\perp^2} 
\gamma_\perp^\nu T^B u_n + \dots \nn  \\ 
&= -i 4 \pi \alpha_s {\bar u}_{\bar n} \gamma_\perp^\mu T^A { \slashed{q}_\perp \over q_\perp^2} \gamma_\perp^\nu T^B u_n \ {\alpha_s \over 2 \pi} \left( C_F - {C_A \over 2} \right)  {h (\epsilon, \mu^2/m^2) \over \eta}  + \dots \,, \\
\text{Fig.~\ref{fig:collinearAD}d} &= - (4 \pi \alpha_s)^2  C_A  \int \dbar^d k  {\iota^\epsilon \mu^{2\epsilon} |\bar n \cdot k|^{-\eta} \nu^\eta  {\bar n} \cdot p_4   \over (k^2 - m^2)  (k+p_4)^2 {\bar n} \cdot k}     {\bar u}_{\bar n} \gamma_\perp^\mu T^A { \slashed{q}_\perp \over q_\perp^2} 
\gamma_\perp^\nu T^B u_n  +\dots \nn  \\ 
&= -i 4 \pi \alpha_s {\bar u}_{\bar n} \gamma_\perp^\mu T^A { \slashed{q}_\perp \over q_\perp^2} \gamma_\perp^\nu T^B u_n \ {\alpha_s \over 2 \pi} {C_A \over 2}  {h (\epsilon, \mu^2/m^2) \over \eta}  + \dots \,,
\end{align}
where
\begin{align}
h(\epsilon, \mu^2/m^2) = e^{\epsilon \gamma_E} \left(\mu^2 \over m^2\right)^\epsilon  \Gamma(\epsilon)  \,.
\end{align}
Here we see an explicit dependence on the IR regulator $m$.
Note that the $C_A$  dependence of the $1/\eta$ pole cancels in the sum for both the V graphs and Wilson line graphs. Upon summing all diagrams in Fig.~\ref{fig:collinearAD}, we find
\begin{align}
\delta V_n &=  { \alpha_s C_F  \over 2\pi} \left[ {g (\epsilon, \mu^2/t)+ h (\epsilon, \mu^2/m^2) \over \eta} \right] \,,\nn \\
\gamma^\nu_{\cO_n} &= { \alpha_s C_F  \over 2\pi} \bigg[ g (\epsilon, \mu^2/t)+ h (\epsilon, \mu^2/m^2) \bigg] = { \alpha_s C_F  \over 2\pi} \text{ln}\left( -t \over m^2 \right) \label{eq:gamman}\,,
\end{align}
where we expanded in $\epsilon$ in the final result for $\gamma^\nu_{\cO_n} $. This result is the same as for the Glauber gluon case up to Casimir scaling.

\subsection{One-Loop Virtual Anomalous Dimension for the Soft Operator}\label{sec:one_loop_soft}
\begin{figure}
\begin{center}
\begin{tabular}{cc}
\fd{3.4cm}{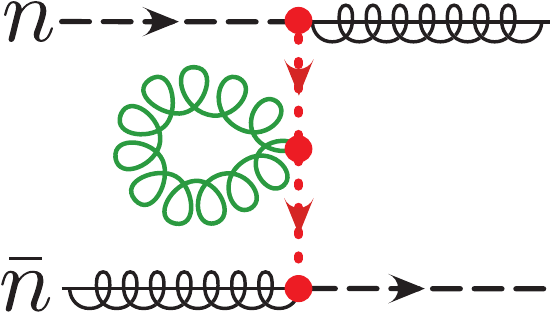} &
\fd{2.6cm}{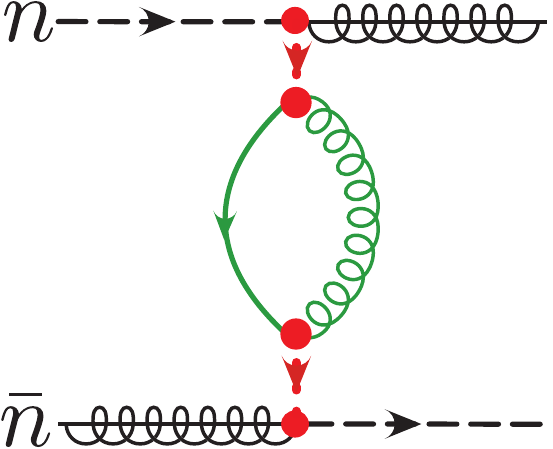} \\[30pt]
a) & b)
\end{tabular}
\vspace{-0.4cm}
\end{center}
\caption{One-loop virtual contributions to the renormalization of the soft operator $\cO_s$. The flower graph is labeled a) and the eye graph is labeled b).
}\label{fig:softAD}
\end{figure}

The result for the anomalous dimension $\gamma^\nu_{\cO_n}$ in Eq.~(\ref{eq:gamman}), along with the relations in Eq.~(\ref{eq:ADrelations}), specify the complete set of anomalous dimensions for our operators. Nonetheless, in this section we explicitly compute the renormalization of the soft operator $\cO_s$, verifying the structure of the operator mixing and the result for the combination $\gamma^{\text{dir}}_{s\nu} + \gamma^{T}_{s\nu}$. 

The relevant diagrams are shown in \Fig{fig:softAD}, which we refer to as the flower graph, and the eye graph. As in the previous section, all integrals can be performed using techniques from~\cite{Rothstein:2016bsq}, so we present only the final results, and again we keep only terms that contribute to the $1/\eta$ pole, as required for the rapidity renormalization. 
For the flower diagram, we find 
\begin{align}
\text{Fig.~\ref{fig:softAD}a}&= - (4 \pi \alpha_s)^2 2 C_F {\bar u}_{\bar n} \gamma_\perp^\mu T^A { \slashed{q}_\perp \over q_\perp^2} \gamma_\perp^\nu T^B u_n  \int \dbar^d k \frac{\iota^\epsilon \mu^{2\epsilon} |2k_z|^{-\eta} \nu^\eta}{(k^2-m^2) n\cdot k \bar n \cdot k}  +\dots \nn \\
&=  -i 4 \pi \alpha_s {\bar u}_{\bar n} \gamma_\perp^\mu T^A { \slashed{q}_\perp \over q_\perp^2} \gamma_\perp^\nu T^B u_n \left[  - {\alpha_s \over \pi} C_F  {h (\epsilon, \mu^2/m^2) \over \eta}   \right]+\dots \,.
\end{align}
For the eye diagram, we find
\begin{align}
\text{Fig.~\ref{fig:softAD}b}&=  -(4\pi\alpha_s)^2 2 C_F  {\bar u}_{\bar n} \gamma_\perp^\mu T^A { \slashed{q}_\perp \over q_\perp^2} \left[ \int \dbar^d k {\iota^\epsilon \mu^{2\epsilon} |2k_z|^{-\eta} \nu^\eta  \slashed{k}_\perp (\slashed{k}_\perp + \slashed{q}_\perp) \slashed{k}_\perp \over  (k^2-m^2) (k + q)^2 n \cdot k {\bar n} \cdot k } \right]  { \slashed{q}_\perp \over q_\perp^2} \gamma_\perp^\nu T^B u_n +\dots \nn \\
&= -i 4 \pi \alpha_s {\bar u}_{\bar n} \gamma_\perp^\mu T^A { \slashed{q}_\perp \over q_\perp^2} \gamma_\perp^\nu T^B u_n  \left[  - {\alpha_s \over \pi} C_F  {g (\epsilon, \mu^2/t) \over \eta}   \right]  +\dots \,. 
\end{align}
These results determine the counterterms and anomalous dimensions as
\begin{align}
\delta V_s &= - {\alpha_s \over \pi} C_F  {h (\epsilon, \mu^2/m^2) \over \eta} \,, \qquad 
\delta V_s^T = - {\alpha_s \over \pi} C_F  {g (\epsilon, \mu^2/t) \over \eta} \,, \nn\\
\gamma_{s \nu}^\text{dir} &=  - {\alpha_s \over \pi} C_F  h (\epsilon, \mu^2/m^2)\,, \quad \gamma_{s \nu}^T = - {\alpha_s \over \pi} C_F  g (\epsilon, \mu^2/t)  \, ,
\end{align}
consistent with those for the collinear sector. In the next section, we will solve the RGE and see that the anomalous dimension fixes the form of the Regge trajectory.

\subsection{Solving the Rapidity RGE}\label{sec:RG_solve}

With the anomalous dimensions in hand, it is now straightforward to achieve amplitude level Reggeization through solving the rapidity RGE.
We have the rapidity anomalous dimensions $\gamma^\nu_{\cO_n}$ for the collinear operator $\cO_n$ and $\gamma^{\text{dir}}_{s\nu} + \gamma^{T}_{s\nu}$ for the soft operator $\cO_s$, which satisfy the required consistency relations in Eq.~(\ref{eq:ADrelations}), This ensures that we can equivalently either run the collinear operators to the soft scale, or the soft operators to the collinear scale. We choose to run the collinear operators to the soft scale. The rapidity RGE is given by
\begin{align}
\nu \frac{d}{d\nu} \cO_n(\nu) =\gamma^\nu_{\cO_n} \cO_n(\nu)\,,
\end{align}
where the argument explicitly denotes the dependence on the $\nu$ scale (the $\mu$ scale does not enter our analysis). Since the anomalous dimension is independent of $\nu$, the solution is
\begin{align}
\cO_n\big(\sqrt{-t}\big)=\left( \frac{s}{-t}  \right)^{-\frac12 \gamma^\nu_{\cO_n} } \cO_n\big(\sqrt{s}\big)\,,
\end{align}
with an analogous expression for the $\bn$-collinear sector. Upon substituting the evolved collinear operators into the forward scattering operator, we find
\begin{align}\label{eq:evolvedscattering}
\cO_{n\nbar}
&=\left( \frac{s}{-t}  \right)^{-\frac{\alpha_s(\mu)C_F}{2\pi}  \log\left( \frac{-t}{m^2}  \right)}    \bar{\cO}_n\big(\sqrt{s}\big) \frac{1}{\Sl{\cP}_\perp}  \cO_s\big(\sqrt{-t}\big)   \frac{1}{\Sl{\cP}_\perp} \cO_{\bar n}\big(\sqrt{s}\big) \,,
\end{align}
which is the one-loop Reggeization of the quark. We emphasize again that we have not decomposed this result into amplitudes of definite signature. At LL order, $\log(s/|t|)$ and $\log(-s/|t|)$ are equivalent, and only differ at NLL order. The one-loop Regge trajectory for the quark is given by the exponent in \Eq{eq:evolvedscattering}:
\begin{equation}
\omega_q=-\frac{\alpha_s(\mu)C_F}{2\pi}  \log\left( \frac{-t}{m^2}  \right)\,,
\end{equation}
which agrees with the known result~\cite{Fadin:1976nw,Bogdan:2002sr}. Here it emerges directly from the rapidity renormalization of operators in the SCET subleading power Lagrangian. The one-loop quark Regge trajectory is identical to that for the gluon up to Casimir scaling, $C_A \to  C_F$. In a physical cross section, the dependence on the IR cutoff $m$ is cancelled by real emission diagrams, leading to an IR finite result. In \Sec{sec:quark_BFKL}, we will consider Reggeization at the cross section level for $q\bar q \to \gamma \gamma$, which will lead to the IR finite BFKL equation.

\subsection{Glauber Boxes}\label{sec:glauber_box}

\begin{figure}
\begin{center}
\begin{tabular}{cccc}
\hspace{-0.4cm} \fd{3.5cm}{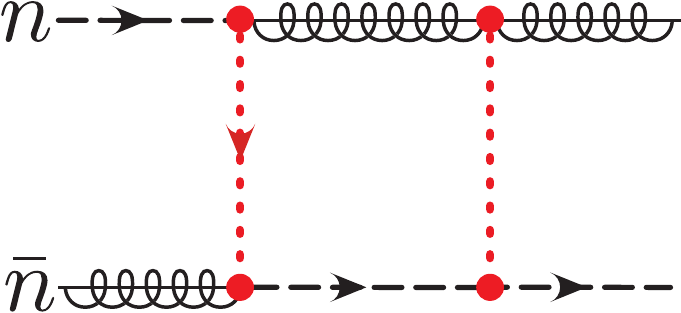}  &
\hspace{0.15cm}\fd{3.5cm}{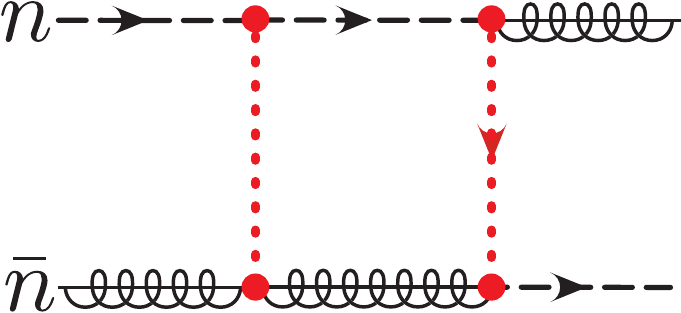} &
 \hspace{0.15cm}\fd{3.5cm}{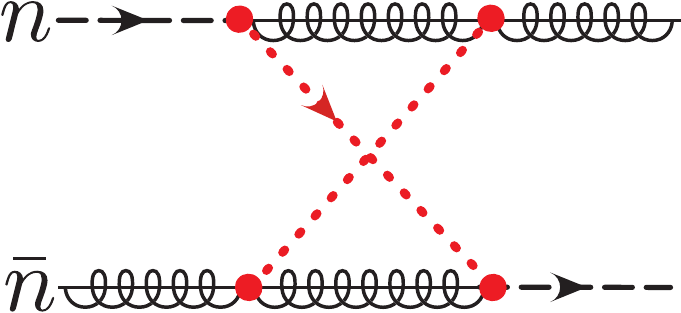} &
\hspace{0.15cm}\fd{3.5cm}{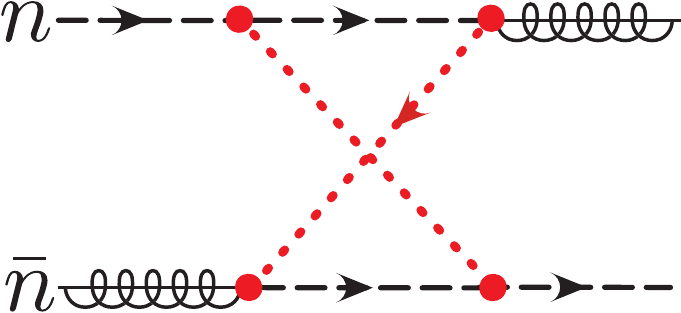}  
\\[25pt] 
a)  & b) &
  c) & d)
\end{tabular}
\vspace{-0.5cm}
\end{center}
\caption{Graphs contributing to the $\bar 6$ and $15$ color structures of the t-channel exchange. The cross box diagrams labeled c) and d) vanish with our regulator. 
}\label{fig:boxes}
\end{figure}

So far, in this section we have focused on the structure of the rapidity divergent contributions, which lead to the Reggeization of the $3$ color channel. At $\cO(\alpha_s^2)$ there are also non-vanishing contributions to the $\bar 6$ and $15$ color channels, which are known in the literature~\cite{Bogdan:2002sr}. In this section, we show that these are reproduced in a very simple manner in our framework by the simultaneous exchange of a Glauber quark and a Glauber gluon, as shown in \Fig{fig:boxes}.

As discussed in detail in~\cite{Rothstein:2016bsq}, the box graphs with Glauber scaling for the loop momentum require the rapidity regulator $|2k^z|^{-\eta} \nu^\eta$ to make them well defined (but are independent of $\eta$ as $\eta \to 0$). In particular, the Glauber cross box diagram vanishes due to having poles in $k^0$ on the same side of the contour. This is crucial since the box and cross box diagrams have different color factors, and thus illustrates the nontrivial mapping between the calculations in the EFT defined with our regulator, and full QCD. The ability to reproduce the known results for the $\bar 6$ and $15$ channels therefore provides a non-trivial test of the regulator, and of the EFT simultaneously involving quark and gluon Glauber operators.

Since the Glauber cross boxes shown in Fig.~\ref{fig:boxes}c and Fig.~\ref{fig:boxes}d vanish, we only compute the boxes shown in Fig.~\ref{fig:boxes}a and Fig.~\ref{fig:boxes}b. The $k^0$ and $k^z$ integrations are the same as for the box graphs with only Glauber gluons considered in~\cite{Rothstein:2016bsq}, while the $k_\perp$ integration is modified by the presence of the Glauber quark. Employing the results for the integrals in~\cite{Rothstein:2016bsq}, we find
\begin{align}
\text{Fig.~\ref{fig:boxes}a}   &= -\delta^{AB}\delta_{ij}   2 \pi^2 \alpha_s^2 \bar{u}_{\bar n}  \gamma_\perp^\mu  \left[ \left( -i \over 4\pi \right)\int  {\dbar^{d-2} k_\perp \slashed{k}_\perp (-i \pi) \over \vec{k}_\perp^2  (\vec{k}_\perp +\vec{q}_\perp)^2}    \right] \gamma_\perp^\nu u_n  \,, \\
\text{Fig.~\ref{fig:boxes}b} &=    \delta^{AB} \delta_{ij}  2 \pi^2 \alpha_s^2 \bar{u}_{\bar n}  \gamma_\perp^\mu \left[ \left( -i \over 4\pi \right) \int {\dbar^{d-2} k_\perp(\slashed{k}_\perp + \slashed{q}_\perp) (-i \pi) \over \vec{k}_\perp^2  (\vec{k}_\perp +\vec{q}_\perp)^2}   \right] \gamma_\perp^\nu u_n \,.
\end{align}
Just like for the exchange of two Glauber gluons, these box diagrams yield ``$i\pi$" factors that are characteristic of Glauber loops. Here we have simplified the color structure as $(T^D T^A T^C)_{ij} f^{BCD} = i\delta^{AB} \delta_{ij}/4 $. The sum of the diagrams is
\begin{align}
\text{Fig.~\ref{fig:boxes}a}    + \text{Fig.~\ref{fig:boxes}b}   &=   \left[ -i 4 \pi \alpha_s \bar{u}_{\bar n}  \gamma_\perp^\mu { \slashed{q}_\perp \over q_\perp^2} \gamma_\perp^\nu u_n \right]   \delta^{AB} \delta_{ij} { \alpha_s \over 4\pi} \left[ -{1 \over \epsilon} - \log { \mu^2 \over -t} \right]   (-i \pi) \, . \label{eq:boxes}
\end{align}
From Eq.~(\ref{eq:boxes}) we find a nonzero contribution to the color amplitude $\cC$ in the decomposition of Eqs.~(\ref{eq:Colordecomp}-\ref{eq:Colordecomp2}), and thus the contributions to the $\bar 6$ and $15$ color structures are
\begin{align}
\cM_{\bar 6} = \cM_{15} =  \left[ -i 4 \pi \alpha_s \bar{u}_{\bar n}  \gamma_\perp^\mu { \slashed{q}_\perp \over q_\perp^2} \gamma_\perp^\nu u_n \right] { \alpha_s \over 4\pi} \left[ -{1 \over \epsilon} - \log { \mu^2 \over -t} \right]   (-i \pi) \, ,
\end{align}
which agrees with the results of~\cite{Bogdan:2002sr} upon accounting for conventions.

\section{BFKL for $q\bar q\to \gamma \gamma$}\label{sec:quark_BFKL}

In this section we consider the application of Glauber quark operators for $q\bar q\to \gamma \gamma$ forward scattering. In QED, fermion Reggeization in the process $e^+e^- \to \gamma \gamma$ was studied in~\cite{Sen:1982xv}. Here we will follow the framework laid out in~\cite{Rothstein:2016bsq}, where the BFKL equation was derived from the rapidity renormalization of Glauber gluon operators at the cross section level. With Glauber operators in the effective theory, one can no longer factorize soft and collinear dynamics to all orders. However, with any fixed number of Glauber exchanges, the factorization is still possible, and therefore one can consider an expansion in the number of Glauber operator insertions. The first term in this expansion has a single Glauber gluon on either side of the cut and is referred to as the Low-Nussinov Pomeron approximation. This was used in~\cite{Rothstein:2016bsq} to derive the BFKL equation at LL order.

Unlike for the gluon BFKL, where one must consider an arbitrary number of Glauber operator insertions, for the case of quark Reggeization, the Glauber quark operators have an explicit power suppression, and therefore cannot be iteratively inserted. Instead, we must consider a single quark Glauber operator insertion on either side of the cut plus an arbitrary number of Glauber gluon operator insertions with $\cL_G^{(0)}$. To proceed, one must therefore still expand in the number of leading power Glauber gluon exchanges. To LL accuracy the situation simplifies significantly, and we only need to consider the factorization of the forward scattering matrix element with a single quark Glauber insertion on either side of the cut. Following \cite{Rothstein:2016bsq}, we can write the transition matrix element as
\begin{align}
T^q_{(1,1)} = \int d^2 q_\perp d^2 q^\prime_\perp C^q_n(q_\perp, p^-) S^q(q_\perp, q_\perp^\prime) C^q_{\bar n} (q^\prime_\perp, p^{\prime +}) \,,
\end{align}
where $C^q_n(q_\perp, p^-)$ and $C^q_{\bar n} (q^\prime_\perp, p^{\prime +})$ are squared collinear matrix elements and $S^q(q_\perp, q_\perp^\prime)$ is a squared soft matrix element. The subscript $(1,1)$ indicates that there is a single quark Glauber exchange on either side of the cut and the $q$ superscript distinguishes these matrix elements from the matrix elements of operators of $\cL_G^{(0)}$ describing Glauber gluon exchange from \cite{Rothstein:2016bsq}. In evaluating the matrix elements above, large logs arise due to the interplay of collinear modes whose natural rapidity scale is $\sqrt{s}$ and soft modes whose natural rapidity scale is $\sqrt{-t}$. We will resum these logs by considering the renormalization of the transition amplitude $T^q_{(1,1)}$ at LL order, and we will find that the resulting evolution equation is the same as the BFKL equation~\cite{Kuraev:1977fs,Balitsky:1978ic} up to Casimir scaling.

\begin{figure}
\begin{center}
\begin{tabular}{ccc}
\fd{3.2cm}{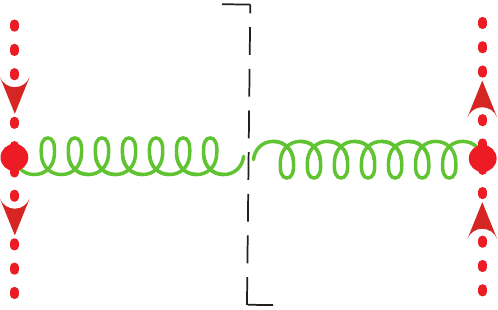}  &
\hspace{1.0cm}\fd{2.6cm}{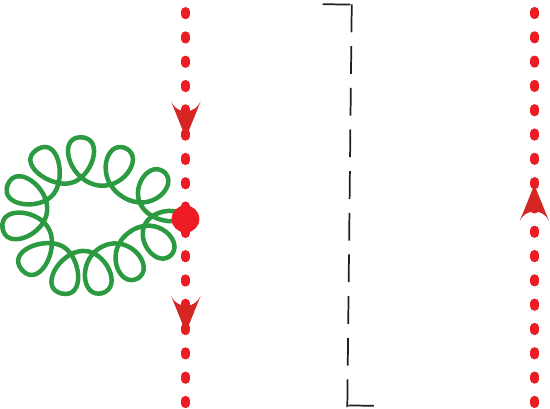} &
\hspace{1.0cm}\fd{2.6cm}{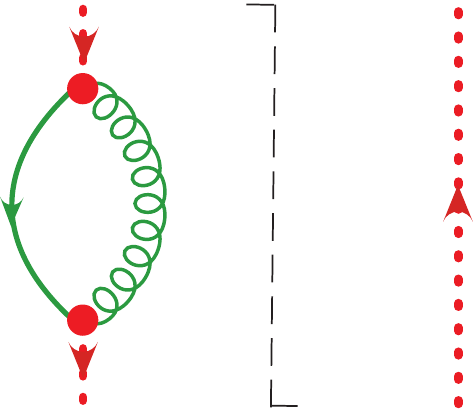} 
\\[30pt]
a)  & \hspace{1.3cm} b)  & \hspace{1.2cm} c)
\end{tabular}
\vspace{-0.5cm}
\end{center}
\caption{Graphs contributing to the LL order evolution of the soft function $S(q_\perp, q_\perp^\prime)$. The real contribution is labeled a), and the virtual contributions are labeled b) and c). The black dashed line represents the final state cut.
}\label{fig:BFKLgraphs}
\end{figure}

\subsection{BFKL Equation for the Soft Function}
Let us choose the rapidity scale in the renormalized transition matrix element  $T^q_{(1,1)}(\nu)$ to be $\nu = \sqrt{s}$, and consider the running of the soft function from $\nu= \sqrt{-t}$ to $\nu=\sqrt{s}$ to resum the large logs. This requires the one-loop real and virtual diagrams shown in Fig.~\ref{fig:BFKLgraphs}. In addition to these diagrams, there are also diagrams involving a Glauber gluon, and real soft quarks crossing the cut, coming from a power suppressed SCET Lagrangian. It is straightforward to show that such contributions are not rapidity divergent, which is expected, since the analogous virtual graphs are not associated with the Reggeization of the quark. For the calculations in this section we drop the mass regulator since IR divergences will cancel between the real and virtual contributions, and we set $d=4$ since only rapidity divergences are relevant for our analysis.

We define the soft function as
\begin{align}\label{eq:BFKLsoft}
S^q(q_\perp,q^\prime_\perp) = -{(2\pi)^4 \over V_2} {\delta^{ii'} \delta^{j j'} \over 
q^\mu_\perp q^{\prime \nu}_\perp \gamma^{\{ \mu}_{\alpha \bar{\alpha}} \gamma_{\beta \bar{\beta}}^{ \dagger \nu\} }} \sum_X \langle 0 | \cO_{s \alpha \bar{\alpha}} ^{ij}(q_\perp, q^\prime_\perp)  | X \rangle  \langle X | \cO_{s \beta \bar{\beta}}^{ \dagger i^\prime j^\prime}(q_\perp, q^\prime_\perp) |0 \rangle \,,
\end{align}
where the volume factor is $V_2 =(2\pi)^2 \delta^2(0) $, the color indices $i,j,i^\prime, j^\prime$ and fermionic indices $\alpha, {\bar \alpha}, \beta , {\bar \beta}$ have been made explicit, and for normalization we divide out by  $- q^\mu_\perp q^{\prime \nu}_\perp \gamma^{\{ \mu}_{\alpha \bar{\alpha}} \gamma_{\beta \bar{\beta}}^{ \dagger \nu\} }  = -\frac12 \{ \slashed{q}_\perp ^\prime \slashed{q}^\dagger_\perp + \slashed{q}_\perp \slashed{q}^{\prime \dagger}_\perp  \}$. 

We now compute the tree level and one-loop real and virtual contributions to the soft function. At tree level, the matrix element of the soft operator and the soft function obtained from squaring it are
\begin{align}\label{eq:BFKLtree}
\langle 0 | \cO_s^{ij}  | 0 \rangle = - i 4\pi \alpha_s \slashed{q}_\perp \delta^2(\vec{q}_\perp + \vec{q}^\prime_\perp ) \delta^{ij} \,, \qquad S^q_0(q_\perp,q^\prime_\perp) = (4\pi \alpha_s)^2 \delta^{ii} (2\pi)^2 \delta^2(\vec{q}_\perp + \vec{q}^\prime_\perp ) \, .
\end{align}

For the $\cO(\alpha_s)$ real contribution shown in Fig.~\ref{fig:BFKLgraphs}a, we compute the square of the one-gluon Feynman rule from the Fadin-Sherman vertex. Upon summing over gluon polarizations in Feynman gauge, we find
\begin{align}
{(2\pi)^4 \over V_2} \langle 0 | \cO_S^{ij}  | g \rangle  \langle g | \cO_S ^{ij \dagger}   | 0 \rangle = -(4\pi \alpha_s)^3 2C_F \delta^{ii} (2\pi)^2 \delta^2(\vec{q}_\perp + \vec{q}^\prime_\perp + \vec{k}_\perp ) { \{ \slashed{q}_\perp ^\prime \slashed{q}^\dagger_\perp + \slashed{q}_\perp \slashed{q}^{\prime \dagger}_\perp  \}  \over n \cdot k {\bar n} \cdot k}+\cdots\,,
\end{align}
where we have dropped the term having $\gamma^\mu_{\alpha \bar{\alpha}\perp} \gamma^{\dagger \mu}_{\beta \bar{\beta} \perp }$, which is rapidity finite. Using this result in Eq.~(\ref{eq:BFKLsoft}) we find the contribution to the soft function
\begin{align}
S_{1}^{q,\text{real}} = {\alpha_s C_F \over \pi^2 } \Gamma \left[ \eta \over 2 \right] \int {d^2k_\perp \over (\vec{k}_\perp - \vec{q}_\perp)^2 } S^q_{0}(k_\perp, q^\prime_\perp) +\cdots \, ,
\end{align}
where we have included the integral over phase space and identified the tree-level soft function. The ellipses denote rapidity finite contributions that will not play a role in the rapidity renormalization.

For the virtual corrections, we have the same flower and eye graphs appearing in the analysis for quark Reggeization in Sec.~\ref{sec:one_loop_soft}. As before, we keep only rapidity divergent contributions. The flower graph, appearing in Fig.~\ref{fig:BFKLgraphs}b, is given by
\begin{align}
\fd{1.0cm}{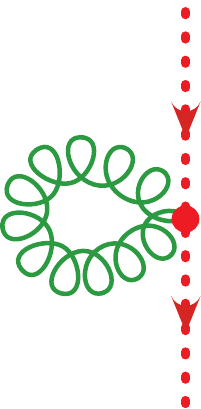} = -2(4 \pi \alpha_s)^2  C_F \delta_{ij} \int \dbar^4 k { w^2 |2 k_z|^{-\eta} \nu^{\eta} \slashed{q}_\perp \over k^2~ n \cdot k~ \bar{n} \cdot k}  \delta^2(\vec{q}_\perp + \vec{q}^\prime_\perp ) + \dots \, ,
\end{align}
where the ellipses denote rapidity finite terms. The eye graph, appearing in Fig.~\ref{fig:BFKLgraphs}c, is given by
\begin{align}
\fd{0.85cm}{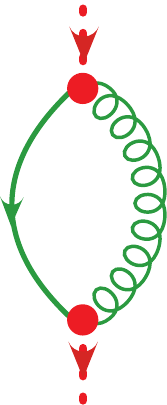} &=- 2(4\pi \alpha_s)^2  C_F \delta_{ij} \int \dbar^4 k { w^2 |2 k_z|^{-\eta} \nu^{\eta} \slashed{k}_\perp (\slashed{k} + \slashed{q}_\perp) \slashed{k}_\perp \over k^2~ (k+q_\perp)^2 ~n \cdot k~ \bar{n} \cdot k} \delta^2(\vec{q}_\perp + \vec{q}^\prime_\perp ) +\cdots  \\
&=2 (4\pi \alpha_s)^2 C_F \delta_{ij}  \int \dbar^4 k { w^2 |2 k_z|^{-\eta} \nu^{\eta} \over n \cdot k~ \bar{n} \cdot k} \left[  {\slashed{q}_\perp \over (k+q_\perp)^2}   + {q_\perp^2 \slashed{k} \over k^2 (k+q_\perp)^2} \right]  \delta^2(\vec{q}_\perp + \vec{q}^\prime_\perp ) +\cdots \,, \nn
\end{align}
where in the second line we dropped integrands that are odd in $k$. Note that the first term in the square brackets cancels the flower graph. The total virtual contribution is then
\begin{align}
\fd{1cm}{figures_BFKL/fermionic_flower_BFKL_oneside_low.pdf} \ \ +  \ \fd{0.85cm}{figures_BFKL/fermionic_eye_BFKL_oneside_low.pdf} \ = i 4\pi \alpha_s^2 C_F \delta_{ij} \Gamma \left[ \eta \over 2 \right] \int \dbar^2 k_\perp {\vec{q}_\perp^{\,\,2} \slashed{q}_\perp  \over \vec{k}_\perp^{\,2} (\vec{k}_\perp - \vec{q}_\perp)^2  }  \delta^2(\vec{q}_\perp + \vec{q}^\prime_\perp ) +\cdots\, .
\end{align}
We combine this result with the tree-level matrix element in Eq.~(\ref{eq:BFKLtree}) to obtain the squared matrix element. Hence the one-loop virtual contribution to the soft function is 
\begin{align}
S_{1}^{q,\text{virtual}}= -  {\alpha_s C_F \over 2 \pi^2}  \Gamma \left[ \eta \over 2 \right] \int d^2 k_\perp {\vec{q}_\perp^{\,\,2}  \over \vec{k}_\perp^{\,2} (\vec{k}_\perp - \vec{q}_\perp)^2  } S^q_0(q_\perp, q^\prime_\perp) +\cdots  \,. 
\end{align}

These results for the real and virtual corrections, $S_{1}^{q,\text{real}}$ and $S_{1}^{q,\text{virtual}}$, to the bare soft function are the same as in the gluon case up to Casimir scaling. Hence the rest of the analysis towards deriving the BFKL follows that of~\cite{Rothstein:2016bsq}, and we refer the reader there for further details. Let us mention a few key steps and then present the final evolution equation.
The rapidity divergence is multiplicatively renormalized with a $k_\perp$ convolution by a standard SCET soft function counterterm to cancel the $1/\eta$ divergence. Then the rapidity renormalization group follows from the $\nu$-independence of the bare soft function. The resulting RGE for $S^q(q_\perp, q^\prime_\perp)$ is precisely the leading log BFKL up to Casimir scaling:
\begin{align}
\nu \frac{d}{d\nu} S^q(q_\perp, q'_\perp,\nu)= \frac{2 C_F \, \alpha_s(\mu)}{\pi^2} \int d^2 k_\perp \left[  \frac{S^q(k_\perp, q'_\perp,\nu)}{(\vec k_\perp-\vec q_\perp)^2} -\frac{\vec{q}_\perp^{\,\,2} S^q(q_\perp, q'_\perp,\nu)}{2\vec k_\perp^2(\vec k_\perp-\vec q_\perp)^2} \right] \, .
\end{align}
Note that unlike the amplitude level Reggeization, the BFKL equation is IR finite due to the cancellation between the real and virtual emissions. 

Just as in \cite{Rothstein:2016bsq}, the rapidity RGE consistency,
\be
	0=\nu\frac{d}{d\nu} T^q_{(1,1)}  \implies 0 = \gamma_{S^q} + \gamma_{C_n^q} + \gamma_{C_\bn^q} = \gamma_{S^q} + 2\gamma_{C_n^q}\,,
\ee 
also implies a BFKL equation for the $n$-collinear function
\begin{align}
\nu \frac{d}{d\nu} C_n^q(q_\perp, p^-,\nu)= -\frac{ C_F \, \alpha_s(\mu)}{\pi^2} \int d^2 k_\perp \left[  \frac{C_n^q(k_\perp, p^-,\nu)}{(\vec k_\perp-\vec q_\perp)^2} -\frac{\vec{q}_\perp^{\,\,2} C_n^q(q_\perp, p^-,\nu)}{2\vec k_\perp^2(\vec k_\perp-\vec q_\perp)^2} \right] \,, 
\end{align}
and an analogous BFKL equation for $C^q_\bn$ with $(n,p^-,q_\perp)\leftrightarrow (\bn,p^{\prime +},q^\prime_\perp)$.

\section{Conclusions} \label{sec:conc}

In this paper we derived operators describing the exchange of Glauber quarks in the Regge limit, within the framework of the SCET.  
These Glauber quark operators describe certain soft and collinear gluon emissions to all orders in $\alpha_s$, and, for the case of a single soft gluon emission, reproduce the classic result of Fadin and Sherman~\cite{Fadin:1976nw,Fadin:1977jr}. From the rapidity renormalization of the Glauber quark operators, we derived the LL Reggeization of the quark and the LL BFKL equation for $q\bar q \to \gamma \gamma$. The rapidity renormalization gives rise to an interesting structure involving operator mixing between the $T$-product of two $\cO(\sqrt{\lambda})$ operators describing soft-collinear scattering, and an $\cO(\lambda)$ operator describing collinear-collinear scattering. We also showed that rapidity finite diagrams involving simultaneous Glauber quark and Glauber gluon exchanges quite simply reproduce known results in the $\bar 6$ and $15$ color channels, showing the consistency of our regulator. These results give a first view of the structure of the EFT for forward scattering in SCET at subleading power. 

There are a number of interesting directions for future study. In particular, it will be important to extend the study of Reggeization through renormalization group evolution to derive the two-loop Regge trajectory, both for the quark and the gluon. It is known that the two-loop quark Regge trajectory is related to the two-loop gluon Regge trajectory by Casimir scaling, $C_A \to C_F$~\cite{Bogdan:2002sr}, and it would be interesting to derive this property directly from the structure of Glauber operators, and to understand at what loop order it fails. Furthermore, now that the effective theory describes both quark and gluon Glauber exchanges, the structure of the higher logarithmic corrections for quantum numbers corresponding to compound Reggeon states can be studied using techniques in the effective theory. Finally, we have studied the subset of operators responsible for quark Reggeization at LL order, and it would be interesting to derive the complete set of power suppressed operators in the EFT for forward scattering, such as those describing subleading power corrections to the Regge trajectory of the gluon.

\vspace{0.2cm}
Note added: As this paper was being finalized, Ref.~\cite{Nefedov:2017qzc} appeared, which studies $\gamma \gamma \to q \bar q$ amplitudes at one-loop in the Regge limit by constructing the quark Reggeization terms in the effective action formalism of Lipatov~\cite{Lipatov:1995pn}. In the SCET language this corresponds to formulating an auxiliary field Lagrangian for the offshell Glauber quarks, while using the full QCD Lagrangian for other fields (without defining EFT fields for the $n$-collinear, soft and $\bar n$-collinear sectors).  Since having distinct fields for these sectors enables their factorization properties to be easily determined and studied, such as in our BFKL calculation, we believe there are certain advantages to our approach.  It would be interesting to make a more explicit comparison between these formalisms.

\chapter{Conclusions}\label{sec:conclusions_thesis}

In this thesis I have presented recent developments on the study of QCD beyond leading power. 

I have given a broad overview of the subject, highlighting the key ideas arising in the study of massless gauge theories beyond leading power, as well as its challenges and applications. 

Using and extending the framework of SCET, I have explored the ingredients of factorization beyond leading power constructing subleading hard scattering operator and radiative jet and soft functions. 
Combining these ingredients, I have motivated the reasons why the resummation at subleading power is structurally different from leading power already at leading logarithmic accuracy. 
I have shown that in order to overcome the challenges to achieve resummation at subleading power, the introduction of several new ideas was required.
This involve the introduction of subleading power jet and soft functions that don't arise through matching calculation, but that are generated through mixing effects in the RG evolution of subleading power of jet and soft functions. It also includes the derivation and solution of new RG equations that play a universal role in the resummation of observables at subleading power. 
Thanks to these findings, I have presented the first resummation of collinear and soft logarithms beyond leading power for an event shape observable in QCD.
While the results presented in \chap{subRGE} and published in \cite{Moult:2018jjd} are derived for Higgs thrust, many of the results derived in this work constitute universal elements of renormalization and resummation for massless gauge theories at subleading power. As we have shown in \cite{Moult:2019uhz}, the same framework allows the resummation of soft fermion emissions in $\cN=1$ SUSY QCD. As an other example, the $\theta$-soft function and its mixing under RG evolution with subleading power soft functions, as well as the solution of the subleading RGE equations derived in our work was later employed in the literature to resum threshold logarithms beyond leading power for Drell-Yan and Higgs production at the LHC \cite{Beneke:2018gvs,Beneke:2019mua}.

I have explained how the calculation of perturbative power corrections can be used to improve state-of-the-art fixed order calculations. In particular, the improvement of slicing methods, as $q_T$-subtraction~\cite{Catani:2007vq} or $N$-jettiness subtraction~\cite{Boughezal:2015aha,Gaunt:2015pea}, for the calculation of fully differential distributions at hadron colliders via the inclusion analytic power corrections will be crucial for pushing these methods to higher loops and multiplicity. Going forward, it will also be interesting to study the impact that analytically calculated power corrections can have on the improvement of matching to parton showers, as for example those based on $\Tau_0$ resummation like \texttt{GENEVA}~\cite{Alioli:2012fc,Alioli:2015toa}, for NNLO and N3LO cross sections.

I have illustrated and solved the subtleties related to the regularization of rapidity divergences beyond leading power for which I have introduced the pure rapidity regulator and its corresponding $\overline{\rm MS}$-like scheme, which handles rapidity divergences while maintaining the homogeneity of the power expansion.
As another example of how the study of tools for the systematic expansions of QCD in its infrared limits can benefit the precision program at modern and future colliders, I have presented a new method to employ cutting edge multiloop techniques for the computation of the expansions of cross sections in the collinear limit. I have shown how this method can be used to obtain state-of-the-art predictions for the infrared behavior of QCD and to construct systematically improvable analytic approximations of differential distributions at the LHC.

Finally, I have developed a formalism for the systematic treatment and resummation of processes mediated by the exchange of a Glauber quark. This provides and example of factorization, renormalization and resummation of effective field theory ingredients for the study of the Regge limit of amplitudes and the cross sections beyond leading power. 
As an application, I have shown how to obtain the quark reggeization via the rapidity renormalizion group equations of Glauber quark operators as well as how the BFKL resummation of small-$x$ logarithms for $q\bar{q} \to \gamma \gamma$ arises from the rapidity renormalizion group evolution of the Glauber quark soft function.
A natural direction for future work is the extension of the SCET framework  for forward scattering processes to include hard scattering operators. This would allow a systematic way of resumming high energy logarithms in hard scattering processes such as Higgs production in gluon and vector boson fusion and study their impact on cross sections, especially at future colliders. It would also be interesting to study the behavior of Glauber modes in the presence of subleading power operators and its implication for factorization or for the appearance of factorization breaking effects.

Going forward, another aspects that should be analyzed further is the study of processes where subleading power contributions are naturally enhanced, as for example the resummation of differential distributions in low-mass Drell-Yan~\cite{Bacchetta:2019tcu}. Also, the analysis of how subleading power contributions can be enhanced, or suppressed, via different choices of experimental cuts deserves further investigation.
Moreover, the application of QCD beyond leading power is crucial in the study processes that cannot be described with only leading power operators and factorization. The case of forward scattering mediated by a fermionic exchange presented in \chap{qregge} and published in \cite{Moult:2017xpp} is one example, but there is plenty of other cases, as for example processes due to soft quark emissions, Drell-Yan angular coefficients~\cite{Bacchetta:2019qkv,Ebert:fiducial}, transverse momentum distributions of quarkonia in peculiar spin configurations \cite{Fleming:2019pzj}, and spin asymmetries in Drell-Yan and SIDIS~\cite{Schweitzer:2010tt,Zhou:2010ui,Bacchetta:2019qkv}.

In this thesis a variety of first steps have been taken to initiate the study of gauge theories beyond leading power. With the range of new findings that have been achieved and the even larger variety of interesting questions that have appeared, this field is likely to continue to grow in the years ahead.

\chapter{Acknowledgments}

My time at MIT has been really rewarding and enjoyable and for this I am extremely grateful to many people.

First of all, I want to thank Iain. Yes, he has thaught me an infinite number of concepts in physics and everything I know about SCET (and EFTs in general), but that's not the main reason I want to thank him. I am particularly thankful to him for the many examples of leadership, for pushing me when I needed to be pushed, to give me time when I needed a break,  to be present when I needed support, to direct me when I needed guidance, and to give me freedom when I was able to experiment and try new ideas.

During my time at MIT I also had the pleasure and priviledge to work and interact with many great collaborators.
I want to thank in particular the postdocs I've mainly collaborated with. Ian, for answering all of my questions during the first projects we did together, while at the same time treating me like a peer when I was proposing new ideas, for all the great discussions about physics, life, and the future that we had in these 5 years of work together, as well as for the memorable memories of breakfast with unsalted tortellini and the Hello Kitty wine.
Markus, for all the great time and good physics (as well as steaks and beers) we had in the three years we overlapped at MIT and for being the German that every Italian needs.
Bernhard, for introducing me to the world of multiloop amplitudes and for giving me a different prospective on many aspects of our field.
I am thankful to have had the possibilty to work with great physicists like HuaXing, Frank, Kai, Lais, Patrick, and Mikhail, as well as the pleasure to tutor students like Cyuan-Han and Arindam. I am deeply grateful to Stefano Forte, without his guidance all of this would not even have begun.

I want to thank the Italian community at MIT. Being part of the MITaly both as a member and then as a president has contributed to carve a small piece of home, 6000 km away from it. Thanks for the dozens of events, parties, activities and dinners that we have organized and enjoyed together and for the friendships that are lasting even now that we are scattered around the world.

The list of people in the MIT community that made my time at MIT so enjoyable is too long to be included here, but I want to at least mention Ryuji, Yong, Sarah, Jasmine, Andrew, Hongwan, Lina, and all the people that have passed through the CTP in these years, my friends throughout the physics department, Masa and all the friends from the Japanese association for the joyful nights spent together drinking mojitos, sake and grappa, the people of the MIT FC, Spain@MIT and MITaly soccer teams for the countless matches played and trophies won. 
I want also to thank Scott at the CTP, Cathy in the physics department, Janka at the international student office and all the people in the MIT administration, that softened the impact of everything that happened outside of physics in these complicated years, pandemic included.

Mi riservo l'uso dell'italiano per gli ultimi, ma non meno importanti, ringraziamenti. 
Grazie mamma e grazie pap\`a per avermi fatto comprendere fin da piccolo che i confini geografici sono l\`i per essere attraversati, di farmi rendere conto che c'\`e sempre un contesto, un palcoscenico, un risultato pi\`u grande cui aspirare e di avermi insegnato l'importanza non solo di raggiungere i risultati, ma di meritarseli. Grazie per avermi insegnato tante cose per affrontare il mondo e di non aver avuto paura di lasciarmi andare, rendendomi orgoglioso del percorso intrapeso.
Chiudo ringraziando chi pi\`u di tutti mi \`e stata vicina e mi ha accompagnato e sostenuto durante il mio percorso ad MIT. Grazie amore per questo e per ogni momento di questi undici anni insieme.

\appendix
\chapter{Additional Tools for Subleading Power Factorization}

\section{BPS Identities}\label{app:BPS_identities}

In this appendix we collect a number of basic identities related to the BPS transformation. Given an operator $\hat{O}$ we denote the BPS transformed operator by $\BPS[\hat{O}]$. Using the fundamental representation for the ultrasoft Wilson lines $Y_n$ we have
\begin{align}
	&\BPS[W_n] = Y_n W_n^{(0)} Y_n^\dagger \,, &&\BPS[W_n^\dagger] = Y_n W_n^{(0)\dagger} Y_n^\dagger\,, \nn\\
	&\BPS[\chi_n] = Y_n \chi^{(0)}_n \,,	&&\BPS[\hat{O}_{us}] = \hat{O}_{us}  \,,\nn
\end{align}
\be
	\BPS[\cD_i] = \BPS[ W_n^\dagger D_i W_n] = Y_n W_n^{(0)\dagger} Y_n^\dagger\, \BPS[D_i]\, Y_n W_n^{(0)} Y_n^\dagger \,.
\ee
Ghost particles $c_n$, transform under BPS field redefinitions according to
\begin{align}
\BPS[c_n] = Y_n c_n Y_n^\dagger\,.
\end{align}
Other useful relations ($A,B$ are generic operators) are the following:
\begin{align}
	&[Y_n,\cP_\perp^\mu] = 0 \implies \cP_\perp^\mu = Y_n \cP_\perp^\mu Y_n^\dagger\,, \nn\\
	&Y_n^\dagger in\cdot D_{us} Y_n = i n \cdot \partial_{us}\,, \qquad &&Y_n^\dagger gn \cdot A_{us} Y_n = i n \cdot \partial_{us} - Y_n^\dagger in\cdot \partial_{us} Y_n \,,\nn\\
    &[ Y_n  A \,Y_n^\dagger\,,\, B] = Y_n\,[A\,,\,Y^\dagger_n B \,Y_n]\, Y_n^\dagger\,,\qquad
    &&[ Y_n A \, Y_n^\dagger, \, Y_n \,B\, Y_n^\dagger] = Y_n \,[A, B] \,Y^\dagger_n \label{eq:commextr}\,.
\end{align}
Using these relations we can compute the BPS field redefinition of the derivative operators appearing in the Lagrangian
\begin{align}
	\BPS[iD^\mu_{n \perp}] &\equiv \BPS[P_\perp^\mu +g A^\mu_{n \perp}] =  Y_n iD^{(0)\mu}_{n \perp} Y_n^\dagger\,,\nn \\
	\BPS[i\cD^\mu_{n\perp}] &\equiv \BPS[W_n^\dagger iD^\mu_{n \perp} W_n] = Y_n i\cD^{(0) \mu}_{n \perp} Y_n^\dagger\,,\nn \\
	\BPS[i\cD^\mu_{n}] &= Y_n \left( i\cD^{(0) \mu}_{n \perp} + \frac{n^\mu}{2} \cPbar \right) Y_n^\dagger + \frac{\bar n^\mu}{2}Y_n W_n^{(0)\dagger} \, ( Y_n^\dagger in\cdot \partial_{us}Y_n + g n\cdot A^{(0)}_n)\,  W_n^{(0)} Y_n^\dagger \nn \\
	&= Y_n \left( i\cD^{(0) \mu}_{n \perp} + \frac{n^\mu}{2} \cPbar \right) Y_n^\dagger + \frac{\bar n^\mu}{2}Y_n W_n^{(0)\dagger} \, ( Y_n^\dagger [in\cdot \partial_{us}Y_n] + in\cdot \partial_{us} + g n\cdot A^{(0)}_n)\,  W_n^{(0)} Y_n^\dagger \nn \\
	&= Y_n i\cD^{(0) \mu}_{n } Y_n^\dagger + \frac{\bar n^\mu}{2}Y_n W_n^{(0)\dagger} \, Y_n^\dagger [in\cdot \partial_{us}Y_n]  W_n^{(0)} Y_n^\dagger\,.
\end{align}
A less trivial calculation is how $i\cD^\mu_{ns}$ transforms under BPS field redefinition. We have
\begin{align}
	\BPS[i\cD^\mu_{ns}] &\equiv \BPS[W_n^\dagger iD^\mu_{ns} W_n] = \BPS[i\cD_n^\mu] + \frac{\bn^\mu}{2} Y_n W_n^{(0)\dagger} Y_n^\dagger\, gn \cdot A_{us} \, Y_n W_n^{(0)} Y_n^\dagger  \nn\\ 
	&= \BPS[i\cD_n^\mu] - \frac{\bn^\mu}{2} Y_n W_n^{(0)\dagger} Y_n^\dagger\, in \cdot \partial_{us} \, Y_n W_n^{(0)} Y_n^\dagger + \frac{\bn^\mu}{2} Y_n W_n^{(0)\dagger} in\cdot \partial_{us} W_n^{(0)} Y_n^\dagger \nn\\
	&= Y_n \left( i\cD^{(0) \mu}_{n \perp} + \frac{n^\mu}{2} \cPbar  + \frac{\bn^\mu}{2} in\cdot \cD^{0}_n \right) Y_n^\dagger \equiv Y_n i\cD^{(0) \mu}_{n } Y_n^\dagger \,.
\end{align}
The most important thing to note here is that 
\be
	\BPS[i\cD^\mu_{n \perp}] = Y_n i\cD^{(0)\mu}_{n \perp} Y_n^\dagger\,,\label{eq:Dperptransf}
\ee
but
\be
	\BPS[i\cD^\mu_{n}] \neq Y_n i\cD^{(0)\mu}_{n} Y_n^\dagger \,. \label{eq:Dstransf}
\ee

\section{Feynman Rules for Subleading Power Matching}\label{app:expand_gluon}

In this appendix we summarize for convenience several useful Feynman rules used in the text, both from the Higgs effective theory, and from SCET. 

The Feynman rules in the Higgs effective theory with
\begin{align}
O^{\text{hard}}=G^{\mu \nu}G_{\mu \nu} H\,,
\end{align}
are well known, and are given by
\begin{align}\label{eq:feynrule_eft}
\fd{2.25cm}{figures/heft_2g_low} &=-4i\delta^{ab} (p_1 \cdot p_2 g^{\rho \delta}  -p_1^\rho p_2^\delta )\,,  
\end{align}

\noindent\begin{minipage}{.5\linewidth}
\begin{align}
\hspace{3cm}\fd{3cm}{figures/heft_3g_low} \nn
\end{align}
\end{minipage}%
\hspace{-3cm}
\begin{minipage}{.6\linewidth}
\begin{align}
=&-4g f^{deg}(p_1^\rho g^{\delta \lambda} -p_1^\lambda g^{\rho \delta}) \nn \\
& -4g  f^{ged}(p_3^\rho g^{\delta \lambda} -p_3^\delta g^{\lambda \rho}) \nn \\
&-4g  f^{egd}(p_2^\lambda g^{\rho\delta} -p_2^\delta g^{\lambda \rho})\,, 
\end{align}
\end{minipage}

\noindent\begin{minipage}{.1\linewidth}
\begin{align}
\hspace{3.5cm}\fd{3.8cm}{figures/heft_4g_low} \nn
\end{align}
\end{minipage}%
\hspace{-2cm}
\begin{minipage}{.7\linewidth}
\begin{align}
=&4ig^2 ( f^{adf}f^{aeg}+f^{aef}f^{adg} )g^{\delta \rho} g^{\lambda \sigma} \nn \\
&  +4ig^2 (f^{ade}f^{afg}+f^{adg}f^{afe}) g^{\delta \lambda} g^{\rho \sigma} \nn \\
&+4ig^2 (f^{ade} f^{agf}+ f^{adf} f^{age})  g^{\delta \sigma} g^{\rho \lambda}\,. 
\end{align}
\end{minipage}\\

Before presenting the subleading power Feynman rules in SCET, we begin by briefly reviewing the Lagrangian, and gauge fixing for the collinear gluons. The gauge covariant derivatives that we will use to write the Lagrangian are defined by
\begin{align}
iD^\mu_n &= i\partial^\mu_n +g A^\mu_n\,, \qquad
 i \partial^\mu_n = \frac{\bn^\mu}{2} n \cdot \partial + \frac{n^\mu}{2} \overline{\cP} + \cP_\perp^\mu\,, \nn \\
 iD^\mu_{ns} &=i D^\mu_n +\frac{\bn^\mu}{2}gn \cdot A_{us}\,,\qquad
i\partial^\mu_{ns}=i \partial^\mu_n +\frac{\bn^\mu}{2} gn\cdot A_{us}\,,
\end{align}
and
\begin{align}
iD_{us}^\mu=i\partial^\mu+gA^\mu_{us}\,,
\end{align}
and their gauge invariant versions are given by
\begin{align}
i\cD^\mu_{n}=W_n^\dagger iD^\mu_{n} W_n\,, \nn \\
i\cD^\mu_{n\perp}=W_n^\dagger iD^\mu_{n\perp} W_n= \cP^\mu_{n\perp}+gB^\mu_{n\perp}\,, \nn \\
i \cD^\mu_{ns}=W_n^\dagger iD^\mu_{ns} W_n\,.
\end{align}
The leading power SCET Lagrangian can be written as
\begin{align} \label{eq:leadingLag_2}
\cL^{(0)} &= \cL^{(0)}_{n \xi} + \cL^{(0)}_{n g} +  \cL^{(0)}_{us}\,, 
\end{align}
where~\cite{Bauer:2001yt}
\begin{align}
\cL^{(0)}_{n \xi} &= \bar{\xi}_n\big(i n \cdot D_{ns} + i \slashed{D}_{n \perp} W_n \frac{1}{\overline{\cP}_n} W_n^\dagger i \slashed{D}_{n \perp} \big)  \frac{\slashed{\bar{n}}}{2} \xi_n\,, \\
\cL^{(0)}_{n g} &= \frac{1}{2 g^2} \tr \big\{ ([i D^\mu_{ns}, i D^\nu_{ns}])^2\big\} + \frac{1}{\alpha} \tr \big\{ ([i \partial^\mu_{ns},A_{n \mu}])^2\big\}+2 \tr \big\{\bar{c}_n [i \partial_\mu^{ns}, [i D^\mu_{ns},c_n]]\big\} \,, \nn
\end{align}
and the ultrasoft Lagrangian, $\cL^{(0)}_{us}$, is simply the QCD Lagrangian. We have used a covariant gauge with gauge fixing parameter $\alpha$ for the collinear gluons.

The $\mathcal{O}(\lambda)$ Lagrangian can be written 
\begin{align}
\cL^{(1)}={\cal L}_{\chi_n}^{(1)}+{\cal L}_{A_n}^{(1)}+{\cal L}_{\chi_n q_{us}}^{(1)} \,,
\end{align}
where \cite{Chay:2002vy,Pirjol:2002km,Manohar:2002fd,Bauer:2003mga}
\begin{align}\label{eq:sublagcollq_2}
{\cal L}_{\chi_n}^{(1)} &= \bar \chi_n \Big(
i \slashed{D}_{us\perp}\frac{1}{ \bar \cP} 
i \slashed{\cal D}_{n\perp}
+i \slashed{\cal D}_{n\perp}\, \frac{1}{\bar \cP} 
i \slashed{D}_{us\perp} 
\Big)\frac{\slashed{\bar{n}}}{2} \chi_n \ ,
 \\ 
{\cal L}_{A_n}^{(1)}&= \frac{2}{g^2}\text{Tr}\Big(
\big[ i {\cal D}_{ns}^\mu,i {\cal D}_{n\perp}^\nu \big]\big[
i {\cal D}_{ns\mu},iD^\perp_{us\,\nu} 
\big]
\Big)  
+ 2 \frac{1}{\alpha} \text{Tr} \left(   [i D_{us\perp}^\mu, A_{n\perp \mu}] [i\partial^\nu_{ns},A_{n\nu}] \right) \nn \\
&+2 \text{Tr}   \left( \bar c_n [iD_{us\perp}^\mu, [iD^\perp_{n\mu}, c_n  ]]  \right)          +2 \text{Tr}   \left( \bar c_n [\cP_\perp^\mu,[ W_n iD^\perp_{us\,\mu} W_n^\dagger, c_n   ]]  \right)    \,, \nn \\
{\cal L}_{\chi_n q_{us}}^{(1)} 
&= \bar{\chi}_n  g \slashed{\cB}_{n\perp} q_{us}+\text{h.c..}\nn
\end{align}

Finally, the $\cO(\lambda^2)$ Lagrangian can be written as \cite{Pirjol:2002km,Manohar:2002fd,Bauer:2003mga}
\begin{align}
	{\cal L}^{(2)} &= {\cal L}_{\chi_n}^{(2)}  + {\cal L}_{A_n}^{(2)} 
	+ {\cal L}_{\chi_n q_{us}}^{(2)} \,,
\end{align}
where
\begin{align}
	\cL_{\xi_n q_{us}}^{(2)}&=  \bar \chi_n \frac{\Sl \bn}{2} [ W_n^\dagger in\cdot D W_n]  q_{us}    + \bar \chi_n \frac{\Sl \bn}{2}  i\Sl \cD_{n\perp}    \frac{1}{\overline{\cP}}   ig \Sl \cB_{n\perp}  q_{us}+ \text{h.c.} \,, \\
	\cL_{n\xi}^{(2)}&=  \bar \chi_n   \left( i\Sl D_{us\perp} \frac{1}{\overline{\cP}}  i\Sl D_{us\perp}-  i\Sl \cD_{n\perp}   \frac{i\bn \cdot D_{us}}{(\overline{\cP})^2}   i\Sl \cD_{n\perp}   \right)    \frac{\Sl \bn}{2} \chi_n\,, \nn\\
	\cL_{ng}^{(2)}&=\frac{1}{g^2} \text{Tr} \left(  [  i\cD^\mu_{ns} , iD_{us}^{\perp \nu}   ]    [ i \cD_{ns\mu}   ,i D^\perp_{us\nu}   ] \right)   +\frac{1}{g^2} \text{Tr} \left(  [ iD^\mu_{us\perp}  ,i D^\nu_{us\perp}   ]    [ i\cD^\perp_{n\mu}  , i\cD^\perp_{n\nu}   ] \right) \nn \\
	&+ \frac{1}{g^2} \text{Tr} \left(  [ i\cD_{ns\mu}  , i n\cdot \cD_{ns}  ]    [  i\cD_{ns\mu} , i\bn \cdot D_{us}   ] \right)+\frac{1}{g^2} \text{Tr} \left(  [ iD^\mu_{us\perp}  ,  i\cD^\perp_{n\nu}  ]    [  i\cD^\perp_{n\mu}  ,  iD_{us \perp}^\nu  ] \right)\,, \nn\\
	\cL_{gf}^{(2)}&= \frac{1}{\alpha} \text{Tr} \left(   [ i D^\mu_{us\perp}, A_{n\perp \mu}]   [i D^\nu_{us\perp}, A_{n\perp \nu}] \right)   + \frac{1}{\alpha} \text{Tr} \left(   [i \bn \cdot D_{us},n \cdot A_n]   [i \partial^\mu_{ns}, A_{n\mu}] \right) \nn \\
	&+ 2 \text{Tr} \left(  \bar c_n [iD^\mu_{us\perp},[ W_n iD^\perp_{us\mu}W_n^\dagger ,c_n]] \right)+ \text{Tr} \left( \bar c_n [i \bn \cdot D_{us} ,[ i n\cdot D_{ns},c_n]] \right) \nn \\
	&+ \text{Tr} \left( \bar c_n [\overline{\cP},[W_n i \bn \cdot D_{us} W_n^\dagger,c_n]] \right)\,. \nn
\end{align}

Using these Lagrangians, one can derive the required Feynman rules for the calculations described in the text. The $\mathcal{O}(\lambda)$ Feynman rule for the emission of a ultrasoft gluon from a collinear gluon in a general covariant gauge, specified by a gauge fixing parameter $\alpha$, is given by
\begin{align}
\fd{2.5cm}{figures/soft_lam1_scet_low}=&-g f^{abc}\left[   g_{\perp}^{\nu\rho}\left(  \left(  1-\frac{1}{\alpha} \right)p^\mu_n-   \left(  1+\frac{1}{\alpha} \right)n\cdot p_s \frac{\bar n^\mu}{2}  -\frac{p_n^2 \bar n^\mu}{\bar n \cdot p_n}  \right)  \right. \nn \\
&-2g^{\mu \nu}p^\rho_{n\perp} +g^{\mu \rho}_\perp\left(   \left(  1-\frac{1}{\alpha} \right)p_n^\nu    -\frac{p_n^2 \bar n^\nu}{\bar n \cdot p_n}  \right) \nn \\
&\left.+\left( \bar n^\mu p^\nu_n+ \bar n^\nu p^\mu_n +\frac{1}{2}\bar n^\mu \bar n^\nu n\cdot p_s    \right)   \frac{p^\rho_{n\perp}}{\bar n \cdot p_n}\right]\,,
\end{align}
and the $\mathcal{O}(\lambda)$ propagator correction to the gluon propagator is given by
\begin{align}
\fd{2.0cm}{figures/feynman_rule_subleading_prop_gluon_low}&=-4i \delta^{ab}g^{\mu \nu} q_\perp \cdot q_{r\perp}+2i(1-\frac{1}{\alpha}) \delta^{ab}\left[  q_{r\perp}^\mu q^\nu +q^{\mu}q_{r\perp}^\nu  \right]\,.
\end{align}

For the matching calculation for the operators involving an ultrasoft derivative in \Sec{sec:us_deriv}, we also needed the $\mathcal{O}(\lambda^2)$ corrections to the propagator, which is given by
\begin{align}\label{eq:lam2_gluon_prop}
\fd{2.0cm}{figures/feynman_rule_subsubleading_prop_gluon_low}&=-i \delta^{ab} q_r^\perp \cdot q_r^\perp g^{\mu \nu}_\perp +i \delta^{ab} \left(  1-\frac{1}{\alpha}  \right) q_{r\perp}^\mu q_{r\perp}^\nu \nn \\
&+\frac{i}{2}\delta^{ab}\left(  1-\frac{1}{\alpha}  \right)(q_\perp^\mu n^\nu \bar n \cdot q_r +q_\perp^\nu n^\mu \bar n \cdot q_r ) +\cdots \,, 
\end{align}
where the dots indicate the other tensor components in the light cone basis, which are not relevant for the current discussion.
For simplicity, the matching was performed using a $\perp$ polarized gluon. In the $n$-collinear sector, the leading power hard scattering operator produces only $\bar n$, and $\perp$ polarized gluons. Therefore, only the $\perp-\perp$ and $n-\perp$ components of the propagator are needed.
In the matching, the $\perp-\perp$ term vanishes since it proportional to the residual $\perp$ momentum, which is set to zero, and the $n-\perp$ term vanishes for a $\perp$ polarized gluon, due to the gluons equation of motion, $q_\perp \cdot \epsilon_\perp=0$.

At $\mathcal{O}(\lambda^2)$, the individual propagator and emission factors are sufficiently complicated that it is also convenient to give the complete result for the matrix element
\begin{align} 
\fd{3cm}{figures/soft_lam2_scet_low}&=\left .\langle 0| T \{\cB^\nu_{n\perp}(0), \cL^{(2)}_{A_n}\} | \epsilon_n, p_n; \epsilon_s,p_s \rangle \right |_{\alpha=1}=\nn\\
&=-i f^{abc} \epsilon_{n\mu} \frac{2\epsilon_{s\rho} p_{s\sigma}}{\nbar \cdot p_n\, n\cdot p_s} \left(  g^{\mu \rho}_\perp g^{\sigma \nu}_\perp -    g^{\mu \sigma}_\perp g^{\rho \nu}_\perp  \right)\,,
\end{align}
where we have restricted to $\alpha=1$ for simplicity.

Since we have also matched to operators involving collinear quarks, we also summarize the subleading power  Feynman rules involving collinear quark. The Feynman rules for the correction to a collinear quark propagator are given by
\be
\fd{3cm}{figures/feynman_rule_subleading_prop_low}=i \frac{\Sl{\bar n}}{2}\frac{2p_\perp \cdot p_{r\perp}}{\bar n \cdot p}\,,
\ee
\be
\fd{3cm}{figures/feynman_rule_subsubleading_prop_low}=i \frac{\Sl{\bar n}}{2}\frac{p_{r\perp}^2}{\bar n\cdot p}\,,
\ee
and the Feynman rules for the emission of a collinear gluon are given by
\be
\fd{2cm}{figures/feynman_rule_collinear_emission_low}=i g T^a \left( n_\mu +\frac{\gamma^\perp_\mu \Sl{p}_\perp}{\bar n \cdot p}+\frac{\Sl{p}_\perp^{'} \gamma_\mu^\perp}{\bar n \cdot p'}- \frac{\Sl{p}_\perp \Sl{p}_\perp^{'}}{\bar n\cdot p \bar n \cdot p'}\bar n_\mu   \right) \frac{\Sl{\bar n}}{2}\,,
\ee
\begin{align}
&\fd{2cm}{figures/feynman_rule_collinear_emission_sub_low}= \\
 &i g T^a \left( \frac{\gamma^\perp_\mu \Sl{p}_{r\perp}}{\bar n \cdot p}+\frac{\Sl{p}_{r\perp}^{'} \gamma_\mu^\perp}{\bar n \cdot p'}+        \frac{\Sl{p}_{r\perp} \Sl{p}_{\perp}}{\bar n\cdot q \bar n \cdot p}\bar n_\mu                    -\frac{\Sl{p}_\perp^{'} \Sl{p}_{r\perp}^{'}}{\bar n\cdot q \bar n \cdot p'}\bar n_\mu-                    \frac{\Sl{p}_{r\perp}^{'} \Sl{p}_\perp}{\bar n\cdot q \bar n \cdot p'}\bar n_\mu+                     \frac{\Sl{p}_\perp^{'} \Sl{p}_{r\perp}}{\bar n\cdot q \bar n \cdot p'}\bar n_\mu \right) \frac{\Sl{\bar n}}{2}\,,\nn
\end{align}
\begin{align}
&\fd{2cm}{figures/feynman_rule_collinear_emission_sub_sub_low}=\\
	&ig T^a \left(\frac{\bar n ^\mu p_{r\perp}^2}{\bar n \cdot p}     -\frac{\bar n ^\mu p_{r\perp}^{'2}}{\bar n \cdot p'}     -\frac{\gamma^\mu_\perp \Sl{p}_\perp  \bar n \cdot p_r }{(\bar n \cdot p)^2}     -\frac{ \Sl{p}^{'}_\perp  \gamma^\mu_\perp \bar n \cdot p_r}{(\bar n \cdot p')^2}     -\frac{\bar n^\mu \Sl{p}^{'}_\perp \Sl{p}_\perp \bar n \cdot p_r}{\bar n \cdot q(\bar n \cdot p)^2}     +\frac{\bar n^\mu \Sl{p}^{'}_\perp \Sl{p}_\perp \bar n \cdot p_r}{\bar n \cdot q(\bar n \cdot p')^2}     \right)    \frac{\Sl{\bar n}}{2}\,.\nn
\end{align}
We can see that each term in the power suppressed collinear Lagrangian insertions are proportional to either $p_{r\perp}$, or $\bar n \cdot p_r$. At tree level, and in the absence of ultrasoft particles, one can use RPI to set all these terms to zero. This was used extensively to simplify our matching calculations.

For convenience we also give the expansion of the Wilson lines and collinear gluon field to two emissions. The collinear Wilson lines are defined by
\begin{align}
W_n =\left[  \sum\limits_{\text{perms}} \exp \left(  -\frac{g}{\bar \cP} \bar n \cdot A_n(x)  \right) \right]\,.
\end{align}
Expanded to two gluons with incoming momentum $k_1$ and $k_2$, we have
\begin{align}
W_n &=1-\frac{gT^a \bar n \cdot A_{nk}^a}{\bar n \cdot k}  +g^2 \left[   \frac{T^a T^b}{\bar n \cdot k_1 (\bar n \cdot k_1+\bar n \cdot k_2)} +\frac{T^b T^a}{\bar n \cdot k_2(\bar n \cdot k_1+\bar n \cdot k_2)}  \right]    \frac{\bar n \cdot A_{nk1}^a   \bar n \cdot A_{nk2}^b  }{2!}\,, \nn \\
W^\dagger_n& =1+\frac{gT^a \bar n \cdot A_{nk}^a}{\bar n \cdot k}  +g^2 \left[   \frac{T^a T^b}{\bar n \cdot k_1 (\bar n \cdot k_1+\bar n \cdot k_2)} +\frac{T^b T^a}{\bar n \cdot k_2(\bar n \cdot k_1+\bar n \cdot k_2)}  \right]    \frac{\bar n \cdot A_{nk1}^a   \bar n \cdot A_{nk2}^b  }{2!}\,.
\end{align}
The collinear gluon field is defined as
\begin{align}
\cB^\mu_{n\perp}=\frac{1}{g}\left[   W_n^\dagger i D^\mu_{n\perp}W_n \right]\,.
\end{align}
Expanded to two gluons, both with incoming momentum, we find
\begin{align}
g\cB^\mu_{n\perp}&=g\left(   A^{\mu a}_{\perp k} T^a -k^\mu_\perp \frac{\bar n \cdot A^a_{nk} T^a}{\bar n \cdot k} \right)+g^2(T^a T^b-T^b T^a) \frac{\bar n \cdot A^a_{nk1} A^{\mu b}_{\perp k2} }{\bar n \cdot k_1}  \\
&+g^2 (k^\mu_{1\perp} +k^\mu_{2\perp}) \left(   \frac{T^a T^b}{\bar n \cdot k_1 (\bar n \cdot k_1+\bar n \cdot k_2)}+\frac{T^b T^a}{\bar n \cdot k_2 (\bar n \cdot k_1+\bar n \cdot k_2)} \right)\frac{\bar n \cdot A_{n k1}^a \bar n \cdot A^b_{n k2}}{2!} \,. \nn
\end{align}
In both cases, at least one of the gluons in the two gluon expansion is not transversely polarized. Such terms can therefore be eliminated in matching calculations by choosing particular polarizations, as was done in the text. For the soft Wilson lines, we have
\begin{align}
S_n &=1-\frac{gT^a n \cdot A_{sk}^a}{ n \cdot k}  +g^2 \left[   \frac{T^a T^b}{n \cdot k_1 } +\frac{T^b T^a}{n \cdot k_2}  \right]    \frac{ n \cdot A_{sk_1}^a   n \cdot A_{sk_2}^b  }{2 n \cdot( k_1+ k_2)}+\cdots\,, \nn \\[3mm]
S^\dagger_n& =1+\frac{gT^a n \cdot A_{sk}^a}{ n \cdot k}  +g^2 \left[   \frac{T^a T^b}{n \cdot k_1} +\frac{T^b T^a}{ n \cdot k_2}  \right]    \frac{ n \cdot A_{sk_1}^a    n \cdot A_{s k_2}^b  }{2 n \cdot( k_1+ k_2)}+\cdots\,.
\end{align}
and
\begin{align}
g\cB^\mu_{s(n)\perp}&=g\left(   A^{\mu a}_{\perp k} T^a -k^\mu_\perp \frac{ n \cdot A^a_{sk} T^a}{ n \cdot k} \right)+g^2(T^a T^b-T^b T^a) \frac{ n \cdot A^a_{sk_1} A^{\mu b}_{\perp k_2} }{ n \cdot k_1}  \\
&+g^2 (k^\mu_{1\perp} +k^\mu_{2\perp}) \left(   \frac{T^a T^b}{ n \cdot k_1 }+\frac{T^b T^a}{n \cdot k_2 } \right)\frac{ n \cdot A_{s k_1}^a  n \cdot A^b_{s k_2}}{2n \cdot( k_1+ k_2)} \,. \nn
\end{align}
When evaluating diagrams involving the soft Glauber operators, the following combination is also useful
\begin{align}
S_n^\dagger S_{\bar n}&=1+   gT^a\left(\frac{ n \cdot A_{sk}^a}{ n \cdot k}  - \frac{ \bar n \cdot A_{sk}^a}{\bar  n \cdot k}\right) -g^2 T^a T^b \frac{n\cdot A_s^a}{n\cdot k} \frac{\bar n \cdot A_s^b}{\bar n \cdot k}\nn \\
&+g^2 \left[   \frac{T^a T^b}{n \cdot k_1 } +\frac{T^b T^a}{n \cdot k_2}  \right]    \frac{ n \cdot A_{sk_1}^a   n \cdot A_{sk_2}^b  }{2n \cdot( k_1+ k_2)}+g^2 \left[   \frac{T^a T^b}{\bar n \cdot k_1} +\frac{T^b T^a}{\bar n \cdot k_2}  \right]    \frac{ \bar  n \cdot A_{s k_1}^a  \bar n \cdot A_{s k_2}^b  }{2\bar n \cdot( k_1+ k_2)}\,.
\end{align}

\section{Fierzing for Radiative Jet Functions}\label{app:Fierzing}

In this appendix we collect some details related to the color and Dirac structure of the radiative jet functions. To obtain scalar jet and soft functions, one must factorize in Dirac and color space.  Factorization in Dirac and color space can be achieved using the SCET Fierz relation \cite{Lee:2004ja}
\begin{align}\label{eq:SCET_fierz}
\big(\delta^{\a'\a}\delta^{i'i} \big)\big( \delta^{\b\b'}\delta^{jj'}\big) 
&= \frac{1}{2}\sum\limits_{k=1}^6 (F_k^{\bar n})^{\b\a}_{ji} \otimes (F_k^n)^{\a'\b'}_{i'j'} \nn\\
&=\frac{1}{2}\Big[
\frac{\Sl{\bar n}}{2N_C} \otimes \frac{\Sl{n}}{2} 
- \frac{\Sl{\bar n}\gamma^5}{2N_C} \otimes \frac{\Sl{n}\gamma^5}{2} 
- \frac{\Sl{\bar n}\gamma^\alpha_\perp}{2N_C} \otimes \frac{\Sl{n}\gamma^\perp_\alpha}{2} \nn \\
&\hspace{1cm}
+ \Sl{\bar n}T^a \otimes \frac{\Sl{n}T^a}{2} 
- \Sl{\bar n}\gamma^5T^a \otimes \frac{\Sl{n}\gamma^5T^a}{2} 
- \Sl{\bar n}\gamma^\alpha_\perp T^a \otimes \frac{\Sl{n}\gamma^\perp_\alpha T^a}{2} 
\Big]\,.
\end{align}
In the text, we primarily focused on the convolution structure. Here we consider the derivation of the Dirac and color structure. 

Consider first the soft quark radiative jet function. Since we are only interested in the Dirac and color structure, it is notationally simplest to ignore all the measurement functions, and just consider the vacuum matrix element of the fields. For the radiative jet function involving a soft quark, we have from \Sec{sec:sub_cross_quark}
\begin{align}
\cM^{(2)}_{\psi_{us}}= \int d^4x d^4y~ &\langle 0 | \left[ \bar \chi_n(x) Y_n^\dagger(x) \gamma^\mu_\perp Y_{\bar n}(x) \chi_{\bar n}(x)   \right]    \left[ \bar \psi_{us(n)}(y) g \Sl{\cB}_{n\perp}(y) \chi_n(y)   \right] \nn \\
&\hspace{1cm}\cdot \left[  \bar \chi_n(z) g\Sl{\cB}_{n\perp}(z)   \psi_{us(n)}(z) \right] 
\left[  \bar \chi_{\bar n} (0) Y_{\bar n}^\dagger(0) \gamma^\mu_\perp Y_n(0) \chi_n(0)  \right] |0\rangle\,,
\end{align}
which will enter the expression for the cross section.
Applying the Fierz relation of \Eq{eq:SCET_fierz} three times, we obtain the factorized form of the matrix element
\begin{align}
\cM^{(2)}_{\psi_{us}}=\int d^4x d^4y~&\langle 0 | \left[  \bar \chi_{\bar n} (0) F^n_k \chi_{\bar n}(x)  \right] |0\rangle \langle 0 | \left[  \bar \chi_{n} (x) F^{\bar n}_{k'} \chi_{n}(0)  \right]  \left[ \bar \chi_n(z) g \Sl{\cB}_{n\perp}(z) F^{\bar n}_l g  \Sl{\cB}_{n\perp}(y) \chi_n(y)  \right]   |0\rangle  \nn \\
&\hspace{-0.9cm}\cdot\langle 0 |   \tr \left[ F^n_{k'} Y_n^\dagger (x) \gamma^\mu_\perp Y_{\bar n}(x) F^{\bar n}_k Y_{\bar n}^\dagger(0) \gamma^\mu_\perp Y_n(0)   \right]  \left[ \bar \psi_{us(n)}(y) F^n_l \psi_{us(n)}(z)   \right] |0\rangle\,.
\end{align}
The symmetries of the soft and collinear sectors can then be used to set $k=k'=l=1$. We can then further simplify the structure of the $n$-collinear matrix element by applying another Fierz relation. We then obtain
\begin{align}
\cM^{(2)}_{\psi_{us}}=\int d^4x d^4y&\langle 0 | \left[  \bar \chi_{\bar n} (0) \frac{\Sl n}{2} \chi_{\bar n}(x)  \right] |0\rangle  \langle 0 |   \tr \left[ Y_n^\dagger (x) Y_{\bar n}(x)  Y_{\bar n}^\dagger(0)  Y_n(0)   \right]  \left[ \bar \psi_{us(n)}(y) \frac{\Sl n}{2} \psi_{us(n)}(z)   \right] |0\rangle \nn \\
&\cdot \langle 0 | \left[  \bar \chi_{n} (x) \frac{\Sl {\bar n}}{2} \chi_n(y) \right]     \tr \left[ g \cB_{n\perp}(z)  \cdot g  \cB_{n\perp}(y)     \right]     \left[  \bar \chi_n(z)   \frac{\Sl {\bar n}}{2} \chi_n(0) \right]   |0\rangle 
\,.
\end{align}
Reinstating the time ordering and measurement function, this corresponds to the result given in \Sec{sec:sub_cross_quark}.

For the case of the radiative jet function involving a gluon emission, from \Sec{sec:sub_cross_gluon}, we must consider the factorization of the matrix element
\begin{align}
\cM^{(2)}_{\cB_{us}}=\langle 0 |    \left[ \bar \chi_n(x) Y_n^\dagger(x) \gamma^\mu_\perp Y_{\bar n}(x) \chi_{\bar n}(x)   \right] &   \bar \chi_n  \left[  T^a \gamma_{\perp \mu} \frac{1}{\bar \cP}  i \Sl \partial_{us\perp} - i  {\overleftarrow{\Sl \partial}}_{us\perp} \frac{1}{\bar \cP} T^a \gamma_{\perp \mu}   \right]   \frac{\Sl \bn}{2} \chi_n (y) \nn \\
&\cdot\left[  \bar \chi_{\bar n} (0) Y_{\bar n}^\dagger(0) \gamma^\mu_\perp Y_n(0) \chi_n(0)  \right]             |0 \rangle\,,
\end{align}
we can flip the action of the ultrasoft derivative onto the $\cB_{us}$ field, and apply twice the Fierz relation of \Eq{eq:SCET_fierz} to obtain
\begin{align}
\cM^{(2)}_{\cB_{us}}=&\langle 0 | \left[ \bar \chi_{\bar n}(0) \frac{\Sl n}{2}  \chi_{\bar n}(x)   \right] |0 \rangle \langle 0 | \left[ \bar \chi_n(x) F_{\bar n}^k \chi_n(y) \right] \frac{1}{\bar \cP}   \left[ \bar \chi_n(y) F_{\bar n}^{k'} \chi_n(0) \right]|0 \rangle \nn \\
&\hspace{2cm}\cdot\langle 0 | \tr \left[   (-i \Sl{\partial}_\perp g\Sl{\cB}_{us(n) \perp} (y))  \frac{\Sl{\bar n}}{2} F_n^k Y_n^\dagger (x) Y_{\bar n}(x) \frac{\Sl{\bar n}}{2} Y_{\bar n}^\dagger (0) Y_n(0) F_n^{k'} \right]\,.
\end{align}
The tree level contribution is with $k=k'=1$. Without having a closed quark loop from which the soft gluon field is emitted, the only possibilities are $k, k'= \frac{\Sl n}{2},  \frac{\Sl n}{2} T^a$. The color neutrality of the vacuum in the collinear sector implies that one would have to contract the color indices between $k$ and $k'$. It seems that one could potentially get this configuration from a fermion bubble type diagram, but this is first at two loops, and should not contribute to LL. (i.e. we know it doesn't from the explicit result, which doesn't have an $n_f$). Simplifying the Dirac structure, we then obtain
\begin{align}
\cM^{(2)}_{\cB_{us}}=&\langle 0 | \left[ \bar \chi_{\bar n}(0) \frac{\Sl n}{2}  \chi_{\bar n}(x)   \right] |0 \rangle \langle 0 | \left[ \bar \chi_n(x)  \frac{\Sl{\bar n}}{2} \chi_n(y) \right] \frac{1}{\bar \cP}   \left[ \bar \chi_n(y)  \frac{\Sl{\bar n}}{2} \chi_n(0) \right]|0 \rangle \nn \\
&\hspace{2cm}\cdot\langle 0 | \tr \left[   (-i \partial_\perp \cdot g\cB_{us(n) \perp} (y))  Y_n^\dagger (x) Y_{\bar n}(x)  Y_{\bar n}^\dagger (0) Y_n(0) \right]|0\rangle\,.
\end{align}
Reinstating the time ordering and measurement function, this corresponds to the result given in \Sec{sec:sub_cross_gluon}.

\section{Higher-Order Plus Distributions}
\label{app:plus_distr}

Subleading power corrections often involve divergences of the form
\begin{equation} \label{eq:power_law_n}
  \frac{1}{(1-z)^{a + \eta}} \,,\qquad a \in \mathbb{N}
\,.\end{equation}
In \sec{distribution} we encountered the two cases $a=2$ and $a=3$, which were treated using integration by parts to relate them to the case $a=1$, where one can use the relation
\begin{align} \label{eq:power_law_1_reg}
 \frac{1}{(1-z)^{1+\eta}} &
 = -\frac{\delta(1-z)}{\eta} + \biggl[ \frac{1}{(1-z)^{1+\eta}} \biggr]_+^1
\nn\\*&
 = -\frac{\delta(1-z)}{\eta} + \cL_0(1-z) - \eta \cL_1(1-z) + \cO(\eta^2)
\,.\end{align}
Here $\cL_n(x) = \bigl[\ln^n x/x\bigr]_+^1$ is defined in terms of standard plus distributions, which regulate functions $g(x)$ with support $x\ge0$ diverging less than $1/x^2$ as $x\to0$.
The defining properties of such plus distributions are
\begin{align} \label{eq:plus_dist_n1}
 \bigl[ g(x) \bigr]_+^1 &= g(x) \,,\qquad x \ne 0
\,,\nn\\*
 \int_0^1 \df x \, \bigl[ g(x) \bigr]_+^1 &= 0
\,.\end{align}
One can also treat the power-law divergences in \eq{power_law_n} similar to \eq{power_law_1_reg} using higher-order plus distributions.
We therefore generalize \eq{plus_dist_n1} as
\begin{align} \label{eq:plus_dist_n}
 \bigl[ g(x) \bigr]_{+(a)}^1 &= g(x) \,,\qquad x \ne 0
\,,\nn\\
 \int_0^1 \df x \, x^k \, \bigl[ g(x) \bigr]_{+(a)}^1 &= 0 \,,\hspace{1.2cm} \forall\, k < a
\,,\end{align}
where $g(x)$ has support $x\ge0$ and diverges less than $1/x^{1+a}$ as $x\to0$.
For $a=1$, this naturally reduces to \eq{plus_dist_n1}.
For $a=2$, one obtains the $++$ distributions used e.g.\ in \refcite{Mateu:2012nk}.

The distributions defined in \eq{plus_dist_n} can be integrated against any test function $f(x)$ that is at least $a{-}1$-times differentiable at $x=0$.
To be specific, consider the example integral
\begin{align} \label{eq:plus_dist_n_example}
 &\int_0^{x_0} \df x \, f(x) \bigl[ g(x) \bigr]_{+(a)}^1
\nn\\*&
 = \int_0^{x_0} \df x \, \biggl[ f(x) - \sum_{k=0}^{a-1} \frac{f^{(k)}(0)}{k!} x^k \biggr] \bigl[ g(x) \bigr]_{+(a)}^1
 + \sum_{k=0}^{a-1} \frac{f^{(k)}(0)}{k!} \int_0^{x_0} \df x \, x^k \bigl[ g(x) \bigr]_{+(a)}^1
\nn\\*&
 = \int_0^{x_0} \df x \, \biggl[ f(x) - \sum_{k=0}^{a-1} \frac{f^{(k)}(0)}{k!} x^k \biggr] g(x)
 - \sum_{k=0}^{a-1} \frac{f^{(k)}(0)}{k!} \int_{x_0}^1 \df x \, x^k g(x)
\,,\end{align}
where we assume $x_0 > 0$ and $f^{(k)}(0)$ is the $k$-th derivative of $f(x)$ at $x=0$.
In \eq{plus_dist_n_example}, we used that the term in square brackets in the first integral behaves as $\cO(x^a)$ and thus cancels the divergent behavior of $g(x)$ as $x\to0$, which allows us to drop the plus prescription in the first integral in the last line.
In the second integral, we used \eq{plus_dist_n} to change the integration bounds from $[0,x_0]$ to $[x_0,1]$. In the latter interval, $g(x)$ is regular and the plus prescription can be dropped.

The power-law divergence in \eq{power_law_n} can be regularized in terms of the higher-order plus distributions in \eq{plus_dist_n} as
\begin{align} \label{eq:power_law_n_reg}
 \frac{1}{(1-z)^{a+\eta}} &
 = \biggl[ \frac{1}{(1-z)^{a+\eta}} \biggr]_{+(a)}^1 + \sum_{k=0}^{a-1} \frac{(-1)^k}{k!} \frac{\delta^{(k)}(1-z)}{1+k-a-\eta}
\,, \qquad a \in \mathbb{N}
\,.\end{align}
This result can be verified by integrating both sides against a test function $(1-z)^m$ with $m<a$, and treating $\eta$ as in dimensional regularization to render all integrals finite.
In \eq{power_law_n_reg}, $\delta^{(k)}(1-z)$ is the $k$-th derivative on $\delta(1-z)$, which thus induces a sign $(-1)^k$ in an integral over $z$ and picks out the $k$-th derivative of any test function it acts on.
Note that only the $k=a-1$ term in \eq{power_law_n_reg} diverges for $\eta\to0$,
\begin{align}
 \frac{1}{(1-z)^{a+\eta}} &
 = -\frac{1}{\eta} \frac{(-1)^{a-1}}{(a-1)!} \delta^{(a-1)}(1-z)
 + \biggl[ \frac{1}{(1-z)^{a}} \biggr]_{+(a)}^1 + \sum_{k=0}^{a-2} \frac{(-1)^k}{k!} \frac{\delta^{(k)}(1-z)}{1+k-a}
 + \cO(\eta)
\,,\end{align}
so irrespective of the power $a$, any power law divergence $(1-z)^{-a-\eta}$ has exactly one single pole.

\chapter{Solution to Subleading Power RG Mixing Equation in Momentum Space}\label{sec:inversefourier}

In \Sec{sec:solution} we have shown that in the leading log approximation, and in the case when $\Gamma^{(0)}_{11}=\Gamma^{(0)}_{22}$, the solution to the subleading power RG mixing equation in position space is \Eq{eq:FLL}. Here we provide additional details on the transformation of this result back to momentum space. In position space the logarithms for the boundary condition are minimized by the choice $\mu_0=\mu_y$. For thrust at subleading power there are no distributions, and the logarithms have a simple correspondence between position and momentum space without subtleties. This is analogous to the situation between position space and cumulative thrust at leading power. To derive an exact relation for the Fourier transform we note that 
\begin{align} \label{eq:FTNLP}
 \int\! \frac{dy}{2\pi}\, e^{i k y}\, (iy)^{-1-\epsilon} = \frac{ \theta(k) k^\epsilon}{\Gamma(1+\epsilon)} \,,
\end{align}
where branch cuts are defined by $y=y-i0$.
Defining $e^{-\epsilon\gamma_E}/\Gamma(1+\epsilon) = \sum_{k=0}^\infty e_k\, \epsilon^k$, we have $e_{0}=1$, $e_{1}=0$, $e_{2}=-\zeta_2/2$, $e_3=\zeta_3/2$, etc. Expanding \Eq{eq:FTNLP} in $\epsilon$ leads to the identity we need to connect the subleading power logarithms in position and momentum space,
\be\label{eq:fourier}
	\int \frac{dy}{2\pi} e^{iky}\frac{\log^n(iye^{\gamma_E} \mu^p)}{i(y-i0)} = (-1)^n \sum_{j = 0}^{n} \frac{n!}{j!}\, e_{n-j}\, \log^j \Bigl(\frac{k}{\mu^p}\Bigr)\, \theta(k)\,.
\ee
Keeping only the LL term on the RHS gives the simple correspondence $\log^n(iye^{\gamma_E}\mu^p)/(iy) \to (-1)^n \log^n(k/\mu^p)\, \theta(k)$. 
To see how this works in an explicit example, we can rewrite the resummed position space result in \Eq{eq:FLL} as
\begin{align}\label{eq:Fposspace}
	\tilde F^{(2)\LL}_{\delta}(y,\mu)
 &= \tilde U_{\delta\theta}^{F,{\rm LL}}(y,\mu,\mu_0) \tilde F_{\theta}^{(2)}(y,\mu_0)
  = A \frac{\left(e^{\gamma_E}iy\mu_0^p \right)^{\omega}}{i(y-i0)} \\
 &= \frac{A}{i(y-i0)} e^{\omega \log\left(e^{\gamma_E}iy \mu_0^p\right)}
   = A \sum_{n=0}^\infty \frac{1}{n!} \omega^n \frac{\log^n\left(e^{\gamma_E}iy\mu_0^p\right)}{i(y-i0)} \,,\nn
\end{align}
where $A \equiv -\frac{\gamma^{(0)}_{12}}{2\beta_0}\log r\exp\Bigl[\frac{p\pi\Gamma^{(0)}_{11}}{\beta_0^2\alpha_s(\mu_0)} \left(\frac{1}{r} - 1 + \log r \right) \Bigr]$ and  $\omega\equiv -\frac{\Gamma^{(0)}_{11}}{2\beta_0}\log(r) $ are dimensionless $y$ independent expressions, where here $r=\alpha_s(\mu)/\alpha_s(\mu_0)$. Using \Eq{eq:fourier} we have
\begin{align}
	F^{(2)\LL}_{\delta}(k,\mu) &= \int \frac{dy}{2\pi} e^{iky} \tilde F^{(2)\LL}_{\delta}(y,\mu)  = A \sum_{n=0}^{\infty} \frac{1}{n!} \omega^n  \int \frac{dy}{2\pi} e^{iky} \frac{\log^n\left(e^{\gamma_E}iy\mu_0^p \right)}{i(y-i0)}  
  \nn \\
	&= A \sum_{n=0}^{\infty} \sum_{j= 0}^{n} \omega^n (-1)^n\frac{e_{n-j}}{j!} \log^j\Bigl(\frac{k}{\mu_0^p}\Bigr) \theta(k)  
  \,.
\end{align}
Here all the terms with $j < n$ are subleading logs, therefore at LL order we can keep just the $j = n$ term to give
\begin{align}\label{eq:Fmomspace}
	F^{(2)\LL}_{\delta}(k,\mu) &= A \sum_{n=0}^{\infty} \frac{(-\omega)^n}{n!} e_{0} \log^n(k) \theta(k) = A e^{-\omega\log(k)} \theta(k)   \\
	&= -\theta(k) \frac{\gamma^{(0)}_{12}}{2\beta_0}\log r\exp\left[\frac{p\pi\Gamma^{(0)}_{11}}{\beta_0^2\alpha_s(\mu_0)} \left(\frac{1}{r} - 1 + \log r \right) \right] \left(\frac{k}{\mu_0^p}\right)^{\frac{\Gamma^{(0)}_{11}}{2\beta_0}\log(r) } \nn\\
	&\equiv \theta(k) U_{\delta \theta}^\LL(k,\mu,\mu_0) \,.\nn
\end{align}
Note that this is simply obtained from the starting result in \Eq{eq:Fposspace} by taking $iye^{\gamma_E}\to 1/k$ everywhere, except for in the explicit prefactor $1/(y-i0)\to \theta(k)$. 
\Eq{eq:Fmomspace} is the LL solution to the subleading RG mixing equation in momentum space which was quoted in the main text in \Eq{eq:Umomspace}.

\chapter{Leading Logarithms for Thrust from Collinear Limits of Amplitudes}\label{sec:LLfromCollinear}

In this Appendix we explain how to obtain the LP LL series for thrust using only the information from collinear limits of scattering amplitudes. The NLP case, which is the focus of this paper, is similar. However, here we present the LP case in detail since this approach to obtaining the LL series is not traditional. The key idea is that the infrared scale dependence should cancel out in a physical cross section. Just as in the NLP analysis leading to \Eq{eq:constraints_final}, consistency at LP implies that the LL term can be obtained from loop corrections to the amplitude for a single collinear emission encoded in coefficients $d_{hc,2N}^{(0)}$, 
\begin{align}
\frac{1}{\sigma_0}\frac{\df\sigma^{(0,N)}}{\df\tau}
&= d_{hc,2N}^{(0)} \frac{\log^{2N - 1} \tau}{\tau} +\cdots
\,.\end{align}
We will work this out explicitly for the first two loop orders below.

Here, as in the text, we take thrust for Higgs decay in pure glue QCD as an example. We write the NLO cumulant at LP as
\begin{align}
  \label{eq:cumulantNLO}
  R^{(0,1)}(\tau) & = \frac{1}{\sigma_0}\int_0^\tau d\tau' \frac{d\sigma^{(0,1)}}{d\tau'} 
\nn\\
& =\frac{\alpha_s}{4 \pi} \frac{C_A}{\e^2} \left(c_h \left( \frac{\mu^2}{Q^2} \right)^\e + c_c \left( \frac{\mu^2}{\tau Q^2} \right)^\e  + c_s \left( \frac{\mu^2}{\tau^2 Q^2} \right)^\e\right) + {\cal O}\left(\frac{1}{\e} \right) \,, 
\end{align}
where we have separated the contribution between hard virtual corrections $c_h$, collinear corrections $c_c$, and soft corrections $c_s$. For a physical cross section both the divergent terms and the LL $\mu$ dependence should cancel. In particular, they should cancel between the $1/\e^2$ terms in Eq.~\eqref{eq:cumulantNLO}. There is no cancellation between the expansion of the $1/\e^2$ terms and the ${\cal O}(1/\e)$ terms. That's why we don't need to write down the ${\cal O}(1/\e)$ terms explicitly, at least for LL. It then follows that
\begin{align}
  \label{eq:NLOconsistency}
  c_h = - \frac{1}{2} c_c \,, \qquad c_s  = - \frac{1}{2} c_c \,.
\end{align}
Substituting the relation in Eq.~\eqref{eq:NLOconsistency} into Eq.~\eqref{eq:cumulantNLO}, we find
\begin{align}
  \label{eq:NLOLL}
  R^{(0,1)}(\tau) = - \frac{1}{2} \frac{\alpha_s}{4 \pi} \, C_A c_c \log^2 \tau  + \text{subleading logs}\,.
\end{align}
That is, the leading logarithm at NLO is uniquely determined by the contribution from the hard collinear splitting. Specifically, at NLO for thrust, the collinear corrections to the cumulant can be written as
\begin{align}
  \label{eq:RcNLO}
  R_{c,\text{LL}}^{(0,1)}(\tau) & = 2 \frac{\alpha_s \mu^{2 \e}}{4 \pi} \int_0^{\tau Q^2} \frac{ds}{Q^2} \int_0^1 dz \frac{e^{\e \gamma_E} [s z (1-z) ]^{-\e}}{\Gamma(1-\e)} C_A P_{gg,\text{LL}}^{(0,0)}
\nn\\
& = \frac{\alpha_s}{4 \pi}  \frac{8 C_A}{\e^2} \left(\frac{\mu^2}{\tau Q^2}\right)^\e+ {\cal O}\left( \frac{1}{\e} \right) \,,
\end{align}
where $P_{gg,\text{LL}}^{(0,0)}$ is introduced in Eq.~\eqref{eq:split_def}.
Therefore $c_c = 8$, and $R^{(0,1)}(\tau) = - \frac{\alpha_s}{\pi} C_A \log^2\tau + \text{subleading logs}$. 

At NNLO, there are several combinations of different modes, but the idea is similar. We write the cumulant as
\begin{align}
  \label{eq:RNNLO}
  R^{(0,2)} (\tau) & =\left( \frac{\alpha_s}{4 \pi}  \right)^2 \frac{C_A^2}{\e^4}
\left(c_{hh} \left( \frac{\mu^4}{Q^4} \right)^{\e} + c_{hc} \left( \frac{\mu^4}{\tau Q^4} \right)^\e  + (c_{cc} + c_{hs}) \left( \frac{\mu^4}{\tau^2 Q^4} \right)^\e + c_{cs} \left( \frac{\mu^4}{\tau^3 Q^4}  \right)^\e
\right. \nn\\
&
\left. + c_{ss} \left( \frac{\mu^4}{\tau^4 Q^4} \right)^\e \right) + {\cal O}\left(\frac{1}{\e^3} \right) \,, 
\end{align}
Here $c_{hh}$ denotes hard modes contributions from pure virtual diagrams, $c_{hc}$ denotes real-virtual contributions with virtual hard mode and real collinear mode, $c_{cc}$ denotes  both real-virtual or double real contributions with virtual or real collinear modes, $c_{hs}$ denotes real-virtual contributions with virtual hard mode and real soft mode, and finally $c_{ss}$ denotes real-virtual or double real contributions with virtual or real soft modes. 
Demanding that all the poles and $\mu$ dependence from expanding the $1/\e^4$ terms cancel, we find
\begin{gather}
  \label{eq:NNLOconsistency}
  c_{hc} = -4 c_{hh} \,, \qquad c_{cc} + c_{hs} = 6 c_{hh} \,, \qquad c_{cs} =  - 4 c_{hh} \,, \qquad  c_{ss} = c_{hh} \,.
\end{gather}
We then find
\begin{align}
  \label{eq:RNNLOhc}
  R^{(0,2)}(\tau) = - \left( \frac{\alpha_s}{4 \pi}  \right)^2 \frac{C_A^2}{4} c_{hc} \log^4 \tau + \text{subleading logs} \,.
\end{align}
Specifically, the real-virtual collinear corrections to the cumulant is given by
\begin{align}
  \label{eq:RVcollinear}
  R_{RVc,\text{LL}}^{(0,2)}(\tau) & = 2 \frac{\alpha_s \mu^{2 \e}}{4 \pi} \int_0^{\tau Q^2} \frac{ds}{Q^2} \int_0^1 dz \frac{e^{\e \gamma_E} [s z (1-z) ]^{-\e}}{\Gamma(1-\e)} C_A P_{gg,\text{LL}}^{(0,0)} \left( F_\text{dipole} + F_R \right) \,,
\end{align}
where we have separated the corrections into the dipole term and the remainder term, see Eq.~\eqref{eq:dipole} and \eqref{eq:remainder}. The dipole term gives
\begin{align}
  \label{eq:Rdipole}
  R_{RVc,\text{dipole,LL}}^{(0,2)}(\tau) & = \left(\frac{\alpha_s}{4 \pi} \right)^2 \left[ - \frac{24 C_A^2}{\e^4} \left(\frac{\mu^4}{\tau Q^4} \right)^\e -  \frac{8 C_A^2}{\e^4} \left(\frac{\mu^4}{\tau^2 Q^4} \right)^\e \right]+ {\cal O}\left( \frac{1}{\e^3} \right)\,.
\end{align}
And the remainder term gives
\begin{align}
  \label{eq:Rremainder}
  R_{RVc,R,\text{LL}}^{(0,2)}(\tau) & = \left(\frac{\alpha_s}{4 \pi} \right)^2 \left[- \frac{8 C_A^2}{\e^4}  \left(\frac{\mu^4}{\tau Q^4} \right)^\e + \frac{4 C_A^2}{\e^4}  \left(\frac{\mu^4}{\tau^2 Q^4} \right)^\e \right] + {\cal O}\left( \frac{1}{\e^3} \right)\,.
\end{align}
Adding the dipole and remainder terms, we find that the hard-collinear coefficient is $c_{hc} = -32$, and the NNLO cumulant is
\begin{align}
  \label{eq:RNNLOres}
  R^{(0,2)}(\tau) = \left( \frac{\alpha_s}{4 \pi} \right)^2 8 \, C_A^2 \log^4 \tau + \text{subleading logs} \,.
\end{align}
This is the correct leading logarithm for thrust. We see explicitly that both the dipole term and the remainder term contribute to thrust at LL. The analysis above can be straightforwardly carried out to all orders in $\alpha_s$.

\chapter{NLP \texorpdfstring{\boldmath $q_T$}{qT} spectra with Generalized Pure Rapidity Regulators}
\label{app:master_formula_c}

In \secs{master_formula}{master_formula_2}, we derived master formulas for the NLP correction to the $q_T$ spectrum using the $\eta$ regulator and the pure rapidity regulator, respectively.
In \sec{upsilonreg}, we also introduced a class of homogeneous rapidity regulators spanned by a parameter $c \neq 1$.
Here, we give the master formulas for this regulator for generic $c\ne1$.
In this regulator, the soft contribution is scaleless and vanishes, similar to the pure rapidity regulator.
Thus, one only needs to consider the $n$-collinear and $\bn$-collinear limits.

The derivation of the $n$-collinear expansion proceeds similar to the calculation shown in \sec{collinear_master}.
One can also obtain it from the result for the pure rapidity regulator, \eq{sigma_ncoll_NLP_LL_vita}, using the replacement
\begin{align}
 &\upsilon^\eta \biggl|\frac{k^-}{k^+}\biggr|^{-\eta/2} = \upsilon^\eta q_T^\eta |k^-|^{-\eta}
 \nn\\*
\quad\to\quad
 &\upsilon^{(1-c)\eta/2} \biggl|\frac{k^-}{\nu}\biggr|^{-\eta/2} \biggl|\frac{k^+}{\nu}\biggr|^{-c \eta/2}
 = \biggl[\upsilon \Bigl(\frac{\nu}{q_T} \Bigr)^{\frac{1+c}{1-c}}\biggr]^{(1-c)\eta/2} q_T^{(1-c)\eta/2} |k^-|^{-(1-c)\eta/2}
\,.\end{align}
Thus, in \eq{sigma_ncoll_NLP_LL_vita} one has to shift $\eta \to (1-c)\eta/2$ and $\upsilon \to \upsilon (\nu/q_T)^{\frac{1+c}{1-c}}$, giving
\begin{align} \label{eq:sigma_ncoll_NLP_LL_vita_c}
 \frac{\df\sigma^{(2),\text{LL}}_n}{\df Q^2 \df Y \df q_T^2} &
 = \frac{1}{(4\pi)^2} \frac{q_T^2}{2Q^2}
   \frac{1}{2 x_a x_b \Ecm^4}\,w^2\biggl(\frac{2}{(1-c)\smallupsilon} - \ln\frac{Q e^Y}{q_T} + \frac{1+c}{1-c} \ln\frac{\nu}{q_T} + \ln(\upsilon) \biggr)
   \nn\\&\quad \times  \biggl\{
   f_a(x_a) f_b(x_b) \Bigl[ \altMp{2}_n(1) - 2 \altMp{0}_n(1) \Bigr]
   + f_a(x_a) \, x_b f'_b(x_b) \Bigl[ \altM{0}_n(1) + 2 {\altMp{0}_n}(1) \Bigr]
   \nn\\*&\qquad
   + x_a f'_a(x_a) f_b(x_b) \Bigl[ \altM{0}_n(1) - \altM{2}_n(1) \Bigr]
   - 2 x_a f'_a(x_a) \, x_b f'_b(x_b)  \altM{0}_n(1)
   \biggr\}
\,.\end{align}
This result is well-defined for all $c\ne1$, whereas one encounters two explicit poles as $c\to1$.
This behavior is expected because for $c=1$ the regulator depends on the boost-invariant product $k^+ k^- = q_T^2$ and therefore does not regulate rapidity divergences, as explained at the end of \sec{upsilonreg}.
For $c=-1$ we recover the result of pure rapidity regularization of \eq{sigma_ncoll_NLP_LL_vita}.
In this case, the $\nu$ dependence in the regulator \eq{vita_regulator} cancels,
which is reflected by the vanishing of the coefficient of $\ln(\nu/q_T)$ in \eq{sigma_ncoll_NLP_LL_vita_c}.

In the $\bn$-collinear limit, the regulator for arbitrary $c\ne1$ is obtained from the pure rapidity regulator through
\begin{align}
 &\upsilon^\eta \biggl|\frac{k^-}{k^+}\biggr|^{-\eta/2} = \upsilon^\eta q_T^{-\eta} |k^+|^{\eta}
 \nn\\*
\quad\to\quad
 &\upsilon^{(1-c)\eta/2} \biggl|\frac{k^-}{\nu}\biggr|^{-\eta/2} \biggl|\frac{k^+}{\nu}\biggr|^{-c \eta/2}
 = \biggl[\upsilon \Bigl(\frac{\nu}{q_T} \Bigr)^{\frac{1+c}{1-c}}\biggr]^{(1-c)\eta/2} q_T^{-(1-c)\eta/2} |k^+|^{(1-c)\eta/2}
\,.\end{align}
Thus, in \eq{sigma_nbarcoll_NLP_LL_vita} one has to shift $\eta \to (c-1)\eta/2$ and $\upsilon \to \upsilon (\nu/q_T)^{\frac{1+c}{1-c}}$, giving
\begin{align} \label{eq:sigma_nbarcoll_NLP_LL_vita_c}
 \frac{\df\sigma^{(2),\text{LL}}_{\bn}}{\df Q^2 \df Y \df q_T^2} &
 = \frac{1}{(4\pi)^2} \frac{q_T^2}{2Q^2}
   \frac{1}{2 x_a x_b \Ecm^4}\,w^2\biggl(\frac{2}{(c-1)\smallupsilon} - \ln\frac{Q e^{-Y}}{q_T} - \frac{1+c}{1-c} \ln\frac{\nu}{q_T}  - \ln(\upsilon) \biggr)
   \\&\quad \times  \biggl\{
   f_a(x_a) f_b(x_b) \Bigl[ \altMp{2}_{\bn}(1) - 2 \altMp{0}_{\bn}(1) \Bigr]
   + f_a(x_a) \, x_b f'_b(x_b) \Bigl[ \altM{0}_{\bn}(1) - \altM{2}_{\bn}(1) \Bigr]
   \nn\\*&\qquad
   + x_a f'_a(x_a) f_b(x_b)  \Bigl[ \altM{0}_{\bn}(1) + 2 {\altMp{0}_{\bn}}(1) \Bigr]
   - 2 x_a f'_a(x_a) \, x_b f'_b(x_b)  \altM{0}_{\bn}(1)
   \biggr\}
\nn\,.\end{align}
Summing \eqs{sigma_ncoll_NLP_LL_vita_c}{sigma_nbarcoll_NLP_LL_vita_c}, the poles in $\smallupsilon$ precisely cancel,
and the dependence on $c$, $\bigupsilon$ and $e^Y$ cancels as well to yield a pure logarithm in $\ln(Q/q_T)$.
As for the pure rapidity regulator, this cancellation has to occur between the two collinear sectors,
since the soft sector does not give a contribution.

The NLP NLL result is identical to that in pure rapidity regularization, which is given by \eq{sigma_coll_NLP_NLL} upon dropping all regulator-dependent pieces, as explained in \sec{master_formula_2}. This provides another check of our regularization procedure.

\chapter{Integrating Out Glauber Quarks via Auxiliary Lagrangians}
In this appendix we will use the formalism of auxiliary fields to build the fermionic Glauber Lagrangians of \chap{qregge} and \refcite{Moult:2017xpp} by integrating out Glauber modes at the path integral level. 
Here, we will focus only on Glauber modes, therefore ignoring the integration of other offshell modes such as hard and hard-collinear modes. Hard and hard-collinear modes can be integrated out to give rise to the collinear and soft Wilson Lines via auxiliary Lagrangians by following the techniques presented in Appendix A of \refscite{Bauer:2001yt,Bauer:2002nz}.

\section{Deriving the Glauber Quark Auxiliary Lagrangians}
Let us introduce the auxiliary Glauber quark fields we aim to integrate out.
\begin{table}[h]
\centering
	\begin{tabular}{lcccl}
		\toprule
		Glauber mode &&Fields && Momentum scaling \\
		\midrule
		$n-s$ (left-moving)&&$\psi_{ns}^G$, $\varphi_{ns}^G$ && $p^\mu \sim (\lambda^2, \lambda,\lambda)$ \\
		$\bn-s$ (right-moving)&&$\psi_{\bn s}^G$, $\varphi_{\bn s}^G$ && $p^\mu \sim (\lambda, \lambda^2,\lambda)$ \\
		$n-\bn$ (central) &&$\psi_{n\bn}^G $ && $p^\mu \sim (\lambda^2, \lambda^2,\lambda)$ \\
		\bottomrule
	\end{tabular}
\end{table}
We find that introducing a single Glauber quark auxiliary field is not sufficient as there are Glauber regions that have different momentum scaling which induce different interacting auxiliary Lagrangians.
We label the fields $n-s$, $n-\bn$, $\bn-s$ according to their momentum scaling.\footnote{For these three Glauber regions we follow the naming convention of \refcite{Rothstein:2016bsq}. 
Note, however, that they are sometimes refereed in the literature also as central, left-, and right-moving Glaubers, see for example \cite{Boer:2017hqr}.}
Note that with this notation $\psi_{n\bn}^G \equiv \psi_{\bn n}^G$.

The exchange of a Glauber quark between two collinear operators $\cO_n$ and $\cO_\bn$, with no soft emission, have momentum scaling $p_{n\bn}^G \sim (\lambda^2,\lambda^2,\lambda)$.
By expanding the kinetic term of the QCD Lagrangian, a quark $\psi_{n\bn}^G$ with such momentum has a (leading) kinetic term of the form
\be
	\cL_\text{kin.}^{n\bn} = \bar \psi_{n\bn}^G \Sl{\cP}_\perp \psi_{n\bn}^G\,.
\ee
This Glauber quark field $\psi_{n\bn}^G$ couples to the collinear operators via the interaction vertex
\be\label{eq:Lcnbn}
	\cL_\text{c}^{n\bn} = g\bar \psi_{n\bn}^G  ( \Sl{\cB}_\perp^n \chi_n + \Sl{\cB}_\perp^\bn \chi_\bn ) + \text{h.c.}\,.
\ee
Given the definition of the collinear operators $\cO_n$ and $\cO_\bn$ of \eq{coll_ops} we recognize that we can rewrite \eq{Lcnbn} as
\be
	\cL_\text{c}^{n\bn} = g\bar \psi_{n\bn}^G  ( \cO_n + \cO_\bn)  + \text{h.c.}\,.
\ee
Because of momentum conservation in the $n$ and $\bn$ direction, a $\psi_{n\bn}^G$ can't emit a soft gluon, hence there is no interaction with soft fields\footnote{In principle momentum conservation allows an interaction of the form $\bar \psi_s gA^\mu_s \psi_{n\bn}^G$, but this operator is subleading with respect to the operator $\bar \psi_s gA^\mu_s \psi_{ns}^G$.}. 
We have therefore determined the kinetic and interacting Lagrangians for $\psi_{n\bn}^G$, i.e. the $n-\bn$ (central) Glauber mode.

Now let us move to the case of $n-s$ and $\bn-s$ Glaubers, which shows up in the interaction of collinear particles with soft ones via a Glauber exchange.
The interaction of a Glauber quark with soft fields starts with the exchange of Glauber modes with momentum $p_{ns}^G \sim (\lambda^2,\lambda,\lambda)$ and $p_{\bn s}^G \sim (\lambda,\lambda^2,\lambda)$.
To study these modes let's consider a QCD quark field $\Psi(p)$ and take the limit  $p \to p_{ns}^G \sim (\lambda^2,\lambda,\lambda)$. We can define the Glauber $n-s$ quark fields $\psi_{ns}^G$ and $\varphi_{ns}^G$ to be 
\be
	\psi_{ns}^G \equiv P_n \Psi(p_{ns}^G) \,,\quad \varphi_{ns}^G \equiv P_{\bn} \Psi(p_{ns}^G) \implies \Psi(p_{ns}^G) = (\psi_{ns}^G +\varphi_{ns}^G)\,.
\ee
The decomposition is complete since $P_n + P_\bn \equiv \frac{\nslash \bnslash}{4} + \frac{\bnslash \nslash}{4} = \id$ and $P_n,P_\bn$ are ortogonal projectors. By definition these spinors satisfy the projection relations
\be\label{eq:spinorrel}
	\nslash \psi_{ns}^G = 0 \,, \qquad \bnslash \varphi_{ns}^G = 0\,,
\ee
and all the other relations that are derived from \eq{spinorrel} like
\be
	P_{n} \psi_{ns}^G = \psi_{ns}^G \,,~ P_{\bn} \psi_{ns}^G = \overline{\psi}_{ns}^G P_{n} = 0 \,, \qquad P_{\bn} \varphi_{ns}^G = \varphi_{ns}^G \,,~ P_{n} \varphi_{ns}^G = \overline{\varphi}_{ns}^G P_{\bn} = 0\,.
\ee
We now want to study the interaction of $\cO_n \equiv \Sl{\cB}^\perp_n \chi_n$ with a Glauber quark. The Lagrangian that mediates this interaction is clearly $g\overline{\Psi}(p) \Sl{\cB}^\perp_n \chi_n$ in the limit $p \to p_{ns}^G$. Now let's use the spinor relations in this limit to simplify the Lagrangian
\begin{align}
	\overline{\Psi}(p_{ns}^G) \Sl{\cB}^\perp_n \chi_n &=  (\overline{\psi}_{ns}^G +\overline{\varphi}_{ns}^G) \Sl{\cB}^\perp_n \chi_n =  (\overline{\psi}_{ns}^G +\overline{\varphi}_{ns}^G) \Sl{\cB}^\perp_n P_n \chi_n = (\overline{\psi}_{ns}^G +\overline{\varphi}_{ns}^G) P_n \Sl{\cB}^\perp_n  \chi_n \nn \\
	&= \overline{\varphi}_{ns}^G  \Sl{\cB}^\perp_n  \chi_n\,.
\end{align}
Therefore, we see that $\cO_n$ couples to the $\varphi$ component of $\Psi$.\\
Now, let's study the free Lagrangian for $\psi_{ns}^G$ and $\varphi_{ns}^G$.
We start from the dirac lagrangian $\overline{\Psi} i\slashed{\partial} \Psi $ and we expand it in the $n-s$ Glauber limit 
\be\label{eq:freequarklagr_withphi}
	\cL^G_\text{kin} = (\overline{\psi} + \overline{\varphi}) \left[\frac{\bnslash}{2} n \cdot \partial +\frac{\nslash}{2} \bn \cdot \cP + \Sl{\cP}_\perp \right] (\psi + \varphi) = \overline{\psi}\frac{\bnslash}{2} n \cdot \partial \psi +  \overline{\varphi}\frac{\nslash}{2} \bn \cdot \cP \varphi + \overline{\psi} \Sl{\cP}_\perp\varphi +  \overline{\varphi}\Sl{\cP}_\perp \psi \,,
\ee
where we suppressed the $._{ns}^G$ labels for conciseness.

If we power count the Lagrangian $\cL^G_\text{kin}$ in \eq{freequarklagr_withphi} we realize that the propagator for both $\psi_{ns}^G$ is subleading with respect to the kinetic mixing terms and the propagator of $\varphi_{ns}^G$, therefore these fields propagate through
\be\label{eq:Gkineticlagr}
	\cL_\text{kin}^G = \overline{\psi}_{ns}^G \Sl{\cP}_\perp\varphi_{ns}^G  + \overline{\varphi}_{ns}^G \Sl{\cP}_\perp\psi_{ns}^G + \overline{\varphi}_{ns}^G \frac{\nslash}{2} \bn \cdot \cP \varphi_{ns}^G \equiv \overline{\Psi}_n 
	\matrixx{0}{\Sl{\cP}_\perp}{\Sl{\cP}_\perp}{\frac{\nslash}{2} \bn \cdot \cP } \Psi_n\,,
\ee
where we define 
\be 
	\Psi_n \equiv \vectorr{\psi_{ns}^G}{\varphi_{ns}^G}\,.
\ee
If we look at the interaction with one soft gluon, then by momentum conservation we need one $n-s$ and one $\bn-s$ Glauber
\begin{align}
	p^{G}_{n s} + p^{G}_{\bn s} &= p_s\,, \nn \\
	 (\lambda^2,\lambda,\lambda) + (\lambda,\lambda^2,\lambda) &= (\lambda, \lambda, \lambda)\,.
\end{align}
therefore we start by expanding the QCD vertex
\begin{align}
	 \cL^G_{s} &= \bar{\Psi}(p_{ns}^G) g\Sl{A}_s \Psi(p_{\bn s}^G) + \text{h.c.} = g(\overline{\psi}_{ns}^G + \overline{\varphi}_{ns}^G) \left[\frac{\bnslash}{2} n \cdot A_s +\frac{\nslash}{2} \bn \cdot A_s + \Sl{A}_s^\perp \right] (\psi_{\bn s}^G + \varphi_{\bn s}^G) + \text{h.c.} \nn \\
	 &= \overline{\varphi}_{ns}^G \frac{\nslash}{2} g\bn \cdot A_s\psi_{\bn s}^G + \overline{\psi
	 }_{ns}^G \frac{\bnslash}{2} n \cdot A_s\varphi_{\bn s}^G+ \overline{\psi}_{ns}^G g\Sl{A}_s^\perp \psi_{\bn s}^G + \overline{\varphi}_{ns}^G g\Sl{A}_s^\perp \varphi_{\bn s}^G +\text{h.c.}\,,
\end{align}
where in the last step we used $\bnslash \psi_{\bn s}^G = \bar{\psi}_{ns}^G \nslash = 0$ to simplify the result.
We are now finally able to obtain the full Glauber quark Lagrangians
\begin{align}\label{eq:GQLagr}
	\cL^G_\text{kin} &= \overline{\psi}_{ns}^G \Sl{\cP}_\perp\varphi_{ns}^G  + \overline{\varphi}_{ns}^G \Sl{\cP}_\perp\psi_{ns}^G + \overline{\varphi}_{ns}^G\frac{\nslash}{2} \bn \cdot \cP \varphi_{ns}^G + (n \leftrightarrow \bn) + \cL_\text{kin.}^{n\bn} \\
	\cL^G_{c} &=  \bar{O}_n \varphi_{n s}^G +\bar{O}_\bn \varphi_{\bn s}^G + \text{h.c.} + \cL_\text{c}^{n\bn} \nn \\
	\cL^G_{s} &=  \overline{\varphi}_{ns}^G \frac{\nslash}{2} g\bn \cdot A_s\psi_{\bn s}^G + \overline{\psi}_{ns}^G \frac{\bnslash}{2} n \cdot A_s\varphi_{\bn s}^G+ \overline{\psi}_{ns}^G g\Sl{A}_s^\perp \psi_{\bn s}^G + \overline{\varphi}_{ns}^G g\Sl{A}_s^\perp \varphi_{\bn s}^G +\text{h.c.}\nn\,.
\end{align}

\section{Integrating Out Glauber Quarks}
Having derived the kinetic and interaction Lagrangians for the auxiliary Glauber quark fields we can proceed to integrating them out.
First of all we observe that we can rewrite the Lagrangians in \eq{GQLagr} in matrix form by defining the Glauber field multiplet $\cG$ and the collinear multiplet $\cC$ as
\be
	\cG \equiv \left(\begin{array}{c} \psi_{ns}^G \\ \varphi_{n s}^G \\ \psi_{n\bn}^G \\ \psi_{\bn s}^G \\  \varphi_{\bn s}^G \end{array}\right) \equiv \vectorrr{\Psi_n}{\psi_{n\bn}^G}{\Psi_\bn}\, , \qquad \cC \equiv \left(\begin{array}{c} 0 \\ \cO_n \\ \cO_n + \cO_\bn \\ 0 \\ \cO_\bn \end{array}\right)\,,
\ee
so that the Glauber quark Lagrangian \eq{GQLagr} can be rewritten as
\begin{align}
	\cL_q^G &= \cL^G_\text{kin} + \cL^G_{c} + \cL^G_{s} = \bar{\cG} 
	\left(\begin{array}{ccccc} 
		0 &\Sl{\cP}_\perp & 0 & g\Sl{A}_s^\perp & \frac{\bnslash}{2} n \cdot A_s \\
		\Sl{\cP}_\perp & \frac{\nslash}{2} \bn \cdot \cP & 0 & \frac{\nslash}{2} g\bn \cdot A_s & g\Sl{A}_s^\perp \\
 		0 & 0 &\Sl{\cP}_\perp & 0 & 0 \\
		g\Sl{A}_s^\perp & \frac{\nslash}{2} g\bn \cdot A_s & 0 & 0 &\Sl{\cP}_\perp  \\
		\frac{\bnslash}{2} n \cdot A_s & g\Sl{A}_s^\perp &0  &\Sl{\cP}_\perp & \frac{\bnslash}{2} n \cdot \cP  
	\end{array}\right) \cG ~ + ~ g\bar{\cC}\cdot\cG + g\bar{\cG}\cdot \cC \nn \\
	&\equiv \bar{\cG} K \cG + g\bar{\cG}\cdot \cC + g\bar\cC \cdot \cG\,.
\end{align}
This step was not strictly necessary, but it is very conveniently since it allows the integration of all Glauber fields simultaneously.
Integrating them out consists in integrating out the multiplet $\cG$ itself. Since $\cG$ doesn't a have source term, to integrate it out at the path integral level it is sufficient to solve the e.o.m. for it
\be\label{eq:eom_G}
	0 = \frac{\delta \cL_q^G}{\delta\bar{\cG} } = K \cG + g\cC \implies \boxed{\cG = -K^{-1} g\cC}\,.
\ee
where 
\be
	K^{-1} = 
	\left(\begin{array}{ccccc} 
		0 &\frac{1}{\Sl{\cP}_\perp} & 0 & 0 & 0 \\
		\frac{1}{\Sl{\cP}_\perp} & 0 & 0 & 0 & -\frac{1}{\Sl{\cP}_\perp}g\Sl{A}_s^\perp\frac{1}{\Sl{\cP}_\perp} \\
 		0 & 0 &\frac{1}{\Sl{\cP}_\perp} & 0 & 0 \\
		0 & 0 & 0 & 0 &\frac{1}{\Sl{\cP}_\perp}  \\
		0 & -\frac{1}{\Sl{\cP}_\perp}g\Sl{A}_s^\perp\frac{1}{\Sl{\cP}_\perp} &0  &\frac{1}{\Sl{\cP}_\perp} & 0  
	\end{array}\right)\,,
\ee
so that \eq{eom_G} in components reads
\begin{align}\label{eq:eom_G_components}
	\psi_{ns}^G &= -\frac{1}{\Sl{\cP}_\perp} g\cO_n \\
	\varphi_{ns}^G &= \frac{1}{\Sl{\cP}_\perp}g\Sl{A}_s^\perp\frac{1}{\Sl{\cP}_\perp} g\cO_\bn \\
	\psi_{n\bn}^G &= -\frac{1}{\Sl{\cP}_\perp} g(\cO_n + \cO_\bn) \\
	\psi_{\bn s}^G &= -\frac{1}{\Sl{\cP}_\perp} g\cO_\bn \\
	\varphi_{\bn s}^G &= \frac{1}{\Sl{\cP}_\perp}g\Sl{A}_s^\perp\frac{1}{\Sl{\cP}_\perp} g\cO_n\,.
\end{align}
By plugging the result of \eq{eom_G} in $\cL_G$ and using that
\be 
	\cO_n^\dagger \frac{1}{\Sl{\cP}_\perp} \cO_n = \cO_\bn^\dagger \frac{1}{\Sl{\cP}_\perp} \cO_\bn = 0   \,,
\ee	
we get 
\be\label{eq:LGfinal}
	\cL_G = -g\bar{\cC} K^{-1} g\cC = -4\pi \alpha_s\cO_n^\dagger \frac{1}{\Sl{\cP}_\perp} (\Sl{\cP}_\perp -g \Sl{A}^s_\perp) \frac{1}{\Sl{\cP}_\perp} \cO_\bn + \text{h.c.}\,.
\ee
As desired the Lagrangian $\cL_G$ of \eq{LGfinal} matches the result for the three rapidity sector Lagrangian $\cO_{\bar n n}$ derived in \sec{nn_ops}, up to soft Wilson lines that would arise from integrating out also hard-collinear modes.

Let us finish this appendix by showing how we can make use of the equation of motion for the auxiliary Glauber quark fields to also derive the two-rapidity Glauber quark Lagrangian $\cO_{ns}$ of \sec{ns_ops} from the interaction of Glauber quarks with soft fields.
In QCD the interaction between a collinear $\cO_n$ and a soft Glauber operator $\cO_s  \equiv \Sl{\cB}^s_\perp\psi_s$ is given by the T-product 
\be
	T\Big[\underbrace{\bar{\Psi} g\Sl{A}}_\text{collinear} \Psi(p^G_{ns}) ,\bar{\Psi}(p^G_{ns}) \underbrace{g\Sl{A} \Psi}_\text{soft}\Big] \,,
\ee
which matches onto the T-product of auxiliary Lagrangian EFT operators
\be
	T\Big[g\bar{\cO}_n \varphi_{ns}^G,g\bar{\Psi}_{ns}^G \cO_s \Big]\,.
\ee
The fact that the kinetic mixing is dominant for the Glauber quark field, implies that
\be 
	T[\varphi_{ns}^G,\bar\varphi_{ns}^G] \ll T[\varphi_{ns}^G,\bar\psi_{ns}^G]\,.
\ee
Hence, the relevant interaction between the Glauber quark and the soft operator $\cO_s$ is obtained via the $\psi_{ns}^G$ component of the Glauber quark field multiplet, i.e. from the interaction Lagrangian term
\be\label{eq:Gss_lagr}
	\cL^G_{ss} = \bar\psi_{ns}^G g\Sl{\cB}^s_\perp\psi_s + \bar\psi_s g\Sl{\cB}^s_\perp\psi_{ns}^G\,.
\ee
We can now plug the solution of the equations of motion for $\psi_{ns}^G$ of \eq{eom_G_components} to get
\be
	\cL^G_{ss} = -g\bar{\chi}_n \Sl{\cB}_\perp^n\frac{1}{\Sl{\cP}_\perp} g\Sl{\cB}^s_\perp\psi_s + \text{h.c.} = -4\pi \alpha_s \cO_n^\dagger \frac{1}{\Sl{\cP}_\perp} \cO_s  + \text{h.c.} \,,
\ee
which matches the $\cO_{ns}$ Lagrangian obtained by direct calculation in the bottom-up approach of \chap{qregge} in \eq{Onsmatching}.

\begin{singlespace}
\bibliography{main}
\bibliographystyle{plain}
\end{singlespace}

\end{document}